\documentclass[11pt]{article}
\usepackage{amsfonts}
\usepackage{amssymb}
\usepackage{amsmath}
\usepackage{geometry}
\usepackage{appendix}

\DeclareMathSizes{10.95}{10}{7}{6}

\begin{document}

\title{Statistical Field Theory and Neural Structures Dynamics V: Synthesis
and extensions}
\author{Pierre Gosselin\thanks{%
Pierre Gosselin : Institut Fourier, UMR 5582 CNRS-UGA, Universit\'{e}
Grenoble Alpes, BP 74, 38402 St Martin d'H\`{e}res, France.\ E-Mail:
Pierre.Gosselin@univ-grenoble-alpes.fr} \and A\"{\i}leen Lotz\thanks{%
A\"{\i}leen Lotz: Cerca Trova, BP 114, 38001 Grenoble Cedex 1, France.\
E-mail: a.lotz@cercatrova.eu}}
\maketitle

\begin{abstract}
We present a unified field-theoretic framework for the dynamics of activity
and connectivity in interacting neuronal systems. Building upon previous
works (\cite{GLr}, \cite{GL}, \cite{GLs}, \cite{GLt}, \cite{GLw}), where a
field approach to activity--connectivity dynamics, formation of collective
states and effective fields of collective states were successively
introduced, the present paper synthesizes and extends these results toward a
general description of multiple hierarchical collective structures. Starting
with the dynamical system representing collective states in terms of
connections, activity levels, and internal frequencies, we analyze its
stability, emphasizing the possibility of transitions between
configurations. Then, turning to the field formalism of collective states,
we extend this framework to include substructures (subobjects) participating
in larger assemblies while retaining intrinsic properties. We define
activation classes describing compatible or independent activity patterns
between objects and subobjects, and study stability conditions arising from
their alignment or mismatch. The global system is described as the
collection of landscapes of coexisting and interacting collective states,
each characterized both by continuous (activity, frequency) and discrete
(class) variables. A corresponding field formalism is developed, with an
action functional incorporating both internal dynamics and interaction
terms. This nonlinear field model captures cascading transitions between
collective states and the formation of composite structures, providing a
coherent theoretical basis for emergent neuronal assemblies and their mutual
couplings.
\end{abstract}

\section{Introduction}

In this series of papers, we developed a field-theoretic framework for
modeling the dynamics of connectivities in interacting spiking neurons. In 
\cite{GLr}, we introduced a two-field model describing both neural activity
and the evolving connectivity between points along a thread. These
interacting fields capture the mutual influences between neuronal activity
and connectivity, expanding the formalism of \cite{GL} and providing a
unified microscopic description through an action functional. Minimization
of this functional yields background fields representing collective
configurations of the system and determining equilibrium states for activity
and connectivity. These equilibria, shaped by internal parameters and
external stimuli, condition fluctuations and signal propagation.

In \cite{GLs}, we showed how repeated activations can modify the
connectivity background through interference effects, leading to transient
states of enhanced connectivity between points. These states, which can be
reactivated by perturbations, behave as partial neuronal assemblies.

In \cite{GLt}, we derived an effective theory for the connectivity field by
integrating out the degrees of freedom of individual neurons.
Self-interactions within the connectivity field generate internal dynamics
that alter background states locally, producing persistent shifts or new
patterns of connectivity. These collective modifications emerge as
higher-order structures built upon the background field.

Finally,\cite{GLw} reoriented the analysis toward the collective structures
themselves. Individual neuronal dynamics become secondary to the behavior of
emerging and interacting assemblies. We introduce an effective field
formalism to describe their activations, transitions between internal
states, and mutual interactions. The state space of this formalism is
broader than in the initial framework, as a separate field is assigned to
each possible structure, with the states of these fields representing single
or multiple activations of the corresponding structure. The action
functional captures both the intrinsic characteristics of these states and
their dynamic interactions. This formalism allows for transitions of a given
structure between distinct states, induced either by interactions with other
structures or by external perturbations. Consequently, a single structure
may undergo activation or deactivation, which in turn can trigger or
suppress the activation of others. In this context, state transitions of a
given structure and the associated modifications in terms of connectivities
and activity frequencies can facilitate or inhibit interactions with other
collective states, resulting in synchronization or desynchronization with
other structures. Moreover, interactions between collective states may
result in the assembly of several states into larger composite sets. An
effective formalism, based on integrating out intermediate degrees of
freedom, characterizes the binding of multiple structures through indirect
couplings mediated by intermediate collective states.

The present work synthesizes and extends these different studies along
several directions. First, we revisit the various results concerning the
description of collective states in terms of connections, activity levels,
and the internal frequencies of these activities. We gather these different
elements in a compact form as a dynamical system describing the
characteristics of a collective state, and then we study the stability of
such a system, confirming the necessity of investigating the notion of
transition. We then summarize the extended formalism of collective state
fields, as well as its main results in terms of interaction, transitions,
and activation--deactivation of states. In particular, we study the notion
of subobjects --- states participating in larger entities but possessing
their own intrinsic characteristics. Building upon these results, we develop
a global study concerning collective states, or objects, and their
subobjects. We focus particularly on the compatibility between the activity
of one or several subobjects and that of a global object. The activity of a
subobject (its levels of connectivity, the proper frequencies of its
activity) may arise from the activity of the global object, but may also be
independent from it. This constitutes an advantage of the field formalism:
several states can overlap, which models the fact that one cell, or several,
may be engaged in different processes due to their large numbers of synapses
and dendrites.

This leads us to decompose the possible states of an object and its
subobjects into different classes. The class of an object is defined by its
activation state and by the induced activation states of all its subobjects,
allowing for continuous modifications or possible permutations (the same
global or local activity may be obtained by locally modifying the role of
each cell composing the object, for example through permutation or global
spatial transformation). Each object may exist in different states of
frequency or connectivity, and thus there exists a collection of classes for
each object. Each subobject has its own collection of activation classes. An
object will be stable if the activation classes of its subobjects coincide
with that of the global object; however, instability may arise if the
activation classes of the subobjects fail to recombine to produce the
activation class of the object. We thus reach a global description of the
system. This corresponds to considering the entire set of possible
landscapes of collective states --- i.e., of objects that may overlap
spatially --- and of their subobjects, themselves possessing subobjects, and
so on. A state of the system is described by continuous data (levels of
activation, frequencies, etc.) for the objects and subobjects, but also by
discrete data, namely the classes to which the different elements belong. A
dynamical description of such landscapes requires the consideration of a
global field whose arguments represent the possible collective states, their
classes, and their subobjects. We construct an action functional for this
global field, incorporating the expected characteristics of interaction
terms, and study the corresponding interaction mechanisms. At a macroscopic
level, this description evokes a nonlinear field model inducing cascading
modifications and transitions of representations, some of which induce or
replace others.

This work is organized as follows. Section 2 presents a literature review.
In Section 3, we recall the field-theoretic formalism for activities and
connectivities. We recall the effective action and present the static and
dynamic background fields of the system. We extend the previous results by
studying the stability of collective states with enhanced connectivities.
Sections 4, 5, and 6 recall the field formalism for collective states. We
present the associated operator formalism and the mechanism of transitions
between several states, including the possibilities for integrating states
into larger structures, and the description of the corresponding bound
states. Section 7 builds on this formalism to focus on systems composed of
collective states and their substates, also defined as objects and their
subobjects. The corresponding collection of collective states is now seen as
a whole in which the characteristics of each part are interdependent.
Section 8 translates this approach to the level of state space. The state
space of an object and those of its subobjects are related by certain
restriction maps. In Section 9, we describe the state space for such a
system as the tensor product of the object and subobject state spaces. We
present the restriction maps in this description and the compatibility
conditions for such a system of states to be stable. Based on this
description, Section 10 develops the field formalism for the
object/subobject system. The fields incorporate the characteristics of both
objects and their subobjects. These characteristics include discrete classes
describing the possible independent activations of subobjects within a given
object. Section 11 describes the interactions and transitions of such
fields. Section 12 investigates the possibility of a global field
encompassing the entire possible collection of objects/subobjects. In
Section 13, we derive the classical action of such a field. Section 14
computes the propagator for this classical action, and Section 15 presents
the effective action series expansion and the mechanisms of transitions it
can describe. Section 16 investigates the connected correlation functions
and the associated transitions. Section 17 builds on these results to
present the general form of transitions. Section 18 concludes by discussing
the possibilities of a macro-effective description derived from the
formalism.

\section{Litterature review}

Several branches of the literature are related to our work. At the
macroscopic scale, our approach shares common goals with the literature on
mean fields or neural fields. Neural fields model large populations of
neurons as homogeneous structures, with individual neurons indexed by
spatial coordinates. These models are employed to describe various patterns
of brain activity. Following the seminal works of Wilson, Cowan, and Amari 
\cite{1,1b,2,2b,3,4,5,6,7,8,9,10}, neural field dynamics are typically
investigated in the continuum limit, with neural activity represented by a
macroscopic variable---the population-averaged firing rate. Mean-field
theory has been extended in various ways and has found a wide range of
applications, including the study of traveling wave solutions \cite{11,12b}
and stochastic perturbations in firing rates \cite{13b,13c}. Extensions have
also incorporated the influence of neural network topology on spatial
arrangements of activity \cite{14,15}.

Despite their success, mean-field approaches aggregate the degrees of
freedom of underlying processes, neglecting interaction delays,
heterogeneity in connectivity, and the true microscopic structure.
Consequently, emergent behaviors are not directly captured.

In comparison, our statistical field theory model maintains explicit
micro-level interactions while retaining the macroscopic continuity of
neural fields. We assign spatial coordinates to neurons and derive
continuous dynamical equations, but our fields---complex-valued functionals
inspired by statistical field theory \cite{16}---encode microscopic
information at the collective level. The field translation of a micro-model
involves activities and connectivities by introducing two coupled
statistical fields, one for activities and one for connectivity, thereby
incorporating all possible distributions for these variables. This allows
the study of the emergence of multiple collective states of activity and
connectivity rather than considering averaged approximations. Our formalism,
in fact, operates on a much larger state space than the mean-field approach
and accounts for the multiple aspects of collective states. It bridges the
microscopic and macroscopic scales, allowing for richer emergent phenomena,
including phase transitions and dynamic coupling between neuronal activity
and connectivity.

Some prior attempts have extended mean-field theory using statistical field
theory tools \cite{17,17a,17b,17c,17d,17e,18,19,20,20a,20b,20c,20d,21,22,23}%
. In these works, neural activity is represented as a field, and an
effective action is formulated for collective activity, going beyond the
mean-field approximation. However, these remain collective-level models,
correcting the mean field description, that are not derived from a
microscopic description. Closer to our approach are partition-function or
effective-action methods \cite{24,25,26,26a}, though these either rely on
simplified assumptions or impose a priori forms of the effective action.

At the mesoscopic scale, another body of research examines the role of
connectivities and plasticity in systems of interacting neurons. This
literature lies at the intersection of network theory and neural field
modeling \cite{27}. For instance, Jirsa and collaborators \cite%
{28,29,29a,29b,30,31} analyzed the stability of resting states and the
impact of inhomogeneous connectivities, revealing that symmetry breaking can
lead to functional subspaces, reminiscent of the background states studied
here. However, these states are imposed rather than dynamically emerging.

More advanced diagrammatic approaches use connectivity tensors and
propagators to represent brain connectivity patterns \cite{32,33,34}. These
models, using graphs expansions, allow an exploration of connectivity
patterns beyond phenomenological models. Nevertheless, they treat
connectivity as a derived quantity, not as a dynamical field in itself,
unlike our formulation, which explicitly models the dynamics of
connectivities.

At the microscopic scale, models from dynamical systems and computational
neuroscience study the interactions of individual neurons \cite{35,36,37}.

These models account for oscillatory behaviors, spiking regimes, and the
emergence of assemblies through local rules such as Hebbian plasticity. They
form the foundation for understanding phenomena such as the binding problem
or polychronization \cite{38,39,40,41,42,43,44,45,46,47,48}.

However, while such models capture local emergent effects, they lack an
analytical framework to describe collective behaviors. Our approach builds a
bridge between these detailed microscopic models and the collective-level
formulations of neural field theory. It is based on one of these frameworks 
\cite{49}, originally designed for polychronization, that we extended into a
field-theoretic model for the dynamics of connectivity functions.

Our research also relates to experimental and theoretical studies on
engrams---sets of connected neurons associated with memory traces. Recent
works \cite{50,51,52,53} support the engram hypothesis, emphasizing their
persistence, reactivation, and distributed nature. Additional studies \cite%
{54,55,56,56a,56b,56c,56d,56e,56f,56g,56h,56i,56j,56k,56l,56m,56n,57}
explore the influence of neuronal excitability and overlapping engrams.
These findings align with our theoretical interpretation of connectivity
states as dynamic interacting assemblies, whose timing and overlap govern
their association. Other works emphasize regulatory and homeostatic
mechanisms stabilizing these structures \cite{58,59,60,61,62,63,64,65,66},
which parallel our notion of connectivity and activity potentials
stabilizing collective states.

A final line of research concerns numerical simulations of neural networks
and the emergence of collective assemblies. Large-scale spiking network
simulations \cite{67,68,69,70,71,72} have demonstrated that structured
patterns such as oscillations, assemblies, and wave propagation can arise
spontaneously from simple local rules. These models often rely on
computational frameworks. However, despite their descriptive richness, they
typically lack a theoretical framework explaining how these emergent
assemblies arise from microscopic dynamics or how they relate to effective
macroscopic variables. This gap underscores the need for a field-theoretic
approach like ours, which provides an analytical basis for connecting
micro-level interactions, emergent assemblies, and their collective
dynamics.\bigskip

\section{From a field model for activities and connections to an effective
model for collective states}

Starting with a field model describing a large number of interacting neurons
along with their connectivities, we describe how the study of this model
leads to an effective theory for collective states of cells activated in
groups. The microfoundations and the method of translating these foundations
into a field formalism are presented in Appendix 1.

\subsection{System of activities and connectivities}

In \cite{GLr,GLs,GLt,GLw}, we develop a statistical field formalism to
describe both cell and connectivity dynamics for a large number of
interacting neurons. This decription relies on two fields, $\Psi $ for
cells, and $\Gamma $ for connectivities. The field action for the system is
written at the classical level as: 
\begin{eqnarray}
S_{full} &=&S\left( \Psi \right) +S\left( \Gamma \right)  \label{Sf} \\
&=&-\frac{1}{2}\Psi ^{\dagger }\left( \theta ,Z,\omega \right) \nabla \left( 
\frac{\sigma _{\theta }^{2}}{2}\nabla -\omega ^{-1}\left( J,\theta
,Z,\left\vert \Psi \right\vert ^{2}\right) \right) \Psi \left( \theta
,Z\right) +V\left( \Psi \right)  \notag \\
&&+\frac{1}{2\eta ^{2}}\left( S_{\Gamma }^{\left( 0\right) }+S_{\Gamma
}^{\left( 1\right) }+S_{\Gamma }^{\left( 2\right) }+S_{\Gamma }^{\left(
3\right) }+S_{\Gamma }^{\left( 4\right) }\right) +U\left( \left\{ \left\vert
\Gamma \left( \theta ,Z,Z^{\prime },C,D\right) \right\vert ^{2}\right\}
\right)  \notag
\end{eqnarray}%
The first part in the right hand side of (\ref{Sf}):%
\begin{equation}
S\left( \Psi \right) =-\frac{1}{2}\Psi ^{\dagger }\left( \theta ,Z,\omega
\right) \nabla \left( \frac{\sigma _{\theta }^{2}}{2}\nabla -\omega
^{-1}\left( J,\theta ,Z,\left\vert \Psi \right\vert ^{2}\right) \right) \Psi
\left( \theta ,Z\right) +V\left( \Psi \right)  \label{Sp}
\end{equation}%
depends on a field $\Psi $, encompassng the neural activity of a large set
of neurons spread over a thread described by coordinate $Z$. This action
describes the dynamics for neural activity $\omega \left( J,\theta
,Z,\left\vert \Psi \right\vert ^{2}\right) $ within the thread. The current $%
J$ represents the external signals that condition the states for $S\left(
\Psi \right) $. To derive an effective action for the field $\Gamma $, we
have to integrate the degrees of freedom of the neurons field $\Psi $. This
is done by computing the effective action for $\Psi $ and to compute the
associated background field. The effective action and the background field
are presented in \cite{GL} as well as the equations for the corresponding
activities.

At classical level, activities satisfy:%
\begin{equation}
\omega ^{-1}\left( \theta ,Z,\left\vert \Psi \right\vert ^{2}\right)
=G\left( \int \frac{\kappa }{N}\frac{\omega \left( J,\theta -\frac{%
\left\vert Z-Z_{1}\right\vert }{c},Z_{1},\Psi \right) T\left( Z,\theta
,Z_{1},\theta -\frac{\left\vert Z-Z_{1}\right\vert }{c}\right) }{\omega
\left( J,\theta ,Z,\left\vert \Psi \right\vert ^{2}\right) }\left\vert \Psi
\left( \theta -\frac{\left\vert Z-Z_{1}\right\vert }{c},Z_{1}\right)
\right\vert ^{2}dZ_{1}\right)  \label{M}
\end{equation}%
where $T\left( Z,\theta ,Z_{1},\theta -\frac{\left\vert Z-Z_{1}\right\vert }{%
c}\right) $ is the average connectivity between $Z$ and $Z_{1}$ depending on
the connectivity state of the sstm:%
\begin{eqnarray*}
T\left( Z,\theta ,Z_{1},\theta -\frac{\left\vert Z-Z_{1}\right\vert }{c}%
\right) &=&\left\langle T\left\vert \Gamma \left( T,\hat{T},\theta
,Z,Z_{1}\right) \right\vert ^{2}\right\rangle \\
&=&\int T\left\vert \Gamma \left( T,\hat{T},\theta ,Z,Z_{1}\right)
\right\vert ^{2}dTd\hat{T}
\end{eqnarray*}%
By the classical level, we mean that computations are performed only using $%
S\left( \Psi \right) $, without considering the perturbative corrections
arising from the full statistical interactions encompassed in the effective
action. Some details of these corrections can be found in Appendix 1, since
they are relevant for the existence of nonlinear stable propagation of
oscillating activities. The potential $V\left( \Psi \right) $ induces a
certain stabilization of neural activities, limiting them around some
average $\left\vert \Psi _{0}\left( Z\right) \right\vert ^{2}$. Thus, it
regulates the overall activity.

We choose:%
\begin{equation*}
V\left( \Psi \right) =\frac{1}{2}\left( \left\vert \Psi \left( Z\right)
\right\vert ^{2}-\int T\left( Z^{\prime },Z_{1}\right) \left\vert \Psi
_{0}\left( Z\right) \right\vert ^{2}dZ_{1}\right) ^{2}
\end{equation*}

This potential loosely constrains the number of firing neurons at every time
in each zone of the thread. This limitation is not constant but depends on
the connectivities between points.

The second part in the right hand side of (\ref{Sf}):%
\begin{equation}
S\left( \Gamma \right) =\frac{1}{2\eta ^{2}}\left( S_{\Gamma }^{\left(
0\right) }+S_{\Gamma }^{\left( 1\right) }+S_{\Gamma }^{\left( 2\right)
}+S_{\Gamma }^{\left( 3\right) }+S_{\Gamma }^{\left( 4\right) }\right)
+U\left( \left\{ \left\vert \Gamma \left( \theta ,Z,Z^{\prime },C,D\right)
\right\vert ^{2}\right\} \right)  \label{Sg}
\end{equation}

where $S_{\Gamma }^{\left( 1\right) }$, $S_{\Gamma }^{\left( 2\right) }$, $%
S_{\Gamma }^{\left( 3\right) }$, $S_{\Gamma }^{\left( 4\right) }$ are given
by: 
\begin{equation}
S_{\Gamma }^{\left( 1\right) }=\int \Gamma ^{\dag }\left( T,\hat{T},\theta
,Z,Z^{\prime },C,D\right) \nabla _{T}\left( \frac{\sigma _{T}^{2}}{2}\nabla
_{T}+O_{T}\right) \Gamma \left( T,\hat{T},\theta ,Z,Z^{\prime },C,D\right)
\label{wgDd}
\end{equation}%
\begin{equation}
S_{\Gamma }^{\left( 2\right) }=\int \Gamma ^{\dag }\left( T,\hat{T},\theta
,Z,Z^{\prime },C,D\right) \nabla _{\hat{T}}\left( \frac{\sigma _{\hat{T}}^{2}%
}{2}\nabla _{\hat{T}}+O_{\hat{T}}\right) \Gamma \left( T,\hat{T},\theta
,Z,Z^{\prime },C,D\right)
\end{equation}%
\begin{equation}
S_{\Gamma }^{\left( 3\right) }=\Gamma ^{\dag }\left( T,\hat{T},\theta
,Z,Z^{\prime },C,D\right) \nabla _{C}\left( \frac{\sigma _{C}^{2}}{2}\nabla
_{C}+O_{C}\right) \Gamma \left( T,\hat{T},\theta ,Z,Z^{\prime },C,D\right)
\label{wgQd}
\end{equation}%
\begin{equation}
S_{\Gamma }^{\left( 4\right) }=\Gamma ^{\dag }\left( T,\hat{T},\theta
,Z,Z^{\prime },C,D\right) \nabla _{D}\left( \frac{\sigma _{D}^{2}}{2}\nabla
_{D}+O_{D}\right) \Gamma \left( T,\hat{T},\theta ,Z,Z^{\prime },C,D\right)
\end{equation}

The connectivty field $\Gamma $ depends on several variables, accounting for
the dynamics of the connectivities between neurons located at different
points $\left( Z,Z^{\prime }\right) $ of the thread. Variables $C$ and $D$
represents the accumulation of input and output spikes which
influence--depending on their synchronization--the modification of
connectivity $\hat{T}$. In turn, this modifies the connectivity $T$. The
operators $O_{T}$, $O_{\hat{T}}$, $O_{C}$, $O_{D}$ represent, in terms of
fields, the dynamics for these variables. Their precise formula are given in
Appendix 1, and their interpretations are provided in Appendix 3. The
difference between $T$ and $\hat{T}$ measures the difference of scale
between modifications in activities and modifications in connectivities. \
The interpretation of these formulas is summarized in Appendix 3.

In (\ref{flt}), we included a potential:%
\begin{equation*}
U\left( \left\{ \left\vert \Gamma \left( \theta ,Z,Z^{\prime },C,D\right)
\right\vert ^{2}\right\} \right) =U\left( \int T\left\vert \Gamma \left( T,%
\hat{T},\theta ,Z,Z^{\prime },C,D\right) \right\vert ^{2}dTd\hat{T}\right)
\end{equation*}%
that models the constraint on the number of possible active connections in
the system.

The action $S\left( \Gamma \right) $ describes the collective and individual
dynamics of the connectivities $\Gamma $. This action represents the
long-term behavior of the system. In a first approximation, it depends on
the particular states defined by $S\left( \Psi \right) $.\bigskip

\subsection{Study of the background states}

The system described by (\ref{Sf}) is studied by first finding the
background fields for $\Psi $ and $\Gamma $. They are obtained by minimizing
the effective actions of the system. These background fields enable the
derivation of effective actions that, in turn, provide insights into the
system's dynamics in a given background field, as well as the emergence of
some types of states depending on external or internal conditions.

The background fields for $\Psi $ and $\Gamma $ are not independent, since
both fields are involved in the two parts of the action functional. This
reflects the fact that neural activity is conditioned by the system of
connections, while neuronal activities modify the form of the
connectivities. However, the search for the background fields can be
decomposed into two successive parts, at least to first approximation.

Given that neuronal dynamics take place on a faster timescale than the
dynamics of the connectivities, we first solve for the background field
equation derived from the saddle point equation for $S\left( \Psi \right) $.
This expresses the background field $\Psi $ and the activity $\omega \left(
J,\theta ,Z,\left\vert \Psi \right\vert ^{2}\right) $ as functionals of the
connectivity field $\Gamma $. This is done at both static and dynamical
levels. We show, in particular, that certain nonlinear stable oscillations
in activity exist, their stability being induced by the modification of the
background field in which the oscillations occur. Some details are given in
appendix 3.

\subsubsection{Effective actions}

The background fields minimize the effective actions of the system. In a
first approximation, the effective actions are the classical actions defined
above. However, when considering the dynamical aspects of the background
field of $\Psi $, it is necessary to include the perturbative corrections to
the action functional of $\Psi \left( \theta ,Z\right) $. Moreover, when
considering the connectivities, it is possible to reduce the action
functional $\sum_{i=1}^{4}S_{\Gamma }^{\left( i\right) }$ to an effective
action obtained by projecting the field $\Gamma \left( T,\hat{T},\theta
,Z,Z^{\prime },C,D\right) $ onto a reduced field $\Gamma \left( T,\theta
,Z,Z^{\prime }\right) $ describing directly the level of connectivity. This
field is decribed by an effective action given below.

\paragraph{Effective action for $\Psi $ and activities}

The background field $\Psi \left( \theta ,Z\right) $ satisfies the saddle
point equation for the "classical" effective action. This effective action
modifies the classical action:%
\begin{equation}
-\frac{1}{2}\Psi ^{\dagger }\left( \theta ,Z\right) \left( \nabla _{\theta
}\left( \frac{\sigma ^{2}}{2}\nabla _{\theta }-\omega ^{-1}\left( J,\theta
,Z,\left\vert \Psi \right\vert ^{2}\right) \right) \right) \Psi \left(
\theta ,Z\right)  \label{nmS}
\end{equation}%
by perturbative corrections (see Appendix 2) and writes:%
\begin{equation}
\Gamma \left( \Psi ^{\dagger },\Psi \right) \simeq \int \Psi ^{\dagger
}\left( \theta ,Z\right) \left( -\nabla _{\theta }\left( \frac{\sigma
_{\theta }^{2}}{2}\nabla _{\theta }-\omega ^{-1}\left( J\left( \theta
\right) ,\theta ,Z,\mathcal{G}_{0}+\left\vert \Psi \right\vert ^{2}\right)
\right) \delta \left( \theta _{f}-\theta _{i}\right) +\Gamma _{p}\left( \Psi
^{\dagger },\Psi \right) \right) \Psi \left( \theta ,Z\right)  \label{nmr}
\end{equation}%
where $\Gamma _{p}\left( \Psi ^{\dagger },\Psi \right) $ is a corrective
perturbative term depending on the successive\ derivatives of the field (the
constants $a_{j}$ are derived in (\cite{GL})):%
\begin{eqnarray}
\Gamma _{p}\left( \Psi ^{\dagger },\Psi \right) &=&\int \sum_{\substack{ %
j\geqslant 1  \\ m\geqslant 1}}\sum_{\substack{ \left( p_{l}^{i}\right)
_{m\times j}  \\ p_{l}+\sum_{i}p_{l}^{i}\geqslant 2}}\frac{a_{j}}{j!}\frac{%
\delta ^{\sum_{l}p_{l}}\left[ -\frac{1}{2}\left( \nabla _{\theta }\left( 
\frac{\sigma _{\theta }^{2}}{2}\nabla _{\theta }-\omega ^{-1}\left(
\left\vert \Psi \left( \theta ,Z\right) \right\vert ^{2}\right) \right)
\right) \right] }{\prod\limits_{l=1}^{j}\prod\limits_{k_{l}^{i}=1}^{p_{l}}%
\delta \left\vert \Psi \left( \theta ^{\left( l\right) },Z_{_{l}}\right)
\right\vert ^{2}}\Psi \left( \theta ,Z\right) \\
&&\times \left( \prod\limits_{l=1}^{j}\Psi ^{\dagger }\left( \theta
_{f}^{\left( l\right) },Z_{l}\right) \right) \prod\limits_{i=1}^{m}\left[ 
\frac{\delta ^{\sum_{l}p_{l}^{i}}\left[ \hat{S}_{cl,\theta }\left( \Psi
^{\dagger },\Psi \right) \right] }{\prod\limits_{l=1}^{j}\prod%
\limits_{k_{l}^{i}=1}^{p_{l}^{i}}\delta \left\vert \Psi \left( \theta
^{\left( l\right) },Z_{_{l}}\right) \right\vert ^{2}}\right] \left(
\prod\limits_{l=1}^{j}\Psi \left( \theta _{i}^{\left( l\right)
},Z_{l}\right) \right)  \notag
\end{eqnarray}%
The term $\mathcal{G}_{0}$ is a function of $Z$ and represents the two
points free Green function (see Appendix 2 and \cite{GL}).

In presence of a source term $J\left( \theta ,Z\right) $, the function $%
\omega \left( J\left( \theta \right) ,\theta ,Z,\mathcal{\bar{G}}%
_{0}+\left\vert \Psi \right\vert ^{2}\right) $ is the solution of:%
\begin{eqnarray}
\omega ^{-1}\left( J,\theta ,Z,\left\vert \Psi \right\vert ^{2}\right)
&=&G\left( J\left( \theta ,Z\right) +\int \frac{\kappa }{N}\frac{\omega
\left( J,\theta -\frac{\left\vert Z-Z_{1}\right\vert }{c},Z_{1},\Psi \right)
T\left( Z,\theta ,Z_{1},\theta -\frac{\left\vert Z-Z_{1}\right\vert }{c}%
\right) }{\omega \left( J,\theta ,Z,\left\vert \Psi \right\vert ^{2}\right) }%
\right.  \label{Dyn} \\
&&\times \left. \left( \mathcal{\bar{G}}_{0}\left( 0,Z_{1}\right)
+\left\vert \Psi \left( \theta -\frac{\left\vert Z-Z_{1}\right\vert }{c}%
,Z_{1}\right) \right\vert ^{2}\right) dZ_{1}\right)  \notag
\end{eqnarray}%
which is the classical activity equation, up to the inclusion of the Green
function $\mathcal{\bar{G}}_{0}\left( 0,Z_{1}\right) $. The Green function
is defined in Appendix 1.

\subsubsection{Background for $\Gamma $}

Once the saddle point for $\Psi $ is obtained, we can, in a second step
study the saddle-point equations for $\Gamma $. We consider a quasi-static
approximation. This amounts to considering neural activities as quite-static
along the thread and replacing them by their averages. Appendix 3 shows that
we can reduce the action functionals for connectivities to an effective
action depending directly on $T$:%
\begin{eqnarray}
S &&\left( \Gamma _{0}\right) =\int \Gamma _{0}^{\dag }\left( T,\theta
,Z,Z^{\prime }\right) \left( \frac{\sigma _{T}^{2}}{2}\nabla _{T}^{2}-\frac{1%
}{2\sigma _{T}^{2}}\left( \left( \frac{1}{\tau \omega }\left( T-\left\langle
T\right\rangle \right) \right) \left\vert \Psi \left( \theta ,Z\right)
\right\vert ^{2}\right) ^{2}-\frac{1}{2\tau \omega \left( Z\right) }\right)
\Gamma _{0}\left( T,\theta ,Z,Z^{\prime }\right)  \label{Ftn} \\
&&-\Gamma _{0}^{\dag }\left( T,\theta ,Z,Z^{\prime }\right) \left(
a_{C}\left( Z\right) +a_{D}\left( Z\right) +\frac{\rho h_{C}\left( \omega
\left( \theta ,Z\right) \right) \left\vert \bar{\Psi}\left( \theta
,Z,Z^{\prime }\right) \right\vert ^{2}}{\omega \left( \theta ,Z,\left\vert
\Psi \right\vert ^{2}\right) }\right) \Gamma _{0}\left( T,\theta
,Z,Z^{\prime }\right)  \notag
\end{eqnarray}%
with the associated saddle point:%
\begin{equation*}
\frac{\delta S\left( \Gamma _{0}\right) }{\delta \Gamma _{0}\left( T,\theta
,Z,Z^{\prime }\right) }=\frac{\delta S\left( \Gamma _{0}\right) }{\delta
\Gamma _{0}^{\dag }\left( T,\theta ,Z,Z^{\prime }\right) }=0
\end{equation*}

We show in this Appendix that several--indeed an infinite number--of static
background fields are possible, corresponding to different overall
environment for neural activities. It is possible to study the modifications
of the background connectivities, by considering the effect of
signal-induced activities. At the static level, shifting the static level of
activity in some zone modifies the background values. Dynamically,
considering the diffusions of waves of activity may results in the binding
of groups of cells that collectively exhibit enhanced activity and
connectivity. This shows that, given a background state, some shift to an
activated group is possible, the resulting state constituting a modified
background for activities. This static version of transitions serves as a
starting point for studying the effective action for connectivities. It is
summarized in appendix 3.

\subsubsection{Static background}

These activations may depend on external signals, or more fundamentally, on
internal interactions between connectivity states. In the static case, this
effective approach encompasses the averaged cellular activity through
indirect interactions terms between distant connections: modifications in
connections enhance the activities in several regions, which then propagate
and, in turn, induce modified--i.e. activated--connections. We show that
some large groups of activated connections may emerge. This corresponds to
sets of connected cells whose connections and activities are mutually bound.
The quasi-static solutions for the field of activities are:

\begin{equation}
\left\vert \Psi \left( Z\right) \right\vert ^{2}=\frac{2T\left( Z\right)
\left\langle \left\vert \Psi _{0}\left( Z^{\prime }\right) \right\vert
^{2}\right\rangle _{Z}}{\left( 1+\sqrt{1+4\left( \frac{\lambda \tau \nu
c-T\left( Z\right) }{\left( \frac{1}{\tau _{D}\alpha _{D}}+\frac{1}{\tau
_{C}\alpha _{C}}+\Omega \right) T\left( Z\right) -\frac{1}{\tau _{D}\alpha
_{D}}\lambda \tau \nu c}\right) ^{2}T\left( Z\right) \left\langle \left\vert
\Psi _{0}\left( Z^{\prime }\right) \right\vert ^{2}\right\rangle _{Z}}%
\right) }  \label{fld}
\end{equation}%
where:%
\begin{equation*}
T\left( Z\right) =\left\langle T\left( Z,Z^{\prime }\right) \right\rangle
_{Z^{\prime }}
\end{equation*}%
is the average of $T\left( Z,Z^{\prime }\right) $ over $Z^{\prime }$.

The static activities are defined by:%
\begin{equation*}
\omega ^{-1}\left( Z,\left\vert \Psi \right\vert ^{2}\right) =G\left( \int 
\frac{\kappa }{N}\frac{\omega \left( J,\Psi \right) T\left( Z,Z_{1}\right) }{%
\omega \left( J,Z,\left\vert \Psi \right\vert ^{2}\right) }\left\vert \Psi
\left( Z_{1}\right) \right\vert ^{2}dZ_{1}\right)
\end{equation*}%
These solutions are defined for a given average level of connectivities,
which are given by:

\begin{equation*}
T\left( Z,Z^{\prime }\right) =\frac{\lambda \tau \exp \left( -\frac{%
\left\vert Z-Z^{\prime }\right\vert }{\nu c}\right) }{1+\frac{\alpha
_{D}\omega h_{D}}{\alpha _{C}\omega ^{\prime }h_{C}}\frac{\frac{1}{\tau _{C}}%
+\alpha _{C}\omega ^{\prime }\left\vert \Psi \left( Z^{\prime }\right)
\right\vert ^{2}}{\frac{1}{\tau _{D}}+\alpha _{D}\omega \left\vert \Psi
\left( Z\right) \right\vert ^{2}}}=\lambda \tau \hat{T}\left( Z,Z^{\prime
}\right)
\end{equation*}%
with $h_{D}\left( D\right) \simeq D$ and $h_{C}\left( C\right) \simeq C$ and:%
\begin{eqnarray}
C &\rightarrow &\left\langle C\left( \theta \right) \right\rangle =\frac{%
\alpha _{C}\frac{\omega \left( Z^{\prime }\right) \left\vert \Psi \left(
Z^{\prime }\right) \right\vert ^{2}}{\omega \left( Z\right) }}{\frac{1}{\tau
_{C}\omega }+\alpha _{C}\frac{\omega \left( Z^{\prime }\right) \left\vert
\Psi \left( Z^{\prime }\right) \right\vert ^{2}}{\omega \left( Z\right) }} \\
D &\rightarrow &\left\langle D\left( \theta \right) \right\rangle =\frac{%
\alpha _{D}\omega \left( Z\right) \left\vert \Psi \left( Z\right)
\right\vert ^{2}}{\frac{1}{\tau _{D}}+\alpha _{D}\omega \left( Z\right)
\left\vert \Psi \left( Z\right) \right\vert ^{2}}
\end{eqnarray}%
Approximate multiple solutions are presented in Appendix 3 and we show that
cells may organise in groups of connected elements.

\subsubsection{Dynamical aspects}

Coming back to the activities in such connected groups, we use the previous
results on $S\left( \Psi \right) $ to compute the dynamic activities for
such connectd groups. We find that an emerging group above the background
has its own system of average connections, activities and frequencies for
oscillations of these activities. Such a group is localized and involves a
certain number of cells. Cells may participate in different collective
states. This is presented in appendix 4. The dynamics of collective states
for $\Gamma $ thus depend on the cell states and their activations by
signals. In turn, the states for $\Gamma $ arising from the minimization of $%
S\left( \Gamma \right) $, given some surrounding neural activity, will shape
these activities. Some collective states, along with compatible stable
internal time-dependent activities, may arise. These activities are
characterized by their amplitudes and characeristic frquencies. These
collective states are those that would arise from the simultaneous
minimization of $S=S\left( \Psi \right) +S\left( \Gamma \right) $.

\paragraph{Dynamical activities}

The dynamical solutions to the saddle-point equation for (\ref{nmr}),
together with (\ref{Dyn}), are obtained by expansion around the static
solutions for the background state. We showed, in a first approximation (see
Appendix 4 for an account) that the dynamical activities satisfy a wave
equation:%
\begin{equation}
f\Omega \left( \theta ,Z\right) =J\left( \theta ,Z\right) +\left( \frac{\hat{%
f}_{1}}{\omega \left( \theta ,Z\right) }+N_{1}\right) \nabla _{\theta
}\Omega \left( \theta ,Z\right) +\left( \frac{\hat{f}_{3}}{\omega \left(
\theta ,Z\right) }-N_{2}\right) \nabla _{\theta }^{2}\Omega \left( \theta
,Z\right) +\frac{c^{2}\hat{f}_{3}}{\omega \left( \theta ,Z\right) }\nabla
_{Z}^{2}\Omega \left( \theta ,Z\right)
\end{equation}%
where we defined:%
\begin{eqnarray}
\hat{f}_{1} &=&\frac{W^{\prime }\left( 1\right) -W\left( 1\right) }{c}\Gamma
_{1}\text{, }\hat{f}_{3}=\frac{\left( W\left( 1\right) -W^{\prime }\left(
1\right) \right) \Gamma _{2}}{c^{2}}  \label{ctf} \\
\Gamma _{1} &=&\frac{\kappa }{NX_{r}}\int \left\vert Z-Z_{1}\right\vert
T\left( Z,Z_{1}\right) \mathcal{\bar{G}}_{0}\left( 0,Z_{1}\right) dZ_{1} 
\notag \\
\Gamma _{2} &=&\frac{\kappa }{2NX_{r}}\int \left( Z-Z_{1}\right) ^{2}T\left(
Z,Z_{1}\right) \mathcal{\bar{G}}_{0}\left( 0,Z_{1}\right) dZ_{1}  \notag
\end{eqnarray}%
and:%
\begin{eqnarray*}
N_{1} &=&\frac{\Psi _{0}\left( Z\right) }{U^{\prime \prime }\left(
X_{0}\right) \omega ^{2}\left( J\left( \theta \right) ,\theta ,Z,\mathcal{G}%
_{0}\right) } \\
N_{2} &=&\frac{\Gamma _{1}\Psi _{0}\left( Z\right) }{U^{\prime \prime
}\left( X_{0}\right) \omega ^{2}\left( J\left( \theta \right) ,\theta ,Z,%
\mathcal{G}_{0}\right) }
\end{eqnarray*}%
those two last coefficients arising from the perturbative expansion of the
effective action (\ref{nmr}). They represent the backreaction of the system
to the fluctuations of activities, and stabilize the system allowing for the
possibility of stable traveling waves solutions.

The dynamic for activities is closed by accounting for the fluctuations of
the field of activities. To the local linear approximation we find for $%
\delta \Psi \left( \theta ,Z\right) $ (see Appendix 5 for the derivation):%
\begin{equation}
\delta \Psi \left( \theta ,Z\right) \simeq \frac{\nabla _{\theta }\omega
\left( \theta ,Z,\mathcal{G}_{0}+\left\vert \Psi _{0}\right\vert ^{2}\right) 
}{U^{\prime \prime }\left( X_{0}\right) \omega ^{2}\left( J\left( \theta
\right) ,\theta ,Z,\mathcal{G}_{0}+\left\vert \Psi _{0}\right\vert
^{2}\right) }\Psi _{0}\simeq \frac{\nabla _{\theta }\omega \left( \theta ,Z,%
\mathcal{G}_{0}\right) }{U^{\prime \prime }\left( X_{0}\right) \omega
^{2}\left( J\left( \theta \right) ,\theta ,Z,\mathcal{G}_{0}\right) }\Psi
_{0}  \label{bgd}
\end{equation}

The values of connectivities to the quasi-static approximation are given by:%
\begin{equation*}
\left\langle T\left( Z,Z^{\prime }\right) \right\rangle =\frac{\lambda \tau
\exp \left( -\frac{\left\vert Z-Z^{\prime }\right\vert }{\nu c}\right) }{1+%
\frac{\alpha _{D}\omega h_{D}}{\alpha _{C}\omega ^{\prime }h_{C}}\frac{\frac{%
1}{\tau _{C}}+\alpha _{C}\omega ^{\prime }\left\vert \Psi \left( \theta -%
\frac{\left\vert Z-Z^{\prime }\right\vert }{c},Z^{\prime }\right)
\right\vert ^{2}}{\frac{1}{\tau _{D}}+\alpha _{D}\omega \left\vert \Psi
\left( \theta ,Z\right) \right\vert ^{2}}}=\lambda \tau \left\langle \hat{T}%
\left( Z,Z^{\prime }\right) \right\rangle
\end{equation*}%
where the quantities $h_{D}$ and $h_{C}$\ depend on the accumulation of
input and outpout spikes with averages $\left\langle C\left( \theta \right)
\right\rangle $ and $\left\langle D\left( \theta \right) \right\rangle $: 
\begin{eqnarray}
C &\rightarrow &\left\langle C\left( \theta \right) \right\rangle =\frac{%
\alpha _{C}\frac{\omega ^{\prime }\left\vert \Psi \left( \theta -\frac{%
\left\vert Z-Z^{\prime }\right\vert }{c},Z^{\prime }\right) \right\vert ^{2}%
}{\omega }}{\frac{1}{\tau _{C}\omega }+\alpha _{C}\frac{\omega ^{\prime
}\left\vert \Psi \left( \theta -\frac{\left\vert Z-Z^{\prime }\right\vert }{c%
},Z^{\prime }\right) \right\vert ^{2}}{\omega }}=\frac{\alpha _{C}\omega
^{\prime }\left\vert \Psi \left( \theta -\frac{\left\vert Z-Z^{\prime
}\right\vert }{c},Z^{\prime }\right) \right\vert ^{2}}{\frac{1}{\tau _{C}}%
+\alpha _{C}\omega ^{\prime }\left\vert \Psi \left( \theta -\frac{\left\vert
Z-Z^{\prime }\right\vert }{c},Z^{\prime }\right) \right\vert ^{2}}\equiv
C\left( \theta \right) \\
D &\rightarrow &\left\langle D\left( \theta \right) \right\rangle =\frac{%
\alpha _{D}\omega \left\vert \Psi \left( \theta ,Z\right) \right\vert ^{2}}{%
\frac{1}{\tau _{D}}+\alpha _{D}\omega \left\vert \Psi \left( \theta
,Z\right) \right\vert ^{2}}\equiv D\left( \theta \right)
\end{eqnarray}

\subsection{Background field for groups wth signal-enhanced connectivity,
stability.}

Considering the static background, the solutions for activities depend on
the average local connectivities. Conversely, some signals may modify both
activities and connectivities. We showed in \cite{GLr} that some signals can
activate additional connectivities, among some sets of cells. We also showed
that these cell states can be regarded as an isolated group. This provides a
first indication that collective states may emerge from signals and
interactions. We present the results here by first assuming quasi-static
solutions for activities, and then by including dynamical activities.

\subsubsection{Static connectivities and activities}

The connectivities between cells activated by certain signals are enhanced
and may form a stable state. Such a state is locked by the cells' enhanced
activities $\omega _{M}$. The connections between points $\left\{
Z_{M}^{\left( \varepsilon \right) }\right\} _{\epsilon }$ are given by:

\begin{equation}
T\left( Z_{M}^{\left( \varepsilon _{1}\right) },Z_{M}^{\left( \varepsilon
_{2}\right) }\right) =\frac{\lambda \tau \exp \left( -\frac{\left\vert
Z_{M}^{\left( \varepsilon _{1}\right) }-Z_{M}^{\left( \varepsilon
_{2}\right) }\right\vert }{\nu c}\right) \left\vert \Psi \left( \theta -%
\frac{\left\vert Z_{M}^{\left( \varepsilon _{1}\right) }-Z_{M}^{\left(
\varepsilon _{2}\right) }\right\vert }{c},Z_{M}^{\left( \varepsilon
_{2}\right) }\right) \right\vert ^{2}h_{C}}{\left\vert \Psi \left( \theta -%
\frac{\left\vert Z_{M}^{\left( \varepsilon _{1}\right) }-Z_{M}^{\left(
\varepsilon _{2}\right) }\right\vert }{c},Z_{M}^{\left( \varepsilon
_{2}\right) }\right) \right\vert ^{2}h_{C}+\frac{\left( \frac{1}{\alpha
_{C}\tau _{C}}+\omega _{M}\left\vert \Psi \left( \theta -\frac{\left\vert
Z_{M}^{\left( \varepsilon _{1}\right) }-Z_{M}^{\left( \varepsilon
_{2}\right) }\right\vert }{c},Z_{M}^{\left( \varepsilon _{2}\right) }\right)
\right\vert ^{2}\right) \alpha _{D}h_{D}}{\frac{1}{\tau _{D}}+\alpha
_{D}\omega _{M}\left\vert \Psi \left( \theta ,Z_{M}^{\left( \varepsilon
_{1}\right) }\right) \right\vert ^{2}}}
\end{equation}%
In a lon-run static perspective, it becomes:%
\begin{equation*}
T\left( Z_{M}^{\left( \varepsilon _{1}\right) },Z_{M}^{\left( \varepsilon
_{2}\right) }\right) =\frac{\lambda \tau \exp \left( -\frac{\left\vert
Z_{M}^{\left( \varepsilon _{1}\right) }-Z_{M}^{\left( \varepsilon
_{2}\right) }\right\vert }{\nu c}\right) \left\vert \Psi _{0}\left(
Z_{M}^{\left( \varepsilon _{2}\right) }\right) \right\vert ^{2}h_{C}}{%
\left\vert \Psi _{0}\left( Z_{M}^{\left( \varepsilon _{2}\right) }\right)
\right\vert ^{2}h_{C}+\left( \frac{1}{\alpha _{C}\tau _{C}}+\omega
_{M}\left\vert \Psi _{0}\left( Z_{M}^{\left( \varepsilon _{2}\right)
}\right) \right\vert ^{2}\right) \frac{\alpha _{D}h_{D}}{\frac{1}{\tau _{D}}%
+\alpha _{D}\omega _{M}\left\vert \Psi _{0}\left( Z_{M}^{\left( \varepsilon
_{1}\right) }\right) \right\vert ^{2}}}
\end{equation*}%
where the corresponding activities satify:

\begin{eqnarray}
\omega _{M}^{-1}\left( Z_{M}^{\left( \varepsilon _{1}\right) },\left\vert
\Psi \right\vert ^{2}\right) &\simeq &G\left( \frac{\kappa }{N}%
\sum_{Z_{M}^{\left( \varepsilon _{2}\right) }}T\left( Z_{M}^{\left(
\varepsilon _{1}\right) },Z_{M}^{\left( \varepsilon _{2}\right) }\right)
\left\vert \Psi _{0}\left( Z_{M}^{\left( \varepsilon _{2}\right) }\right)
\right\vert ^{2}\right)  \label{nh} \\
&\simeq &G\left( C\frac{\left\vert \Psi _{0}\left( Z_{M}\right) \right\vert
^{4}h_{C}}{\left\vert \Psi _{0}\left( Z_{M}\right) \right\vert
^{2}h_{C}+\left( \frac{1}{\alpha _{C}\tau _{C}}+\omega _{M}\left\vert \Psi
_{0}\left( Z_{M}\right) \right\vert ^{2}\right) \frac{\alpha _{D}h_{D}}{%
\frac{1}{\tau _{D}}+\alpha _{D}\omega _{M}\left\vert \Psi _{0}\left(
Z_{M}\right) \right\vert ^{2}}}\right)  \notag
\end{eqnarray}%
where:%
\begin{equation*}
C=\frac{\kappa \lambda \tau }{N\left( \sharp \left\{ Z_{M}^{\left(
\varepsilon _{1}\right) }\right\} \right) }\sum_{Z_{M}^{\left( \varepsilon
_{1}\right) },Z_{M}^{\left( \varepsilon _{2}\right) }}\exp \left( -\frac{%
\left\vert Z_{M}^{\left( \varepsilon _{1}\right) }-Z_{M}^{\left( \varepsilon
_{2}\right) }\right\vert }{\nu c}\right)
\end{equation*}%
and $\left\vert \Psi _{0}\left( Z_{M}\right) \right\vert ^{2}$ is the
average of $\left\vert \Psi _{0}\left( Z_{M}^{\left( \varepsilon _{2}\right)
}\right) \right\vert ^{2}$ over $\left\{ Z_{M}^{\left( \varepsilon
_{2}\right) }\right\} $ and the values of $\left\vert \Psi _{0}\left(
Z_{M}^{\left( \varepsilon _{2}\right) }\right) \right\vert ^{2}$ are given
by (\ref{fld}). The point $Z_{M}$ is the average of the $\left\{
Z_{M}^{\left( \varepsilon _{2}\right) }\right\} $. We also show in \cite{GLr}
how signals may activate, combine, or desactivate such groups. We now
consider the dynamical oscillations of cell activity for these states.

\subsubsection{Dynamic activities}

We can look for specific dynamical activities within such a group of
connected elements. Dynamicaly, these activities are defined by specific
frequencies.

In a state with $S^{2}$ $=\left\{ \left( Z,Z^{\prime }\right) \right\} $, we
aim at finding the possible activity frequencies $\omega \left( \theta
,Z,\left\vert \Psi \right\vert ^{2}\right) $ associated with the state $%
\Delta T$. \ We consider the state $\left\{ \Delta T\left( Z,Z^{\prime
}\right) \right\} $ as a system with its own associated activity $\Delta
\omega \left( \theta ,Z,\left\vert \Psi \right\vert ^{2}\right) $ in the
given background field $\left\vert \Psi \right\vert ^{2}$.

To find $\omega \left( \theta ,Z,\left\vert \Psi \right\vert ^{2}\right) $,
we start with the defining equation with source:%
\begin{eqnarray}
&&\omega ^{-1}\left( J,\theta ,Z,\left\vert \Psi \right\vert ^{2}\right)
\label{qg} \\
&=&G\left( J\left( \theta ,Z\right) +\int \frac{\kappa }{N}\frac{\omega
\left( J,\theta -\frac{\left\vert Z-Z_{1}\right\vert }{c},Z_{1},\Psi \right)
T\left( Z,\theta ,Z_{1},\theta -\frac{\left\vert Z-Z_{1}\right\vert }{c}%
\right) }{\omega \left( J,\theta ,Z,\left\vert \Psi \right\vert ^{2}\right) }%
\left\vert \Psi \left( \theta -\frac{\left\vert Z-Z_{1}\right\vert }{c}%
,Z_{1}\right) \right\vert ^{2}dZ_{1}\right)  \notag
\end{eqnarray}

We show in \cite{GL} that $\omega ^{-1}\left( \theta ,Z,\left\vert \Psi
\right\vert ^{2}\right) $ decomposes into a static part and a variable part.
The solution for the specific activties of the group is:

\begin{eqnarray}
&&\omega \left( \theta ,Z_{i},\mathbf{T},\left\vert \Psi \right\vert
^{2}\right)  \label{cTv} \\
&=&\overline{\omega }\left( \mathbf{Z},\mathbf{T},\left\vert \Psi
\right\vert ^{2}\right)  \notag \\
&&+A\left( Z_{1}\right) \left( 1,\left( 1-\mathbf{\hat{T}}\exp \left(
-i\Upsilon _{p}\left( \mathbf{\hat{T}}\right) \frac{\left\vert \mathbf{%
\Delta Z}\right\vert }{c}\right) \right) ^{-1}\hat{T}_{1}\left( \mathbf{Z}%
\right) \exp \left( -i\Upsilon _{p}\left( \mathbf{\hat{T}}\right) \frac{%
\left\vert \mathbf{\Delta Z}_{1}\right\vert }{c}\right) \right) ^{t}\exp
\left( i\Upsilon _{p}\left( \mathbf{\hat{T}}\right) \theta \right)  \notag
\end{eqnarray}%
The static part for a group of activated states 
\begin{equation*}
\overline{\omega }\left( \mathbf{Z},\mathbf{T},\left\vert \Psi \right\vert
^{2}\right)
\end{equation*}%
is a vector, whose component $i$ satisfies for each point $Z_{i}$ of this
group:%
\begin{equation}
\overline{\omega }^{-1}\left( Z_{i},\left\vert \Psi \right\vert ^{2}\right)
=G\left( \sum_{j}\frac{\kappa }{N}\frac{\overline{\omega }\left( Z_{j},\Psi
\right) T\left( Z_{i},Z_{j}\right) }{\overline{\omega }\left(
Z_{i},\left\vert \Psi \right\vert ^{2}\right) }\left\vert \Psi \left(
Z_{j}\right) \right\vert ^{2}\right)  \label{dM}
\end{equation}%
which corresponds to (\ref{nh}) if we replace $\left\vert \Psi \left(
Z_{j}\right) \right\vert ^{2}$. The second contribution in (\ref{CTv})
describes an osicllatory behavior for all elements of the groups with a set
of possible frequencs, with $\mathbf{\hat{T}}$ the matrix with elements $%
\hat{T}\left( Z_{i},Z_{j}\right) $ defined by:%
\begin{equation*}
\hat{T}\left( Z_{i},Z_{j}\right) =\frac{\kappa }{N}\frac{T\left(
Z_{i},Z_{j}\right) \overline{\omega }\left( J,Z_{j},\Psi \right) \left\vert
\Psi _{0}\left( Z_{j}\right) \right\vert ^{2}}{G^{-1}\left( \overline{\omega 
}^{-1}\left( J,Z_{i},\left\vert \Psi \right\vert ^{2}\right) \right) -%
\overline{\omega }^{-1}\left( J,Z_{i},\left\vert \Psi \right\vert
^{2}\right) }
\end{equation*}%
and $\overline{\omega }\left( \theta ,\mathbf{Z},\left\vert \Psi \right\vert
^{2}\right) $ and $\hat{T}_{1}\left( \mathbf{Z}\right) $\ are vectors with
coordinates $\overline{\omega }\left( J,\theta ,Z_{i},\left\vert \Psi
\right\vert ^{2}\right) $ and $\hat{T}\left( Z_{1},Z_{j}\right) $
respectively. The point $Z_{1}$ is an arbitrary point chosen in the group
and $A\left( Z_{1}\right) $ is the amplitude of $\omega $ at some given $%
Z_{1}$. The frequencies $\Upsilon _{p}\left( \mathbf{T}\right) $ belong to a
discrete set satisfying the equation:%
\begin{equation}
\det \left( 1-\hat{T}\left( Z_{i},Z_{j}\right) \exp \left( -i\Upsilon _{p}%
\frac{\left\vert Z_{i}-Z_{j}\right\vert }{c}\right) \right) =0  \label{drc}
\end{equation}%
The possible osillatory activities associated to the assembly is thus
described by the sets:%
\begin{equation*}
\left\{ \left\{ A\left( Z_{i}\right) \right\} _{i=1,...n},\Upsilon
_{p}\left( \left\{ \hat{T}\left( Z_{i},Z_{j}\right) \right\} \right)
\right\} _{p}
\end{equation*}%
where $p$ refers to the frequencies $\Upsilon $ and the amplitudes solve:%
\begin{equation*}
A\left( Z_{i}\right) =\sum_{j\neq i}A\left( Z_{j}\right) \hat{T}\left(
Z_{i},Z_{j}\right) \exp \left( -i\Upsilon _{p}\left( \left\{ \hat{T}\left(
Z_{i},Z_{j}\right) \right\} \right) \frac{\left\vert Z_{i}-Z_{j}\right\vert 
}{c}\right)
\end{equation*}

\subsubsection{Collective states fluctuations, stability}

The stability of such states is studied by considering fluctuations around
the saddle-point equations that define them:%
\begin{equation*}
\frac{\delta }{\delta \Psi }\Gamma \left( \Psi ^{\dagger },\Psi \right) =0
\end{equation*}%
\begin{equation*}
\frac{\delta S\left( \Gamma _{0}\right) }{\delta \Gamma _{0}\left( T,\theta
,Z,Z^{\prime }\right) }=\frac{\delta S\left( \Gamma _{0}\right) }{\delta
\Gamma _{0}^{\dag }\left( T,\theta ,Z,Z^{\prime }\right) }=0
\end{equation*}%
together with the dynamical equation for activities (\ref{qg}).

We show in Appendix 5 that this yields two equations for the fluctuations of
the background field and activity $\frac{\delta \Psi \left( \theta ,Z\right) 
}{\Psi \left( \theta ,Z\right) }$, $\frac{\delta \omega ^{-1}\left( \theta
_{1},Z_{1}\right) }{\omega ^{-1}\left( \theta _{1},Z_{1}\right) }$ and one
for the fluctuations $\frac{\delta \check{T}\left( \theta _{1},Z_{1}\right) 
}{\check{T}\left( \theta _{1},Z_{1}\right) }$ of an "effective" connectivity 
$\check{T}\left( Z,\theta ,Z_{1},\theta -\frac{\left\vert Z-Z_{1}\right\vert 
}{c}\right) $ given by:%
\begin{equation*}
\check{T}\left( Z,\theta ,Z_{1},\theta -\frac{\left\vert Z-Z_{1}\right\vert 
}{c}\right) =\frac{\left( \omega ^{-1}\left( Z\right) \right) ^{2}T\left(
Z,\theta ,Z_{1},\theta -\frac{\left\vert Z-Z_{1}\right\vert }{c}\right)
\left( 1+\frac{\alpha _{D}h_{D}}{\alpha _{C}h_{C}}\frac{T\left( Z,\theta
,Z_{1},\theta -\frac{\left\vert Z-Z_{1}\right\vert }{c}\right) }{\lambda
\tau \exp \left( -\frac{\left\vert Z-Z^{\prime }\right\vert }{\nu c}\right) }%
\frac{\omega ^{-1}\left( Z\right) }{\left( \omega ^{\prime }\left(
Z_{1}\right) \right) ^{-1}}\right) }{\left( \left( \omega ^{\prime }\left(
Z_{1}\right) \right) ^{-1}\right) ^{2}\left( G^{-1}\omega ^{-1}\left(
Z\right) -\omega ^{-1}\left( Z\right) \left( G^{-1}\right) ^{\prime }\left(
\omega ^{-1}\left( Z\right) \right) +\tilde{T}_{2}\left( Z,\theta \right)
\right) }
\end{equation*}%
with:%
\begin{eqnarray*}
\tilde{T}_{2}\left( Z,\theta \right) &=&\frac{\kappa }{N}\int dZ_{1}\frac{%
\alpha _{D}h_{D}}{\alpha _{C}h_{C}}\frac{T\left( Z,\theta ,Z_{1},\theta -%
\frac{\left\vert Z-Z_{1}\right\vert }{c}\right) }{\lambda \tau \exp \left( -%
\frac{\left\vert Z-Z^{\prime }\right\vert }{\nu c}\right) } \\
&&\times \frac{\left( \omega ^{-1}\left( Z\right) \right) ^{2}T\left(
Z,\theta ,Z_{1},\theta -\frac{\left\vert Z-Z_{1}\right\vert }{c}\right)
\left( \mathcal{\bar{G}}_{0}\left( 0,Z_{1}\right) +\left\vert \Psi \left(
\theta -\frac{\left\vert Z-Z_{1}\right\vert }{c},Z_{1}\right) \right\vert
^{2}\right) }{\left( \left( \omega ^{\prime }\left( Z_{1}\right) \right)
^{-1}\right) ^{2}}
\end{eqnarray*}%
and where we used the abreviations:%
\begin{equation}
\omega \left( Z\right) \rightarrow \omega \left( J,\theta ,Z,\left\vert \Psi
\right\vert ^{2}\right)  \label{Df}
\end{equation}%
\begin{equation}
\omega ^{\prime }\left( Z_{1}\right) \rightarrow \omega \left( J,\theta -%
\frac{\left\vert Z-Z_{1}\right\vert }{c},Z_{1},\left\vert \Psi \right\vert
^{2}\right)  \label{Dh}
\end{equation}

The equation for the fluctuations of the background field is:%
\begin{eqnarray}
&&0=D_{\theta }\frac{\delta \Psi \left( \theta ,Z\right) }{\Psi \left(
\theta ,Z\right) }+\frac{1}{2}\nabla _{\theta }\left( \delta \omega
^{-1}\left( \theta ,Z\right) \right)  \label{Fln} \\
&&+M_{\omega }^{\left( 1,r\right) }\left( \left( \theta ,Z\right) ,\left(
\theta _{1},Z_{1}\right) \right) \left( \frac{\delta \omega ^{-1}\left(
\theta _{1},Z_{1}\right) }{\omega ^{-1}\left( \theta _{1},Z_{1}\right) }+%
\frac{\delta \check{T}\left( \theta _{1},Z_{1}\right) }{\check{T}\left(
\theta _{1},Z_{1}\right) }\right)  \notag \\
&&+M_{\omega }^{\left( 2,r\right) }\left( \left( \theta ,Z\right) ,\left(
\theta _{1},Z_{1}\right) \right) \nabla _{\theta }\left( \frac{\delta \omega
^{-1}\left( \theta _{1},Z_{1}\right) }{\omega ^{-1}\left( \theta
_{1},Z_{1}\right) }+\frac{\delta \check{T}\left( \theta _{1},Z_{1}\right) }{%
\check{T}\left( \theta _{1},Z_{1}\right) }\right)  \notag \\
&&+M_{\Psi }^{\left( 1,r\right) }\left( \left( \theta ,Z\right) ,\left(
\theta _{1},Z_{1}\right) \right) \frac{\delta \Psi \left( \theta
_{1},Z_{1}\right) }{\Psi \left( \theta _{1},Z_{1}\right) }+M_{\Psi }^{\left(
2,r\right) }\left( \left( \theta ,Z\right) ,\left( \theta _{1},Z_{1}\right)
\right) \nabla _{\theta }\frac{\delta \Psi \left( \theta _{1},Z_{1}\right) }{%
\Psi \left( \theta _{1},Z_{1}\right) }  \notag \\
&&+M_{\omega }^{\left( 1,a\right) }\left( \left( \theta ,Z\right) ,\left(
\theta _{1},Z_{1}\right) \right) \left( \frac{\delta \omega ^{-1}\left(
\theta _{1},Z_{1}\right) }{\omega ^{-1}\left( \theta _{1},Z_{1}\right) }+%
\frac{\delta \check{T}\left( \theta _{1},Z_{1}\right) }{\check{T}\left(
\theta _{1},Z_{1}\right) }\right)  \notag \\
&&+M_{\omega }^{\left( 2,a\right) }\left( \left( \theta ,Z\right) ,\left(
\theta _{1},Z_{1}\right) \right) \nabla _{\theta }\left( \frac{\delta \omega
^{-1}\left( \theta _{1},Z_{1}\right) }{\omega ^{-1}\left( \theta
_{1},Z_{1}\right) }+\frac{\delta \check{T}\left( \theta _{1},Z_{1}\right) }{%
\check{T}\left( \theta _{1},Z_{1}\right) }\right)  \notag \\
&&+\bar{N}_{\omega }^{\left( 1\right) }\left( \theta ,Z\right) \frac{\delta
\omega ^{-1}\left( \theta ,Z\right) }{\omega ^{-1}\left( \theta ,Z\right) }+%
\bar{N}_{\omega }^{\left( 2\right) }\left( \theta ,Z\right) \nabla _{\theta }%
\frac{\delta \omega ^{-1}\left( \theta ,Z\right) }{\omega ^{-1}\left( \theta
,Z\right) }  \notag \\
&&+M_{\Psi }^{\left( 1,a\right) }\left( \left( \theta ,Z\right) ,\left(
\theta _{1},Z_{1}\right) \right) \frac{\delta \Psi \left( \theta
_{1},Z_{1}\right) }{\Psi \left( \theta _{1},Z_{1}\right) }+M_{\Psi }^{\left(
2,a\right) }\left( \left( \theta ,Z\right) ,\left( \theta _{1},Z_{1}\right)
\right) \nabla _{\theta }\frac{\delta \Psi \left( \theta _{1},Z_{1}\right) }{%
\Psi \left( \theta _{1},Z_{1}\right) }  \notag
\end{eqnarray}%
where the operator:%
\begin{eqnarray*}
&&D_{\theta }=\frac{1}{2}\left( -\frac{\sigma _{\theta }^{2}}{2}\nabla
_{\theta }+\omega ^{-1}\left( J\left( \theta \right) ,\theta ,Z\right)
\right) \nabla _{\theta } \\
&&+\left( A\left( \Psi \left( \theta ,Z\right) \right) -\frac{\sigma
_{\theta }^{2}}{2}\right) \nabla _{\theta }+U^{\prime \prime }\left(
\left\vert \Psi \left( \theta ,Z\right) \right\vert ^{2}\right) \left\vert
\Psi \left( \theta ,Z\right) \right\vert ^{2}
\end{eqnarray*}%
describes the propagation of activities within the collective state defined
by $\Psi \left( \theta ,Z\right) $, $\omega ^{-1}\left( J\left( \theta
\right) ,\theta ,Z\right) $ and the background connectivities. The various
terms in (\ref{Fln}) are derived in Appendix 5 and depend primarily on the
derivatives:%
\begin{equation*}
\prod\limits_{\substack{ i=1  \\ i\neq k}}^{m+1}\int \left\{ \frac{1}{2}%
\frac{\delta ^{\sum_{l}p_{l}^{i}}\left( \nabla _{\theta }\omega ^{-1}\left(
\theta _{i},Z_{i},\left\vert \Psi \right\vert ^{2}\right) \right) }{%
\prod\limits_{l=1}^{j}\delta ^{\sum_{l}p_{l}^{i}}\left\vert \Psi \left(
\theta ^{\left( l\right) },Z_{_{l}}\right) \right\vert ^{2}}\left\vert \Psi
\left( \theta _{i},Z_{i}\right) \right\vert ^{2}\right\}
\end{equation*}%
The fluctuation equation for the activity $\frac{\delta \omega ^{-1}}{\omega
^{-1}}$ is:%
\begin{eqnarray*}
&&\delta \omega ^{-1}\left( \theta ,Z,\left\vert \Psi \right\vert ^{2}\right)
\\
&=&\int \check{T}\left( Z,\theta ,Z_{1},\theta -\frac{\left\vert
Z-Z_{1}\right\vert }{c}\right) \left( \mathcal{\bar{G}}_{0}\left(
0,Z_{1}\right) +\left\vert \Psi \left( \theta -\frac{\left\vert
Z-Z_{1}\right\vert }{c},Z_{1}\right) \right\vert ^{2}\right) \delta \omega
^{-1}\left( \theta -\frac{\left\vert Z-Z_{1}\right\vert }{c},Z_{1},\Psi
\right) dZ_{1} \\
&&-\int dZ_{1}\omega ^{-1}\left( \theta -\frac{\left\vert Z-Z_{1}\right\vert 
}{c},Z_{1},\Psi \right) \frac{\check{T}\left( Z,\theta ,Z_{1},\theta -\frac{%
\left\vert Z-Z_{1}\right\vert }{c}\right) }{1+\frac{\alpha _{D}h_{D}}{\alpha
_{C}h_{C}}\frac{T\left( Z,\theta ,Z_{1},\theta -\frac{\left\vert
Z-Z_{1}\right\vert }{c}\right) }{\lambda \tau \exp \left( -\frac{\left\vert
Z-Z^{\prime }\right\vert }{\nu c}\right) }\frac{\omega ^{-1}\left( \theta
,Z,\left\vert \Psi \right\vert ^{2}\right) }{\omega ^{-1}\left( \theta -%
\frac{\left\vert Z-Z_{1}\right\vert }{c},Z_{1},\Psi \right) }}\delta
\left\vert \Psi \left( \theta -\frac{\left\vert Z-Z_{1}\right\vert }{c}%
,Z_{1}\right) \right\vert ^{2}
\end{eqnarray*}%
and the fluctuations in the connections are given by:%
\begin{eqnarray*}
&&\frac{\delta \check{T}\left( \theta ^{\left( l\right) }-\frac{\left\vert
Z^{\left( 1\right) }-Z^{\left( l\right) }\right\vert }{c},Z^{\left( 1\right)
},Z^{\left( l\right) }\right) }{\check{T}\left( \theta ^{\left( l\right) }-%
\frac{\left\vert Z^{\left( 1\right) }-Z^{\left( 2\right) }\right\vert }{c}%
,Z^{\left( 1\right) },Z^{\left( l\right) }\right) } \\
&=&\frac{\partial _{\frac{\delta \omega _{0}}{\omega _{0}}\left( J,\theta
^{\left( 1\right) }-\frac{\left\vert Z^{\left( 1\right) }-Z^{\left( 2\right)
}\right\vert }{c},Z^{\left( 2\right) }\right) }\check{T}\left( \theta
^{\left( 1\right) }-\frac{\left\vert Z^{\left( 1\right) }-Z^{\left( 2\right)
}\right\vert }{c},Z^{\left( 1\right) },Z^{\left( 2\right) },\omega
_{0}\right) }{\check{T}\left( \theta ^{\left( 1\right) }-\frac{\left\vert
Z^{\left( 1\right) }-Z^{\left( 2\right) }\right\vert }{c},Z^{\left( 1\right)
},Z^{\left( 2\right) },\omega _{0}\right) } \\
&&\times \left( \frac{\delta \omega _{0}\left( J,\theta ^{\left( 1\right) }-%
\frac{\left\vert Z^{\left( 1\right) }-Z^{\left( 2\right) }\right\vert }{c}%
,Z^{\left( 2\right) }\right) }{\omega _{0}\left( J,\theta ^{\left( 1\right)
}-\frac{\left\vert Z^{\left( 1\right) }-Z^{\left( 2\right) }\right\vert }{c}%
,Z^{\left( 2\right) }\right) }-\frac{\delta \omega _{0}\left( J,\theta
^{\left( 1\right) },Z^{\left( 1\right) }\right) }{\omega _{0}\left( J,\theta
^{\left( 1\right) },Z^{\left( 1\right) }\right) }\right) +\frac{\delta 
\check{T}_{i}\left( \theta ^{\left( 1\right) }-\frac{\left\vert Z^{\left(
1\right) }-Z^{\left( 2\right) }\right\vert }{c},Z^{\left( 1\right)
},Z^{\left( 2\right) },\omega _{0}\right) }{\check{T}\left( \theta ^{\left(
1\right) }-\frac{\left\vert Z^{\left( 1\right) }-Z^{\left( 2\right)
}\right\vert }{c},Z^{\left( 1\right) },Z^{\left( 2\right) },\omega
_{0}\right) }
\end{eqnarray*}%
This system of three equations is studied in Appendix 5. We show that, to
the first approximation of local interactions, this system can be reduced to
a differential equation for the field fluctuations $\delta \Psi $. After
change of variable on the field:%
\begin{equation*}
\delta \Psi \left( \theta ,Z\right) \rightarrow \int \exp \left( -\frac{%
\theta -\theta _{0}}{c}\right) \delta \Psi \left( \theta _{0},Z\right)
d\theta _{0}
\end{equation*}%
and considering a background state with internal frequency $\Upsilon $, this
equation writes:%
\begin{eqnarray*}
&&0=\left( 1-\left( \frac{i\Upsilon }{c\eta ^{2}}-\frac{1}{\nu c}\left( 1-%
\frac{1}{\eta ^{2}}\right) \right) ^{2}\nabla _{Z}^{2}\right) \left(
D_{\theta }+\Lambda _{G}\right) \delta \Psi \left( \theta ,Z\right) \\
&&-\Lambda _{H}\frac{\left( \frac{i\Upsilon }{c\eta ^{2}}-\frac{1}{\nu c}%
\left( 1-\frac{1}{\eta ^{2}}\right) \right) ^{2}\nabla _{Z}^{2}}{1+\frac{%
\alpha _{D}h_{D}}{\alpha _{C}h_{C}}\frac{\omega ^{-1}\left( \theta
,Z,\left\vert \Psi \right\vert ^{2}\right) }{\omega ^{-1}\left( \theta -%
\frac{\left\vert Z-Z_{1}\right\vert }{c},Z_{1},\Psi \right) }\frac{\check{T}%
}{\lambda \tau \exp \left( -\frac{\left\vert Z-Z^{\prime }\right\vert }{\nu c%
}\right) }}\delta \Psi \left( \theta ,Z\right)
\end{eqnarray*}%
with:%
\begin{equation*}
\Lambda _{G}=c\frac{\frac{\omega _{0}^{-1}\left( J,\left( \theta ^{\left(
l\right) },Z_{l}\right) \right) }{\mathcal{\bar{G}}_{0}\left( 0,Z\right)
\left( 1+T_{0}\left( \theta ^{\left( l\right) },Z_{l}\right) \right) D}}{%
\left( 1-\frac{\omega _{0}^{-1}\left( J,\left( \theta ^{\left( l\right)
},Z_{l}\right) \right) }{\mathcal{\bar{G}}_{0}\left( 0,Z\right) \left(
1+T_{0}\left( \theta ^{\left( l\right) },Z_{l}\right) \right) D}\right) ^{2}}%
\left\langle \frac{\left\vert \Psi \left( \theta _{i},Z_{i}\right)
\right\vert ^{2}}{G^{i}\left( \left( \theta _{i},Z_{i}\right) \right) }%
\right\rangle \left\langle \left\vert \Psi \left( \theta ^{\left( l\right)
},Z_{l}\right) \right\vert ^{2}\right\rangle
\end{equation*}%
and:%
\begin{equation*}
\Lambda _{H}=-c\frac{\left( \frac{\omega _{0}^{-1}\left( J,\left( \theta
^{\left( l\right) },Z_{l}\right) \right) }{\mathcal{\bar{G}}_{0}\left(
0,Z\right) \left( 1+T_{0}\left( \theta ^{\left( l\right) },Z_{l}\right)
\right) D}\right) ^{2}}{\left( 1-\frac{\omega _{0}^{-1}\left( J,\left(
\theta ^{\left( l\right) },Z_{l}\right) \right) }{\mathcal{\bar{G}}%
_{0}\left( 0,Z\right) \left( 1+T_{0}\left( \theta ^{\left( l\right)
},Z_{l}\right) \right) D}\right) ^{2}}\left\langle \frac{\left\vert \Psi
\left( \theta _{i},Z_{i}\right) \right\vert ^{2}}{G^{i}\left( \left( \theta
_{i},Z_{i}\right) \right) }\right\rangle \left\langle \left\vert \Psi \left(
\theta ^{\left( l\right) },Z_{l}\right) \right\vert ^{2}\right\rangle
\end{equation*}

Projecting this equation onto the space of fluctuations with wave vector $%
k_{Z}$, we show that the condition for the perturbation to be stable can be
written, for $k_{Z}>>1$, as:

\begin{eqnarray*}
k_{Z}^{2} &<&\alpha +\alpha \frac{1-\frac{\omega _{0}^{-1}\left( J,\left(
\theta ^{\left( l\right) },Z_{l}\right) \right) }{\mathcal{\bar{G}}%
_{0}\left( 0,Z\right) \left( 1+T_{0}\left( \theta ^{\left( l\right)
},Z_{l}\right) \right) D}}{\frac{\omega _{0}^{-1}\left( J,\left( \theta
^{\left( l\right) },Z_{l}\right) \right) }{\mathcal{\bar{G}}_{0}\left(
0,Z\right) \left( 1+T_{0}\left( \theta ^{\left( l\right) },Z_{l}\right)
\right) D}\left\langle \frac{\left\vert \Psi \left( \theta _{i},Z_{i}\right)
\right\vert ^{2}}{G^{i}\left( \left( \theta _{i},Z_{i}\right) \right) }%
\right\rangle \left\langle \left\vert \Psi \left( \theta ^{\left( l\right)
},Z_{l}\right) \right\vert ^{2}\right\rangle c} \\
&&\times \left( U^{\prime \prime }\left( \left\vert \Psi \left( \theta
,Z\right) \right\vert ^{2}\right) \left\vert \Psi \left( \theta ,Z\right)
\right\vert ^{2}+\frac{\left( \frac{1}{2}\omega ^{-1}\left( J\left( \theta
\right) ,\theta ,Z\right) +\left( A\left( \Psi \left( \theta ,Z\right)
\right) -\frac{\sigma _{\theta }^{2}}{2}\right) \right) ^{2}}{\left( \sigma
_{\theta }^{2}+4\frac{\Lambda _{G}}{\alpha }\left( 1-\frac{\omega
_{0}^{-1}\left( J,\left( \theta ^{\left( l\right) },Z_{l}\right) \right) }{%
\mathcal{\bar{G}}_{0}\left( 0,Z\right) \left( 1+T_{0}\left( \theta ^{\left(
l\right) },Z_{l}\right) \right) D}\right) \right) }\right)
\end{eqnarray*}

Thus, any collective state becomes unstable for perturbations with a squared
wave vector amplitude $k_{Z}^{2}$ sufficiently large, i.e. for highly
fluctuating perturbations. However, the higher the average level of
activity, the lower $\omega _{0}^{-1}\left( J,\left( \theta ^{\left(
l\right) },Z_{l}\right) \right) $, and the larger $k_{Z}^{2}$ must be to
destabilize the collective state.

On the contrary, the higher $\left\langle \frac{\left\vert \Psi \left(
\theta _{i},Z_{i}\right) \right\vert ^{2}}{G^{i}\left( \left( \theta
_{i},Z_{i}\right) \right) }\right\rangle $, the smaller $k_{Z}^{2}$ must be
to induce destibilization. This condition implies that the higher the number
of activated cells, the further the perturbation will propagate throughout
the system.

\subsection{Towards field model for collective states}

The derivation of collective states in connectivities, along with the
dynamical description of their activities, arose from the minimization of $%
S\left( \Psi \right) +S\left( \Gamma \right) $ in the system as a whole. It
showed the possibility of quasi-stable clusters, especially when these
clusters are induced by external signals binding their elements. These
clusters are described by their internal parameters, average level of
connections and activations, and characteristic frequencies of signals
throughout the structure. However, this description remains partly exogenous
and static. It thus led us to consider an extended point of view. Since the
system is described by these composite states, it is appropriate to adopt an
effective field theory for collective states.

The fundamental entities then become the states of any possible group of
cells, represented as large tensor products of individual states for the
connectivity field, along with the induced activities characterized by their
possible frequencies. Such collective states should thus be labeled not only
by their connectivities but also by their possible averages and frequencies.

\subsection{Effective action for $\Gamma $ above background field}

The possibility of connected groups is then examined within a formalism of
effective action as an intrinsic property of connectivities. When a static
background field is selected, we can consider the effective action for the
connectivities above this given background to inspect their dynamical
aspects. For later purposes, we write this effective action here:%
\begin{eqnarray}
&&\hat{S}\left( \Delta \Gamma \left( T,\hat{T},\theta ,Z,Z^{\prime }\right)
\right) \\
&=&-\Delta \Gamma ^{\dag }\left( T,\hat{T},\theta ,Z,Z^{\prime }\right)
\left( \nabla _{T}\left( \nabla _{T}+\frac{\left( \Delta T-\Delta
\left\langle T\right\rangle \right) -\lambda \left( \Delta \hat{T}-\Delta
\left\langle \hat{T}\right\rangle \right) }{\tau \omega _{0}\left( Z\right) }%
\left\vert \Psi \left( \theta ,Z\right) \right\vert ^{2}\right) \right)
\Delta \Gamma \left( T,\hat{T},\theta ,Z,Z^{\prime }\right)  \notag \\
&&-\Delta \Gamma ^{\dag }\left( T,\hat{T},\theta ,Z,Z^{\prime }\right)
\nabla _{\hat{T}}\left( \nabla _{\hat{T}}+\left\vert \bar{\Psi}_{0}\left(
Z,Z^{\prime }\right) \right\vert ^{2}\left( \Delta \hat{T}-\left( \Delta
\left\langle \hat{T}\right\rangle \right) \right) \right) \Delta \Gamma
\left( T,\hat{T},\theta ,Z,Z^{\prime }\right)  \notag \\
&&+U_{\Delta \Gamma }\left( \left\Vert \Delta \Gamma \left( Z,Z^{\prime
}\right) \right\Vert ^{2}\right)  \notag
\end{eqnarray}%
This effective action allows us to derive a model for activated connections
above the background field and to study the possibility of emerging groups
of connected states activated in groups.

Some details are given in Appendix 6. The field $\Delta \Gamma \left( T,\hat{%
T},\theta ,Z,Z^{\prime }\right) $ represents the modification of the field
above the background, $\Delta \left\langle T\right\rangle $ and $\Delta
\left\langle \hat{T}\right\rangle $ are the average for these connectivities
when they are activated. The conditions for this activation are given in
appendix 6, along with the form of the associated state.

An activated state consists at first in a collection of values:%
\begin{equation*}
\left( \left\{ \Delta \hat{T},\Delta T,Z,Z^{\prime }\right\} ,\Delta
\left\langle \hat{T}\right\rangle ,\Delta \left\langle T\right\rangle \right)
\end{equation*}%
describing the values of connections between the locations $\left(
Z,Z^{\prime }\right) $ within the spatial extension of the state. The value
of $\Delta \left\langle \hat{T}\right\rangle $ describes the average level
of connections within the group and $\omega _{0}\left( Z\right) $ is the
static value of activity at location $Z$.

The activated state can be written:%
\begin{equation}
\Delta \Gamma =\prod\limits_{Z,Z^{\prime }}\left\vert \Delta T\left(
Z,Z^{\prime }\right) ,\Delta \hat{T}\left( Z,Z^{\prime }\right)
\right\rangle \equiv \left\vert \Delta \mathbf{T,}\Delta \mathbf{\hat{T}}%
\right\rangle
\end{equation}%
However, as shown in the previous section, such a state should also be
described by the level of cell's activity and the frequencies of their
fluctuations. We should thus describe a state not only by the values of
connections but also by the indicices refering to the possible frequencies
and averages.

We define:

\begin{eqnarray*}
\left[ \mathbf{\Delta T}\right] &=&\left( \Delta \mathbf{T,}\Delta \mathbf{%
\hat{T}}\right) \\
\left\langle \left[ \mathbf{\Delta T}\right] \right\rangle _{p}^{\alpha }
&=&\left( \left\langle \Delta \mathbf{T}\right\rangle _{p}^{\alpha
},\left\langle \Delta \mathbf{\hat{T}}\right\rangle _{p}^{\alpha }\right)
\end{eqnarray*}%
with $\left( \left\langle \Delta \mathbf{T}\right\rangle _{p}^{\alpha
},\left\langle \Delta \mathbf{\hat{T}}\right\rangle _{p}^{\alpha }\right) $
the set of averages connectvities and variation of connectivities in the
group. The index $\alpha $ refers to multiple possibilities for the state.
For each possibility a sequence of frequencies $\left( \Upsilon _{p}^{\alpha
}\right) $ satisfying:%
\begin{equation}
\det \left( 1-\Delta T\left( Z_{i},Z_{j}\right) \exp \left( -i\Upsilon _{p}%
\frac{\left\vert Z_{i}-Z_{j}\right\vert }{c}\right) \right) =0
\end{equation}%
are compatible. The activities are given by a static part $\overline{\Delta
\omega }\left( Z,\mathbf{T},\left\vert \Psi \right\vert ^{2}\right) $ and
variable part:%
\begin{equation*}
\Delta \omega _{p}^{\alpha }\left( \theta ,Z,\mathbf{\Delta T}\right) =%
\overline{\Delta \omega }\left( Z,\mathbf{T},\left\vert \Psi \right\vert
^{2}\right) +\left( \mathbf{N}_{p}^{\alpha }\right) ^{-1}\Delta \mathbf{%
\omega }_{0}\exp \left( -i\Upsilon _{p}^{\alpha }\frac{\left\vert \Delta 
\mathbf{Z}_{i}\right\vert }{c}\right)
\end{equation*}%
where $\left[ \mathbf{N}_{p}^{\alpha }\right] _{\left( Z_{i},Z_{j}\right) }$
is a matrix depending on the frequencies detailed in Appendix 6. The
activated state can be written:%
\begin{equation}
\Delta \Gamma =\prod\limits_{Z,Z^{\prime }}\left\vert \Delta T\left(
Z,Z^{\prime }\right) ,\Delta \hat{T}\left( Z,Z^{\prime }\right) ,\alpha
\left( Z,Z^{\prime }\right) ,p\left( Z,Z^{\prime }\right) \right\rangle
\equiv \left\vert \mathbf{\alpha },\mathbf{p},S^{2}\right\rangle  \label{st}
\end{equation}%
with detailed formulas in the same Appendix.

Such a group, as a whole, has effective action: 
\begin{eqnarray}
&&\hat{S}\left( \Delta \Gamma \left( T,\hat{T},\theta ,Z,Z^{\prime }\right)
\right) \\
&=&-\Delta \Gamma ^{\dag }\left( T,\hat{T},\theta ,Z,Z^{\prime }\right)
\left( \nabla _{T}^{2}+\nabla _{\hat{T}}^{2}-\frac{1}{2}\left( \left[ 
\mathbf{\Delta T}\right] \mathbf{-}\left\langle \left[ \mathbf{\Delta T}%
\right] \right\rangle _{p}^{\alpha }\right) ^{t}\mathbf{A}_{p}^{\alpha
}\left( \left[ \mathbf{\Delta T}\right] \mathbf{-}\left\langle \left[ 
\mathbf{\Delta T}\right] \right\rangle _{p}^{\alpha }\right) \right) \Delta
\Gamma \left( T,\hat{T},\theta ,Z,Z^{\prime }\right)  \notag \\
&&+C\left( Z,Z^{\prime }\right) \left\Vert \Delta \Gamma \left( T,\hat{T}%
,\theta ,Z,Z^{\prime }\right) \right\Vert ^{2}  \notag
\end{eqnarray}%
However, to describe full the possibilities of dynamical entities, we have
to go one step further and consider a field model in which the argument of
the fields are the whole states (\ref{st}) and the field action functional
describes the interactions between such global states.

\section{Field theory for collective states}

In \cite{GLw}, we introduced the field formalism for collective states,
based on the initial developments of the model. The advantage of adopting
this approach is that it simplifies the analysis of transitions between such
states. Considering the desciption of states in the previous section, the
field formalism allows to consider the activation or desactivation of states
through interaction potentials. This makes the transitions mechanisms, such
as the fission or the merger of several states, more tractable.

This section presents the main elements of the model.

\subsection{Action functional for states}

Starting with states (\ref{St}) $\left\vert \mathbf{\alpha },\mathbf{p}%
,S^{2}\right\rangle $ and their conjugates $\left\langle \mathbf{\alpha },%
\mathbf{p},S^{2}\right\vert $, we can build a field-theoretic description.
First, we rewrite the action functional for this state: we consider the same
action as (\ref{TRM}), but summing only over points $\left( Z,Z^{\prime
}\right) \in S^{2}$ belonging to the collective state with spatial extension 
$S^{2}$.

When considering collective states directly on their own, we saw that the
corresponding action for such state is:

\begin{eqnarray}
&&\hat{S}\left( \left\vert \mathbf{\alpha },\mathbf{p},S^{2}\right\rangle
\right)  \label{Cn} \\
&=&\left\langle \mathbf{\alpha },\mathbf{p},S^{2}\right\vert \left( -\nabla
_{T}^{2}-\nabla _{\hat{T}}^{2}+\frac{1}{2}\left( \left[ \mathbf{\Delta T}%
\right] \mathbf{-}\left\langle \left[ \mathbf{\Delta T}\right] \right\rangle
_{p}^{\alpha }\right) ^{t}\mathbf{A}_{p}^{\alpha }\left( \left[ \mathbf{%
\Delta T}\right] \mathbf{-}\left\langle \left[ \mathbf{\Delta T}\right]
\right\rangle _{p}^{\alpha }\right) +C\right) \left\vert \mathbf{\alpha },%
\mathbf{p},S^{2}\right\rangle  \notag
\end{eqnarray}%
with:%
\begin{equation}
\left\vert \mathbf{\alpha },\mathbf{p},S^{2}\right\rangle
=\prod\limits_{Z,Z^{\prime }}\left\vert \Delta T\left( Z,Z^{\prime }\right)
,\Delta \hat{T}\left( Z,Z^{\prime }\right) ,\alpha \left( Z,Z^{\prime
}\right) ,p\left( Z,Z^{\prime }\right) \right\rangle  \label{snt}
\end{equation}%
and:%
\begin{equation*}
\left\langle \mathbf{\alpha },\mathbf{p},S^{2}\right\vert =\left\vert 
\mathbf{\alpha },\mathbf{p},S^{2}\right\rangle ^{\dag }
\end{equation*}%
In (\ref{Cn}), we have defined:%
\begin{equation}
\left[ \mathbf{\Delta T}\right] \mathbf{-}\left\langle \left[ \mathbf{\Delta
T}\right] \right\rangle _{p}^{\alpha }=\left( 
\begin{array}{c}
\mathbf{\Delta T}-\left\langle \mathbf{\Delta T}\right\rangle _{p}^{\alpha }
\\ 
\mathbf{\Delta \hat{T}}-\left\langle \mathbf{\Delta \hat{T}}\right\rangle
_{p}^{\alpha }%
\end{array}%
\right)
\end{equation}%
and the matrix $\mathbf{A}_{p}^{\alpha }$ is defined by (an account of the
derivation is given in Appendix 6):%
\begin{equation*}
\mathbf{A}_{p}^{\alpha }=\left( 
\begin{array}{cc}
\left( \frac{1}{\tau \omega _{0}\left( Z\right) }\right) ^{2}+\left( \mathbf{%
M}_{1}^{\alpha }\left( Z,Z^{\prime }\right) \right) ^{2} & -\lambda \left( 
\frac{1}{\tau \omega _{0}\left( Z\right) }\right) ^{2}+D\left( Z,Z^{\prime
}\right) \mathbf{M}_{2}^{\alpha }\left( Z,Z^{\prime }\right) \\ 
-\lambda \left( \frac{1}{\tau \omega _{0}\left( Z\right) }\right)
^{2}+D\left( Z,Z^{\prime }\right) \mathbf{M}_{1}^{\alpha }\left( Z,Z^{\prime
}\right) & \left( \frac{\lambda }{\tau \omega _{0}\left( Z\right) }\right)
^{2}+D^{2}\left( Z,Z^{\prime }\right) +\left( \mathbf{M}_{2}^{\alpha }\left(
Z,Z^{\prime }\right) \right) ^{2}%
\end{array}%
\right)
\end{equation*}%
with:%
\begin{equation*}
D\left( Z,Z^{\prime }\right) =\frac{\rho \left( C\left( \theta \right)
\left\vert \Psi _{0}\left( Z\right) \right\vert ^{2}\omega _{0}\left(
Z\right) +D\left( \theta \right) \hat{T}\left\vert \Psi _{0}\left( Z^{\prime
}\right) \right\vert ^{2}\omega _{0}\left( Z^{\prime }\right) \right) }{%
\omega _{0}\left( Z\right) }
\end{equation*}%
and $\mathbf{M}^{\alpha }\left( Z,Z^{\prime }\right) $ is a two components
vector:%
\begin{equation*}
\mathbf{M}^{\alpha }\left( Z,Z^{\prime }\right) =\frac{\rho }{\omega
_{0}\left( Z\right) }\left( D\left( \theta \right) \left\langle \hat{T}%
\right\rangle \left\vert \Psi _{0}\left( Z^{\prime }\right) \right\vert
^{2}A\left\vert Z-Z^{\prime }\right\vert \left( \nabla _{\mathbf{\Delta T}%
_{\left( Z_{1},Z_{1}^{\prime }\right) }}\left( \Delta \omega \left(
Z,\left\langle \mathbf{\Delta T}\right\rangle \right) \right) _{\left(
\left\langle \Delta \mathbf{T}_{\left( Z_{1},Z_{1}^{\prime }\right)
}\right\rangle ^{\alpha }\right) }\right) \right)
\end{equation*}%
The constant $C$ arising in (\ref{Cn}) is given by:%
\begin{equation*}
C=\sum_{\left( Z,Z^{\prime }\right) }\mathbf{C}\left( Z,Z^{\prime }\right)
\end{equation*}%
where $\mathbf{C}\left( Z,Z^{\prime }\right) $ is defined by:%
\begin{equation}
\mathbf{C}\left( Z,Z^{\prime }\right) =\frac{\tau \omega _{0}\left( Z\right) 
}{2}+\frac{\rho \left( C\left( \theta \right) \left\vert \Psi _{0}\left(
Z\right) \right\vert ^{2}\omega _{0}\left( Z\right) +D\left( \theta \right)
\left\vert \Psi _{0}\left( Z^{\prime }\right) \right\vert ^{2}\omega
_{0}\left( Z^{\prime }\right) \right) }{2\omega _{0}\left( Z\right) }
\label{LC}
\end{equation}%
Ultimately, given our hypothesis that:%
\begin{equation*}
\left\Vert \Delta \mathbf{\hat{T}}-\left\langle \Delta \mathbf{\hat{T}}%
\right\rangle \right\Vert <<\left\Vert \Delta \mathbf{T}-\left\langle \Delta 
\mathbf{T}\right\rangle ^{\alpha }\right\Vert
\end{equation*}%
the field action simplifies as: 
\begin{eqnarray}
&&\hat{S}\left( \left\vert \mathbf{\alpha },\mathbf{p},S^{2}\right\rangle
\right)  \label{DFc} \\
&=&\left\langle \mathbf{\alpha },\mathbf{p},S^{2}\right\vert \left( -\nabla
_{T}^{2}+\frac{1}{2}\left( \left[ \mathbf{\Delta T}\right] \mathbf{-}%
\left\langle \left[ \mathbf{\Delta T}\right] \right\rangle _{p}^{\alpha
}\right) ^{t}\mathbf{A}_{p}^{\alpha }\left( \left[ \mathbf{\Delta T}\right] 
\mathbf{-}\left\langle \left[ \mathbf{\Delta T}\right] \right\rangle
_{p}^{\alpha }\right) +C\right) \left\vert \mathbf{\alpha },\mathbf{p}%
,S^{2}\right\rangle +U\left( \left\vert \mathbf{\alpha },\mathbf{p}%
,S^{2}\right\rangle \right)  \notag
\end{eqnarray}%
where the matrix $\mathbf{A}^{\alpha }$ is defined by:%
\begin{equation*}
\mathbf{A}^{\alpha }=\sqrt{\mathbf{D}^{2}+\left( \mathbf{M}^{\alpha }\right)
^{t}\mathbf{M}^{\alpha }}
\end{equation*}%
and the matrix $\mathbf{D}$ is diagonal with elements $D\left( Z,Z^{\prime
}\right) $.

The state (\ref{snt}) $\left\vert \mathbf{\alpha },\mathbf{p}%
,S^{2}\right\rangle $ minimizing (\ref{Cn})\ is given by formula (See
Appendix 6):%
\begin{eqnarray}
&&\left\vert \mathbf{\alpha },\mathbf{p},S^{2}\right\rangle  \label{Stn} \\
&=&\exp \left( -\frac{1}{2}\left( \left[ \mathbf{\Delta T}\right] \mathbf{-}%
\left\langle \left[ \mathbf{\Delta T}\right] \right\rangle _{p}^{\alpha
}\right) ^{t}\mathbf{A}_{p}^{\alpha }\left( \left[ \mathbf{\Delta T}\right] 
\mathbf{-}\left\langle \left[ \mathbf{\Delta T}\right] \right\rangle
_{p}^{\alpha }\right) \right) H_{p}\left( \frac{1}{2}\left( \left[ \mathbf{%
\Delta T}\right] \mathbf{-}\left\langle \left[ \mathbf{\Delta T}\right]
\right\rangle _{p}^{\alpha }\right) ^{t}\mathbf{A}_{p}^{\alpha }\left( \left[
\mathbf{\Delta T}\right] \mathbf{-}\left\langle \left[ \mathbf{\Delta T}%
\right] \right\rangle _{p}^{\alpha }\right) \right)  \notag \\
&&\times H_{p}\left( \left( \left[ \mathbf{\Delta T}^{\prime }\right] 
\mathbf{-}\left\langle \left[ \mathbf{\Delta T}\right] \right\rangle
_{p}^{\alpha }\right) _{1}^{t}\left( \mathbf{D}_{p}^{\alpha }\right)
_{1}\left( \left[ \mathbf{\Delta T}^{\prime }\right] \mathbf{-}\left\langle %
\left[ \mathbf{\Delta T}\right] \right\rangle _{p}^{\alpha }\right)
_{1}^{t}\right)  \notag \\
&&H_{p-\delta }\left( \left( \left[ \mathbf{\Delta T}^{\prime }\right] 
\mathbf{-}\left\langle \left[ \mathbf{\Delta T}\right] \right\rangle
_{p}^{\alpha }\right) _{2}^{t}\left( \mathbf{D}_{p}^{\alpha }\right)
_{2}\left( \left[ \mathbf{\Delta T}^{\prime }\right] \mathbf{-}\left\langle %
\left[ \mathbf{\Delta T}\right] \right\rangle _{p}^{\alpha }\right)
_{2}^{t}\right)  \notag
\end{eqnarray}%
where the functions $H_{p}$ are the Hermite polynomials.

\subsection{States with given connectivities}

Rather than the states $\left\vert \mathbf{\alpha },\mathbf{p}%
,S^{2}\right\rangle $, we can consider states:%
\begin{equation}
\left\vert \Delta \mathbf{T},\Delta \mathbf{\hat{T},\alpha },\mathbf{p}%
,S^{2}\right\rangle  \label{gns}
\end{equation}%
that have a given values of $\Delta \mathbf{T}$ and $\Delta \mathbf{\hat{T}}$
of connectivity. They are found by diagonalizing the operator arising in (%
\ref{DFc}) (assuming that $\nabla _{T}^{2}$ is preserved by the
diagonalization):%
\begin{eqnarray}
&&\left( -\nabla _{T}^{2}+\frac{1}{2}\left( \left[ \mathbf{\Delta T}\right] 
\mathbf{-}\left\langle \left[ \mathbf{\Delta T}\right] \right\rangle
_{p}^{\alpha }\right) ^{t}\mathbf{A}_{p}^{\alpha }\left( \left[ \mathbf{%
\Delta T}\right] \mathbf{-}\left\langle \left[ \mathbf{\Delta T}\right]
\right\rangle _{p}^{\alpha }\right) +\mathbf{C}\right)  \label{PRc} \\
&\rightarrow &\left( -\nabla _{\bar{T}}^{2}+\frac{1}{2}\left( \left[ \mathbf{%
\Delta \bar{T}}\right] \mathbf{-}\left\langle \left[ \mathbf{\Delta \bar{T}}%
\right] \right\rangle _{p}^{\alpha }\right) ^{t}\mathbf{D}_{p}^{\alpha
}\left( \left[ \mathbf{\Delta \bar{T}}\right] \mathbf{-}\left\langle \left[ 
\mathbf{\Delta \bar{T}}\right] \right\rangle _{p}^{\alpha }\right) +\mathbf{C%
}\right)  \notag \\
&=&\frac{1}{2}\mathbf{D}_{p}^{\alpha }\mathbf{A}^{-}\left( \alpha
,p,S^{2}\right) \mathbf{A}^{+}\left( \alpha ,p,S^{2}\right) +\mathbf{C} 
\notag
\end{eqnarray}%
where we defined the operator:%
\begin{eqnarray}
\mathbf{A}_{D}^{-}\left( \alpha ,p,S^{2}\right) &=&\frac{1}{2}\left( \sqrt{%
\mathbf{\bar{D}}_{S^{2}}^{\alpha }}\Delta \mathbf{\bar{T}}-\frac{1}{\sqrt{%
\mathbf{\bar{D}}_{S^{2}}^{\alpha }}}\nabla _{\left( \left[ \mathbf{\bar{T}}%
\right] \right) _{S^{2}}}^{2}\right)  \label{CRN} \\
\mathbf{A}_{D}^{+}\left( \alpha ,p,S^{2}\right) &=&\frac{1}{2}\left( \sqrt{%
\mathbf{\bar{D}}_{S^{2}}^{\alpha }}\Delta \left[ \mathbf{\bar{T}}\right] +%
\frac{1}{\sqrt{\mathbf{\bar{D}}_{S^{2}}^{\alpha }}}\nabla _{\left( \left[ 
\mathbf{\bar{T}}\right] \right) _{S^{2}}}^{2}\right)  \notag
\end{eqnarray}%
where $\Delta \mathbf{\bar{T}}_{p}^{\alpha }$ is seen as an operator of
multiplicatn b $\Delta \mathbf{\bar{T}}_{p}^{\alpha }$. Then, we can rewrite:%
\begin{equation*}
\Delta \mathbf{\bar{T}}_{p}^{\alpha }=\left( \sqrt{\mathbf{\bar{D}}%
_{S^{2}}^{\alpha }}\right) ^{-1}\left( \mathbf{A}_{D}^{-}\left( \alpha
,p,S^{2}\right) +\mathbf{A}_{D}^{+}\left( \alpha ,p,S^{2}\right) \right)
\end{equation*}%
and:%
\begin{equation*}
\nabla _{\left( \mathbf{\bar{T}}\right) _{S^{2}}}^{2}=\left( \sqrt{\mathbf{%
\bar{D}}_{S^{2}}^{\alpha }}\right) \left( \mathbf{A}_{D}^{+}\left( \alpha
,p,S^{2}\right) -\mathbf{A}_{D}^{-}\left( \alpha ,p,S^{2}\right) \right)
\end{equation*}%
The operators $\mathbf{A}_{D}^{\pm }\left( \alpha ,p,S^{2}\right) $\ satisfy
the following commutation relation:%
\begin{equation*}
\left[ \left( \mathbf{A}_{D}^{-}\right) _{i}\left( \alpha ,p,S^{2}\right)
,\left( \mathbf{A}_{D}^{+}\right) _{j}\left( \alpha ^{\prime },p^{\prime
},S^{\prime 2}\right) \right] =\delta \left( \underline{S^{2}}-S^{\prime
2}\right) \delta \left( \left( \alpha ,p\right) -\left( \alpha ,p\right)
^{\prime }\right) \delta _{i,j}
\end{equation*}%
where $\delta $ is the Dirac function.

Coming back to the initial variables, this defines operators $\mathbf{A}%
^{-}\left( \alpha ,p,S^{2}\right) $ and $\mathbf{A}^{+}\left( \alpha
,p,S^{2}\right) $ and:%
\begin{equation}
\Delta \mathbf{T}_{p}^{\alpha }=\left( \sqrt{\mathbf{A}_{p}^{\alpha }}%
\right) ^{-1}\left( \mathbf{A}^{-}\left( \alpha ,p,S^{2}\right) +\mathbf{A}%
^{+}\left( \alpha ,p,S^{2}\right) \right)  \label{GNv}
\end{equation}%
and:%
\begin{equation}
\nabla _{\left( \mathbf{\bar{T}}\right) _{S^{2}}}^{2}=\left( \sqrt{\mathbf{A}%
_{p}^{\alpha }}\right) \left( \mathbf{A}^{+}\left( \alpha ,p,S^{2}\right) -%
\mathbf{A}^{-}\left( \alpha ,p,S^{2}\right) \right)  \label{Gnvt}
\end{equation}%
The states (\ref{gns}) $\left\vert \Delta \mathbf{T},\Delta \mathbf{\hat{T}%
,\alpha },\mathbf{p},S^{2}\right\rangle $ are the eigenstates of operators $%
\Delta \mathbf{T}$ and $\Delta \mathbf{\hat{T}}$ with given eigenvalues $%
\Delta \mathbf{T}$ and $\Delta \mathbf{\hat{T}}$ of connectivity. They are
not solutions of the saddle-point equations, which present possible
fluctuations in $\Delta \mathbf{T}$ and $\Delta \mathbf{\hat{T}}$ around
their averages. Instead, they represent statistical combinations of
generalized states defined by $\mathbf{\alpha },\mathbf{p}$. The states $%
\left\vert \Delta \mathbf{T},\Delta \mathbf{\hat{T},\alpha },\mathbf{p}%
,S^{2}\right\rangle $ are obtained by series of successive actions of the
components $\mathbf{A}_{i}^{+}\left( \alpha ,p,S^{2}\right) $ of $\mathbf{A}%
_{i}^{+}\left( \alpha ,p,S^{2}\right) $ on the minimum $\left\vert
vac\right\rangle $

The connection between $\left\vert \Delta \mathbf{T},\Delta \mathbf{\hat{T}%
,\alpha },\mathbf{p},S^{2}\right\rangle $ and the saddle-point solutions is
as follows. Using (\ref{GNv}), the states $\left\vert \Delta \mathbf{T}%
,\Delta \mathbf{\hat{T},\alpha },\mathbf{p},S^{2}\right\rangle $ satisfy the
eigenvalue equation:%
\begin{equation*}
\left( \sqrt{\mathbf{A}_{p}^{\alpha }}\right) ^{-1}\left( \mathbf{A}%
^{-}\left( \alpha ,p,S^{2}\right) +\mathbf{A}^{+}\left( \alpha
,p,S^{2}\right) \right) \left\vert \Delta \mathbf{T},\Delta \mathbf{\hat{T}%
,\alpha },\mathbf{p},S^{2}\right\rangle =\left[ \Delta \mathbf{T}%
_{p}^{\alpha }\right] \left\vert \Delta \mathbf{T},\Delta \mathbf{\hat{T}%
,\alpha },\mathbf{p},S^{2}\right\rangle
\end{equation*}%
whose solution writes:

\begin{eqnarray}
&&\left\vert \Delta \mathbf{T}_{p}^{\alpha },\mathbf{\alpha },\mathbf{p}%
,S^{2}\right\rangle  \label{LCT} \\
&=&\exp \left( -\left( \left( \Delta \mathbf{T}_{p}^{\alpha }\right) ^{t}%
\mathbf{A}_{p}^{\alpha }\Delta \mathbf{T}_{p}^{\alpha }+2\left( \Delta 
\mathbf{T}_{p}^{\alpha }\right) ^{t}\sqrt{\mathbf{A}_{p}^{\alpha }}\mathbf{%
\hat{A}}^{+}\left( \mathbf{\alpha },\mathbf{p},S^{2}\right) \right) +\frac{1%
}{2}\mathbf{\hat{A}}^{+}\left( \mathbf{\alpha },\mathbf{p},S^{2}\right) .%
\mathbf{\hat{A}}^{+}\left( \mathbf{\alpha },\mathbf{p},S^{2}\right) \right)
\left\vert Vac\right\rangle  \notag
\end{eqnarray}%
This states are mixed states. Their given values of connections hides some
unstability that leads to transitions towards other states.

Formula (\ref{LCT}) indicates that we should generally consider mixed
states, so that instability should be accounted for by allowing for
transitions between states. This possibility can be described within a field
theoretic framework for collectiv states.

\subsection{Field action functional}

Before developing a field formalism for collective states we simplify the
notations. First, we replace the notations for extra connections by removing
the relative notation induced by $\Delta $:%
\begin{equation*}
\Delta \mathbf{T\rightarrow T},\Delta \mathbf{\hat{T}\rightarrow \hat{T}}
\end{equation*}%
so that the extra connections states are simply considered as connections.
This implies that we replace:%
\begin{eqnarray*}
\mathbf{\Delta T}-\left\langle \mathbf{\Delta T}\right\rangle _{p}^{\alpha }
&\rightarrow &\mathbf{T}-\left\langle \mathbf{T}\right\rangle _{p}^{\alpha }
\\
\mathbf{\Delta \hat{T}}-\left\langle \mathbf{\Delta \hat{T}}\right\rangle
_{p}^{\alpha } &\rightarrow &\mathbf{\hat{T}}-\left\langle \mathbf{\hat{T}}%
\right\rangle _{p}^{\alpha }
\end{eqnarray*}%
Ultimately we gather both variables in one single variable:%
\begin{equation*}
\left[ \Delta \mathbf{T}\right] =\left( \Delta \mathbf{T,}\Delta \mathbf{%
\hat{T}}\right) \rightarrow \mathbf{T}
\end{equation*}

In \cite{GLw}, we extended the previous action for general states defined by
functions $\underline{\gamma }\left( \mathbf{\bar{T},\alpha },\mathbf{p}%
,S^{2},\theta \right) $. This amounts to consider mixed states:%
\begin{equation*}
\sum_{\left\{ \mathbf{\alpha },\mathbf{p},S^{2}\right\} }\underline{\gamma }%
\left( \mathbf{\bar{T},\alpha },\mathbf{p},S^{2},\theta \right) \left\vert 
\mathbf{\alpha },\mathbf{p},S^{2}\right\rangle
\end{equation*}%
or in a continuous case:%
\begin{equation*}
\int \underline{\gamma }\left( \mathbf{\bar{T},\alpha },\mathbf{p}%
,S^{2},\theta \right) \left\vert \mathbf{\alpha },\mathbf{p}%
,S^{2}\right\rangle
\end{equation*}%
In our context it amounts to describe statistical combinations of states $%
\left\vert \mathbf{\alpha },\mathbf{p},S^{2}\right\rangle $ corresponding to
assume that each collective state may be activated

We now replace $\gamma \left( \mathbf{\bar{T},\alpha },\mathbf{p}%
,S^{2},\theta \right) $ by a field $\underline{\Gamma }\left( \mathbf{%
T,\alpha },\mathbf{p},S^{2},\theta \right) $ which is a field whose $%
\underline{\gamma }\left( \mathbf{\bar{T},\alpha },\mathbf{p},S^{2},\theta
\right) $ are the realizations. To describe the system as a whole we
consider that set of multi-component field $\left\{ \underline{\Gamma }%
\left( \mathbf{T,\alpha },\mathbf{p},S^{2},\theta \right) \right\} $.
Including the constants $\mathbf{C}$ defined in (\ref{LC})\ and the
potential $U$ we are led to the action functional: 
\begin{eqnarray}
&&S\left( \left\{ \underline{\Gamma }\left( \mathbf{T},\mathbf{\alpha },%
\mathbf{p},S^{2},\theta \right) \right\} \right)  \label{FCz} \\
&=&\sum_{\left\{ \mathbf{\alpha },\mathbf{p},S^{2}\right\} }\underline{%
\Gamma }^{\dag }\left( \mathbf{T},\mathbf{\alpha },\mathbf{p},S^{2},\theta
\right) \left( -\nabla _{\Delta \mathbf{T}}^{2}+\frac{1}{2}\left( \Delta 
\mathbf{T}_{p}^{\alpha }\right) _{S^{2}}^{t}\mathbf{A}_{S^{2}}^{\alpha
}\left( \Delta \mathbf{T}_{p}^{\alpha }\right) +\mathbf{C}\right) \underline{%
\Gamma }\left( \Delta \mathbf{T},\mathbf{\alpha },\mathbf{p},S^{2},\theta
\right)  \notag \\
&&+U\left( \left\Vert \underline{\Gamma }\left( \left( \Delta \mathbf{T}%
_{p}^{\alpha }\right) _{S^{2}},\mathbf{\alpha },\mathbf{p},S^{2},\theta
\right) \right\Vert ^{2}\right)  \notag
\end{eqnarray}%
where:%
\begin{equation*}
\left( \Delta \mathbf{T}_{p}^{\alpha }\right) _{S^{2}}=\mathbf{T}%
_{S^{2}}-\left\langle \mathbf{T}_{p}^{\alpha }\right\rangle _{S^{2}}
\end{equation*}%
\begin{equation*}
\mathbf{C}=\sum_{\left( Z,Z^{\prime }\right) \in S\times S}C\left(
Z,Z^{\prime }\right)
\end{equation*}%
This action allows to extend (\ref{Cn}) to products of states representing
multiple states activation of the same group.

Note that $S\left( \underline{\Gamma }\left( \mathbf{\bar{T},\alpha },%
\mathbf{p},S^{2},\theta \right) \right) $ is similar to $S\left( \underline{%
\gamma }\left( \mathbf{\bar{T},\alpha },\mathbf{p},S^{2},\theta \right)
\right) $ but with the difference that $\underline{\Gamma }\left( \mathbf{%
\bar{T},\alpha },\mathbf{p},S^{2},\theta \right) $ is a random variable with
realizations $\underline{\gamma }\left( \mathbf{\bar{T},\alpha },\mathbf{p}%
,S^{2},\theta \right) $. This extension allows to consider activation and
desactivation of stats.

The dynamics of the system will thus be encompassed in the partition
function:%
\begin{equation}
\int \exp \left( -S\left( \underline{\Gamma }\left( \mathbf{\bar{T},\alpha },%
\mathbf{p},S^{2},\theta \right) \right) \right) D\left\{ \underline{\Gamma }%
\left( \mathbf{T},\mathbf{\alpha },\mathbf{p},S^{2},\theta \right) \right\}
\label{Pf}
\end{equation}%
Moreover, since $S\left( \underline{\Gamma }\left( \mathbf{\bar{T},\alpha },%
\mathbf{p},S^{2},\theta \right) \right) $ describes the whole set of
possible states, it depends on the collection $\left\{ \underline{\Gamma }%
\left( \mathbf{T},\mathbf{\alpha },\mathbf{p},S^{2},\theta \right) \right\} $%
. This will allow to include interactions between collective states in (\ref%
{Pf}).

Note also that the action $S\left( \left\{ \underline{\Gamma }\left( \mathbf{%
T},\mathbf{\alpha },\mathbf{p},S^{2},\theta \right) \right\} \right) $ is
similar to $\hat{S}\left( \Delta \Gamma \left( T,\hat{T},\theta ,Z,Z^{\prime
}\right) \right) $ where the dynamics for $\hat{T}$ has been neglected but
with the replacement:%
\begin{equation*}
\Delta \Gamma \left( T,\hat{T},\theta ,Z,Z^{\prime }\right) \rightarrow 
\underline{\Gamma }\left( \mathbf{T},\mathbf{\alpha },\mathbf{p}%
,S^{2},\theta \right)
\end{equation*}%
where $S$ is the spatial extension of the collective state. This replacement
stresses that the fundamental object are now the states made of the set of
activated interacting connections and producing the activities $\Delta
\omega _{p}^{\alpha }\left( \theta ,\mathbf{Z}\right) $.

\subsubsection{Operators description}

Alternatively, using (\ref{GNv}) and (\ref{Gnvt}), the field action $S\left( 
\underline{\Gamma }\left( \mathbf{T,\alpha },\mathbf{p},S^{2},\theta \right)
\right) $ can be considered as a matrix element of an operator. Actually,
given (\ref{DRT}), the free part of the action (\ref{FCz}) can be written:%
\begin{eqnarray}
&&S_{f}\left( \underline{\Gamma }\left( \mathbf{T,\alpha },\mathbf{p}%
,S^{2},\theta \right) \right) \\
&=&\int \left\langle \left( \Delta \mathbf{\bar{T}}_{p}^{\alpha }\right)
^{\prime },\mathbf{\alpha },\mathbf{p},S^{2}\right\vert \underline{\Gamma }%
^{\dag }\left( \mathbf{T}^{\prime },\mathbf{\alpha },\mathbf{p},S^{2}\right)
\notag \\
&&\times \left( -\frac{1}{2}\nabla _{\left( \mathbf{T}\right) _{S^{2}}}^{2}+%
\frac{1}{2}\left( \Delta \mathbf{T}_{p}^{\alpha }\right) _{S^{2}}^{t}\mathbf{%
A}_{S^{2}}^{\alpha }\left( \Delta \mathbf{T}_{p}^{\alpha }\right) +\mathbf{C}%
\right) \underline{\Gamma }\left( \mathbf{T},\mathbf{\alpha },\mathbf{p}%
,S^{2}\right) \left\vert \Delta \mathbf{\bar{T}}_{p}^{\alpha },\mathbf{%
\alpha },\mathbf{p},S^{2}\right\rangle d\mathbf{T}d\mathbf{T}^{\prime } 
\notag \\
&=&\left\langle \underline{\Gamma }\left( \mathbf{T,}\boldsymbol{\alpha },%
\mathbf{p},S^{2}\right) \right\vert \left( \mathbf{\bar{D}}_{S^{2}}^{\alpha
}\left( \mathbf{A}^{+}\left( \alpha ,p,S^{2}\right) \mathbf{A}^{-}\left(
\alpha ,p,S^{2}\right) +\frac{1}{2}+\mathbf{C}\right) \right) \left\vert 
\underline{\Gamma }\left( \mathbf{T,}\boldsymbol{\alpha },\mathbf{p}%
,S^{2}\right) \right\rangle  \notag
\end{eqnarray}%
Equivalently, using the initial variables:%
\begin{equation*}
S_{f}\left( \underline{\Gamma }\left( \mathbf{T,\alpha },\mathbf{p}%
,S^{2},\theta \right) \right) =\left\langle \underline{\Gamma }\left( 
\mathbf{T,}\boldsymbol{\alpha },\mathbf{p},S^{2}\right) \right\vert \left(
\left( \mathbf{\hat{A}}^{+}\left( \alpha ,p,S^{2}\right) \right) ^{t}\mathbf{%
A}_{S^{2}}^{\alpha }\mathbf{\hat{A}}^{-}\left( \alpha ,p,S^{2}\right) +\frac{%
1}{2}+\mathbf{C}\right) \left\vert \underline{\Gamma }\left( \mathbf{T,}%
\boldsymbol{\alpha },\mathbf{p},S^{2}\right) \right\rangle
\end{equation*}%
with:%
\begin{equation*}
\mathbf{\hat{A}}^{\pm -}\left( \alpha ,p,S^{2}\right) =O\mathbf{A}^{\pm
-}\left( \alpha ,p,S^{2}\right)
\end{equation*}%
with:%
\begin{eqnarray*}
\mathbf{\bar{D}}_{S^{2}}^{\alpha } &=&O^{-1}\left( \mathbf{A}%
_{S^{2}}^{\alpha }\right) O \\
\left( \Delta \mathbf{\bar{T}}_{p}^{\alpha }\right) _{S^{2}} &=&O^{-1}\left(
\Delta \mathbf{T}_{p}^{\alpha }\right) _{S^{2}}
\end{eqnarray*}%
Thus, integrating over the degrees of freedom for $\underline{\Gamma }\left(
\Delta \mathbf{T},\mathbf{\alpha },\mathbf{p},S^{2},\theta \right) $ is
equivalent to compute transition elements of operators.

Considering the potential $U\left( \left\Vert \underline{\Gamma }\left(
\left( \Delta \mathbf{T}_{p}^{\alpha }\right) _{S^{2}},\mathbf{\alpha },%
\mathbf{p},S^{2},\theta \right) \right\Vert ^{2}\right) $ in the action (\ref%
{FCz}),\ it is a matrix element between tensor products of states. Actually,
starting with a series expansion for $U$:%
\begin{eqnarray}
&&U\left( \left\Vert \underline{\Gamma }\left( \left( \Delta \mathbf{T}%
_{p}^{\alpha }\right) _{S^{2}},\mathbf{\alpha },\mathbf{p},S^{2},\theta
\right) \right\Vert ^{2}\right)  \label{PL} \\
&=&\sum_{k}\int \prod\limits_{l=1}^{k}\underline{\Gamma }^{\dag }\left(
\left( \Delta \mathbf{T}_{p}^{\alpha }\right) _{S^{2}}^{\left( l\right) },%
\mathbf{\alpha },\mathbf{p},S^{2},\theta \right) \hat{U}_{k}\left( \left(
\left( \Delta \mathbf{T}_{p}^{\alpha }\right) _{S^{2}}^{\left( l\right)
}\right) _{l=1...k}\right) \prod\limits_{l=1}^{k}\underline{\Gamma }\left(
\left( \Delta \mathbf{T}_{p}^{\alpha }\right) _{S^{2}}^{\left( l\right) },%
\mathbf{\alpha },\mathbf{p},S^{2},\theta \right)  \notag
\end{eqnarray}%
this writes;%
\begin{eqnarray*}
&&U\left( \left\Vert \underline{\Gamma }\left( \left( \Delta \mathbf{T}%
_{p}^{\alpha }\right) _{S^{2}},\mathbf{\alpha },\mathbf{p},S^{2},\theta
\right) \right\Vert ^{2}\right) \\
&=&\sum_{k}\int \prod\limits_{l=1}^{k}\left\langle \underline{\Gamma }\left(
\left( \Delta \mathbf{T}_{p}^{\alpha }\right) _{S^{2}}^{\left( l\right) }%
\mathbf{,}\boldsymbol{\alpha },\mathbf{p},S^{2}\right) \right\vert \hat{U}%
_{k}\left( \left( \left( \Delta \mathbf{T}_{p}^{\alpha }\right)
_{S^{2}}^{\left( l\right) }\right) _{l=1...k}\right)
\prod\limits_{l=1}^{k}\left\vert \underline{\Gamma }\left( \left( \Delta 
\mathbf{T}_{p}^{\alpha }\right) _{S^{2}}^{\left( l\right) }\mathbf{,}%
\boldsymbol{\alpha },\mathbf{p},S^{2}\right) \right\rangle
\end{eqnarray*}%
so that the potential is the matrix elements of the operator:%
\begin{equation*}
\hat{U}=\sum_{k}\int \prod\limits_{l=1}^{k}\left\vert \left( \Delta \mathbf{T%
}_{p}^{\alpha }\right) _{S^{2}}^{\left( l\right) },\mathbf{\alpha },\mathbf{p%
},S^{2}\right\rangle \hat{U}_{k}\left( \left( \left( \Delta \mathbf{T}%
_{p}^{\alpha }\right) _{S^{2}}^{\left( l\right) }\right) _{l=1...k}\right)
\prod\limits_{l=1}^{k}\left\langle \left( \Delta \mathbf{T}_{p}^{\alpha
}\right) _{S^{2}}^{\left( l\right) },\mathbf{\alpha },\mathbf{p}%
,S^{2}\right\vert
\end{equation*}%
between tensor product of states. This operator can be written in terms of
creation annihilation operators by a change of basis. Defining:%
\begin{eqnarray*}
&&\bar{U}_{mn}\left( \mathbf{\alpha },\mathbf{p},S^{2}\right)
=\sum_{k}\left( \left\langle \mathbf{\alpha },\mathbf{p},S^{2}\right\vert
\right) ^{\otimes m}\hat{U}_{k}\left( \left( \left( \Delta \mathbf{T}%
_{p}^{\alpha }\right) _{S^{2}}^{\left( l\right) }\right) _{l=1...k}\right)
\left( \left\vert \mathbf{\alpha },\mathbf{p},S^{2}\right\rangle \right)
^{\otimes n} \\
&=&\sum_{k}\int \hat{U}_{k}\left( \left( \left( \Delta \mathbf{T}%
_{p}^{\alpha }\right) _{S^{2}}^{\left( l\right) }\right) _{l=1...k}\right) \\
&&\times \left( \prod\limits_{l=1}^{k}\left\langle \left( \Delta \mathbf{T}%
_{p}^{\alpha }\right) _{S^{2}}^{\left( l\right) },\mathbf{\alpha },\mathbf{p}%
,S^{2}\right\vert \left( \left\vert \mathbf{\alpha },\mathbf{p}%
,S^{2}\right\rangle \right) ^{\otimes n}\right) \left(
\prod\limits_{l=1}^{k}\left\langle \left( \Delta \mathbf{T}_{p}^{\alpha
}\right) _{S^{2}}^{\left( l\right) },\mathbf{\alpha },\mathbf{p}%
,S^{2}\right\vert \left( \left\vert \mathbf{\alpha },\mathbf{p}%
,S^{2}\right\rangle \right) ^{\otimes m}\right) ^{\dag }
\end{eqnarray*}%
We can thus replace:%
\begin{equation*}
\hat{U}=\sum_{m,n}\left( \left\vert \mathbf{\alpha },\mathbf{p}%
,S^{2}\right\rangle \right) ^{\otimes n}\bar{U}_{mn}\left( \mathbf{\alpha },%
\mathbf{p},S^{2}\right) \left( \left\langle \mathbf{\alpha },\mathbf{p}%
,S^{2}\right\vert \right) ^{\otimes m}
\end{equation*}%
and the operator becomes:%
\begin{equation*}
\hat{U}=\sum_{m,n}\bar{U}_{mn}\left( \mathbf{\alpha },\mathbf{p}%
,S^{2}\right) \left( \mathbf{\hat{A}}^{+}\left( \mathbf{\alpha },\mathbf{p}%
,S^{2}\right) \right) ^{m}\left( \mathbf{\hat{A}}^{-}\left( \mathbf{\alpha },%
\mathbf{p},S^{2}\right) \right) ^{n}
\end{equation*}%
As a consequence, field action $S\left( \left\{ \underline{\gamma }\left( 
\mathbf{T},\mathbf{\alpha },\mathbf{p},S^{2},\theta \right) \right\} \right) 
$ has the same content as the operator:%
\begin{eqnarray}
\mathbf{S} &=&\left( \mathbf{\hat{A}}^{+}\left( \alpha ,p,S^{2}\right)
\right) ^{t}\mathbf{A}_{S^{2}}^{\alpha }\mathbf{\hat{A}}^{-}\left( \alpha
,p,S^{2}\right) +\frac{1}{2}+\mathbf{C}  \label{RT} \\
&&+\sum_{m,n}\bar{U}_{mn}\left( \mathbf{\alpha },\mathbf{p},S^{2}\right)
\left( \mathbf{\hat{A}}^{+}\left( \mathbf{\alpha },\mathbf{p},S^{2}\right)
\right) ^{m}\left( \mathbf{\hat{A}}^{-}\left( \mathbf{\alpha },\mathbf{p}%
,S^{2}\right) \right) ^{n}  \notag
\end{eqnarray}%
and this operator will compute the same transitions between states as the
integration over the field degrees of freedom of the field $\underline{%
\Gamma }\left( \Delta \mathbf{T},\mathbf{\alpha },\mathbf{p},S^{2},\theta
\right) $ in the partition function (\ref{Pf}).

\section{Interactions between collective states}

So far, we have examined fields describing independent collective states. In
this section, we introduce interaction terms and explore their implications
for transitions.

\subsection{Principle}

Previous mechanism translates in term of fields by considering $n$
multi-components fields corresponding to the structures:%
\begin{eqnarray}
&&S\left( \left\{ \underline{\Gamma }\left( \Delta \mathbf{T},\mathbf{\alpha 
},\mathbf{p},S_{i}^{2},\theta \right) \right\} \right)  \label{Nd} \\
&=&-\underline{\Gamma }^{\dag }\left( \Delta \mathbf{T},\mathbf{\alpha },%
\mathbf{p},S_{i}^{2},\theta \right) \left( -\nabla _{\mathbf{T}}^{2}+\frac{1%
}{2}\left( \Delta \mathbf{T}_{p}^{\alpha }\right) _{S_{i}^{2}}^{t}\mathbf{A}%
_{S_{i}^{2}}^{\alpha }\left( \Delta \mathbf{T}_{p}^{\alpha }\right)
_{S_{i}^{2}}+\mathbf{C}\right) \underline{\Gamma }\left( \Delta \mathbf{T},%
\mathbf{\alpha },\mathbf{p},S_{i}^{2},\theta \right)  \notag \\
&&+U\left( \left\Vert \underline{\Gamma }\left( \Delta \mathbf{T},\mathbf{%
\alpha },\mathbf{p},S_{i}^{2},\theta \right) \right\Vert ^{2}\right)  \notag
\end{eqnarray}%
The set $S_{i}^{2}$ characterizes the structure localization along with its
possible states. The multi-components labelled by $\mathbf{\alpha },\mathbf{p%
}$ transcribes the possible averages and frequencies.

The full action for the system described above should be a sum of individual
actions:%
\begin{equation*}
\sum_{i}S\left( \left\{ \underline{\Gamma }\left( \Delta \mathbf{T},\mathbf{%
\alpha },\mathbf{p},S_{i}^{2},\theta \right) \right\} \right) +V\left(
\left\{ \underline{\Gamma }\left( \Delta \mathbf{T},\mathbf{\alpha },\mathbf{%
p},\left( \cup S_{i}\right) \times \left( \cup S_{i}\right) ,\theta \right)
\right\} \right)
\end{equation*}%
with additional interaction terms accounting for trnstns between various
states. To describe dynamically this transition in terms of fields, we
showed in (P.4 ) that we can add to the action a term of the form:%
\begin{equation*}
\sum_{nn^{\prime }}\sum_{\substack{ k=1...n  \\ l=1,...,n^{\prime }}}%
\sum_{\left\{ S_{k},S_{l}\right\} _{\substack{ l=1,...,n^{\prime }  \\ %
k=1...n }}}\prod_{l}\underline{\Gamma }^{\dag }\left( \mathbf{T}_{l}^{\prime
},\mathbf{\alpha }_{l}^{\prime },\mathbf{p}_{l}^{\prime },S_{l}^{\prime
2}\right) V\left( \left\{ \left\vert \mathbf{T}_{l}^{\prime },\mathbf{\alpha 
}_{l}^{\prime },\mathbf{p}_{l}^{\prime },S_{l}^{\prime 2}\right\rangle
\right\} ,\left\{ \left\vert \mathbf{T}_{l},\mathbf{\alpha }_{k},\mathbf{p}%
_{k},S_{k}^{2}\right\rangle \right\} \right) \prod_{k}\underline{\Gamma }%
\left( \mathbf{T}_{k},\mathbf{\alpha }_{k},\mathbf{p}_{k},S_{k}^{2}\right)
\end{equation*}%
where:%
\begin{equation*}
V\left( \left\{ \left\vert \mathbf{T}_{l}^{\prime },\mathbf{\alpha }%
_{l}^{\prime },\mathbf{p}_{l}^{\prime },S_{l}^{\prime 2}\right\rangle
\right\} ,\left\{ \left\vert \mathbf{T}_{k},\mathbf{\alpha }_{k},\mathbf{p}%
_{k},S_{k}^{2}\right\rangle \right\} \right) =V\left( \left\{ \mathbf{T}%
_{l}^{\prime },\mathbf{\alpha }_{l}^{\prime },\mathbf{p}_{l}^{\prime
},S_{l}^{\prime 2}\right\} ,\left\{ \mathbf{T}_{k},\mathbf{\alpha }_{k},%
\mathbf{p}_{k},S_{k}^{2}\right\} \right)
\end{equation*}%
and the action for interacting structures becomes:%
\begin{eqnarray}
S &=&\sum_{S}\underline{\Gamma }^{\dag }\left( \mathbf{T},\mathbf{\alpha },%
\mathbf{p},S^{2}\right) \left( -\frac{1}{2}\nabla _{\left( \mathbf{\hat{T}}%
\right) _{S^{2}}}^{2}+\frac{1}{2}\left( \Delta \mathbf{T}_{p}^{\alpha
}\right) _{S^{2}}^{t}\mathbf{A}_{S^{2}}^{\alpha }\left( \Delta \mathbf{T}%
_{p}^{\alpha }\right) +\mathbf{C}\right) \underline{\Gamma }\left( \mathbf{T}%
,\mathbf{\alpha },\mathbf{p},S^{2}\right)  \label{RCN} \\
&&+\sum_{nn^{\prime }}\sum_{\substack{ k=1...n  \\ l=1,...,n^{\prime }}}%
\sum_{\left\{ S_{k},S_{l}\right\} _{\substack{ l=1,...,n^{\prime }  \\ %
k=1...n }}}\prod_{l}\underline{\Gamma }^{\dag }\left( \mathbf{T}_{l}^{\prime
},\mathbf{\alpha }_{l}^{\prime },\mathbf{p}_{l}^{\prime },S_{l}^{\prime
2}\right)  \notag \\
&&\times V\left( \left\{ \left\vert \mathbf{T}^{\prime },\mathbf{\alpha }%
_{l}^{\prime },\mathbf{p}_{l}^{\prime },S_{l}^{\prime 2}\right\rangle
\right\} ,\left\{ \left\vert \mathbf{T},\mathbf{\alpha }_{k},\mathbf{p}%
_{k},S_{k}^{2}\right\rangle \right\} \right) \prod_{k}\underline{\Gamma }%
\left( \mathbf{T}_{k},\mathbf{\alpha }_{k},\mathbf{p}_{k},S_{k}^{2}\right) 
\notag
\end{eqnarray}%
allowing for transitions between sets of several collective states. The form
of $V$ is conditionned by frequencies of oscillation:%
\begin{equation*}
V\left( \left\{ \mathbf{T}_{l}^{\prime },\mathbf{\alpha }_{l}^{\prime },%
\mathbf{p}_{l}^{\prime },S_{l}^{\prime 2}\right\} ,\left\{ \mathbf{T}_{k},%
\mathbf{\alpha }_{k},\mathbf{p}_{k},S_{k}^{2}\right\} \right) =V\left(
\Upsilon _{l}^{\mathbf{p}_{l}^{\prime }}\left( \mathbf{T}_{l}^{\prime
}\right) ,\Upsilon _{k}^{\mathbf{p}_{k}}\left( \mathbf{T}_{k}\right) \right)
\end{equation*}%
and models the results of the first part of this article, transitions depend
both on initial states characteritics and that of the merged ones.

\subsection{Interactions between subobjects}

We also consider the possibility of activation by a substructure. This is a
particular case of interaction where the activation of some substructure
induces the full structure activation. To describe this type of transition,
the term:%
\begin{equation*}
\underline{\Gamma }^{\dag }\left( \mathbf{T},\mathbf{\alpha },\mathbf{p}%
,S^{2}\right) \left( -\frac{1}{2}\nabla _{\left( \mathbf{\hat{T}}\right)
_{S^{2}}}^{2}+\frac{1}{2}\left( \Delta \mathbf{T}_{p}^{\alpha }\right)
_{S^{2}}^{t}\mathbf{A}_{S^{2}}^{\alpha }\left( \Delta \mathbf{T}_{p}^{\alpha
}\right) +\mathbf{C}\right) \underline{\Gamma }\left( \mathbf{T},\mathbf{%
\alpha },\mathbf{p},S^{2}\right)
\end{equation*}%
is generalized by including free action terms for each substructures plus
interaction terms between these substructures, including the full one:%
\begin{eqnarray}
&&\sum_{S_{1}\subseteq S}\underline{\Gamma }^{\dag }\left( \mathbf{T},%
\mathbf{\alpha },\mathbf{p},S_{1}^{2}\right) \left( -\frac{1}{2}\nabla
_{\left( \mathbf{\hat{T}}\right) _{S_{1}^{2}}}^{2}+\frac{1}{2}\left( \Delta 
\mathbf{T}_{p}^{\alpha }\right) _{S_{1}^{2}}^{t}\mathbf{A}%
_{S_{1}^{2}}^{\alpha }\left( \Delta \mathbf{T}_{p}^{\alpha }\right) +\mathbf{%
C}\right) \underline{\Gamma }\left( \mathbf{T},\mathbf{\alpha },\mathbf{p}%
,S_{1}^{2}\right)  \label{NTBJ} \\
&&+\sum_{n}\sum_{S_{1},...,S_{n}\subseteq S}\sum_{\left( \alpha
_{1},p_{1}\right) ,...,\left( \alpha _{n},p_{n}\right) }\sum_{k}V_{k}\left(
\left( \left\{ \mathbf{T}_{i},\mathbf{\alpha }_{i},\mathbf{p}_{i}\right\}
,\left( S_{i}\right) ^{2}\right) \right) \prod\limits_{i\leqslant k}%
\underline{\Gamma }^{\dag }\left( \mathbf{T}_{i},\mathbf{\alpha }_{i},%
\mathbf{p}_{i}\right) \prod\limits_{k+1\leqslant i\leqslant n}\underline{%
\Gamma }\left( \mathbf{T}_{i},\mathbf{\alpha }_{i},\mathbf{p}%
_{i},S_{i}^{2}\right)  \notag
\end{eqnarray}

The potential $V_{k}\left( \left( \left\{ \mathbf{T}_{i},\mathbf{\alpha }%
_{i},\mathbf{p}_{i}\right\} ,\left( S_{i}\right) ^{2}\right) \right) $
induces transition from some state with $k$ substructures towards a state
with $n-k$ subtructures. It may include the transition from one or several
subsets to the full activated structure. This situation is depicted by a
potential of the type:%
\begin{equation*}
V_{k}\left( \left( \left\{ \mathbf{T}_{i},\mathbf{\alpha }_{i},\mathbf{p}%
_{i},\left( S_{i}\right) ^{2}\right\} \right) \right) =V_{k}\left( \left(
\left\{ \mathbf{T}_{i},\mathbf{\alpha }_{i},\mathbf{p}_{i},\left(
S_{i}\right) ^{2}\right\} _{i\leqslant k},\left\{ \mathbf{T}_{i},\mathbf{%
\alpha }_{i},\mathbf{p}_{i},\left( S_{i}\right) ^{2}\right\} _{k+1\leqslant
i\leqslant n}\right) \right)
\end{equation*}%
the group $\mathbf{Z}$ of possible states is defined, at least partly, by
initial background.

\subsection{Operator formalism for interactions}

We have seen in (\ref{RT}) the operator formulation for the dynamic of one
type of structure:%
\begin{eqnarray}
\mathbf{S} &=&\left( \mathbf{\hat{A}}^{+}\left( \alpha ,p,S^{2}\right)
\right) ^{t}\mathbf{A}_{S^{2}}^{\alpha }\mathbf{\hat{A}}^{-}\left( \alpha
,p,S^{2}\right) +\frac{1}{2}+\mathbf{C} \\
&&+\sum_{m,n}\bar{U}_{mn}\left( \mathbf{\alpha },\mathbf{p},S^{2}\right)
\left( \mathbf{\hat{A}}^{+}\left( \mathbf{\alpha },\mathbf{p},S^{2}\right)
\right) ^{m}\left( \mathbf{\hat{A}}^{-}\left( \mathbf{\alpha },\mathbf{p}%
,S^{2}\right) \right) ^{n}  \notag
\end{eqnarray}

As for the field version, we can consider interaction potential between
different structures. The operator counterpart of the potential term in (\ref%
{RCN}) writes in terms of annihilation and creation operators:%
\begin{eqnarray}
&&\hat{V}=\sum_{n,n^{\prime }}\sum_{\left\{ S_{k},S_{l}\right\} _{\substack{ %
l=1,...,n^{\prime }  \\ k=1...n}}}\sum_{\left\{ m_{l}^{\prime
},m_{k}\right\} }\prod_{l=1}^{n^{\prime }}\left( \mathbf{\hat{A}}^{+}\left( 
\mathbf{\alpha }_{l}^{\prime },\mathbf{p}_{l}^{\prime },S_{l}^{\prime
2}\right) \right) ^{m_{l}^{\prime }}  \label{NT} \\
&&\times V_{n,n^{\prime }}\left( \left\{ \mathbf{\alpha }_{l}^{\prime },%
\mathbf{p}_{l}^{\prime },S_{l}^{\prime 2},m_{l}^{\prime }\right\}
_{l\leqslant n^{\prime }},\left\{ \mathbf{\alpha }_{k},\mathbf{p}%
_{k},S_{k}^{2},m_{k}\right\} _{l\leqslant n}\right) \prod_{k=1}^{n}\left( 
\mathbf{\hat{A}}^{-}\left( \mathbf{\alpha }_{k},\mathbf{p}%
_{k},S_{k}^{2}\right) \right) ^{m_{k}}  \notag
\end{eqnarray}%
Th prssn fr $V_{n,n^{\prime }}$ s fnd n P 4. Thus the corresponding operator
is to $S\left( \left\{ \underline{\Gamma }\left( \mathbf{T},\mathbf{\alpha },%
\mathbf{p},S^{2},\theta \right) \right\} \right) $ is:%
\begin{eqnarray}
&&\mathbf{S}=\sum_{S\times S}\mathbf{\bar{D}}_{S^{2}}^{\alpha }\left( 
\mathbf{A}^{+}\left( \alpha ,p,S^{2}\right) \mathbf{A}^{-}\left( \alpha
,p,S^{2}\right) +\frac{1}{2}\right)  \label{SP} \\
&&+\sum_{m,n}\bar{U}_{mn}\left( \mathbf{\alpha },\mathbf{p},S^{2}\right)
\left( \mathbf{\hat{A}}^{+}\left( \mathbf{\alpha },\mathbf{p},S^{2}\right)
\right) ^{m}\left( \mathbf{\hat{A}}^{-}\left( \mathbf{\alpha },\mathbf{p}%
,S^{2}\right) \right) ^{n}+\hat{V}  \notag
\end{eqnarray}

The advantage of this formulation is to directly translate the dynamics in
terms of creation and destruction of structures, describing the transition
resulting from such operators. It also allows straightforward computations
at the lowest order of approximation, presenting a direct interpretation as
transitions of structures.

In the sequel we will simplify the notation:%
\begin{equation*}
V_{n,n^{\prime }}\left( \left\{ \mathbf{\alpha }_{l}^{\prime },\mathbf{p}%
_{l}^{\prime },S_{l}^{\prime 2},m_{l}^{\prime }\right\} _{l\leqslant
n^{\prime }},\left\{ \mathbf{\alpha }_{k},\mathbf{p}_{k},S_{k}^{2},m_{k}%
\right\} _{l\leqslant n}\right) \rightarrow V_{n,n^{\prime }}\left( \left\{ 
\mathbf{\alpha }_{l}^{\prime },\mathbf{p}_{l}^{\prime },S_{l}^{\prime
2},m_{l}^{\prime }\right\} ,\left\{ \mathbf{\alpha }_{k},\mathbf{p}%
_{k},S_{k}^{2},m_{k}\right\} \right)
\end{equation*}

\subsection{External sources action}

An external source modifies the state of a given structur. Using the wave
equation for activity presented previously, in presence of source, the
activty satisfies the following equation:%
\begin{equation}
f\Omega \left( \theta ,Z\right) =J\left( \theta ,Z\right) +\left( \frac{\hat{%
f}_{1}}{\omega \left( \theta ,Z\right) }+N_{1}\right) \nabla _{\theta
}\Omega \left( \theta ,Z\right) +\left( \frac{\hat{f}_{3}}{\omega \left(
\theta ,Z\right) }-N_{2}\right) \nabla _{\theta }^{2}\Omega \left( \theta
,Z\right) +\frac{c^{2}\hat{f}_{3}}{\omega \left( \theta ,Z\right) }\nabla
_{Z}^{2}\Omega \left( \theta ,Z\right)  \label{WQ}
\end{equation}%
with:%
\begin{equation*}
\Omega \left( \theta ,Z\right) =\omega \left( J,\theta ,Z\right) -\omega
_{0}\left( J,\theta ,Z\right)
\end{equation*}%
measure the deviation of activity from some background, usually considered
as static.

The coefficients of the equation are given in Appendix 4. In first
approximation the solutions are of the form:%
\begin{equation*}
\Omega \left( \theta ,Z\right) =\bar{\Omega}\left( \theta ,Z\right) +\Omega
_{J}\left( \theta ,Z\right)
\end{equation*}%
where $\bar{\Omega}\left( \theta ,Z\right) $ is solution of (\ref{WQ}) with $%
J\left( \theta ,Z\right) =0$ and $\Omega _{J}\left( \theta ,Z\right) $ is
the activity induced by the external source. The formula for $\Omega _{J}$
was obtained in \cite{GLs} and an account of the derivation is given in
Appendix 8.

When the source is composed of oscillating signals located at some points $%
Z_{i}$:%
\begin{equation*}
a\left( Z_{i},\theta \right) \propto a\exp \left( i\varpi \theta \right)
\end{equation*}%
An estimation of $\Omega _{J}\left( \theta ,Z\right) $ at the lowest order
was given in \cite{GLs}. We obtain:

\begin{equation}
\Omega _{J}\left( \theta ,Z\right) =S\left( Z,\varpi \right) \exp \left(
i\varpi \theta \right) \sum_{i}\exp \left( i\frac{\varpi \left\vert
Z_{i}-Z_{0}\right\vert }{c\left\vert Z-Z_{0}\right\vert }\right)  \label{ntf}
\end{equation}%
where $Z_{0}\in \left\{ Z_{i}\right\} $ is the closest point to $Z$ and $%
S\left( Z,\varpi \right) $ is given in Appendix 8. the contribution $\Omega
_{J}\left( \theta ,Z\right) $ induces interferences. As a consequence, for a
large number of points $Z_{i}$:%
\begin{equation*}
\sum_{i}\exp \left( i\frac{\varpi \left\vert Z_{i}-Z_{0}\right\vert }{%
c\left\vert Z-Z_{0}\right\vert }\right) \simeq 0
\end{equation*}%
except for the maxima of interferences with magnitude:%
\begin{equation*}
\frac{a\exp \left( -\left\vert Z-Z_{0}\right\vert \right) }{c\sqrt{\left(
1+2\alpha \left\vert Z-Z_{0}\right\vert \right) ^{2}+\left( \frac{\varpi }{c}%
\right) ^{2}}}
\end{equation*}%
Note that for these maxima and $N$ large:%
\begin{equation*}
a=\sum_{i}\exp \left( i\frac{\varpi \left\vert Z_{i}-Z_{0}\right\vert }{%
c\left\vert Z-Z_{0}\right\vert }\right)
\end{equation*}%
is proportional to $N$ so that $a>>1$.

If $\varpi $ is equal to some internal frequency $\Upsilon _{p}$, when the
source is switched off, the state may experience a transition $\left(
\Upsilon _{p}^{\alpha }\right) \rightarrow \left( \Upsilon _{p^{\prime
}}^{\alpha ^{\prime }}\right) $. This occurs when the initial state has been
damped upon switching off the source. This is modeled by:%
\begin{equation*}
\left( \mathbf{\hat{A}}^{+}\left( \mathbf{\alpha }^{\prime },\mathbf{p}%
^{\prime },S^{2}\right) \right) \left( \mathbf{\hat{A}}^{-}\left( \mathbf{%
\alpha },\mathbf{p},S^{2}\right) \right) J\left( \left( \Upsilon _{p^{\prime
}}^{\alpha ^{\prime }}\right) \right)
\end{equation*}%
If the state $\left( \mathbf{\alpha },\mathbf{p},S^{2}\right) $ is damping,
then the oprtr descrbing the trnstn is:%
\begin{equation*}
\left( \mathbf{\hat{A}}^{+}\left( \mathbf{\alpha }^{\prime },\mathbf{p}%
^{\prime },S^{2}\right) \right) \left( \mathbf{\hat{A}}^{+}\left( \mathbf{%
\alpha },\mathbf{p},S^{2}\right) \right) \left( \mathbf{\hat{A}}^{-}\left( 
\mathbf{\alpha },\mathbf{p},S^{2}\right) \right) J\left( \left( \Upsilon
_{p^{\prime }}^{\alpha ^{\prime }}\right) \right)
\end{equation*}

Which possible realization will occur depends on the coefficient:%
\begin{equation*}
\frac{\hat{f}_{1}}{\omega \left( \theta ,Z\right) }+N_{1}
\end{equation*}%
When this coefficient is negative, the oscillations $\bar{\Omega}\left(
\theta ,Z\right) $ are dampening. Moreover, since $N_{1}$ is proportionl to $%
\frac{1}{\left( \omega \left( \theta ,Z\right) \right) ^{2}}$, the
introduction of the srce $\Omega _{J}\left( \theta ,Z\right) $ increass $%
\omega \left( \theta ,Z\right) $ and lowrs $\frac{\hat{f}_{1}}{\omega \left(
\theta ,Z\right) }+N_{1}$. As a consequence, a stable internl oscillation
with frequenc $\left( \Upsilon _{p}^{\alpha }\right) $ mays switch to the
dampening zone, so that in the long run:%
\begin{equation*}
\Omega \left( \theta ,Z\right) =\Omega _{J}\left( \theta ,Z\right)
\end{equation*}%
switching off the srce, leadn to the oscillatn equation:%
\begin{equation}
f\Omega \left( \theta ,Z\right) =\left( \frac{\hat{f}_{1}}{\omega \left(
\theta ,Z\right) }+N_{1}\right) \nabla _{\theta }\Omega \left( \theta
,Z\right) +\left( \frac{\hat{f}_{3}}{\omega \left( \theta ,Z\right) }%
-N_{2}\right) \nabla _{\theta }^{2}\Omega \left( \theta ,Z\right) +\frac{%
c^{2}\hat{f}_{3}}{\omega \left( \theta ,Z\right) }\nabla _{Z}^{2}\Omega
\left( \theta ,Z\right)
\end{equation}

If the source frequency is an internal frequency $\left( \Upsilon
_{p^{\prime }}^{\alpha ^{\prime }}\right) ${}, then the structure will
oscillate at this frequency, realizing the transition from $\left( \Upsilon
_{p}^{\alpha }\right) $\ to $\left( \Upsilon _{p^{\prime }}^{\alpha ^{\prime
}}\right) $. However, if the oscillation $\varpi $ is not an internal
frequency $\left( \Upsilon _{p^{\prime }}^{\alpha ^{\prime }}\right) $, the
entire activation may shut down or decay towards some close internal
frequency. This possibility can be understood as follows: having been bound
by the source, the various cells constituting the structure continue to
interact, but at some internal frequency different from that of the source.

To minimize the modification, the internal frequency $\left( \Upsilon
_{p^{\prime }}^{\alpha ^{\prime }}\right) ${}should be close to $\varpi $.
This implies that the initial state $\left( \mathbf{\alpha },\mathbf{p}%
,S^{2}\right) $ can be preserved, with the addition of a new state $\left( 
\mathbf{\alpha }^{\prime },\mathbf{p}^{\prime },S^{2}\right) $.
Alternatively, it may also transform into another structure with a different
spatial support. This possibility of modification in spatial support has
been studied in \cite{GLw}, where it is described by a modification of the
"free" dynamics for single structures and is given by the operator (\ref{Nd}%
) and inclding the intrinsic displacement. The system is described by:%
\begin{equation}
-\underline{\Gamma }^{\dag }\left( \Delta \mathbf{T},\mathbf{\alpha },%
\mathbf{p},S_{i}^{2},\theta \right) \left( -\frac{1}{2}\frac{\delta ^{2}}{%
\delta \left( S_{i}^{2}\right) ^{2}}-\bar{U}\left( S_{i}^{2}\right) \frac{%
\delta }{\delta S_{i}^{2}}-\nabla _{\mathbf{T}}^{2}+\frac{1}{2}\left( \Delta 
\mathbf{T}_{p}^{\alpha }\right) _{S_{i}^{2}}^{t}\mathbf{A}%
_{S_{i}^{2}}^{\alpha }\left( \Delta \mathbf{T}_{p}^{\alpha }\right)
_{S_{i}^{2}}+\mathbf{C}\right) \underline{\Gamma }\left( \Delta \mathbf{T},%
\mathbf{\alpha },\mathbf{p},S_{i}^{2},\theta \right)  \label{CVS}
\end{equation}%
The second-order operator:%
\begin{equation}
-\frac{1}{2}\frac{\delta ^{2}}{\delta \left( S_{i}^{2}\right) ^{2}}-\bar{U}%
\left( S_{i}^{2}\right) \frac{\delta }{\delta S_{i}^{2}}  \label{sps}
\end{equation}%
represents the modifications to the spatial extension of the structure,
including oscillations and possible trends driven by $\bar{U}\left(
S_{i}^{2}\right) $. Such modification operators should induce the
displacement of the structure, leading to changes in its internal
characteristics, amplitudes, and frequencies.

The transition from $\left( \Upsilon _{p}^{\alpha }\right) $ in $S^{2}$
towards $\left( \Upsilon _{p^{\prime }}^{\alpha ^{\prime }}\right) $ in $%
\left( S^{\prime }\right) ^{2}$ should be described as a transition from one
state to another:%
\begin{equation*}
\left( \mathbf{\hat{A}}^{+}\left( \mathbf{\alpha }^{\prime },\mathbf{p}%
^{\prime },\left( S^{\prime }\right) ^{2}\right) \right) \left( \mathbf{\hat{%
A}}^{-}\left( \mathbf{\alpha },\mathbf{p},S^{2}\right) \right)
\end{equation*}%
However, since the transition requires the transformation of the support,
the parameter to consider is rather:%
\begin{equation}
\mathcal{G}\left( S^{\prime },S\right) \left( \mathbf{\hat{A}}^{+}\left( 
\mathbf{\alpha }^{\prime },\mathbf{p}^{\prime },\left( S^{\prime }\right)
^{2}\right) \right) \left( \mathbf{\hat{A}}^{-}\left( \mathbf{\alpha },%
\mathbf{p},S^{2}\right) \right)  \label{Trd}
\end{equation}%
where $\mathcal{G}\left( S^{\prime },S\right) $ is the Green function of (%
\ref{sps}).

\subsection{Structure-structure interaction}

A collective state may modify another state of a given structure by acting
as an external source. This can be obtained by an estimation of the second
term in the RHS of (\ref{ntf}) at the lowest order for some oscillating
signals $a\left( Z_{i},\theta \right) \propto a\exp \left( i\varpi \theta
\right) $, with:%
\begin{equation*}
\varpi =\Upsilon _{p^{\prime }}^{\alpha ^{\prime }}
\end{equation*}%
since the source is now a collective state. It yields:

\begin{equation}
\omega \left( J,\theta ,Z\right) =\omega _{0}\left( J,\theta ,Z\right)
+S\left( Z,\Upsilon _{p^{\prime }}^{\alpha ^{\prime }}\right) \exp \left(
i\Upsilon _{p^{\prime }}^{\alpha ^{\prime }}\theta \right) \sum_{i}\exp
\left( i\frac{\Upsilon _{p^{\prime }}^{\alpha ^{\prime }}\left\vert
Z_{i}-Z_{0}\right\vert }{c\left\vert Z-Z_{0}\right\vert }\right)  \label{Lfr}
\end{equation}%
and this terms induces interferences. As a consequence, for a large number
of points $Z_{i}$:%
\begin{equation*}
\sum_{i}\exp \left( i\frac{\Upsilon _{p^{\prime }}^{\alpha ^{\prime
}}\left\vert Z_{i}-Z_{0}\right\vert }{c\left\vert Z-Z_{0}\right\vert }%
\right) \simeq 0
\end{equation*}%
except for the maxima of interferences with magnitude:%
\begin{equation*}
\frac{a\exp \left( -\left\vert Z-Z_{0}\right\vert \right) }{c\sqrt{\left(
1+2\alpha \left\vert Z-Z_{0}\right\vert \right) ^{2}+\left( \frac{\varpi }{c}%
\right) ^{2}}}
\end{equation*}%
Note that for these maxima and for $N$ large, the amplitude for constructive
interferences:%
\begin{equation*}
a=\sum_{i}\exp \left( i\frac{\varpi \left\vert Z_{i}-Z_{0}\right\vert }{%
c\left\vert Z-Z_{0}\right\vert }\right)
\end{equation*}%
is proportional to $N$, so that $a>>1$.

\begin{equation*}
S\left( Z,\Upsilon _{p^{\prime }}^{\alpha ^{\prime }}\right) \exp \left(
i\Upsilon _{p^{\prime }}^{\alpha ^{\prime }}\theta \right)
\end{equation*}%
Written in terms of collective states, (\ref{Lfr}) becomes:%
\begin{equation}
\omega \left( S^{2}\right) =\omega \left( \mathbf{\alpha },\mathbf{p}%
,S^{2}\right) +S\left( Z,\Upsilon _{p^{\prime }}^{\alpha ^{\prime }}\right)
\exp \left( i\Upsilon _{p^{\prime }}^{\alpha ^{\prime }}\theta \right)
\sum_{i}\exp \left( i\frac{\Upsilon _{p^{\prime }}^{\alpha ^{\prime
}}\left\vert Z_{i}-Z_{0}\right\vert }{c\left\vert Z-Z_{0}\right\vert }\right)
\label{stc}
\end{equation}%
If $\Upsilon _{p^{\prime }}^{\alpha ^{\prime }}${} is one of the $\Upsilon
_{p}$,{} the state may experience a transition $\left( \Upsilon _{p}^{\alpha
}\right) \rightarrow \left( \Upsilon _{p^{\prime }}^{\alpha ^{\prime
}}\right) $ when the source switches off,. This happens when the initial
state has dampened. This is modeled by the term:%
\begin{equation}
\mathbf{\hat{A}}^{+}\left( \mathbf{\alpha }_{1}^{\prime },\mathbf{p}%
_{1}^{\prime },\left( S_{1}\right) ^{2}\right) \left( \mathbf{\hat{A}}%
^{+}\left( \mathbf{\alpha }_{2}^{\prime },\mathbf{p}_{2}^{\prime },\left(
S_{2}\right) ^{2}\right) \right) \left( \mathbf{\hat{A}}^{-}\left( \mathbf{%
\alpha }_{1},\mathbf{p}_{1},S_{1}^{2}\right) \right) \left( \mathbf{\hat{A}}%
^{-}\left( \mathbf{\alpha }_{2},\mathbf{p}_{2},S_{2}^{2}\right) \right)
\label{Trn}
\end{equation}%
if the state $\left( \mathbf{\alpha }_{i},\mathbf{p}_{i},S_{i}^{2}\right) $
are dampening, or:%
\begin{eqnarray}
&&\mathbf{\hat{A}}^{+}\left( \mathbf{\alpha }_{1}^{\prime },\mathbf{p}%
_{1}^{\prime },\left( S_{1}\right) ^{2}\right) \left( \mathbf{\hat{A}}%
^{+}\left( \mathbf{\alpha }_{2}^{\prime },\mathbf{p}_{2}^{\prime },\left(
S_{2}\right) ^{2}\right) \right) \left( \mathbf{\hat{A}}^{+}\left( \mathbf{%
\alpha }_{1},\mathbf{p}_{1},S_{1}^{2}\right) \right) \left( \mathbf{\hat{A}}%
^{+}\left( \mathbf{\alpha }_{2},\mathbf{p}_{2},S_{2}^{2}\right) \right) 
\notag \\
&&\times \left( \mathbf{\hat{A}}^{-}\left( \mathbf{\alpha }_{1},\mathbf{p}%
_{1},S_{1}^{2}\right) \right) \left( \mathbf{\hat{A}}^{-}\left( \mathbf{%
\alpha }_{2},\mathbf{p}_{2},S_{2}^{2}\right) \right)  \label{Trt}
\end{eqnarray}%
otherwise. When interactions imply the transition for space, the equivalent
of (\ref{Trd}) becomes:%
\begin{equation}
\mathcal{G}\left( S_{2}^{\prime },S_{2}\right) \mathcal{G}\left(
S_{1}^{\prime },S_{1}\right) \mathbf{\hat{A}}^{+}\left( \mathbf{\alpha }%
_{1}^{\prime },\mathbf{p}_{1}^{\prime },\left( S_{1}^{\prime }\right)
^{2}\right) \left( \mathbf{\hat{A}}^{+}\left( \mathbf{\alpha }_{2}^{\prime },%
\mathbf{p}_{2}^{\prime },\left( S_{2}^{\prime }\right) ^{2}\right) \right)
\left( \mathbf{\hat{A}}^{-}\left( \mathbf{\alpha }_{1},\mathbf{p}%
_{1},S_{1}^{2}\right) \right) \left( \mathbf{\hat{A}}^{-}\left( \mathbf{%
\alpha }_{2},\mathbf{p}_{2},S_{2}^{2}\right) \right)  \label{Trh}
\end{equation}

\section{State space for collective states and their subobjects}

In P4 we studied the interaction of subobjects among a global one through
action (\ref{NTBJ}). Here we are rather interested in the merging of
subobjects into integrated ones, as well as the stability of such systems
and their possibilities of dissociations. We consder this case in the next
paragraphs and deduce the set up needed to study an object along with its
subparts. We also study the conditions for stability of such composites. We
consider firt the integration of an object into a larger one, and second,
the composite object from several structures due to interactions.

\subsection{Integration into higher object}

This situation can be modelled by including independent action functional
terms fer subobjects. It corresponds to add contributions:%
\begin{eqnarray}
&&\sum_{S_{i}\subseteq S}\underline{\Gamma }^{\dag }\left( \mathbf{T}_{i},%
\mathbf{\alpha }_{i},\mathbf{p}_{i},\left( S_{i}\right) ^{2}\right) \left( -%
\frac{1}{2}\nabla _{\left( \mathbf{\hat{T}}\right) _{S_{i}^{2}}}^{2}+\frac{1%
}{2}\left( \Delta \mathbf{T}_{p}^{\alpha }\right) _{S_{i}^{2}}^{t}\mathbf{A}%
_{S_{i}^{2}}^{\alpha }\left( \Delta \mathbf{T}_{p}^{\alpha }\right) +\mathbf{%
C}_{i}\right) \underline{\Gamma }\left( \mathbf{T}_{i},\mathbf{\alpha }_{i},%
\mathbf{p}_{i},\left( S_{i}\right) ^{2}\right)  \label{TRCMH} \\
&&V\left( \left( \left\{ \mathbf{T}_{i},\mathbf{\alpha }_{i},\mathbf{p}%
_{i}\right\} ,\left( S_{i}\right) ^{2}\right) ,\mathbf{T},\mathbf{\alpha },%
\mathbf{p},S^{2}\right) \left\vert \underline{\Gamma }\left( \mathbf{T}_{i},%
\mathbf{\alpha }_{i},\mathbf{p}_{i},S_{i}^{2}\right) \right\vert ^{2}  \notag
\end{eqnarray}%
which amounts to consider the subobjects as independent, with the
possibility to switch towards a state of n ntgrtd bjct. The potential has to
transcribe the interactions between the objects at stake that could merge as
consequence of these interactions. In P4, we studied the details of the
mechanisms of interactions. Each potential subobjects sends signals to the
others depending on their characteristics amplitudes and frequencies. This
may result in a bound global objects whose frequencies are combinations of
the sbbjcts frqncs. Appendix 7 recalls the principle and results of this
mechanisms, while some details and dynamical arguments for these transitions
are presented in \cite{GLw}. In short the potential $V$ should impose the
relation between frequencies and activts of the full object and its subpart
which has the form: 
\begin{equation}
\Upsilon _{\left( i,l_{i}\right) }=\sum_{i,j\neq i}\left[ \left[ M\right] %
\right] _{i,j}\frac{\gamma _{i,k_{i}}+\gamma _{j,l_{j}}}{2}\pm \sqrt{\left(
\sum_{i,j\neq i}\frac{\left[ \left[ M\right] \right] _{i,j}}{\sum_{i,j\neq i}%
\left[ \left[ M\right] \right] _{i,j}}\frac{\gamma _{i,k_{i}}-\gamma
_{j,l_{j}}}{2}\right) ^{2}+\sum_{i,j\neq i}\left[ \left[ M\right] \right]
_{i,j}}  \label{Rl}
\end{equation}%
The matrices $\left[ M\right] $ are detailed in Appendix 7.

The transition between such state is induced by the potential $V\left(
\left( \left\{ \mathbf{T}_{i},\mathbf{\alpha }_{i},\mathbf{p}_{i}\right\}
,\left( S_{i}\right) ^{2}\right) ,\mathbf{T},\mathbf{\alpha },\mathbf{p}%
,S^{2}\right) $ and can be represented by:%
\begin{equation*}
\prod\limits_{i}\left\vert \mathbf{T}_{i},\mathbf{\alpha }_{i},\mathbf{p}%
_{i},\left( S_{i}\right) ^{2}\right\rangle \overset{V\left( \left( \left\{ 
\mathbf{T}_{i},\mathbf{\alpha }_{i},\mathbf{p}_{i}\right\} ,\left(
S_{i}\right) ^{2}\right) ,\mathbf{T},\mathbf{\alpha },\mathbf{p}%
,S^{2}\right) }{{\Large \rightarrow }}\left\vert \mathbf{T},\mathbf{\alpha },%
\mathbf{p},S^{2}\right\rangle
\end{equation*}%
with an amplitude of transition between initial and final states of the form:%
\begin{eqnarray*}
&&\prod\limits_{i}G\left( \left( \mathbf{T}_{i},\mathbf{\alpha }_{i},\mathbf{%
p}_{i},\left( S_{i}\right) ^{2}\right) _{i},\mathbf{T}_{i},\mathbf{\alpha }%
_{i},\mathbf{p}_{i},\left( S_{i}\right) ^{2}\right) \\
&&\ast V\left( \left( \left\{ \mathbf{T}_{i},\mathbf{\alpha }_{i},\mathbf{p}%
_{i}\right\} ,\left( S_{i}\right) ^{2}\right) ,\mathbf{T},\mathbf{\alpha },%
\mathbf{p},S^{2}\right) \ast G\left( \mathbf{T},\mathbf{\alpha },\mathbf{p}%
,S^{2},\left( \mathbf{T},\mathbf{\alpha },\mathbf{p},S^{2}\right) _{f}\right)
\end{eqnarray*}%
where the stars stand for convolutions.

\subsection{Transformations from independent structures to composite ones}

By translating the dynamics in terms of creation and destruction of
structures, the advantge of operator formalism, is to read directly which
terms induce transitions between structures and thus to understand the
dynamical mechanisms of transitions. Consider the action for several
structures in which the possibility of transitions are included:

\begin{eqnarray}
&&S^{\left( O\right) }\left( \left( \mathbf{\alpha },\mathbf{p},S^{2}\right)
\right)  \label{SPR} \\
&=&\sum_{S\times S}\mathbf{\bar{D}}_{S^{2}}^{\alpha }\left( \mathbf{A}%
^{+}\left( \alpha ,p,S^{2}\right) \mathbf{A}^{-}\left( \alpha
,p,S^{2}\right) +\frac{1}{2}\right)  \notag \\
&&+\sum_{m,n}\bar{U}_{mn}\left( \mathbf{\alpha },\mathbf{p},S^{2}\right)
\left( \mathbf{\hat{A}}^{+}\left( \mathbf{\alpha },\mathbf{p},S^{2}\right)
\right) ^{m}\left( \mathbf{\hat{A}}^{-}\left( \mathbf{\alpha },\mathbf{p}%
,S^{2}\right) \right) ^{n}+\hat{V}  \notag
\end{eqnarray}%
This formulation shows the instablity of a state since the form of
interaction always allows apriori for transitions. However, some change of
basis makes possible to integrate out the overall results of these
interactions and to reveal the appearance of resulting stable dressed
structures, having included the action of some structures considered as
auxiliary in this perspective. This change of basis is similar to the
effective action formalism but is more precise in the present approach.

\subsubsection{Transformation of $S^{\left( O\right) }$\ and emergence of
composed structures}

Starting with operators describing transitions between structures, the idea
is to perform a transformation that modifies $S^{\left( O\right) }\left(
\left( \mathbf{\alpha },\mathbf{p},S^{2}\right) \right) $. The
transformation is performed through an operator $\exp \left( -F\right) $,
with $F$ to be determined in order, at least in first approximation, to
diagonalize partially (\ref{SPR}) and cancel the interaction terms between
two types of structures. This terms will be replaced by an effective
interaction terms between a subset of remaining bound structures.

\paragraph{Interaction terms}

Technically, we divide the structures into two sets. The first one labelled
by indices $k$ and $l$ describes the strctrs for which we aim at finding an
effective description. The second set labelled by indices $c$ and $d$
corresponds to structures that will be integrated out to produce effective
interactions in the remaining subset. The interaction between the subsets
takes the form: 
\begin{eqnarray*}
&&\sum_{n,n^{\prime }}\sum_{\left\{ S_{k/c},S_{l/d}\right\} _{\substack{ %
l/d=1,...,n^{\prime }  \\ k/c=1...n}}}\sum_{\left\{ m_{l/d}^{\prime
},m_{k/c}\right\} }\prod_{l/d=1}^{n^{\prime }}\left( \mathbf{\hat{A}}%
^{+}\left( \mathbf{\alpha }_{l/d}^{\prime },\mathbf{p}_{l/d}^{\prime
},S_{l/d}^{\prime 2}\right) \right) ^{m_{l/d}^{\prime }} \\
&&V_{n,n^{\prime }}\left( \left\{ \mathbf{\alpha }_{l/d}^{\prime },\mathbf{p}%
_{l/d}^{\prime },S_{l/d}^{\prime 2},m_{l/d}^{\prime }\right\} ,\left\{ 
\mathbf{\alpha }_{k/c},\mathbf{p}_{k/c},S_{k/c}^{2},m_{k/c}\right\} \right)
\prod_{k/c=1}^{n}\left( \mathbf{\hat{A}}^{-}\left( \mathbf{\alpha }_{k/c},%
\mathbf{p}_{k/c},S_{k/c}^{2}\right) \right) ^{m_{k}}
\end{eqnarray*}%
where indices $l/d$ or $k/c$ indicate that the structures can be of either
type. Our goal is to integrate the crossed interactions:%
\begin{equation}
\prod\limits_{l=1}^{n^{\prime }}\left( \mathbf{\hat{A}}^{+}\left( \mathbf{%
\alpha }_{l}^{\prime },\mathbf{p}_{l}^{\prime },S_{l}^{\prime 2}\right)
\right) ^{m_{l}^{\prime }}V_{n,n^{\prime }}\left( \left\{ \mathbf{\alpha }%
_{l}^{\prime },\mathbf{p}_{l}^{\prime },S_{l}^{\prime 2},m_{l}^{\prime
}\right\} ,\left\{ \mathbf{\alpha }_{c},\mathbf{p}_{c},S_{c}^{2},m_{c}\right%
\} \right) \prod_{k/c=1}^{n}\left( \mathbf{\hat{A}}^{-}\left( \mathbf{\alpha 
}_{c},\mathbf{p}_{c},S_{c}^{2}\right) \right) ^{m_{c}}+\left( \left(
l,c\right) \leftrightarrow \left( d,k\right) \right)  \label{SNR}
\end{equation}%
to obtain an effective action for structures $S_{k}^{2}$, $S_{l}^{\prime 2}$.

\paragraph{Transformation operator}

To integrate the crossed interactions and obtain the required effective
action, we consider the following transformation: 
\begin{equation}
\left( S^{\left( O\right) }\left( \left( \mathbf{\alpha },\mathbf{p}%
,S^{2}\right) \right) \right) ^{\prime }=\exp \left( -F\right) S^{\left(
O\right) }\left( \left( \mathbf{\alpha },\mathbf{p},S^{2}\right) \right)
\exp \left( F\right)  \label{TRS}
\end{equation}%
where $F$ will be found to cancel the interaction term (\ref{SNR}) after
transformation. Doing so modifies $S^{\left( O\right) }\left( \left( \mathbf{%
\alpha },\mathbf{p},S^{2}\right) \right) $ to a descrption of composed,
stable structures.

The derivation is performd in \cite{GLw} and we find to the lowest order
that:

\begin{eqnarray*}
F &=&\sum_{nn^{\prime }}\sum_{\substack{ k=1...n  \\ l=1,...,n^{\prime }}}%
\sum_{\left\{ S_{k},S_{l}\right\} _{\substack{ l=1,...,n^{\prime }  \\ %
k=1...n }}}\prod_{l=1}^{n^{\prime }}\prod\limits_{s=1}^{m_{l}^{\prime }}%
\mathbf{A}^{+}\left( \mathbf{\alpha }_{l}^{\prime },\mathbf{p}_{l}^{\prime
},S_{l}^{\prime 2}\right) \\
&&\times F\left( \left\{ \mathbf{\alpha }_{l}^{\prime },\mathbf{p}%
_{l}^{\prime },S_{l}^{\prime 2},m_{l}^{\prime }\right\} ,\left\{ \mathbf{%
\alpha }_{k},\mathbf{p}_{k},S_{k}^{2},m_{k}\right\} \right)
\prod_{k=1}^{n}\prod\limits_{s=1}^{m_{k}}\mathbf{A}^{-}\left( \mathbf{\alpha 
}_{k},\mathbf{p}_{k},S_{k}^{2}\right)
\end{eqnarray*}%
with:%
\begin{equation}
F\left( \left\{ \mathbf{\alpha }_{l}^{\prime },\mathbf{p}_{l}^{\prime
},S_{l}^{\prime 2},m_{l}^{\prime }\right\} ,\left\{ \mathbf{\alpha }_{k},%
\mathbf{p}_{k},S_{k}^{2},m_{k}\right\} \right) =-\frac{V_{n,n^{\prime
}}\left( \left\{ \mathbf{\alpha }_{l}^{\prime },\mathbf{p}_{l}^{\prime
},S_{l}^{\prime 2},m_{l}^{\prime }\right\} ,\left\{ \mathbf{\alpha }_{k},%
\mathbf{p}_{k},S_{k}^{2},m_{k}\right\} \right) }{\sum_{l=1}^{n^{\prime
}}m_{l}^{\prime }\mathbf{\bar{D}}_{S_{l}^{\prime 2}}^{\mathbf{\alpha }%
_{l}^{\prime }}-\sum_{k=1}^{n}m_{k}\mathbf{\bar{D}}_{S_{k}^{2}}^{\mathbf{%
\alpha }_{k}}}  \label{CMT}
\end{equation}

\subsubsection{Effective structures}

After transformation, the operator version of the action writes:%
\begin{eqnarray}
&&\left( S^{\left( O\right) }\right) ^{\prime }=\sum_{S\times S}\mathbf{\bar{%
D}}_{S^{2}}^{\alpha }\left( \mathbf{A}^{+}\left( \alpha ,p,S^{2}\right) 
\mathbf{A}^{-}\left( \alpha ,p,S^{2}\right) +\frac{1}{2}\right)  \label{FST}
\\
&&+\sum_{n,n^{\prime }}\sum_{\left\{ S_{k},S_{l}\right\} _{\substack{ %
l=1,...,n^{\prime }  \\ k=1...n}}}\sum_{\left\{ m_{l}^{\prime
},m_{k}\right\} }\prod_{l=1}^{n^{\prime }}\left( \mathbf{\hat{A}}^{+}\left( 
\mathbf{\alpha }_{l}^{\prime },\mathbf{p}_{l}^{\prime },S_{l}^{\prime
2}\right) \right) ^{m_{l}^{\prime }}  \notag \\
&&\times V_{n,n^{\prime }}\left( \left\{ \mathbf{\alpha }_{l}^{\prime },%
\mathbf{p}_{l}^{\prime },S_{l}^{\prime 2},m_{l}^{\prime }\right\} ,\left\{ 
\mathbf{\alpha }_{k},\mathbf{p}_{k},S_{k}^{2},m_{k}\right\} \right)
\prod_{k=1}^{n}\left( \mathbf{\hat{A}}^{-}\left( \mathbf{\alpha }_{k},%
\mathbf{p}_{k},S_{k}^{2}\right) \right) ^{m_{k}}  \notag \\
&&+\frac{1}{2}\prod_{L=1}^{n^{\prime }}\prod\limits_{s^{\prime
}=1}^{M_{L}^{\prime }}\mathbf{A}^{+}\left( \mathbf{\alpha }_{L}^{\prime },%
\mathbf{p}_{L}^{\prime },S_{L}^{\prime 2}\right) W\left( \left\{ \mathbf{%
\alpha }_{L}^{\prime },\mathbf{p}_{L}^{\prime },S_{L}^{\prime
2},M_{L}^{\prime }\right\} ,\left\{ \mathbf{\alpha }_{K},\mathbf{p}%
_{K},S_{K}^{2},M_{K}\right\} \right)
\prod\limits_{K=1}^{n}\prod\limits_{s=1}^{M_{k}^{\prime }}\mathbf{A}%
^{-}\left( \mathbf{\alpha }_{K},\mathbf{p}_{K},S_{K}^{2}\right)  \notag
\end{eqnarray}%
where:

\begin{eqnarray}
&&W\left( \left\{ \mathbf{\Lambda }_{L}^{\prime },M_{L}^{\prime }\right\}
,\left\{ \mathbf{\Lambda }_{K},M_{K}\right\} \right)  \label{CMR} \\
&=&-\sum_{P_{K},P_{L}}\sum_{\left\{ \epsilon _{d}^{\prime }\right\} ,\left\{
\epsilon _{c}\right\} }\sum_{\left\{ \delta _{k}\right\} ,\left\{ \delta
_{l}^{\prime }\right\} }\prod \left( \epsilon _{d}^{\prime }!\epsilon
_{c}!\right) ^{2}\prod\limits_{\bar{k},\bar{l},k,l}\left( -1\right) ^{\delta
_{l}^{\prime }}C_{m_{l}^{\prime }+\delta _{l}^{\prime }}^{\delta
_{l}^{\prime }}C_{m_{k}+\delta _{k}}^{\delta _{k}}  \notag \\
&&\times \mathbf{\bar{D}}_{S_{k}^{2}}^{\mathbf{\alpha }_{k}}\mathbf{\bar{D}}%
_{S_{k}^{2}}^{\mathbf{\alpha }_{l}^{\prime }}\mathbf{\bar{D}}_{S_{c}^{2}}^{%
\mathbf{\alpha }_{c}}\mathbf{\bar{D}}_{S_{d}^{2}}^{\mathbf{\alpha }%
_{d}^{\prime }}\delta \left( \mathbf{\Lambda }_{k}-\mathbf{\bar{\Lambda}}_{%
\bar{l}}^{\prime }\right) \delta \left( \mathbf{\Lambda }_{l}^{\prime }-%
\mathbf{\bar{\Lambda}}_{\bar{k}}\right) \delta \left( \mathbf{\Lambda }_{c}-%
\mathbf{\bar{\Lambda}}_{\bar{d}}^{\prime }\right) \delta \left( \mathbf{%
\Lambda }_{d}^{\prime }-\mathbf{\bar{\Lambda}}_{\bar{c}}\right)  \notag \\
&&\times \frac{V^{\left( 2\right) }\left( \left\{ \left( \mathbf{\Lambda }%
_{l\cup d}^{\prime },m_{l}^{\prime }+\delta _{l}^{\prime },\epsilon
_{d}^{\prime }\right) ,\left( \mathbf{\Lambda }_{k\cup c},m_{k}+\delta
_{k},\epsilon _{c}\right) ,\left( \mathbf{\bar{\Lambda}}_{\bar{l}\cup \bar{d}%
}^{\prime },\bar{m}_{^{\bar{l}}}^{\prime }+\delta _{k},\epsilon _{d}\right)
,\left( \mathbf{\bar{\Lambda}}_{\bar{k}\cup \bar{c}},\bar{m}_{^{\bar{k}%
}}+\delta _{l}^{\prime },\epsilon _{c}^{\prime }\right) \right\} \right) }{%
\sum_{l=1}^{n^{\prime }}\left( m_{l}^{\prime }+\delta _{l^{\prime }}^{\prime
}\right) \mathbf{\bar{D}}_{S_{l}^{\prime 2}}^{\mathbf{\alpha }_{l}^{\prime
}}-\sum_{k=1}^{n}\left( m_{k}+\delta _{k}\right) \mathbf{\bar{D}}%
_{S_{k}^{2}}^{\mathbf{\alpha }_{k}}+\sum_{d=1}^{p^{\prime }}\epsilon
_{d}^{\prime }\mathbf{\bar{D}}_{S_{d}^{\prime 2}}^{\mathbf{\alpha }%
_{d}^{\prime }}-\sum_{c=1}^{p}\epsilon _{c}\mathbf{\bar{D}}_{S_{c}^{2}}^{%
\mathbf{\alpha }_{c}}}  \notag
\end{eqnarray}%
where:%
\begin{eqnarray*}
\mathbf{\Lambda }_{k} &=&\left( \mathbf{\alpha }_{k},\mathbf{p}%
_{k},S_{k}^{2}\right) \\
\mathbf{\Lambda }_{l}^{\prime } &=&\left( \mathbf{\alpha }_{l}^{\prime },%
\mathbf{p}_{l}^{\prime },S_{l}^{\prime 2}\right)
\end{eqnarray*}%
and:%
\begin{eqnarray*}
\mathbf{\bar{\Lambda}}_{\bar{k}} &=&\left( \mathbf{\bar{\alpha}}_{\bar{k}},%
\mathbf{\bar{p}}_{\bar{k}},\bar{S}_{\bar{k}}^{2}\right) \\
\mathbf{\bar{\Lambda}}_{\bar{l}}^{\prime } &=&\left( \mathbf{\bar{\alpha}}_{%
\bar{l}}^{\prime },\mathbf{\bar{p}}_{\bar{l}}^{\prime },\bar{S}_{\bar{l}%
}^{\prime 2}\right)
\end{eqnarray*}%
and:%
\begin{eqnarray*}
&&V^{\left( 2\right) }\left( \left\{ \left( \mathbf{\Lambda }_{l\cup
d}^{\prime },m_{l}^{\prime }+\delta _{l}^{\prime },\epsilon _{d}^{\prime
}\right) ,\left( \mathbf{\Lambda }_{k\cup c},m_{k}+\delta _{k},\epsilon
_{c}\right) ,\left( \mathbf{\bar{\Lambda}}_{\bar{l}\cup \bar{d}}^{\prime },%
\bar{m}_{^{\bar{l}}}^{\prime }+\delta _{k},\epsilon _{d}\right) ,\left( 
\mathbf{\bar{\Lambda}}_{\bar{k}\cup \bar{c}},\bar{m}_{^{\bar{k}}}+\delta
_{l}^{\prime },\epsilon _{c}^{\prime }\right) \right\} \right) \\
&=&V\left( \left\{ \mathbf{\Lambda }_{l}^{\prime },m_{l}^{\prime }+\delta
_{l}^{\prime }\right\} \cup \left\{ \mathbf{\Lambda }_{d}^{\prime },\epsilon
_{d}^{\prime }\right\} ,\left\{ \mathbf{\Lambda }_{k},m_{k}+\delta
_{k}\right\} \cup \left\{ \mathbf{\Lambda }_{c},\epsilon _{c}\right\} \right)
\\
&&\times V\left( \left\{ \mathbf{\bar{\Lambda}}_{\bar{l}}^{\prime },\bar{m}%
_{^{\bar{l}}}^{\prime }+\delta _{k}\right\} \cup \left\{ \mathbf{\bar{\Lambda%
}}_{\bar{d}}^{\prime },\epsilon _{c}\right\} ,\left\{ \mathbf{\bar{\Lambda}}%
_{\bar{k}},\bar{m}_{^{\bar{k}}}+\delta _{l}^{\prime }\right\} \cup \left\{ 
\mathbf{\bar{\Lambda}}_{\bar{c}},\epsilon _{d}^{\prime }\right\} \right)
\end{eqnarray*}%
with $P_{K},P_{L}$ are partitions of $\left\{ \mathbf{\alpha }_{K},\mathbf{p}%
_{K},S_{K}^{2},M_{K}\right\} $, $\left\{ \mathbf{\alpha }_{L}^{\prime },%
\mathbf{p}_{L}^{\prime },S_{L}^{\prime 2},M_{L}^{\prime }\right\} $:%
\begin{eqnarray*}
\left\{ \mathbf{\alpha }_{L}^{\prime },\mathbf{p}_{L}^{\prime
},S_{L}^{\prime 2},M_{L}^{\prime }\right\} &=&\left\{ \mathbf{\alpha }%
_{l}^{\prime },\mathbf{p}_{l}^{\prime },S_{l}^{\prime 2},m_{l^{\prime
}}^{\prime }\right\} \cup \left\{ \mathbf{\bar{\alpha}}_{\bar{l}}^{\prime },%
\mathbf{\bar{p}}_{\bar{l}}^{\prime },\bar{S}_{\bar{l}}^{\prime 2},\bar{m}_{%
\bar{l}^{\prime }}^{\prime }\right\} \\
\left\{ \mathbf{\alpha }_{K},\mathbf{p}_{K},S_{K}^{2},M_{K}\right\}
&=&\left\{ \mathbf{\alpha }_{k},\mathbf{p}_{k},S_{k}^{2},m_{k}\right\} \cup
\left\{ \mathbf{\bar{\alpha}}_{\bar{k}},\mathbf{\bar{p}}_{\bar{k}},\bar{S}_{%
\bar{k}}^{2},\bar{m}_{\bar{k}}\right\}
\end{eqnarray*}%
In the commutator (\ref{CMR}), we sum over all of these possible partitions.

Remark that an effective action:%
\begin{equation*}
W\left( \left\{ \mathbf{\alpha }_{L}^{\prime },\mathbf{p}_{L}^{\prime
},S_{L}^{\prime 2},M_{L}^{\prime }\right\} ,\left\{ \mathbf{\alpha }_{K},%
\mathbf{p}_{K},S_{K}^{2},M_{K}\right\} \right)
\end{equation*}%
the structures $\left\{ \mathbf{\alpha }_{c},\mathbf{p}_{c},S_{c}^{2}\right%
\} $ $\left\{ \mathbf{\bar{\alpha}}_{\bar{d}}^{\prime },\mathbf{\bar{p}}_{%
\bar{d}}^{\prime },\bar{S}_{\bar{d}}^{\prime 2}\right\} $ have been
integrated and do not appear anymore in the interaction. They have glued
structures $\left\{ \mathbf{\alpha }_{l}^{\prime },\mathbf{p}_{l}^{\prime
},S_{l}^{\prime 2}\right\} $ and $\left\{ \mathbf{\alpha }_{k},\mathbf{p}%
_{k},S_{k}^{2}\right\} $ even if this ones were not interacting initially
that is, even if:%
\begin{equation}
V_{n,n^{\prime }}\left( \left\{ \mathbf{\alpha }_{l}^{\prime },\mathbf{p}%
_{l}^{\prime },S_{l}^{\prime 2},m_{l}^{\prime }\right\} _{l\leqslant
n^{\prime }},\left\{ \mathbf{\alpha }_{k},\mathbf{p}_{k},S_{k}^{2},m_{k}%
\right\} _{l\leqslant n}\right) =0  \label{VZ}
\end{equation}%
Depending on the form of the resulting interaction (\ref{FST}), some new
combined structures may appear.

\subsubsection{Bound states}

Assuming that condition (\ref{VZ}) is satisfied, we can describe the
combined structures by computing the eigenstates of (\ref{FST}) with lowest
eigenvalues. These states satisfy:%
\begin{equation}
\left( S^{\left( O\right) }\right) \left\vert \prod\limits_{K}\left( \mathbf{%
\alpha }_{K},\mathbf{p}_{K},S_{K}^{2}\right) \right\rangle =\eta _{\left(
M_{K}\right) }\left\vert \prod\limits_{K}\left( \mathbf{\alpha }_{K},\mathbf{%
p}_{K},S_{K}^{2}\right) \right\rangle  \label{Gt}
\end{equation}%
and define boundd stbl stts. Such state is written as a series involving the 
$n$ types of sructures:%
\begin{equation}
\left\vert \prod\limits_{K}\left( \left( \mathbf{\alpha }_{K},\mathbf{p}%
_{K},S_{K}^{2},M_{K}\right) \right) \right\rangle =\sum_{\left( M_{K}\right)
}A\left( \left( \mathbf{\alpha }_{K},\mathbf{p}_{K},S_{K}^{2},M_{K}\right)
\right) \prod\limits_{K=1}^{n}\prod\limits_{s=1}^{M_{K}}\mathbf{A}^{+}\left( 
\mathbf{\alpha }_{K},\mathbf{p}_{K},S_{K}^{2}\right)
\prod\limits_{K}\left\vert Vac\right\rangle _{K}  \label{Bs}
\end{equation}%
The coefficients can be computed recursively to satisfy the eigenstate
equation (\ref{Gt}).

This formula can be read both ways. We can consider also that a "large"
state can be decomposed into a sequence of products of subobjcts. This point
of view is adopted in an other context, in field theory with large number of
states (\cite{G}).

As a consequence, we can read (\ref{Bs}) as a transition equation:%
\begin{equation*}
\left\vert BS\right\rangle \rightarrow \prod\limits_{K}\left\vert \left(
\left( \mathbf{\alpha }_{K},\mathbf{p}_{K},S_{K}^{2},M_{K}\right) \right)
\right\rangle
\end{equation*}%
or:%
\begin{equation*}
\prod\limits_{K}\left\vert \left( \left( \mathbf{\alpha }_{K},\mathbf{p}%
_{K},S_{K}^{2},M_{K}\right) \right) \right\rangle \rightarrow \left\vert
BS\right\rangle
\end{equation*}%
where $\left\vert BS\right\rangle $ is the state defined in (\ref{Bs}).

\section{Systems objects/subobjects}

Until now we have inspected the possible mechanism that would allow some
collective states, or objects to integrate into higher structures, and thus
become some subobjects from this structure. In this section we change the
point of view and consider the consistency, i.e. the condition of stability
for a global structure, or object to coincide with its possible subpart or
subobjects.

The inspection of (\ref{NTBJ}) and (\ref{TRCMH}) and \ (\ref{Bs}) shows
that, in our formalism, while inspecting objects and subobjects, we have to
consider both independent fields for subobjects and the full objects.
Moreover, the term:%
\begin{equation}
\sum_{S_{1}\subseteq S}\underline{\Gamma }^{\dag }\left( \mathbf{T},\mathbf{%
\alpha },\mathbf{p},S_{1}^{2}\right) \left( -\frac{1}{2}\nabla _{\left( 
\mathbf{\hat{T}}\right) _{S_{1}^{2}}}^{2}+\frac{1}{2}\left( \Delta \mathbf{T}%
_{p}^{\alpha }\right) _{S_{1}^{2}}^{t}\mathbf{A}_{S_{1}^{2}}^{\alpha }\left(
\Delta \mathbf{T}_{p}^{\alpha }\right) +\mathbf{C}\right) \underline{\Gamma }%
\left( \mathbf{T},\mathbf{\alpha },\mathbf{p},S_{1}^{2}\right)  \label{sbj}
\end{equation}%
involved in the action functional (\ref{NTBJ}), shows that we have to
consider the field of the full object to be evaluated on each subobjects
extensions. Such terms are necessary to assess some possible stability. This
stability can be reached, at the level of the suboject if this evaluation
corresponds to some minima of the subobject action functional.

If the restriction on some subobject of the full field does not match with
consistency conditions, the corresponding state will be unstable. This
implies that those field are not independent. Said differently, this
corresponds to require that the activation of a subobject as part of a
global object should match with its own stable configurations arising from
the saddle point equations of the subobject action functional. As a
consequence, we should consider as whole, both an object and its subobject.
Since a subobject may itself include some subobjects, the full description
has also to include sequences of potential subobjects.

In terms of states, i.e. functional of fields, this means that we will
consider tensor products of subobjects state spaces. Element of such spaces
are inded b the connectns $\mathbf{T}_{i}$ and the characteristc parameters $%
\mathbf{\alpha }_{i}$ and $\mathbf{p}_{i}$. This description in terms of
state-spaces is necessary to detail the relations and consistency conditions
between objects and subobjects. We thus consider this description, and come
back later to the field-level.

\subsection{Sequences of subobjects}

We consider the space of all possible collective states. with possible
inclusions. Thus one object is not only given by a spatial extensn $\Sigma $
with its characteristic properties in terms of states, but also as a
sequence including all its possible included subobjects and their
characteristics, themselves including potential subsubject... At the level
of spatial extension, a possible state is thus a collection: 
\begin{equation}
\left\{ \left\{ \Sigma _{p,,...,1}\right\} ...%
\begin{array}{c}
\rightarrow \\ 
\rightarrow \\ 
\rightarrow%
\end{array}%
\left\{ \Sigma _{1}\right\} 
\begin{array}{c}
\rightarrow \\ 
\rightarrow \\ 
\rightarrow%
\end{array}%
\Sigma \right\}  \label{Sn}
\end{equation}%
of sequences of sets of subobjcts and their subbjcts along with their
possible internal frequencis. The internal frequencies of subobjects differ
from the frequencies of the object. Note that references to the inclusion of
spatial extensions of subobjects and objects can be discarded and replaced
with an arbitrary mapping from an object to an other object.

This generalization accounts for the fact that the notion of an object may
not be defined by spatial inclusion but rather by certain relationships
between frequencies and activities.

Such sequence (\ref{Sn}) is not sufficient to define some relations of
subobjects since it does not include the label describing average
connections and frequencies of activities, but it suggests that, to define
some stable relation between objects and subobjects, some constraints should
exist between the characteristics of a given object and that of its
subobjects. If these characteristics are completly disconnected, we may
expect that no stable relation exists between the objects and subobjects
except they may be binded temporarily by some interaction term that could be
switched off, or canceled by some other structure interacting with the
object or one of its subobjects. Such situation corresponds to our previous
decscrition where subobjects were described solely by their own field. In
such situation, the larger objects may cover the spatial extension of some
other objects but those "smaller" objects act and interact as, at least
partly, independent structures.

\subsection{State space and restrictions for object subobjects}

\subsubsection{State spaces}

For a collectiv state defined by its spatial extension $\Sigma $, its set of
intrinsic characteristics are the averages connectivits $\left\langle \Delta 
\mathbf{T}\left( \Sigma \times \Sigma \right) \right\rangle _{a}$, their
variations $\left\langle \Delta \mathbf{\hat{T}}\left( \Sigma \times \Sigma
\right) \right\rangle _{a}$, the index $a$ accounts for the multiple
possible states. To such variables are associatd the set of intrinsic
frequencies $\left\{ \gamma _{a,p}\right\} $, where $p$ represnts the
various frequencies, such as harmonics associated to $a$, alng with the
average amplituds $\Delta \omega _{a}$. This corresponds to the set:

\begin{equation*}
\left( \left\langle \Delta \mathbf{T}\left( \Sigma \times \Sigma \right)
\right\rangle _{a},\left\langle \Delta \mathbf{\hat{T}}\left( \Sigma \times
\Sigma \right) \right\rangle _{a},\Delta \omega _{a},\left\{ \gamma
_{a,p}\right\} _{p}\right) _{a}
\end{equation*}%
To this state, we associate the set of states:%
\begin{equation}
F\left( \Sigma \right) =H\left( \left( \left\langle \Delta \mathbf{T}\left(
\Sigma \times \Sigma \right) \right\rangle _{a},\left\langle \Delta \mathbf{%
\hat{T}}\left( \Sigma \times \Sigma \right) \right\rangle _{a},\Delta \omega
_{a},\left\{ \gamma _{a,p}\right\} _{p}\right) _{a}\right)  \label{Stp}
\end{equation}%
generatd by linear combinations and tensor products of states:%
\begin{equation}
\left\vert \left\langle \Delta T\left( Z_{a_{i}},Z_{b_{j}}\right)
\right\rangle ,\left\langle \Delta \hat{T}_{ij}\left(
Z_{a_{i}},Z_{b_{j}}\right) \right\rangle ,\alpha \left(
Z_{a_{i}},Z_{b_{j}}\right) ,p\left( Z_{a_{i}},Z_{b_{j}}\right) \right\rangle
\label{gr}
\end{equation}%
these states have the form (\ref{Stn}).

This descibes a state with variables connectivities $\Delta T\left(
Z_{a_{i}},Z_{b_{j}}\right) $ and connectivities variations $\Delta \hat{T}%
_{ij}\left( Z_{a_{i}},Z_{b_{j}}\right) $ and defined by parameters $\alpha
\left( Z_{a_{i}},Z_{b_{j}}\right) $ describing the average connectivities $%
\left\langle \Delta T\left( Z_{a_{i}},Z_{b_{j}}\right) \right\rangle $ and
connectivities variations $\left\langle \Delta \hat{T}_{ij}\left(
Z_{a_{i}},Z_{b_{j}}\right) \right\rangle $ of the state. The parameters $%
p\left( Z_{a_{i}},Z_{b_{j}}\right) $ represent the frequencies of activities
within the state.

In this context, as in the discussion before (\ref{FCz}), linear
combinations of generators (\ref{gr}):%
\begin{eqnarray*}
&&\int g\left( \left\{ \left\langle \Delta T\left(
Z_{a_{i}},Z_{b_{j}}\right) \right\rangle ,\left\langle \Delta \hat{T}%
_{ij}\left( Z_{a_{i}},Z_{b_{j}}\right) \right\rangle ,\alpha \left(
Z_{a_{i}},Z_{b_{j}}\right) ,p\left( Z_{a_{i}},Z_{b_{j}}\right) \right\}
\right) \\
&&\left\vert \left\langle \Delta T\left( Z_{a_{i}},Z_{b_{j}}\right)
\right\rangle ,\left\langle \Delta \hat{T}_{ij}\left(
Z_{a_{i}},Z_{b_{j}}\right) \right\rangle ,\alpha \left(
Z_{a_{i}},Z_{b_{j}}\right) ,p\left( Z_{a_{i}},Z_{b_{j}}\right) \right\rangle
\end{eqnarray*}%
have to be understood as statistical combinations of large number of states (%
\ref{gr}), ths combinatn describing the global activity for the object.

A dependent subobject is defined by: 
\begin{equation*}
\Sigma _{l}\subset \Sigma
\end{equation*}%
but also by assuming that the characteristics of the object built on $\Sigma 
$ restricted to $\Sigma _{l}$ belongs to the characteristics of the object
built on $\Sigma _{l}$, so that the restriction:%
\begin{equation*}
\left( \mathbf{a}_{i},\mathbf{p}_{i},\left( \cup \Sigma \right) \times
\left( \cup \Sigma \right) \right) \rightarrow \left( \mathbf{a}_{i},\mathbf{%
p}_{i},\left( \cup \Sigma _{a}\right) \times \left( \cup \Sigma _{a}\right)
\right)
\end{equation*}%
belongs to the set of caracteristics for $\Sigma _{l}$.

In this case, we, as in (\ref{Stp}): 
\begin{equation}
F\left( \Sigma _{l}\right) =H\left( \left\langle \Delta \mathbf{T}\left(
\Sigma \times \Sigma \right) \right\rangle _{a_{l}},\left\langle \Delta 
\mathbf{\hat{T}}\left( \Sigma \times \Sigma \right) \right\rangle
_{a_{l}},\Delta \omega _{a_{l}},\left\{ \gamma _{a_{l},p_{l}}\right\}
_{p_{l}}\right) _{a_{l}}  \label{Str}
\end{equation}%
the state space for a subobject $\Sigma _{l}$ and we expect that the
restriction of an element of $F\left( \Sigma \right) $ belongs to $F\left(
\Sigma _{l}\right) $. \ This is detailed below.

\subsubsection{Restrictions between sets of spatial extension and
characteristics}

The spatial extension of object and subobjects, a priori overlap, or at
least are related through maps $\Sigma _{l}\rightarrow \Sigma $. If both
structures do not activate independently, there should be some consitency
conditions between the characteristic frequencies, or any other parameter
characteristc of these structurs. Thus, there should be a map:%
\begin{equation}
F\left( \Sigma \right) \rightarrow F\left( \Sigma _{l}\right)  \label{MPn}
\end{equation}%
mapping at least partly between the states of the full object and those of
its subparts. This map should define a presheaf over the site defined by the
collective state considered and its subobjects. This presheaf property
translates precisely that restrictions of the collective characteristics are
themselves characteritics of the subobjects.

The site consists in:%
\begin{equation*}
S^{c}=\left( \Delta \mathbf{T,}\Delta \mathbf{\hat{T},a}_{i},\mathbf{p}%
_{i},\Sigma \times \Sigma ,\prod \left( \Sigma _{l}\times \Sigma _{l},\left(
\Delta \mathbf{T,}\Delta \mathbf{\hat{T},a}_{i},\mathbf{p}_{i}\right)
_{l}\right) \right)
\end{equation*}%
and the presheaf consists in states over this non-connected space.

However, describing the relations between collective states does not reduce
to maps between sets. We should rather describe the state space for the
system and try to define on these state spaces the equivalent of the maps
between sets. This would correspond to lifting of the inclusion map to the
level of restriction maps of a presheaf.

\section{Description of states in term of functionals and restrictns maps}

In field formalism, the states are described functionals:%
\begin{equation*}
G\left( \underline{\Gamma }_{\Sigma }\left( \Delta \mathbf{T}\left( \Sigma
\right) \mathbf{,}\Delta \mathbf{\hat{T}}\left( \Sigma \right) \mathbf{,}%
a,\gamma _{a}\right) \right)
\end{equation*}%
of multi-component fields $\underline{\Gamma }_{\Sigma }\left( \Delta 
\mathbf{T}\left( \Sigma \right) \mathbf{,}\Delta \mathbf{\hat{T}}\left(
\Sigma \right) \mathbf{,}a,\gamma _{a}\right) $. Here $a,\gamma _{a}$
denotes the possible components for average connectivities and frequency
belonging to $\mathbf{a}_{i},\mathbf{p}_{i}$. The map (\ref{MPn}) can be
translated on these states.

\subsection{Base space and restriction maps}

To describe the states as field functionials, we define first the base space
over $\Sigma $:%
\begin{equation*}
L\left( \Sigma \times \Sigma \right) =\left\{ 
\mathbb{R}
^{\Sigma \times \Sigma }\mathbf{,%
\mathbb{R}
^{\Sigma \times \Sigma },}\left( \left\langle \Delta \mathbf{T}\left( \Sigma
\times \Sigma \right) \right\rangle _{a},\left\langle \Delta \mathbf{\hat{T}}%
\left( \Sigma \times \Sigma \right) \right\rangle _{a},\Delta \omega
_{a},\left\{ \gamma _{a,p}\right\} _{p}\right) _{a}\right\}
\end{equation*}%
This describes the variables on which field object depend: the
connectivities withn the object, but also the possible state of the objects,
in terms of averages, as well as frequencies.

Since we will consider functions on $L\left( \Sigma \times \Sigma \right) $
that are square integrable, we can consider rather:%
\begin{equation*}
L\left( \Sigma \times \Sigma \right) =\left\{ S^{\Sigma \times \Sigma }%
\mathbf{,}S^{\Sigma \times \Sigma }\mathbf{,}\left( \left\langle \Delta 
\mathbf{T}\left( \Sigma \times \Sigma \right) \right\rangle
_{a},\left\langle \Delta \mathbf{\hat{T}}\left( \Sigma \times \Sigma \right)
\right\rangle _{a},\Delta \omega _{a},\left\{ \gamma _{a,p}\right\}
_{p}\right) _{a}\right\}
\end{equation*}%
where $S$ is a two dimensional sphere.

Elements of $L\left( \Sigma \times \Sigma \right) $ have the form: 
\begin{equation*}
\left\{ \Delta \mathbf{T}\left( \Sigma \times \Sigma \right) \mathbf{,}%
\Delta \mathbf{\hat{T}}\left( \Sigma \times \Sigma \right) \mathbf{,}\left(
\left\langle \Delta \mathbf{T}\left( \Sigma \times \Sigma \right)
\right\rangle _{a},\left\langle \Delta \mathbf{\hat{T}}\left( \Sigma \times
\Sigma \right) \right\rangle _{a},\Delta \omega _{a},\left\{ \gamma
_{a,p}\right\} _{p}\right) _{a}\right\}
\end{equation*}%
In addition to the characteristics of the possible collective states over $%
\Sigma $, this include any possible values for connections $\Delta \mathbf{T}
$ and $\Delta \mathbf{\hat{T}}$\ over $\Sigma $ since the dynamics are
described by a field depending of these variables. The previous restriction
on the object-states with respect to a subobject expresses in terms of $%
L\left( \Sigma \times \Sigma \right) $ as: 
\begin{eqnarray}
&&\left\{ \left( \left( \left\langle \Delta \mathbf{T}\left( \Sigma \times
\Sigma \right) \right\rangle _{a}\right) _{\Sigma _{l}},\left( \left\langle
\Delta \mathbf{\hat{T}}\left( \Sigma \times \Sigma \right) \right\rangle
_{a}\right) _{\Sigma _{l}},\Delta \omega _{a},\left\{ \gamma _{a,p}\right\}
_{p}\right) _{a}\right\}  \label{Nc} \\
&\subset &\left\{ \left( \left\langle \Delta \mathbf{T}\left( \Sigma
_{l}\times \Sigma _{l}\right) \right\rangle _{a_{l}},\left\langle \Delta 
\mathbf{\hat{T}}\left( \Sigma _{l}\times \Sigma _{l}\right) \right\rangle
_{a_{l}},\Delta \omega _{a_{l}},\left\{ \gamma _{a_{l},p_{l}}\right\}
_{p_{l}}\right) _{al}\right\}  \notag
\end{eqnarray}%
with:%
\begin{equation*}
\left( \left\langle \Delta \mathbf{T}\left( \Sigma \times \Sigma \right)
\right\rangle _{a}\right) _{\Sigma _{l}},\left( \left\langle \Delta \mathbf{%
\hat{T}}\left( \Sigma \times \Sigma \right) \right\rangle _{a}\right)
_{\Sigma _{l}}
\end{equation*}%
standing for the restrictions of the values of:%
\begin{equation*}
\left( \left\langle \Delta \mathbf{T}\left( \Sigma \times \Sigma \right)
\right\rangle _{a}\right) _{\Sigma _{l}},\left( \left\langle \Delta \mathbf{%
\hat{T}}\left( \Sigma \times \Sigma \right) \right\rangle _{a}\right)
\end{equation*}%
to $\Sigma _{l}$. Condition (\ref{Nc}) is not be satisfied in general. This
condition is satisfied only if some characteristics are shared by $\Sigma $
and $\Sigma _{l}$ which implies a dependnt subobject relation, represented
by a restriction $P_{l}$ when it exists: 
\begin{equation*}
P_{l}:L\left( \Sigma \times \Sigma \right) \rightarrow L\left( \Sigma
_{l}\times \Sigma _{l}\right)
\end{equation*}%
\begin{eqnarray}
P_{l} &:&\left\{ 
\mathbb{R}
^{\Sigma \times \Sigma }\mathbf{,%
\mathbb{R}
^{\Sigma \times \Sigma },}\left( \left\langle \Delta \mathbf{T}\left( \Sigma
\times \Sigma \right) \right\rangle _{a},\left\langle \Delta \mathbf{\hat{T}}%
\left( \Sigma \times \Sigma \right) \right\rangle _{a},\Delta \omega
_{a},\left\{ \gamma _{a,p}\right\} _{p}\right) _{a}\right\}  \label{pl} \\
&\rightarrow &\left\{ 
\mathbb{R}
^{\Sigma _{l}\times \Sigma _{l}}\mathbf{,%
\mathbb{R}
^{\Sigma _{l}\times \Sigma _{l}},}\left( \left( \left\langle \Delta \mathbf{T%
}\left( \Sigma \times \Sigma \right) \right\rangle _{a}\right) _{\Sigma
_{l}},\left( \left\langle \Delta \mathbf{\hat{T}}\left( \Sigma \times \Sigma
\right) \right\rangle _{a}\right) _{\Sigma _{l}},\Delta \omega
_{a_{l}},\left\{ \gamma _{a_{l},p_{l}}\right\} _{p_{l}}\right) _{al}\right\}
\notag
\end{eqnarray}

In the sequel we write: 
\begin{eqnarray*}
&&\left\{ \Delta \mathbf{T}\left( \Sigma \times \Sigma \right) \mathbf{,}%
\Delta \mathbf{\hat{T}}\left( \Sigma \times \Sigma \right) \mathbf{,}\left(
\left\langle \Delta \mathbf{T}\left( \Sigma \times \Sigma \right)
\right\rangle _{a},\left\langle \Delta \mathbf{\hat{T}}\left( \Sigma \times
\Sigma \right) \right\rangle _{a},\Delta \omega _{a},\left\{ \gamma
_{a,p}\right\} _{p}\right) _{a}\right\} \\
&=&\left\{ \Delta \mathbf{T}\left( \Sigma \times \Sigma \right) \mathbf{,}%
\Delta \mathbf{\hat{T}}\left( \Sigma \times \Sigma \right) \mathbf{,}\left(
a,\gamma _{a}\right) \right\}
\end{eqnarray*}

\subsection{Maps between subobject and object field realizations}

Recall that a field is a random variable whose realizations are functions
(in general fields and their realisations are written using the same
notation). The map (\ref{pl}) induces a map of fields $\Pi _{l}$ from $%
\underline{\Gamma }_{\Sigma _{l}}$ to $\underline{\Gamma }_{\Sigma }$ that
transforms a realization of $\underline{\Gamma }_{\Sigma _{l}}$ in a
realization if $\underline{\Gamma }_{\Sigma }$

The map $P_{l}$ induces the map between fields: 
\begin{equation*}
\Pi _{l}\underline{\Gamma }_{\Sigma _{l}}\left( \Delta \mathbf{T}\left(
\Sigma \times \Sigma \right) \mathbf{,}\Delta \mathbf{\hat{T}}\left( \Sigma
\times \Sigma \right) \mathbf{,}\left( a,\gamma _{a}\right) \right) =%
\underline{\Gamma }_{\Sigma _{l}}\left( P_{l}\left( \Delta \mathbf{T}\left(
\Sigma \times \Sigma \right) \mathbf{,}\Delta \mathbf{\hat{T}}\left( \Sigma
\times \Sigma \right) \mathbf{,}\left( a,\gamma _{a}\right) \right) \right)
\end{equation*}%
which transforms a realization of $\underline{\Gamma }_{\Sigma _{l}}$ in a
realization of $\Pi _{l}\underline{\Gamma }_{\Sigma _{l}}$ written $%
\underline{\Gamma }_{\Sigma }$.

\subsection{Maps between object state and subbjct state}

By duality, $\Pi _{l}$ induces a map between the state spaces $H\left(
\Sigma \right) $ and $H\left( \Sigma _{l}\right) $:%
\begin{eqnarray*}
\hat{\Pi}_{l} &:&H\left( \Sigma \right) \rightarrow H\left( \Sigma
_{l}\right) \\
\hat{\Pi}_{l}G\left( \underline{\Gamma }_{\Sigma _{l}}\right) &=&G\left( \Pi
_{l}\underline{\Gamma }_{\Sigma _{l}}\right)
\end{eqnarray*}

In terms of functionals, when this map exists, it is defined by integration
along fibers. Writing a state as a series expansion:%
\begin{equation*}
G\left( \underline{\Gamma }_{\Sigma }\right) =\sum_{r}\sum_{\left( a,\gamma
_{a}\right) }\int g_{r}\left( \left( \Delta \mathbf{T}\left( \Sigma \times
\Sigma \right) \mathbf{,}\Delta \mathbf{\hat{T}}\left( \Sigma \times \Sigma
\right) \mathbf{,}\left( a,\gamma _{a}\right) \right) ^{r}\right) \underline{%
\Gamma }_{\Sigma }^{r}\left( \left( \Delta \mathbf{T}\left( \Sigma \times
\Sigma \right) \mathbf{,}\Delta \mathbf{\hat{T}}\left( \Sigma \times \Sigma
\right) \mathbf{,}\left( a,\gamma _{a}\right) \right) ^{r}\right)
\end{equation*}%
then:%
\begin{eqnarray}
&&\hat{\Pi}_{l}G\left( \underline{\Gamma }_{\Sigma _{l}}\right)  \label{Srn}
\\
&=&\sum_{r}\sum_{\left( a,\gamma _{a}\right) }\int g_{r}\left( \left( \Delta 
\mathbf{T}\left( \Sigma \times \Sigma \right) \mathbf{,}\Delta \mathbf{\hat{T%
}}\left( \Sigma \times \Sigma \right) \mathbf{,}\left( a,\gamma _{a}\right)
\right) ^{r}\right) \underline{\Gamma }_{\Sigma _{l}}^{r}\left( \left(
\Delta \mathbf{T}_{l}\left( \Sigma _{l}\times \Sigma _{l}\right) \mathbf{,}%
\Delta \mathbf{\hat{T}}_{l}\left( \Sigma _{l}\times \Sigma _{l}\right) 
\mathbf{,}\left( a,\gamma _{a}\right) \right) ^{r}\right)  \notag \\
&=&\sum_{r}\sum_{\left( a,\gamma _{a}\right) }\int \hat{\Pi}_{l}g_{r}\left(
\left( \Delta \mathbf{T}_{l}\left( \Sigma _{l}\times \Sigma _{l}\right) 
\mathbf{,}\Delta \mathbf{\hat{T}}_{l}\left( \Sigma _{l}\times \Sigma
_{l}\right) \mathbf{,}\left( a,\gamma _{a}\right) \right) ^{r}\right) 
\underline{\Gamma }_{\Sigma _{l}}^{r}\left( \left( \Delta \mathbf{T}%
_{l}\left( \Sigma _{l}\times \Sigma _{l}\right) \mathbf{,}\Delta \mathbf{%
\hat{T}}_{l}\left( \Sigma _{l}\times \Sigma _{l}\right) \mathbf{,}\left(
a,\gamma _{a}\right) \right) ^{r}\right)  \notag
\end{eqnarray}%
where $\hat{\Pi}_{l}g_{r}$ is is the integration:%
\begin{equation*}
\hat{\Pi}_{l}g_{r}\left( \left( \Delta \mathbf{T}_{l}\left( \Sigma
_{l}\times \Sigma _{l}\right) \mathbf{,}\Delta \mathbf{\hat{T}}_{l}\left(
\Sigma _{l}\times \Sigma _{l}\right) \mathbf{,}\left( a,\gamma _{a}\right)
\right) ^{r}\right) =\int_{F_{\Sigma /\Sigma _{l}}}g_{r}\left( \left( \Delta 
\mathbf{T}\left( \Sigma \times \Sigma \right) \mathbf{,}\Delta \mathbf{\hat{T%
}}\left( \Sigma \times \Sigma \right) \mathbf{,}\left( a,\gamma _{a}\right)
\right) ^{r}\right)
\end{equation*}%
over the fiber $F_{\Sigma /\Sigma _{l}}$ defined by:%
\begin{equation*}
F_{\Sigma /\Sigma _{l}}=\left\{ \Delta \mathbf{T}\left( \Sigma \times \Sigma
\right) \mathbf{,}\Delta \mathbf{\hat{T}}\left( \Sigma \times \Sigma \right)
\right\} /\left\{ \Delta \mathbf{T}_{l}\left( \Sigma _{l}\times \Sigma
_{l}\right) \mathbf{,}\Delta \mathbf{\hat{T}}_{l}\left( \Sigma _{l}\times
\Sigma _{l}\right) \right\}
\end{equation*}

\section{Description of states in terms of series of tensor products}

In this section, we detail the previous state spaces. This allows to refine
the previous description by including some symetry groups. In the
coreespondences induces by maps between objects and subobjects, we should
not require strictly that two states are directly related by restriction
maps, but that they correspond up to some permutation or global continuous
symetry group. This corresponds to consider that a collective state bear
some information that, to some extent, does not rely on a rigid spatial
extension. What matters are rather the coherence between its different
elements.

\subsection{Basic formalism}

The states can be described a sries of tensor products. We rewrite:%
\begin{equation*}
F\left( \Sigma \right) =\underset{\alpha \left( Z_{a},Z_{b}\right)
,p_{\left\{ \alpha \left( Z_{a},Z_{b}\right) \right\} }}{\cup }%
\prod\limits_{\left( Z_{a},Z_{b}\right) \subset \Sigma \times \Sigma }\left(
S^{2}\times S^{2}\times \left\{ \left\langle \Delta T\left(
Z_{a},Z_{b}\right) \right\rangle ,\left\langle \Delta \hat{T}\left(
Z_{a},Z_{b}\right) \right\rangle ,\alpha \left( Z_{a},Z_{b}\right)
,p_{\left\{ \alpha \left( Z_{a},Z_{b}\right) \right\} }\right\} \right)
\end{equation*}

This is fibered above $\Sigma \times \Sigma $ with the possibly of non
constant fiber:%
\begin{equation*}
S^{2}\times S^{2}\times \left\{ \left\langle \Delta T\left(
Z_{a},Z_{b}\right) \right\rangle ,\left\langle \Delta \hat{T}\left(
Z_{a},Z_{b}\right) \right\rangle ,\alpha \left( Z_{a},Z_{b}\right)
,p_{\left\{ \alpha \left( Z_{a},Z_{b}\right) \right\} }\right\}
\end{equation*}%
only $p_{\left\{ \alpha \left( Z_{a},Z_{b}\right) \right\} }$ is constant,
since it depends on the overall connectivities on $\Sigma \times \Sigma $.
If the number of possible $\alpha \left( Z_{a},Z_{b}\right) $ is constant,
this is the trivial sheaf over $\Sigma \times \Sigma $ with fiber $%
S^{2}\times N$ or possibly $S^{2}\times 
\mathbb{Z}
$.

A simple state is an element of the dual space of some test space\
functions: 
\begin{equation*}
D=L\left( \left\{ \underline{\Gamma }:F\left( \Sigma \right) \rightarrow 
\mathbb{C}
\right\} ,%
\mathbb{C}
\right)
\end{equation*}%
where $L$ are linear functions from the test functions $\left\{ \underline{%
\Gamma }:F\left( \Sigma \right) \rightarrow 
\mathbb{C}
\right\} $ to $%
\mathbb{C}
$.

Not every functions are allowed, since the average should be satisfied on
the fibers:%
\begin{equation*}
\frac{\int_{\left( Z_{a},Z_{b}\right) }\Delta T\left( Z_{a},Z_{b}\right)
\left\vert \underline{\Gamma }\right\vert ^{2}}{\int_{\left(
Z_{a},Z_{b}\right) }\left\vert \underline{\Gamma }\right\vert ^{2}}%
=\left\langle \Delta T\left( Z_{a},Z_{b}\right) \right\rangle
\end{equation*}

Including subobjects leads to consider the spaces:%
\begin{equation*}
D_{l}=L\left( \left\{ \underline{\Gamma }_{l}:F\left( \Sigma _{l}\right)
\rightarrow 
\mathbb{C}
\right\} ,%
\mathbb{C}
\right)
\end{equation*}%
and a sequence of state defines some presheaf $\mathbf{D}$ on $\left\{
\Sigma _{l}\rightarrow \Sigma \right\} $ if the matching condition are valid.

In this case, the state of space is the sum of tensor power of $\mathbf{S}$:%
\begin{equation*}
H_{\Sigma }=\sum \mathbf{D}^{\otimes n}
\end{equation*}%
As a consequence $H_{\Sigma }$ is the sum of tensorial power of a presheaf
of functions over $\left\{ \Sigma _{l}\rightarrow \Sigma \right\} $, i.e.
presheaf of functional space over $\left\{ \Sigma _{l}\rightarrow \Sigma
\right\} $. Equivalently, this is a space of functions over the space of
sections of some presheaf over $\left\{ \Sigma _{l}\rightarrow \Sigma
\right\} $.

\subsection{States including symmetries}

\subsubsection{Restriction maps}

The previous description has to be modified if we consider that the action
for the collective state present some symmetry group. Actually, a collective
state for a set of $n^{2}$ connection is described by the vector with $%
2n^{2} $ coordinates:%
\begin{equation*}
\left( \Delta T_{11}\mathbf{,}\Delta \hat{T}_{11},...\right)
\end{equation*}%
The action includes laplacian terms and some potentials. If the background
surrounding the emerging state presents some homogeneity, at least locally,
there should be some symetry groups in the action. The states should be
described by:%
\begin{equation*}
\Sigma ^{c}=\left( \left( S^{2}\right) ^{n^{2}}\times \left( S^{2}\right)
^{n^{2}}/G_{\left( \mathbf{a}_{i},\mathbf{p}_{i}\right) }\mathbf{,}\left( 
\mathbf{a}_{i},\mathbf{p}_{i},\left( \cup \Sigma \right) \times \left( \cup
\Sigma \right) \right) /G_{\left( \mathbf{a}_{i},\mathbf{p}_{i}\right)
}\right)
\end{equation*}%
with $n^{2}$ the number of cells involved in the bound state and with $%
G_{\left( \mathbf{a}_{i},\mathbf{p}_{i}\right) }$ a subgroup of the linear
transformations $SO_{n^{2}}$ depending on $\left( \mathbf{a}_{i},\mathbf{p}%
_{i}\right) $. Thus, we define:%
\begin{eqnarray*}
&&F\left( \Sigma _{G}\right) \\
&\rightarrow &\underset{\alpha \left( Z_{a},Z_{b}\right) ,p_{\left\{ \alpha
\left( Z_{a},Z_{b}\right) \right\} }}{\cup } \prod\limits_{\left(
Z_{a},Z_{b}\right) \subset \Sigma \times \Sigma }\left\{ S^{2}\times
S^{2}\times \left\{ \left\langle \Delta T\left( Z_{a},Z_{b}\right)
\right\rangle ,\left\langle \Delta \hat{T}\left( Z_{a},Z_{b}\right)
\right\rangle ,\alpha \left( Z_{a},Z_{b}\right) ,p_{\left\{ \alpha \left(
Z_{a},Z_{b}\right) \right\} }\right\} \right\} /G_{\Sigma ,\left\{ \alpha
\left( Z_{a},Z_{b}\right) \right\} }
\end{eqnarray*}%
where $G_{\Sigma ,\left\{ \alpha \left( Z_{a},Z_{b}\right) \right\} }$ is
some symmetry group allowing for reducing the degrees of freedom.

Defining $\left( G_{\Sigma ,\left\{ \alpha \left( Z_{a},Z_{b}\right)
\right\} }\right) _{\Sigma _{l}}$ the restriction of $G_{\Sigma ,\left\{
\alpha \left( Z_{a},Z_{b}\right) \right\} }$ to $\Sigma _{l}$, the
symmetries imply the condition: $\left( G_{\Sigma ,\left\{ \alpha \left(
Z_{a},Z_{b}\right) \right\} }\right) _{\Sigma _{l}}\subset G_{\Sigma
_{l},\left\{ \alpha \left( Z_{a_{l}},Z_{b_{l}}\right) \right\} }$ for the
restriction (\ref{pl}) to be defined component by component:%
\begin{eqnarray}
&&P_{l,\alpha \left( Z_{a},Z_{b}\right) ,p_{\left\{ \alpha \left(
Z_{a},Z_{b}\right) \right\} }} \\
&:&\prod\limits_{\left( Z_{a},Z_{b}\right) \subset \Sigma \times \Sigma
}\left\{ S^{2}\times S^{2}\times \left\{ \left\langle \Delta T\left(
Z_{a},Z_{b}\right) \right\rangle ,\left\langle \Delta \hat{T}\left(
Z_{a},Z_{b}\right) \right\rangle ,\alpha \left( Z_{a},Z_{b}\right)
,p_{\left\{ \alpha \left( Z_{a},Z_{b}\right) \right\} }\right\} \right\}
/G_{\Sigma ,\left\{ \alpha \left( Z_{a},Z_{b}\right) \right\} }\rightarrow 
\notag \\
&&\prod\limits_{\left( Z_{a},Z_{b}\right) \subset \Sigma _{l}\times \Sigma
_{l}}\left\{ S^{2}\times S^{2}\times \left\{ \left( \left\langle \Delta 
\mathbf{T}\left( \Sigma \times \Sigma \right) \right\rangle _{a}\right)
_{\Sigma _{l}},\left( \left\langle \Delta \mathbf{\hat{T}}\left( \Sigma
\times \Sigma \right) \right\rangle _{a}\right) _{\Sigma _{l}},\alpha \left(
Z_{a_{l}},Z_{b_{l}}\right) ,p_{\left\{ \alpha \left(
Z_{a_{l}},Z_{b_{l}}\right) \right\} }\right\} \right\} /G_{\Sigma
_{l},\left\{ \alpha \left( Z_{a_{l}},Z_{b_{l}}\right) \right\} }  \notag
\end{eqnarray}%
\noindent

\subsubsection{Functional description}

As before, if the restrictions exist, it lifts to fields and functionals:%
\begin{equation*}
\Pi _{l}\underline{\Gamma }_{\Sigma _{l}}\left( \left( \Delta \mathbf{T}%
\left( \Sigma \times \Sigma \right) \mathbf{,}\Delta \mathbf{\hat{T}}\left(
\Sigma \times \Sigma \right) \mathbf{,}\left( a,\gamma _{a}\right) \right)
/G\right) =\underline{\Gamma }_{\Sigma _{l}}\left( P_{l}\left( \left( \Delta 
\mathbf{T}\left( \Sigma \times \Sigma \right) \mathbf{,}\Delta \mathbf{\hat{T%
}}\left( \Sigma \times \Sigma \right) \mathbf{,}\left( a,\gamma _{a}\right)
\right) /G\right) \right)
\end{equation*}%
wher the notation stands fr component equation $/G_{\Sigma ,\left\{ \alpha
\left( Z_{a},Z_{b}\right) \right\} }$. By duality: 
\begin{eqnarray*}
\hat{\Pi}_{l} &:&H\left( \Sigma \right) \rightarrow H\left( \Sigma
_{l}\right) \\
\hat{\Pi}_{l}G\left( \underline{\Gamma }_{\Sigma _{l}}\right) &=&G\left( \Pi
_{l}\underline{\Gamma }_{\Sigma _{l}}\right)
\end{eqnarray*}

In terms of functionals, this map is defined again by integration along the
fibers. States are series expansion:%
\begin{equation}
G\left( \underline{\Gamma }_{\Sigma }\right) =\sum_{r}\sum_{\left( a,\gamma
_{a}\right) }\int g_{r}\left( \left( \Delta \mathbf{T}\left( \Sigma \times
\Sigma \right) \mathbf{,}\Delta \mathbf{\hat{T}}\left( \Sigma \times \Sigma
\right) \mathbf{,}\left( a,\gamma _{a}\right) \right) ^{r}\right) \Psi
_{\Sigma }^{r}\left( \left( \Delta \mathbf{T}\left( \Sigma \times \Sigma
\right) \mathbf{,}\Delta \mathbf{\hat{T}}\left( \Sigma \times \Sigma \right) 
\mathbf{,}\left( a,\gamma _{a}\right) \right) ^{r}\right)  \label{Srns}
\end{equation}%
where integrals are taken over $\prod\limits_{\left( Z_{a},Z_{b}\right)
\subset \Sigma \times \Sigma }\left( S^{2}\times S^{2}\right) /G_{\Sigma
,\left\{ \alpha \left( Z_{a},Z_{b}\right) \right\} }$

Then:%
\begin{eqnarray*}
\hat{\Pi}_{l}G\left( \underline{\Gamma }_{\Sigma _{l}}\right)
&=&\sum_{r}\sum_{\left( a,\gamma _{a}\right) }\int \hat{\Pi}_{l}g_{r}\left(
\left( \Delta \mathbf{T}_{l}\left( \Sigma _{l}\times \Sigma _{l}\right) 
\mathbf{,}\Delta \mathbf{\hat{T}}_{l}\left( \Sigma _{l}\times \Sigma
_{l}\right) \mathbf{,}\left( a,\gamma _{a}\right) \right) ^{r}\right) \\
&&\times \underline{\Gamma }_{\Sigma _{l}}^{r}\left( \left( \Delta \mathbf{T}%
_{l}\left( \Sigma _{l}\times \Sigma _{l}\right) \mathbf{,}\Delta \mathbf{%
\hat{T}}_{l}\left( \Sigma _{l}\times \Sigma _{l}\right) \mathbf{,}\left(
a,\gamma _{a}\right) \right) ^{r}\right)
\end{eqnarray*}%
where $\hat{\Pi}_{l}g_{r}$ is obtained through integratn along the fibers:%
\begin{equation*}
\hat{\Pi}_{l}g_{r}\left( \left( \Delta \mathbf{T}_{l}\left( \Sigma
_{l}\times \Sigma _{l}\right) \mathbf{,}\Delta \mathbf{\hat{T}}_{l}\left(
\Sigma _{l}\times \Sigma _{l}\right) \mathbf{,}\left( a,\gamma _{a}\right)
\right) ^{r}\right) =\int_{F_{\Sigma /\Sigma _{l}}}g_{r}\left( \left( \Delta 
\mathbf{T}\left( \Sigma \times \Sigma \right) \mathbf{,}\Delta \mathbf{\hat{T%
}}\left( \Sigma \times \Sigma \right) \mathbf{,}\left( a,\gamma _{a}\right)
\right) ^{r}\right)
\end{equation*}%
\begin{equation*}
F_{\Sigma /\Sigma _{l}}=\left( \left\{ \Delta \mathbf{T}\left( \Sigma \times
\Sigma \right) \mathbf{,}\Delta \mathbf{\hat{T}}\left( \Sigma \times \Sigma
\right) \right\} /G\right) /\left( \left\{ \Delta \mathbf{T}_{l}\left(
\Sigma _{l}\times \Sigma _{l}\right) \mathbf{,}\Delta \mathbf{\hat{T}}%
_{l}\left( \Sigma _{l}\times \Sigma _{l}\right) \right\} /G\right)
\end{equation*}

\subsubsection{Description in terms of tensor products}

In this case, the space of sections over these reduced space accounts for
these symetries. Defining test functions:%
\begin{equation*}
\left\{ \underline{\Gamma }:F\left( \Sigma _{G}\right) \rightarrow 
\mathbb{C}
\right\}
\end{equation*}%
where:%
\begin{equation*}
F\left( \Sigma _{G}\right) =\underset{\alpha \left( Z_{a},Z_{b}\right)
,p_{\left\{ \alpha \left( Z_{a},Z_{b}\right) \right\} }}{\cup }F_{\left\{
\alpha \left( Z_{a},Z_{b}\right) \right\} }\left( \Sigma \right) /G_{\Sigma
,\left\{ \alpha \left( Z_{a},Z_{b}\right) \right\} }
\end{equation*}%
nd:%
\begin{equation*}
F_{\left\{ \alpha \left( Z_{a},Z_{b}\right) \right\} }\left( \Sigma \right)
= \left[ \prod\limits_{\left( Z_{a},Z_{b}\right) \subset \Sigma \times
\Sigma }\left\{ S^{2}\times \left\{ \left\langle \Delta T\left(
Z_{a},Z_{b}\right) \right\rangle ,\left\langle \Delta \hat{T}\left(
Z_{a},Z_{b}\right) \right\rangle ,\alpha \left( Z_{a},Z_{b}\right)
,p_{\left\{ \alpha \left( Z_{a},Z_{b}\right) \right\} }\right\} \right\} %
\right]
\end{equation*}%
Here, we have to consider separately the spaces $F_{\left\{ \alpha \left(
Z_{a},Z_{b}\right) \right\} }\left( \Sigma \right) $ as disconnectd fibrs
since the symtrs group depend on the averages, i.e. on the choice of $\alpha
\left( Z_{a},Z_{b}\right) $. The state space is thus:%
\begin{equation*}
D=L\left( \left\{ \underline{\Gamma }:F\left( \Sigma _{G}\right) \rightarrow 
\mathbb{C}
\right\} ,%
\mathbb{C}
\right)
\end{equation*}

The space of sections for subject will be defined similarly:%
\begin{equation*}
D_{l}=L\left( \left\{ \underline{\Gamma }:\underset{\alpha \left(
Z_{a_{l}},Z_{b_{l}}\right) ,p_{\left\{ \alpha \left(
Z_{a_{l}},Z_{b_{l}}\right) \right\} }}{\cup }F_{\left\{ \alpha \left(
Z_{a_{l}},Z_{b_{l}}\right) \right\} }\left( \Sigma _{l}\right) /G_{\Sigma
_{l},\left\{ \alpha \left( Z_{a_{l}},Z_{b_{l}}\right) \right\} }\rightarrow 
\mathbb{C}
\right\} ,%
\mathbb{C}
\right)
\end{equation*}%
with:%
\begin{equation*}
F\left( \Sigma _{lG_{l}}\right) =\underset{\alpha \left(
Z_{a_{l}},Z_{b_{l}}\right) ,p_{\left\{ \alpha \left(
Z_{a_{l}},Z_{b_{l}}\right) \right\} }}{\cup }F_{\left\{ \alpha \left(
Z_{a_{l}},Z_{b_{l}}\right) \right\} }\left( \Sigma _{l}\right) /G_{\Sigma
_{l},\left\{ \alpha \left( Z_{a_{l}},Z_{b_{l}}\right) \right\} }
\end{equation*}%
The various section spaces ths depnds on diffrnt grps dependng on the symtrs
of sbbjcts. These symmetries should imply non-trivial cohomological changes
and conditn the matching conditions between objects.

The state of space is the sum of tensor powers of $\mathbf{D}$:%
\begin{equation*}
H_{\Sigma }=\sum \mathbf{D}^{\otimes n}
\end{equation*}%
and, as before, the projections $P_{l}$ define some presheaf $\mathbf{D}$ on 
$\left\{ \Sigma _{l}\rightarrow \Sigma _{l}\right\} $ if the matching
condition are valid. Here again, the previous descriptn, implies tht $%
H_{\Sigma }$ is the sum of tensorial power of a presheaf of functions
satifying some symetry conditions over $\left\{ \Sigma _{l}\rightarrow
\Sigma \right\} $ , i.e. presheaf of functional space over $\left\{ \Sigma
_{l}\rightarrow \Sigma \right\} $, i.e. a space of functions over the space
of sections of some presheaf over $\left\{ \Sigma _{l}\rightarrow \Sigma
\right\} $.

\section{Field and action functionals for objcts and sbbjects systems}

To describe an object along with its dependent subobjects we should consider
that a configuration includes the full object configurations, and some
independent activations for the subobjects. This means that we should
consider a fld:%
\begin{eqnarray*}
&&\underline{\Gamma }\left( \left[ s\in F\left( \Sigma \right) ,\left[
s_{l}\in F\left( \Sigma _{l}\right) \right] _{l},\left[ s_{lk}\in F\left(
\Sigma _{lk}\right) \right] _{lk},..\right] ,\right. \\
&&\left[ s_{l}^{\left( l\right) }\in F\left( \Sigma _{l}\right) ,\left[
s_{lk}^{\left( l\right) }\in F\left( \Sigma _{lk}\right) \right] _{lk},\left[
s_{lkm}^{\left( l\right) }\in F\left( \Sigma _{lkm}\right) \right] _{lkm},...%
\right] _{l}, \\
&&\left. \left[ s_{lk}^{lk}\in F\left( \Sigma _{lk}\right) ,s_{lkm}^{\left(
lk\right) }\in F\left( \Sigma _{lkm}\right) ,..\right] _{lk},\left[
s_{lkm}^{\left( lkm\right) }\in F\left( \Sigma _{lkm}\right) ,...\right]
_{lkm},..\right)
\end{eqnarray*}%
Alternatively, we replace the object $\Sigma \rightarrow \Sigma _{G}${} and
the subobjects $\Sigma _{l}\rightarrow \Sigma _{lG}$, etc., if certain
symmetries are present. The description of the field object then depends on
all sequences of possible arrows:%
\begin{equation*}
\left\{ ..\rightarrow \Sigma _{lk}\rightarrow \Sigma _{l}\rightarrow \Sigma
\right\}
\end{equation*}%
for the considered object, or:%
\begin{equation*}
\rightarrow \Sigma _{lk}\rightarrow \Sigma _{l}
\end{equation*}%
for sequences starting with some subobject, and so on if we consider
sequences starting at subobjects of subobjects, etc...

To these arrows are associated sequences:%
\begin{equation*}
\left\{ s,s_{l},s_{lk}..\right\}
\end{equation*}%
and:%
\begin{equation*}
\left\{ s_{l}^{\left( l\right) },s_{lk}^{\left( l\right) }..\right\}
\end{equation*}%
with:%
\begin{equation}
s\in \left\{ \Delta \mathbf{T}\left( \Sigma \times \Sigma \right) \mathbf{,}%
\Delta \mathbf{\hat{T}}\left( \Sigma \times \Sigma \right) \mathbf{,}\left(
a,\gamma _{a}\right) \right\}  \label{cd}
\end{equation}%
\begin{equation}
s_{l}\text{ and }s_{l}^{\left( l\right) }\in \Delta \mathbf{T}\left( \Sigma
_{l}\times \Sigma _{l}\right) \mathbf{,}\Delta \mathbf{\hat{T}}\left( \Sigma
_{l}\times \Sigma _{l}\right) \mathbf{,}\left( a_{l},\gamma _{a}\right)
\label{ct}
\end{equation}%
so that the full object is described by considering all activations of the
object and its subobjects, as well as all possible independent activations
of the subobjects along with their sequences of subobject activtions. This
framework allows the system to be analyzed as a whole, with supplementary
activity contributed by subobjcts, or alternatively, to isolate specific
subobjcts and their hierarchical set of subobjcts derived from them.{}

We write the field as:%
\begin{equation*}
\underline{\Gamma }\left( \left[ s,\left[ s_{l}\right] _{l},,\left[ s_{lk}%
\right] _{lk},..\right] ,\left[ s_{l}^{\left( l\right) },\left[
s_{lk}^{\left( l\right) }\right] _{lk},\left[ s_{lkm}^{\left( l\right) }%
\right] _{lkm},...\right] _{l},\left[ s_{lk}^{lk},\left[ s_{lkm}^{\left(
lk\right) }..\right] _{lkm}\right] _{lk},\left[ s_{lkm}^{\left( lkm\right)
},...\right] _{lkm},..\right)
\end{equation*}%
Alternatively, we can consider:%
\begin{equation*}
\underline{\Gamma }\left( s\in F\left( \Sigma \right) ,...,\left[ s_{l}\in
F\left( \Sigma _{l}\right) .\right] _{l},\left[ s_{lk}\in F\left( \Sigma
_{lk}\right) \right] _{lk},.\right)
\end{equation*}%
alng with:%
\begin{equation*}
\underline{\Gamma }\left( \left[ s_{l}^{\left( l\right) }\in F\left( \Sigma
_{l}\right) ,\left[ s_{lk}^{\left( l\right) }\in F\left( \Sigma _{lk}\right) %
\right] _{lk},...\right] _{l}\right) \text{, }\underline{\Gamma }\left( %
\left[ s_{lk}^{lk}\in F\left( \Sigma _{lk}\right) \right] _{lk},\left[
s_{lkm}^{\left( lk\right) }\in F\left( \Sigma _{lkm}\right) \right]
,..\right) \text{,..}
\end{equation*}%
This amounts to consider the subobject as partly independent element to
account at least for independent activations of some part of the overall
structre. Both possibilities generates the same state spaces as series of
tensors, except that the first possibility already includes the series
expansion in terms of subobject states as basic state. This may seem
overdetermined and redundant, but takes into account that the full object
should be described not only as entangled subparts but also directly as the
exposition of the whole set of configurations and relations.

The full action for the system should thus be:%
\begin{eqnarray}
&&\underline{\Gamma }^{\dag }\left( \left[ s,\left[ s_{l}\right] _{l},,\left[
s_{lk}\right] _{lk},..\right] ,\left[ s_{l}^{\left( l\right) },\left[
s_{lk}^{\left( l\right) }\right] _{lk},\left[ s_{lkm}^{\left( l\right) }%
\right] _{lkm},...\right] _{l},\left[ s_{lk}^{lk},\left[ s_{lkm}^{\left(
lk\right) }..\right] _{lkm}\right] _{lk},\left[ s_{lkm}^{\left( lkm\right)
},...\right] _{lkm},..\right)  \label{Tn} \\
&&\times \left( -\frac{1}{2}\nabla _{\left( \mathbf{T}\right) _{\Sigma
^{2}}}^{2}+\frac{1}{2}\left( \Delta \mathbf{T}_{p}^{\alpha }\right) _{\Sigma
^{2}}^{t}\mathbf{A}_{\Sigma ^{2}}^{\alpha }\left( \Delta \mathbf{T}%
_{p}^{\alpha }\right) +\mathbf{C}\right) +\sum_{s}\left( -\frac{1}{2}\nabla
_{\left( \mathbf{T}\right) _{\Sigma _{s}^{2}}}^{2}+\frac{1}{2}\left( \Delta 
\mathbf{T}_{p}^{\alpha }\right) _{_{\Sigma _{s}^{2}}}^{t}\mathbf{A}%
_{_{\Sigma _{s}^{2}}}^{\alpha }\left( \Delta \mathbf{T}_{p}^{\alpha }\right)
+\mathbf{C}\right)  \notag \\
&&\times \underline{\Gamma }\left( \left[ s,\left[ s_{l}\right] _{l},,\left[
s_{lk}\right] _{lk},..\right] ,\left[ s_{l}^{\left( l\right) },\left[
s_{lk}^{\left( l\right) }\right] _{lk},\left[ s_{lkm}^{\left( l\right) }%
\right] _{lkm},...\right] _{l},\left[ s_{lk}^{lk},\left[ s_{lkm}^{\left(
lk\right) }..\right] _{lkm}\right] _{lk},\left[ s_{lkm}^{\left( lkm\right)
},...\right] _{lkm},..\right)  \notag
\end{eqnarray}%
and should also include the sum over all possible subobject contributions,
subobject of subobjects:%
\begin{eqnarray}
&&\underline{\Gamma }^{\dag }\left( \left[ s,\left[ s_{l}\right] _{l},,\left[
s_{lk}\right] _{lk},..\right] ,\left[ s_{l}^{\left( l\right) },\left[
s_{lk}^{\left( l\right) }\right] _{lk},\left[ s_{lkm}^{\left( l\right) }%
\right] _{lkm},...\right] _{l},\left[ s_{lk}^{lk},\left[ s_{lkm}^{\left(
lk\right) }..\right] _{lkm}\right] _{lk},\left[ s_{lkm}^{\left( lkm\right)
},...\right] _{lkm},..\right)  \label{Td} \\
&&\times \sum_{st}\left( -\frac{1}{2}\nabla _{\left( \mathbf{T}\right)
_{\Sigma _{st}^{2}}}^{2}+\frac{1}{2}\left( \Delta \mathbf{T}_{p}^{\alpha
}\right) _{_{\Sigma _{st}^{2}}}^{t}\mathbf{A}_{_{\Sigma _{st}^{2}}}^{\alpha
}\left( \Delta \mathbf{T}_{p}^{\alpha }\right) +\mathbf{C}\right)  \notag \\
&&\times \underline{\Gamma }\left( \left[ s,\left[ s_{l}\right] _{l},,\left[
s_{lk}\right] _{lk},..\right] ,\left[ s_{l}^{\left( l\right) },\left[
s_{lk}^{\left( l\right) }\right] _{lk},\left[ s_{lkm}^{\left( l\right) }%
\right] _{lkm},...\right] _{l},\left[ s_{lk}^{lk},\left[ s_{lkm}^{\left(
lk\right) }..\right] _{lkm}\right] _{lk},\left[ s_{lkm}^{\left( lkm\right)
},...\right] _{lkm},..\right)  \notag
\end{eqnarray}%
Remark tht there is some redundency in the presence of: 
\begin{equation*}
-\frac{1}{2}\nabla _{\left( \mathbf{T}\right) _{\Sigma _{s}^{2}}}^{2}+\frac{1%
}{2}\left( \Delta \mathbf{T}_{p}^{\alpha }\right) _{_{\Sigma _{s}^{2}}}^{t}%
\mathbf{A}_{_{\Sigma _{s}^{2}}}^{\alpha }\left( \Delta \mathbf{T}%
_{p}^{\alpha }\right) +\mathbf{C}
\end{equation*}%
that accounts for global activation but also fr independent activation. This
corresponds to a partial independency of subobjects. Satisfying consistency
conditions due to the restriction map, some states peculiar to the
subobjects may nevertheless activte. This activtion may ultimately modify
the whl state.

\subsection{Stability and presheaf property}

The sum (\ref{Tn}) nd (\ref{Td}) shows that the minimal value of the action
functionl is reached for field configurations such that a full object as a
whole describd by $s$ has to select one common minimum to its subobjects to
minimize the action. This corresponds to systems that satisfy some presheaf
property quoted above. We can refine this property by considering states
that satify the presheaf, or sheaf, property.

Considering an object plus its subobjects, along with their subobjects:%
\begin{equation}
\begin{array}{c}
\begin{array}{c}
\rightarrow \\ 
\rightarrow%
\end{array}%
\Sigma _{ab}^{c}/G_{ab}\rightarrow \\ 
\begin{array}{c}
\rightarrow \\ 
\rightarrow%
\end{array}%
\Sigma _{ab^{\prime }}^{c}/G_{ab^{\prime }}\rightarrow%
\end{array}%
\Sigma _{a}^{c}/G_{a}\rightarrow \Sigma ^{c}/G  \label{Cv}
\end{equation}%
along with restrictions:%
\begin{equation*}
F\left( \Sigma ^{c}/G\right) \rightarrow 
\begin{array}{c}
F\left( \Sigma _{ab}^{c}/G_{ab}\right) 
\begin{array}{c}
\rightarrow \\ 
\rightarrow%
\end{array}
\\ 
F\left( \Sigma _{ab^{\prime }}^{c}/G_{ab^{\prime }}\right) 
\begin{array}{c}
\rightarrow \\ 
\rightarrow%
\end{array}%
\end{array}%
\end{equation*}%
the condition for sheaf implies that subobjects with states matching on some
intersection should be extended to an object whose subobjects are given by
restrictns. Such object should be stable with respect to perturbation
signals. The same applies to subobjects considered independently. The
presheaf property is less constraining and imposes that the states of
subobjects match on the spatial extension of their common subobjects.

Not considering sheaf or presheaf condition makes this stability
problematic. Other configurations are unstable. One object could be
dislocated into several subobjects that could not rearrange in the initial
object. If not even presheaf property is satisfied, the object contains
subojects only in a constrained manner. Their existence and conditions as
part of the dominant object is only maintained by this object.
Perturbations, dislocations should divide the initial assembly as
independent parts with different caracteristics. \ Actually, the stability
of object witht presheaf condition is possible only if the system is bound
by some interaction terms added to (\ref{Td}):%
\begin{eqnarray}
&&\underline{\Gamma }^{\dag }\left( \left[ s,\left[ s_{l}\right] _{l},,\left[
s_{lk}\right] _{lk},..\right] ,\left[ s_{l}^{\left( l\right) },\left[
s_{lk}^{\left( l\right) }\right] _{lk},\left[ s_{lkm}^{\left( l\right) }%
\right] _{lkm},...\right] _{l},\left[ s_{lk}^{lk},\left[ s_{lkm}^{\left(
lk\right) }..\right] _{lkm}\right] _{lk},\left[ s_{lkm}^{\left( lkm\right)
},...\right] _{lkm},..\right) \\
&&\times V\left( \left[ s,\left[ s_{l}\right] _{l},,\left[ s_{lk}\right]
_{lk},..\right] \right)  \notag \\
&&\times \underline{\Gamma }\left( \left[ s,\left[ s_{l}\right] _{l},,\left[
s_{lk}\right] _{lk},..\right] ,\left[ s_{l}^{\left( l\right) },\left[
s_{lk}^{\left( l\right) }\right] _{lk},\left[ s_{lkm}^{\left( l\right) }%
\right] _{lkm},...\right] _{l},\left[ s_{lk}^{lk},\left[ s_{lkm}^{\left(
lk\right) }..\right] _{lkm}\right] _{lk},\left[ s_{lkm}^{\left( lkm\right)
},...\right] _{lkm},..\right)  \notag
\end{eqnarray}%
with a minimum corresponding to the characteristics of the full subobject.

However, if the object interacts with another field $\underline{\Gamma }%
^{\prime }$ describing an other distinct object, the interaction term can be
expressed as follows:%
\begin{equation*}
V^{\prime }\left( \underline{\Gamma }^{\dag }\underline{\Gamma }^{\prime
}+\left( \underline{\Gamma }^{\prime }\right) ^{\dag }\underline{\Gamma }%
\right)
\end{equation*}%
with:%
\begin{equation*}
V^{\prime }=V^{\prime }\left( \left[ s_{l}^{\left( l\right) },\left[
s_{lk}^{\left( l\right) }\right] _{lk},\left[ s_{lkm}^{\left( l\right) }%
\right] _{lkm},...\right] _{l},\left[ s_{l}^{\left( l\right) },\left[
s_{lk}^{\left( l\right) }\right] _{lk},\left[ s_{lkm}^{\left( l\right) }%
\right] _{lkm},...\right] _{l}^{\prime }\right)
\end{equation*}%
This term involves certain subobjects from the two different objects. As
explained in \cite{GLw}, such a potential may induce the subobjects of each
distinct objct\ to bind in different configurations.

\subsection{Presheaf properties and transitions}

The states of an object satisfying the presheaf property do not prevent a
subobject from activating other configurations independently, which should
be tracked by a subfield. These independent activations may be responsible
for binding with other structures and inducing transitions.

Conversely, in order to bind, several subobjects must reach a stable
configuration within the space defined by a larger object. This can be
achieved if the subobject's states can be extended as part of the larger
structure. This means that such an extension is possible if certain
conditions for a sheaf, not just a presheaf, are satisfied. As a result, due
to interactions, some subobjects that were initially bound by some
interactions may break into pieces, with no possibility of restoring the
original configuration.

On the other hand, some subobjects, even if they do not define a global
state, could at least overlap locally in a more or less stable way.
Moreover, even if a presheaf or sheaf condition is satisfied, some
non-equivalent configurations may still be reached by a system
object/subobjects. No continuous dynamics should allow for a transition from
one configuration to another. This discontinuity is measured by some \v{C}%
ech cohomology for the presheaf of states over the covering (\ref{Cv}).

\subsection{Description of states for objects/subobjects in terms of class}

The states induced by the field (\ref{Td}) are given by (a series of) tensor
products of distributions over test functions:

\begin{equation*}
\left[ s,\left[ s_{l}\right] _{l},,\left[ s_{lk}\right] _{lk},..\right] ,%
\left[ s_{l}^{\left( l\right) },\left[ s_{lk}^{\left( l\right) }\right]
_{lk},\left[ s_{lkm}^{\left( l\right) }\right] _{lkm},...\right] _{l},\left[
s_{lk}^{lk},\left[ s_{lkm}^{\left( lk\right) }..\right] _{lkm}\right] _{lk},%
\left[ s_{lkm}^{\left( lkm\right) },...\right] _{lkm},..
\end{equation*}%
that is product of functions of the form:%
\begin{eqnarray*}
&&\left[ \psi \left( s\right) ,\left[ \psi _{l}\left( s_{l}\right) ..\right]
_{l},\left[ \psi _{lk}\left( s_{lk}\right) \right] _{lk},..\right] \\
&&\left[ \psi _{l}^{\left( l\right) }\left( s_{l}^{\left( l\right) }\right) ,%
\left[ \psi _{l}^{\left( l\right) }\left( s_{lk}^{\left( l\right) }\right)
,...\right] _{lk},..\right] _{l},\left[ \psi _{lk}^{\left( lk\right) }\left(
s_{lk}^{lk}\right) ,\left[ \psi _{lkm}^{\left( lk\right) }\left(
s_{lkm}^{\left( lk\right) }\right) \right] _{lkm}..\right] _{lk},\left[ \psi
_{lkm}^{lkm}\left( s_{lkm}^{\left( lkm\right) }\right) ,...\right] _{lkm},..
\end{eqnarray*}%
Similary to (\ref{Srn}) and(\ref{Srns}), the distributions over such
functions are given by the product of sequences:%
\begin{eqnarray}
&&\left[ g\left( s\right) ,\left[ g_{l}\left( s_{l}\right) \right] _{l},%
\left[ g_{lk}\left( s_{lk}\right) \right] _{lk},..\right] ,  \label{Cfg} \\
&&\left[ g_{l}^{\left( l\right) }\left( s_{l}^{\left( l\right) }\right) ,%
\left[ g_{l}^{\left( l\right) }\left( s_{lk}^{\left( l\right) }\right) %
\right] _{lk},..\right] _{l},\left[ g_{lk}^{\left( lk\right) }\left(
s_{lk}^{lk}\right) ,g_{lkm}^{\left( lk\right) }\left( s_{lkm}^{\left(
lk\right) }\right) ..\right] _{lk},\left[ g_{lkm}^{lkm}\left(
s_{lkm}^{\left( lkm\right) }\right) ,...\right] _{lkm},..  \notag
\end{eqnarray}%
Each of these functions has restrictions defined by integration along fiber
bundles. These components describe sections over $\Sigma $ and encompass all
possible elements of the covering, consisting of $\Sigma _{l}$.

Consider first states $g_{l}\left( s_{l}\right) $ and $g_{k}\left(
s_{k}\right) $ that do not match on $\Sigma _{lk}$, up to a symetry
transformation on $\Sigma _{lk}$. This amounts to the activation on some $%
g_{lk}\left( s_{lk}\right) $%
\begin{equation*}
g_{lk}\left( s_{lk}\right) =\frac{\prod\limits_{lk}g_{l}\left( s_{l}\right) 
}{\prod\limits_{kl}g_{k}\left( s_{k}\right) }
\end{equation*}%
where $\prod\limits_{lk}$ and $\prod\limits_{kl}$ are the restriction
applications. Given that these restrictions are projction, there is no
reason that $g_{lk}\left( s_{lk}\right) $ can be extended to $\Sigma _{l}$
and $\Sigma _{k}$. This activation measures the discrepency between the
local states $g_{l}\left( s_{l}\right) $ and $g_{k}\left( s_{k}\right) $.
This means that the activation $g_{lk}\left( s_{lk}\right) $ is an
activation of a state over $\Sigma _{lk}$, describing a different state from
the states $g_{l}\left( s_{l}\right) $ and $g_{k}\left( s_{k}\right) $. Such
state should be stable, only if $g_{l}\left( s_{l}\right) $, $g_{k}\left(
s_{k}\right) $ and $g_{lk}\left( s_{lk}\right) $ correspond to some states
with localized minimal action. Such states $g_{lk}\left( s_{lk}\right) $
corresponding to some activation should not be independent.

On the contrary, when states $g_{l}\left( s_{l}\right) $ and $g_{k}\left(
s_{k}\right) $ match on $\Sigma _{lk}$, up to a symetry transformation on $%
\Sigma _{lk}$ that is:%
\begin{equation}
g_{lk}\left( s_{lk}\right) =\frac{\hat{g}_{l}\left( s_{l}\right) }{\hat{g}%
_{k}\left( s_{k}\right) }  \label{Mt}
\end{equation}%
can be transformed to a trivial state, the function $1$, corresponding to
the vacuum, i.e. no activation in terms of tensor prodcts. This means that
the subobjects $g_{l}\left( s_{l}\right) $ and $g_{k}\left( s_{k}\right) $
may correspond to some global object wth restriction $\frac{g_{l}\left(
s_{l}\right) }{\hat{g}_{l}\left( s_{l}\right) }$ and $\frac{g_{k}\left(
s_{k}\right) }{\hat{g}_{k}\left( s_{k}\right) }$. The subsets can be
twisted, potentially continuously, to define a global object. In this case,
the local activations $\left\{ g_{l}\left( s_{l}\right) \right\} $ are those
of a global state

We can classify the collections of subobjects states into inequivalent
classes. Actually, given consistency in compositions of restrictions:%
\begin{eqnarray*}
\prod\limits_{lkm}\frac{\prod\limits_{lk}g_{l}\left( s_{l}\right) }{%
\prod\limits_{kl}g_{k}\left( s_{k}\right) }\frac{\prod\limits_{km}g_{k}%
\left( s_{k}\right) }{\prod\limits_{mk}g_{m}\left( s_{m}\right) }\frac{%
\prod\limits_{ml}g_{m}\left( s_{m}\right) }{\prod\limits_{lm}g_{l}\left(
s_{l}\right) } &=&\frac{\prod\limits_{lkm}g_{l}\left( s_{l}\right) }{%
\prod\limits_{lkm}g_{k}\left( s_{k}\right) }\frac{\prod\limits_{lkm}g_{k}%
\left( s_{k}\right) }{\prod\limits_{lkm}g_{m}\left( s_{m}\right) }\frac{%
\prod\limits_{lkm}g_{m}\left( s_{m}\right) }{\prod\limits_{lkm}g_{l}\left(
s_{l}\right) } \\
&\rightarrow &1
\end{eqnarray*}%
we have the relation: 
\begin{equation}
\prod\limits_{lkm}\left[ g_{lk}\left( s_{lk}\right) \right]
_{lk}\prod\limits_{lkm}\left[ g_{km}\left( s_{km}\right) \right]
_{km}\prod\limits_{lkm}\left[ g_{ml}\left( s_{ml}\right) \right] _{m}=1
\label{Cln}
\end{equation}%
Thus, when the presheaf hypothesis is satisfied, and if we assume that
consistency conditions between states are met, we can define cohomology
classes over $\Sigma _{lk}$. The various elements form cycles, i.e. over
intersections $\Sigma _{lk}$ that define each some cohomology class. The
same reasoning applies to the to the $g_{lkm...}\left( s_{lkm...}\right) $
so that the states of the objects: 
\begin{equation*}
\left[ s,\left[ s_{l}\right] _{l},,\left[ s_{lk}\right] _{lk},..\right] ,%
\left[ s_{l}^{\left( l\right) },\left[ s_{lk}^{\left( l\right) }\right]
_{lk},\left[ s_{lkm}^{\left( l\right) }\right] _{lkm},...\right] _{l},\left[
s_{lk}^{lk},\left[ s_{lkm}^{\left( lk\right) }..\right] _{lkm}\right] _{lk},%
\left[ s_{lkm}^{\left( lkm\right) },...\right] _{lkm},..
\end{equation*}
that are consisitent over intersections, are classified between sequences of
inequivalent classes of cech cohomology. The subobjects that do not satisfy
the consistencies correspond to some independent activations, with their
sequence of subobject being themselves consistent or not.

Configurations (\ref{Cfg}) should thus be labelled by sequences of cech
cohomology classes, both for the initial objects describing global and local
activation, but also for all possible subobjects.

The fields should thus be described as defined on some presheaf (or sheaf)
over the spatial extension of the objects characterized by some given cech
cohomology classes. The transitions between states with different classes,
should arise as global discontinuous transitions rather than continuous
modifications.

States depend on possible internal frequencies of object and subobjects.
Moreover presence of some symetries implies that the sections of the
presheaves defining these states should represent these symetries. The
classes characterizing a state thus depend on these representations and on
the frequencies, and states belonging to different classes present different
type of frequencies.

If a collection of states $\left\{ s_{l}^{\left( l\right) }\right\} $, i.e.
several subobjects with spatial extension $\Sigma _{l}$ and independent
activation define a given cohomology class, this implies that such states
match on their intersections $\Sigma _{lk}$, but cannot be gathered as a
global object activation. This may be stable but may interact destructibely
with the global object. This represents activations that do not fit with the
range of possible frequencies and amplitude allowed as a global objct.

\section{Interactions and transitions}

Operator formalism fits with the description of equivalence class of states.
Actually, the potential (\ref{NT}):%
\begin{eqnarray}
&&\hat{V}=\sum_{n,n^{\prime }}\sum_{\left\{ S_{k},S_{l}\right\} _{\substack{ %
l=1,...,n^{\prime }  \\ k=1...n}}}\sum_{\left\{ m_{l}^{\prime
},m_{k}\right\} }\prod_{l=1}^{n^{\prime }}\left( \mathbf{\hat{A}}^{+}\left( 
\mathbf{\alpha }_{l}^{\prime },\mathbf{p}_{l}^{\prime },S_{l}^{\prime
2}\right) \right) ^{m_{l}^{\prime }} \\
&&\times V_{n,n^{\prime }}\left( \left\{ \mathbf{\alpha }_{l}^{\prime },%
\mathbf{p}_{l}^{\prime },S_{l}^{\prime 2},m_{l}^{\prime }\right\}
_{l\leqslant n^{\prime }},\left\{ \mathbf{\alpha }_{k},\mathbf{p}%
_{k},S_{k}^{2},m_{k}\right\} _{l\leqslant n}\right) \prod_{k=1}^{n}\left( 
\mathbf{\hat{A}}^{-}\left( \mathbf{\alpha }_{k},\mathbf{p}%
_{k},S_{k}^{2}\right) \right) ^{m_{k}}  \notag
\end{eqnarray}%
provides a mechanism of transition induced by interactions. Actually,
consider a state (\ref{Cfg}) composed of eigenstates of some product of
operators $\mathbf{\hat{A}}^{+}\left( \mathbf{\alpha }_{l}^{\prime },\mathbf{%
p}_{l}^{\prime },S_{l}^{\prime 2}\right) \mathbf{\hat{A}}^{-}\left( \mathbf{%
\alpha }_{k},\mathbf{p}_{k},S_{k}^{2}\right) $ describing collctive state.

The action of interaction operator:%
\begin{equation*}
\left( \mathbf{\hat{A}}^{+}\left( \mathbf{\alpha }_{l}^{\prime },\mathbf{p}%
_{l}^{\prime },S_{l}^{\prime 2}\right) \right) ^{m_{l}^{\prime }}\left( 
\mathbf{\hat{A}}^{-}\left( \mathbf{\alpha }_{k},\mathbf{p}%
_{k},S_{k}^{2}\right) \right) ^{m_{k}}
\end{equation*}%
switches directly a state (\ref{Cfg}) into a state:%
\begin{eqnarray}
&&\left[ g\left( s\right) ,\left[ g_{l}\left( s_{l}\right) \right] _{l},%
\left[ g_{lk}\left( s_{lk}\right) \right] _{lk},..\right] ^{\prime }, \\
&&\left[ g_{l}^{\left( l\right) }\left( s_{l}^{\left( l\right) }\right) ,%
\left[ g_{l}^{\left( l\right) }\left( s_{lk}^{\left( l\right) }\right) %
\right] _{lk},..\right] _{l}^{\prime },\left[ g_{lk}^{\left( lk\right)
}\left( s_{lk}^{lk}\right) ,g_{lkm}^{\left( lk\right) }\left(
s_{lkm}^{\left( lk\right) }\right) ..\right] _{lk}^{\prime },\left[
g_{lkm}^{lkm}\left( s_{lkm}^{\left( lkm\right) }\right) ,...\right]
_{lkm}^{\prime },..  \notag
\end{eqnarray}%
Where the "prime superscript denotes the modification of all or part of the
states:%
\begin{equation*}
\left[ g_{l}\left( s_{l}\right) \right] _{l}\rightarrow \left[ g_{l}\left(
s_{l}\right) \right] _{l}^{\prime }\text{, }\left[ g_{l}^{\left( l\right)
}\left( s_{lk}^{\left( l\right) }\right) \right] _{lk}^{\prime }
\end{equation*}%
for instance. Such a modification corresponds to our previous description in
terms of the modification of the class of the state, apart from some
deformation that occurs once the new class of the state is reached.

The operators $\mathbf{\hat{A}}^{+}\left( \mathbf{\alpha }_{l}^{\prime },%
\mathbf{p}_{l}^{\prime },S_{l}^{\prime 2}\right) $, $\mathbf{\hat{A}}%
^{-}\left( \mathbf{\alpha }_{l}^{\prime },\mathbf{p}_{l}^{\prime
},S_{l}^{\prime 2}\right) $ should be decomposed into a part that modifies
the class and a part that does not:%
\begin{equation*}
\mathbf{\hat{A}}^{\pm }\left( \mathbf{\alpha }_{l}^{\prime },\mathbf{p}%
_{l}^{\prime },S_{l}^{\prime 2}\right) =\mathbf{\hat{A}}_{C}^{\pm }\left( 
\mathbf{\alpha }_{l}^{\prime },\mathbf{p}_{l}^{\prime },S_{l}^{\prime
2}\right) +\mathbf{\hat{A}}_{D}^{\pm }\left( \mathbf{\alpha }_{l}^{\prime },%
\mathbf{p}_{l}^{\prime },S_{l}^{\prime 2}\right)
\end{equation*}

Here, the subscript $C$ denotes the component associated with the change of
class, and $D$ corresponds to the deformation. This decomposition is not
unique and can be defined such that $\mathbf{\hat{A}}_{C}^{\pm }\left( 
\mathbf{\alpha }_{l}^{\prime },\mathbf{p}_{l}^{\prime },S_{l}^{\prime
2}\right) $ indicates a particular basis among the possible modifications
that transform the class.

\section{Possibility for field of fields: global field}

To sum up, a general description should take into account, not only the
fields describing several interacting collective states or objects, but also
their sub-collective states or subobject. This description was given in the
previous sections. However, this framework remains disagregated, since we
considered multiple fields, one for each objects/subobjects systems.
Moreover, since all possibilities of fields of objects should be considered,
a unified picture should encompass, all possible types of object/subobject
system. We should thus define a global field whose variables are the
characteristics of all possible systems of objects/subobject. Among these
systems, some global landscapes, i.e. involving all cells in particular
configurations of objects/subobjects should also be considered. The global
field thus also includes also, as variables, all the possible landscapes, or
collections of landscapes with reduced spatial extensions. Said differently,
the global field should be defined over the spaces of presheaves over an
infinite (non-countable) collection of some kind of sites $\left\{ \Sigma
_{l}\rightarrow \Sigma \right\} $.

\subsection{Global field}

The collection of fields for objects/subobjects:%
\begin{equation*}
\underline{\Gamma }\left( \left[ s,\left[ s_{l}\right] _{l},,\left[ s_{lk}%
\right] _{lk},..\right] ,\left[ s_{l}^{\left( l\right) },\left[
s_{lk}^{\left( l\right) }\right] _{lk},\left[ s_{lkm}^{\left( l\right) }%
\right] _{lkm},...\right] _{l},\left[ s_{lk}^{lk},\left[ s_{lkm}^{\left(
lk\right) }..\right] _{lkm}\right] _{lk},\left[ s_{lkm}^{\left( lkm\right)
},...\right] _{lkm},..\right)
\end{equation*}%
describes the system above the cover (\ref{Cv}):%
\begin{equation}
\begin{array}{c}
\begin{array}{c}
\rightarrow \\ 
\rightarrow%
\end{array}%
\Sigma _{ab}^{c}/G_{ab}\rightarrow \\ 
\begin{array}{c}
\rightarrow \\ 
\rightarrow%
\end{array}%
\Sigma _{ab^{\prime }}^{c}/G_{ab^{\prime }}\rightarrow%
\end{array}%
\Sigma _{a}^{c}/G_{a}\rightarrow \Sigma ^{c}/G  \label{Cvr}
\end{equation}%
Thus, there is one field for each possible cover of the system.
Consequently, an infinite number of fields describes the system. These
fields can be considered collectively, potentially gathered into a single
field. We should introduce a field ${\Huge \Lambda }$ that depends on the
covering (\ref{Cvr}) in addition to the local coordinates:

\begin{equation}
\left( \left[ s,\left[ s_{l}\right] _{l},,\left[ s_{lk}\right] _{lk},..%
\right] ,\left[ s_{l}^{\left( l\right) },\left[ s_{lk}^{\left( l\right) }%
\right] _{lk},\left[ s_{lkm}^{\left( l\right) }\right] _{lkm},...\right]
_{l},\left[ s_{lk}^{lk},\left[ s_{lkm}^{\left( lk\right) }..\right] _{lkm}%
\right] _{lk},\left[ s_{lkm}^{\left( lkm\right) },...\right] _{lkm},..\right)
\label{lc}
\end{equation}%
Thus, the field should be defined over the set of states (\ref{Cvr}),
endowed with presheaves over each subobject:%
\begin{equation}
\begin{array}{c}
\begin{array}{c}
\rightarrow \\ 
\rightarrow%
\end{array}%
\frac{\mathbf{S}_{c}}{\Sigma _{ab}^{c}/G_{ab}}\frac{\overset{r_{ca}}{%
\leftarrow }}{\rightarrow } \\ 
\begin{array}{c}
\rightarrow \\ 
\rightarrow%
\end{array}%
\frac{\mathbf{S}_{c^{\prime }}}{\Sigma _{ab^{\prime }}^{c}/G_{ab^{\prime }}}%
\rightarrow \frac{\overset{r_{c^{\prime }a}}{\leftarrow }}{\rightarrow }%
\end{array}%
\frac{\mathbf{S}_{a}}{\Sigma _{a}^{c}/G_{a}}\frac{\overset{r_{a}}{\leftarrow 
}}{\rightarrow }\frac{\mathbf{S}}{\Sigma ^{c}/G}
\end{equation}

Here we use the notation $\frac{\mathbf{S}}{\Sigma ^{c}/G}$ to describe the
presheaf (or sheaf) $\mathbf{S}$ over $\Sigma ^{c}/G$.

The arrows denote, as before, the inclusion or any map between a subobject
and an object. The left arrow $\overset{r_{a}}{\leftarrow }$ represents the
restriction map of the presheaf's sections. Here, $\mathbf{S}_{a}$, $\mathbf{%
S}_{c}$, $\mathbf{S}_{c^{\prime }}$ denote presheaves over the subjects.
These presheaves are independent of the restriction from the preasheaf $%
\mathbf{S}$ and imply the possibility of independent activations. Each of
these independent activation is stable in itself but may be responsible for
transitions and potential transformations of the full object.

This multiplicity of activations implies that the object and its subobjects
are defined locally by coordinates (\ref{lc}). Each of these coordinates
representing states of connectivities, average connectivities, and
frequencies of activities given by (\ref{cd}) and (\ref{ct}) for the
corresponding subbjects. However, as mentioned before, the presheaves are
defined by inequivalent cohomology classes. With this, we write the system
of classes and coordinates as:

\begin{equation}
\begin{array}{c}
\begin{array}{c}
\rightarrow \\ 
\rightarrow%
\end{array}%
\frac{\mathbf{C}_{c}\mathbf{,S}_{\mathbf{C}_{c}}}{\Sigma _{ab}^{c}/G_{ab}}%
\frac{\overset{r_{ca}}{\leftarrow }}{\rightarrow } \\ 
\begin{array}{c}
\rightarrow \\ 
\rightarrow%
\end{array}%
\frac{\mathbf{C}_{c^{\prime }}\mathbf{,S}_{\mathbf{C}_{c^{\prime }}}}{\Sigma
_{ab^{\prime }}^{c}/G_{ab^{\prime }}}\rightarrow \frac{\overset{r_{c^{\prime
}a}}{\leftarrow }}{\rightarrow }%
\end{array}%
\frac{\mathbf{C}_{a}\mathbf{,S}_{\mathbf{C}_{a}}}{\Sigma _{a}^{c}/G_{a}}%
\frac{\overset{r_{a}}{\leftarrow }}{\rightarrow }\frac{\mathbf{C,S}_{\mathbf{%
C}}}{\Sigma ^{c}/G}
\end{equation}

Here, $\left( \mathbf{C,S}_{\mathbf{C}}\right) $, $\left( \mathbf{C}_{a}%
\mathbf{,S}_{\mathbf{C}_{a}}\right) $, ... represent the cohomology classes
of the considered presheaves along with their local coordinates (\ref{lc}).
We should also consider restricted objects. For $\left( \frac{\mathbf{C,S}_{%
\mathbf{C}}}{\Sigma ^{c}/G}\right) ^{^{\prime }}\overset{n}{\rightarrow }%
\frac{\mathbf{C,S}_{\mathbf{C}}}{\Sigma ^{c}/G}$, we have the restricted
object:%
\begin{equation}
\begin{array}{c}
\begin{array}{c}
\rightarrow \\ 
\rightarrow%
\end{array}%
\left( \frac{\mathbf{C}_{c}\mathbf{,S}_{\mathbf{C}_{c}}}{\Sigma
_{ab}^{c}/G_{ab}}\right) ^{\prime }\frac{\overset{r_{ca}}{\leftarrow }}{%
\rightarrow } \\ 
\begin{array}{c}
\rightarrow \\ 
\rightarrow%
\end{array}%
\left( \frac{\mathbf{C}_{c^{\prime }}\mathbf{,S}_{\mathbf{C}_{c^{\prime }}}}{%
\Sigma _{ab^{\prime }}^{c}/G_{ab^{\prime }}}\right) ^{\prime }\rightarrow 
\frac{\overset{r_{c^{\prime }a}}{\leftarrow }}{\rightarrow }%
\end{array}%
\left( \frac{\mathbf{C}_{a}\mathbf{,S}_{\mathbf{C}_{a}}}{\Sigma
_{a}^{c}/G_{a}}\right) ^{\prime }\frac{\overset{r_{a}}{\leftarrow }}{%
\rightarrow }\left( \frac{\mathbf{C,S}_{\mathbf{C}}}{\Sigma ^{c}/G}\right)
^{\prime }
\end{equation}%
with:%
\begin{equation*}
\left( \frac{\mathbf{C}_{a}\mathbf{,S}_{\mathbf{C}_{a}}}{\Sigma
_{a}^{c}/G_{a}}\right) ^{\prime }:\left( \frac{\mathbf{C}_{a}\mathbf{,S}_{%
\mathbf{C}_{a}}}{\Sigma _{a}^{c}/G_{a}}\right) \underset{\overset{n}{%
\rightarrow }\times \rightarrow }{\times }\left( \frac{\mathbf{C,S}_{\mathbf{%
C}}}{\Sigma ^{c}/G}\right) ^{\prime }
\end{equation*}%
and by restriction maps

As a consequence, the field should depend on the set of variables $\frac{%
\mathbf{C,S}_{\mathbf{C}}}{\Sigma /G_{\Sigma }}$ and writes:

\begin{equation*}
{\Huge \Lambda }\left[ \frac{\mathbf{C,S}_{\mathbf{C}}}{\Sigma /G_{\Sigma }}%
\right]
\end{equation*}

where $\mathbf{C}$ describes the class of states for the covering $\Sigma
/G_{\Sigma }$, and the $\mathbf{S}_{\mathbf{C}}${} are the coordinates in
the considered class (\ref{lc}). This corresponds to chose a class and local
coordinates for each subobject of $\Sigma /G_{\Sigma }$.

To connect this point of view with the previous description, we sum the
elements encompassd in $\frac{\mathbf{C,S}_{\mathbf{C}}}{\Sigma /G_{\Sigma }}
$. Recall that:%
\begin{equation}
s\in \left\{ \Delta \mathbf{T}\left( \Sigma \times \Sigma \right) \mathbf{,}%
\Delta \mathbf{\hat{T}}\left( \Sigma \times \Sigma \right) \mathbf{,}\left(
a,\gamma _{a}\right) \right\}
\end{equation}%
and:%
\begin{equation}
s_{l}\text{ and }s_{l}^{\left( l\right) }\in \Delta \mathbf{T}\left( \Sigma
_{l}\times \Sigma _{l}\right) \mathbf{,}\Delta \mathbf{\hat{T}}\left( \Sigma
_{l}\times \Sigma _{l}\right) \mathbf{,}\left( a_{l},\gamma _{a}\right)
\end{equation}%
so that the local coordinates represent the relative connections $\Delta 
\mathbf{T}$ and $\Delta \mathbf{\hat{T}}$ for the object and subobjects.
This implies the choice of vectors of $\left\langle \mathbf{T}\right\rangle $%
, $\left\langle \mathbf{\hat{T}}\right\rangle $ given by the states $\left(
a_{l},\gamma _{a}\right) $. The set of data $\left( a_{l},\gamma _{a}\right) 
$, written as $\left( \mathbf{\alpha }_{k},\mathbf{p}_{k}\right) $ for the
object and subobject, are discrete data describing the collective state. It
depends on the class of the presheaves describing the state for the object
and its subobjects. We can thus describe more precisely $\frac{\mathbf{C,S}_{%
\mathbf{C}}}{\Sigma /G_{\Sigma }}$:%
\begin{eqnarray*}
\frac{\mathbf{C}}{\Sigma /G_{\Sigma }} &\rightarrow &\frac{\mathbf{C,}\left( 
\mathbf{\alpha }_{k},\mathbf{p}_{k}\right) _{\mathbf{C}}}{\Sigma /G_{\Sigma }%
} \\
\mathbf{S}_{\mathbf{C}} &\rightarrow &\left( \left[ \Delta \mathbf{T},\Delta 
\mathbf{\hat{T},}\left[ \Delta \mathbf{T}_{l},\Delta \mathbf{\hat{T}}_{l}%
\right] _{l},\left[ \Delta \mathbf{T}_{lk},\Delta \mathbf{\hat{T}}_{lk}%
\right] _{lk},..\right] ,\left[ \Delta \mathbf{T}_{l}^{\left( l\right)
},\Delta \mathbf{\hat{T}}_{l}^{\left( l\right) },\left[ \Delta \mathbf{T}%
_{lk}^{\left( l\right) },\Delta \mathbf{\hat{T}}_{lk}^{\left( l\right) }%
\right] _{lk},..\right] _{l},..\right)
\end{eqnarray*}

Thus $\frac{\mathbf{C,S}_{\mathbf{C}}}{\Sigma /G_{\Sigma }}$ encompasses a
large set of data: for a given discrete class, it represents one of the
potential average connectivities for all connectivities of an object or
subobject and the field ${\Huge \Lambda }\left[ \frac{\mathbf{C,S}_{\mathbf{C%
}}}{\Sigma /G_{\Sigma }}\right] $ stands for:%
\begin{eqnarray}
&&{\Huge \Lambda }\left[ \frac{\mathbf{C}}{\Sigma /G_{\Sigma }},\left( \left[
s,\left[ s_{l}\right] _{l},,\left[ s_{lk}\right] _{lk},..\right] ,\left[
s_{l}^{\left( l\right) },\left[ s_{lk}^{\left( l\right) }\right] _{lk},\left[
s_{lkm}^{\left( l\right) }\right] _{lkm},...\right] _{l},\left[ s_{lk}^{lk},%
\left[ s_{lkm}^{\left( lk\right) }..\right] _{lkm}\right] _{lk},\left[
s_{lkm}^{\left( lkm\right) },...\right] _{lkm},..\right) \right]  \label{Fr}
\\
&=&{\Huge \Lambda }\left[ \frac{\mathbf{C,}\left( \mathbf{\alpha }_{k},%
\mathbf{p}_{k}\right) _{\mathbf{C}}}{\Sigma /G_{\Sigma }},\left( \left[
\Delta \mathbf{T},\Delta \mathbf{\hat{T},}\left[ \Delta \mathbf{T}%
_{l},\Delta \mathbf{\hat{T}}_{l}\right] _{l},\left[ \Delta \mathbf{T}%
_{lk},\Delta \mathbf{\hat{T}}_{lk}\right] _{lk},..\right] ,\left[ \Delta 
\mathbf{T}_{l}^{\left( l\right) },\Delta \mathbf{\hat{T}}_{l}^{\left(
l\right) },\left[ \Delta \mathbf{T}_{lk}^{\left( l\right) },\Delta \mathbf{%
\hat{T}}_{lk}^{\left( l\right) }\right] _{lk},..\right] _{l},..\right) %
\right]  \notag
\end{eqnarray}%
and takes values in the space of sections of spaces belonging to the class $%
\mathbf{C}$ for $\Sigma /G_{\Sigma }$.

As a consequence, the states built by this field are functionals whose
arguments are sections of presheaves over $\Sigma /G_{\Sigma }$ constrained
by their class. These functional are labeled by these classes and two states
described by different classes cannot be linkes by some continuous
transformation.

Remark that, in the field:%
\begin{equation*}
{\Huge \Lambda }\left[ \frac{\mathbf{C,S}_{\mathbf{C}}}{\Sigma /G_{\Sigma }}%
\right]
\end{equation*}%
the covering $\Sigma /G_{\Sigma }${} is itself variable. Its variations
describe the entire set of possible objects together with their subobjects.
The set of fields has been replaced by a single field with
infinite-dimensional arguments. This field can be considered as being
defined over the category with objects $\frac{\mathbf{C}}{\Sigma /G_{\Sigma }%
}$, and the maps are inclusions or, more generally, maps between the objects.

As said in the introduction, objects are not predetermined, so that each
object should be considered, including those whose spatial extension is
maximal, i.e. some collective states involving the all space of clls. Since
there are some relations of inclusion between objects, we can consider these
states of maximal extension, as some "terminal" states, each of them definin
a global surrounding for the objects that are in a relation of inclusion
with these terminal states. Even if those included objects have their own
dynamics, interactions etc... one may expect such global surroundings to
impact the objects they include, acting as colective states for these
collective local colective stats.

If the field ${\Huge \Lambda }\left[ \frac{\mathbf{C,S}_{\mathbf{C}}}{\Sigma
/G_{\Sigma }}\right] $ encompasses both states for some object and all its
subobjects, these states can also be localized over some particular
subobject. Actually, those states are sections of presheaf of functionals
whose arguments are sections (in given classes) over some spatial support.
Localized states corresponds to sections whose support is restricted to that
of a particular subobject. To include this localization in the dynamics of
the system, some constraints should appear in the interactions terms of the
effective action (section 13). This remark also applies to the terminal
objects. Some constraints in the interaction processes may restrict the
states involved in the interaction to some particular subobject, but the
form of the interaction should nethertheless depend on the global
surrounding created by the trminal object. This does \ not mean that we
should restrict the domain of the field ${\Huge \Lambda }$ to the terminal
objects, since we can also aim at considering particular and independent
activations of every kind of object.

\subsection{Operators description}

From operatr point of view, field ${\Huge \Lambda }$ seen itself as an
operator:%
\begin{equation*}
{\Huge \Lambda }={\Huge A}^{{\Large +}}\left[ \frac{\mathbf{C,S}_{\mathbf{C}}%
}{\Sigma /G_{\Sigma }}\right] +{\Huge A}\left[ \frac{\mathbf{C}^{\prime }%
\mathbf{,S}_{\mathbf{C}}^{\prime }}{\Sigma ^{\prime }/G_{\Sigma }^{\prime }}%
\right]
\end{equation*}%
As before, the creation and annihilation operators should be expanded as a
series (over some infinite-dimensional functional space) of operators
describing both the characteristics $\left( \mathbf{\alpha }_{k},\mathbf{p}%
_{k}\right) $ of a particular state, but also the class $\mathbf{C}$.
Moreover, these characteristics are not independent. This implies the
decomposition:%
\begin{eqnarray}
{\Huge \Lambda } &=&{\Huge A}^{{\Large +}}\left[ \frac{\mathbf{C,S}_{\mathbf{%
C}}}{\Sigma /G_{\Sigma }}\right] +{\Huge A}\left[ \frac{\mathbf{C}^{\prime }%
\mathbf{,S}_{\mathbf{C}}^{\prime }}{\Sigma ^{\prime }/G_{\Sigma }^{\prime }}%
\right]  \label{pt} \\
&=&\int \mathcal{D}\left[ \frac{\mathbf{C,}\left( \mathbf{\alpha }_{k},%
\mathbf{p}_{k}\right) _{\mathbf{C}}}{\Sigma /G_{\Sigma }}\right]
\sum_{A}\left( \prod \mathbf{\hat{A}}_{\mathbf{C}}^{+}\mathbf{\hat{A}}%
_{\left( \mathbf{\alpha }_{k},\mathbf{p}_{k}\right) _{\mathbf{C}}}^{+}+\prod 
\mathbf{\hat{A}}_{\mathbf{C}}^{-}\mathbf{\hat{A}}_{\left( \mathbf{\alpha }%
_{k},\mathbf{p}_{k}\right) _{\mathbf{C}}}^{-}\right)  \notag
\end{eqnarray}%
The sum $\sum_{A}$ being over the possible activations of objct and
subobject states, and the products being performed over this activation. By
analogy with field theory, contributions:%
\begin{equation*}
\left( \prod \mathbf{\hat{A}}_{\mathbf{C}}^{+}\mathbf{\hat{A}}_{\left( 
\mathbf{\alpha }_{k},\mathbf{p}_{k}\right) _{\mathbf{C}}}^{+}+\prod \mathbf{%
\hat{A}}_{\mathbf{C}}^{-}\mathbf{\hat{A}}_{\left( \mathbf{\alpha }_{k},%
\mathbf{p}_{k}\right) _{\mathbf{C}}}^{-}\right)
\end{equation*}%
would correspond to the creation annihilation of a single state. The
operator:%
\begin{equation*}
\mathbf{\hat{A}}_{\mathbf{C}}^{+}\mathbf{\hat{A}}_{\left( \mathbf{\alpha }%
_{k},\mathbf{p}_{k}\right) _{\mathbf{C}}}^{+}
\end{equation*}%
describes both the creation $\mathbf{\hat{A}}_{\mathbf{C}}^{+}$ of a
cohomology class $\mathbf{C}$, while $\mathbf{\hat{A}}_{\left( \mathbf{%
\alpha }_{k},\mathbf{p}_{k}\right) _{\mathbf{C}}}^{+}${} describes the
creation of characteristics $\left( \mathbf{\alpha }_{k},\mathbf{p}%
_{k}\right) _{\mathbf{C}}${}, depending on the class $\mathbf{C}$. The
perturbation:

The operator:%
\begin{equation*}
\mathbf{\hat{A}}_{\mathbf{C}}^{-}\mathbf{\hat{A}}_{\left( \mathbf{\alpha }%
_{k},\mathbf{p}_{k}\right) _{\mathbf{C}}}^{-}
\end{equation*}%
describes the destruction of such state. The product:%
\begin{equation*}
\prod \mathbf{\hat{A}}_{\mathbf{C}}^{+}\mathbf{\hat{A}}_{\left( \mathbf{%
\alpha }_{k},\mathbf{p}_{k}\right) _{\mathbf{C}}}^{+}+\prod \mathbf{\hat{A}}%
_{\mathbf{C}}^{-}\mathbf{\hat{A}}_{\left( \mathbf{\alpha }_{k},\mathbf{p}%
_{k}\right) _{\mathbf{C}}}^{-}
\end{equation*}%
refers to the product over subobjects.

\subsection{States}

The states are functional of ${\Huge \Lambda }\left[ \frac{\mathbf{C,S}_{%
\mathbf{C}}}{\Sigma /G_{\Sigma }}\right] $ of the form:%
\begin{equation}
\left\vert H\right\rangle =\sum_{n}\sum_{\mathbf{C}_{i},\Sigma
_{i}/G_{\Sigma _{i}}}\int \prod\limits_{1}^{n}H\left( \frac{\mathbf{C}_{i}%
\mathbf{,S}_{\mathbf{C}_{i}}}{\Sigma _{i}/G_{\Sigma _{i}}}\right) {\Huge %
\Lambda }\left[ \frac{\mathbf{C}_{i}\mathbf{,S}_{\mathbf{C}_{i}}}{\Sigma
_{i}/G_{\Sigma _{i}}}\right] d\mathbf{S}_{\mathbf{C}_{i}}  \label{Hm}
\end{equation}

A state describes combinations of functionals over any arrangement of
products of collective states, representing the activation of multiple
transitory structures. The system should be characterized by an action
functional, whose vertices describe possible states, interactions, and
transitions. We should also consider states of the form:%
\begin{equation*}
\left\langle H\right\vert =\sum_{n}\sum_{\mathbf{C}_{i},\Sigma
_{i}/G_{\Sigma _{i}}}\int \prod\limits_{1}^{n}H\left( \frac{\mathbf{C}_{i}%
\mathbf{,S}_{\mathbf{C}_{i}}}{\Sigma _{i}/G_{\Sigma _{i}}}\right) {\Huge %
\Lambda }^{\dag }\left[ \frac{\mathbf{C}_{i}\mathbf{,S}_{\mathbf{C}_{i}}}{%
\Sigma _{i}/G_{\Sigma _{i}}}\right] d\mathbf{S}_{\mathbf{C}_{i}}
\end{equation*}%
and dynamic transitions should occur between states of the form $\left\vert
H\right\rangle $ and $\left\langle H^{\prime }\right\vert $, the first
corresponding to an initial state and the second to a final state.

Alternatively, we can consider states within the operator formalism. Recall
that states with a given value $\Delta \mathbf{T}_{p}^{\alpha }${} were
described by the state (\ref{LCT}):%
\begin{eqnarray*}
&&\left\vert \Delta \mathbf{T}_{p}^{\alpha },\mathbf{\alpha },\mathbf{p}%
,S^{2}\right\rangle \\
&=&\exp \left( -\left( \left( \Delta \mathbf{T}_{p}^{\alpha }\right) ^{t}%
\mathbf{A}_{p}^{\alpha }\Delta \mathbf{T}_{p}^{\alpha }+2\left( \Delta 
\mathbf{T}_{p}^{\alpha }\right) ^{t}\sqrt{\mathbf{A}_{p}^{\alpha }}\mathbf{%
\hat{A}}^{+}\left( \mathbf{\alpha },\mathbf{p},S^{2}\right) \right) +\frac{1%
}{2}\mathbf{\hat{A}}^{+}\left( \mathbf{\alpha },\mathbf{p},S^{2}\right) .%
\mathbf{\hat{A}}^{+}\left( \mathbf{\alpha },\mathbf{p},S^{2}\right) \right)
\left\vert Vac\right\rangle
\end{eqnarray*}%
We can generalize the state (\ref{LCT}) for an object by including a
substate, where the formalism (\ref{pt}) allows us to define a state $%
\left\vert \mathbf{S}_{\mathbf{C}}\right\rangle $ with a given value of $%
\mathbf{S}_{\mathbf{C}}$:%
\begin{eqnarray}
&&\left\vert \mathbf{S}_{\mathbf{C}}\right\rangle  \label{stn} \\
&=&\exp \left( -\left( \mathbf{S}_{\mathbf{C}}^{T}\mathbf{\nu }\left( \frac{%
\mathbf{C}}{\Sigma /G_{\Sigma }}\right) \mathbf{S}_{\mathbf{C}}+2\mathbf{S}_{%
\mathbf{C}}^{T}\sqrt{\mathbf{\nu }\left( \frac{\mathbf{C}}{\Sigma /G_{\Sigma
}}\right) }{\Huge A}^{{\Large +}}\left[ \frac{\mathbf{C,S}_{\mathbf{C}}}{%
\Sigma /G_{\Sigma }}\right] \right) +\frac{1}{2}{\Huge A}^{{\Large +}}\left[ 
\frac{\mathbf{C,S}_{\mathbf{C}}}{\Sigma /G_{\Sigma }}\right] .{\Huge A}^{%
{\Large +}}\left[ \frac{\mathbf{C,S}_{\mathbf{C}}}{\Sigma /G_{\Sigma }}%
\right] \right) \left\vert Vac\right\rangle  \notag
\end{eqnarray}%
where $\mathbf{C}$ corresponds to the elements describing the state, that is
the subobjects and the coordinates $\left\vert \mathbf{S}_{\mathbf{C}%
}\right\rangle $.

\section{Classical action}

The field ${\Huge \Lambda }$ should be described by an effective action,
including interactions and accounting for transitions. This effective action
should be derived from the action of the collective states. We begin with
the quadratic part (\ref{Td}).

\subsection{Qudratic action}

The quadratic part applies to both the objects and their subobjects (\ref{Td}%
): 
\begin{eqnarray}
&&\underline{\Gamma }^{\dag }\left( \left[ s,\left[ s_{l}\right] _{l},,\left[
s_{lk}\right] _{lk},..\right] ,\left[ s_{l}^{\left( l\right) },\left[
s_{lk}^{\left( l\right) }\right] _{lk},\left[ s_{lkm}^{\left( l\right) }%
\right] _{lkm},...\right] _{l},\left[ s_{lk}^{lk},\left[ s_{lkm}^{\left(
lk\right) }..\right] _{lkm}\right] _{lk},\left[ s_{lkm}^{\left( lkm\right)
},...\right] _{lkm},..\right)  \label{Tg} \\
&&\times \sum_{st}\left( -\frac{1}{2}\nabla _{\left( \mathbf{T}\right)
_{\Sigma _{st}^{2}}}^{2}+\frac{1}{2}\left( \Delta \mathbf{T}_{p}^{\alpha
}\right) _{_{\Sigma _{st}^{2}}}^{t}\mathbf{A}_{_{\Sigma _{st}^{2}}}^{\alpha
}\left( \Delta \mathbf{T}_{p}^{\alpha }\right) +\mathbf{C}\right)  \notag \\
&&\times \underline{\Gamma }\left( \left[ s,\left[ s_{l}\right] _{l},,\left[
s_{lk}\right] _{lk},..\right] ,\left[ s_{l}^{\left( l\right) },\left[
s_{lk}^{\left( l\right) }\right] _{lk},\left[ s_{lkm}^{\left( l\right) }%
\right] _{lkm},...\right] _{l},\left[ s_{lk}^{lk},\left[ s_{lkm}^{\left(
lk\right) }..\right] _{lkm}\right] _{lk},\left[ s_{lkm}^{\left( lkm\right)
},...\right] _{lkm},..\right)  \notag
\end{eqnarray}%
should be replaced by a quadratic action for ${\Huge \Lambda }$. The action
for the field ${\Huge \Lambda }$ is non-local, as it involves interactions
between possible coverings. At the lowest order, we consider the 'quadratic'
action:

\begin{equation}
{\huge S}\left( {\Huge \Lambda ,\Lambda }^{{\Large \dag }}\right) {\huge =}%
{\Huge \Lambda }^{{\Large \dag }}\left[ \frac{\mathbf{C,S}_{\mathbf{C}}}{%
\Sigma /G_{\Sigma }}\right] {\Large S}_{1,1}\left[ \frac{\mathbf{C,S}_{%
\mathbf{C}}}{\Sigma /G_{\Sigma }},\frac{\mathbf{C}^{\prime }\mathbf{,S}_{%
\mathbf{C}}^{\prime }}{\Sigma ^{\prime }/G_{\Sigma }^{\prime }}\right] 
{\Huge \Lambda }\left[ \frac{\mathbf{C}^{\prime }\mathbf{,S}_{\mathbf{C}%
}^{\prime }}{\Sigma ^{\prime }/G_{\Sigma }^{\prime }}\right]  \label{DRT}
\end{equation}

In this descrption, the points are the classes $\frac{\mathbf{C,S}_{\mathbf{C%
}}}{\Sigma /G_{\Sigma }}$, the potential objects with their potential
subobjects, aolng with the space of states. Not all possible objects are
admissible.

There are conditions depending on some background, described in the previous
part. However, as in part 4, the states can evolve continuously, including
in their spatial extension. This may result in a slow, continuous
modification, staying in the same discrete class $C$, while, once certain
values of the parameters are reached, the state is locked in an interaction,
inducing a transition toward a new class or deactivation.

We should not consider that object are repertory of representations, or only
as merely approximations. Rather, they should be considered as some emerging
states, with rearrangement, spatial displacement, sometimes experiencing
fast abrupt transitions, then fading away. The dynamics, even if
signal-induced, should ultimately rely on the characteristic of the system.
Several states should be activated, displaced, interact with resulting
transitions, giving rise to different states and inducing chain reactions

The kernel ${\Large S}_{1,1}\left[ \frac{\mathbf{C,S}_{\mathbf{C}}}{\Sigma
/G_{\Sigma }},\frac{\mathbf{C}^{\prime }\mathbf{,S}_{\mathbf{C}}^{\prime }}{%
\Sigma ^{\prime }/G_{\Sigma }^{\prime }}\right] $ describes transitions and
dynamics of states. We can postulate the form of the quadratic action, by
using the operator arising in (\ref{Td}):%
\begin{equation*}
\left( -\frac{1}{2}\nabla _{\left( \mathbf{T}\right) _{\Sigma
_{st}^{2}}}^{2}+\frac{1}{2}\left( \Delta \mathbf{T}_{p}^{\alpha }\right)
_{_{\Sigma _{st}^{2}}}^{t}\mathbf{A}_{_{\Sigma _{st}^{2}}}^{\alpha }\left(
\Delta \mathbf{T}_{p}^{\alpha }\right) +\mathbf{C}\right)
\end{equation*}%
and (\ref{cd}) and (\ref{ct}). The operator $-\frac{1}{2}\nabla _{\left( 
\mathbf{T}\right) _{\Sigma _{st}^{2}}}^{2}$ corresponds to:%
\begin{equation*}
-\frac{1}{2}\left( \frac{\delta ^{2}}{\delta \left[ s,\left[ s_{l}\right]
_{l},,\left[ s_{lk}\right] _{lk},..\right] ^{2}}+\frac{\delta ^{2}}{\left[
s_{l}^{\left( l\right) },\left[ s_{lk}^{\left( l\right) }\right] _{lk},\left[
s_{lkm}^{\left( l\right) }\right] _{lkm},...\right] _{l}^{2}}+...\right)
\rightarrow -\frac{1}{2}\frac{\delta ^{2}}{\delta \mathbf{S}_{\mathbf{C}}^{2}%
}
\end{equation*}%
This is the local part of the quadratic action. The non-local part includes
the variation of classes:%
\begin{equation*}
-\frac{1}{2}\frac{\Delta ^{2}}{\Delta \mathbf{C}^{2}}
\end{equation*}%
This corresponds to the second-order variation with respect to the discrete
variable $\mathbf{C}$ without changing $\Sigma /G_{\Sigma }$.

The operator $\frac{1}{2}\left( \Delta \mathbf{T}_{p}^{\alpha }\right)
_{_{\Sigma _{st}^{2}}}^{t}\mathbf{A}_{_{\Sigma _{st}^{2}}}^{\alpha }\left(
\Delta \mathbf{T}_{p}^{\alpha }\right) $ is replaced by:%
\begin{equation*}
\frac{1}{2}\left( \mathbf{S}_{\mathbf{C}}^{T}\mathbf{\nu }\left( \frac{%
\mathbf{C}}{\Sigma /G_{\Sigma }}\right) \mathbf{S}_{\mathbf{C}}+\mathbf{S}_{%
\mathbf{C}^{\prime }}^{\prime T}\mathbf{\nu }\left( \frac{\mathbf{C}^{\prime
}}{\Sigma /G_{\Sigma }}\right) \mathbf{S}_{\mathbf{C}^{\prime }}^{\prime
}\right)
\end{equation*}%
where the metric $\mathbf{\nu }\left( \frac{\mathbf{C}}{\Sigma /G_{\Sigma }}%
\right) $ is local and replaces $\mathbf{A}_{_{\Sigma _{st}^{2}}}^{\alpha }$%
{}.

Including the operator $-\frac{\delta }{\delta \Sigma ^{2}}$ for the
modification of spatial support, the quadratic part becomes:

\begin{equation*}
-\frac{1}{2}\frac{\delta }{\delta \Sigma ^{2}}-\frac{1}{2}\frac{\delta ^{2}}{%
\delta \mathbf{S}_{\mathbf{C}}^{2}}+\frac{1}{2}\left( \mathbf{S}_{\mathbf{C}%
}^{T}\mathbf{\nu }\left( \frac{\mathbf{C}}{\Sigma /G_{\Sigma }}\right) 
\mathbf{S}_{\mathbf{C}}+\mathbf{S}_{\mathbf{C}^{\prime }}^{\prime T}\mathbf{%
\nu }\left( \frac{\mathbf{C}^{\prime }}{\Sigma /G_{\Sigma }}\right) \mathbf{S%
}_{\mathbf{C}^{\prime }}^{\prime }\right) -\frac{1}{2}\frac{\Delta ^{2}}{%
\Delta \mathbf{C}^{2}}
\end{equation*}%
Including also the possibility of quadratic terms, where we include some
off-diagonal terms: ${\Large \hat{S}}_{1,1}\left[ \frac{\mathbf{C,S}_{%
\mathbf{C}}}{\Sigma /G_{\Sigma }},\frac{\mathbf{C}^{\prime }\mathbf{,S}_{%
\mathbf{C}}^{\prime }}{\Sigma ^{\prime }/G_{\Sigma }^{\prime }}\right] $,
and we write:%
\begin{eqnarray}
&&{\Large S}_{1,1}\left[ \frac{\mathbf{C,S}_{\mathbf{C}}}{\Sigma /G_{\Sigma }%
},\frac{\mathbf{C}^{\prime }\mathbf{,S}_{\mathbf{C}}^{\prime }}{\Sigma
^{\prime }/G_{\Sigma }^{\prime }}\right]  \label{th} \\
&=&-\frac{1}{2}\frac{\delta }{\delta \Sigma ^{2}}-\frac{1}{2}\frac{\delta
^{2}}{\delta \mathbf{S}_{\mathbf{C}}^{2}}+\frac{1}{2}\left( \mathbf{S}_{%
\mathbf{C}}^{T}\mathbf{\nu }\left( \frac{\mathbf{C}}{\Sigma /G_{\Sigma }}%
\right) \mathbf{S}_{\mathbf{C}}+\mathbf{S}_{\mathbf{C}^{\prime }}^{\prime T}%
\mathbf{\nu }\left( \frac{\mathbf{C}^{\prime }}{\Sigma /G_{\Sigma }}\right) 
\mathbf{S}_{\mathbf{C}^{\prime }}^{\prime }\right) -\frac{1}{2}\frac{\Delta
^{2}}{\Delta \mathbf{C}^{2}}+{\Large \hat{S}}_{1,1}\left[ \frac{\mathbf{C,S}%
_{\mathbf{C}}}{\Sigma /G_{\Sigma }},\frac{\mathbf{C}^{\prime }\mathbf{,S}_{%
\mathbf{C}}^{\prime }}{\Sigma ^{\prime }/G_{\Sigma }^{\prime }}\right] 
\notag
\end{eqnarray}%
The operator $-\frac{1}{2}\frac{\delta }{\delta \Sigma ^{2}}$ is local in
appearance, since the modification of $\Sigma $ induces certain transitions
or modifications of the possible classes $\mathbf{C}$. We will inspect the
impact of this modification while considering the propagators.

\subsection{Interaction terms}

Interaction terms in (\ref{NTBJ})$:$%
\begin{equation*}
\sum_{n}\sum_{S_{1},...,S_{n}\subseteq S}\sum_{\left( \alpha
_{1},p_{1}\right) ,...,\left( \alpha _{n},p_{n}\right) }\sum_{k}V_{k}\left(
\left( \left\{ \mathbf{T}_{i},\mathbf{\alpha }_{i},\mathbf{p}_{i}\right\}
,\left( S_{i}\right) ^{2}\right) \right) \prod\limits_{i\leqslant k}%
\underline{\Gamma }^{\dag }\left( \mathbf{T}_{i},\mathbf{\alpha }_{i},%
\mathbf{p}_{i}\right) \prod\limits_{k+1\leqslant i\leqslant n}\underline{%
\Gamma }\left( \mathbf{T}_{i},\mathbf{\alpha }_{i},\mathbf{p}%
_{i},S_{i}^{2}\right)
\end{equation*}%
correspond by extension to interactions:%
\begin{equation*}
\sum_{k}\prod \underline{\Gamma }^{\dag }\left( \mathbf{S}_{i}\right)
V_{k}\left( \left( \left\{ \mathbf{S}_{i}\right\} \right) \right) \prod 
\underline{\Gamma }\left( \mathbf{S}_{i}\right)
\end{equation*}%
where:%
\begin{equation}
\mathbf{S}_{i}=\left( \left( \left[ s,\left[ s_{l}\right] _{l},,\left[ s_{lk}%
\right] _{lk},..\right] \right) _{i},\left( \left[ s_{l}^{\left( l\right) },%
\left[ s_{lk}^{\left( l\right) }\right] _{lk},\left[ s_{lkm}^{\left(
l\right) }\right] _{lkm},...\right] _{l}\right) _{i},\left( \left[
s_{lk}^{lk},\left[ s_{lkm}^{\left( lk\right) }..\right] _{lkm}\right] _{lk},%
\left[ s_{lkm}^{\left( lkm\right) },...\right] _{lkm}\right) _{i},..\right)
\label{crd}
\end{equation}

Thus the full classical action is given by:

\begin{eqnarray}
&&{\Huge S}\left( {\Huge \Lambda }^{{\Large \dag }},{\Huge \Lambda }\right)
\label{sfr} \\
&=&\sum_{r,r^{\prime }}\int \prod\limits_{r=1}^{v}{\Huge \Lambda }^{{\Large %
\dag }}\left[ \left( \frac{\mathbf{C,S}_{\mathbf{C}}}{\Sigma /G_{\Sigma }}%
\right) _{r}\right] \frac{1}{\nu !}\frac{1}{\nu ^{\prime }!}{\Large S}_{\nu
,\nu ^{\prime }}\left[ \left\{ \left( \frac{\mathbf{C,S}_{\mathbf{C}}}{%
\Sigma /G_{\Sigma }}\right) _{r}\right\} ,\left\{ \left( \frac{\mathbf{C}%
^{\prime }\mathbf{,S}_{\mathbf{C}}^{\prime }}{\Sigma ^{\prime }/G_{\Sigma
}^{\prime }}\right) _{r^{\prime }}\right\} \right] \prod\limits_{r^{\prime
}=1}^{v^{\prime }}{\Huge \Lambda }\left[ \left( \frac{\mathbf{C}^{\prime }%
\mathbf{,S}_{\mathbf{C}}^{\prime }}{\Sigma ^{\prime }/G_{\Sigma }^{\prime }}%
\right) _{r^{\prime }}\right]  \notag
\end{eqnarray}%
The coefficients:%
\begin{equation}
{\Large S}_{\nu ,\nu ^{\prime }}\left[ \left\{ \left( \frac{\mathbf{C,S}_{%
\mathbf{C}}}{\Sigma /G_{\Sigma }}\right) _{r}\right\} ,\left\{ \left( \frac{%
\mathbf{C}^{\prime }\mathbf{,S}_{\mathbf{C}}^{\prime }}{\Sigma ^{\prime
}/G_{\Sigma }^{\prime }}\right) _{r^{\prime }}\right\} \right]  \label{vr}
\end{equation}%
include the $V_{k}\left( \left( \left\{ \mathbf{s}_{i}\right\} \right)
\right) $ terms tht describs interactions between subobjects of several
objects $\left\{ \mathbf{s}_{i}\right\} $and $\left\{ \mathbf{s}_{j}\right\} 
$.

The vertices (\ref{vr}) represent the modification of objects from state $%
\left\{ \left( \frac{\mathbf{C}^{\prime }\mathbf{,S}_{\mathbf{C}}^{\prime }}{%
\Sigma ^{\prime }/G_{\Sigma }^{\prime }}\right) _{r^{\prime }}\right\} ${}
to the state $\left\{ \left( \frac{\mathbf{C,S}_{\mathbf{C}}}{\Sigma
/G_{\Sigma }}\right) _{r}\right\} $.

We have seen that two types of processes are possible. We details them in
the next paragraph.

\subsubsection{Gathering and rearagmn}

Given the results presented in \cite{GLw} (see also Appendix 7 for some
details), several states may gather to constitute a new one. The feasibility
of such processes depends on the potential internal frequencies of the
initial state and the final one. The frequencies of incoming states and
those of outgoing states should satisfy certain relations to ensure the
modification

In general, for several states to interact and recompose to produce new
states, specific relations between their frequencies are required:%
\begin{equation*}
\delta \left( f\left( \left\{ \Upsilon _{\left( \mathbf{p}_{k}\right)
_{r^{\prime }}^{\prime }}\right\} _{r^{\prime }},\left\{ \Upsilon _{\left( 
\mathbf{p}_{k}\right) _{r}}\right\} _{r}\right) \right)
\end{equation*}%
if the recomposition of $\left\{ \Upsilon _{\left( \mathbf{p}_{k}\right)
_{r^{\prime }}^{\prime }}\right\} _{r^{\prime }}$ in $\left\{ \Upsilon
_{\left( \mathbf{p}_{k}\right) _{r}}\right\} _{r}$ is global, or if the
objects recompose within groups:%
\begin{equation*}
\sum_{\substack{ f_{1},f_{2}  \\ r_{1}+r_{2}=r  \\ r_{1}^{\prime
}+r_{2}^{\prime }=r^{\prime }}}\delta \left( f_{1}\left( \left\{ \Upsilon
_{\left( \mathbf{p}_{k}\right) _{r_{1}^{\prime }}^{\prime }}\right\}
_{r_{1}^{\prime }},\left\{ \Upsilon _{\left( \mathbf{p}_{k}\right)
_{r_{1}}}\right\} _{r_{1}}\right) \right) \delta \left( f_{2}\left( \left\{
\Upsilon _{\left( \mathbf{p}_{k}\right) _{r_{2}^{\prime }}^{\prime
}}\right\} _{r_{2}^{\prime }},\left\{ \Upsilon _{\left( \mathbf{p}%
_{k}\right) _{r_{2}}}\right\} _{r_{2}}\right) \right)
\end{equation*}%
and similar formula stand for decomposition in several groups.

In the process of transition, the constraint should be affected by
coefficients measuring the strength of transition:%
\begin{equation*}
V\left( f\left( \left\{ \Upsilon _{\left( \mathbf{p}_{k}\right) _{r^{\prime
}}^{\prime }}\right\} _{r^{\prime }},\left\{ \Upsilon _{\left( \mathbf{p}%
_{k}\right) _{r}}\right\} _{r}\right) \right)
\end{equation*}%
and%
\begin{equation*}
V\left( f_{1}\left( \left\{ \Upsilon _{\left( \mathbf{p}_{k}\right)
_{r_{1}^{\prime }}^{\prime }}\right\} _{r_{1}^{\prime }},\left\{ \Upsilon
_{\left( \mathbf{p}_{k}\right) _{r_{1}}}\right\} \right) ,f_{2}\left(
\left\{ \Upsilon _{\left( \mathbf{p}_{k}\right) _{r_{2}^{\prime }}^{\prime
}}\right\} _{r_{2}^{\prime }},\left\{ \Upsilon _{\left( \mathbf{p}%
_{k}\right) _{r_{2}}}\right\} _{r_{2}}\right) \right)
\end{equation*}%
respectively.

The frequencies $\left\{ \Upsilon _{\left( \mathbf{p}_{k}\right) _{r^{\prime
}}^{\prime }}\right\} _{r^{\prime }}${}and $\left\{ \Upsilon _{\left( 
\mathbf{p}_{k}\right) _{r}}\right\} _{r}$ are potential frequencies for
objects and subobjects of the resulting and incoming object respectively.
The functions $f$ or $f_{i}$ implement relations between initial and
compound njcts f th typ (\ref{Rl}).\ Th various terms corresponds to the sum
over different groups of composed objects.

To express these contributions in terms of the variable $\left( \frac{%
\mathbf{C,S}_{\mathbf{C}}}{\Sigma /G_{\Sigma }}\right) _{r}$, we write:%
\begin{equation*}
\Upsilon \left[ \left( \frac{\mathbf{C,S}_{\mathbf{C}}}{\Sigma /G_{\Sigma }}%
\right) _{r}\right]
\end{equation*}%
the frequencies for the state $\left( \frac{\mathbf{C,S}_{\mathbf{C}}}{%
\Sigma /G_{\Sigma }}\right) _{r}$.

This formulation allows us to present interactions from another point of
view. Snce frqncy depnd n th spprt $\Sigma /G_{\Sigma }$ we can consider
that the resulting shape of the final object is fluctuating and can adapt to
produce frequencies satisfying the constraints imposed by the incoming
objects. The rationale for that is condition (\ref{thr}) in appendix 7. that
conditions the activation of a given connection in a collective state to the
background field in which both objects stand. As a conseqnce, the support of
the composed objct may be smaller than the union of support.

Morover, the notion of supprt corresponds to some minimization of a global
action. As a consequence, the gathered objct may also corrsponds to a larger
structures. The frequency of the composed structure should be equal to some
internal possibility of the composed objct, and this condition corresponds
to the constraint given by the functions $f$, $f_{i}$.

As a consequence, the shape of the object's support becomes a fluctuating
dynamical variable, allowing for changes in frequencies, classes, and
connectivities. The shape of the composed structure should be defined so
that the composition of frequency corresponds to internal frequencies.

The shape of the composed structure should be defined such that the
composition of frequencies corresponds to internal frequencies. This shape
can be selected by the amplitudes $V$ and we consider it is defined by the
differential of potential for connections:%
\begin{eqnarray}
V &\rightarrow &\Delta U=U\left( \left\{ \Upsilon \left[ \left( \frac{%
\mathbf{C,S}_{\mathbf{C}}}{\Sigma /G_{\Sigma }}\right) _{r^{\prime }}\right]
\right\} _{r^{\prime }}\right) -U\left( \left\{ \left( \frac{\mathbf{C,S}_{%
\mathbf{C}}}{\Sigma /G_{\Sigma }}\right) _{r}\right\} \right)  \label{Pn} \\
V_{2} &\rightarrow &\Delta U_{12}=U\left( \left\{ \Upsilon \left[ \left( 
\frac{\mathbf{C,S}_{\mathbf{C}}}{\Sigma /G_{\Sigma }}\right) _{r_{1}^{\prime
}}\right] \right\} _{r_{1}^{\prime }}\right) +U\left( \left\{ \Upsilon \left[
\left( \frac{\mathbf{C,S}_{\mathbf{C}}}{\Sigma /G_{\Sigma }}\right)
_{r_{2}^{\prime }}\right] \right\} _{r_{2}^{\prime }}\right)  \notag \\
&&-U\left( \left\{ \Upsilon \left[ \left( \frac{\mathbf{C,S}_{\mathbf{C}}}{%
\Sigma /G_{\Sigma }}\right) _{r_{1}}\right] \right\} _{r_{1}}\right)
-U\left( \left\{ \Upsilon \left[ \left( \frac{\mathbf{C,S}_{\mathbf{C}}}{%
\Sigma /G_{\Sigma }}\right) _{r_{2}}\right] \right\} _{r_{2}}\right)  \notag
\end{eqnarray}%
The system favors the shape with the lowest potential, that is, the shape
with minimal connections given its own strctr.

The coefficients of interactions should therefore represent processes
written in terms of frequencies including all possible support satisfying
the constraint. As a consequence, (\ref{vr}) should include some series
expansion of relations:%
\begin{eqnarray*}
&&\sum_{f}V\left( f\left( \left\{ \Upsilon \left[ \left( \frac{\mathbf{C,S}_{%
\mathbf{C}}}{\Sigma /G_{\Sigma }}\right) _{r^{\prime }}\right] \right\}
_{r^{\prime }},\left\{ \Upsilon \left[ \left( \frac{\mathbf{C,S}_{\mathbf{C}}%
}{\Sigma /G_{\Sigma }}\right) _{r}\right] \right\} _{r}\right) \right)
\delta \left( f\left( \left\{ \Upsilon \left[ \left( \frac{\mathbf{C,S}_{%
\mathbf{C}}}{\Sigma /G_{\Sigma }}\right) _{r^{\prime }}\right] \right\}
_{r^{\prime }},\left\{ \Upsilon \left[ \left( \frac{\mathbf{C,S}_{\mathbf{C}}%
}{\Sigma /G_{\Sigma }}\right) _{r}\right] \right\} _{r}\right) \right) \\
&&+\sum_{\substack{ f_{1},f_{2}  \\ r_{1}+r_{2}=r  \\ r_{1}^{\prime
}+r_{2}^{\prime }=r^{\prime }}}\delta \left( f_{1}\left( \left\{ \Upsilon %
\left[ \left( \frac{\mathbf{C,S}_{\mathbf{C}}}{\Sigma /G_{\Sigma }}\right)
_{r_{1}^{\prime }}\right] \right\} _{r_{1}^{\prime }},\left\{ \Upsilon \left[
\left( \frac{\mathbf{C,S}_{\mathbf{C}}}{\Sigma /G_{\Sigma }}\right) _{r_{1}}%
\right] \right\} _{r_{1}}\right) \right) \\
&&\times V_{2}\left( f_{1}\left( \left\{ \Upsilon \left[ \left( \frac{%
\mathbf{C,S}_{\mathbf{C}}}{\Sigma /G_{\Sigma }}\right) _{r_{1}^{\prime }}%
\right] \right\} _{r_{1}^{\prime }},\left\{ \Upsilon \left[ \left( \frac{%
\mathbf{C,S}_{\mathbf{C}}}{\Sigma /G_{\Sigma }}\right) _{r_{1}}\right]
\right\} _{r_{1}}\right) ,f_{2}\left( \left\{ \Upsilon \left[ \left( \frac{%
\mathbf{C,S}_{\mathbf{C}}}{\Sigma /G_{\Sigma }}\right) _{r_{2}^{\prime }}%
\right] \right\} _{r_{2}^{\prime }},\left\{ \Upsilon \left[ \left( \frac{%
\mathbf{C,S}_{\mathbf{C}}}{\Sigma /G_{\Sigma }}\right) _{r_{2}}\right]
\right\} _{r_{2}}\right) \right) \\
&&\times \delta \left( f_{2}\left( \left\{ \Upsilon \left[ \left( \frac{%
\mathbf{C,S}_{\mathbf{C}}}{\Sigma /G_{\Sigma }}\right) _{r_{2}^{\prime }}%
\right] \right\} _{r_{2}^{\prime }},\left\{ \Upsilon \left[ \left( \frac{%
\mathbf{C,S}_{\mathbf{C}}}{\Sigma /G_{\Sigma }}\right) _{r_{2}}\right]
\right\} _{r_{2}}\right) \right) +...
\end{eqnarray*}%
where the functn:%
\begin{equation*}
\delta \left( f\left( \left\{ \Upsilon \left[ \left( \frac{\mathbf{C,S}_{%
\mathbf{C}}}{\Sigma /G_{\Sigma }}\right) _{r^{\prime }}\right] \right\}
_{r^{\prime }},\left\{ \Upsilon \left[ \left( \frac{\mathbf{C,S}_{\mathbf{C}}%
}{\Sigma /G_{\Sigma }}\right) _{r}\right] \right\} _{r}\right) \right)
\end{equation*}%
This imposes the constraints on the frequencies between states, and the
integration over $\left( \Sigma /G_{\Sigma }\right) $ considers all possible
shapes that satisfy the constraints on frequencies.\ Potential $V$, $V_{2}$
ar given by (\ref{Pn}).

\subsubsection{Distant interaction and transitions}

Several objects can also interact at some distance and behave as sources
activating other objects. Given (\ref{stc}), (\ref{Trn}), (\ref{Trt}), (\ref%
{Trh}), the distant interactions between two objects where:

\begin{equation}
\mathbf{\hat{A}}^{+}\left( \mathbf{\alpha }_{1}^{\prime },\mathbf{p}%
_{1}^{\prime },\left( S_{1}\right) ^{2}\right) \left( \mathbf{\hat{A}}%
^{+}\left( \mathbf{\alpha }_{2}^{\prime },\mathbf{p}_{2}^{\prime },\left(
S_{2}\right) ^{2}\right) \right) \left( \mathbf{\hat{A}}^{-}\left( \mathbf{%
\alpha }_{1},\mathbf{p}_{1},S_{1}^{2}\right) \right) \left( \mathbf{\hat{A}}%
^{-}\left( \mathbf{\alpha }_{2},\mathbf{p}_{2},S_{2}^{2}\right) \right)
\end{equation}%
\begin{equation}
\mathbf{\hat{A}}^{+}\left( \mathbf{\alpha }_{1}^{\prime },\mathbf{p}%
_{1}^{\prime },\left( S_{1}\right) ^{2}\right) \left( \mathbf{\hat{A}}%
^{+}\left( \mathbf{\alpha }_{2}^{\prime },\mathbf{p}_{2}^{\prime },\left(
S_{2}\right) ^{2}\right) \right) \left( \mathbf{\hat{A}}^{+}\left( \mathbf{%
\alpha }_{1},\mathbf{p}_{1},S_{1}^{2}\right) \right) \left( \mathbf{\hat{A}}%
^{+}\left( \mathbf{\alpha }_{2},\mathbf{p}_{2},S_{2}^{2}\right) \right)
\left( \mathbf{\hat{A}}^{-}\left( \mathbf{\alpha }_{1},\mathbf{p}%
_{1},S_{1}^{2}\right) \right) \left( \mathbf{\hat{A}}^{-}\left( \mathbf{%
\alpha }_{2},\mathbf{p}_{2},S_{2}^{2}\right) \right)
\end{equation}%
or:%
\begin{equation}
\mathbf{\hat{A}}^{+}\left( \mathbf{\alpha }_{1}^{\prime },\mathbf{p}%
_{1}^{\prime },\left( S_{1}^{\prime }\right) ^{2}\right) \left( \mathbf{\hat{%
A}}^{+}\left( \mathbf{\alpha }_{2}^{\prime },\mathbf{p}_{2}^{\prime },\left(
S_{2}^{\prime }\right) ^{2}\right) \right) \left( \mathbf{\hat{A}}^{-}\left( 
\mathbf{\alpha }_{1},\mathbf{p}_{1},S_{1}^{2}\right) \right) \left( \mathbf{%
\hat{A}}^{-}\left( \mathbf{\alpha }_{2},\mathbf{p}_{2},S_{2}^{2}\right)
\right)
\end{equation}%
depending respectively if the states $\left( \mathbf{\alpha }_{i},\mathbf{p}%
_{i},S_{i}^{2}\right) $ are damping, or not, or if some transition is
involved.

Such interactions can be generalized to more than one structures, depending
on the fact that sme initial structures are preserved or not, other
activatd... This leads to the interaction terms:%
\begin{equation*}
{\Huge \Lambda }^{{\Large \dag }}\left[ \left( \frac{\mathbf{C,S}_{\mathbf{C}%
}}{\Sigma /G_{\Sigma }}\right) _{r}\right] \frac{1}{\nu !}\frac{1}{\nu
^{\prime }!}{\Large S}_{\nu ,\nu ^{\prime }}\left[ \left\{ \left( \frac{%
\mathbf{C,S}_{\mathbf{C}}}{\Sigma /G_{\Sigma }}\right) _{r}\right\} ,\left\{
\left( \frac{\mathbf{C}^{\prime }\mathbf{,S}_{\mathbf{C}}^{\prime }}{\Sigma
^{\prime }/G_{\Sigma }^{\prime }}\right) _{r^{\prime }}\right\} \right]
\prod\limits_{r^{\prime }=1}^{v^{\prime }}{\Huge \Lambda }\left[ \left( 
\frac{\mathbf{C}^{\prime }\mathbf{,S}_{\mathbf{C}}^{\prime }}{\Sigma
^{\prime }/G_{\Sigma }^{\prime }}\right) _{r^{\prime }}\right]
\end{equation*}%
or in the operator formalism:%
\begin{equation*}
{\Huge A}^{{\Large \dag }}\left[ \left( \frac{\mathbf{C,S}_{\mathbf{C}}}{%
\Sigma /G_{\Sigma }}\right) _{r}\right] \frac{1}{\nu !}\frac{1}{\nu ^{\prime
}!}{\Large S}_{\nu ,\nu ^{\prime }}\left[ \left\{ \left( \frac{\mathbf{C,S}_{%
\mathbf{C}}}{\Sigma /G_{\Sigma }}\right) _{r}\right\} ,\left\{ \left( \frac{%
\mathbf{C}^{\prime }\mathbf{,S}_{\mathbf{C}}^{\prime }}{\Sigma ^{\prime
}/G_{\Sigma }^{\prime }}\right) _{r^{\prime }}\right\} \right]
\prod\limits_{r^{\prime }=1}^{v^{\prime }}{\Huge A}\left[ \left( \frac{%
\mathbf{C}^{\prime }\mathbf{,S}_{\mathbf{C}}^{\prime }}{\Sigma ^{\prime
}/G_{\Sigma }^{\prime }}\right) _{r^{\prime }}\right]
\end{equation*}%
The process involved is different from gathering or breaking objects. The
distant interactions may induce transitions, change of support without the
compatibility constraint. The condition on support reduces to the condition
that the support allows for activation of an object. A chang of initial
objct creating some interferences, with coherent range:%
\begin{equation*}
\Sigma ^{I}\left( \Upsilon \left[ \left( \frac{\mathbf{C,S}_{\mathbf{C}}}{%
\Sigma /G_{\Sigma }}\right) _{r}\right] \right)
\end{equation*}%
we can consider that an object:%
\begin{equation*}
\left( \frac{\mathbf{C,S}_{\mathbf{C}}}{\Sigma /G_{\Sigma }}\right) _{r}
\end{equation*}%
can be activated when:%
\begin{equation*}
\left( \Sigma /G_{\Sigma }\right) ^{\prime }\cap \left( \cup _{r}\Sigma
^{I}\left( \Upsilon \left[ \left( \frac{\mathbf{C,S}_{\mathbf{C}}}{\Sigma
/G_{\Sigma }}\right) _{r}\right] \right) \right) \neq \emptyset
\end{equation*}

This arises if the $\Sigma ^{\prime }$ are, at least partly in the
constructive interaction range of the signals sent by the objct. This non
restrictive condition arises from the fact that the objects considered,
while transitory, present some stability. Only a part of them is activated
by signals, so that ultimately, the total object is either activated, or
not, which is captured by the notion of amplitudes of transitions, that are
probabilities. The coefficients of transitions are thus:%
\begin{eqnarray*}
&&{\Large S}_{\nu ,\nu ^{\prime }}\left[ \left\{ \left( \frac{\mathbf{C,S}_{%
\mathbf{C}}}{\Sigma /G_{\Sigma }}\right) _{r}\right\} ,\left\{ \left( \frac{%
\mathbf{C}^{\prime }\mathbf{,S}_{\mathbf{C}}^{\prime }}{\Sigma ^{\prime
}/G_{\Sigma }^{\prime }}\right) _{r^{\prime }}\right\} \right] \\
&=&V\left( \left\{ \left( \frac{\mathbf{C,S}_{\mathbf{C}}}{\Sigma /G_{\Sigma
}}\right) _{r}\right\} ,\left\{ \left( \frac{\mathbf{C}^{\prime }\mathbf{,S}%
_{\mathbf{C}}^{\prime }}{\Sigma ^{\prime }/G_{\Sigma }^{\prime }}\right)
_{r^{\prime }}\right\} \right) H\left( \left( \Sigma /G_{\Sigma }\right)
^{\prime }\cap \left( \cup _{r}\Sigma ^{I}\left( \Upsilon \left[ \left( 
\frac{\mathbf{C,S}_{\mathbf{C}}}{\Sigma /G_{\Sigma }}\right) _{r}\right]
\right) \right) \right)
\end{eqnarray*}%
where $H$ is the indicator function, equal to $1$ if the support is not
empty.

The coefficient can be taken is:%
\begin{eqnarray*}
&&V\left( \left\{ \left( \frac{\mathbf{C,S}_{\mathbf{C}}}{\Sigma /G_{\Sigma }%
}\right) _{r}\right\} ,\left\{ \left( \frac{\mathbf{C}^{\prime }\mathbf{,S}_{%
\mathbf{C}}^{\prime }}{\Sigma ^{\prime }/G_{\Sigma }^{\prime }}\right)
_{r^{\prime }}\right\} \right) \\
&=&\exp \left( 1-\frac{Vol\left( \left( \Sigma /G_{\Sigma }\right) ^{\prime
}\cap \left( \cup _{r}\Sigma ^{I}\left( \Upsilon \left[ \left( \frac{\mathbf{%
C,S}_{\mathbf{C}}}{\Sigma /G_{\Sigma }}\right) _{r}\right] \right) \right)
\right) }{Vol\left( \left( \Sigma /G_{\Sigma }\right) ^{\prime }\right) }%
\right)
\end{eqnarray*}%
where $Vol$\ is the volume of the supprt.

\section{Propagator}

The transitions between states are computed through the propagator which is
obtained by considering the quadratic action only.

\subsection{Diagonal part}

The first contribution to the quadratic action is the "diagonal part":%
\begin{equation*}
{\Huge \Lambda }^{{\Large \dag }}\left[ \frac{\mathbf{C,S}_{\mathbf{C}}}{%
\Sigma /G_{\Sigma }}\right] \mathit{G}^{-1}\left[ \frac{\mathbf{C,S}_{%
\mathbf{C}}}{\Sigma /G_{\Sigma }},\frac{\mathbf{C}^{\prime }\mathbf{,S}_{%
\mathbf{C}}^{\prime }}{\Sigma ^{\prime }/G_{\Sigma }^{\prime }}\right] 
{\Huge \Lambda }\left[ \frac{\mathbf{C}^{\prime }\mathbf{,S}_{\mathbf{C}%
}^{\prime }}{\Sigma ^{\prime }/G_{\Sigma }^{\prime }}\right]
\end{equation*}%
This expression describes the "free" propagation of a structure and its
subobjects. This diagonal propagator is obtained by considering the inverse
of the operator (\ref{th}), excluding the off diagonal term ${\Large \hat{S}}%
_{1,1}\left[ \frac{\mathbf{C,S}_{\mathbf{C}}}{\Sigma /G_{\Sigma }},\frac{%
\mathbf{C}^{\prime }\mathbf{,S}_{\mathbf{C}}^{\prime }}{\Sigma ^{\prime
}/G_{\Sigma }^{\prime }}\right] $:%
\begin{eqnarray*}
&&\mathit{G}^{-1}\left[ \frac{\mathbf{C,S}_{\mathbf{C}}}{\Sigma /G_{\Sigma }}%
,\frac{\mathbf{C}^{\prime }\mathbf{,S}_{\mathbf{C}}^{\prime }}{\Sigma
^{\prime }/G_{\Sigma }^{\prime }}\right] \\
&=&-\frac{1}{2}\frac{\delta }{\delta \Sigma ^{2}}-\frac{1}{2}\frac{\delta
^{2}}{\delta \mathbf{S}_{\mathbf{C}}^{2}}+\frac{1}{2}\left( \mathbf{S}_{%
\mathbf{C}}^{T}\mathbf{\nu }\left( \frac{\mathbf{C}}{\Sigma /G_{\Sigma }}%
\right) \mathbf{S}_{\mathbf{C}}+\mathbf{S}_{\mathbf{C}^{\prime }}^{\prime T}%
\mathbf{\nu }\left( \frac{\mathbf{C}^{\prime }}{\Sigma /G_{\Sigma }}\right) 
\mathbf{S}_{\mathbf{C}^{\prime }}^{\prime }-\frac{1}{2}\frac{\Delta ^{2}}{%
\Delta \mathbf{C}^{2}}\right)
\end{eqnarray*}

\subsubsection{Formula without support modifcation}

Disregarding initially the modification of spatial support $-\frac{1}{2}%
\frac{\delta }{\delta \Sigma ^{2}}$, the Green's function for a transition
from $\mathbf{S}_{\mathbf{C}}${} to $\mathbf{S}_{\mathbf{C}}^{\prime }${}
describes the oscillations around the state:

\begin{eqnarray}
&&G\left( \frac{\mathbf{C,S}_{\mathbf{C}}}{\Sigma /G_{\Sigma }},\frac{%
\mathbf{C}^{\prime }\mathbf{,S}_{\mathbf{C}}^{\prime }}{\Sigma ^{\prime
}/G_{\Sigma }^{\prime }}\right)  \label{Gr} \\
&=&\left( -\frac{1}{2}\frac{\delta }{\delta \Sigma ^{2}}-\frac{1}{2}\frac{%
\delta ^{2}}{\delta \mathbf{S}_{\mathbf{C}}^{2}}+\frac{1}{2}\left( \mathbf{S}%
_{\mathbf{C}}^{T}\mathbf{\nu }\left( \frac{\mathbf{C}}{\Sigma /G_{\Sigma }}%
\right) \mathbf{S}_{\mathbf{C}}+\mathbf{S}_{\mathbf{C}^{\prime }}^{\prime T}%
\mathbf{\nu }\left( \frac{\mathbf{C}^{\prime }}{\Sigma /G_{\Sigma }}\right) 
\mathbf{S}_{\mathbf{C}^{\prime }}^{\prime }-\frac{1}{2}\frac{\Delta ^{2}}{%
\Delta \mathbf{C}^{2}}\right) \right) ^{-1}  \notag
\end{eqnarray}%
We assume that the kernel of the operator $-\frac{1}{2}\frac{\Delta ^{2}}{%
\Delta \mathbf{C}^{2}}${} behaves like the kernel of some Laplacian.
Consequently, the associated Green's function depends on some squared
distance between the classes. Note that the discrete part of the state
depends on the class $\mathbf{C}$ but also on various characteristics of the
subobjects, such as frequencies, average activities, encompassd in $\left( 
\mathbf{\alpha }_{k},\mathbf{p}_{k}\right) _{\mathbf{C}}${}. For the
discrete part, we assume:

\begin{eqnarray}
&&G_{-\frac{1}{2}\frac{\Delta ^{2}}{\Delta \mathbf{C}^{2}}}\left( \frac{%
\mathbf{C,S}_{\mathbf{C}}}{\Sigma /G_{\Sigma }},\frac{\mathbf{C}^{\prime }%
\mathbf{,S}_{\mathbf{C}}^{\prime }}{\Sigma ^{\prime }/G_{\Sigma }^{\prime }}%
\right)  \label{Gp} \\
&=&\exp \left( -\left( \frac{\mathbf{C,}\left( \mathbf{\alpha }_{k},\mathbf{p%
}_{k}\right) _{\mathbf{C}}}{\Sigma /G_{\Sigma }}-\frac{\mathbf{C,}\left( 
\mathbf{\alpha }_{k},\mathbf{p}_{k}\right) _{\mathbf{C}}^{\prime }}{\Sigma
^{\prime }/G_{\Sigma }^{\prime }}\right) ^{t}\mathcal{G}^{-1}\left( \frac{%
\mathbf{C,}\left( \mathbf{\alpha }_{k},\mathbf{p}_{k}\right) _{\mathbf{C}}}{%
\Sigma /G_{\Sigma }}\frac{\mathbf{C,}\left( \mathbf{\alpha }_{k},\mathbf{p}%
_{k}\right) _{\mathbf{C}}^{\prime }}{\Sigma ^{\prime }/G_{\Sigma }^{\prime }}%
\right) \left( \frac{\mathbf{C,}\left( \mathbf{\alpha }_{k},\mathbf{p}%
_{k}\right) _{\mathbf{C}}}{\Sigma /G_{\Sigma }}-\frac{\mathbf{C,}\left( 
\mathbf{\alpha }_{k},\mathbf{p}_{k}\right) _{\mathbf{C}}^{\prime }}{\Sigma
^{\prime }/G_{\Sigma }^{\prime }}\right) \right)  \notag
\end{eqnarray}

where $\frac{\mathbf{C,}\left( \mathbf{\alpha }_{k},\mathbf{p}_{k}\right) _{%
\mathbf{C}}}{\Sigma /G_{\Sigma }}${} nd $\frac{\mathbf{C,}\left( \mathbf{%
\alpha }_{k},\mathbf{p}_{k}\right) _{\mathbf{C}}^{\prime }}{\Sigma ^{\prime
}/G_{\Sigma }^{\prime }}${} encompass the discrete part of the states $\frac{%
\mathbf{C,S}_{\mathbf{C}}}{\Sigma /G_{\Sigma }}$ and $\frac{\mathbf{C}%
^{\prime }\mathbf{,S}_{\mathbf{C}}^{\prime }}{\Sigma ^{\prime }/G_{\Sigma
}^{\prime }}${} respectively. The function $\mathcal{G}^{-1}\left( \frac{%
\mathbf{C,}\left( \mathbf{\alpha }_{k},\mathbf{p}_{k}\right) _{\mathbf{C}}}{%
\Sigma /G_{\Sigma }}\frac{\mathbf{C,}\left( \mathbf{\alpha }_{k},\mathbf{p}%
_{k}\right) _{\mathbf{C}}^{\prime }}{\Sigma ^{\prime }/G_{\Sigma }^{\prime }}%
\right) $ represents the metric for state transitions.

Combining (\ref{Gr}) and (\ref{Gp}) we obtain the following expression for
the Green's function:

\begin{eqnarray}
&&G\left( \frac{\mathbf{C,S}_{\mathbf{C}}}{\Sigma /G_{\Sigma }},\frac{%
\mathbf{C}^{\prime }\mathbf{,S}_{\mathbf{C}}^{\prime }}{\Sigma ^{\prime
}/G_{\Sigma }^{\prime }}\right)  \label{Tr} \\
&=&\delta \left( \Sigma ^{\prime }/G_{\Sigma }^{\prime },\Sigma /G_{\Sigma
}\right) \exp \left( -\left( \frac{\mathbf{S}_{\mathbf{C}}-\mathbf{S}_{%
\mathbf{C}}^{^{\prime }}}{\mathbf{\nu }\left( \frac{\mathbf{C}}{\Sigma
/G_{\Sigma }}\right) }\right) ^{2}-\left( \left( \mathbf{S}_{\mathbf{C}}^{T}%
\mathbf{\nu }\left( \frac{\mathbf{C}}{\Sigma /G_{\Sigma }}\right) \mathbf{S}%
_{\mathbf{C}}+\mathbf{S}_{\mathbf{C}^{\prime }}^{\prime T}\mathbf{\nu }%
\left( \frac{\mathbf{C}^{\prime }}{\Sigma /G_{\Sigma }}\right) \mathbf{S}_{%
\mathbf{C}^{\prime }}^{\prime }\right) \right) \right)  \notag \\
&&\times \exp \left( -\left( \frac{\mathbf{C,}\left( \mathbf{\alpha }_{k},%
\mathbf{p}_{k}\right) _{\mathbf{C}}}{\Sigma /G_{\Sigma }}-\frac{\mathbf{C,}%
\left( \mathbf{\alpha }_{k},\mathbf{p}_{k}\right) _{\mathbf{C}}^{\prime }}{%
\Sigma ^{\prime }/G_{\Sigma }^{\prime }}\right) ^{t}\mathcal{G}^{-1}\left( 
\frac{\mathbf{C,}\left( \mathbf{\alpha }_{k},\mathbf{p}_{k}\right) _{\mathbf{%
C}}}{\Sigma /G_{\Sigma }}\frac{\mathbf{C,}\left( \mathbf{\alpha }_{k},%
\mathbf{p}_{k}\right) _{\mathbf{C}}^{\prime }}{\Sigma ^{\prime }/G_{\Sigma
}^{\prime }}\right) \left( \frac{\mathbf{C,}\left( \mathbf{\alpha }_{k},%
\mathbf{p}_{k}\right) _{\mathbf{C}}}{\Sigma /G_{\Sigma }}-\frac{\mathbf{C,}%
\left( \mathbf{\alpha }_{k},\mathbf{p}_{k}\right) _{\mathbf{C}}^{\prime }}{%
\Sigma ^{\prime }/G_{\Sigma }^{\prime }}\right) \right)  \notag
\end{eqnarray}%
Practically, the first term in the Green's function can be expressed as:

\begin{equation}
\delta \left( \mathbf{C-C}^{\prime }\right) \prod\limits_{\left\{ \Delta 
\mathbf{T}_{p}^{\alpha }\right\} \rightarrow \Sigma /G_{\Sigma }}\exp \left(
-\left( \frac{\left\{ \Delta \mathbf{T}_{p}^{\alpha }\right\} -\left\{
\Delta \mathbf{T}_{p}^{\alpha }\right\} ^{\prime }}{\mathbf{\nu }\left( 
\frac{\mathbf{C}}{\Sigma /G_{\Sigma }}\right) }\right) ^{2}-\left( \mathbf{%
\nu }\left( \frac{\mathbf{C}}{\Sigma /G_{\Sigma }}\right) \right) ^{2}\left(
\left( \left\{ \Delta \mathbf{T}_{p}^{\alpha }\right\} \right) ^{2}+\left(
\left\{ \Delta \mathbf{T}_{p}^{\alpha }\right\} ^{\prime }\right)
^{2}\right) \right)
\end{equation}%
where the product is over the set \{$\left\{ \Delta \mathbf{T}_{p}^{\alpha
}\right\} $, defining the sheaf over the state $\Sigma /G_{\Sigma }${} and
the subobjects\footnote{%
Rcl tht gvn (\ref{crd}), th $S_{i}$ r sqncs 
\begin{equation*}
\left( \left( \left[ s,\left[ s_{l}\right] _{l},,\left[ s_{lk}\right]
_{lk},..\right] \right) _{i},\left( \left[ s_{l}^{\left( l\right) },\left[
s_{lk}^{\left( l\right) }\right] _{lk},\left[ s_{lkm}^{\left( l\right) }%
\right] _{lkm},...\right] _{l}\right) _{i},\left( \left[ s_{lk}^{lk},\left[
s_{lkm}^{\left( lk\right) }..\right] _{lkm}\right] _{lk},\left[
s_{lkm}^{\left( lkm\right) },...\right] _{lkm}\right) _{i},..\right)
\end{equation*}%
tht r thm f th fr $\left\{ \Delta \mathbf{T}_{p}^{\alpha }\right\} $.} where 
$\mathbf{S}_{\mathbf{C}}$ is a set of vctor, one per subobjct, i.e. elment
of the site, with associated variances $\mathbf{\nu }\left( \frac{\mathbf{C}%
}{\Sigma /G_{\Sigma }}\right) $.

Remark that the Green functions are associated to the transitions of
products of states:%
\begin{equation*}
\prod\limits_{\mathbf{C}}\left\vert \mathbf{S}_{\mathbf{C}}\right\rangle
\end{equation*}%
where $\left\vert \mathbf{S}_{\mathbf{C}}\right\rangle $ is given by (\ref%
{stn}). The transition writes:%
\begin{equation*}
\prod\limits_{\mathbf{C}^{\prime }}\left\langle \mathbf{S}_{\mathbf{C}%
^{\prime }}^{\prime }\right\vert \prod\limits_{\mathbf{C}}\left\vert \mathbf{%
S}_{\mathbf{C}}\right\rangle =\sum_{\mathbf{C}^{\prime },\mathbf{C}%
}\prod\limits_{\mathbf{C}^{\prime }}\prod\limits_{\mathbf{C}}G\left( \frac{%
\mathbf{C,S}_{\mathbf{C}}}{\Sigma /G_{\Sigma }},\frac{\mathbf{C}^{\prime }%
\mathbf{,S}_{\mathbf{C}}^{\prime }}{\Sigma ^{\prime }/G_{\Sigma }^{\prime }}%
\right)
\end{equation*}%
whr the summation is ovr all pairs $\left( \mathbf{C}^{\prime },\mathbf{C}%
\right) $.

\subsubsection{Including support modification}

Considering also continuous displacement operators,the term $-\frac{1}{2}%
\frac{\delta }{\delta \Sigma ^{2}}$, corresponds to amplitudes that could
modify the class of the bjct by altering the support. This should modify the
local coordinates and class, but this case will be included in off diagonal
terms of the propagator.

We assume that the propagator associated with the continuous change of
support is:%
\begin{equation*}
\exp \left[ -\int \left( \Sigma /G_{\Sigma }-\Sigma ^{\prime }/G_{\Sigma
}^{\prime }\right) ^{t}\mathcal{F}\left( \frac{\mathbf{C}}{\Sigma /G_{\Sigma
}},\frac{\mathbf{C}^{\prime }}{\Sigma ^{\prime }/G_{\Sigma }^{\prime }}%
\right) \left( \Sigma /G_{\Sigma }-\Sigma ^{\prime }/G_{\Sigma }^{\prime
}\right) \right]
\end{equation*}%
whr $\mathcal{F}\left( \frac{\mathbf{C}}{\Sigma /G_{\Sigma }},\frac{\mathbf{C%
}^{\prime }}{\Sigma ^{\prime }/G_{\Sigma }^{\prime }}\right) $ is a metric
depending on the initial and final state reached by the object. The vector $%
\left( \Sigma /G_{\Sigma }-\Sigma ^{\prime }/G_{\Sigma }^{\prime }\right) $
represents the displacment of all points in the support from $\Sigma
/G_{\Sigma }$ to $\Sigma ^{\prime }/G_{\Sigma }^{\prime }$.

The matrix and the summation should define a distance between objects and
betwn subobjects. In the displacement, the class should be kept constant,
which necessitates the inclusion of the factor $\delta \left( \mathbf{C-C}%
^{\prime }\right) $. This ensures that the propagation respects the class
structure in transitions.

The full form for the propagator in the first approximation should thus be
expressed as:%
\begin{eqnarray}
&&\mathit{G}\left( \frac{\mathbf{C,S}_{\mathbf{C}}}{\Sigma /G_{\Sigma }},%
\frac{\mathbf{C}^{\prime }\mathbf{,S}_{\mathbf{C}}^{\prime }}{\Sigma
^{\prime }/G_{\Sigma }^{\prime }}\right)  \label{gn} \\
&=&\exp \left( -\left( \frac{\mathbf{S}_{\mathbf{C}}-\mathbf{S}_{\mathbf{C}%
}^{^{\prime }}}{\mathbf{\nu }\left( \frac{\mathbf{C}}{\Sigma /G_{\Sigma }}%
\right) }\right) ^{2}-\left( \left( \mathbf{S}_{\mathbf{C}}^{T}\mathbf{\nu }%
\left( \frac{\mathbf{C}}{\Sigma /G_{\Sigma }}\right) \mathbf{S}_{\mathbf{C}}+%
\mathbf{S}_{\mathbf{C}^{\prime }}^{\prime T}\mathbf{\nu }\left( \frac{%
\mathbf{C}^{\prime }}{\Sigma /G_{\Sigma }}\right) \mathbf{S}_{\mathbf{C}%
^{\prime }}^{\prime }\right) \right) -\frac{1}{2}\left( \mathbf{C}-\mathbf{C}%
^{\prime }\right) ^{2}\right)  \notag \\
&&\times \exp \left( -\left( \frac{\mathbf{C,}\left( \mathbf{\alpha }_{k},%
\mathbf{p}_{k}\right) _{\mathbf{C}}}{\Sigma /G_{\Sigma }}-\frac{\mathbf{C,}%
\left( \mathbf{\alpha }_{k},\mathbf{p}_{k}\right) _{\mathbf{C}}^{\prime }}{%
\Sigma ^{\prime }/G_{\Sigma }^{\prime }}\right) ^{t}\mathcal{G}^{-1}\left( 
\frac{\mathbf{C,}\left( \mathbf{\alpha }_{k},\mathbf{p}_{k}\right) _{\mathbf{%
C}}}{\Sigma /G_{\Sigma }}\frac{\mathbf{C,}\left( \mathbf{\alpha }_{k},%
\mathbf{p}_{k}\right) _{\mathbf{C}}^{\prime }}{\Sigma ^{\prime }/G_{\Sigma
}^{\prime }}\right) \left( \frac{\mathbf{C,}\left( \mathbf{\alpha }_{k},%
\mathbf{p}_{k}\right) _{\mathbf{C}}}{\Sigma /G_{\Sigma }}-\frac{\mathbf{C,}%
\left( \mathbf{\alpha }_{k},\mathbf{p}_{k}\right) _{\mathbf{C}}^{\prime }}{%
\Sigma ^{\prime }/G_{\Sigma }^{\prime }}\right) \right)  \notag \\
&&\times \exp \left[ -\int \left( \Sigma /G_{\Sigma }-\Sigma ^{\prime
}/G_{\Sigma }^{\prime }\right) ^{t}\mathcal{F}\left( \frac{\mathbf{C}}{%
\Sigma /G_{\Sigma }},\frac{\mathbf{C}^{\prime }}{\Sigma ^{\prime }/G_{\Sigma
}^{\prime }}\right) \left( \Sigma /G_{\Sigma }-\Sigma ^{\prime }/G_{\Sigma
}^{\prime }\right) \right]  \notag
\end{eqnarray}

\subsection{Off diagonal terms}

\subsubsection{Form of the coefficents}

For off diagonal terms, the kernel ${\Large S}_{1,1}\left[ \frac{\mathbf{C,S}%
_{\mathbf{C}}}{\Sigma /G_{\Sigma }},\frac{\mathbf{C}^{\prime }\mathbf{,S}_{%
\mathbf{C}}^{\prime }}{\Sigma ^{\prime }/G_{\Sigma }^{\prime }}\right] $ and
the associated terms involving transtns between structures:%
\begin{equation*}
{\Huge \Lambda }^{{\Large \dag }}\left[ \frac{\mathbf{C,S}_{\mathbf{C}}}{%
\Sigma /G_{\Sigma }}\right] {\Large S}_{1,1}\left[ \frac{\mathbf{C,S}_{%
\mathbf{C}}}{\Sigma /G_{\Sigma }},\frac{\mathbf{C}^{\prime }\mathbf{,S}_{%
\mathbf{C}}^{\prime }}{\Sigma ^{\prime }/G_{\Sigma }^{\prime }}\right] 
{\Huge \Lambda }\left[ \frac{\mathbf{C}^{\prime }\mathbf{,S}_{\mathbf{C}%
}^{\prime }}{\Sigma ^{\prime }/G_{\Sigma }^{\prime }}\right]
\end{equation*}%
can be also considered, in the operator formalism a series of
creation/annihilation, replacing the field ${\Huge \Lambda }$ by an operator
(\ref{pt}), this leads to: 
\begin{eqnarray*}
&&{\Huge A}^{{\Large \dag }}\left[ \frac{\mathbf{C,S}_{\mathbf{C}}}{\Sigma
/G_{\Sigma }}\right] {\Large S}_{1,1}\left[ \frac{\mathbf{C,S}_{\mathbf{C}}}{%
\Sigma /G_{\Sigma }},\frac{\mathbf{C}^{\prime }\mathbf{,S}_{\mathbf{C}%
}^{\prime }}{\Sigma ^{\prime }/G_{\Sigma }^{\prime }}\right] {\Huge A}\left[ 
\frac{\mathbf{C}^{\prime }\mathbf{,S}_{\mathbf{C}}^{\prime }}{\Sigma
^{\prime }/G_{\Sigma }^{\prime }}\right] \\
&=&\sum_{\frac{\mathbf{C,}\left( \mathbf{\alpha }_{k},\mathbf{p}_{k}\right)
_{\mathbf{C}}}{\Sigma /G_{\Sigma }},\frac{\mathbf{C}^{\prime }\mathbf{,}%
\left( \mathbf{\alpha }_{k},\mathbf{p}_{k}\right) _{\mathbf{C}}^{\prime }}{%
\Sigma ^{\prime }/G_{\Sigma }^{\prime }}}\prod \left( \mathbf{\hat{A}}_{%
\mathbf{C}}^{+}\mathbf{\hat{A}}_{\left( \mathbf{\alpha }_{k},\mathbf{p}%
_{k}\right) _{\mathbf{C}}}^{+}+\mathbf{\hat{A}}_{\mathbf{C}}^{-}\mathbf{\hat{%
A}}_{\left( \mathbf{\alpha }_{k},\mathbf{p}_{k}\right) _{\mathbf{C}%
}}^{-}\right) \\
&&\mathcal{S}\left[ \frac{\mathbf{C,}\left( \mathbf{\alpha }_{k},\mathbf{p}%
_{k}\right) _{\mathbf{C}}}{\Sigma /G_{\Sigma }},\frac{\mathbf{C}^{\prime }%
\mathbf{,}\left( \mathbf{\alpha }_{k},\mathbf{p}_{k}\right) _{\mathbf{C}%
}^{\prime }}{\Sigma ^{\prime }/G_{\Sigma }^{\prime }}\right]
\prod_{p^{\prime }}\left( \mathbf{\hat{A}}_{\mathbf{C}^{\prime }}^{+}\mathbf{%
\hat{A}}_{\left( \mathbf{\alpha }_{k},\mathbf{p}_{k}\right) _{\mathbf{C}%
^{\prime }}^{\prime }}^{+}+\mathbf{\hat{A}}_{\mathbf{C}^{\prime }}^{-}%
\mathbf{\hat{A}}_{\left( \mathbf{\alpha }_{k},\mathbf{p}_{k}\right) _{%
\mathbf{C}^{\prime }}^{\prime }}^{-}\right)
\end{eqnarray*}%
Recall that $\mathbf{C}$ represents the possible classes of states over $%
\Sigma /G_{\Sigma }${}, while $\left( \mathbf{\alpha }_{k},\mathbf{p}%
_{k}\right) _{\mathbf{C}}${} serves as a basis for generators within the
class $\mathbf{C}_{p}$. These are labeled by the possible configurations $%
\mathbf{\alpha }_{k}${} and the state parameters $\mathbf{p}_{k}$. Together,
they form the coordinates within the class and the associated state among
the classes for the object. The summation over the variabels $\frac{\mathbf{%
C,}\left( \mathbf{\alpha }_{k},\mathbf{p}_{k}\right) _{\mathbf{C}}}{\Sigma
/G_{\Sigma }}$ is written as a functional integral of the frm: 
\begin{equation*}
\int \mathcal{D}\left[ \frac{\mathbf{C,}\left( \mathbf{\alpha }_{k},\mathbf{p%
}_{k}\right) _{\mathbf{C}}}{\Sigma /G_{\Sigma }}\right]
\end{equation*}%
This integral accounts for all possible variations and contributions arising
from the parameters $\left( \mathbf{\alpha }_{k},\mathbf{p}_{k}\right) _{%
\mathbf{C}}${}within the class $\mathbf{C}$.

The choice of the operator formalism allows us to emphasize the irreversible
nature of certain transitions from unstable to stable states. This captures
the dynamics where transitions between states are not symmetric and can
reflect a directional evolution.

Note that this products $\prod_{p^{\prime }}\mathbf{\hat{A}}_{\mathbf{C}%
^{\prime }}^{-}\mathbf{\hat{A}}_{\left( \mathbf{\alpha }_{k},\mathbf{p}%
_{k}\right) _{\mathbf{C}^{\prime }}^{\prime }}^{-}$ are consistent with the
quadratic action (\ref{DRT}), since the field ${\Huge \Lambda }\left[ \frac{%
\mathbf{C}^{\prime }\mathbf{,S}_{\mathbf{C}}^{\prime }}{\Sigma ^{\prime
}/G_{\Sigma }^{\prime }}\right] $ describing all possible activations of an
object's state along with subobject activation, corresponds in terms of
states of products of creation annihilations on the state.

The lowest order contribution is the diagonal part:%
\begin{equation}
\sum_{\frac{\mathbf{C,}\left( \mathbf{\alpha }_{k},\mathbf{p}_{k}\right) _{%
\mathbf{C}}}{\Sigma /G_{\Sigma }}}\left[ \prod \mathbf{\hat{A}}_{\mathbf{C}%
}^{+}\mathbf{\hat{A}}_{\left( \mathbf{\alpha }_{k},\mathbf{p}_{k}\right) _{%
\mathbf{C}}}^{+}\mathcal{S}\left[ \frac{\mathbf{C,}\left( \mathbf{\alpha }%
_{k},\mathbf{p}_{k}\right) _{\mathbf{C}}}{\Sigma /G_{\Sigma }},\frac{\mathbf{%
C}^{\prime }\mathbf{,}\left( \mathbf{\alpha }_{k},\mathbf{p}_{k}\right) _{%
\mathbf{C}}^{\prime }}{\Sigma ^{\prime }/G_{\Sigma }^{\prime }}\right] 
\mathbf{\hat{A}}_{\mathbf{C}^{\prime }}^{-}\mathbf{\hat{A}}_{\left( \mathbf{%
\alpha }_{k},\mathbf{p}_{k}\right) _{\mathbf{C}^{\prime }}}^{-}+C_{\left[ 
\mathbf{C}\right] }\right]  \label{fd}
\end{equation}%
which consist in a sum of terms (\ref{Td}) describing the internal
interctions for all coverings (\ref{Cv}) or (\ref{Cvr}). The matrices: 
\begin{equation*}
\mathcal{S}\left[ \frac{\mathbf{C,}\left( \mathbf{\alpha }_{k},\mathbf{p}%
_{k}\right) _{\mathbf{C}}}{\Sigma /G_{\Sigma }},\frac{\mathbf{C}^{\prime }%
\mathbf{,}\left( \mathbf{\alpha }_{k},\mathbf{p}_{k}\right) _{\mathbf{C}%
}^{\prime }}{\Sigma ^{\prime }/G_{\Sigma }^{\prime }}\right]
\end{equation*}%
are in general, the expression of coefficents of:%
\begin{eqnarray*}
&&\left( -\frac{1}{2}\nabla _{\left( \mathbf{T}\right) _{\Sigma ^{2}}}^{2}+%
\frac{1}{2}\left( \Delta \mathbf{T}_{p}^{\alpha }\right) _{\Sigma ^{2}}^{t}%
\mathbf{A}_{\Sigma ^{2}}^{\alpha }\left( \Delta \mathbf{T}_{p}^{\alpha
}\right) +\mathbf{C}\right) +\sum_{s}\left( -\frac{1}{2}\nabla _{\left( 
\mathbf{T}\right) _{\Sigma _{s}^{2}}}^{2}+\frac{1}{2}\left( \Delta \mathbf{T}%
_{p}^{\alpha }\right) _{_{\Sigma _{s}^{2}}}^{t}\mathbf{A}_{_{\Sigma
_{s}^{2}}}^{\alpha }\left( \Delta \mathbf{T}_{p}^{\alpha }\right) +\mathbf{C}%
\right) \\
&=&\mathbf{\hat{A}}_{\mathbf{C}_{\alpha ,p}}^{+}\mathbf{\hat{A}}_{\mathbf{S}%
_{\mathbf{C}\alpha ,p}}^{+}\sqrt{\mathbf{A}_{\Sigma ^{2}}^{\alpha }}\mathbf{%
\hat{A}}_{\mathbf{C}_{\alpha ,p}}^{-}\mathbf{\hat{A}}_{\mathbf{S}_{\mathbf{C}%
\alpha ,p}}^{-}+\left( \mathbf{C-}\frac{1}{2}\left\vert \Sigma
^{2}\right\vert \right) +\sum_{s}\left( \mathbf{\hat{A}}_{\mathbf{C}_{\alpha
,p,s}}^{+}\mathbf{\hat{A}}_{\mathbf{S}_{\mathbf{C}\alpha ,p,s}}^{+}\sqrt{%
\mathbf{A}_{\Sigma _{s}^{2}}^{\alpha }}\mathbf{\hat{A}}_{\mathbf{C}_{\alpha
,p}}^{-}\mathbf{\hat{A}}_{\mathbf{S}_{\mathbf{C}\alpha ,p}}^{-}+\mathbf{C-}%
\frac{1}{2}\left\vert \Sigma _{s}^{2}\right\vert \right)
\end{eqnarray*}%
in terms of creation annihilation, so that they are the collection of
matrices:%
\begin{equation*}
\sqrt{\mathbf{A}_{\Sigma ^{2}}^{\alpha }},\sqrt{\mathbf{A}_{\Sigma
_{s}^{2}}^{\alpha }}
\end{equation*}%
The $C_{\left[ p\right] }$ are the collection of $\mathbf{C-}\frac{1}{2}%
\left\vert \Sigma ^{2}\right\vert ,\mathbf{C-}\frac{1}{2}\left\vert \Sigma
_{s}^{2}\right\vert $.

Discontinuous transitions are thus determined by the interactions terms.
When a state with charateristics:%
\begin{equation*}
\frac{\mathbf{C,}\left( \mathbf{\alpha }_{k},\mathbf{p}_{k}\right) _{\mathbf{%
C}}^{\prime }}{\Sigma ^{\prime }/G_{\Sigma }^{\prime }}
\end{equation*}%
is such that these characteristics fits with some interaction potential, a
change of discrete class may arise. This corresponds to an interaction term:%
\begin{eqnarray*}
&&\mathcal{S}\left[ \frac{\mathbf{C,}\left( \mathbf{\alpha }_{k},\mathbf{p}%
_{k}\right) _{\mathbf{C}}}{\Sigma /G_{\Sigma }},\frac{\mathbf{C}^{\prime }%
\mathbf{,}\left( \mathbf{\alpha }_{k},\mathbf{p}_{k}\right) _{\mathbf{C}%
}^{\prime }}{\Sigma ^{\prime }/G_{\Sigma }^{\prime }}\right] \\
&=&\sum_{\frac{\mathbf{C,}\left( \mathbf{\alpha }_{k},\mathbf{p}_{k}\right)
_{\mathbf{C}}}{\Sigma /G_{\Sigma }},\frac{\mathbf{C}^{\prime }\mathbf{,}%
\left( \mathbf{\alpha }_{k},\mathbf{p}_{k}\right) _{\mathbf{C}}^{\prime }}{%
\Sigma ^{\prime }/G_{\Sigma }^{\prime }}}\prod \mathbf{\hat{A}}_{\mathbf{C}%
}^{+}\mathbf{\hat{A}}_{\left( \mathbf{\alpha }_{k},\mathbf{p}_{k}\right) _{%
\mathbf{C}}}^{+}\delta \left( \Upsilon _{\left( \mathbf{p}_{k}\right)
^{\prime }}-\Upsilon _{\mathbf{p}_{k}}\right) \prod_{p^{\prime }}\mathbf{%
\hat{A}}_{\mathbf{C}^{\prime }}^{-}\mathbf{\hat{A}}_{\left( \mathbf{\alpha }%
_{k},\mathbf{p}_{k}\right) _{\mathbf{C}^{\prime }}^{\prime }}^{-}
\end{eqnarray*}

\subsection{Propagator including off diagonal terms}

Including off diagonal terms in the Green function leads to consider the
following modification of the Green functions as a series expansion:%
\begin{eqnarray}
&&\mathit{\hat{G}}\left( \frac{\mathbf{C,S}_{\mathbf{C}}}{\Sigma /G_{\Sigma }%
},\frac{\mathbf{C}^{\prime }\mathbf{,S}_{\mathbf{C}}^{\prime }}{\Sigma
^{\prime }/G_{\Sigma }^{\prime }}\right)  \label{Pg} \\
&=&\left( \mathit{G}^{-1}\left( \frac{\mathbf{C,S}_{\mathbf{C}}}{\Sigma
/G_{\Sigma }},\frac{\mathbf{C}^{\prime }\mathbf{,S}_{\mathbf{C}}^{\prime }}{%
\Sigma ^{\prime }/G_{\Sigma }^{\prime }}\right) +{\Large F}\left[ \frac{%
\mathbf{C,S}_{\mathbf{C}}}{\Sigma /G_{\Sigma }},\frac{\mathbf{C}^{\prime }%
\mathbf{,S}_{\mathbf{C}}^{\prime }}{\Sigma ^{\prime }/G_{\Sigma }^{\prime }}%
\right] \right) ^{-1}  \notag \\
&=&\mathit{G}\left( \frac{\mathbf{C,S}_{\mathbf{C}}}{\Sigma /G_{\Sigma }},%
\frac{\mathbf{C}^{\prime }\mathbf{,S}_{\mathbf{C}}^{\prime }}{\Sigma
^{\prime }/G_{\Sigma }^{\prime }}\right)  \notag \\
&&-\mathit{G}\left( \frac{\mathbf{C,S}_{\mathbf{C}}}{\Sigma /G_{\Sigma }},%
\frac{\mathbf{C}^{\prime }\mathbf{,S}_{\mathbf{C}}^{\prime }}{\Sigma
^{\prime }/G_{\Sigma }^{\prime }}\right) \ast {\Large F}\left[ \frac{\mathbf{%
C,S}_{\mathbf{C}}}{\Sigma /G_{\Sigma }},\frac{\mathbf{C}^{\prime }\mathbf{,S}%
_{\mathbf{C}}^{\prime }}{\Sigma ^{\prime }/G_{\Sigma }^{\prime }}\right]
\ast \mathit{G}\left( \frac{\mathbf{C,S}_{\mathbf{C}}}{\Sigma /G_{\Sigma }},%
\frac{\mathbf{C}^{\prime }\mathbf{,S}_{\mathbf{C}}^{\prime }}{\Sigma
^{\prime }/G_{\Sigma }^{\prime }}\right)  \notag \\
&&+...  \notag
\end{eqnarray}%
where $\mathit{G}\left( \frac{\mathbf{C,S}_{\mathbf{C}}}{\Sigma /G_{\Sigma }}%
,\frac{\mathbf{C}^{\prime }\mathbf{,S}_{\mathbf{C}}^{\prime }}{\Sigma
^{\prime }/G_{\Sigma }^{\prime }}\right) $ is defined by (\ref{gn}) and the
general formula for ${\Large F}\left[ \frac{\mathbf{C,S}_{\mathbf{C}}}{%
\Sigma /G_{\Sigma }},\frac{\mathbf{C}^{\prime }\mathbf{,S}_{\mathbf{C}%
}^{\prime }}{\Sigma ^{\prime }/G_{\Sigma }^{\prime }}\right] $ is (\ref{fd}):%
\begin{eqnarray*}
&&\mathit{G}\left( \frac{\mathbf{C,S}_{\mathbf{C}}}{\Sigma /G_{\Sigma }},%
\frac{\mathbf{C}^{\prime }\mathbf{,S}_{\mathbf{C}}^{\prime }}{\Sigma
^{\prime }/G_{\Sigma }^{\prime }}\right) \ast {\Large F}\left[ \frac{\mathbf{%
C,S}_{\mathbf{C}}}{\Sigma /G_{\Sigma }},\frac{\mathbf{C}^{\prime }\mathbf{,S}%
_{\mathbf{C}}^{\prime }}{\Sigma ^{\prime }/G_{\Sigma }^{\prime }}\right]
\ast \mathit{G}\left( \frac{\mathbf{C,S}_{\mathbf{C}}}{\Sigma /G_{\Sigma }},%
\frac{\mathbf{C}^{\prime }\mathbf{,S}_{\mathbf{C}}^{\prime }}{\Sigma
^{\prime }/G_{\Sigma }^{\prime }}\right) \\
&=&\sum_{\frac{\mathbf{C}_{1}\mathbf{,}\left( \mathbf{\alpha }_{k},\mathbf{p}%
_{k}\right) _{\mathbf{C}_{1}}}{\Sigma /G_{\Sigma }},\frac{\mathbf{C}%
_{1}^{\prime }\mathbf{,}\left( \mathbf{\alpha }_{k},\mathbf{p}_{k}\right) _{%
\mathbf{C}_{1}^{\prime }}}{\Sigma /G_{\Sigma }}}\prod \mathcal{F}\left[ 
\frac{\mathbf{C}_{1}\mathbf{,}\left( \mathbf{\alpha }_{k},\mathbf{p}%
_{k}\right) _{\mathbf{C}_{1}}}{\Sigma _{1}/G_{\Sigma _{1}}},\frac{\mathbf{C}%
^{\prime }\mathbf{,}\left( \mathbf{\alpha }_{k},\mathbf{p}_{k}\right) _{%
\mathbf{C}^{\prime }}}{\Sigma /G_{\Sigma }}\right] \\
&&\times \mathit{G}\left( \frac{\mathbf{C}_{1}^{\prime }\mathbf{,S}_{\mathbf{%
C}_{1}}^{\prime }}{\Sigma _{1}^{\prime }/G_{\Sigma _{1}^{\prime }}},\frac{%
\mathbf{C}_{1}\mathbf{,S}_{\mathbf{C}_{1}}}{\Sigma _{1}/G_{\Sigma _{1}}}%
\right) \mathbf{\hat{A}}_{\mathbf{C}_{1}}^{+}\mathbf{\hat{A}}_{\left( 
\mathbf{\alpha }_{k},\mathbf{p}_{k}\right) _{\mathbf{C}_{1}}}^{+}\mathbf{%
\hat{A}}_{\mathbf{C}_{1}^{\prime }}^{-}\mathbf{\hat{A}}_{\left( \mathbf{%
\alpha }_{k},\mathbf{p}_{k}\right) _{\mathbf{C}_{1}^{\prime }}}^{-}\mathit{G}%
\left( \frac{\mathbf{C}^{\prime }\mathbf{,S}_{\mathbf{C}}^{\prime }}{\Sigma
^{\prime }/G_{\Sigma }^{\prime }},\frac{\mathbf{C,S}_{\mathbf{C}}}{\Sigma
/G_{\Sigma }}\right) \\
&=&\sum_{\frac{\mathbf{C}_{1}\mathbf{,}\left( \mathbf{\alpha }_{k},\mathbf{p}%
_{k}\right) _{\mathbf{C}_{1}}}{\Sigma /G_{\Sigma }},\frac{\mathbf{C}%
_{1}^{\prime }\mathbf{,}\left( \mathbf{\alpha }_{k},\mathbf{p}_{k}\right) _{%
\mathbf{C}_{1}^{\prime }}}{\Sigma /G_{\Sigma }}}\prod \mathit{G}\left( \frac{%
\mathbf{C,S}_{\mathbf{C}}}{\Sigma /G_{\Sigma }},\frac{\mathbf{C}_{1}\mathbf{%
,S}_{\mathbf{C}_{1}}}{\Sigma _{1}/G_{\Sigma _{1}}}\right) \mathcal{F}\left[ 
\frac{\mathbf{C}_{1}\mathbf{,}\left( \mathbf{\alpha }_{k},\mathbf{p}%
_{k}\right) _{\mathbf{C}_{1}}}{\Sigma /G_{\Sigma }},\frac{\mathbf{C}^{\prime
}\mathbf{,}\left( \mathbf{\alpha }_{k},\mathbf{p}_{k}\right) _{\mathbf{C}%
^{\prime }}}{\Sigma /G_{\Sigma }}\right] \mathit{G}\left( \frac{\mathbf{C}%
^{\prime }\mathbf{,S}_{\mathbf{C}}^{\prime }}{\Sigma ^{\prime }/G_{\Sigma
}^{\prime }},\frac{\mathbf{C,S}_{\mathbf{C}}}{\Sigma /G_{\Sigma }}\right)
\end{eqnarray*}%
where the product $\prod $ is for the object $\frac{\mathbf{C,S}_{\mathbf{C}}%
}{\Sigma /G_{\Sigma }}$ and the whole set of subobjects.

\section{Effective action}

\subsection{Principle}

The computation of transitions of the system are performed by transitions
amplitudes of the form:%
\begin{equation*}
\left\langle H^{\prime }\right\vert \exp \left( -{\Huge S}\left( {\Huge %
\Lambda }^{{\Large \dag }},{\Huge \Lambda }\right) \right) \left\vert
H\right\rangle
\end{equation*}%
where ${\Huge S}\left( {\Huge \Lambda }^{{\Large \dag }},{\Huge \Lambda }%
\right) $ is given by (\ref{sfr}). These transitions involve series of
graphs describing intermediate transitions between initial and final states.

The interactions within the whole system and the transitions can be
described by an effective action of the system. Its form is similar to the
classical action and has the following form:

\begin{eqnarray}
{\huge S}^{\left( ef\right) }\left( {\Huge \Lambda ,\Lambda }^{{\Large \dag }%
}\right) &{\huge =}&{\Huge \Lambda }^{{\Large \dag }}\left[ \frac{\mathbf{C,S%
}_{\mathbf{C}}}{\Sigma /G_{\Sigma }}\right] {\Large F}\left[ \frac{\mathbf{%
C,S}_{\mathbf{C}}}{\Sigma /G_{\Sigma }},\frac{\mathbf{C}^{\prime }\mathbf{,S}%
_{\mathbf{C}}^{\prime }}{\Sigma ^{\prime }/G_{\Sigma }^{\prime }}\right] 
{\Huge \Lambda }\left[ \frac{\mathbf{C}^{\prime }\mathbf{,S}_{\mathbf{C}%
}^{\prime }}{\Sigma ^{\prime }/G_{\Sigma }^{\prime }}\right]  \label{Fc} \\
&&+\sum_{v}\left( \prod\limits_{r=1}^{v}{\Huge \Lambda }_{r}^{{\Large \dag }}%
\left[ \left( \frac{\mathbf{C,S}_{\mathbf{C}}}{\Sigma /G_{\Sigma }}\right)
_{r}\right] \right) {\Large F}\left[ \left\{ \left( \frac{\mathbf{C,S}_{%
\mathbf{C}}}{\Sigma /G_{\Sigma }}\right) _{r}\right\} ,\left\{ \left( \frac{%
\mathbf{C}^{\prime }\mathbf{,S}_{\mathbf{C}}^{\prime }}{\Sigma ^{\prime
}/G_{\Sigma }^{\prime }}\right) _{r^{\prime }}\right\} \right] \left(
\prod\limits_{r^{\prime }=1}^{v^{\prime }}{\Huge \Lambda }_{r^{\prime }}%
\left[ \left( \frac{\mathbf{C}^{\prime }\mathbf{,S}_{\mathbf{C}}^{\prime }}{%
\Sigma ^{\prime }/G_{\Sigma }^{\prime }}\right) _{r^{\prime }}\right] \right)
\notag
\end{eqnarray}%
corresponding to the series of operators in the operator formalism:%
\begin{eqnarray*}
&&{\Huge A}^{{\Large \dag }}\left[ \frac{\mathbf{C,S}_{\mathbf{C}}}{\Sigma
/G_{\Sigma }}\right] {\Large F}\left[ \frac{\mathbf{C,S}_{\mathbf{C}}}{%
\Sigma /G_{\Sigma }},\frac{\mathbf{C}^{\prime }\mathbf{,S}_{\mathbf{C}%
}^{\prime }}{\Sigma ^{\prime }/G_{\Sigma }^{\prime }}\right] {\Huge A}\left[ 
\frac{\mathbf{C}^{\prime }\mathbf{,S}_{\mathbf{C}}^{\prime }}{\Sigma
^{\prime }/G_{\Sigma }^{\prime }}\right] \\
&&+\sum_{v}\left( \prod\limits_{r=1}^{v}{\Huge A}_{r}^{{\Large +}}\left[
\left( \frac{\mathbf{C,S}_{\mathbf{C}}}{\Sigma /G_{\Sigma }}\right) _{r}%
\right] \right) {\Large F}\left[ \left\{ \left( \frac{\mathbf{C,S}_{\mathbf{C%
}}}{\Sigma /G_{\Sigma }}\right) _{r}\right\} ,\left\{ \left( \frac{\mathbf{C}%
^{\prime }\mathbf{,S}_{\mathbf{C}}^{\prime }}{\Sigma ^{\prime }/G_{\Sigma
}^{\prime }}\right) _{r^{\prime }}\right\} \right] \left(
\prod\limits_{r^{\prime }=1}^{v^{\prime }}{\Huge A}_{r^{\prime }}\left[
\left( \frac{\mathbf{C}^{\prime }\mathbf{,S}_{\mathbf{C}}^{\prime }}{\Sigma
^{\prime }/G_{\Sigma }^{\prime }}\right) _{r^{\prime }}\right] \right)
\end{eqnarray*}%
This effective action can be considered as a classical action, in which the
various integrations have been integratd out by path integrals. The
functionals:%
\begin{equation*}
{\Large F}\left[ \left\{ \left( \frac{\mathbf{C,S}_{\mathbf{C}}}{\Sigma
/G_{\Sigma }}\right) _{r}\right\} ,\left\{ \left( \frac{\mathbf{C}^{\prime }%
\mathbf{,S}_{\mathbf{C}}^{\prime }}{\Sigma ^{\prime }/G_{\Sigma }^{\prime }}%
\right) _{r^{\prime }}\right\} \right]
\end{equation*}%
compute the connected correlation functions. For a set of classes $\mathbf{C}%
_{r}$, $\mathbf{C}_{r^{\prime }}^{\prime }$, the connected correlations are
functions:%
\begin{equation*}
{\Large F}_{\left\{ \mathbf{C}_{r}\right\} ,\left\{ \mathbf{C}_{r^{\prime
}}^{\prime }\right\} }\left( \left\{ \mathbf{S}_{\mathbf{C}_{r}}\right\}
,\left\{ \mathbf{S}_{\mathbf{C}_{r^{\prime }}}^{\prime }\right\} \right)
\end{equation*}%
that are invariant with respct to the action of:%
\begin{equation*}
\prod \left( G_{\Sigma _{r}}\right) \prod \left( G_{\Sigma _{r^{\prime
}}^{\prime }}^{\prime }\right)
\end{equation*}%
Technically, the coeffcients:%
\begin{equation}
{\Large F}\left[ \left\{ \left( \frac{\mathbf{C,S}_{\mathbf{C}}}{\Sigma
/G_{\Sigma }}\right) _{r}\right\} ,\left\{ \left( \frac{\mathbf{C}^{\prime }%
\mathbf{,S}_{\mathbf{C}}^{\prime }}{\Sigma ^{\prime }/G_{\Sigma }^{\prime }}%
\right) _{r^{\prime }}\right\} \right]  \label{cfn}
\end{equation}%
are renormalized series of graphs involving the propagators $\mathit{G}%
\left( \left( \frac{\mathbf{C,S}_{\mathbf{C}}}{\Sigma /G_{\Sigma }}\right)
_{i},\left( \frac{\mathbf{C}^{\prime }\mathbf{,S}_{\mathbf{C}}^{\prime }}{%
\Sigma ^{\prime }/G_{\Sigma }^{\prime }}\right) _{i}\right) $ of
intermediate transitions as well as constraint betwn various intermediate
states. These propagators connect the coefficients (\ref{vr}) of the
classical action. Thus, coefficients (\ref{cfn}) gather sums and succession
of simple transition They are obtaind by computing product and convolutions
of contributions:%
\begin{equation*}
\prod {\Large S}_{\nu ,\nu ^{\prime }}\left[ \left\{ \left( \frac{\mathbf{C,S%
}_{\mathbf{C}}}{\Sigma /G_{\Sigma }}\right) _{r}\right\} ,\left\{ \left( 
\frac{\mathbf{C}^{\prime }\mathbf{,S}_{\mathbf{C}}^{\prime }}{\Sigma
^{\prime }/G_{\Sigma }^{\prime }}\right) _{r^{\prime }}\right\} \right]
\end{equation*}%
that arr linked by propagators under some condition.

\subsection{Interpretation of some terms}

We provide the interpretation for several terms of the effective action,
including activation or desactivation of subbjct states, and transition
induced by interactions.

\subsubsection{Activation, disactivation of subobjects}

\paragraph{Activation of subobject}

Assume that a state of a subobject is activated through some external signal:%
\begin{equation*}
\left[ s_{r}^{\left( r\right) },\left[ s_{rk}^{\left( r\right) }\right]
_{rk},\left[ s_{rkm}^{\left( r\right) }\right] _{rkm},...\right] ^{\left(
1\right) }\rightarrow \left[ s_{r}^{\left( r\right) },\left[ s_{rk}^{\left(
r\right) }\right] _{rk},\left[ s_{rkm}^{\left( r\right) }\right] _{rkm},...%
\right] ^{\left( 2\right) }
\end{equation*}%
Ths induces that the subobjects transition towards:%
\begin{eqnarray*}
&&\left\{ \left\{ \left[ s_{l}^{\left( l\right) },\left[ s_{lk}^{\left(
l\right) }\right] _{lk},\left[ s_{lkm}^{\left( l\right) }\right] _{lkm},...%
\right] \right\} _{l\neq r},\left[ s_{r}^{\left( r\right) },\left[
s_{rk}^{\left( r\right) }\right] _{rk},\left[ s_{rkm}^{\left( r\right) }%
\right] _{rkm},...\right] ^{\left( 1\right) }\right\} \\
&\rightarrow &\left\{ \left\{ \left[ s_{l}^{\left( l\right) },\left[
s_{lk}^{\left( l\right) }\right] _{lk},\left[ s_{lkm}^{\left( l\right) }%
\right] _{lkm},...\right] \right\} _{l\neq r},\left[ s_{r}^{\left( r\right)
},\left[ s_{rk}^{\left( r\right) }\right] _{rk},\left[ s_{rkm}^{\left(
r\right) }\right] _{rkm},...\right] ^{\left( 2\right) }\right\}
\end{eqnarray*}%
If this set of indepedent activation satisfies the condition (\ref{Cln}) to
define a global state transition:%
\begin{equation*}
\left[ s,\left[ s_{l}\right] _{l},,\left[ s_{lk}\right] _{lk},..\right]
\rightarrow \left[ s,\left[ s_{l}\right] _{l},,\left[ s_{lk}\right] _{lk},..%
\right] ^{\left( 2\right) }
\end{equation*}%
with the following action, where integrals are included:%
\begin{equation*}
\int D\left( \Sigma /G_{\Sigma }\right) D\left( \Sigma ^{\prime }/G_{\Sigma
}^{\prime }\right) D\left( \mathbf{C,S}_{\mathbf{C}}\right) D\left( \mathbf{C%
}^{\prime }\mathbf{,S}_{\mathbf{C}}^{\prime }\right) {\Huge \Lambda }^{%
{\Large \dag }}\left[ \frac{\mathbf{C,S}_{\mathbf{C}}}{\Sigma /G_{\Sigma }}%
\right] {\Large F}\left[ \frac{\mathbf{C,S}_{\mathbf{C}}}{\Sigma /G_{\Sigma }%
},\frac{\mathbf{C}^{\prime }\mathbf{,S}_{\mathbf{C}}^{\prime }}{\Sigma
/G_{\Sigma }}\right] {\Huge \Lambda }\left[ \frac{\mathbf{C}^{\prime }%
\mathbf{,S}_{\mathbf{C}}^{\prime }}{\Sigma /G_{\Sigma }}\right]
\end{equation*}%
with:%
\begin{equation*}
{\Large F}\left[ \frac{\mathbf{C,S}_{\mathbf{C}}}{\Sigma /G_{\Sigma }},\frac{%
\mathbf{C}^{\prime }\mathbf{,S}_{\mathbf{C}}^{\prime }}{\Sigma ^{\prime
}/G_{\Sigma }^{\prime }}\right] =\delta \left( \left( \left( \frac{\mathbf{%
C,S}_{\mathbf{C}}}{\Sigma /G_{\Sigma }}\right) -\frac{\mathbf{C}^{\prime }%
\mathbf{,S}_{\mathbf{C}}^{\prime }}{\Sigma ^{\prime }/G_{\Sigma }^{\prime }}%
\right) /\left( \frac{\mathbf{C,S}_{\mathbf{C}}}{\Sigma /G_{\Sigma }}\right)
_{l}\right) {\Large F}\left[ \left( \frac{\mathbf{C,S}_{\mathbf{C}}}{\Sigma
/G_{\Sigma }}\right) _{l}\right]
\end{equation*}%
which imposes the constraint that $\frac{\mathbf{C,S}_{\mathbf{C}}}{\Sigma
/G_{\Sigma }}$ and $\frac{\mathbf{C}^{\prime }\mathbf{,S}_{\mathbf{C}%
}^{\prime }}{\Sigma ^{\prime }/G_{\Sigma }^{\prime }}$ coincides except on
the subobject $\left( \frac{\mathbf{C,S}_{\mathbf{C}}}{\Sigma /G_{\Sigma }}%
\right) _{l}$ \ where $l$ is any label, integer or not. The kernel has
support on $\left( \frac{\mathbf{C,S}_{\mathbf{C}}}{\Sigma /G_{\Sigma }}%
\right) _{l}$.

If the creation of a new state:%
\begin{equation*}
\left[ s,\left[ s_{l}\right] _{l},,\left[ s_{lk}\right] _{lk},..\right]
^{\left( 2\right) }
\end{equation*}%
preserves the initial state, the corresponding terms writes:%
\begin{equation*}
{\Huge \Lambda }^{{\Large \dag }}\left[ \frac{\mathbf{C,S}_{\mathbf{C}}}{%
\Sigma /G_{\Sigma }}\right] {\Huge \Lambda }^{{\Large \dag }}\left[ \frac{%
\mathbf{C}^{\prime }\mathbf{,S}_{\mathbf{C}}^{\prime }}{\Sigma ^{\prime
}/G_{\Sigma }^{\prime }}\right] {\Large F}\left[ \frac{\mathbf{C,S}_{\mathbf{%
C}}}{\Sigma /G_{\Sigma }},\frac{\mathbf{C}^{\prime }\mathbf{,S}_{\mathbf{C}%
}^{\prime }}{\Sigma ^{\prime }/G_{\Sigma }^{\prime }}\right] {\Huge \Lambda }%
\left[ \frac{\mathbf{C}^{\prime }\mathbf{,S}_{\mathbf{C}}^{\prime }}{\Sigma
^{\prime }/G_{\Sigma }^{\prime }}\right]
\end{equation*}

In \cite{GLr} we showed that structures, here subobjects, can be activated
by some signal if their frequencies satisfy some constraints depending on
the frequencies and amplitude of the external signal. We write the external
source as: 
\begin{equation*}
J\left( \left\{ f_{i}\right\} \right)
\end{equation*}%
and the conditions to activate $\left( \frac{\mathbf{C,S}_{\mathbf{C}}}{%
\Sigma /G_{\Sigma }}\right) _{l}$ has the form:%
\begin{equation*}
\delta \left( f\left( \Upsilon _{\mathbf{p}}\left( \left( \frac{\mathbf{C,S}%
_{\mathbf{C}}}{\Sigma /G_{\Sigma }}\right) _{l}\right) ,\left\{
f_{i}\right\} \right) \right)
\end{equation*}%
The two possibilities of activation gather in terms written as:%
\begin{eqnarray*}
&&{\Huge \Lambda }^{{\Large \dag }}\left[ \frac{\mathbf{C,S}_{\mathbf{C}}}{%
\Sigma /G_{\Sigma }}\right] {\Huge \Lambda }^{{\Large \dag }}\left[ \frac{%
\mathbf{C}^{\prime }\mathbf{,S}_{\mathbf{C}}^{\prime }}{\Sigma ^{\prime
}/G_{\Sigma }^{\prime }}\right] {\Large F}\left[ \frac{\mathbf{C,S}_{\mathbf{%
C}}}{\Sigma /G_{\Sigma }},\frac{\mathbf{C}^{\prime }\mathbf{,S}_{\mathbf{C}%
}^{\prime }}{\Sigma ^{\prime }/G_{\Sigma }^{\prime }}\right] {\Huge \Lambda }%
\left[ \frac{\mathbf{C}^{\prime }\mathbf{,S}_{\mathbf{C}}^{\prime }}{\Sigma
^{\prime }/G_{\Sigma }^{\prime }}\right] \\
&&+{\Huge \Lambda }^{{\Large \dag }}\left[ \left( \frac{\mathbf{C,S}_{%
\mathbf{C}}}{\Sigma /G_{\Sigma }}\right) _{l}\right] \delta \left( f\left(
\Upsilon _{\mathbf{p}}\left( \left( \frac{\mathbf{C,S}_{\mathbf{C}}}{\Sigma
/G_{\Sigma }}\right) _{l}\right) ,\left\{ f_{i}\right\} \right) \right)
J\left( \left\{ f_{i}\right\} \right)
\end{eqnarray*}

Such terms induce the emergence of a new class for the initial object, along
with new states with different activities and connections. This new states
may coexist with the initial object.

\paragraph{Desactivation of objects}

Assume that:%
\begin{equation*}
\left[ s,\left[ s_{l}\right] _{l},,\left[ s_{lk}\right] _{lk},..\right]
\rightarrow \left[ s,\left[ s_{l}\right] _{l}^{\prime },,\left[ s_{lk}\right]
_{lk},..\right]
\end{equation*}%
and that the state obtained does not satisfies the preseahf condition. As
explained before, this implies some instability, i.e. a non minimal
effective action. As a consequence, in this case, either 
\begin{equation*}
\left[ s,\left[ s_{l}\right] _{l},,\left[ s_{lk}\right] _{lk},..\right]
\rightarrow vac
\end{equation*}%
or:%
\begin{equation*}
\left[ s,\left[ s_{l}\right] _{l},,\left[ s_{lk}\right] _{lk},..\right]
\rightarrow \left[ s^{\prime },\left[ s_{l}\right] _{l}^{\prime },,\left[
s_{lk}\right] _{lk},..\right]
\end{equation*}%
which is a new state, i.e. a new object. This corresponds to a change of
class with effective vertices:%
\begin{equation*}
{\Huge \Lambda }^{{\Large \dag }}\left[ \frac{\mathbf{C,S}_{\mathbf{C}}}{%
\Sigma /G_{\Sigma }}\right] {\Large F}\left[ \frac{\mathbf{C,S}_{\mathbf{C}}%
}{\Sigma /G_{\Sigma }},\frac{\mathbf{C}^{\prime }\mathbf{,S}_{\mathbf{C}%
}^{\prime }}{\Sigma /G_{\Sigma }}\right] {\Huge \Lambda }\left[ \frac{%
\mathbf{C}^{\prime }\mathbf{,S}_{\mathbf{C}}^{\prime }}{\Sigma /G_{\Sigma }}%
\right] +{\Huge \Lambda }^{{\Large \dag }}\left[ \frac{\mathbf{C,S}_{\mathbf{%
C}}}{\Sigma /G_{\Sigma }}\right] {\Large F}\left[ \frac{\mathbf{C,S}_{%
\mathbf{C}}}{\Sigma /G_{\Sigma }}\right]
\end{equation*}%
where the second term describes the desactivation of the initial state.

\subsubsection{Transitions}

As seen previously, the transitions of some object due to interactions (\ref%
{Trn}), (\ref{Trt}), (\ref{Trh}) includes transitions of subobjects. They
are described by terms:%
\begin{eqnarray}
&&{\Huge \Lambda }^{{\Large \dag }}\left[ \frac{\mathbf{C}_{1}^{\prime }%
\mathbf{,S}_{\mathbf{C}_{1}}^{\prime }}{\Sigma _{1}/G_{\Sigma _{1}}}\right] 
{\Huge \Lambda }^{{\Large \dag }}\left[ \frac{\mathbf{C}_{2}^{\prime }%
\mathbf{,S}_{\mathbf{C}_{2}}^{\prime }}{\Sigma _{2}/G_{\Sigma _{2}}}\right]
\label{TRn} \\
&&\times {\Large F}\left[ \frac{\mathbf{C}_{1}^{\prime }\mathbf{,S}_{\mathbf{%
C}_{1}}^{\prime }}{\Sigma _{1}/G_{\Sigma _{1}}},\frac{\mathbf{C}_{2}^{\prime
}\mathbf{,S}_{\mathbf{C}_{2}}^{\prime }}{\Sigma _{2}/G_{\Sigma _{2}}},\frac{%
\mathbf{C}_{1}\mathbf{,S}_{\mathbf{C}_{1}}}{\Sigma _{1}/G_{\Sigma _{1}}},%
\frac{\mathbf{C}_{2}\mathbf{,S}_{\mathbf{C}_{2}}}{\Sigma _{2}/G_{\Sigma _{2}}%
}\right] {\Huge \Lambda }\left[ \frac{\mathbf{C}_{1}\mathbf{,S}_{\mathbf{C}%
_{1}}}{\Sigma _{1}/G_{\Sigma _{1}}}\right] {\Huge \Lambda }\left[ \frac{%
\mathbf{C}_{2}\mathbf{,S}_{\mathbf{C}_{2}}}{\Sigma _{2}/G_{\Sigma _{2}}}%
\right]  \notag
\end{eqnarray}%
if the state defined by $\mathbf{\alpha },\mathbf{p},S^{2}$ is damping, or:%
\begin{eqnarray}
&&{\Huge \Lambda }^{{\Large \dag }}\left[ \frac{\mathbf{C}_{1}^{\prime }%
\mathbf{,S}_{\mathbf{C}_{1}}^{\prime }}{\Sigma _{1}/G_{\Sigma _{1}}}\right] 
{\Huge \Lambda }^{{\Large \dag }}\left[ \frac{\mathbf{C}_{2}^{\prime }%
\mathbf{,S}_{\mathbf{C}_{2}}^{\prime }}{\Sigma _{2}/G_{\Sigma _{2}}}\right] 
{\Huge \Lambda }^{\dag }\left[ \frac{\mathbf{C}_{1}\mathbf{,S}_{\mathbf{C}%
_{1}}}{\Sigma _{1}/G_{\Sigma _{1}}}\right] {\Huge \Lambda }^{\dag }\left[ 
\frac{\mathbf{C}_{2}\mathbf{,S}_{\mathbf{C}_{2}}}{\Sigma _{2}/G_{\Sigma _{2}}%
}\right]  \label{TRt} \\
&&\times {\Large F}\left[ \frac{\mathbf{C}_{1}^{\prime }\mathbf{,S}_{\mathbf{%
C}_{1}}^{\prime }}{\Sigma _{1}/G_{\Sigma _{1}}},\frac{\mathbf{C}_{2}^{\prime
}\mathbf{,S}_{\mathbf{C}_{2}}^{\prime }}{\Sigma _{2}/G_{\Sigma _{2}}},\frac{%
\mathbf{C}_{1}\mathbf{,S}_{\mathbf{C}_{1}}}{\Sigma _{1}/G_{\Sigma _{1}}},%
\frac{\mathbf{C}_{2}\mathbf{,S}_{\mathbf{C}_{2}}}{\Sigma _{2}/G_{\Sigma _{2}}%
}\right] {\Huge \Lambda }\left[ \frac{\mathbf{C}_{1}\mathbf{,S}_{\mathbf{C}%
_{1}}}{\Sigma _{1}/G_{\Sigma _{1}}}\right] {\Huge \Lambda }\left[ \frac{%
\mathbf{C}_{2}\mathbf{,S}_{\mathbf{C}_{2}}}{\Sigma _{2}/G_{\Sigma _{2}}}%
\right]  \notag
\end{eqnarray}%
otherwise.

When interaction between states implies some transition, the equivalent of (%
\ref{Trh}) becoms:%
\begin{eqnarray}
&&{\Huge \Lambda }^{{\Large \dag }}\left[ \frac{\mathbf{C}_{1}^{\prime }%
\mathbf{,S}_{\mathbf{C}_{1}}^{\prime }}{\Sigma _{1}^{\prime }/G_{\Sigma
_{1}^{\prime }}}\right] {\Huge \Lambda }^{{\Large \dag }}\left[ \frac{%
\mathbf{C}_{2}^{\prime }\mathbf{,S}_{\mathbf{C}_{2}}^{\prime }}{\Sigma
_{2}^{\prime }/G_{\Sigma _{2}^{\prime }}}\right]  \label{TRh} \\
&&\times {\Large F}\left[ \frac{\mathbf{C}_{1}^{\prime }\mathbf{,S}_{\mathbf{%
C}_{1}}^{\prime }}{\Sigma _{1}^{\prime }/G_{\Sigma _{1}^{\prime }}},\frac{%
\mathbf{C}_{2}^{\prime }\mathbf{,S}_{\mathbf{C}_{2}}^{\prime }}{\Sigma
_{2}^{\prime }/G_{\Sigma _{2}^{\prime }}},\frac{\mathbf{C}_{1}\mathbf{,S}_{%
\mathbf{C}_{1}}}{\Sigma _{1}/G_{\Sigma _{1}}},\frac{\mathbf{C}_{2}\mathbf{,S}%
_{\mathbf{C}_{2}}}{\Sigma _{2}/G_{\Sigma _{2}}}\right] {\Huge \Lambda }\left[
\frac{\mathbf{C}_{1}\mathbf{,S}_{\mathbf{C}_{1}}}{\Sigma _{1}/G_{\Sigma _{1}}%
}\right] {\Huge \Lambda }\left[ \frac{\mathbf{C}_{2}\mathbf{,S}_{\mathbf{C}%
_{2}}}{\Sigma _{2}/G_{\Sigma _{2}}}\right]  \notag
\end{eqnarray}%
The coefficients matrices: 
\begin{equation*}
{\Large F}\left[ \frac{\mathbf{C}_{1}^{\prime }\mathbf{,S}_{\mathbf{C}%
_{1}}^{\prime }}{\Sigma _{1}/G_{\Sigma _{1}}},\frac{\mathbf{C}_{2}^{\prime }%
\mathbf{,S}_{\mathbf{C}_{2}}^{\prime }}{\Sigma _{2}/G_{\Sigma _{2}}},\frac{%
\mathbf{C}_{1}\mathbf{,S}_{\mathbf{C}_{1}}}{\Sigma _{1}/G_{\Sigma _{1}}},%
\frac{\mathbf{C}_{2}\mathbf{,S}_{\mathbf{C}_{2}}}{\Sigma _{2}/G_{\Sigma _{2}}%
}\right]
\end{equation*}%
and:%
\begin{equation*}
{\Large F}\left[ \frac{\mathbf{C}_{1}^{\prime }\mathbf{,S}_{\mathbf{C}%
_{1}}^{\prime }}{\Sigma _{1}^{\prime }/G_{\Sigma _{1}^{\prime }}},\frac{%
\mathbf{C}_{2}^{\prime }\mathbf{,S}_{\mathbf{C}_{2}}^{\prime }}{\Sigma
_{2}^{\prime }/G_{\Sigma _{2}^{\prime }}},\frac{\mathbf{C}_{1}\mathbf{,S}_{%
\mathbf{C}_{1}}}{\Sigma _{1}/G_{\Sigma _{1}}},\frac{\mathbf{C}_{2}\mathbf{,S}%
_{\mathbf{C}_{2}}}{\Sigma _{2}/G_{\Sigma _{2}}}\right]
\end{equation*}%
are general. When transitions only involve subobject, the matrices involve
only coefficients related to this subobject.

\subsection{Derivation of effective action}

The series for the effective action starts with classical action, and
computes the Legendre transform of generatng function of connected
correlation functions.

The effective term:%
\begin{equation*}
{\Large F}_{\nu ,\nu ^{\prime }}\left[ \left\{ \left( \frac{\mathbf{C,S}_{%
\mathbf{C}}}{\Sigma /G_{\Sigma }}\right) _{r}\right\} ,\left\{ \left( \frac{%
\mathbf{C}^{\prime }\mathbf{,S}_{\mathbf{C}}^{\prime }}{\Sigma ^{\prime
}/G_{\Sigma }^{\prime }}\right) _{r^{\prime }}\right\} \right]
\end{equation*}%
arising in (\ref{Fc}) results from integratd procses computed perturbatively
by computing 1P irreductible graphs arising in (\ref{sfr}):%
\begin{eqnarray*}
&&{\Large F}_{\nu ,\nu ^{\prime }}\left[ \left\{ \left( \frac{\mathbf{C,S}_{%
\mathbf{C}}}{\Sigma /G_{\Sigma }}\right) _{r}\right\} ,\left\{ \left( \frac{%
\mathbf{C}^{\prime }\mathbf{,S}_{\mathbf{C}}^{\prime }}{\Sigma ^{\prime
}/G_{\Sigma }^{\prime }}\right) _{r^{\prime }}\right\} \right] \\
&=&\frac{1}{\nu !}\frac{1}{\nu ^{\prime }!}{\Large S}_{\nu ,\nu ^{\prime }}%
\left[ \left\{ \left( \frac{\mathbf{C,S}_{\mathbf{C}}}{\Sigma /G_{\Sigma }}%
\right) _{r}\right\} ,\left\{ \left( \frac{\mathbf{C}^{\prime }\mathbf{,S}_{%
\mathbf{C}}^{\prime }}{\Sigma ^{\prime }/G_{\Sigma }^{\prime }}\right)
_{r^{\prime }}\right\} \right] \\
&&+{\Large S}_{\nu ,\nu _{1}}\left[ \left\{ \left( \frac{\mathbf{C,S}_{%
\mathbf{C}}}{\Sigma /G_{\Sigma }}\right) _{r}\right\} ,\left\{ \left( \frac{%
\mathbf{C}_{1}^{\prime }\mathbf{,S}_{\mathbf{C}_{1}}^{\prime }}{\Sigma
_{1}^{\prime }/G_{\Sigma _{1}}^{\prime }}\right) _{r_{1}}\right\} \right] \\
&&\times \prod\limits_{r_{1}}\mathit{G}\left( \left( \frac{\mathbf{C}%
_{1}^{\prime }\mathbf{,S}_{\mathbf{C}_{1}}^{\prime }}{\Sigma _{1}^{\prime
}/G_{\Sigma _{1}}^{\prime }}\right) _{r_{1}},\left( \frac{\mathbf{C}_{1}%
\mathbf{,S}_{\mathbf{C}_{1}}}{\Sigma _{1}/G_{\Sigma _{1}}}\right)
_{r_{1}}\right) {\Large S}_{\nu _{1},\nu ^{\prime }}\left[ \left\{ \left( 
\frac{\mathbf{C}_{1}\mathbf{,S}_{\mathbf{C}_{1}}}{\Sigma _{1}/G_{\Sigma _{1}}%
}\right) _{r_{1}}\right\} ,\left\{ \left( \frac{\mathbf{C}^{\prime }\mathbf{%
,S}_{\mathbf{C}}^{\prime }}{\Sigma ^{\prime }/G_{\Sigma }^{\prime }}\right)
_{r^{\prime }}\right\} \right] \\
&&+...
\end{eqnarray*}%
where the coefficients ${\Large S}_{\nu ,\nu ^{\prime }}$ are the vertices
of an initial action (\ref{sfr}) and the sum is constrained by restricting
the graphs to the 1PI graphs.

\subsubsection{First and second order in vertices expansions}

At the first order in vertices, the effective action is the classical action:

\begin{equation*}
{\huge S}_{0}^{\left( ef\right) }\left( {\Huge \Lambda ,\Lambda }^{{\Large %
\dag }}\right) ={\Huge S}\left( {\Huge \Lambda }^{{\Large \dag }},{\Huge %
\Lambda }\right)
\end{equation*}

To find the second order in vertices-number corrections, we consider graphs
with 2 vertices:%
\begin{equation*}
{\Large S}_{l_{1}^{\prime },l_{1}}\left[ \left\{ \left( \frac{\mathbf{C,S}_{%
\mathbf{C}}}{\Sigma /G_{\Sigma }}\right) _{i_{1}^{\prime }}^{\prime
}\right\} ,\left\{ \left( \frac{\mathbf{C,S}_{\mathbf{C}}}{\Sigma /G_{\Sigma
}}\right) _{i_{1}}\right\} \right]
\end{equation*}%
and:%
\begin{equation*}
{\Large S}_{l_{2}^{\prime },l_{2}}\left[ \left\{ \left( \frac{\mathbf{C,S}_{%
\mathbf{C}}}{\Sigma /G_{\Sigma }}\right) _{i_{2}^{\prime }}^{\prime
}\right\} ,\left\{ \left( \frac{\mathbf{C,S}_{\mathbf{C}}}{\Sigma /G_{\Sigma
}}\right) _{i_{2}}\right\} \right]
\end{equation*}%
Since propagators are not symetric by time reversal (see Appendix 1 for
example), we consider irreversible processes only. To obtain $1$P
irreductible graph, we have to order the vertices as:%
\begin{equation*}
{\Large S}_{l_{1}^{\prime },l_{1}}\left[ \left\{ \left( \frac{\mathbf{C,S}_{%
\mathbf{C}}}{\Sigma /G_{\Sigma }}\right) _{i_{1}^{\prime }}^{\prime
}\right\} ,\left\{ \left( \frac{\mathbf{C,S}_{\mathbf{C}}}{\Sigma /G_{\Sigma
}}\right) _{i_{1}}\right\} \right] ,{\Large S}_{l_{2}^{\prime },l_{2}}\left[
\left\{ \left( \frac{\mathbf{C,S}_{\mathbf{C}}}{\Sigma /G_{\Sigma }}\right)
_{i_{2}^{\prime }}^{\prime }\right\} ,\left\{ \left( \frac{\mathbf{C,S}_{%
\mathbf{C}}}{\Sigma /G_{\Sigma }}\right) _{i_{2}}\right\} \right]
\end{equation*}

The graphs \ involving:%
\begin{equation*}
\prod\limits_{k=1^{\prime }}^{n^{\prime }}{\Huge \Lambda }^{{\Large \dag }}%
\left[ \left( \frac{\mathbf{C,S}_{\mathbf{C}}}{\Sigma /G_{\Sigma }}\right)
_{k^{\prime }}^{\prime }\right] \prod\limits_{i=1}^{n}{\Huge \Lambda }\left[
\left( \frac{\mathbf{C,S}_{\mathbf{C}}}{\Sigma /G_{\Sigma }}\right) _{k}%
\right]
\end{equation*}%
are obtained by connecting $l_{1}^{\prime }$ states $\left( \frac{\mathbf{C,S%
}_{\mathbf{C}}}{\Sigma /G_{\Sigma }}\right) _{k^{\prime }}^{\prime }$ to $%
{\Large S}_{l_{1}^{\prime },l_{1}}$ and $n^{\prime }-l_{1}^{\prime }$ to $%
{\Large S}_{l_{2}^{\prime },l_{2}}$. Then by connecting $l_{2}$ states $%
\left( \frac{\mathbf{C,S}_{\mathbf{C}}}{\Sigma /G_{\Sigma }}\right) _{k}$ to 
${\Large S}_{l_{2}^{\prime },l_{2}}\ $nd $n-l_{2}$ t ${\Large S}%
_{l_{1}^{\prime },l_{1}}$. Then the $l_{1}-\left( n-l_{2}\right) $ remaining
valences ${\Large S}_{l_{1}^{\prime },l_{1}}$ are connected to the $%
l_{2}^{\prime }-\left( n^{\prime }-l_{1}^{\prime }\right) $ remaining
valences of ${\Large S}_{l_{2}^{\prime },l_{2}}$. To each connection,
propagtor are associatd. These connections imply constraints:%
\begin{equation*}
l_{1}-\left( n-l_{2}\right) =l_{2}^{\prime }-\left( n^{\prime
}-l_{1}^{\prime }\right)
\end{equation*}%
that is:%
\begin{equation*}
l_{1}+l_{2}-\left( l_{1}^{\prime }+l_{2}^{\prime }\right) =n-n^{\prime }
\end{equation*}

Ultimately, the second order contribution to the effective action becomes:%
\begin{eqnarray*}
&&\prod\limits_{k=1^{\prime }}^{n^{\prime }}{\Huge \Lambda }^{{\Large \dag }}%
\left[ \left( \frac{\mathbf{C,S}_{\mathbf{C}}}{\Sigma /G_{\Sigma }}\right)
_{k^{\prime }}^{\prime }\right] {\Large S}_{l_{1}^{\prime },l_{1}}\left[
\left\{ \left( \frac{\mathbf{C,S}_{\mathbf{C}}}{\Sigma /G_{\Sigma }}\right)
_{i_{1}^{\prime }}^{\prime }\right\} ,\left\{ \left( \frac{\mathbf{C,S}_{%
\mathbf{C}}}{\Sigma /G_{\Sigma }}\right) _{i_{1}}\right\} \right] \\
&&\times {\Large S}_{l_{2}^{\prime },l_{2}}\left[ \left\{ \left( \frac{%
\mathbf{C,S}_{\mathbf{C}}}{\Sigma /G_{\Sigma }}\right) _{i_{2}^{\prime
}}^{\prime }\right\} ,\left\{ \left( \frac{\mathbf{C,S}_{\mathbf{C}}}{\Sigma
/G_{\Sigma }}\right) _{i_{2}}\right\} \right] \prod\limits_{i=1}^{n}{\Huge %
\Lambda }\left[ \left( \frac{\mathbf{C,S}_{\mathbf{C}}}{\Sigma /G_{\Sigma }}%
\right) _{k}\right] \\
&&\times \prod\limits_{r^{\prime }}\mathit{G}\left( \left( \frac{\mathbf{C,S}%
_{\mathbf{C}}}{\Sigma /G_{\Sigma }}\right) _{r},\left( \frac{\mathbf{C}%
^{\prime }\mathbf{,S}_{\mathbf{C}}^{\prime }}{\Sigma ^{\prime }/G_{\Sigma
}^{\prime }}\right) _{r^{\prime }}\right)
\end{eqnarray*}

where:%
\begin{eqnarray*}
\left\{ \left( \frac{\mathbf{C,S}_{\mathbf{C}}}{\Sigma /G_{\Sigma }}\right)
_{i_{1}^{\prime }}^{\prime }\right\} &\subset &\left\{ \left( \frac{\mathbf{%
C,S}_{\mathbf{C}}}{\Sigma /G_{\Sigma }}\right) _{k^{\prime }}^{\prime
}\right\} \\
\left\{ \left( \frac{\mathbf{C,S}_{\mathbf{C}}}{\Sigma /G_{\Sigma }}\right)
_{i_{2}}\right\} &\subset &\left\{ \left( \frac{\mathbf{C,S}_{\mathbf{C}}}{%
\Sigma /G_{\Sigma }}\right) _{k}\right\}
\end{eqnarray*}%
and:%
\begin{eqnarray*}
\left( \frac{\mathbf{C,S}_{\mathbf{C}}}{\Sigma /G_{\Sigma }}\right) _{r}
&\in &\left\{ \left( \frac{\mathbf{C,S}_{\mathbf{C}}}{\Sigma /G_{\Sigma }}%
\right) _{i_{2}}\right\} /\left( \left\{ \left( \frac{\mathbf{C,S}_{\mathbf{C%
}}}{\Sigma /G_{\Sigma }}\right) _{i_{1}}\right\} \cap \left\{ \left( \frac{%
\mathbf{C,S}_{\mathbf{C}}}{\Sigma /G_{\Sigma }}\right) _{k}\right\} \right)
\\
\left( \frac{\mathbf{C}^{\prime }\mathbf{,S}_{\mathbf{C}}^{\prime }}{\Sigma
^{\prime }/G_{\Sigma }^{\prime }}\right) _{r^{\prime }} &\in &\left\{ \left( 
\frac{\mathbf{C,S}_{\mathbf{C}}}{\Sigma /G_{\Sigma }}\right) _{i_{2}^{\prime
}}^{\prime }\right\} /\left( \left\{ \left( \frac{\mathbf{C,S}_{\mathbf{C}}}{%
\Sigma /G_{\Sigma }}\right) _{i_{2}^{\prime }}^{\prime }\right\} \cap
\left\{ \left( \frac{\mathbf{C,S}_{\mathbf{C}}}{\Sigma /G_{\Sigma }}\right)
_{k^{\prime }}^{\prime }\right\} \right)
\end{eqnarray*}

\subsubsection{Full series of graphs}

Appendx 9 presents the derivation of the series of graphs. The coefficients: 
\begin{eqnarray*}
&&{\Large F}\left[ \left\{ \left( \frac{\mathbf{C,S}_{\mathbf{C}}}{\Sigma
/G_{\Sigma }}\right) _{n}\right\} ,\left\{ \left( \frac{\mathbf{C}^{\prime }%
\mathbf{,S}_{\mathbf{C}}^{\prime }}{\Sigma ^{\prime }/G_{\Sigma }^{\prime }}%
\right) _{n^{\prime }}\right\} \right] \\
&=&\sum_{\substack{ \left\{ \left( \frac{\mathbf{C}^{\prime }\mathbf{,S}_{%
\mathbf{C}}^{\prime }}{\Sigma ^{\prime }/G_{\Sigma }^{\prime }}\right)
_{n^{\prime }}\right\} =\cup \left( \frac{\mathbf{C,S}_{\mathbf{C}}}{\Sigma
/G_{\Sigma }}\right) _{k_{i,l}^{\prime }\leqslant \left( n_{i}\right)
_{\gamma _{l}}}^{\prime }  \\ \left\{ \left( \frac{\mathbf{C,S}_{\mathbf{C}}%
}{\Sigma /G_{\Sigma }}\right) _{n}\right\} =\cup \left( \frac{\mathbf{C,S}_{%
\mathbf{C}}}{\Sigma /G_{\Sigma }}\right) _{k_{i,l}\leqslant \left(
n_{f}\right) _{\gamma _{l}}}}}\mathit{G}\left( \mathcal{L}^{\left( L\right)
}\right) \left[ \left\{ \left( \frac{\mathbf{C,S}_{\mathbf{C}}}{\Sigma
/G_{\Sigma }}\right) _{k_{i,l}^{\prime }\leqslant \left( n_{i}\right)
_{\gamma _{l}}}^{\prime }\right\} _{i,l},\left\{ \left( \frac{\mathbf{C,S}_{%
\mathbf{C}}}{\Sigma /G_{\Sigma }}\right) _{k_{i,l}\leqslant \left(
n_{f}\right) _{\gamma _{l}}}\right\} _{i,l}\right]
\end{eqnarray*}%
where $\mathit{G}\left( \mathcal{L}^{\left( l\right) }\right) $ is the $L$
loops contribution to the $\left( n,n^{\prime }\right) $ vertices. This
vertices is computed by products of $L$ loops graphs connected with each
other.%
\begin{eqnarray}
&&\mathit{G}\left( \mathcal{L}^{\left( L\right) }\right) \left[ \left\{
\left( \frac{\mathbf{C,S}_{\mathbf{C}}}{\Sigma /G_{\Sigma }}\right)
_{k_{i,l}^{\prime }\leqslant \left( n_{i}\right) _{\gamma _{l}}}^{\prime
}\right\} _{i,l},\left\{ \left( \frac{\mathbf{C,S}_{\mathbf{C}}}{\Sigma
/G_{\Sigma }}\right) _{k_{i,l}\leqslant \left( n_{f}\right) _{\gamma
_{l}}}\right\} _{i,l}\right]  \label{ffC} \\
&=&\prod\limits_{l=1}^{L}\left\{ \mathit{G}\left( L_{p,p^{\prime }}\right)
\right\} \left[ 
\begin{array}{c}
\left\{ \left( \frac{\mathbf{C,S}_{\mathbf{C}}}{\Sigma /G_{\Sigma }}\right)
_{k_{\gamma _{l}}^{\prime }}^{\prime }\right\} _{2\leqslant k_{\gamma
_{l}}^{\prime }\leqslant \left( a_{i}^{\prime }\right) _{\gamma
_{l}}},\left\{ \left( \frac{\mathbf{C,S}_{\mathbf{C}}}{\Sigma /G_{\Sigma }}%
\right) \right\} _{2\leqslant k_{\gamma _{l}}\leqslant \left( a_{i}\right)
_{\gamma _{l}}} \\ 
\left[ \left\{ \left( \frac{\mathbf{C,S}_{\mathbf{C}}}{\Sigma /G_{\Sigma }}%
\right) _{k_{i,l}^{\prime }\leqslant \left( n_{i}\right) _{\gamma
_{l}}}^{\prime }\right\} ,\left\{ \left( \frac{\mathbf{C,S}_{\mathbf{C}}}{%
\Sigma /G_{\Sigma }}\right) _{k_{i,l}^{\prime }\leqslant \left( n_{f}\right)
_{\gamma _{l}}}\right\} \right]%
\end{array}%
\right]  \notag \\
&&C\left[ \left\{ \left\{ \left( \frac{\mathbf{C,S}_{\mathbf{C}}}{\Sigma
/G_{\Sigma }}\right) _{k_{\gamma _{l}}^{\prime }}^{\prime }\right\}
_{2\leqslant k_{\gamma _{l}}^{\prime }\leqslant \left( a_{i}^{\prime
}\right) _{\gamma _{l}}},\left\{ \left( \frac{\mathbf{C,S}_{\mathbf{C}}}{%
\Sigma /G_{\Sigma }}\right) \right\} _{2\leqslant k_{\gamma _{l}}\leqslant
\left( a_{i}\right) _{\gamma _{l}}}\right\} _{l}\right]  \notag
\end{eqnarray}%
The functions $\left\{ \mathit{G}\left( L_{p,p^{\prime }}\right) \right\} $
represent multiple loops, that is processes of transitions between some
states dissociating and then gathering at some points, after transformation.
This describes some internal chnrctn include infinte cascade transitions
between states, i.e. objects, subobjcts. This objects have thus no fixed or
determined form. They are permanently modified. These loops have some
interactions points, that are encompassed in the function:%
\begin{equation*}
\left\{ \mathit{G}\left( L_{p,p^{\prime }}\right) \right\} \left[ 
\begin{array}{c}
\left\{ \left( \frac{\mathbf{C,S}_{\mathbf{C}}}{\Sigma /G_{\Sigma }}\right)
_{k^{\prime }}^{\prime }\right\} _{2\leqslant k^{\prime }\leqslant
a_{i}^{\prime }},\left\{ \left( \frac{\mathbf{C,S}_{\mathbf{C}}}{\Sigma
/G_{\Sigma }}\right) \right\} _{2\leqslant k\leqslant a_{i}} \\ 
\left[ \left\{ \left( \frac{\mathbf{C,S}_{\mathbf{C}}}{\Sigma /G_{\Sigma }}%
\right) _{k_{i}^{\prime }}^{\prime }\right\} ,\left\{ \left( \frac{\mathbf{%
C,S}_{\mathbf{C}}}{\Sigma /G_{\Sigma }}\right) _{k_{i}}\right\} \right]%
\end{array}%
\right]
\end{equation*}%
These loops are connected to external vertices:%
\begin{equation*}
\left[ \left\{ \left( \frac{\mathbf{C,S}_{\mathbf{C}}}{\Sigma /G_{\Sigma }}%
\right) _{k_{i}^{\prime }}^{\prime }\right\} ,\left\{ \left( \frac{\mathbf{%
C,S}_{\mathbf{C}}}{\Sigma /G_{\Sigma }}\right) _{k_{i}}\right\} \right]
\end{equation*}%
describing thus the transition from several objects $\left\{ \left( \frac{%
\mathbf{C,S}_{\mathbf{C}}}{\Sigma /G_{\Sigma }}\right) _{k_{i}}\right\} $
toward other objects $\left\{ \left( \frac{\mathbf{C,S}_{\mathbf{C}}}{\Sigma
/G_{\Sigma }}\right) _{k_{i}^{\prime }}^{\prime }\right\} $. The loops are
also connected with other loops through combinations:%
\begin{equation*}
\left\{ \left( \frac{\mathbf{C,S}_{\mathbf{C}}}{\Sigma /G_{\Sigma }}\right)
_{k^{\prime }}^{\prime }\right\} _{2\leqslant k^{\prime }\leqslant
a_{i}^{\prime }},\left\{ \left( \frac{\mathbf{C,S}_{\mathbf{C}}}{\Sigma
/G_{\Sigma }}\right) \right\} _{2\leqslant k\leqslant a_{i}}
\end{equation*}%
the coefficients of the combinatn being given by the connection tensor:%
\begin{equation*}
C\left[ \left\{ \left\{ \left( \frac{\mathbf{C,S}_{\mathbf{C}}}{\Sigma
/G_{\Sigma }}\right) _{k_{\gamma _{l}}^{\prime }}^{\prime }\right\}
_{2\leqslant k_{\gamma _{l}}^{\prime }\leqslant \left( a_{i}^{\prime
}\right) _{\gamma _{l}}},\left\{ \left( \frac{\mathbf{C,S}_{\mathbf{C}}}{%
\Sigma /G_{\Sigma }}\right) \right\} _{2\leqslant k_{\gamma _{l}}\leqslant
\left( a_{i}\right) _{\gamma _{l}}}\right\} _{l}\right]
\end{equation*}%
Formula for the various objects are given in appendix 9. The connections
between loops implies the transitions induced by objcts interacting.
Ultimately permanent transitions, motions etc... should imply that process
should be modeled as fields in some abstract space.

The functions arising in the effective action are called proper vertices.
They represent the overall interactions allowing transitns between some
states defined by:%
\begin{equation*}
\prod\limits_{k_{i}=1}^{n_{i}}{\Huge \Lambda }\left[ \left( \frac{\mathbf{C,S%
}_{\mathbf{C}}}{\Sigma /G_{\Sigma }}\right) _{k_{i}^{\prime }}\right]
\end{equation*}%
towards states defined by:%
\begin{equation*}
\prod\limits_{k_{i}^{\prime }=1}^{n_{i}^{\prime }}{\Huge \Lambda }^{{\Large %
\dag }}\left[ \left( \frac{\mathbf{C,S}_{\mathbf{C}}}{\Sigma /G_{\Sigma }}%
\right) _{k_{i}^{\prime }}^{\prime }\right]
\end{equation*}%
The inspection of formula for $\left\{ \mathit{G}\left( L_{p,p^{\prime
}}\right) \right\} $ shows bindless activity where subobjects emerge,
interact, disactivate, activate again. This creats some perpetual changing
landscape from which some dynamic object emerge, reaching some dynamic
eqlibrium.

\section{Connected transition functions and transition amplitudes}

\subsection{Connected transition functions}

The connected transition functions compute the transitions of connected
blocks of objects. They are obtained from the effective action, seen as the
series of fundamental processes, and by connecting these fundamental
processes via propagators. The connected transitions are obtained by
connecting several loop products:%
\begin{equation*}
\mathit{G}\left( \mathcal{L}^{\left( L\right) }\right) \left[ \left\{ \left( 
\frac{\mathbf{C,S}_{\mathbf{C}}}{\Sigma /G_{\Sigma }}\right)
_{k_{i,l}^{\prime }\leqslant \left( n_{i}\right) _{\gamma _{l}}}^{\prime
}\right\} _{i,l},\left\{ \left( \frac{\mathbf{C,S}_{\mathbf{C}}}{\Sigma
/G_{\Sigma }}\right) _{k_{i,l}\leqslant \left( n_{f}\right) _{\gamma
_{l}}}\right\} _{i,l}\right]
\end{equation*}%
by dressed propagatrs. These dressed propagators are the two vertices of the
effective action. They are the propagators computed previously, modified by
the interaction and series of graphs. In such a view, the loops' propagators
represent complex processes involving several objects that are transformed
into products of objects. These ones propagate and then involve in other
processes.

\subsection{Dressed propagator}

The dressed propagators obtained by (\ref{Fcr}) with $\sum n_{i}^{\prime }=1$%
, $\sum n_{i}=1$. They write:%
\begin{equation*}
\mathit{\hat{G}}\left( \frac{\mathbf{C,S}_{\mathbf{C}}}{\Sigma /G_{\Sigma }},%
\frac{\mathbf{C}^{\prime }\mathbf{,S}_{\mathbf{C}}^{\prime }}{\Sigma
^{\prime }/G_{\Sigma }^{\prime }}\right) +\mathit{G}_{D}\left( \left[ \left( 
\frac{\mathbf{C,S}_{\mathbf{C}}}{\Sigma /G_{\Sigma }}\right) ^{\prime
},\left( \frac{\mathbf{C,S}_{\mathbf{C}}}{\Sigma /G_{\Sigma }}\right) \right]
\right)
\end{equation*}%
where: 
\begin{equation*}
\mathit{\hat{G}}\left( \frac{\mathbf{C,S}_{\mathbf{C}}}{\Sigma /G_{\Sigma }},%
\frac{\mathbf{C}^{\prime }\mathbf{,S}_{\mathbf{C}}^{\prime }}{\Sigma
^{\prime }/G_{\Sigma }^{\prime }}\right)
\end{equation*}%
is the propagator (\ref{Pg}) s computed with the classical action, and $%
\mathit{G}_{D}\left( \left[ \left( \frac{\mathbf{C,S}_{\mathbf{C}}}{\Sigma
/G_{\Sigma }}\right) ^{\prime },\left( \frac{\mathbf{C,S}_{\mathbf{C}}}{%
\Sigma /G_{\Sigma }}\right) \right] \right) $ is the correction computed
with the effective action. It decomposes in two types of terms:%
\begin{eqnarray}
&&\mathit{G}_{D}\left( \left[ \left( \frac{\mathbf{C,S}_{\mathbf{C}}}{\Sigma
/G_{\Sigma }}\right) ^{\prime },\left( \frac{\mathbf{C,S}_{\mathbf{C}}}{%
\Sigma /G_{\Sigma }}\right) \right] \right) \\
&=&\left( \mathit{G}_{D_{1}}\left( \left[ \left( \frac{\mathbf{C,S}_{\mathbf{%
C}}}{\Sigma /G_{\Sigma }}\right) ^{\prime },\left( \frac{\mathbf{C,S}_{%
\mathbf{C}}}{\Sigma /G_{\Sigma }}\right) \right] \right) +\mathit{G}%
_{D_{2}}\left( \left[ \left( \frac{\mathbf{C,S}_{\mathbf{C}}}{\Sigma
/G_{\Sigma }}\right) ^{\prime },\left( \frac{\mathbf{C,S}_{\mathbf{C}}}{%
\Sigma /G_{\Sigma }}\right) \right] \right) \right)  \notag \\
&&\times \prod\limits_{l=1}^{L}C\left[ \left\{ \left\{ \left( \frac{\mathbf{%
C,S}_{\mathbf{C}}}{\Sigma /G_{\Sigma }}\right) _{k_{\gamma _{l}}^{\prime
}}^{\prime }\right\} _{2\leqslant k_{\gamma _{l}}^{\prime }\leqslant \left(
a_{i}^{\prime }\right) _{\gamma _{l}}},\left\{ \left( \frac{\mathbf{C,S}_{%
\mathbf{C}}}{\Sigma /G_{\Sigma }}\right) \right\} _{2\leqslant k_{\gamma
_{l}}\leqslant \left( a_{i}\right) _{\gamma _{l}}}\right\} _{l}\right] 
\notag
\end{eqnarray}%
\begin{eqnarray*}
&&\mathit{G}_{D_{1}}\left( \left[ \left( \frac{\mathbf{C,S}_{\mathbf{C}}}{%
\Sigma /G_{\Sigma }}\right) ^{\prime },\left( \frac{\mathbf{C,S}_{\mathbf{C}}%
}{\Sigma /G_{\Sigma }}\right) \right] \right) \\
&=&\left\{ \mathit{G}\left( L_{p,p^{\prime }}\right) \right\} \left[ \left\{
\left( \frac{\mathbf{C,S}_{\mathbf{C}}}{\Sigma /G_{\Sigma }}\right)
_{k_{\gamma _{1}}^{\prime }}^{\prime }\right\} _{2\leqslant k_{\gamma
_{1}}^{\prime }\leqslant \left( a_{i}^{\prime }\right) _{\gamma
_{1}}},\left\{ \left( \frac{\mathbf{C,S}_{\mathbf{C}}}{\Sigma /G_{\Sigma }}%
\right) \right\} _{2\leqslant k_{\gamma _{1}}\leqslant \left( a_{i}\right)
_{\gamma _{1}}},\left[ \left( \frac{\mathbf{C,S}_{\mathbf{C}}}{\Sigma
/G_{\Sigma }}\right) ^{\prime },\left( \frac{\mathbf{C,S}_{\mathbf{C}}}{%
\Sigma /G_{\Sigma }}\right) \right] \right] \\
&&\times \prod\limits_{l=2}^{L}\left\{ \mathit{G}\left( L_{p,p^{\prime
}}\right) \right\} \left[ \left\{ \left( \frac{\mathbf{C,S}_{\mathbf{C}}}{%
\Sigma /G_{\Sigma }}\right) _{k_{\gamma _{l}}^{\prime }}^{\prime }\right\}
_{2\leqslant k_{\gamma _{l}}^{\prime }\leqslant \left( a_{i}^{\prime
}\right) _{\gamma _{l}}},\left\{ \left( \frac{\mathbf{C,S}_{\mathbf{C}}}{%
\Sigma /G_{\Sigma }}\right) \right\} _{2\leqslant k_{\gamma _{l}}\leqslant
\left( a_{i}\right) _{\gamma _{l}}}\right]
\end{eqnarray*}%
in which in and out legs belong to the same loop, and:%
\begin{eqnarray*}
&&\mathit{G}_{D_{2}}\left( \left[ \left( \frac{\mathbf{C,S}_{\mathbf{C}}}{%
\Sigma /G_{\Sigma }}\right) ^{\prime },\left( \frac{\mathbf{C,S}_{\mathbf{C}}%
}{\Sigma /G_{\Sigma }}\right) \right] \right) \\
&=&\left\{ \mathit{G}\left( L_{p,p^{\prime }}\right) \right\} \left[ \left\{
\left( \frac{\mathbf{C,S}_{\mathbf{C}}}{\Sigma /G_{\Sigma }}\right)
_{k_{\gamma _{1}}^{\prime }}^{\prime }\right\} _{2\leqslant k_{\gamma
_{1}}^{\prime }\leqslant \left( a_{i}^{\prime }\right) _{\gamma
_{1}}},\left\{ \left( \frac{\mathbf{C,S}_{\mathbf{C}}}{\Sigma /G_{\Sigma }}%
\right) \right\} _{2\leqslant k_{\gamma _{1}}\leqslant \left( a_{i}\right)
_{\gamma _{1}}},\left[ \left( \frac{\mathbf{C,S}_{\mathbf{C}}}{\Sigma
/G_{\Sigma }}\right) ^{\prime }\right] \right] \\
&&\times \left\{ \mathit{G}\left( L_{p,p^{\prime }}\right) \right\} \left[
\left\{ \left( \frac{\mathbf{C,S}_{\mathbf{C}}}{\Sigma /G_{\Sigma }}\right)
_{k_{\gamma _{2}}^{\prime }}^{\prime }\right\} _{2\leqslant k_{\gamma
_{2}}^{\prime }\leqslant \left( a_{i}^{\prime }\right) _{\gamma
_{2}}},\left\{ \left( \frac{\mathbf{C,S}_{\mathbf{C}}}{\Sigma /G_{\Sigma }}%
\right) \right\} _{2\leqslant k_{\gamma _{2}}\leqslant \left( a_{i}\right)
_{\gamma _{2}}},\left[ \left( \frac{\mathbf{C,S}_{\mathbf{C}}}{\Sigma
/G_{\Sigma }}\right) \right] \right] \\
&&\times \prod\limits_{l=3}^{L}\left\{ \mathit{G}\left( L_{p,p^{\prime
}}\right) \right\} \left[ \left\{ \left( \frac{\mathbf{C,S}_{\mathbf{C}}}{%
\Sigma /G_{\Sigma }}\right) _{k_{\gamma _{l}}^{\prime }}^{\prime }\right\}
_{2\leqslant k_{\gamma _{l}}^{\prime }\leqslant \left( a_{i}^{\prime
}\right) _{\gamma _{l}}},\left\{ \left( \frac{\mathbf{C,S}_{\mathbf{C}}}{%
\Sigma /G_{\Sigma }}\right) \right\} _{2\leqslant k_{\gamma _{l}}\leqslant
\left( a_{i}\right) _{\gamma _{l}}}\right]
\end{eqnarray*}%
in which the in and out legs belong to different loop.

\subsection{Connected transition functions}

The connected transition functions are given by convolutions:%
\begin{equation}
\mathit{\hat{G}}\left( \left\{ \left( \frac{\mathbf{C,S}_{\mathbf{C}}}{%
\Sigma /G_{\Sigma }}\right) \right\} _{n},\left\{ \left( \frac{\mathbf{C}%
^{\prime }\mathbf{,S}_{\mathbf{C}}^{\prime }}{\Sigma ^{\prime }/G_{\Sigma
}^{\prime }}\right) \right\} _{n^{\prime }}\right) =\sum_{\substack{ \sum
n_{i}=n+l  \\ \sum n_{i}^{\prime }=n^{\prime }+l}}\sum_{l}\sum_{G_{l}}\prod%
\limits_{i=1}^{l+1}{\Large F}\left[ \left\{ \left( \frac{\mathbf{C,S}_{%
\mathbf{C}}}{\Sigma /G_{\Sigma }}\right) _{k_{i}}\right\} _{n_{i}},\left\{
\left( \frac{\mathbf{C}^{\prime }\mathbf{,S}_{\mathbf{C}}^{\prime }}{\Sigma
^{\prime }/G_{\Sigma }^{\prime }}\right) _{k_{i}^{\prime }}\right\}
_{n_{i}^{\prime }}\right] \mathit{G}\left( G_{l}\right)  \label{ctr}
\end{equation}%
where $G_{l}$ is any simple connected graph \ At each vertex of the graph we
insert effective vertex $\Gamma \left( n_{i},n_{i}^{\prime }\right) $
connected to the incoming line and the outgoing line at one point, and to
each line we associate the propagator:%
\begin{equation*}
\mathit{G}_{D}\left( \left[ \left( \frac{\mathbf{C,S}_{\mathbf{C}}}{\Sigma
/G_{\Sigma }}\right) _{k_{i}}^{\prime },\left( \frac{\mathbf{C,S}_{\mathbf{C}%
}}{\Sigma /G_{\Sigma }}\right) _{k_{i}}\right] \right)
\end{equation*}%
and $\mathit{G}\left( G_{l}\right) $ is the product of these terms along the
graph:%
\begin{equation*}
\mathit{G}\left( G_{l}\right) =\prod \mathit{G}_{D}\left( \left[ \left( 
\frac{\mathbf{C,S}_{\mathbf{C}}}{\Sigma /G_{\Sigma }}\right)
_{k_{i}}^{\prime },\left( \frac{\mathbf{C,S}_{\mathbf{C}}}{\Sigma /G_{\Sigma
}}\right) _{k_{i}}\right] \right)
\end{equation*}%
where the product is computed along the graph.

\subsection{Example with three objects transitions}

There are two types of transitions involving three objects. First, the term:%
\begin{eqnarray*}
&&\mathit{\hat{G}}\left( \left\{ \left( \frac{\mathbf{C,S}_{\mathbf{C}}}{%
\Sigma /G_{\Sigma }}\right) _{1},\left( \frac{\mathbf{C,S}_{\mathbf{C}}}{%
\Sigma /G_{\Sigma }}\right) _{2}\right\} ,\left( \frac{\mathbf{C}^{\prime }%
\mathbf{,S}_{\mathbf{C}}^{\prime }}{\Sigma ^{\prime }/G_{\Sigma }^{\prime }}%
\right) \right) \\
&=&\int \mathit{G}\left( \left( \frac{\mathbf{C,S}_{\mathbf{C}}}{\Sigma
/G_{\Sigma }}\right) _{2},\left( \frac{\mathbf{C,S}_{\mathbf{C}}}{\Sigma
/G_{\Sigma }}\right) _{2}^{\prime }\right) \mathit{G}\left( \left( \frac{%
\mathbf{C,S}_{\mathbf{C}}}{\Sigma /G_{\Sigma }}\right) _{1},\left( \frac{%
\mathbf{C,S}_{\mathbf{C}}}{\Sigma /G_{\Sigma }}\right) _{1}^{\prime }\right)
\\
&&\times {\Large F}\left[ \left\{ \left( \frac{\mathbf{C,S}_{\mathbf{C}}}{%
\Sigma /G_{\Sigma }}\right) _{1}^{\prime },\left( \frac{\mathbf{C,S}_{%
\mathbf{C}}}{\Sigma /G_{\Sigma }}\right) _{2}^{\prime }\right\} ,\left( 
\frac{\mathbf{C}^{\prime }\mathbf{,S}_{\mathbf{C}}^{\prime }}{\Sigma
^{\prime }/G_{\Sigma }^{\prime }}\right) ^{\prime }\right] \mathit{G}\left(
\left( \frac{\mathbf{C}^{\prime }\mathbf{,S}_{\mathbf{C}}^{\prime }}{\Sigma
^{\prime }/G_{\Sigma }^{\prime }}\right) ^{\prime },\left( \frac{\mathbf{C}%
^{\prime }\mathbf{,S}_{\mathbf{C}}^{\prime }}{\Sigma ^{\prime }/G_{\Sigma
}^{\prime }}\right) \right) \\
&&\times D\left[ \left\{ \left( \frac{\mathbf{C,S}_{\mathbf{C}}}{\Sigma
/G_{\Sigma }}\right) _{1}^{\prime },\left( \frac{\mathbf{C,S}_{\mathbf{C}}}{%
\Sigma /G_{\Sigma }}\right) _{2}^{\prime }\right\} ,\left( \frac{\mathbf{C}%
^{\prime }\mathbf{,S}_{\mathbf{C}}^{\prime }}{\Sigma ^{\prime }/G_{\Sigma
}^{\prime }}\right) ^{\prime }\right]
\end{eqnarray*}%
describes the transition of $\left( \frac{\mathbf{C}^{\prime }\mathbf{,S}_{%
\mathbf{C}}^{\prime }}{\Sigma ^{\prime }/G_{\Sigma }^{\prime }}\right) $
towards $2$ objects $\left( \frac{\mathbf{C,S}_{\mathbf{C}}}{\Sigma
/G_{\Sigma }}\right) _{1}$ and $\left( \frac{\mathbf{C,S}_{\mathbf{C}}}{%
\Sigma /G_{\Sigma }}\right) _{2}$.

Symetrically:%
\begin{eqnarray*}
&&\mathit{\hat{G}}\left( \left( \frac{\mathbf{C,S}_{\mathbf{C}}}{\Sigma
/G_{\Sigma }}\right) ,\left\{ \left( \frac{\mathbf{C,S}_{\mathbf{C}}}{\Sigma
/G_{\Sigma }}\right) _{1}^{\prime },\left( \frac{\mathbf{C,S}_{\mathbf{C}}}{%
\Sigma /G_{\Sigma }}\right) _{2}^{\prime }\right\} \right) \\
&=&\int \mathit{G}\left( \left( \frac{\mathbf{C,S}_{\mathbf{C}}}{\Sigma
/G_{\Sigma }}\right) ^{\prime },\left( \frac{\mathbf{C}^{\prime }\mathbf{,S}%
_{\mathbf{C}}^{\prime }}{\Sigma ^{\prime }/G_{\Sigma }^{\prime }}\right)
\right) {\Large F}\left[ \left( \frac{\mathbf{C}^{\prime }\mathbf{,S}_{%
\mathbf{C}}^{\prime }}{\Sigma ^{\prime }/G_{\Sigma }^{\prime }}\right)
,\left\{ \left( \left( \frac{\mathbf{C,S}_{\mathbf{C}}}{\Sigma /G_{\Sigma }}%
\right) _{1}^{\prime }\right) ^{\prime },\left( \left( \frac{\mathbf{C,S}_{%
\mathbf{C}}}{\Sigma /G_{\Sigma }}\right) _{2}^{\prime }\right) ^{\prime
}\right\} \right] \\
&&\times \mathit{G}\left( \left( \left( \frac{\mathbf{C,S}_{\mathbf{C}}}{%
\Sigma /G_{\Sigma }}\right) _{2}^{\prime }\right) ^{\prime },\left( \frac{%
\mathbf{C,S}_{\mathbf{C}}}{\Sigma /G_{\Sigma }}\right) _{2}^{\prime }\right) 
\mathit{G}\left( \left( \left( \frac{\mathbf{C,S}_{\mathbf{C}}}{\Sigma
/G_{\Sigma }}\right) _{1}^{\prime }\right) ^{\prime },\left( \frac{\mathbf{%
C,S}_{\mathbf{C}}}{\Sigma /G_{\Sigma }}\right) _{1}^{\prime }\right) \\
&&\times D\left[ \left\{ \left( \left( \frac{\mathbf{C,S}_{\mathbf{C}}}{%
\Sigma /G_{\Sigma }}\right) _{1}^{\prime }\right) ^{\prime },\left( \frac{%
\mathbf{C,S}_{\mathbf{C}}}{\Sigma /G_{\Sigma }}\right) _{1}^{\prime
}\right\} ,\left( \frac{\mathbf{C}^{\prime }\mathbf{,S}_{\mathbf{C}}^{\prime
}}{\Sigma ^{\prime }/G_{\Sigma }^{\prime }}\right) ^{\prime }\right]
\end{eqnarray*}%
describes the merging of $2$ objects into one structure.

Both mechanisms involve interactions between subobjects and transitions of
subobjects as internal mechanisms.

\subsection{Transition amplitudes}

The effective action (\ref{Fc}) allows the computation of transitions
between unstable states. It starts with the connected amplitudes and gathers
them to produce amplitudes. The transitions are given by products of Green's
functions multiplied by coefficients in (\ref{Fc}) of order higher than 2.
The formula have the form for connected graphs:%
\begin{eqnarray*}
&&\left( \prod\limits_{r}\mathit{G}\left( \left( \frac{\mathbf{C,S}_{\mathbf{%
C}}}{\Sigma /G_{\Sigma }}\right) _{r}^{\left( i\right) },\left( \frac{%
\mathbf{C,S}_{\mathbf{C}}}{\Sigma /G_{\Sigma }}\right) _{r}\right) \right) \\
&&\times \mathit{\hat{G}}\left( \left\{ \left( \frac{\mathbf{C,S}_{\mathbf{C}%
}}{\Sigma /G_{\Sigma }}\right) \right\} _{r},\left\{ \left( \frac{\mathbf{C}%
^{\prime }\mathbf{,S}_{\mathbf{C}}^{\prime }}{\Sigma ^{\prime }/G_{\Sigma
}^{\prime }}\right) \right\} _{r^{\prime }}\right) \left(
\prod\limits_{r^{\prime }}\mathit{G}\left( \left( \frac{\mathbf{C}^{\prime }%
\mathbf{,S}_{\mathbf{C}}^{\prime }}{\Sigma ^{\prime }/G_{\Sigma }^{\prime }}%
\right) _{r^{\prime }},\left( \frac{\mathbf{C,S}_{\mathbf{C}}}{\Sigma
/G_{\Sigma }}\right) _{r^{\prime }}^{\left( f\right) }\right) \right)
\end{eqnarray*}

To obtain the full transition for groups, the formula is decomposed into
amplitudes for connected process:%
\begin{eqnarray*}
&&\sum_{C}\left( \prod\limits_{r}\mathit{G}\left( \left( \frac{\mathbf{C,S}_{%
\mathbf{C}}}{\Sigma /G_{\Sigma }}\right) _{n}^{\left( i\right) },\left( 
\frac{\mathbf{C,S}_{\mathbf{C}}}{\Sigma /G_{\Sigma }}\right) _{n}\right)
\right) \\
&&\times \prod\limits_{r,r^{\prime }}\mathit{\hat{G}}\left( \left\{ \left( 
\frac{\mathbf{C,S}_{\mathbf{C}}}{\Sigma /G_{\Sigma }}\right) \right\}
_{r},\left\{ \left( \frac{\mathbf{C}^{\prime }\mathbf{,S}_{\mathbf{C}%
}^{\prime }}{\Sigma ^{\prime }/G_{\Sigma }^{\prime }}\right) \right\}
_{r^{\prime }}\right) \left( \prod\limits_{r^{\prime }}\mathit{G}\left(
\left( \frac{\mathbf{C}^{\prime }\mathbf{,S}_{\mathbf{C}}^{\prime }}{\Sigma
^{\prime }/G_{\Sigma }^{\prime }}\right) _{n^{\prime }},\left( \frac{\mathbf{%
C,S}_{\mathbf{C}}}{\Sigma /G_{\Sigma }}\right) _{n^{\prime }}^{\left(
f\right) }\right) \right)
\end{eqnarray*}%
where the sum is constrained by:%
\begin{equation*}
C=\left( \cup \left\{ \left( \frac{\mathbf{C,S}_{\mathbf{C}}}{\Sigma
/G_{\Sigma }}\right) \right\} _{r}=\left\{ \left( \frac{\mathbf{C,S}_{%
\mathbf{C}}}{\Sigma /G_{\Sigma }}\right) \right\} _{n},\cup \left\{ \left( 
\frac{\mathbf{C}^{\prime }\mathbf{,S}_{\mathbf{C}}^{\prime }}{\Sigma
^{\prime }/G_{\Sigma }^{\prime }}\right) \right\} _{r^{\prime }}=\left\{
\left( \frac{\mathbf{C}^{\prime }\mathbf{,S}_{\mathbf{C}}^{\prime }}{\Sigma
^{\prime }/G_{\Sigma }^{\prime }}\right) \right\} _{n^{\prime }}\right)
\end{equation*}

Again, the formula is quite genral and may involve inital subobjects. Ths
corresponds to set matrices element associated to other subobjects to $0$.

The transitions matrices should be ranked with respect to their classes,
labeling invariants of the various sheaves $\left( \frac{\mathbf{C,S}_{%
\mathbf{C}}}{\Sigma /G_{\Sigma }}\right) _{r}^{\left( i\right) }$ n st $%
\left( \Sigma /G_{\Sigma }\right) _{r}$. Some topology should be preserved
or broken by interactions. If, during the processes some invariants are
modified, some inequivalent objects or processes are created.

\section{States and transitions}

Recall that state (\ref{stn}): 
\begin{eqnarray}
&&\left\vert \mathbf{S}_{\mathbf{C}}\right\rangle =\left\vert \Delta \mathbf{%
T}_{p}^{\alpha },\mathbf{\alpha },\mathbf{p},S^{2}\right\rangle \\
&=&\exp \left( -\left( \left( \Delta \mathbf{T}_{p}^{\alpha }\right) ^{t}%
\mathbf{A}_{p}^{\alpha }\Delta \mathbf{T}_{p}^{\alpha }+2\left( \Delta 
\mathbf{T}_{p}^{\alpha }\right) ^{t}\sqrt{\mathbf{A}_{p}^{\alpha }}\mathbf{%
\hat{A}}^{+}\left( \mathbf{\alpha },\mathbf{p},S^{2}\right) \right) +\frac{1%
}{2}\mathbf{\hat{A}}^{+}\left( \mathbf{\alpha },\mathbf{p},S^{2}\right) .%
\mathbf{\hat{A}}^{+}\left( \mathbf{\alpha },\mathbf{p},S^{2}\right) \right)
\left\vert Vac\right\rangle  \notag
\end{eqnarray}%
write also:%
\begin{eqnarray*}
&&\prod\limits_{\left\{ \Delta \mathbf{T}_{p}^{\alpha }\right\} \rightarrow
\Sigma /G_{\Sigma }}\exp \left( -\mathbf{\nu }\left( \frac{\mathbf{C}}{%
\Sigma /G_{\Sigma }}\right) \left( \left\{ \Delta \mathbf{T}_{p}^{\alpha
}\right\} -\left\{ \Delta \mathbf{T}_{p}^{\alpha }\right\} ^{\prime }\right)
^{2}+\left( \mathbf{\nu }\left( \frac{\mathbf{C}}{\Sigma /G_{\Sigma }}%
\right) \right) ^{2}\left( \left( \left\{ \Delta \mathbf{T}_{p}^{\alpha
}\right\} \right) ^{2}+\left( \left\{ \Delta \mathbf{T}_{p}^{\alpha
}\right\} ^{\prime }\right) ^{2}\right) \right) \\
&&\times H_{p}\left( \frac{1}{2}\left( \mathbf{\Delta T-}\left\langle 
\mathbf{\Delta T}\right\rangle _{p}^{\alpha }\right) ^{t}\mathbf{A}%
_{p}^{\alpha }\left( \mathbf{\Delta T-}\left\langle \mathbf{\Delta T}%
\right\rangle _{p}^{\alpha }\right) \right)
\end{eqnarray*}%
where $\mathbf{\Delta T}$ is the vector of connections for object and
subobjects. Similarly to (\ref{Stn}) the states:%
\begin{eqnarray}
&&\left\vert \mathbf{S}_{\mathbf{C}}\right\rangle \\
&=&\exp \left( -\left( \mathbf{S}_{\mathbf{C}}^{T}\mathbf{\nu }\left( \frac{%
\mathbf{C}}{\Sigma /G_{\Sigma }}\right) \mathbf{S}_{\mathbf{C}}+2\mathbf{S}_{%
\mathbf{C}}^{T}\sqrt{\mathbf{\nu }\left( \frac{\mathbf{C}}{\Sigma /G_{\Sigma
}}\right) }{\Huge A}^{{\Large +}}\left[ \frac{\mathbf{C,S}_{\mathbf{C}}}{%
\Sigma /G_{\Sigma }}\right] \right) +\frac{1}{2}{\Huge A}^{{\Large +}}\left[ 
\frac{\mathbf{C,S}_{\mathbf{C}}}{\Sigma /G_{\Sigma }}\right] .{\Huge A}^{%
{\Large +}}\left[ \frac{\mathbf{C,S}_{\mathbf{C}}}{\Sigma /G_{\Sigma }}%
\right] \right) \left\vert Vac\right\rangle  \notag
\end{eqnarray}

This can\ also be written in terms of field:%
\begin{equation}
\exp \left( -\frac{1}{2}{\Huge \Lambda }\left[ \frac{\mathbf{C}_{\nu }%
\mathbf{,S}_{\mathbf{C}_{\nu }}}{\Sigma _{\nu }/G_{\Sigma _{\nu }}}\right] 
\mathbf{\nu }\left( \frac{\mathbf{C}}{\Sigma /G_{\Sigma }}\right) {\Huge %
\Lambda }\left[ \frac{\mathbf{C}_{\nu }\mathbf{,S}_{\mathbf{C}_{\nu }}}{%
\Sigma _{\nu }/G_{\Sigma _{\nu }}}\right] \right) H_{p}\left( \frac{1}{2}%
{\Huge \Lambda }\left[ \frac{\mathbf{C}_{\nu }\mathbf{,S}_{\mathbf{C}_{\nu }}%
}{\Sigma _{\nu }/G_{\Sigma _{\nu }}}\right] \mathbf{\nu }\left( \frac{%
\mathbf{C}}{\Sigma /G_{\Sigma }}\right) {\Huge \Lambda }\left[ \frac{\mathbf{%
C}_{\nu }\mathbf{,S}_{\mathbf{C}_{\nu }}}{\Sigma _{\nu }/G_{\Sigma _{\nu }}}%
\right] \right)  \label{bgs}
\end{equation}%
More generally, collective state for the system is defined by countable or
not countable collection of states $\left\{ \frac{\mathbf{C}_{\nu }\mathbf{,S%
}_{\mathbf{C\nu }}}{\Sigma \nu /G_{\Sigma \nu }}\right\} _{\nu }$ and states
have the functional form:%
\begin{equation}
\sum \int H\left( \left\{ \frac{\mathbf{C}_{\nu }\mathbf{,S}_{\mathbf{C}%
_{\nu }}}{\Sigma _{\nu }/G_{\Sigma _{\nu }}}\right\} _{\nu }\right)
\prod\limits_{\nu }{\Huge \Lambda }\left[ \frac{\mathbf{C}_{\nu }\mathbf{,S}%
_{\mathbf{C}_{\nu }}}{\Sigma _{\nu }/G_{\Sigma _{\nu }}}\right] D\left[ 
\frac{\mathbf{C}_{\nu }\mathbf{,S}_{\mathbf{C}_{\nu }}}{\Sigma _{\nu
}/G_{\Sigma _{\nu }}}\right]  \label{bgw}
\end{equation}%
This states can be expressd as combinations of products of states (\ref{bgs}%
). This represents mixed states. Such states should be considered since the
formula for effective action (\ref{ffC}) and connectd transitions (\ref{ctr}%
), leads to consider proceesss where infinite number of structures are
active. Considering terms as (\ref{bgw}) consists in considering that mixed
states have some persistence.

The transitions between such stats are driven by interactions of the form:%
\begin{equation}
\sum \int H\left( \left\{ \frac{\mathbf{C}_{\nu ^{\prime }}^{\prime }\mathbf{%
,S}_{\mathbf{C}_{\nu ^{\prime }}}^{\prime }}{\Sigma _{\nu ^{\prime
}}^{\prime }/G_{\Sigma _{\nu }^{\prime }}}\right\} _{\nu ^{\prime }}\left\{ 
\frac{\mathbf{C}_{\nu }\mathbf{,S}_{\mathbf{C}_{\nu }}}{\Sigma _{\nu
}/G_{\Sigma _{\nu }}}\right\} _{\nu }\right) \prod\limits_{\nu ,\nu ^{\prime
}}{\Huge \Lambda }^{{\Large \dag }}\left[ \frac{\mathbf{C}_{\nu ^{\prime
}}^{\prime }\mathbf{,S}_{\mathbf{C}_{\nu ^{\prime }}}^{\prime }}{\Sigma
_{\nu ^{\prime }}^{\prime }/G_{\Sigma _{\nu }^{\prime }}}\right] {\Huge %
\Lambda }\left[ \frac{\mathbf{C}_{\nu }\mathbf{,S}_{\mathbf{C}_{\nu }}}{%
\Sigma _{\nu }/G_{\Sigma _{\nu }}}\right] D\left[ \frac{\mathbf{C}_{\nu
^{\prime }}^{\prime }\mathbf{,S}_{\mathbf{C}_{\nu ^{\prime }}}^{\prime }}{%
\Sigma _{\nu ^{\prime }}^{\prime }/G_{\Sigma _{\nu }^{\prime }}}\right] D%
\left[ \frac{\mathbf{C}_{\nu }\mathbf{,S}_{\mathbf{C}_{\nu }}}{\Sigma _{\nu
}/G_{\Sigma _{\nu }}}\right]
\end{equation}%
Writting two states as:%
\begin{equation*}
\left\vert H\right\rangle =\sum \int H\left( \left\{ \frac{\mathbf{C}_{\nu }%
\mathbf{,S}_{\mathbf{C}_{\nu }}}{\Sigma _{\nu }/G_{\Sigma _{\nu }}}\right\}
_{\nu }\right) \prod\limits_{\nu }{\Huge \Lambda }\left[ \frac{\mathbf{C}%
_{\nu }\mathbf{,S}_{\mathbf{C}_{\nu }}}{\Sigma _{\nu }/G_{\Sigma _{\nu }}}%
\right] D\left[ \frac{\mathbf{C}_{\nu }\mathbf{,S}_{\mathbf{C}_{\nu }}}{%
\Sigma _{\nu }/G_{\Sigma _{\nu }}}\right]
\end{equation*}%
and:%
\begin{equation*}
\left\langle H^{\prime }\right\vert =\sum \int H\left( \left\{ \frac{\mathbf{%
C}_{\nu ^{\prime }}^{\prime }\mathbf{,S}_{\mathbf{C}_{\nu ^{\prime
}}}^{\prime }}{\Sigma _{\nu ^{\prime }}^{\prime }/G_{\Sigma _{\nu }^{\prime
}}}\right\} _{\nu ^{\prime }}\right) \prod\limits_{\nu ^{\prime }}{\Huge %
\Lambda }^{{\Large \dag }}\left[ \frac{\mathbf{C}_{\nu ^{\prime }}^{\prime }%
\mathbf{,S}_{\mathbf{C}_{\nu ^{\prime }}}^{\prime }}{\Sigma _{\nu ^{\prime
}}^{\prime }/G_{\Sigma _{\nu }^{\prime }}}\right] D\left[ \frac{\mathbf{C}%
_{\nu ^{\prime }}^{\prime }\mathbf{,S}_{\mathbf{C}_{\nu ^{\prime }}}^{\prime
}}{\Sigma _{\nu ^{\prime }}^{\prime }/G_{\Sigma _{\nu }^{\prime }}}\right]
\end{equation*}%
the amplitude has the form:%
\begin{eqnarray*}
\left\langle H\right\vert T\left\vert H^{\prime }\right\rangle &=&\sum \int
H\left( \left\{ \frac{\mathbf{C}_{\nu }\mathbf{,S}_{\mathbf{C}_{\nu }}}{%
\Sigma _{\nu }/G_{\Sigma _{\nu }}}\right\} _{\nu }\right) H\left( \left\{ 
\frac{\mathbf{C}_{\nu ^{\prime }}^{\prime }\mathbf{,S}_{\mathbf{C}_{\nu
^{\prime }}}^{\prime }}{\Sigma _{\nu ^{\prime }}^{\prime }/G_{\Sigma _{\nu
}^{\prime }}}\right\} _{\nu ^{\prime }}\right) \\
&&\times \prod\limits_{\nu ^{\prime }}{\Huge \Lambda }^{{\Large \dag }}\left[
\frac{\mathbf{C}_{\nu ^{\prime }}^{\prime }\mathbf{,S}_{\mathbf{C}_{\nu
^{\prime }}}^{\prime }}{\Sigma _{\nu ^{\prime }}^{\prime }/G_{\Sigma _{\nu
}^{\prime }}}\right] T\prod\limits_{\nu }{\Huge \Lambda }\left[ \frac{%
\mathbf{C}_{\nu }\mathbf{,S}_{\mathbf{C}_{\nu }}}{\Sigma _{\nu }/G_{\Sigma
_{\nu }}}\right] D\left[ \frac{\mathbf{C}_{\nu }\mathbf{,S}_{\mathbf{C}_{\nu
}}}{\Sigma _{\nu }/G_{\Sigma _{\nu }}}\right] D\left[ \frac{\mathbf{C}_{\nu
^{\prime }}^{\prime }\mathbf{,S}_{\mathbf{C}_{\nu ^{\prime }}}^{\prime }}{%
\Sigma _{\nu ^{\prime }}^{\prime }/G_{\Sigma _{\nu }^{\prime }}}\right]
\end{eqnarray*}%
where:%
\begin{eqnarray*}
&&\prod\limits_{\nu ^{\prime }}{\Huge \Lambda }^{{\Large \dag }}\left[ \frac{%
\mathbf{C}_{\nu ^{\prime }}^{\prime }\mathbf{,S}_{\mathbf{C}_{\nu ^{\prime
}}}^{\prime }}{\Sigma _{\nu ^{\prime }}^{\prime }/G_{\Sigma _{\nu }^{\prime
}}}\right] T\prod\limits_{\nu }{\Huge \Lambda }\left[ \frac{\mathbf{C}_{\nu }%
\mathbf{,S}_{\mathbf{C}_{\nu }}}{\Sigma _{\nu }/G_{\Sigma _{\nu }}}\right] \\
&=&\int \prod\limits_{\nu ^{\prime }}{\Huge \Lambda }^{{\Large \dag }}\left[ 
\frac{\mathbf{C}_{\nu ^{\prime }}^{\prime }\mathbf{,S}_{\mathbf{C}_{\nu
^{\prime }}}^{\prime }}{\Sigma _{\nu ^{\prime }}^{\prime }/G_{\Sigma _{\nu
}^{\prime }}}\right] \exp \left( -{\Huge S}\left( {\Huge \Lambda }^{{\Large %
\dag }},{\Huge \Lambda }\right) \right) \prod\limits_{\nu }{\Huge \Lambda }%
\left[ \frac{\mathbf{C}_{\nu }\mathbf{,S}_{\mathbf{C}_{\nu }}}{\Sigma _{\nu
}/G_{\Sigma _{\nu }}}\right] {\Large D{\Huge \Lambda }D}{\Huge \Lambda }^{%
{\Large \dag }}
\end{eqnarray*}%
with ${\Huge S}\left( {\Huge \Lambda }^{{\Large \dag }},{\Huge \Lambda }%
\right) $ defined in (\ref{sfr}).

These integrals are correlation functions, that is products of connected
correlation functions.%
\begin{eqnarray*}
&&\prod\limits_{\nu ^{\prime }}{\Huge \Lambda }^{{\Large \dag }}\left[ \frac{%
\mathbf{C}_{\nu ^{\prime }}^{\prime }\mathbf{,S}_{\mathbf{C}_{\nu ^{\prime
}}}^{\prime }}{\Sigma _{\nu ^{\prime }}^{\prime }/G_{\Sigma _{\nu }^{\prime
}}}\right] T\prod\limits_{\nu }{\Huge \Lambda }\left[ \frac{\mathbf{C}_{\nu }%
\mathbf{,S}_{\mathbf{C}_{\nu }}}{\Sigma _{\nu }/G_{\Sigma _{\nu }}}\right] \\
&=&\sum_{\substack{ \cup _{i}\left\{ \left\{ \left( \frac{\mathbf{C,S}_{%
\mathbf{C}}}{\Sigma /G_{\Sigma }}\right) \right\} _{n_{i}}\right\} =\left\{ 
\frac{\mathbf{C}_{\nu }\mathbf{,S}_{\mathbf{C}_{\nu }}}{\Sigma _{\nu
}/G_{\Sigma _{\nu }}}\right\}  \\ \cup _{i}\left\{ \left\{ \left( \frac{%
\mathbf{C}^{\prime }\mathbf{,S}_{\mathbf{C}}^{\prime }}{\Sigma ^{\prime
}/G_{\Sigma }^{\prime }}\right) \right\} _{n_{i}^{\prime }}\right\} =\left\{ 
\frac{\mathbf{C}_{\nu ^{\prime }}^{\prime }\mathbf{,S}_{\mathbf{C}_{\nu
^{\prime }}}^{\prime }}{\Sigma _{\nu ^{\prime }}^{\prime }/G_{\Sigma _{\nu
}^{\prime }}}\right\} }}\prod\limits_{n_{i},n_{i}^{\prime }}\mathit{\hat{G}}%
\left( \left\{ \left( \frac{\mathbf{C,S}_{\mathbf{C}}}{\Sigma /G_{\Sigma }}%
\right) \right\} _{n_{i}},\left\{ \left( \frac{\mathbf{C}^{\prime }\mathbf{,S%
}_{\mathbf{C}}^{\prime }}{\Sigma ^{\prime }/G_{\Sigma }^{\prime }}\right)
\right\} _{n_{i}^{\prime }}\right)
\end{eqnarray*}%
As a consequence:%
\begin{eqnarray}
&&\left\langle H\right\vert T\left\vert H^{\prime }\right\rangle  \label{Mr}
\\
&=&\sum \int H\left( \left\{ \frac{\mathbf{C}_{\nu }\mathbf{,S}_{\mathbf{C}%
_{\nu }}}{\Sigma _{\nu }/G_{\Sigma _{\nu }}}\right\} _{\nu }\right) H\left(
\left\{ \frac{\mathbf{C}_{\nu ^{\prime }}^{\prime }\mathbf{,S}_{\mathbf{C}%
_{\nu ^{\prime }}}^{\prime }}{\Sigma _{\nu ^{\prime }}^{\prime }/G_{\Sigma
_{\nu }^{\prime }}}\right\} _{\nu ^{\prime }}\right)  \notag \\
&&\times \sum_{\substack{ \cup _{i}\left\{ \left\{ \left( \frac{\mathbf{C,S}%
_{\mathbf{C}}}{\Sigma /G_{\Sigma }}\right) \right\} _{n_{i}}\right\}
=\left\{ \frac{\mathbf{C}_{\nu }\mathbf{,S}_{\mathbf{C}_{\nu }}}{\Sigma
_{\nu }/G_{\Sigma _{\nu }}}\right\}  \\ \cup _{i}\left\{ \left\{ \left( 
\frac{\mathbf{C}^{\prime }\mathbf{,S}_{\mathbf{C}}^{\prime }}{\Sigma
^{\prime }/G_{\Sigma }^{\prime }}\right) \right\} _{n_{i}^{\prime }}\right\}
=\left\{ \frac{\mathbf{C}_{\nu ^{\prime }}^{\prime }\mathbf{,S}_{\mathbf{C}%
_{\nu ^{\prime }}}^{\prime }}{\Sigma _{\nu ^{\prime }}^{\prime }/G_{\Sigma
_{\nu }^{\prime }}}\right\} }}\prod\limits_{n_{i},n_{i}^{\prime }}\mathit{%
\hat{G}}\left( \left\{ \left( \frac{\mathbf{C,S}_{\mathbf{C}}}{\Sigma
/G_{\Sigma }}\right) \right\} _{n_{i}},\left\{ \left( \frac{\mathbf{C}%
^{\prime }\mathbf{,S}_{\mathbf{C}}^{\prime }}{\Sigma ^{\prime }/G_{\Sigma
}^{\prime }}\right) \right\} _{n_{i}^{\prime }}\right) D\left[ \frac{\mathbf{%
C}_{\nu }\mathbf{,S}_{\mathbf{C}_{\nu }}}{\Sigma _{\nu }/G_{\Sigma _{\nu }}}%
\right] D\left[ \frac{\mathbf{C}_{\nu ^{\prime }}^{\prime }\mathbf{,S}_{%
\mathbf{C}_{\nu ^{\prime }}}^{\prime }}{\Sigma _{\nu ^{\prime }}^{\prime
}/G_{\Sigma _{\nu }^{\prime }}}\right]  \notag
\end{eqnarray}

The transition between mixed states contains internal transitions of groups
of objects, these internal transitions combining with each other to compose
new states.

\subsection{Description of transitions}

Coming back to the expanded form of ${\Huge \Lambda }\left[ \frac{\mathbf{C}%
_{\nu }\mathbf{,S}_{\mathbf{C}_{\nu }}}{\Sigma _{\nu }/G_{\Sigma _{\nu }}}%
\right] $: 
\begin{equation*}
{\Huge \Lambda }\left[ \frac{\mathbf{C,}\left( \mathbf{\alpha }_{k},\mathbf{p%
}_{k}\right) _{\mathbf{C}}}{\Sigma /G_{\Sigma }},\left( \left[ \Delta 
\mathbf{T},\Delta \mathbf{\hat{T},}\left[ \Delta \mathbf{T}_{l},\Delta 
\mathbf{\hat{T}}_{l}\right] _{l},\left[ \Delta \mathbf{T}_{lk},\Delta 
\mathbf{\hat{T}}_{lk}\right] _{lk},..\right] ,\left[ \Delta \mathbf{T}%
_{l}^{\left( l\right) },\Delta \mathbf{\hat{T}}_{l}^{\left( l\right) },\left[
\Delta \mathbf{T}_{lk}^{\left( l\right) },\Delta \mathbf{\hat{T}}%
_{lk}^{\left( l\right) }\right] _{lk},..\right] _{l},..\right) \right]
\end{equation*}%
and given (\ref{Mr}), the description of transition is obtained by
considering a cloud of variables:%
\begin{equation*}
S=\left\{ \frac{\mathbf{C,}\left( \mathbf{\alpha }_{k},\mathbf{p}_{k}\right)
_{\mathbf{C}}}{\Sigma /G_{\Sigma }},\left( \left[ \Delta \mathbf{T},\Delta 
\mathbf{\hat{T},}\left[ \Delta \mathbf{T}_{l},\Delta \mathbf{\hat{T}}_{l}%
\right] _{l},\left[ \Delta \mathbf{T}_{lk},\Delta \mathbf{\hat{T}}_{lk}%
\right] _{lk},..\right] ,\left[ \Delta \mathbf{T}_{l}^{\left( l\right)
},\Delta \mathbf{\hat{T}}_{l}^{\left( l\right) },\left[ \Delta \mathbf{T}%
_{lk}^{\left( l\right) },\Delta \mathbf{\hat{T}}_{lk}^{\left( l\right) }%
\right] _{lk},..\right] _{l},..\right) \right\}
\end{equation*}%
Some distinguished values enter the cloud, while some other variables come
out of the cloud.

A mechanism of transition, obtained by summing sequences of elementary
transitions, is graphically represented by a thread of an infinite number of
lines from one initial object to another one. The objects evolve and
displace their support. These lines are sequences:%
\begin{eqnarray}
\left( \frac{\mathbf{C,S}_{\mathbf{C}}}{\Sigma /G_{\Sigma }}\right) _{a_{1}}
&\rightarrow &\left( \frac{\mathbf{C,S}_{\mathbf{C}}}{\Sigma /G_{\Sigma }}%
\right) _{a_{1}}\rightarrow \left( \frac{\mathbf{C,S}_{\mathbf{C}}}{\Sigma
/G_{\Sigma }}\right) _{a_{1}}\rightarrow ...  \label{ld} \\
&\hookrightarrow &\left( \frac{\mathbf{C,S}_{\mathbf{C}}}{\Sigma /G_{\Sigma }%
}\right) _{a_{1}^{\prime }}^{\prime }\rightarrow ...\quad \quad \quad \quad
\quad \quad \quad \quad \quad \quad \quad \quad \frac{{}}{\hookrightarrow
\left( \frac{\mathbf{C,S}_{\mathbf{C}}}{\Sigma /G_{\Sigma }}\right)
_{a_{1}^{\prime }}^{\prime }\rightarrow ..\Rsh }\rightarrow  \notag
\end{eqnarray}%
the downward and upward arrows corresponding to a shift from one line to a
lower or upper line respectively, each new line corresponding to the
creation of a new object. These lines are not independent but are
intertwined via some interactions:%
\begin{equation*}
\begin{array}{c}
\left( \frac{\mathbf{C,S}_{\mathbf{C}}}{\Sigma /G_{\Sigma }}\right)
_{a_{1,0}} \\ 
... \\ 
\left( \frac{\mathbf{C,S}_{\mathbf{C}}}{\Sigma /G_{\Sigma }}\right)
_{a_{k,0}}%
\end{array}%
\left[ 
\begin{array}{c}
\left( \frac{\mathbf{C,S}_{\mathbf{C}}}{\Sigma /G_{\Sigma }}\right)
_{a_{1,1}}\rightarrow \left( \frac{\mathbf{C,S}_{\mathbf{C}}}{\Sigma
/G_{\Sigma }}\right) _{a_{1,2}}\rightarrow \left( \frac{\mathbf{C,S}_{%
\mathbf{C}}}{\Sigma /G_{\Sigma }}\right) _{a_{1,3}}\rightarrow ... \\ 
\downdownarrows \upuparrows \\ 
... \\ 
\left( \frac{\mathbf{C,S}_{\mathbf{C}}}{\Sigma /G_{\Sigma }}\right)
_{a_{k,1}}\rightarrow \left( \frac{\mathbf{C,S}_{\mathbf{C}}}{\Sigma
/G_{\Sigma }}\right) _{a_{k,2}}\rightarrow \left( \frac{\mathbf{C,S}_{%
\mathbf{C}}}{\Sigma /G_{\Sigma }}\right) _{a_{k,3}}\rightarrow ... \\ 
... \\ 
\downdownarrows \upuparrows%
\end{array}%
\right] 
\begin{array}{c}
\left( \frac{\mathbf{C,S}_{\mathbf{C}}}{\Sigma /G_{\Sigma }}\right)
_{a_{1,f}} \\ 
... \\ 
\left( \frac{\mathbf{C,S}_{\mathbf{C}}}{\Sigma /G_{\Sigma }}\right)
_{a_{k,f}}%
\end{array}%
\end{equation*}

where the $\downdownarrows \upuparrows $ represent the branching in (\ref{ld}%
). These dressed lines present some interactions, and the probability of
such a thread gives the amplitudes for transitions.

An effective description should include some fluctuating and interacting
state $S$ of continuous and discrete variables. Then, from this cloud, some
clusters should appear,interact and vanish,. The process of transition
should correspond to some fluctuations in some field or plasma in an
abstract space where points have coordinates $\frac{\mathbf{C,S}_{\mathbf{C}}%
}{\Sigma /G_{\Sigma }}$. Note that the space defined by these points has
variable dimensions, due to the variable parameters involved in the local
coordinates and the dimension of space of classes. The properties of such a
field should be given by some function:%
\begin{equation*}
{\Huge \Lambda }_{cl}\left( \left( \frac{\mathbf{C,S}_{\mathbf{C}}}{\Sigma
/G_{\Sigma }}\right) \right)
\end{equation*}%
that should be the background field of the effective action. Such field
should decribe the potentially stable configurations for each site or
landscape.

\section{Effective macro description}

The classical field ${\Huge \Lambda }_{cl}\left( \left( \frac{\mathbf{C,S}_{%
\mathbf{C}}}{\Sigma /G_{\Sigma }}\right) \right) $ is unknown, however,
given our previous description, an effective approach to the transitions
would be to discard the local coordinates that describe the objects, as well
as the support variables, since both these variables are evolving at some
fast time scale that can be considered as integrated. The dynamics should
thus be between lines of $\mathbf{C,}\left( \mathbf{\alpha }_{k},\mathbf{p}%
_{k}\right) _{\mathbf{C}}$ or $\left( \mathbf{\alpha }_{k},\mathbf{p}%
_{k}\right) _{\mathbf{C}}$ for short:%
\begin{equation*}
\begin{array}{c}
\left( \left( \mathbf{\alpha }_{1},\mathbf{p}_{1}\right) _{\mathbf{C}%
}\right) _{a_{1,0}} \\ 
... \\ 
\left( \left( \mathbf{\alpha }_{k},\mathbf{p}_{k}\right) _{\mathbf{C}%
}\right) _{a_{k,0}}%
\end{array}%
\left[ 
\begin{array}{c}
\left( \left( \mathbf{\alpha }_{k},\mathbf{p}_{k}\right) _{\mathbf{C}%
}\right) _{a_{1,1}}\rightarrow \left( \left( \mathbf{\alpha }_{k},\mathbf{p}%
_{k}\right) _{\mathbf{C}}\right) _{a_{1,2}}\rightarrow \left( \left( \mathbf{%
\alpha }_{k},\mathbf{p}_{k}\right) _{\mathbf{C}}\right)
_{a_{1,3}}\rightarrow ... \\ 
\downdownarrows \upuparrows \\ 
... \\ 
\left( \left( \mathbf{\alpha }_{k},\mathbf{p}_{k}\right) _{\mathbf{C}%
}\right) _{a_{k,1}}\rightarrow \left( \frac{\mathbf{C,S}_{\mathbf{C}}}{%
\Sigma /G_{\Sigma }}\right) _{a_{k,2}}\rightarrow \left( \frac{\mathbf{C,S}_{%
\mathbf{C}}}{\Sigma /G_{\Sigma }}\right) _{a_{k,3}}\rightarrow ... \\ 
... \\ 
\downdownarrows \upuparrows%
\end{array}%
\right] 
\begin{array}{c}
\left( \left( \mathbf{\alpha }_{1},\mathbf{p}_{1}\right) _{\mathbf{C}%
}\right) _{a_{1,f}} \\ 
... \\ 
\left( \left( \mathbf{\alpha }_{k},\mathbf{p}_{k}\right) _{\mathbf{C}%
}\right) _{a_{k,f}}%
\end{array}%
\end{equation*}

These states have frequencies and amplitude $\gamma \left( \left( \mathbf{%
\alpha }_{k},\mathbf{p}_{k}\right) _{\mathbf{C}}\right) $.

Some interactions should preserve the frequencies and classes, some other
modifying these discrete variables.

At the "macro" scale, the dynamics should encompass the result of a cascade
of transitions of constrained representations. As a consequence of the
nonlinear aspects of interactions, some representations should induce the
emergence of several new representations, the amplitudes of the transitions
being functions of the constraints between these representations, their
classes, their frequencies. The emergence of subobjects with classes that do
not match those of the initial objects should be a factor of instability and
transition, with a possible change in the class of a global object.

Since classes ultimately depend on the internal frequencies of objects and
on the symmetries of states, classes can be modified by interactions that
break those symmetries or by perturbations that modify the internal
frequencies through some instability. This possibility has already been
investigated in Section 3.3. Perturbations with large wave vectors, that is,
exhibiting large variability across the spatial extension of an object, may
induce instabilities that can break the object or induce a transition toward
another state---for example, a state with frequencies induced by other
objects with which it interacts\textbf{.}

\section*{Appendix 1 Microfoundations and translation method}

We present an account of the translation techniques that transform a general
class of models with large number of degrees of freedom into a field
formalism. We then present the micro-foundations of our model, and its
translation in terms of fld.

The application of this formalism to our particular case has been presented
in \cite{GL}, \cite{GLr}.

\subsection*{A1.1 Statistical weight and minimization functions for a
classical system}

In a dynamic system with a large number of agents, each agent is
characterized by one or more stochastic dynamic equations. Some of these
equations result from the optimization of one or several objective
functions. Deriving the statistical weight from these equations is
straightforward: it associates, to each trajectory of the group of agents $%
\left\{ T_{i}\right\} $, a probability that is peaked around the set of
optimal trajectories of the system, of the form:%
\begin{equation}
W\left( s\left( \left\{ T_{i}\right\} \right) \right) =\exp \left( -s\left(
\left\{ T_{i}\right\} \right) \right)  \label{wdt}
\end{equation}%
where $s\left( \left\{ T_{i}\right\} \right) $ measures the distance between
the trajectories $\left\{ T_{i}\right\} $ and the optimal ones.

As explained above, this paper studies two types of agents: cells and
connectivities between cells. To remain at a general level in this section,
we rather consider two arbitrary types of agents characterized by
vector-variables $\left\{ \mathbf{A}_{i}\left( t\right) \right\}
_{i=1,...N}, $ and $\left\{ \mathbf{\hat{A}}_{l}\left( t\right) \right\}
_{i=1,...\hat{N}} $ respectively, where $N$ and $\hat{N}$ are the number of
agents of each type, with vectors $\mathbf{A}_{i}\left( t\right) $\ and $%
\mathbf{\hat{A}}_{l}\left( t\right) $ of arbitrary dimension. For such a
system, the statistical weight writes:%
\begin{equation}
W\left( \left\{ \mathbf{A}_{i}\left( t\right) \right\} ,\left\{ \mathbf{\hat{%
A}}_{l}\left( t\right) \right\} \right) =\exp \left( -s\left( \left\{ 
\mathbf{A}_{i}\left( t\right) \right\} ,\left\{ \mathbf{\hat{A}}_{l}\left(
t\right) \right\} \right) \right)  \label{wdh}
\end{equation}

The optimal paths for the system are assumed to be described by the sets of
equations:%
\begin{equation}
\frac{d\mathbf{A}_{i}\left( t\right) }{dt}-\sum_{j,k,l...}f\left( \mathbf{A}%
_{i}\left( t\right) ,\mathbf{A}_{j}\left( t\right) ,\mathbf{A}_{k}\left(
t\right) ,\mathbf{\hat{A}}_{l}\left( t\right) ,\mathbf{\hat{A}}_{m}\left(
t\right) ...\right) =\epsilon _{i}\text{, }i=1...N  \label{gauche}
\end{equation}%
\begin{equation}
\frac{d\mathbf{\hat{A}}_{l}\left( t\right) }{dt}-\sum_{i,j,k...}\hat{f}%
\left( \mathbf{A}_{i}\left( t\right) ,\mathbf{A}_{j}\left( t\right) ,\mathbf{%
A}_{k}\left( t\right) ,\mathbf{\hat{A}}_{l}\left( t\right) ,\mathbf{\hat{A}}%
_{m}\left( t\right) ...\right) =\hat{\epsilon}_{l}\text{, }i=1...\hat{N}
\label{dnw}
\end{equation}%
where the $\epsilon _{i}$ and $\hat{\epsilon}_{i}$ are idiosynchratic random
shocks.\ These equations describe the general dynamics of the two types
agents, including their interactions with other agents. They may\ encompass
the dynamics of optimizing agents where interactions act as externalities so
that this set of equations is the full description of a system of
interacting agents\footnote{%
Expectations of agents could be included by replacing $\frac{d\mathbf{A}%
_{i}\left( t\right) }{dt}$ with $E\frac{d\mathbf{A}_{i}\left( t\right) }{dt}$%
, where $E$ is the expectation operator. This would amount to double some
variables by distinguishing "real variables" and expectations. However, for
our purpose, in the context of a large number of agents, at least in this
work, we discard as much as possible this possibility.}\footnote{%
A generalisation of equations (\ref{gauche}) and (\ref{dnw}), in which
agents interact at different times, and its translation in term of field is
presented in appendix 1.}\textbf{. }

For equations (\ref{gauche}) and (\ref{dnw}), the quadratic deviation at
time $t$ of any trajectory with respect to the optimal one for each type of
agent are:%
\begin{equation}
\left( \frac{d\mathbf{A}_{i}\left( t\right) }{dt}-\sum_{j,k,l...}f\left( 
\mathbf{A}_{i}\left( t\right) ,\mathbf{A}_{j}\left( t\right) ,\mathbf{A}%
_{k}\left( t\right) ,\mathbf{\hat{A}}_{l}\left( t\right) ,\mathbf{\hat{A}}%
_{m}\left( t\right) ...\right) \right) ^{2}  \label{pst}
\end{equation}%
and:%
\begin{equation}
\left( \frac{d\mathbf{\hat{A}}_{l}\left( t\right) }{dt}-\sum_{i,j,k...}\hat{f%
}\left( \mathbf{A}_{i}\left( t\right) ,\mathbf{A}_{j}\left( t\right) ,%
\mathbf{A}_{k}\left( t\right) ,\mathbf{\hat{A}}_{l}\left( t\right) ,\mathbf{%
\hat{A}}_{m}\left( t\right) ...\right) \right) ^{2}  \label{psh}
\end{equation}%
Since the function (\ref{wdh}) involves the deviations for all agents over
all trajectories, the function: 
\begin{equation*}
s\left( \left\{ \mathbf{A}_{i}\left( t\right) \right\} ,\left\{ \mathbf{\hat{%
A}}_{l}\left( t\right) \right\} \right)
\end{equation*}%
is obtained by summing (\ref{pst}) and (\ref{psh}) over all agents, and
integrate over $t$. We thus find:%
\begin{eqnarray}
s\left( \left\{ \mathbf{A}_{i}\left( t\right) \right\} ,\left\{ \mathbf{\hat{%
A}}_{l}\left( t\right) \right\} \right) &=&\int dt\sum_{i}\left( \frac{d%
\mathbf{A}_{i}\left( t\right) }{dt}-\sum_{j,k,l...}f\left( \mathbf{A}%
_{i}\left( t\right) ,\mathbf{A}_{j}\left( t\right) ,\mathbf{A}_{k}\left(
t\right) ,\mathbf{\hat{A}}_{l}\left( t\right) ,\mathbf{\hat{A}}_{m}\left(
t\right) ...\right) \right) ^{2}  \label{prw} \\
&&+\int dt\sum_{l}\left( \frac{d\mathbf{\hat{A}}_{l}\left( t\right) }{dt}%
-\sum_{i,j,k...}\hat{f}\left( \mathbf{A}_{i}\left( t\right) ,\mathbf{A}%
_{j}\left( t\right) ,\mathbf{A}_{k}\left( t\right) ,\mathbf{\hat{A}}%
_{l}\left( t\right) ,\mathbf{\hat{A}}_{m}\left( t\right) ...\right) \right)
^{2}  \notag
\end{eqnarray}%
There is an alternate, more general, form to (\ref{prw}). We can assume that
the dynamical system is originally defined by some equations of type (\ref%
{gauche}) and (\ref{dnw}), plus some objective functions for agents $i$ and $%
l$, and that these agents aim at minimizing respectively:%
\begin{equation}
\sum_{j,k,l...}g\left( \mathbf{A}_{i}\left( t\right) ,\mathbf{A}_{j}\left(
t\right) ,\mathbf{A}_{k}\left( t\right) ,\mathbf{\hat{A}}_{l}\left( t\right)
,\mathbf{\hat{A}}_{m}\left( t\right) ...\right)  \label{glf}
\end{equation}%
and:%
\begin{equation}
\sum_{i,j,k..}\hat{g}\left( \mathbf{A}_{i}\left( t\right) ,\mathbf{A}%
_{j}\left( t\right) ,\mathbf{A}_{k}\left( t\right) ,\mathbf{\hat{A}}%
_{l}\left( t\right) ,\mathbf{\hat{A}}_{m}\left( t\right) ...\right)
\label{gln}
\end{equation}%
In the above equations, the objective functions depend on other agents'
actions seen as externalities\footnote{%
We may also assume intertemporal objectives, see (\cite{GL1}).{}}. The
functions (\ref{glf}) and (\ref{gln}) could themselves be considered as a
measure of the deviation of a trajectory from the optimum. Actually, the
higher the distance, the higher (\ref{glf}) and (\ref{gln}).

Thus, rather than describing the systm by a full system of dynamic
equations, we can consider some ad-hoc equations of type (\ref{gauche}) and (%
\ref{dnw}) and some objective functions (\ref{glf}) and (\ref{gln}) to write
the alternate form of (\ref{prw}) as:%
\begin{eqnarray}
&&s\left( \left\{ \mathbf{A}_{i}\left( t\right) \right\} ,\left\{ \mathbf{%
\hat{A}}_{l}\left( t\right) \right\} \right)  \label{mNZ} \\
&=&\int dt\sum_{i}\left( \frac{d\mathbf{A}_{i}\left( t\right) }{dt}%
-\sum_{j,k,l...}f\left( \mathbf{A}_{i}\left( t\right) ,\mathbf{A}_{j}\left(
t\right) ,\mathbf{A}_{k}\left( t\right) ,\mathbf{\hat{A}}_{l}\left( t\right)
,\mathbf{\hat{A}}_{m}\left( t\right) ...\right) \right) ^{2}  \notag \\
&&+\int dt\sum_{l}\left( \frac{d\mathbf{\hat{A}}_{l}\left( t\right) }{dt}%
-\sum_{i,j,k...}\hat{f}\left( \mathbf{A}_{i}\left( t\right) ,\mathbf{A}%
_{j}\left( t\right) ,\mathbf{A}_{k}\left( t\right) ,\mathbf{\hat{A}}%
_{l}\left( t\right) ,\mathbf{\hat{A}}_{m}\left( t\right) ...\right) \right)
^{2}  \notag \\
&&+\int dt\sum_{i,j,k,l...}\left( g\left( \mathbf{A}_{i}\left( t\right) ,%
\mathbf{A}_{j}\left( t\right) ,\mathbf{A}_{k}\left( t\right) ,\mathbf{\hat{A}%
}_{l}\left( t\right) ,\mathbf{\hat{A}}_{m}\left( t\right) ...\right) +\hat{g}%
\left( \mathbf{A}_{i}\left( t\right) ,\mathbf{A}_{j}\left( t\right) ,\mathbf{%
A}_{k}\left( t\right) ,\mathbf{\hat{A}}_{l}\left( t\right) ,\mathbf{\hat{A}}%
_{m}\left( t\right) ...\right) \right)  \notag
\end{eqnarray}

In the sequel, we will refer to the various terms arising in equation (\ref%
{mNZ}) as the "minimization functions",\textbf{\ i.e}. the functions whose
minimization yield the dynamics equations of the system\footnote{%
A generalisation of equation (\ref{mNZ}), in which agents interact at
different times, and its translation in term of field is presented in
appendix 1 of \cite{GLr}.{}}.

We have shown in \cite{GL1}\cite{GL2}\cite{GL3} that the probabilistic
description of the system (\ref{mNZ}) is equivalent to a statistical field
formalism. In such a formalism, the system is collectively described by a
field that is an element of the Hilbert space of complex functions. The
arguments of\ these functions are the same as those describing an individual
neuron and the connectivity function between two cells. A shortcut of the
translation of systems similar to (\ref{mNZ}) in terms of field, is given in 
\cite{GL4} . The next paragraph gives an account of this method.

\subsection*{A1.2 Translation techniques}

Once the statistical weight $W\left( s\left( \left\{ T_{i}\right\} \right)
\right) $ defined in (\ref{wdt}) iscomputed, it can be translated in terms
of field. To do so, and for each type $\alpha $ of agent, the sets of
trajectories $\left\{ \mathbf{A}_{\alpha i}\left( t\right) \right\} $ are
replaced by a field $\Psi _{\alpha }\left( \mathbf{A}_{\alpha }\right) $, a
random variable whose realizations are complex-valued functions $\Psi $ of
the variables $\mathbf{A}_{\alpha }$\footnote{%
In the following, we will use indifferently the term "field" and the
notation $\Psi $ for the random variable or any of its realization $\Psi $.}%
. The statistical weight for the whole set of fields $\left\{ \Psi _{\alpha
}\right\} $ has the form $\exp \left( -S\left( \left\{ \Psi _{\alpha
}\right\} \right) \right) $. The function $S\left( \left\{ \Psi _{\alpha
}\right\} \right) $\ is called the \emph{fields action functional}. It
represents the interactions among different types of agents. Ultimately, the
expression $\exp \left( -S\left( \left\{ \Psi _{\alpha }\right\} \right)
\right) $ is the statistical weight for the field\footnote{%
In general, one must consider the integral of $\exp \left( -S\left( \left\{
\Psi _{\alpha }\right\} \right) \right) $\ over the configurations $\left\{
\Psi _{\alpha }\right\} $. This integral is the partition function of the
system.} that computes the probability of any realization $\left\{ \Psi
_{\alpha }\right\} $\ of the field.

The form of $S\left( \left\{ \Psi _{\alpha }\right\} \right) $\ is obtained
directly from the classical description of our model. For two types of
agents, we start with expression (\ref{mNZ}). The various minimizations
functions involved in the definition of $s\left( \left\{ \mathbf{A}%
_{i}\left( t\right) \right\} ,\left\{ \mathbf{\hat{A}}_{l}\left( t\right)
\right\} \right) $ will be translated in terms of field and the sum of these
translations will produce finally the action functional $S\left( \left\{
\Psi _{\alpha }\right\} \right) $. The translation method can itself be
divided into two relatively simple processes, but varies slightly depending
on the type of terms that appear in the various minimization functions.

\subsubsection*{A1.2.1 Terms without temporal derivative}

In equation (\ref{mNZ}), the terms that involve indexed variables but no
temporal derivative terms are the easiest to translate.\ They are of the
form:%
\begin{equation*}
\sum_{i}\sum_{j,k,l,m...}g\left( \mathbf{A}_{i}\left( t\right) ,\mathbf{A}%
_{j}\left( t\right) ,\mathbf{A}_{k}\left( t\right) ,\mathbf{\hat{A}}%
_{l}\left( t\right) ,\mathbf{\hat{A}}_{m}\left( t\right) ...\right)
\end{equation*}%
These terms describe the whole set of interactions both among and between
two groups of agents. Here, agents are characterized by their variables $%
\mathbf{A}_{i}\left( t\right) ,\mathbf{A}_{j}\left( t\right) ,\mathbf{A}%
_{k}\left( t\right) $... and $\mathbf{\hat{A}}_{l}\left( t\right) ,\mathbf{%
\hat{A}}_{m}\left( t\right) $... respectively, for instance in our model
firms and investors.

In the field translation, agents of type $\mathbf{A}_{i}\left( t\right) $
and $\mathbf{\hat{A}}_{l}\left( t\right) $ are described by a field $\Psi
\left( \mathbf{A}\right) $ and $\hat{\Psi}\left( \mathbf{\hat{A}}\right) $,
respectively.

In a first step, the variables indexed $i$ such as $\mathbf{A}_{i}\left(
t\right) $ are replaced by variables $\mathbf{A}$ in the expression of $g$.
The variables indexed $j$, $k$, $l$, $m$..., such as $\mathbf{A}_{j}\left(
t\right) $, $\mathbf{A}_{k}\left( t\right) $, $\mathbf{\hat{A}}_{l}\left(
t\right) ,\mathbf{\hat{A}}_{m}\left( t\right) $... are replaced by $\mathbf{A%
}^{\prime },\mathbf{A}^{\prime \prime }$, $\mathbf{\hat{A}}$, $\mathbf{\hat{A%
}}^{\prime }$ , and so on for all the indices in the function. This yields
the expression:

\begin{equation*}
\sum_{i}\sum_{j,k,l,m...}g\left( \mathbf{A},\mathbf{A}^{\prime },\mathbf{A}%
^{\prime \prime },\mathbf{\hat{A},\hat{A}}^{\prime }...\right)
\end{equation*}%
In a second step, each sum is replaced by a weighted integration symbol: 
\begin{eqnarray*}
\sum_{i} &\rightarrow &\int \left\vert \Psi \left( \mathbf{A}\right)
\right\vert ^{2}d\mathbf{A}\text{, }\sum_{j}\rightarrow \int \left\vert \Psi
\left( \mathbf{A}^{\prime }\right) \right\vert ^{2}d\mathbf{A}^{\prime }%
\text{, }\sum_{k}\rightarrow \int \left\vert \Psi \left( \mathbf{A}^{\prime
\prime }\right) \right\vert ^{2}d\mathbf{A}^{\prime \prime } \\
\sum_{l} &\rightarrow &\int \left\vert \hat{\Psi}\left( \mathbf{\hat{A}}%
\right) \right\vert ^{2}d\mathbf{\hat{A}}\text{, }\sum_{m}\rightarrow \int
\left\vert \hat{\Psi}\left( \mathbf{\hat{A}}^{\prime }\right) \right\vert
^{2}d\mathbf{\hat{A}}^{\prime }
\end{eqnarray*}%
which leads to the translation:%
\begin{eqnarray}
&&\sum_{i}\sum_{j}\sum_{j,k...}g\left( \mathbf{A}_{i}\left( t\right) ,%
\mathbf{A}_{j}\left( t\right) ,\mathbf{A}_{k}\left( t\right) ,\mathbf{\hat{A}%
}_{l}\left( t\right) ,\mathbf{\hat{A}}_{m}\left( t\right) ...\right)  \notag
\\
&\rightarrow &\int g\left( \mathbf{A},\mathbf{A}^{\prime },\mathbf{A}%
^{\prime \prime },\mathbf{\hat{A},\hat{A}}^{\prime }...\right) \left\vert
\Psi \left( \mathbf{A}\right) \right\vert ^{2}\left\vert \Psi \left( \mathbf{%
A}^{\prime }\right) \right\vert ^{2}\left\vert \Psi \left( \mathbf{A}%
^{\prime \prime }\right) \right\vert ^{2}\times ...d\mathbf{A}d\mathbf{A}%
^{\prime }d\mathbf{A}^{\prime \prime }...  \label{tln} \\
&&\times \left\vert \hat{\Psi}\left( \mathbf{\hat{A}}\right) \right\vert
^{2}\left\vert \hat{\Psi}\left( \mathbf{\hat{A}}^{\prime }\right)
\right\vert ^{2}\times ...d\mathbf{\hat{A}}d\mathbf{\hat{A}}^{\prime }... 
\notag
\end{eqnarray}%
where the dots stand for the products of square fields and integration
symbols needed.

\subsubsection*{A1.2.2 Terms with temporal derivative}

In equation (\ref{mNZ}), the terms that involve a variable temporal
derivative are of the form:%
\begin{equation}
\sum_{i}\left( \frac{d\mathbf{A}_{i}^{\left( \alpha \right) }\left( t\right) 
}{dt}-\sum_{j,k,l,m...}f^{\left( \alpha \right) }\left( \mathbf{A}_{i}\left(
t\right) ,\mathbf{A}_{j}\left( t\right) ,\mathbf{A}_{k}\left( t\right) ,%
\mathbf{\hat{A}}_{l}\left( t\right) ,\mathbf{\hat{A}}_{m}\left( t\right)
...\right) \right) ^{2}  \label{edr}
\end{equation}%
This particular form represents the dynamics of the $\alpha $-th coordinate
of a variable $\mathbf{A}_{i}\left( t\right) $ as a function of the other
agents.

The method of translation is similar to the above, but the time derivative
adds an additional operation.

In a first step, we translate the terms without derivative inside the
parenthesis:%
\begin{equation}
\sum_{j,k,l,m...}f^{\left( \alpha \right) }\left( \mathbf{A}_{i}\left(
t\right) ,\mathbf{A}_{j}\left( t\right) ,\mathbf{A}_{k}\left( t\right) ,%
\mathbf{\hat{A}}_{l}\left( t\right) ,\mathbf{\hat{A}}_{m}\left( t\right)
...\right)  \label{nts}
\end{equation}%
This type of term has already been translated in the previous paragraph, but
since there is no sum over $i$ in equation (\ref{ntr}), there should be no
integral over $\mathbf{A}$\textbf{,} nor factor $\left\vert \Psi \left( 
\mathbf{A}\right) \right\vert ^{2}$.

The translation of equation (\ref{nts}) is therefore, as before:%
\begin{equation}
\int f^{\left( \alpha \right) }\left( \mathbf{A},\mathbf{A}^{\prime },%
\mathbf{A}^{\prime \prime },\mathbf{\hat{A},\hat{A}}^{\prime }...\right)
\left\vert \Psi \left( \mathbf{A}^{\prime }\right) \right\vert
^{2}\left\vert \Psi \left( \mathbf{A}^{\prime \prime }\right) \right\vert
^{2}d\mathbf{A}^{\prime }d\mathbf{A}^{\prime \prime }\left\vert \hat{\Psi}%
\left( \mathbf{\hat{A}}\right) \right\vert ^{2}\left\vert \hat{\Psi}\left( 
\mathbf{\hat{A}}^{\prime }\right) \right\vert ^{2}d\mathbf{\hat{A}}d\mathbf{%
\hat{A}}^{\prime }  \label{trn}
\end{equation}%
A free variable $\mathbf{A}$ remains, which will be integrated later, when
we account for the external sum $\sum_{i}$. We will call $\Lambda (\mathbf{A}%
)$ the expression obtained:%
\begin{equation}
\Lambda (\mathbf{A})=\int f^{\left( \alpha \right) }\left( \mathbf{A},%
\mathbf{A}^{\prime },\mathbf{A}^{\prime \prime },\mathbf{\hat{A},\hat{A}}%
^{\prime }...\right) \left\vert \Psi \left( \mathbf{A}^{\prime }\right)
\right\vert ^{2}\left\vert \Psi \left( \mathbf{A}^{\prime \prime }\right)
\right\vert ^{2}d\mathbf{A}^{\prime }d\mathbf{A}^{\prime \prime }\left\vert 
\hat{\Psi}\left( \mathbf{\hat{A}}\right) \right\vert ^{2}\left\vert \hat{\Psi%
}\left( \mathbf{\hat{A}}^{\prime }\right) \right\vert ^{2}d\mathbf{\hat{A}}d%
\mathbf{\hat{A}}^{\prime }  \label{bdt}
\end{equation}%
In a second step, we account for the derivative in time by using field
gradients. To do so, and as a rule, we replace :%
\begin{equation}
\sum_{i}\left( \frac{d\mathbf{A}_{i}^{\left( \alpha \right) }\left( t\right) 
}{dt}-\sum_{j}\sum_{j,k...}f^{\left( \alpha \right) }\left( \mathbf{A}%
_{i}\left( t\right) ,\mathbf{A}_{j}\left( t\right) ,\mathbf{A}_{k}\left(
t\right) ,\mathbf{\hat{A}}_{l}\left( t\right) ,\mathbf{\hat{A}}_{m}\left(
t\right) ...\right) \right) ^{2}  \label{inco}
\end{equation}%
by:%
\begin{equation}
\int \Psi ^{\dag }\left( \mathbf{A}\right) \left( -\nabla _{\mathbf{A}%
^{\left( \alpha \right) }}\left( \frac{\sigma _{\mathbf{A}^{\left( \alpha
\right) }}^{2}}{2}\nabla _{\mathbf{A}^{\left( \alpha \right) }}-\Lambda (%
\mathbf{A})\right) \right) \Psi \left( \mathbf{A}\right) d\mathbf{A}
\label{Trll}
\end{equation}%
The variance $\sigma _{\mathbf{A}^{\left( \alpha \right) }}^{2}$ reflects
the probabilistic nature of the model which is hidden behind the field
formalism. This variance represents the characteristic level of uncertainty
of the system's dynamics. It is a parameter of the model. Note also that in (%
\ref{Trll}), the integral over $\mathbf{A}$ reappears at the end, along with
the square of the field $\left\vert \Psi \left( \mathbf{A}\right)
\right\vert ^{2}$.\ This square is split into two terms, $\Psi ^{\dag
}\left( \mathbf{A}\right) $ and $\Psi \left( \mathbf{A}\right) $, with a
gradient operator inserted in between.

\subsubsection*{A1.2.3 Action functional}

The field description is ultimately obtained by summing all the terms
translated above and introducing a time dependency. This sum is called the
action functional. It is the sum of terms of the form (\ref{tln}) and (\ref%
{Trll}), and is denoted $S\left( \Psi ,\Psi ^{\dag }\right) $.

For example, in a system with two types of agents described by two fields $%
\Psi \left( \mathbf{A}\right) $and $\hat{\Psi}\left( \mathbf{\hat{A}}\right) 
$, the action functional has the form:%
\begin{eqnarray}
S\left( \Psi ,\Psi ^{\dag }\right) &=&\int \Psi ^{\dag }\left( \mathbf{A}%
\right) \left( -\nabla _{\mathbf{A}^{\left( \alpha \right) }}\left( \frac{%
\sigma _{\mathbf{A}^{\left( \alpha \right) }}^{2}}{2}\nabla _{\mathbf{A}%
^{\left( \alpha \right) }}-\Lambda _{1}(\mathbf{A})\right) \right) \Psi
\left( \mathbf{A}\right) d\mathbf{A}  \label{notime} \\
&&\mathbf{+}\int \hat{\Psi}^{\dag }\left( \mathbf{\hat{A}}\right) \left(
-\nabla _{\mathbf{\hat{A}}^{\left( \alpha \right) }}\left( \frac{\sigma _{%
\mathbf{\hat{A}}^{\left( \alpha \right) }}^{2}}{2}\nabla _{\mathbf{\hat{A}}%
^{\left( \alpha \right) }}-\Lambda _{2}(\mathbf{\hat{A}})\right) \right) 
\hat{\Psi}\left( \mathbf{\hat{A}}\right) d\mathbf{\hat{A}}  \notag \\
&&+\sum_{m}\int g_{m}\left( \mathbf{A},\mathbf{A}^{\prime },\mathbf{A}%
^{\prime \prime },\mathbf{\hat{A},\hat{A}}^{\prime }...\right) \left\vert
\Psi \left( \mathbf{A}\right) \right\vert ^{2}\left\vert \Psi \left( \mathbf{%
A}^{\prime }\right) \right\vert ^{2}\left\vert \Psi \left( \mathbf{A}%
^{\prime \prime }\right) \right\vert ^{2}\times ...d\mathbf{A}d\mathbf{A}%
^{\prime }d\mathbf{A}^{\prime \prime }...  \notag \\
&&\times \left\vert \hat{\Psi}\left( \mathbf{\hat{A}}\right) \right\vert
^{2}\left\vert \hat{\Psi}\left( \mathbf{\hat{A}}^{\prime }\right)
\right\vert ^{2}\times ...d\mathbf{\hat{A}}d\mathbf{\hat{A}}^{\prime }... 
\notag
\end{eqnarray}%
where the sequence of functions $g_{m}$\ describes the various types of
interactions in the system.

\subsection*{A1.3 Probabilistic description of large set of cells and
connectivity functions}

We describe a dynamic system of a large number of neurons ($N>>1$) and their
connectivity functions. We define their individual equations. Then, we write
a probability density for the configurations of the whole system over time.

\subsubsection*{A1.3.1 Cells Individual dynamics}

We follow the description of \cite{V11} for coupled quadratic
integrate-and-fire (QIF) neurons, but use the additional hypothesis that
each neuron is characterized by its position in some spatial range.

Each neuron's potential $X_{i}\left( t\right) $ satisfies the differential
equation:%
\begin{equation}
\dot{X}_{i}\left( t\right) =\gamma X_{i}^{2}\left( t\right) +J_{i}\left(
t\right)  \label{ptn}
\end{equation}%
for $X_{i}\left( t\right) <X_{p}$, where $X_{p}$ denotes the potential level
of a spike. When $X=X_{p}$, the potential is reset to its resting value $%
X_{i}\left( t\right) =X_{r}<X_{p}$. For the sake of simplicity, following (%
\cite{V11}) we have chosen the squared form $\gamma X_{i}^{2}\left( t\right) 
$ in (\ref{ptn}). However any form $f\left( X_{i}\left( t\right) \right) $\
could be used. The current of signals reaching cell $i$ at time $t$ is
written $J_{i}\left( t\right) $.

Our purpose is to find the system dynamics in terms of the spikes'
frequencies, that is neural activities. First, we consider the time for the $%
n$-th spike of cell $i$, $\theta _{n}^{\left( i\right) }$. This is written
as a function of $n$, $\theta ^{\left( i\right) }\left( n\right) $. Then, a
continuous approximation $n\rightarrow t$ allows to write the spike time
variable as $\theta ^{\left( i\right) }\left( t\right) $. We thus have
replaced:

\begin{equation*}
\theta _{n}^{\left( i\right) }\rightarrow \theta ^{\left( i\right) }\left(
n\right) \rightarrow \theta ^{\left( i\right) }\left( t\right)
\end{equation*}%
The continuous approximation could be removed, but is convenient and
simplifies the notations and computations. We assume now that the timespans
between two spikes are relatively small. The time between two spikes for
cell $i$ is obtained by writing (\ref{ptn}) as:%
\begin{equation*}
\frac{dX_{i}\left( t\right) }{dt}=\gamma X_{i}^{2}\left( t\right)
+J_{i}\left( t\right)
\end{equation*}%
and by inverting this relation to write:%
\begin{equation*}
dt=\frac{dX_{i}}{\gamma X_{i}^{2}+J^{\left( i\right) }\left( \theta ^{\left(
i\right) }\left( n-1\right) \right) }
\end{equation*}%
Integrating the potential between two spikes thus yields:%
\begin{equation*}
\theta ^{\left( i\right) }\left( n\right) -\theta ^{\left( i\right) }\left(
n-1\right) \simeq \int_{X_{r}}^{X_{p}}\frac{dX}{\gamma X^{2}+J^{\left(
i\right) }\left( \theta ^{\left( i\right) }\left( n-1\right) \right) }
\end{equation*}%
Replacing $J^{\left( i\right) }\left( \theta ^{\left( i\right) }\left(
n-1\right) \right) $ by its average value during the small time period $%
\theta ^{\left( i\right) }\left( n\right) -\theta ^{\left( i\right) }\left(
n-1\right) $, we can consider $J^{\left( i\right) }\left( \theta ^{\left(
i\right) }\left( n-1\right) \right) $ as constant in first approximation, so
that:

\begin{eqnarray}
\theta ^{\left( i\right) }\left( n\right) -\theta ^{\left( i\right) }\left(
n-1\right) &\simeq &\frac{\left[ \arctan \left( \sqrt{\frac{\gamma }{%
J^{\left( i\right) }\left( \theta ^{\left( i\right) }\left( n-1\right)
\right) }}X\right) \right] _{X_{r}}^{X_{p}}}{\sqrt{\gamma J^{\left( i\right)
}\left( \theta ^{\left( i\right) }\left( n-1\right) \right) }}  \notag \\
&=&\frac{\arctan \left( \frac{\left( \frac{1}{X_{r}}-\frac{1}{X_{p}}\right) 
\sqrt{\frac{J^{\left( i\right) }\left( \theta ^{\left( i\right) }\left(
n-1\right) \right) }{\gamma }}}{1+\frac{J^{\left( n\right) }\left( \theta
^{\left( n-1\right) }\right) }{\gamma X_{r}X_{p}}}\right) }{\sqrt{\gamma
J^{\left( i\right) }\left( \theta ^{\left( i\right) }\left( n-1\right)
\right) }}  \label{FG}
\end{eqnarray}%
To work at the highest level of generality when possible, we write:%
\begin{equation*}
\theta ^{\left( i\right) }\left( n\right) -\theta ^{\left( i\right) }\left(
n-1\right) \equiv G\left( \theta ^{\left( i\right) }\left( n-1\right) \right)
\end{equation*}%
understood that for computations and numerical approximations we will use
formula (\ref{FG}) for $G$.

For $\gamma <<1$, (\ref{FG}) yields:%
\begin{equation*}
\theta ^{\left( i\right) }\left( n\right) -\theta ^{\left( i\right) }\left(
n-1\right) \equiv G\left( \theta ^{\left( i\right) }\left( n-1\right)
\right) =\frac{X_{p}-X_{r}}{J^{\left( i\right) }\left( \theta ^{\left(
i\right) }\left( n-1\right) \right) }
\end{equation*}%
For $\gamma =O\left( 1\right) $ and for $\gamma $ normalized to $1$ and $%
\frac{J^{\left( n\right) }\left( \theta ^{\left( n-1\right) }\right) }{%
X_{r}X_{p}}<<1$, this is: 
\begin{equation}
\theta ^{\left( i\right) }\left( n\right) -\theta ^{\left( i\right) }\left(
n-1\right) \equiv G\left( \theta ^{\left( i\right) }\left( n-1\right)
\right) =\frac{\arctan \left( \left( \frac{1}{X_{r}}-\frac{1}{X_{p}}\right) 
\sqrt{J^{\left( i\right) }\left( \theta ^{\left( i\right) }\left( n-1\right)
\right) }\right) }{\sqrt{J^{\left( i\right) }\left( \theta ^{\left( i\right)
}\left( n-1\right) \right) }}  \label{spt}
\end{equation}%
The activity or firing rate at $t$, $\omega _{i}\left( t\right) $, is
defined by the inverse time span (\ref{spt}) between two spikes:%
\begin{eqnarray*}
\omega _{i}\left( t\right) &=&\frac{1}{G\left( \theta ^{\left( i\right)
}\left( n-1\right) \right) } \\
&\equiv &F\left( \theta ^{\left( i\right) }\left( n-1\right) \right) =\frac{%
\sqrt{J^{\left( i\right) }\left( \theta ^{\left( i\right) }\left( n-1\right)
\right) }}{\arctan \left( \left( \frac{1}{X_{r}}-\frac{1}{X_{p}}\right) 
\sqrt{J^{\left( i\right) }\left( \theta ^{\left( i\right) }\left( n-1\right)
\right) }\right) }
\end{eqnarray*}%
Since we consider small time intervals between two spikes, we can write:%
\begin{equation}
\theta ^{\left( i\right) }\left( n\right) -\theta ^{\left( i\right) }\left(
n-1\right) \simeq \frac{d}{dt}\theta ^{\left( i\right) }\left( t\right)
-\omega _{i}^{-1}\left( t\right) =\varepsilon _{i}\left( t\right)
\label{dnm}
\end{equation}%
where the white noise perturbation $\varepsilon _{i}\left( t\right) $ for
each period was added to account for any internal uncertainty in the time
span $\theta ^{\left( i\right) }\left( n\right) -\theta ^{\left( i\right)
}\left( n-1\right) $. This white noise is independent from the instantaneous
inverse activity $\omega _{i}^{-1}\left( t\right) $. We assume these $%
\varepsilon _{i}\left( t\right) $ to have variance $\sigma ^{2}$, so that
equation (\ref{dnm}) writes: 
\begin{equation}
\frac{d}{dt}\theta ^{\left( i\right) }\left( t\right) -G\left( \theta
^{\left( i\right) }\left( t\right) ,J^{\left( i\right) }\left( \theta
^{\left( i\right) }\left( t\right) \right) \right) =\varepsilon _{i}\left(
t\right)  \label{dnq}
\end{equation}%
The $\omega _{i}\left( t\right) $ are computed by considering the overall
current which, using the discrete time notation, is given by:%
\begin{equation}
\hat{J}^{\left( i\right) }\left( \left( n-1\right) \right) =J^{\left(
i\right) }\left( \left( n-1\right) \right) +\frac{\kappa }{N}\sum_{j,m}\frac{%
\omega _{j}\left( m\right) }{\omega _{i}\left( n-1\right) }\delta \left(
\theta ^{\left( i\right) }\left( n-1\right) -\theta ^{\left( j\right)
}\left( m\right) -\frac{\left\vert Z_{i}-Z_{j}\right\vert }{c}\right)
T_{ij}\left( \left( n-1,Z_{i}\right) ,\left( m,Z_{j}\right) \right)
\label{crt}
\end{equation}%
The quantity $J^{\left( i\right) }\left( \left( n-1\right) \right) $ denotes
an external current. The term inside the sum is the average current sent to $%
i$ by neuron $j$ during the short time span $\theta ^{\left( i\right)
}\left( n\right) -\theta ^{\left( i\right) }\left( n-1\right) $. The
function $T_{ij}\left( \left( n-1,Z_{i}\right) ,\left( m,Z_{j}\right)
\right) $ is the connectivity function between cells $j$ and $i$. It
measures the level of connectivity between $i$ and $j$. If we consider%
\textbf{\ }$T_{ij}\left( \left( n-1,Z_{i}\right) ,\left( m,Z_{j}\right)
\right) $\textbf{\ }as exogenous, we may assume that (see \cite{GL}): 
\begin{equation*}
T_{ij}\left( \left( n-1,Z_{i}\right) ,\left( m,Z_{j}\right) \right) =T\left(
\left( n-1,Z_{i}\right) ,\left( m,Z_{j}\right) \right)
\end{equation*}%
so that the connectivity function of $Z_{j}$ on $Z_{i}$ only depends on
positions and times. It models the connectivity function as an average
connectivity between local zones of the thread. this transfer function is
typically considered as gaussian or decreasing exponentially with the
distance between neurons, so that the closer the cells, the more connected
they are.

However, in this paper, the connectivity function is a dynamical object
whose dynamic equations are described in the next paragraph.

We can justify the other terms arising in (\ref{crt}): given the distance $%
\left\vert Z_{i}-Z_{j}\right\vert $ between the two cells and the signals'
velocity $c$, signals arrive with a delay $\frac{\left\vert
Z_{i}-Z_{j}\right\vert }{c}$. The spike emitted by cell $j$ at time $\theta
^{\left( j\right) }\left( m\right) $ has thus to satisfy: 
\begin{equation*}
\theta ^{\left( i\right) }\left( n-1\right) <\theta ^{\left( j\right)
}\left( m\right) +\frac{\left\vert Z_{i}-Z_{j}\right\vert }{c}<\theta
^{\left( i\right) }\left( n\right)
\end{equation*}%
to reach cell $i$ during the timespan $\left[ \theta ^{\left( i\right)
}\left( n-1\right) ,\theta ^{\left( i\right) }\left( n\right) \right] $.
This relation must be represented by a step function in the current formula.
However given our approximations, this can be replaced by: 
\begin{equation*}
\delta \left( \theta ^{\left( i\right) }\left( n-1\right) -\theta ^{\left(
j\right) }\left( m\right) -\frac{\left\vert Z_{i}-Z_{j}\right\vert }{c}%
\right)
\end{equation*}%
as in (\ref{crt}). However, this Dirac function must be weighted by the
number of spikes emitted during the rise of the potential. This number is
the ratio $\frac{\omega _{j}\left( m\right) }{\omega _{i}\left( n-1\right) }$
that counts the number of spikes emitted by neuron $j$ towards neuron $i$
between the spikes $n-1$ and $n$ of neuron $i$. Again, this is valid for
relatively small timespans between two spikes. For larger timespans, the
firing rates should be replaced by their average over this period of time.

The sum over $m$ and $i$ is the overall contribution to the current from any
possible spike of the thread, provided it arrives at $i$ during the interval 
$\theta ^{\left( i\right) }\left( n\right) -\theta ^{\left( i\right) }\left(
n-1\right) $ considered. Note that the current (\ref{crt}) is partly an
endogenous variable. It depends on signals external to $i$, but depends also
on $i$ through $\omega _{i}\left( n-1\right) $. This is a consequence of the
intrication between the system's elements.

In the sequel, we will work in the continuous approximation, so that formula
(\ref{crt}) is replaced by:%
\begin{equation}
\hat{J}^{\left( i\right) }\left( t\right) =J^{\left( i\right) }\left(
t\right) +\frac{\kappa }{N}\int \sum_{j}\frac{\omega _{j}\left( s\right) }{%
\omega _{i}\left( t\right) }\delta \left( \theta ^{\left( i\right) }\left(
t\right) -\theta ^{\left( j\right) }\left( s\right) -\frac{\left\vert
Z_{i}-Z_{j}\right\vert }{c}\right) T_{ij}\left( \left( t,Z_{i}\right)
,\left( s,Z_{j}\right) \right) ds  \label{crT}
\end{equation}

Formula (\ref{crT}) shows that the dynamic equation (\ref{dnm}) has to be
coupled with the activity equation:%
\begin{eqnarray}
\omega _{i}\left( t\right) &=&G\left( \theta ^{\left( i\right) }\left(
t\right) ,\hat{J}\left( \theta ^{\left( i\right) }\left( t\right) \right)
\right) +\upsilon _{i}\left( t\right)  \label{cstrt} \\
&=&\frac{\sqrt{\hat{J}^{\left( i\right) }\left( t\right) }}{\arctan \left(
\left( \frac{1}{X_{r}}-\frac{1}{X_{p}}\right) \sqrt{\hat{J}^{\left( i\right)
}\left( t\right) }\right) }+\upsilon _{i}\left( t\right)  \notag
\end{eqnarray}%
and $J^{\left( i\right) }\left( t\right) $ is defined by (\ref{crT}). A
white noise $\upsilon _{i}\left( t\right) $ accounts for the possible
deviations from this relation, due to some internal or external causes for
each cell. We assume that the variances of $\upsilon _{i}\left( t\right) $
are constant, and equal to $\eta ^{2}$, such that $\eta ^{2}<<\sigma ^{2}$.

\subsubsection*{A1.3.2 Connectivity functions dynamics}

We describe the dynamics for the connectivity functions $T_{ij}\left( \left(
n-1,Z_{i}\right) ,\left( m,Z_{j}\right) \right) $ between cells. To do so we
adapt the description of (\cite{IFR}) to our context. In this work, the
connectivity functions depend on some intermediate variables and do not
present any space index. The connectivity between neurons $i$ and $j$
satisifies a differential equation:%
\begin{equation}
\frac{dT_{ij}}{dt}=-\frac{T_{ij}\left( t\right) }{\tau }+\lambda \hat{T}%
_{ij}\left( t\right) \sum_{l}\delta \left( t-\Delta t_{ij}-t_{j}^{l}\right)
\label{spn}
\end{equation}%
where $\hat{T}_{ij}\left( t\right) $ represents the variation in
connectivity, due to the synaptic interactions between the two neurons. The
delay $\Delta t_{ij}$ is the time of arrival at neuton $i$ for a spike of
neuron $j$. The time $t_{j}^{l}$ accounts for time of neuron $j$ spikes. The
sum:%
\begin{equation*}
\sum_{l}\delta \left( t-\Delta t_{ij}-t_{j}^{l}\right)
\end{equation*}%
counts the number of spikes emitted by neuron $j$ and arriving at time $t$
at neuron $i$.

The variation in connectivity satisfies itself an equation:%
\begin{equation}
\frac{d\hat{T}_{ij}}{dt}=\rho \left( 1-\hat{T}_{ij}\left( t\right) \right)
C_{ij}\left( t\right) \sum_{k}\delta \left( t-t_{i}^{k}\right) -\hat{T}%
_{ij}\left( t\right) D_{i}\left( t\right) \sum_{l}\delta \left( t-\Delta
t_{ij}-t_{j}^{l}\right)  \label{SPt}
\end{equation}%
where $C_{ij}\left( t\right) $ and $D_{i}\left( t\right) $\ measure the
cumulated postsynaptic and presynaptic activity. The sum:%
\begin{equation*}
\sum_{k}\delta \left( t-t_{i}^{k}\right)
\end{equation*}%
counts the number of spikes emitted at time $t$. Quantities $C_{ij}\left(
t\right) $ and $D_{i}\left( t\right) $ follow the dynamics:%
\begin{equation}
\frac{dC_{ij}}{dt}=-\frac{C_{ij}\left( t\right) }{\tau _{C}}+\alpha
_{C}\left( 1-C_{ij}\left( t\right) \right) \sum_{l}\delta \left( t-\Delta
t_{ij}-t_{j}^{l}\right)  \label{spr}
\end{equation}%
\begin{equation}
\frac{dD_{i}}{dt}=-\frac{D_{i}\left( t\right) }{\tau _{D}}+\alpha _{C}\left(
1-D_{i}\left( t\right) \right) \sum_{k}\delta \left( t-t_{i}^{k}\right)
\label{spf}
\end{equation}

To translate these equations in our set up, we have to consider connectivity
functions of the form:%
\begin{equation*}
T_{ij}\left( \left( n_{i},Z_{i}\right) ,\left( n_{j},Z_{j}\right) \right)
\end{equation*}%
that include the positions of neurons $i$ and $j$ and the parameter $n_{i}$
and $n_{j}$ which are our counting variables of neurons spikes. However,
equations (\ref{spn}), (\ref{SPt}), (\ref{spr}), (\ref{spf}) include a time
variable.

In our formalism, the time variable $\theta ^{\left( i\right) }\left(
n_{i}\right) $ is the time at which neuron $i$ produces its $n_{i}$-th
spike. We should write classical equations depending on these variables.

Moreover, the number of spikes $\sum_{l}\delta \left( t-\Delta
t_{ij}-t_{j}^{l}\right) $ emitted by cell $j$ at time $t_{j}^{l}$ and the
number of spikes $\sum_{k}\delta \left( t-t_{i}^{k}\right) $ emitted by cell 
$i$ at time $t$ are proportional to $\delta \left( \theta ^{\left( j\right)
}\left( n_{j}\right) -\left( t-\Delta t_{ij}\right) \right) \omega
_{j}\left( n_{j}\right) $ and $\delta \left( \theta ^{\left( i\right)
}\left( n_{i}\right) -t\right) \omega _{j}\left( n_{i}\right) $
respectively. Given the introduction of a spatial indices, we have the
relation:%
\begin{equation*}
\Delta t_{ij}=\frac{\left\vert Z_{i}-Z_{j}\right\vert }{c}
\end{equation*}%
and the first $\delta $ function writes:%
\begin{equation*}
\delta \left( \theta ^{\left( j\right) }\left( n_{j}\right) -\left( t-\Delta
t_{ij}\right) \right) =\delta \left( \theta ^{\left( j\right) }\left(
n_{j}\right) -\left( \theta ^{\left( i\right) }\left( n_{i}\right) -\frac{%
\left\vert Z_{i}-Z_{j}\right\vert }{c}\right) \right) \delta \left( \theta
^{\left( i\right) }\left( n_{i}\right) -t\right)
\end{equation*}

As a consequence, we will write first the connectivity functions from $i$ to 
$j$ as:%
\begin{equation*}
T\left( \left( Z_{i},\theta ^{\left( i\right) }\left( n_{i}\right) ,\omega
_{i}\left( n_{i}\right) \right) ,\left( Z_{j},\theta ^{\left( j\right)
}\left( n_{j}\right) ,\omega _{j}\left( n_{j}\right) \right) \right)
\end{equation*}%
This function, together with the variation in connectivity:%
\begin{equation*}
\hat{T}\left( \left( Z_{i},\theta ^{\left( i\right) }\left( n_{i}\right)
,\omega _{i}\left( n_{i}\right) \right) ,\left( Z_{j},\theta ^{\left(
j\right) }\left( n_{j}\right) ,\omega _{j}\left( n_{j}\right) \right) \right)
\end{equation*}%
along with the auxiliary variables:%
\begin{equation*}
C\left( \left( Z_{i},\theta ^{\left( i\right) }\left( n_{i}\right) ,\omega
_{i}\left( n_{i}\right) \right) ,\left( Z_{j},\theta ^{\left( j\right)
}\left( n_{j}\right) ,\omega _{j}\left( n_{j}\right) \right) \right)
\end{equation*}%
and:%
\begin{equation*}
D\left( \left( Z_{i},\theta ^{\left( i\right) }\left( n_{i}\right) ,\omega
_{i}\left( n_{i}\right) \right) \right)
\end{equation*}%
satisfy the following translations of equations (\ref{spn}), (\ref{SPt}), (%
\ref{spr}), (\ref{spf}): 
\begin{eqnarray}
&&\nabla _{\theta ^{\left( i\right) }\left( n_{i}\right) }T\left( \left(
Z_{i},\theta ^{\left( i\right) }\left( n_{i}\right) ,\omega _{i}\left(
n_{i}\right) \right) ,\left( Z_{j},\theta ^{\left( j\right) }\left(
n_{j}\right) ,\omega _{j}\left( n_{j}\right) \right) \right) \\
&=&-\frac{1}{\tau }T\left( \left( Z_{i},\theta ^{\left( i\right) }\left(
n_{i}\right) ,\omega _{i}\left( n_{i}\right) \right) ,\left( Z_{j},\theta
^{\left( j\right) }\left( n_{j}\right) ,\omega _{j}\left( n_{j}\right)
\right) \right)  \notag \\
&&+\lambda \left( \hat{T}\left( \left( Z_{i},\theta ^{\left( i\right)
}\left( n_{i}\right) ,\omega _{i}\left( n_{i}\right) \right) ,\left(
Z_{j},\theta ^{\left( j\right) }\left( n_{j}\right) ,\omega _{j}\left(
n_{j}\right) \right) \right) \right) \delta \left( \theta ^{\left( i\right)
}\left( n_{i}\right) -\theta ^{\left( j\right) }\left( n_{j}\right) -\frac{%
\left\vert Z_{i}-Z_{j}\right\vert }{c}\right)  \notag
\end{eqnarray}%
where $\hat{T}$ measures the variation of $T$ due to the signals send from $%
j $ to $i$ and the signals emitted by $i$. It satisfies the following
equation: 
\begin{eqnarray}
&&\nabla _{\theta ^{\left( i\right) }\left( n_{i}\right) }\hat{T}\left(
\left( Z_{i},\theta ^{\left( i\right) }\left( n_{i}\right) ,\omega
_{i}\left( n_{i}\right) \right) ,\left( Z_{j},\theta ^{\left( j\right)
}\left( n_{j}\right) ,\omega _{j}\left( n_{j}\right) \right) \right)
\label{VL} \\
&=&\rho \delta \left( \theta ^{\left( i\right) }\left( n_{i}\right) -\theta
^{\left( j\right) }\left( n_{j}\right) -\frac{\left\vert
Z_{i}-Z_{j}\right\vert }{c}\right)  \notag \\
&&\times \left\{ \left( h\left( Z,Z_{1}\right) -\hat{T}\left( \left(
Z_{i},\theta ^{\left( i\right) }\left( n_{i}\right) ,\omega _{i}\left(
n_{i}\right) \right) ,\left( Z_{j},\theta ^{\left( j\right) }\left(
n_{j}\right) ,\omega _{j}\left( n_{j}\right) \right) \right) \right) C\left(
\theta ^{\left( i\right) }\left( n\right) \right) h_{C}\left( \omega
_{i}\left( n_{i}\right) \right) \right.  \notag \\
&&\left. -D\left( \theta ^{\left( i\right) }\left( n\right) \right) \hat{T}%
\left( \left( Z_{i},\theta ^{\left( i\right) }\left( n_{i}\right) ,\omega
_{i}\left( n_{i}\right) \right) ,\left( Z_{j},\theta ^{\left( j\right)
}\left( n_{j}\right) ,\omega _{j}\left( n_{j}\right) \right) \right)
h_{D}\left( \omega _{j}\left( n_{j}\right) \right) \right\}  \notag
\end{eqnarray}%
where $h_{C}$ and $h_{D}$\ are increasing functions. In the set of equations
(\ref{spn}), (\ref{SPt}), (\ref{spr}), (\ref{spf}): 
\begin{eqnarray*}
h_{C}\left( \omega _{i}\left( n_{i}\right) \right) &=&\omega _{i}\left(
n_{i}\right) \\
h_{D}\left( \omega _{j}\left( n_{j}\right) \right) &=&\omega _{j}\left(
n_{j}\right)
\end{eqnarray*}

We depart slightly from (\cite{IFR}) by the introduction of the function $%
h\left( Z,Z_{1}\right) $ (they choose $h\left( Z,Z_{1}\right) =1$), to
implement some loss due to the distance. We may choose for example:%
\begin{equation*}
h\left( Z,Z_{1}\right) =\exp \left( -\frac{\left\vert Z_{i}-Z_{j}\right\vert 
}{\nu c}\right)
\end{equation*}%
where $\nu $ is a parameter. Equation (\ref{VL}) involves two dynamic
factors $C\left( \theta ^{\left( i\right) }\left( n-1\right) \right) $ and $%
D\left( \theta _{i}\left( n-1\right) \right) $. The factor $C\left( \theta
^{\left( i\right) }\left( n-1\right) \right) $ describes the accumulation of
input spikes. It is solution of the differential equation:%
\begin{eqnarray}
\nabla _{\theta ^{\left( i\right) }\left( n-1\right) }C\left( \theta
^{\left( i\right) }\left( n-1\right) \right) &=&-\frac{C\left( \theta
^{\left( i\right) }\left( n-1\right) \right) }{\tau _{C}} \\
&&+\alpha _{C}\left( 1-C\left( \theta ^{\left( i\right) }\left( n-1\right)
\right) \right) \omega _{j}\left( \theta ^{\left( i\right) }\left(
n-1\right) -\frac{\left\vert Z_{i}-Z_{j}\right\vert }{c}\right)  \notag
\end{eqnarray}%
The term $D\left( \theta _{i}\left( n-1\right) \right) $ is proportional to
the accumulation of output spikes and is solution of:%
\begin{equation}
\nabla _{\theta ^{\left( i\right) }\left( n-1\right) }D\left( \theta
^{\left( i\right) }\left( n-1\right) \right) =-\frac{D\left( \theta ^{\left(
i\right) }\left( n-1\right) \right) }{\tau _{D}}+\alpha _{D}\left( 1-D\left(
\theta ^{\left( i\right) }\left( n-1\right) \right) \right) \omega
_{i}\left( n_{i}\right)
\end{equation}%
For the purpose of field translation, we have to change the variables in the
derivatives by the counting variable $n_{i}$ and replace $\nabla _{\theta
^{\left( i\right) }\left( n_{i}\right) }\simeq \omega _{i}\left(
n_{i}\right) \nabla _{n_{i}}$ in the previous dynamics equations. We thus
rewrite the dynamic equations in the following form:

For the connectivity $T$:

\begin{eqnarray}
&&\nabla _{n_{i}}T\left( \left( Z_{i},\theta ^{\left( i\right) }\left(
n_{i}\right) ,\omega _{i}\left( n_{i}\right) \right) ,\left( Z_{j},\theta
^{\left( j\right) }\left( n_{j}\right) ,\omega _{j}\left( n_{j}\right)
\right) \right)  \label{nqp} \\
&=&-\frac{1}{\tau \omega _{i}\left( n_{i}\right) }T\left( \left(
Z_{i},\theta ^{\left( i\right) }\left( n_{i}\right) ,\omega _{i}\left(
n_{i}\right) \right) ,\left( Z_{j},\theta ^{\left( j\right) }\left(
n_{j}\right) ,\omega _{j}\left( n_{j}\right) \right) \right)  \notag \\
&&+\frac{\lambda }{\omega _{i}\left( n_{i}\right) }\left( \hat{T}\left(
\left( Z_{i},\theta ^{\left( i\right) }\left( n_{i}\right) ,\omega
_{i}\left( n_{i}\right) \right) ,\left( Z_{j},\theta ^{\left( j\right)
}\left( n_{j}\right) ,\omega _{j}\left( n_{j}\right) \right) \right) \right)
\delta \left( \theta ^{\left( i\right) }\left( n_{i}\right) -\theta ^{\left(
j\right) }\left( n_{j}\right) -\frac{\left\vert Z_{i}-Z_{j}\right\vert }{c}%
\right)  \notag
\end{eqnarray}%
For the variation in connectivity $\hat{T}$:%
\begin{eqnarray}
&&\nabla _{n_{i}}\hat{T}\left( \left( Z_{i},\theta ^{\left( i\right) }\left(
n_{i}\right) ,\omega _{i}\left( n_{i}\right) \right) ,\left( Z_{j},\theta
^{\left( j\right) }\left( n_{j}\right) ,\omega _{j}\left( n_{j}\right)
\right) \right)  \label{nqd} \\
&=&\frac{\rho }{\omega _{i}\left( n_{i}\right) }\delta \left( \theta
^{\left( i\right) }\left( n_{i}\right) -\theta ^{\left( j\right) }\left(
n_{j}\right) -\frac{\left\vert Z_{i}-Z_{j}\right\vert }{c}\right)  \notag \\
&&\times \left\{ \left( h\left( Z,Z_{1}\right) -\hat{T}\left( \left(
Z_{i},\theta ^{\left( i\right) }\left( n_{i}\right) ,\omega _{i}\left(
n_{i}\right) \right) ,\left( Z_{j},\theta ^{\left( j\right) }\left(
n_{j}\right) ,\omega _{j}\left( n_{j}\right) \right) \right) \right) C\left(
\theta ^{\left( i\right) }\left( n-1\right) \right) h_{C}\left( \omega
_{i}\left( n_{i}\right) \right) \right.  \notag \\
&&\left. -D\left( \theta ^{\left( i\right) }\left( n-1\right) \right) \hat{T}%
\left( \left( Z_{i},\theta ^{\left( i\right) }\left( n_{i}\right) ,\omega
_{i}\left( n_{i}\right) \right) ,\left( Z_{j},\theta ^{\left( j\right)
}\left( n_{j}\right) ,\omega _{j}\left( n_{j}\right) \right) \right)
h_{D}\left( \omega _{j}\left( n_{j}\right) \right) \right\}  \notag
\end{eqnarray}%
and for the auxiliary variables $C$ and $D$:%
\begin{eqnarray}
\nabla _{n_{i}}C\left( \theta ^{\left( i\right) }\left( n-1\right) \right)
&=&-\frac{C\left( \theta ^{\left( i\right) }\left( n-1\right) \right) }{\tau
_{C}\omega _{i}\left( n_{i}\right) }  \label{nqt} \\
&&+\alpha _{C}\left( 1-C\left( \theta ^{\left( i\right) }\left( n-1\right)
\right) \right) \frac{\omega _{j}\left( Z_{j},\theta ^{\left( i\right)
}\left( n-1\right) -\frac{\left\vert Z_{i}-Z_{j}\right\vert }{c}\right) }{%
\omega _{i}\left( n_{i}\right) }  \notag
\end{eqnarray}%
\begin{equation}
\nabla _{n_{i}}D\left( \theta ^{\left( i\right) }\left( n-1\right) \right) =-%
\frac{D\left( \theta ^{\left( i\right) }\left( n-1\right) \right) }{\tau
_{D}\omega _{i}\left( n_{i}\right) }+\alpha _{D}\left( 1-D\left( \theta
^{\left( i\right) }\left( n-1\right) \right) \right)  \label{nqQ}
\end{equation}

Then, to describe the connectivity by a field, we have to describe the
connectivity as a set of vectors depending of a set of double indices $kl$
(replacing $ij$) and interacting with the activities, seen as independent
variables indexed by $i,j...$

We thus describe connectivity by a set of matrices:%
\begin{equation*}
\left( T_{kl}\left( n_{kl}\right) ,\hat{T}_{kl}\left( n_{kl}\right) ,\left(
Z_{kl}\left( n_{kl}\right) =\left( Z_{k,},Z_{l}\right) \right) ,\theta
^{\left( kl\right) }\left( n_{kl}\right) ,\omega _{k}\left( n_{kl}\right)
,\omega _{l}^{\prime }\left( n_{kl}\right) ,C_{kl}\left( n_{kl}\right)
,D_{k}\left( n_{kl}\right) \right)
\end{equation*}%
where $n_{kl}$ is an internal parameter given by the average counting
variable for cells or synapses firing simultaneously at point $Z_{k,}$.

Then, we replace the description (\ref{nqp}), (\ref{nqd}), (\ref{nqt}), (\ref%
{nqQ}) by a set of equations in which connectivities $T_{kl}\left(
n_{kl}\right) $ interact with all pairs of neurons at points $Z_{k,}$ and $%
Z_{l}$ whose average firing rates at time $\theta ^{\left( kl\right) }\left(
n_{kl}\right) $ and $\theta ^{\left( kl\right) }\left( n_{kl}\right) -\frac{%
\left\vert Z_{k}-Z_{l}\right\vert }{c}$ are given by $\omega _{k}\left(
n_{kl}\right) ,\omega _{l}^{\prime }\left( n_{kl}\right) $ respectively. As
a consequence, we replace the notion of connectivity $T_{ij}\left( \left(
n-1,Z_{i}\right) ,\left( m,Z_{j}\right) \right) $ between two specific
neurons $i$ and $j$ by the average connectivity between the two sets of
neurons with identical activities at each extremity of the segment $\left(
Z_{i},Z_{j}\right) $ \ This approximation is justified if we consider that
neurons located at the same place and firing at the same rate can be
considered as closely connected and in average identical. Alternatively,
this can also be justified if we consider one neuron per spatial location
and assume each neuron as a complex entity sending several signals
simultaneously. Under this hypothesis, the average considered are taken over
the multiple activities of the same neuron\footnote{%
See section 5.2.1 for more details about these alternative interpretations.}.

Stated mathematically, the variable $n_{kl}$ is replacd by an average $%
n_{kl}=\bar{n}_{i}$ at a given time $\theta ^{\left( kl\right) }$ and we
assume that in average, connectivity variable $T_{kl}\left( n_{kl}\right) $
interacts with all neurons pairs located at $\left( Z_{k,},Z_{l}\right) $ at
times $\theta ^{\left( i\right) }\left( n_{i}\right) =\theta ^{\left(
kl\right) }\left( n_{kl}\right) $. Writing $\bar{\omega}\left(
Z_{i},n_{i}\right) $ for the average activity, we impose $\bar{\omega}\left(
Z_{i},n_{i}\right) =\omega _{k}\left( n_{kl}\right) $ and $\bar{\omega}%
\left( Z_{j},n_{j}\right) =\omega _{l}^{\prime }\left( n_{kl}\right) $ and $%
\theta ^{\left( j\right) }\left( n_{j}\right) =\theta ^{\left( kl\right)
}\left( n_{kl}\right) -\frac{\left\vert Z_{k}-Z_{l}\right\vert }{c}$
respectively. The densities $T_{kl}\left( n_{kl}\right) $ are thus the set
of all connections between points $Z_{k,}$ and $Z_{l}$ between sets of
synchronized neurons at $Z_{k}$ and synchronized neurons at $Z_{l}$, i.e.
between set of neurons at this points or alternatively between multiple
synapses for one or a few number of cells. In this point of view, we replace 
$\nabla _{\theta ^{\left( i\right) }\left( n_{i}\right) }\simeq \omega
_{i}\left( n_{i}\right) \nabla _{n_{i}}$ by:%
\begin{equation*}
\nabla _{\theta ^{\left( kl\right) }\left( n_{kl}\right) }\simeq \frac{%
\partial n_{kl}}{\partial \theta ^{\left( kl\right) }\left( n_{kl}\right) }%
\nabla _{n_{kl}}=\bar{\omega}\left( Z_{i},n_{i}\right) \nabla _{n_{kl}}
\end{equation*}%
As a consequence, the dynamic equations (\ref{nqp}), (\ref{nqd}), (\ref{nqt}%
), (\ref{nqQ}) are replaced by:%
\begin{eqnarray}
&\nabla _{n_{kl}}T_{kl}\left( n_{kl}\right) =&\left( -\sum_{i,n_{i}}\frac{1}{%
\tau \bar{\omega}\left( Z_{i},n_{i}\right) }T_{kl}\left( n_{kl}\right) +%
\frac{\lambda }{\bar{\omega}\left( Z_{i},n_{i}\right) }\hat{T}_{kl}\left(
n_{kl}\right) \right) \\
&&\times \delta \left( \theta ^{\left( i\right) }\left( n_{i}\right) -\theta
^{\left( kl\right) }\left( n_{kl}\right) \right) \delta \left(
Z_{k}-Z_{i}\right) \delta \left( \omega _{k}\left( n_{kl}\right) -\bar{\omega%
}\left( Z_{i},n_{i}\right) \right)  \notag
\end{eqnarray}%
\begin{eqnarray}
&&\nabla _{n_{kl}}\hat{T}\left( n_{kl}\right) \\
&=&\left( \sum_{i,n_{i}}\left( h\left( Z_{k},Z_{l}\right) -\hat{T}\left(
n_{kl}\right) \right) C_{kl}\left( n_{kl}\right) h_{C}\left( \omega
_{i}\left( n_{i}\right) \right) -\sum_{j,n_{j}}D_{k}\left( n_{kl}\right) 
\hat{T}\left( n_{kl}\right) h_{D}\left( \omega _{j}\left( n_{j}\right)
\right) \right)  \notag \\
&&\times \frac{\rho }{\bar{\omega}\left( Z_{i},n_{i}\right) }\delta \left(
\theta ^{\left( i\right) }\left( n_{i}\right) -\theta ^{\left( j\right)
}\left( n_{j}\right) -\frac{\left\vert Z_{i}-Z_{j}\right\vert }{c}\right)
\delta \left( \theta ^{\left( i\right) }\left( n_{i}\right) -\theta ^{\left(
kl\right) }\left( n_{kl}\right) \right)  \notag \\
&&\times \delta \left( \left( Z_{k,},Z_{l}\right) -\left(
Z_{i,},Z_{j}\right) \right) \delta \left( \omega _{k}\left( n_{kl}\right) -%
\bar{\omega}\left( Z_{i},n_{i}\right) \right) \delta \left( \omega
_{l}\left( n_{kl}\right) -\bar{\omega}\left( Z_{j},n_{j}\right) \right) 
\notag
\end{eqnarray}%
\begin{eqnarray}
\nabla _{n_{kl}}C\left( n_{kl}\right) &=&\left( -\frac{C\left( n_{kl}\right) 
}{\tau _{C}\bar{\omega}\left( Z_{i},n_{i}\right) }+\sum_{j,n_{j}}\alpha
_{C}\left( 1-C_{kl}\left( n_{kl}\right) \right) \frac{\omega _{j}\left(
n_{j}\right) }{\bar{\omega}\left( Z_{i},n_{i}\right) }\right) \\
&&\times \delta \left( \theta ^{\left( i\right) }\left( n_{i}\right) -\theta
^{\left( j\right) }\left( n_{j}\right) -\frac{\left\vert
Z_{i}-Z_{j}\right\vert }{c}\right) \delta \left( \theta ^{\left( i\right)
}\left( n_{i}\right) -\theta ^{\left( kl\right) }\left( n_{kl}\right)
\right) \delta \left( \left( Z_{k,},Z_{l}\right) -\left( Z_{i,},Z_{j}\right)
\right)  \notag \\
&&\times \delta \left( \omega _{k}\left( n_{kl}\right) -\bar{\omega}\left(
Z_{i},n_{i}\right) \right) \delta \left( \omega _{l}\left( n_{kl}\right) -%
\bar{\omega}\left( Z_{j},n_{j}\right) \right)  \notag
\end{eqnarray}%
\begin{eqnarray}
\nabla _{n_{kl}}D_{k}\left( n_{kl}\right) &=&\left( -\frac{D_{k}\left(
n_{kl}\right) }{\tau _{D}\bar{\omega}\left( Z_{i},n_{i}\right) }+\frac{1}{%
\bar{\omega}\left( Z_{i},n_{i}\right) }\sum_{i,n_{i}}\alpha _{D}\left(
1-D_{k}\left( n_{kl}\right) \right) \omega _{i}\left( n_{i}\right) \right) \\
&&\times \delta \left( \theta ^{\left( i\right) }\left( n_{i}\right) -\theta
^{\left( kl\right) }\left( n_{kl}\right) \right) \delta \left(
Z_{k}-Z_{i}\right) \delta \left( \omega _{k}\left( n_{kl}\right) -\bar{\omega%
}\left( Z_{i},n_{i}\right) \right) \delta \left( \omega _{l}\left(
n_{kl}\right) -\bar{\omega}\left( Z_{j},n_{j}\right) \right)  \notag
\end{eqnarray}%
Similarly, note that we can also rewrite the currents equation (\ref{crt})
as:%
\begin{equation*}
\hat{J}^{\left( i\right) }\left( \left( n-1\right) \right) =J^{\left(
i\right) }\left( \left( n-1\right) \right) +\frac{\kappa }{N}\sum_{j,m}\frac{%
\omega _{j}\left( m\right) }{\omega _{i}\left( n-1\right) }\delta \left(
\theta ^{\left( i\right) }\left( n-1\right) -\theta ^{\left( j\right)
}\left( m\right) -\frac{\left\vert Z_{i}-Z_{j}\right\vert }{c}\right)
T_{ij}\left( \left( n-1,Z_{i}\right) ,\left( m,Z_{j}\right) \right)
\end{equation*}%
with:%
\begin{equation}
T_{ij}\left( \left( n_{i},Z_{i}\right) ,\left( m_{j},Z_{j}\right) \right)
=\sum_{kl}T_{kl}\left( n_{kl}\right) \delta \left( \theta ^{\left( i\right)
}\left( n_{i}\right) -\theta ^{\left( kl\right) }\left( n_{kl}\right)
\right) \delta \left( \omega _{k}\left( n_{kl}\right) -\bar{\omega}\left(
Z_{i},n_{i}\right) \right) \delta \left( \omega _{l}\left( n_{kl}\right) -%
\bar{\omega}\left( Z_{j},n_{j}\right) \right)  \label{tkl}
\end{equation}

\subsubsection*{A1.3.3 Probability density for the system}

\paragraph*{A1.3.3.1 Individual neurons}

Due to the stochastic nature of equations (\ref{dnq}) and (\ref{cstrt}), the
dynamics of a single neuron can be described by the probability density $%
P\left( \theta ^{\left( i\right) }\left( t\right) ,\omega _{i}^{-1}\left(
t\right) \right) $ for a path $\left( \theta ^{\left( i\right) }\left(
t\right) ,\omega _{i}^{-1}\left( t\right) \right) $ which is given by, up to
a normalization factor:

\begin{equation}
P\left( \theta ^{\left( i\right) }\left( t\right) ,\omega _{i}^{-1}\left(
t\right) \right) =\exp \left( -A_{i}\right)  \label{dnmcs}
\end{equation}%
where:%
\begin{equation}
A_{i}=\frac{1}{\sigma ^{2}}\int \left( \frac{d}{dt}\theta ^{\left( i\right)
}\left( t\right) -\omega _{i}^{-1}\left( t\right) \right) ^{2}dt+\int \frac{%
\left( \omega _{i}^{-1}\left( t\right) -G\left( \theta ^{\left( i\right)
}\left( t\right) ,\hat{J}\left( \theta ^{\left( i\right) }\left( t\right)
\right) \right) \right) ^{2}}{\eta ^{2}}dt  \label{dnmcszz}
\end{equation}%
(see \cite{GL1} and \cite{GL2}). The integral is taken over a time period
that depends on the time scale of the interactions. Actually, the
minimization of (\ref{dnmcszz})\ imposes both (\ref{dnm}) and (\ref{cstrt}),
so that the probability density is, as expected, centered around these two
conditions, i.e. (\ref{dnm}) and (\ref{cstrt}) are satisfied in mean. A
probability density for the whole system of neurons is obtained by summing $%
S_{i}$ over all agents. We thus define the statistical weight for the cells:%
\begin{equation}
P\left( \left( \theta ^{\left( i\right) }\left( t\right) ,\omega
_{i}^{-1}\left( t\right) ,Z_{i}\right) _{i=1...N}\right) =\exp \left(
-A\right)  \label{Prdn}
\end{equation}%
with:%
\begin{equation}
A=\sum_{i}A_{i}=\sum_{i}\frac{1}{\sigma ^{2}}\int \left( \frac{d}{dt}\theta
^{\left( i\right) }\left( t\right) -\omega _{i}^{-1}\left( t\right) \right)
^{2}dt+\int \frac{\left( \omega _{i}^{-1}\left( t\right) -G\left( \theta
^{\left( i\right) }\left( t\right) ,\hat{J}\left( \theta ^{\left( i\right)
}\left( t\right) \right) \right) \right) ^{2}}{\eta ^{2}}dt  \label{ctnt}
\end{equation}%
and (using (\ref{tkl})):%
\begin{eqnarray*}
\hat{J}^{\left( i\right) }\left( \left( n-1\right) \right) &=&J^{\left(
i\right) }\left( \left( n-1\right) \right) +\frac{\kappa }{N}\sum_{j,m}\frac{%
\omega _{j}\left( m\right) }{\omega _{i}\left( n-1\right) }\delta \left(
\theta ^{\left( i\right) }\left( n-1\right) -\theta ^{\left( j\right)
}\left( m\right) -\frac{\left\vert Z_{i}-Z_{j}\right\vert }{c}\right)
T_{ij}\left( \left( n-1,Z_{i}\right) ,\left( m,Z_{j}\right) \right) \\
&&\times \sum_{kl}T_{kl}\left( n_{kl}\right) \delta \left( \theta ^{\left(
i\right) }\left( n_{i}\right) -\theta ^{\left( kl\right) }\left(
n_{kl}\right) \right) \delta \left( \omega _{k}\left( n_{kl}\right) -\omega
_{i}\left( n_{i}\right) \right) \delta \left( \omega _{l}\left(
n_{kl}\right) -\omega _{j}\left( m\right) \right)
\end{eqnarray*}

\paragraph*{A1.3.3.1 Connectivity functions}

The statistical exponent associated to the connectivity functions is
obtained as in the previous paragraph. We obtain the statistical weight:%
\begin{equation*}
\prod\limits_{k,l}P\left( T_{kl}\left( n_{kl}\right) ,\hat{T}_{kl}\left(
n_{kl}\right) ,\left( Z_{kl}\left( n_{kl}\right) =\left( Z_{k,},Z_{l}\right)
\right) ,\theta ^{\left( kl\right) }\left( n_{kl}\right) ,\omega _{k}\left(
n_{kl}\right) ,\omega _{l}^{\prime }\left( n_{kl}\right) ,C_{kl}\left(
n_{kl}\right) ,D_{k}\left( n_{kl}\right) \right) =\exp \left( -B\right)
\end{equation*}%
where:%
\begin{eqnarray}
B &=&\sum_{kl}\left( \nabla _{n_{kl}}T_{kl}\left( n_{kl}\right)
-B_{kl}^{\left( 1\right) }\right) ^{2}+\left( \nabla _{n_{kl}}\hat{T}\left(
n_{kl}\right) -B_{kl}^{\left( 2\right) }\nabla _{n_{kl}}\right) ^{2}
\label{wgp} \\
&&+\left( C\left( n_{kl}\right) -B_{kl}^{\left( 3\right) }\right)
^{2}+\left( \nabla _{n_{kl}}D_{k}\left( n_{kl}\right) -B_{k}\right) ^{2} 
\notag
\end{eqnarray}%
and:%
\begin{eqnarray}
&&B_{kl}^{\left( 1\right) }=\left( -\sum_{i,n_{i}}\frac{1}{\tau \bar{\omega}%
\left( Z_{i},n_{i}\right) }T_{kl}\left( n_{kl}\right) +\frac{\lambda }{\bar{%
\omega}\left( Z_{i},n_{i}\right) }\hat{T}_{kl}\left( n_{kl}\right) \right)
\label{wgd} \\
&&\times \delta \left( \theta ^{\left( i\right) }\left( n_{i}\right) -\theta
^{\left( kl\right) }\left( n_{kl}\right) \right) \delta \left(
Z_{k}-Z_{i}\right) \delta \left( \omega _{k}\left( n_{kl}\right) -\bar{\omega%
}\left( Z_{i},n_{i}\right) \right)  \notag
\end{eqnarray}%
\begin{eqnarray}
&&B_{kl}^{\left( 2\right) }=\left( \sum_{i,n_{i}}\left( h\left(
Z_{k},Z_{l}\right) -\hat{T}\left( n_{kl}\right) \right) C_{kl}\left(
n_{kl}\right) h_{C}\left( \omega _{i}\left( n_{i}\right) \right)
-\sum_{j,n_{j}}D_{k}\left( n_{kl}\right) \hat{T}\left( n_{kl}\right)
h_{D}\left( \omega _{j}\left( n_{j}\right) \right) \right)  \label{wgt} \\
&&\times \frac{\rho }{\bar{\omega}\left( Z_{i},n_{i}\right) }\delta \left(
\theta ^{\left( i\right) }\left( n_{i}\right) -\theta ^{\left( j\right)
}\left( n_{j}\right) -\frac{\left\vert Z_{i}-Z_{j}\right\vert }{c}\right)
\delta \left( \theta ^{\left( i\right) }\left( n_{i}\right) -\theta ^{\left(
kl\right) }\left( n_{kl}\right) \right)  \notag \\
&&\times \delta \left( \left( Z_{k,},Z_{l}\right) -\left(
Z_{i,},Z_{j}\right) \right) \delta \left( \omega _{k}\left( n_{kl}\right) -%
\bar{\omega}\left( Z_{i},n_{i}\right) \right) \delta \left( \omega
_{l}\left( n_{kl}\right) -\bar{\omega}\left( Z_{j},n_{j}\right) \right) 
\notag
\end{eqnarray}%
\begin{eqnarray}
B_{kl}^{\left( 3\right) } &=&\left( -\frac{C\left( n_{kl}\right) }{\tau _{C}%
\bar{\omega}\left( Z_{i},n_{i}\right) }+\sum_{j,n_{j}}\alpha _{C}\left(
1-C_{kl}\left( n_{kl}\right) \right) \frac{\omega _{j}\left( n_{j}\right) }{%
\bar{\omega}\left( Z_{i},n_{i}\right) }\right)  \label{wgq} \\
&&\times \delta \left( \theta ^{\left( i\right) }\left( n_{i}\right) -\theta
^{\left( j\right) }\left( n_{j}\right) -\frac{\left\vert
Z_{i}-Z_{j}\right\vert }{c}\right) \delta \left( \theta ^{\left( i\right)
}\left( n_{i}\right) -\theta ^{\left( kl\right) }\left( n_{kl}\right)
\right) \delta \left( \left( Z_{k,},Z_{l}\right) -\left( Z_{i,},Z_{j}\right)
\right)  \notag \\
&&\times \delta \left( \omega _{k}\left( n_{kl}\right) -\bar{\omega}\left(
Z_{i},n_{i}\right) \right) \delta \left( \omega _{l}\left( n_{kl}\right) -%
\bar{\omega}\left( Z_{j},n_{j}\right) \right)  \notag
\end{eqnarray}%
\begin{eqnarray}
B_{k} &=&\left( -\frac{D_{k}\left( n_{kl}\right) }{\tau _{D}\bar{\omega}%
\left( Z_{i},n_{i}\right) }+\frac{1}{\bar{\omega}\left( Z_{i},n_{i}\right) }%
\sum_{i,n_{i}}\alpha _{D}\left( 1-D_{k}\left( n_{kl}\right) \right) \omega
_{i}\left( n_{i}\right) \right)  \label{wgc} \\
&&\times \delta \left( \theta ^{\left( i\right) }\left( n_{i}\right) -\theta
^{\left( kl\right) }\left( n_{kl}\right) \right) \delta \left(
Z_{k}-Z_{i}\right) \delta \left( \omega _{k}\left( n_{kl}\right) -\bar{\omega%
}\left( Z_{i},n_{i}\right) \right) \delta \left( \omega _{l}\left(
n_{kl}\right) -\bar{\omega}\left( Z_{j},n_{j}\right) \right)  \notag
\end{eqnarray}

\paragraph*{A1.3.3.3 Probability density for the full system}

The probability for the full system is obtained by the product:%
\begin{eqnarray}
&&\prod\limits_{k,l}P\left( T_{kl}\left( n_{kl}\right) ,\hat{T}_{kl}\left(
n_{kl}\right) ,\left( Z_{kl}\left( n_{kl}\right) =\left( Z_{k,},Z_{l}\right)
\right) ,\theta ^{\left( kl\right) }\left( n_{kl}\right) ,\omega _{k}\left(
n_{kl}\right) ,\omega _{l}^{\prime }\left( n_{kl}\right) ,C_{kl}\left(
n_{kl}\right) ,D_{k}\left( n_{kl}\right) \right)  \label{SSW} \\
&&\times P\left( \left( \theta ^{\left( i\right) }\left( t\right) ,\omega
_{i}^{-1}\left( t\right) ,Z_{i}\right) _{i=1...N}\right)  \notag \\
&=&\exp \left( -B\right) \exp \left( -A\right)  \notag
\end{eqnarray}

\subsection*{A1.4 Field theoretic description of the system}

\subsubsection*{A1.4.1 Translation of formula (\protect\ref{SSW}) in terms
of field theory}

In our context two fields are necessary. The field representing the set of
neurons depends on the three variables $\left( \theta ,Z,\omega \right) $,\
and is denoted $\Psi \left( \theta ,Z,\omega \right) $. The connectivity
functions are characterized by the set of variables $\left( T,\hat{T},\omega
,\omega ^{\prime },\theta ,Z,Z^{\prime },C,D\right) $ and represented by the
field $\Gamma \left( T,\hat{T},\omega ,\omega ^{\prime },\theta ,Z,Z^{\prime
},C,D\right) $.We provide an interpretation of the various fields at the end
of this paragraph.

\paragraph*{A1.4.1.1 Translation of (\protect\ref{ctnt})}

The dynamics of neurons is described by an action functional for the field $%
\Psi \left( \theta ,Z,\omega \right) $ and its associated partition
function. This partition function captures both collective and individual
aspects of the system, enabling the retrieval of correlation functions for
number of neurons.

The field theoretic version of (\ref{dnmcszz}) is obtained using (\ref{ctnt}%
). The correspondence detailed in \cite{GL1}\cite{GL2}) yields an action $%
S\left( \Psi \right) $ for a field $\Psi \left( \theta ,Z,\omega \right) $
and a statistical weight $\exp \left( -\left( S\left( \Psi \right) \right)
\right) $ for each configuration $\Psi \left( \theta ,Z,\omega \right) $ of
this field. The functional $S\left( \Psi \right) $ is decomposed in two
parts corresponding to the two contributions in (\ref{ctnt}).

The first term of (\ref{ctnt}):%
\begin{equation}
\frac{1}{\sigma ^{2}}\int \left( \frac{d}{dt}\theta ^{\left( i\right)
}\left( t\right) -\omega _{i}^{-1}\left( t\right) \right) ^{2}dt  \label{fsT}
\end{equation}%
is a term with temporal derivative. Its form is simple since the function $%
f^{\left( \alpha \right) }$ in (\ref{inco}) depends only on the variable $%
\mathbf{X}_{i}\left( t\right) =\left( \theta ^{\left( i\right) }\left(
t\right) ,\omega _{i}^{-1}\left( t\right) ,Z_{i}\right) $. Actually $%
f^{\left( \theta \right) }\left( \mathbf{X}_{i}\left( t\right) \right)
=\omega _{i}^{-1}\left( t\right) $. Using (\ref{Trll}), the term (\ref{fsT})
is thus replaced by the corresponding quadratic functional in field theory:%
\begin{equation}
-\frac{1}{2}\Psi ^{\dagger }\left( \theta ,Z,\omega \right) \nabla \left( 
\frac{\sigma ^{2}}{2}\nabla -\omega ^{-1}\right) \Psi \left( \theta
,Z,\omega \right)  \label{thtdnmcs}
\end{equation}%
where $\sigma ^{2}$ is the variance of the errors $\varepsilon _{i}$.

The field functional that corresponds to the second term of (\ref{dnmcszz}):%
\begin{equation*}
V=\int \frac{\left( \omega _{i}^{-1}\left( t\right) -G\left( \theta ^{\left(
i\right) }\left( t\right) ,\hat{J}\left( \theta ^{\left( i\right) }\left(
t\right) \right) \right) \right) ^{2}}{\eta ^{2}}dt
\end{equation*}
is obtained by expanding the formula (\ref{crT}) for the current induced by
all $j$:

\begin{eqnarray}
V &=&\frac{1}{2\eta ^{2}}\int dt\sum_{i}\left( \omega _{i}^{-1}\left(
t\right) \right.  \label{Cpnt} \\
&&-\left. G\left( J\left( \theta ^{\left( i\right) }\left( t\right)
,Z_{i}\right) +\frac{\kappa }{N}\int ds\sum_{j}\frac{\omega _{j}\left(
s\right) T_{ij}\left( \left( t,Z_{i}\right) ,s,Z_{j}\right) }{\omega
_{i}\left( t\right) }\delta \left( \theta ^{\left( i\right) }\left( t\right)
-\theta ^{\left( j\right) }\left( s\right) -\frac{\left\vert
Z_{i}-Z_{j}\right\vert }{c}\right) \right) \right) ^{2}  \notag
\end{eqnarray}%
with $\eta <<1$, which is the constraint (\ref{cstrt}) imposed
stochastically. Its corresponding potential in field theory is obtained
straightforwardly by using the translation (\ref{tln}):%
\begin{equation}
\frac{1}{2\eta ^{2}}\int \left\vert \Psi \left( \theta ,Z,\omega \right)
\right\vert ^{2}\left( \omega ^{-1}-G\left( J\left( \theta ,Z\right) +\int 
\frac{\kappa }{N}\frac{\omega _{1}T\left( Z,\theta ,Z_{1},\theta -\frac{%
\left\vert Z-Z_{1}\right\vert }{c}\right) }{\omega }\left\vert \Psi \left(
\theta -\frac{\left\vert Z-Z_{1}\right\vert }{c},Z_{1},\omega _{1}\right)
\right\vert ^{2}dZ_{1}d\omega _{1}\right) \right) ^{2}  \label{ptntl}
\end{equation}%
and $T\left( Z,\theta ,Z_{1},\theta -\frac{\left\vert Z-Z_{1}\right\vert }{c}%
\right) $ is obtained by the translation of the term (\ref{tkl}):%
\begin{eqnarray*}
&&\sum_{kl}T_{kl}\left( n_{kl}\right) \delta \left( \theta ^{\left( i\right)
}\left( n_{i}\right) -\theta ^{\left( kl\right) }\left( n_{kl}\right)
\right) \delta \left( \omega _{k}\left( n_{kl}\right) -\omega _{i}\left(
n_{i}\right) \right) \delta \left( \omega _{l}\left( n_{kl}\right) -\omega
_{j}\left( m\right) \right) \\
&\rightarrow &\int T\left\vert \Gamma \left( T,\hat{T},\hat{\omega},\hat{%
\omega}^{\prime },\hat{\theta},\hat{Z},\hat{Z}^{\prime },C,D\right)
\right\vert ^{2}\delta \left( \theta -\hat{\theta}\right) \delta \left( \hat{%
\omega}-\omega \right) \delta \left( \hat{\omega}-\omega _{1}\right) \delta
\left( \left( \hat{Z},\hat{Z}^{\prime }\right) -\left( Z,Z_{1}\right) \right)
\\
&=&\int T\left\vert \Gamma \left( T,\hat{T},\omega ,\omega _{1},\theta
,Z,Z_{1},C,D\right) \right\vert ^{2}dTd\hat{T}dCdD\equiv T\left( Z,\theta
,Z_{1},\theta -\frac{\left\vert Z-Z_{1}\right\vert }{c}\right)
\end{eqnarray*}%
To simplify, we will write in the sequel:%
\begin{equation}
T\left( Z,\theta ,Z_{1},\theta -\frac{\left\vert Z-Z_{1}\right\vert }{c}%
\right) =\int T\left\vert \Gamma \left( T,\hat{T},\omega ,\omega _{1},\theta
,Z,Z_{1},C,D\right) \right\vert ^{2}dTd\hat{T}dCdD\equiv T\left( Z,\theta
,Z_{1}\right)  \label{tpr}
\end{equation}%
which represents the average connectivity between points $Z$ and $Z_{1}$ in
state $\Gamma \left( T,\hat{T},\omega ,\omega _{1},\theta
,Z,Z_{1},C,D\right) $.

The field action is then the sum of (\ref{thtdnmcs}) and (\ref{ptntl}):%
\begin{eqnarray}
S &=&-\frac{1}{2}\Psi ^{\dagger }\left( \theta ,Z,\omega \right) \nabla
\left( \frac{\sigma _{\theta }^{2}}{2}\nabla -\omega ^{-1}\right) \Psi
\left( \theta ,Z,\omega \right)  \label{lfS} \\
&&+\frac{1}{2\eta ^{2}}\int \left\vert \Psi \left( \theta ,Z,\omega \right)
\right\vert ^{2}\left( \omega ^{-1}-G\left( J\left( \theta ,Z\right) +\int 
\frac{\kappa }{N}\frac{\omega _{1}}{\omega }\left\vert \Psi \left( \theta -%
\frac{\left\vert Z-Z_{1}\right\vert }{c},Z_{1},\omega _{1}\right)
\right\vert ^{2}T\left( Z,\theta ,Z_{1}\right) dZ_{1}d\omega _{1}\right)
\right) ^{2}  \notag
\end{eqnarray}%
$\allowbreak $

\paragraph*{A1.4.1.1 Translation for connectivity dynamics (\protect\ref{wgp}%
)}

The translation of the four action terms describing the connectivity
dynamics (\ref{wgd}), (\ref{wgt}), (\ref{wgq}) and (\ref{wgc}) in (\ref{wgp}%
) is straightforward. We obtain four contributions:%
\begin{equation}
S_{\Gamma }^{\left( 1\right) }=\int \Gamma ^{\dag }\left( T,\hat{T},\omega
_{\Gamma },\omega _{\Gamma }^{\prime },\theta ,Z,Z^{\prime },C,D\right)
\nabla _{T}\left( \frac{\sigma _{T}^{2}}{2}\nabla _{T}+O_{T}^{\omega
}\right) \Gamma \left( T,\hat{T},\omega _{\Gamma },\omega _{\Gamma }^{\prime
},\theta ,Z,Z^{\prime },C,D\right)  \label{wgD}
\end{equation}%
\begin{equation}
S_{\Gamma }^{\left( 2\right) }=\int \Gamma ^{\dag }\left( T,\hat{T},\omega
_{\Gamma },\omega _{\Gamma }^{\prime },\theta ,Z,Z^{\prime },C,D\right)
\nabla _{\hat{T}}\left( \frac{\sigma _{\hat{T}}^{2}}{2}\nabla _{\hat{T}}+O_{%
\hat{T}}^{\omega }\right) \Gamma \left( T,\hat{T},\omega _{\Gamma },\omega
_{\Gamma }^{\prime },\theta ,Z,Z^{\prime },C,D\right)  \label{wgT}
\end{equation}%
\begin{equation}
S_{\Gamma }^{\left( 3\right) }=\int \Gamma ^{\dag }\left( T,\hat{T},\omega
_{\Gamma },\omega _{\Gamma }^{\prime },\theta ,Z,Z^{\prime },C,D\right)
\nabla _{C}\left( \frac{\sigma _{C}^{2}}{2}\nabla _{C}+O_{C}^{\omega
}\right) \Gamma \left( T,\hat{T},\omega _{\Gamma },\omega _{\Gamma }^{\prime
},\theta ,Z,Z^{\prime },C,D\right)  \label{wgQ}
\end{equation}%
\begin{equation}
S_{\Gamma }^{\left( 4\right) }=\int \Gamma ^{\dag }\left( T,\hat{T},\omega
_{\Gamma },\omega _{\Gamma }^{\prime },\theta ,Z,Z^{\prime },C,D\right)
\nabla _{D}\left( \frac{\sigma _{D}^{2}}{2}\nabla _{D}+O_{D}^{\omega
}\right) \Gamma \left( T,\hat{T},\omega _{\Gamma },\omega _{\Gamma }^{\prime
},\theta ,Z,Z^{\prime },C,D\right)  \label{wgC}
\end{equation}%
with:%
\begin{eqnarray}
O_{C}^{\omega } &=&\left( \frac{C}{\tau _{C}\bar{\omega}}-\frac{\alpha
_{C}\left( 1-C\right) \int \omega ^{\prime }\left\vert \Psi \left( \theta -%
\frac{\left\vert Z-Z^{\prime }\right\vert }{c},Z^{\prime },\omega ^{\prime
}\right) \right\vert ^{2}d\omega ^{\prime }}{\bar{\omega}}\right) \delta
\left( \left( \omega _{\Gamma },\omega _{\Gamma }^{\prime }\right) -\left( 
\bar{\omega},\bar{\omega}^{\prime }\right) \right)  \label{Gm} \\
O_{D}^{\omega } &=&\frac{D}{\tau _{D}\bar{\omega}}-\frac{\alpha _{D}\left(
1-D\right) \int \omega \left\vert \Psi \left( \theta ,Z,\omega \right)
\right\vert ^{2}d\omega }{\bar{\omega}}\delta \left( \left( \omega _{\Gamma
},\omega _{\Gamma }^{\prime }\right) -\left( \bar{\omega},\bar{\omega}%
^{\prime }\right) \right)  \notag \\
O_{\hat{T}}^{\omega } &=&-\frac{\rho }{\bar{\omega}}\left( \left( h\left(
Z,Z^{\prime }\right) -\hat{T}\right) C\int \left\vert \Psi \left( \theta
,Z,\omega \right) \right\vert ^{2}h_{C}\left( \omega \right) d\omega \right.
\notag \\
&&\left. -D\hat{T}\int \left\vert \Psi \left( \theta -\frac{\left\vert
Z-Z^{\prime }\right\vert }{c},Z^{\prime },\omega ^{\prime }\right)
\right\vert ^{2}h_{D}\left( \omega ^{\prime }\right) d\omega ^{\prime
}\right) \delta \left( \left( \omega _{\Gamma },\omega _{\Gamma }^{\prime
}\right) -\left( \bar{\omega},\bar{\omega}^{\prime }\right) \right)  \notag
\\
O_{T}^{\omega } &=&-\left( -\frac{1}{\tau \bar{\omega}}T+\frac{\lambda }{%
\bar{\omega}}\hat{T}\right) \delta \left( \left( \omega _{\Gamma },\omega
_{\Gamma }^{\prime }\right) -\left( \bar{\omega},\bar{\omega}^{\prime
}\right) \right)  \notag
\end{eqnarray}%
Here:%
\begin{eqnarray*}
\bar{\omega} &=&\frac{\int \omega \left\vert \Psi \left( \theta ,Z,\omega
\right) \right\vert ^{2}d\omega }{\int \left\vert \Psi \left( \theta
,Z,\omega \right) \right\vert ^{2}d\omega } \\
\bar{\omega}^{\prime } &=&\frac{\int \omega ^{\prime }\left\vert \Psi \left(
\theta -\frac{\left\vert Z-Z^{\prime }\right\vert }{c},Z^{\prime },\omega
^{\prime }\right) \right\vert ^{2}d\omega ^{\prime }}{\int \left\vert \Psi
\left( \theta -\frac{\left\vert Z-Z^{\prime }\right\vert }{c},Z^{\prime
},\omega ^{\prime }\right) \right\vert ^{2}d\omega ^{\prime }}
\end{eqnarray*}

\subsubsection*{A1.4.2 Full action for the system}

The full action for the system is obtained by gathering the different terms:%
\begin{equation}
-\frac{1}{2}\int \Psi ^{\dagger }\left( \theta ,Z,\omega \right) \nabla
\left( \frac{\sigma _{\theta }^{2}}{2}\nabla -\omega ^{-1}\right) \Psi
\left( \theta ,Z,\omega \right) +\frac{1}{2\eta ^{2}}\left( S_{\Gamma
}^{\left( 1\right) }+S_{\Gamma }^{\left( 2\right) }+S_{\Gamma }^{\left(
3\right) }+S_{\Gamma }^{\left( 4\right) }\right)  \label{FCN}
\end{equation}%
with $S_{\Gamma }^{\left( 1\right) }$, $S_{\Gamma }^{\left( 2\right) }$, $%
S_{\Gamma }^{\left( 3\right) }$, $S_{\Gamma }^{\left( 4\right) }$ given by (%
\ref{wgD}), (\ref{wgT}), (\ref{wgQ}), (\ref{wgC}).

\subsubsection*{A1.4.3 Remark: interpretation of the various field}

The action functional depends on two fields: $\Psi \left( \theta ,Z,\omega
\right) $ and $\Gamma \left( T,\hat{T},\omega _{\Gamma },\omega _{\Gamma
}^{\prime },\theta ,Z,Z^{\prime },C,D\right) $. These two abstract
quantities will enable us to derive the dynamic state of the entire system
and subsequently study transitions between different states. However, the
squared modulus of the two functions can be interpreted in terms of
statistical distribution, depending on the chosen description. If we
consider a system of simple cells spread along the thread, the function $%
\left\vert \Psi \left( \theta ,Z,\omega \right) \right\vert ^{2}$ measures
at time $\theta $, the density of active cells at point $Z$ with activity $%
\omega $. In the perspective of complex cells with multiple axons and
dendrites, we can consider that one cell stands at $Z$, and $\left\vert \Psi
\left( \theta ,Z,\omega \right) \right\vert ^{2}$ measures for that cell the
density of axons with activity $\omega $. A similar interpretation works for 
$\Gamma \left( T,\hat{T},\omega _{\Gamma },\omega _{\Gamma }^{\prime
},\theta ,Z,Z^{\prime },C,D\right) $. In the perspective of system of simple
cells "accumulated" in the neighborhood of $Z$, $\left\vert \Gamma \left( T,%
\hat{T},\omega _{\Gamma },\omega _{\Gamma }^{\prime },\theta ,Z,Z^{\prime
},C,D\right) \right\vert ^{2}$ measures the density of connections of value $%
T$ \ (and auxiliary variabls $\hat{T}$, $C,D$) between the set of cells
located at points $Z$ and $Z^{\prime }$ with activity $\omega _{\Gamma }$
and $\omega _{\Gamma }^{\prime }$. In the context of complex cells, it
describes the density of connections with strength $T$ between sets of axons
nd dendrites of cells with activity $\omega _{\Gamma },\omega _{\Gamma
}^{\prime }$.

\subsubsection*{A1.4.4 Projection on dependent activity states and effective
action}

We have shown in (\cite{GL}) that some simplifications arise in the action
functional. Using the fact that $\eta ^{2}<<1$, and noting that in this
case, field configurations $\Psi \left( \theta ,Z,\omega \right) $ such that:%
\begin{equation*}
\omega ^{-1}-G\left( J\left( \theta ,Z\right) +\int \frac{\kappa }{N}\frac{%
\omega _{1}}{\omega }\left\vert \Psi \left( \theta -\frac{\left\vert
Z-Z_{1}\right\vert }{c},Z_{1},\omega _{1}\right) \right\vert ^{2}T\left(
Z,\theta ,Z_{1}\right) dZ_{1}d\omega _{1}\right) \neq 0
\end{equation*}%
have negligible statistical weight, we can simplify (\ref{lfS}) and restrict
the fields to those of the form: 
\begin{equation}
\Psi \left( \theta ,Z\right) \delta \left( \omega ^{-1}-\omega ^{-1}\left(
J,\theta ,Z,\left\vert \Psi \right\vert ^{2}\right) \right)  \label{prt}
\end{equation}%
where $\omega ^{-1}\left( J,\theta ,Z,\Psi \right) $ satisfies:%
\begin{eqnarray*}
\omega ^{-1}\left( J,\theta ,Z,\left\vert \Psi \right\vert ^{2}\right)
&=&G\left( J\left( \theta ,Z\right) +\int \frac{\kappa }{N}\frac{\omega
_{1}T\left( Z,\theta ,Z_{1},\theta -\frac{\left\vert Z-Z_{1}\right\vert }{c}%
\right) }{\omega \left( J,\theta ,Z,\left\vert \Psi \right\vert ^{2}\right) }%
\left\vert \Psi \left( \theta -\frac{\left\vert Z-Z_{1}\right\vert }{c}%
,Z_{1},\omega _{1}\right) \right\vert ^{2}dZ_{1}d\omega _{1}\right) \\
&=&G\left( J\left( \theta ,Z\right) +\int \frac{\kappa }{N}\frac{\omega
_{1}T\left( Z,\theta ,Z_{1},\theta -\frac{\left\vert Z-Z_{1}\right\vert }{c}%
\right) }{\omega \left( J,\theta ,Z,\left\vert \Psi \right\vert ^{2}\right) }%
\left\vert \Psi \left( \theta -\frac{\left\vert Z-Z_{1}\right\vert }{c}%
,Z_{1}\right) \right\vert ^{2}\right. \\
&&\times \left. \delta \left( \omega _{1}^{-1}-\omega ^{-1}\left( J,\theta -%
\frac{\left\vert Z-Z_{1}\right\vert }{c},Z_{1},\left\vert \Psi \right\vert
^{2}\right) \right) dZ_{1}d\omega _{1}\right)
\end{eqnarray*}%
The last equation simplifies to yield:%
\begin{eqnarray}
&&\omega ^{-1}\left( J,\theta ,Z,\left\vert \Psi \right\vert ^{2}\right)
\label{qf} \\
&=&G\left( J\left( \theta ,Z\right) +\int \frac{\kappa }{N}\frac{\omega
\left( J,\theta -\frac{\left\vert Z-Z_{1}\right\vert }{c},Z_{1},\Psi \right)
T\left( Z,\theta ,Z_{1},\theta -\frac{\left\vert Z-Z_{1}\right\vert }{c}%
\right) }{\omega \left( J,\theta ,Z,\left\vert \Psi \right\vert ^{2}\right) }%
\left\vert \Psi \left( \theta -\frac{\left\vert Z-Z_{1}\right\vert }{c}%
,Z_{1}\right) \right\vert ^{2}dZ_{1}\right)  \notag
\end{eqnarray}%
The configurations $\Psi \left( \theta ,Z,\omega \right) $ that minimize the
potential (\ref{ptntl}) can now be considered: the field $\Psi \left( \theta
,Z,\omega \right) $ is projected on the subspace (\ref{prt}) of functions of
two variables, and we can therefore replace in (\ref{ptntl}):%
\begin{equation}
\omega \rightarrow \omega \left( J,\theta ,Z,\left\vert \Psi \right\vert
^{2}\right)  \label{DF}
\end{equation}%
\begin{equation}
\omega ^{\prime }\rightarrow \omega \left( J,\theta -\frac{\left\vert
Z-Z^{\prime }\right\vert }{c},Z^{\prime },\left\vert \Psi \right\vert
^{2}\right)  \label{DH}
\end{equation}%
The "classical" effective action becomes (see appendix 1):%
\begin{equation}
-\frac{1}{2}\Psi ^{\dagger }\left( \theta ,Z\right) \left( \nabla _{\theta
}\left( \frac{\sigma ^{2}}{2}\nabla _{\theta }-\omega ^{-1}\left( J,\theta
,Z,\left\vert \Psi \right\vert ^{2}\right) \right) \right) \Psi \left(
\theta ,Z\right)  \label{nmR}
\end{equation}%
with $\omega ^{-1}\left( J,\theta ,Z,\left\vert \Psi \right\vert ^{2}\right) 
$ given by equation (\ref{qf}). As in (\cite{GL}) we add to this action a
stabilization potential $V\left( \Psi \right) $ ensuring an average activity
of the system. The precise form of this potential is irrelevant here, but we
assume that it has a minimum $\Psi _{0}\left( \theta ,Z\right) $.

The projection on dependent activity also applies to connectivity action
terms. We can thus replace $\Gamma \left( T,\hat{T},\omega _{\Gamma },\omega
_{\Gamma }^{\prime },\theta ,Z,Z^{\prime },C,D\right) $ by $\Gamma \left( T,%
\hat{T},\theta ,Z,Z^{\prime },C,D\right) $ and the action becomes:%
\begin{eqnarray}
S_{full} &=&-\frac{1}{2}\Psi ^{\dagger }\left( \theta ,Z,\omega \right)
\nabla \left( \frac{\sigma _{\theta }^{2}}{2}\nabla -\omega ^{-1}\left(
J,\theta ,Z,\left\vert \Psi \right\vert ^{2}\right) \right) \Psi \left(
\theta ,Z\right) +V\left( \Psi \right)  \label{flt} \\
&&+\frac{1}{2\eta ^{2}}\left( S_{\Gamma }^{\left( 0\right) }+S_{\Gamma
}^{\left( 1\right) }+S_{\Gamma }^{\left( 2\right) }+S_{\Gamma }^{\left(
3\right) }+S_{\Gamma }^{\left( 4\right) }\right) +U\left( \left\{ \left\vert
\Gamma \left( \theta ,Z,Z^{\prime },C,D\right) \right\vert ^{2}\right\}
\right)  \notag
\end{eqnarray}%
with $S_{\Gamma }^{\left( 1\right) }$, $S_{\Gamma }^{\left( 2\right) }$, $%
S_{\Gamma }^{\left( 3\right) }$, $S_{\Gamma }^{\left( 4\right) }$ now given
by: 
\begin{equation}
S_{\Gamma }^{\left( 1\right) }=\int \Gamma ^{\dag }\left( T,\hat{T},\theta
,Z,Z^{\prime },C,D\right) \nabla _{T}\left( \frac{\sigma _{T}^{2}}{2}\nabla
_{T}+O_{T}\right) \Gamma \left( T,\hat{T},\theta ,Z,Z^{\prime },C,D\right)
\label{wGD}
\end{equation}%
\begin{equation}
S_{\Gamma }^{\left( 2\right) }=\int \Gamma ^{\dag }\left( T,\hat{T},\theta
,Z,Z^{\prime },C,D\right) \nabla _{\hat{T}}\left( \frac{\sigma _{\hat{T}}^{2}%
}{2}\nabla _{\hat{T}}+O_{\hat{T}}\right) \Gamma \left( T,\hat{T},\theta
,Z,Z^{\prime },C,D\right)  \label{wGT}
\end{equation}%
\begin{equation}
S_{\Gamma }^{\left( 3\right) }=\Gamma ^{\dag }\left( T,\hat{T},\theta
,Z,Z^{\prime },C,D\right) \nabla _{C}\left( \frac{\sigma _{C}^{2}}{2}\nabla
_{C}+O_{C}\right) \Gamma \left( T,\hat{T},\theta ,Z,Z^{\prime },C,D\right)
\label{wGQ}
\end{equation}%
\begin{equation}
S_{\Gamma }^{\left( 4\right) }=\Gamma ^{\dag }\left( T,\hat{T},\theta
,Z,Z^{\prime },C,D\right) \nabla _{D}\left( \frac{\sigma _{D}^{2}}{2}\nabla
_{D}+O_{D}\right) \Gamma \left( T,\hat{T},\theta ,Z,Z^{\prime },C,D\right)
\label{wGC}
\end{equation}%
where:%
\begin{eqnarray}
O_{C} &=&\frac{C}{\tau _{C}\omega \left( J,\theta ,Z,\left\vert \Psi
\right\vert ^{2}\right) }-\frac{\alpha _{C}\left( 1-C\right) \omega \left(
J,\theta -\frac{\left\vert Z-Z^{\prime }\right\vert }{c},Z^{\prime
},\left\vert \Psi \right\vert ^{2}\right) \left\vert \Psi \left( \theta -%
\frac{\left\vert Z-Z^{\prime }\right\vert }{c},Z^{\prime }\right)
\right\vert ^{2}}{\omega \left( J,\theta ,Z,\left\vert \Psi \right\vert
^{2}\right) }  \label{DP} \\
O_{D} &=&\frac{D}{\tau _{D}\omega \left( J,\theta ,Z,\left\vert \Psi
\right\vert ^{2}\right) }-\alpha _{D}\left( 1-D\right) \left\vert \Psi
\left( \theta ,Z\right) \right\vert ^{2}  \notag \\
O_{\hat{T}} &=&-\frac{\rho }{\omega \left( J,\theta ,Z,\left\vert \Psi
\right\vert ^{2}\right) }\left( \left( h\left( Z,Z^{\prime }\right) -\hat{T}%
\right) C\left\vert \Psi \left( \theta ,Z\right) \right\vert ^{2}h_{C}\left(
\omega \left( J,\theta ,Z,\left\vert \Psi \right\vert ^{2}\right) \right)
\right.  \notag \\
&&\left. -D\hat{T}\left\vert \Psi \left( \theta -\frac{\left\vert
Z-Z^{\prime }\right\vert }{c},Z^{\prime }\right) \right\vert ^{2}h_{D}\left(
\omega \left( J,\theta -\frac{\left\vert Z-Z^{\prime }\right\vert }{c}%
,Z^{\prime },\left\vert \Psi \right\vert ^{2}\right) \right) \right)  \notag
\\
O_{T} &=&-\left( -\frac{1}{\tau \omega \left( J,\theta ,Z,\left\vert \Psi
\right\vert ^{2}\right) }T+\frac{\lambda }{\omega \left( J,\theta
,Z,\left\vert \Psi \right\vert ^{2}\right) }\hat{T}\right)  \notag
\end{eqnarray}%
In these equations, the averages $\bar{\omega}$ and $\bar{\omega}^{\prime }$
have been replaced by\ $\omega \left( J,\theta ,Z,\left\vert \Psi
\right\vert ^{2}\right) $ and $\omega \left( J,\theta -\frac{\left\vert
Z-Z^{\prime }\right\vert }{c},Z^{\prime },\left\vert \Psi \right\vert
^{2}\right) $ as a consequence of the projection.

In (\ref{flt}), we added a potential:%
\begin{equation}
U\left( \left\{ \left\vert \Gamma \left( \theta ,Z,Z^{\prime },C,D\right)
\right\vert ^{2}\right\} \right) =U\left( \int T\left\vert \Gamma \left( T,%
\hat{T},\theta ,Z,Z^{\prime },C,D\right) \right\vert ^{2}dTd\hat{T}\right)
\label{pcn}
\end{equation}%
that models the constraint about the number of active connections in the
system.

\section*{Appendix 2 \ Effective action for neurons field $\Psi $}

The perturbative expansion of the path integral for the field action (\ref%
{Sp}) modifies the activities equation. We computed this effective action,
written $\Gamma \left( \Psi ^{\dagger },\Psi \right) $, in (\cite{GL}). It
is not equal to $S\left( \Psi ^{\dagger },\Psi \right) $ defined in (\ref{Sp}%
) since the dependency of $\omega ^{-1}\left( J,\theta ,Z,\left\vert \Psi
\right\vert ^{2}\right) $ in $\left\vert \Psi \right\vert ^{2}$ introduces
self interaction terms.

The computation of the effective action is obtained perturbatively by
expanding the interaction terms and computing the associated grphs. This is
obtained by first computing the propagator associated to the "free action":%
\begin{equation}
-\Psi ^{\dagger }\left( \theta ,Z\right) \nabla _{\theta }\left( \frac{%
\sigma ^{2}}{2}\nabla _{\theta }-\omega ^{-1}\left( J,\theta ,Z,0\right)
\right) \Psi \left( \theta ,Z\right)  \label{cw}
\end{equation}%
where $\omega ^{-1}\left( J,\theta ,Z,0\right) $ is the inverse frequency
given by \ $\omega ^{-1}\left( J,\theta ,Z,\Psi \right) =G\left( J\left(
\theta ,Z\right) \right) $ for $\Psi \equiv 0$.

Adding a potential $V\left( \Psi \right) $, this action decomposes as:%
\begin{equation}
-\frac{1}{2}\Psi ^{\dagger }\left( \theta ,Z\right) \left( \nabla _{\theta
}\left( \frac{\sigma ^{2}}{2}\nabla _{\theta }-\omega ^{-1}\left( J,\theta
,Z,0\right) \right) \right) +\frac{1}{2}\Psi ^{\dagger }\left( \theta
,Z\right) \nabla _{\theta }\omega ^{-1}\left( J,\theta ,Z,\left\vert \Psi
\right\vert ^{2}\right) \Psi \left( \theta ,Z\right) +V\left( \Psi \right)
\label{dcp}
\end{equation}%
and the two last terms in the previous expression are considered
perturbatively in the computation of the graphs.

For an external current decomposed in a static and dynamic part $\bar{J}%
\left( Z\right) +J\left( Z,\theta \right) $, the expression for $G\left(
J\left( \theta ,Z\right) \right) $ can be approximated by $G\left( \bar{J}%
\left( Z\right) \right) $, so that: 
\begin{equation*}
\omega ^{-1}\left( J,\theta ,Z,0\right) =G\left( \bar{J}\left( Z\right)
+J\left( \theta \right) \right) \simeq G\left( \bar{J}\left( Z\right) \right)
\end{equation*}%
Given our choice for the function $G$, we have:%
\begin{equation}
\omega ^{-1}\left( J,\theta ,Z,0\right) \simeq \frac{\arctan \left( \left( 
\frac{1}{X_{r}}-\frac{1}{X_{p}}\right) \sqrt{\bar{J}\left( Z\right) }\right) 
}{\sqrt{\bar{J}\left( Z\right) }}=\frac{1}{\bar{X}_{r}\left( Z\right) }
\label{frz}
\end{equation}%
The propagator associated to (\ref{cw}) $\mathcal{G}_{0}\left( \theta
,\theta ^{\prime },Z,Z^{\prime }\right) $ can be directly computed. It
satifies:%
\begin{equation*}
-\nabla _{\theta }\left( \frac{\sigma ^{2}}{2}\nabla _{\theta }-\omega
^{-1}\left( J,\theta ,Z,0\right) \right) \mathcal{G}_{0}\left( \theta
,\theta ^{\prime },Z,Z^{\prime }\right) =\delta \left( Z-Z^{\prime }\right)
\delta \left( \theta -\theta ^{\prime }\right)
\end{equation*}%
and we find:%
\begin{equation}
\mathcal{G}_{0}\left( \theta ,\theta ^{\prime },Z,Z^{\prime }\right) =\delta
\left( Z-Z^{\prime }\right) \frac{\exp \left( -\Lambda _{1}\left( Z\right)
\left( \theta -\theta ^{\prime }\right) \right) }{\Lambda \left( Z\right) }%
H\left( \theta -\theta ^{\prime }\right)  \label{rg}
\end{equation}%
where: 
\begin{eqnarray*}
\Lambda \left( Z\right) &=&\sqrt{\frac{\pi }{2}}\sqrt{\left( \frac{1}{\sigma
^{2}\bar{X}_{r}\left( Z\right) }\right) ^{2}+\frac{2\alpha }{\sigma ^{2}}} \\
\Lambda _{1}\left( Z\right) &=&\sqrt{\left( \frac{1}{\sigma ^{2}\bar{X}%
_{r}\left( Z\right) }\right) ^{2}+\frac{2\alpha }{\sigma ^{2}}}-\frac{1}{%
\sigma ^{2}\bar{X}_{r}\left( Z\right) }
\end{eqnarray*}%
and $H$ is the heaviside function:%
\begin{eqnarray*}
H\left( \theta -\theta ^{\prime }\right) &=&0\text{ for }\theta -\theta
^{\prime }<0 \\
&=&1\text{ for }\theta -\theta ^{\prime }>0
\end{eqnarray*}%
For the sake of simplicity, we often discard the factor $\delta \left(
Z-Z^{\prime }\right) $ and write $\mathcal{G}_{0}\left( \theta ,\theta
^{\prime },Z\right) $ for $\mathcal{G}_{0}\left( \theta ,\theta ^{\prime
},Z,Z^{\prime }\right) $. In the sequel, for the sake of simplicity, the
dependency in $Z$ of $\bar{X}_{r}\left( Z\right) $, $\Lambda \left( Z\right) 
$, $\Lambda _{1}\left( Z\right) $ will be implicit, so that we will write: 
\begin{equation*}
\bar{X}_{r}\left( Z\right) \equiv \bar{X}_{r}\text{, }\Lambda \left(
Z\right) \equiv \Lambda \text{, }\Lambda _{1}\left( Z\right) \equiv \Lambda
_{1}
\end{equation*}

The particular form of this propagator simplifies the computations of the
perturbative expansion of the effective action. Actually, the heaviside
function $H\left( \theta -\theta ^{\prime }\right) $ in (\ref{rg}) excludes
the loop diagramms and we have to consider only tree graphs.

\subsubsection*{A2.1 Graphs expansion and two point Green functions}

Having found the propagator $\mathcal{G}_{0}\left( \theta ,\theta ^{\prime
},Z\right) $ associated to (\ref{cw}), we can compute the graphs associated
to the decomposition (\ref{dcp}). The two points Green function is shown to
be equal to be the inverse of the operator:%
\begin{equation}
-\frac{1}{2}\nabla _{\theta }\frac{\sigma _{\theta }^{2}}{2}\nabla _{\theta
}+\frac{1}{2}\left[ \frac{\delta \left[ \Psi ^{\dagger }\left( \theta
^{\prime },Z\right) \nabla _{\theta }\left( \omega ^{-1}\left( J,\theta
,Z,\left\vert \Psi \right\vert ^{2}\right) \Psi \left( \theta ,Z\right)
\right) \right] }{\delta \left\vert \Psi \right\vert ^{2}}\right] 
_{\substack{ \theta ^{\prime }=\theta  \\ \left\vert \Psi \left( \theta
,Z\right) \right\vert ^{2}=\mathcal{G}_{0}\left( 0,Z\right) }}+\left[ \frac{%
\delta \left[ V\left( \Psi \right) \right] }{\delta \left\vert \Psi
\right\vert ^{2}}\right] _{\substack{ \left\vert \Psi \left( \theta
,Z\right) \right\vert ^{2}  \\ =\mathcal{G}_{0}\left( 0,Z\right) }}
\label{grt}
\end{equation}%
where:%
\begin{equation*}
\mathcal{G}_{0}\left( 0,Z\right) =\mathcal{G}_{0}\left( \theta ,\theta
,Z\right) =\frac{\exp \left( -\Lambda _{1}\left( Z\right) \left( \theta
-\theta ^{\prime }\right) \right) }{\Lambda \left( Z\right) }H\left( \theta
-\theta ^{\prime }\right)
\end{equation*}%
and $\left\vert \Psi \right\vert ^{2}\left[ \frac{\delta }{\delta \left\vert
\Psi \right\vert ^{2}}\right] $ is a shorthand for:%
\begin{equation}
\int dZ^{\prime }\left\vert \Psi \left( \theta -\frac{\left\vert Z-Z^{\prime
}\right\vert }{c},Z^{\prime }\right) \right\vert ^{2}\times \frac{\delta }{%
\delta \left( \left\vert \Psi \left( \theta -\frac{\left\vert Z-Z^{\prime
}\right\vert }{c},Z^{\prime }\right) \right\vert ^{2}\right) }  \label{sht}
\end{equation}

\subsubsection*{A2.2 Effective action at the lowest order}

The effective order at the lowest order is derived directly from (\ref{grt}%
). We showed that:%
\begin{equation}
\Gamma _{0}\left( \Psi ^{\dagger },\Psi \right) =\Psi ^{\dagger }\left(
\theta ,Z\right) \left[ \frac{\delta \left[ S_{cl}\left( \Psi ^{\dagger
},\Psi \right) \right] }{\delta \left\vert \Psi \right\vert ^{2}}\right] 
_{\substack{ \left\vert \Psi \left( \theta ,Z\right) \right\vert ^{2}  \\ =%
\mathcal{G}_{0}\left( 0,Z\right) }}\Psi \left( \theta ,Z\right)
\label{fctnbs}
\end{equation}%
with:%
\begin{eqnarray}
&&S_{cl}\left( \Psi ^{\dagger },\Psi \right) =-\frac{1}{2}\Psi ^{\dagger
}\left( \theta ,Z\right) \left( \nabla _{\theta }\left( \frac{\sigma
_{\theta }^{2}}{2}\nabla _{\theta }-\omega ^{-1}\left( J,\theta
,Z,\left\vert \Psi \right\vert ^{2}\right) \right) \right) \Psi \left(
\theta ,Z\right)  \label{scL} \\
&&+\alpha \int \left\vert \Psi \left( \theta ^{\left( i\right)
},Z_{i}\right) \right\vert ^{2}+V\left( \Psi \right)  \notag
\end{eqnarray}%
and the brackets notation given in equation (\ref{sht}). Alternatively,
formula (\ref{fctnbs}) can also be written:

\begin{eqnarray}
\Gamma _{0}\left( \Psi ^{\dagger },\Psi \right) &=&-\frac{1}{2}\Psi
^{\dagger }\left( \theta ,Z\right) \left( \nabla _{\theta }\frac{\sigma ^{2}%
}{2}\left( \nabla _{\theta }-\left( \omega ^{-1}\left( \bar{J},Z,\mathcal{G}%
_{0}\right) +\frac{\delta \left[ \omega ^{-1}\left( \bar{J},Z,\mathcal{G}%
_{0}\right) \right] }{\delta \mathcal{G}_{0}\left( 0,Z\right) }\mathcal{G}%
_{0}\left( \theta ^{\prime },\theta ,Z\right) \right) \right) \right)
_{\theta ^{\prime }=\theta }\Psi \left( \theta ,Z\right)  \notag \\
&&+\alpha \int \left\vert \Psi \left( \theta ,Z\right) \right\vert ^{2}+\Psi
^{\dagger }\left( \theta ,Z\right) \left[ \frac{\delta \left[ V\left( \Psi
\right) \right] }{\delta \left\vert \Psi \right\vert ^{2}}\right] 
_{\substack{ \left\vert \Psi \left( \theta ,Z\right) \right\vert ^{2}  \\ =%
\mathcal{G}_{0}\left( 0,Z\right) }}\Psi \left( \theta ,Z\right)  \label{Rf}
\end{eqnarray}%
where $\omega ^{-1}\left( \bar{J},Z,\mathcal{G}_{0}\right) $ is the static
inversed frequency defined as the solution of the equation: 
\begin{equation}
\omega ^{-1}\left( \bar{J}\left( Z\right) ,Z,\mathcal{G}_{0}\right) =G\left( 
\bar{J}\left( Z\right) +\int \frac{\kappa }{N}\frac{\omega \left( \bar{J}%
,Z_{1},\mathcal{G}_{0}\right) }{\omega \left( \bar{J},Z,\mathcal{G}%
_{0}\right) }\mathcal{G}_{0}\left( 0,Z_{1}\right) T\left( Z,\theta
,Z_{1}\right) dZ_{1}\right)  \label{cp}
\end{equation}%
In formula (\ref{cp}), $\bar{J}\left( Z\right) $ is the average over time of 
$J\left( \theta ,Z\right) $. As a consequence, $\omega ^{-1}\left( \bar{J}%
\left( Z\right) ,Z,\mathcal{G}_{0}\right) $ solves:%
\begin{equation}
\omega ^{-1}\left( \bar{J}\left( Z\right) ,Z,\mathcal{G}_{0}\right) =G\left( 
\bar{J}\left( Z\right) +\int \frac{\kappa }{N}\frac{\omega ^{-1}\left( \bar{J%
}\left( Z_{1}\right) ,Z_{1},\mathcal{G}_{0}\right) }{\omega ^{-1}\left( \bar{%
J}\left( Z\right) ,Z,\mathcal{G}_{0}\right) }T\left( Z,\theta ,Z_{1}\right) 
\mathcal{G}_{0}\left( 0,Z_{1}\right) dZ_{1}\right)  \label{Cp}
\end{equation}

The perturbative corrections for the effective action are found by adding
the $1$PI graphs with $2n$ external points with $n\geqslant 2$. The
corrections are ordered by the number of vertices involved in them.

\subsubsection*{A2.3 Effective action at higher orders}

We showed that the effective action is a series of corrections to the
classical effective action (\ref{fctnbs}):%
\begin{equation*}
\Gamma \left( \Psi ^{\dagger },\Psi \right) =\sum_{n=2}^{\infty }\int \left(
\prod\limits_{l=1}^{n}\Psi ^{\dagger }\left( \theta _{f}^{\left( l\right)
},Z_{l}\right) \right) S_{n}\left( \left( \theta _{f}^{\left( l\right)
},\theta _{i}^{\left( l\right) },Z_{l}\right) \right) \left(
\prod\limits_{l=1}^{n}\Psi \left( \theta _{i}^{\left( l\right)
},Z_{l}\right) \right)
\end{equation*}%
The contribution $S_{n}\left( \left( \theta _{f}^{\left( l\right) },\theta
_{i}^{\left( l\right) },Z_{l}\right) \right) $ for a given $n$ is the $n$
points effective vertex. It is the sum of one-particle irreducible graphs ($%
1 $PI) with $n$ lines labelled by their position $Z_{l}$, $l=1,...,n$ \ plus
their starting and ending points $\left( \theta _{f}^{\left( l\right)
},\theta _{i}^{\left( l\right) }\right) $.

To build the series of graphs, we first consider the $l$ points vertices:%
\begin{eqnarray*}
\hat{V}_{2l}\left( \left\{ \left( \theta ^{\left( k_{i}\right)
},Z_{k_{i}}\right) \right\} _{i=1,...,l}\right) &=&\frac{1}{l!}\left[ \frac{%
\delta ^{l}\left[ \int \Psi ^{\dagger }\left( \theta ,Z\right) \nabla
_{\theta }\omega ^{-1}\left( J,\theta ,Z\right) \Psi \left( \theta ,Z\right)
dZd\theta +V\left( \Psi \right) \right] }{\prod\limits_{i=1}^{l}\delta
\left\vert \Psi \left( \theta ^{\left( k_{i}\right) },Z_{k_{i}}\right)
\right\vert ^{2}}\right] _{\left\vert \Psi \left( \theta ,Z\right)
\right\vert ^{2}=\mathcal{G}_{0}\left( 0,Z\right) } \\
&=&\frac{1}{l!}\left[ \frac{\delta ^{l}\left[ S_{cl}\left( \Psi ^{\dagger
},\Psi \right) \right] }{\prod\limits_{i=1}^{l}\delta \left\vert \Psi \left(
\theta ^{\left( k_{i}\right) },Z_{k_{i}}\right) \right\vert ^{2}}\right]
_{\left\vert \Psi \left( \theta ,Z\right) \right\vert ^{2}=\mathcal{G}%
_{0}\left( 0,Z\right) }
\end{eqnarray*}%
for $l=2,...,n$ and $\left( \theta ^{\left( k_{i}\right) },Z_{k_{i}}\right)
\in \left\{ \left( \theta _{i},Z_{i}\right) \right\} _{i=1,...,n}$ with $%
\theta _{i}\in \left[ \theta _{i}^{\left( i\right) },\theta _{f}^{\left(
i\right) }\right] $ and $\left( \theta ^{\left( k_{i}\right)
},Z_{k_{i}}\right) \neq \left( \theta ^{\left( k_{j}\right)
},Z_{k_{j}}\right) $ for $i\neq j$.

These vertices are represented graphically by associating a point $\left(
\theta ,Z\right) _{\hat{V}}$ to each vertex $\hat{V}$, which differs from
all the $\left( \theta ^{\left( i\right) },Z_{i}\right) $ and $\left( \theta
,Z\right) _{\hat{V}}\neq \left( \theta ,Z\right) _{\hat{V}^{\prime }}$ when $%
\hat{V}\neq \hat{V}^{\prime }$. We draw $l$ lines from $\left( \theta
,Z\right) _{\hat{V}}$ ending at the points $\left( \theta ^{\left(
k_{i}\right) },Z_{k_{i}}\right) $.

We then consider the series of graphs $l=2,...,n$, with an arbitrary number
of vertices $\hat{V}_{2l}\left( \left\{ \left( \theta ^{\left( k_{i}\right)
},Z_{k_{i}}\right) \right\} _{i=1,...,l}\right) $, joining the points $%
\left( \theta ^{\left( k_{i}\right) },Z_{k_{i}}\right) $. To each internal
segment between $\theta $ and $\theta ^{\prime }$ at position $Z$, we
associate a propagator $\mathcal{G}_{0}\left( \theta ,\theta ^{\prime
},Z\right) $ defined in (\ref{rg}).\ The effective vertex $S_{n}\left(
\left( \theta _{f}^{\left( l\right) },\theta _{i}^{\left( l\right)
},Z_{l}\right) \right) $\ is obtained by summing the $n$ lines-$1$PI graphs.

Defining:%
\begin{eqnarray}
\hat{S}_{cl}\left( \Psi ^{\dagger },\Psi \right) &\equiv &S_{cl}\left( 
\mathcal{G}_{0}\left( 0,Z\right) +\left\vert \Psi \right\vert ^{2}\right)
\label{fth} \\
&\equiv &-\frac{1}{2}\left( \left( \nabla _{\theta }\left( \frac{\sigma
_{\theta }^{2}}{2}\nabla _{\theta }-\omega ^{-1}\left( \mathcal{G}_{0}\left(
0,Z\right) +\left\vert \Psi \left( \theta ,Z\right) \right\vert ^{2}\right)
\right) \right) \left( \mathcal{G}_{0}\left( \theta ^{\prime },\theta
,Z\right) +\Psi ^{\dagger }\left( \theta ^{\prime },Z\right) \Psi \left(
\theta ,Z\right) \right) \right) _{\theta ^{\prime }=\theta }  \notag \\
&&+\alpha \int \left( \mathcal{G}_{0}\left( 0,Z\right) \mathcal{+}\left\vert
\Psi \left( \theta ,Z\right) \right\vert ^{2}\right) +V\left( \left( 
\mathcal{G}_{0}\left( 0,Z\right) \mathcal{+}\left\vert \Psi \left( \theta
,Z\right) \right\vert ^{2}\right) \right)  \notag
\end{eqnarray}%
with $V\left( \left( \mathcal{G}_{0}\left( 0,Z\right) \mathcal{+}\left\vert
\Psi \left( \theta ,Z\right) \right\vert ^{2}\right) \right) $ given by:%
\begin{equation}
V\left( \mathcal{G}_{0}\left( 0,Z\right) \mathcal{+}\left\vert \Psi \left(
\theta ,Z\right) \right\vert ^{2}\right) =\int \left( \mathcal{G}_{0}\left(
0,Z\right) \mathcal{+}\left\vert \Psi \left( \theta ,Z\right) \right\vert
^{2}\right) U_{0}\left( \int \mathcal{G}_{0}\left( 0,Z^{\prime }\right)
+\left\vert \Psi \left( \theta -\frac{\left\vert Z-Z^{\prime }\right\vert }{c%
},Z^{\prime }\right) \right\vert ^{2}\right)  \label{ptd}
\end{equation}%
the effective action writes as a series expansion:

\begin{eqnarray}
\Gamma \left( \Psi ^{\dagger },\Psi \right) &=&\hat{S}_{cl}\left( \Psi
^{\dagger },\Psi \right) +\sum_{\substack{ j\geqslant 2  \\ m\geqslant 2}}%
\sum _{\substack{ \left( p_{l}^{i}\right) _{m\times j}  \\ %
\sum_{i}p_{l}^{i}\geqslant 2}}\int \left( \prod\limits_{l=1}^{j}\Psi
^{\dagger }\left( \theta _{f}^{\left( l\right) },Z_{l}\right) \right)  \notag
\\
&&\times \prod\limits_{i=1}^{m}\left[ \underset{\prod\limits_{l}\left[
\theta _{i}^{\left( l\right) },\theta _{f}^{\left( l\right) }\right]
^{p_{l}^{i}}}{\int }\frac{\delta ^{\sum_{l}p_{l}^{i}}\left[ \hat{S}%
_{cl}\left( \Psi ^{\dagger },\Psi \right) \right] }{\prod\limits_{l=1}^{j}%
\prod\limits_{k_{l}^{i}=1}^{p_{l}^{i}}\delta \left\vert \Psi \left( \theta
^{\left( k_{l}^{i}\right) },Z_{_{l}}\right) \right\vert ^{2}}%
\prod\limits_{l=1}^{j}\prod\limits_{k_{l}^{i}=1}^{p_{l}^{i}}d\theta ^{\left(
k_{l}^{i}\right) }\right]  \notag \\
&&\times \frac{\prod\limits_{l=1}^{j}\exp \left( -\Lambda _{1}\left( \theta
_{f}^{\left( l\right) }-\theta _{i}^{\left( l\right) }\right) \right) }{%
m!\prod\limits_{k}\left( \sharp _{j,m,k}\left( \left( p_{l}^{i}\right)
\right) \right) !\Lambda ^{\sum_{i,l}p_{l}^{i}}}\left(
\prod\limits_{l=1}^{j}\Psi \left( \theta _{i}^{\left( l\right)
},Z_{l}\right) \right)  \label{ffc}
\end{eqnarray}

\subsubsection*{A2.4 Local approximation}

In the local approximation, this effective action corrects (\ref{Sp}) by a
series expansion in field:

\begin{equation}
\Gamma \left( \Psi ^{\dagger },\Psi \right) \simeq \int \Psi ^{\dagger
}\left( \theta ,Z\right) \left( -\nabla _{\theta }\left( \frac{\sigma
_{\theta }^{2}}{2}\nabla _{\theta }-\omega ^{-1}\left( J\left( \theta
\right) ,\theta ,Z,\mathcal{G}_{0}+\left\vert \Psi \right\vert ^{2}\right)
\right) \delta \left( \theta _{f}-\theta _{i}\right) +\Gamma _{p}\left( \Psi
^{\dagger },\Psi \right) \right) \Psi \left( \theta ,Z\right)  \label{gmf}
\end{equation}%
where $\Gamma _{p}\left( \Psi ^{\dagger },\Psi \right) $ is a corrective
perturbative term depending on the successive\ derivatives of the field (the
constants $a_{j}$ are derived in (\cite{GL})):%
\begin{eqnarray}
\Gamma _{p}\left( \Psi ^{\dagger },\Psi \right) &=&\int \sum_{\substack{ %
j\geqslant 1  \\ m\geqslant 1}}\sum_{\substack{ \left( p_{l}^{i}\right)
_{m\times j}  \\ p_{l}+\sum_{i}p_{l}^{i}\geqslant 2}}\frac{a_{j}}{j!}\frac{%
\delta ^{\sum_{l}p_{l}}\left[ -\frac{1}{2}\left( \nabla _{\theta }\left( 
\frac{\sigma _{\theta }^{2}}{2}\nabla _{\theta }-\omega ^{-1}\left(
\left\vert \Psi \left( \theta ,Z\right) \right\vert ^{2}\right) \right)
\right) \right] }{\prod\limits_{l=1}^{j}\prod\limits_{k_{l}^{i}=1}^{p_{l}}%
\delta \left\vert \Psi \left( \theta ^{\left( l\right) },Z_{_{l}}\right)
\right\vert ^{2}}\Psi \left( \theta ,Z\right)  \label{mG} \\
&&\times \left( \prod\limits_{l=1}^{j}\Psi ^{\dagger }\left( \theta
_{f}^{\left( l\right) },Z_{l}\right) \right) \prod\limits_{i=1}^{m}\left[ 
\frac{\delta ^{\sum_{l}p_{l}^{i}}\left[ \hat{S}_{cl,\theta }\left( \Psi
^{\dagger },\Psi \right) \right] }{\prod\limits_{l=1}^{j}\prod%
\limits_{k_{l}^{i}=1}^{p_{l}^{i}}\delta \left\vert \Psi \left( \theta
^{\left( l\right) },Z_{_{l}}\right) \right\vert ^{2}}\right] \left(
\prod\limits_{l=1}^{j}\Psi \left( \theta _{i}^{\left( l\right)
},Z_{l}\right) \right)  \notag
\end{eqnarray}%
The term $\mathcal{G}_{0}$ is a function of $Z$ and represents a two points
free Green function (see (\cite{GL})).

The previous equation (\ref{gmf}) defines an effective activity that can be
identified as:%
\begin{equation}
\omega _{e}^{-1}\left( J\left( \theta \right) ,\theta ,Z,\mathcal{G}%
_{0}+\left\vert \Psi \right\vert ^{2}\right) =\omega ^{-1}\left( J\left(
\theta \right) ,\theta ,Z,\mathcal{G}_{0}+\left\vert \Psi \right\vert
^{2}\right) +\int^{\theta }\Omega \left( \theta ,Z\right)  \label{fft}
\end{equation}%
where $\omega \left( J\left( \theta \right) ,\theta ,Z,\mathcal{\bar{G}}%
_{0}+\left\vert \Psi \right\vert ^{2}\right) $ is the solution of:%
\begin{eqnarray*}
\omega ^{-1}\left( J,\theta ,Z,\left\vert \Psi \right\vert ^{2}\right)
&=&G\left( J\left( \theta ,Z\right) +\int \frac{\kappa }{N}\frac{\omega
\left( J,\theta -\frac{\left\vert Z-Z_{1}\right\vert }{c},Z_{1},\Psi \right)
T\left( Z,\theta ,Z_{1},\theta -\frac{\left\vert Z-Z_{1}\right\vert }{c}%
\right) }{\omega \left( J,\theta ,Z,\left\vert \Psi \right\vert ^{2}\right) }%
\right. \\
&&\times \left. \left( \mathcal{\bar{G}}_{0}\left( 0,Z_{1}\right)
+\left\vert \Psi \left( \theta -\frac{\left\vert Z-Z_{1}\right\vert }{c}%
,Z_{1}\right) \right\vert ^{2}\right) dZ_{1}\right)
\end{eqnarray*}%
Which is the classical activity equation, up to the inclusion of the Green
function $\mathcal{\bar{G}}_{0}\left( 0,Z_{1}\right) $.

The second term $\int^{\theta }\Omega \left( \theta ,Z\right) $ in (\ref{fft}%
) represents corrections due to the interactions. Using (\ref{mG}), we can
find its expression as a series expansion in terms of activities and field.
The computations of these corrections to the classical equation are
presented in \cite{GL} and confirm the possibility of traveling wave
solutions. At the lowest order, we find:%
\begin{equation*}
\int^{\theta }\Omega \left( \theta ,Z\right) =\frac{1}{4}\int \int^{\theta }%
\frac{\delta \left( \nabla _{\theta }\omega ^{-1}\left( J\left( \theta
\right) ,\theta ,Z,\mathcal{G}_{0}+\left\vert \Psi \right\vert ^{2}\right)
\right) }{\delta \left\vert \Psi \left( \theta ^{\left( l\right)
},Z_{_{l}}\right) \right\vert ^{2}}\times \frac{\delta \left( \nabla
_{\theta }\omega ^{-1}\left( J\left( \theta \right) ,\theta ,Z,\mathcal{G}%
_{0}+\left\vert \Psi \right\vert ^{2}\right) \right) }{\delta \left\vert
\Psi \left( \theta ^{\left( l\right) },Z_{_{l}}\right) \right\vert ^{2}}%
\left\vert \Psi \left( \theta ^{\left( l\right) },Z_{_{l}}\right)
\right\vert ^{2}
\end{equation*}%
This correction will impact the traveling wave solutions for activits (see
Appendix).

\section*{Appendix 3 \ Connectivty field action and Static background for
connectivities}

\subsection*{A3.1 Effective action functional for connectivities}

The connectivity betwn tw pnt $Z,Z^{\prime }$ are described by variables $%
C,D,T,\hat{T}$ (see \cite{GLr}). The connectivity between the two pnts is
measured by $T$ while $\hat{T}$ represents the variation of $T$. Following 
\cite{IFR} we consider that $\hat{T}$ is a function of accumulation of
output spikes and input spikes. The higher the number of output and input
spikes, the lower $\hat{T}$, modeling that those variables should present
some delay to be correlated. The difference between $T$ and $\hat{T}$
measures the difference of scale between the modification in activities and
modification in connectivities. The connectivities are thus described by the
field $\Gamma \left( T,\hat{T},\theta ,Z,Z^{\prime },C,D\right) $.

The field model for connections writes:%
\begin{equation*}
S_{\Gamma }=S_{\Gamma }^{\left( 1\right) }+S_{\Gamma }^{\left( 2\right)
}+S_{\Gamma }^{\left( 3\right) }+S_{\Gamma }^{\left( 4\right) }
\end{equation*}

where $S_{\Gamma }^{\left( 1\right) }$, $S_{\Gamma }^{\left( 2\right) }$, $%
S_{\Gamma }^{\left( 3\right) }$, $S_{\Gamma }^{\left( 4\right) }$ are given
by (\ref{wGD}), (\ref{wGT}), (\ref{wGQ}), (\ref{wGC}) and the four operators 
$O_{C}$, $O_{D}$, $O_{\hat{T}}$, $O_{T}$ defined in (\ref{DP}) are the field
transltn of the dynamics for the variables $C,D,T,\hat{T}$. \ The term $%
-O_{C}$ includes a term:%
\begin{equation*}
\omega \left( J,\theta -\frac{\left\vert Z-Z^{\prime }\right\vert }{c}%
,Z^{\prime },\left\vert \Psi \right\vert ^{2}\right) \left\vert \Psi \left(
\theta -\frac{\left\vert Z-Z^{\prime }\right\vert }{c},Z^{\prime }\right)
\right\vert ^{2}
\end{equation*}%
that computes the number of spikes incoming from other cells while the term $%
-\frac{C}{\tau _{C}\omega \left( J,\theta ,Z,\left\vert \Psi \right\vert
^{2}\right) }$ is an attenuation factor in the accumulation of these spikes.
The term $-O_{D}$ is similar but is computed from the spikes emtted by one
cell.

The term $-O_{\hat{T}}$ computes the difference accumulated input spikes
minus those that are emitted. The input spikes reiforce the connections,
while the simultaneous output decorelates the two cells. The dynamics for $%
\hat{T}$ is determined by these two effect.

The term$-O_{T}$ is the field translation of the dynamics for $T$. The first
term is the attenuation of past effects $\frac{1}{\tau \omega \left(
J,\theta ,Z,\left\vert \Psi \right\vert ^{2}\right) }T$, while the second
trm is proportional to $\hat{T}$ and represents the presnt variation of
connectivity due to the accumulation of output and input spikes.

In \cite{GLr} we show that the background field $\Gamma \left( T,\hat{T}%
,\theta ,Z,Z^{\prime },C,D\right) $ can be decomposed in first approximation
as a product:%
\begin{equation*}
\Gamma \left( T,\hat{T},\theta ,Z,Z^{\prime },C,D\right) =\Gamma _{1}\left(
Z,Z^{\prime },C\right) \Gamma _{2}\left( Z,Z^{\prime },D\right) \Gamma
\left( T,\hat{T},\theta ,Z,Z^{\prime }\right)
\end{equation*}%
where $\Gamma _{1}\left( Z,Z^{\prime },C\right) $ and $\Gamma _{2}\left(
Z,Z^{\prime },D\right) $ satisfy:%
\begin{equation}
\left( \frac{\sigma _{C}^{2}}{2}\nabla _{C}^{2}-\frac{1}{2\sigma _{C}^{2}}%
\left( a_{C}\left( Z\right) \left( C-C\left( \theta \right) \right) \right)
^{2}-\frac{1}{2}a_{C}\left( Z\right) +\lambda _{1}\left( Z\right) \right)
\Gamma _{1}\left( Z,Z^{\prime },C\right) =0  \label{GMN}
\end{equation}%
and:%
\begin{equation}
\left( \frac{\sigma _{D}^{2}}{2}\nabla _{D}^{2}-\frac{1}{2\sigma _{D}^{2}}%
\left( a_{D}\left( Z\right) \left( D-D\left( \theta \right) \right) \right)
^{2}-\frac{1}{2}a_{D}\left( Z\right) +\lambda _{2}\left( Z,Z^{\prime
}\right) \right) \Gamma _{2}\left( Z,Z^{\prime },D\right) =0  \label{GMT}
\end{equation}

\bigskip with:%
\begin{eqnarray}
a_{C}\left( Z\right) &=&\frac{1}{\tau _{C}\omega }+\alpha _{C}\frac{\omega
^{\prime }\left\vert \Psi \left( \theta -\frac{\left\vert Z-Z^{\prime
}\right\vert }{c},Z^{\prime },\omega ^{\prime }\right) \right\vert ^{2}}{%
\omega }  \label{CD} \\
a_{D}\left( Z\right) &=&\frac{1}{\tau _{D}\omega }+\alpha _{D}\left\vert
\Psi \left( \theta ,Z\right) \right\vert ^{2}  \notag
\end{eqnarray}%
and the averages $\left\langle C\left( \theta \right) \right\rangle $ and $%
\left\langle D\left( \theta \right) \right\rangle $ are defined by the
average of $C$ and $D$ in theie background states: 
\begin{eqnarray}
C &\rightarrow &\left\langle C\left( \theta \right) \right\rangle =\frac{%
\alpha _{C}\frac{\omega ^{\prime }\left\vert \Psi \left( \theta -\frac{%
\left\vert Z-Z^{\prime }\right\vert }{c},Z^{\prime }\right) \right\vert ^{2}%
}{\omega }}{\frac{1}{\tau _{C}\omega }+\alpha _{C}\frac{\omega ^{\prime
}\left\vert \Psi \left( \theta -\frac{\left\vert Z-Z^{\prime }\right\vert }{c%
},Z^{\prime }\right) \right\vert ^{2}}{\omega }}=\frac{\alpha _{C}\omega
^{\prime }\left\vert \Psi \left( \theta -\frac{\left\vert Z-Z^{\prime
}\right\vert }{c},Z^{\prime }\right) \right\vert ^{2}}{\frac{1}{\tau _{C}}%
+\alpha _{C}\omega ^{\prime }\left\vert \Psi \left( \theta -\frac{\left\vert
Z-Z^{\prime }\right\vert }{c},Z^{\prime }\right) \right\vert ^{2}}\equiv
C\left( \theta \right) \\
D &\rightarrow &\left\langle D\left( \theta \right) \right\rangle =\frac{%
\alpha _{D}\omega \left\vert \Psi \left( \theta ,Z\right) \right\vert ^{2}}{%
\frac{1}{\tau _{D}}+\alpha _{D}\omega \left\vert \Psi \left( \theta
,Z\right) \right\vert ^{2}}\equiv D\left( \theta \right)
\end{eqnarray}%
The solutions $\Gamma _{1}\left( Z,Z^{\prime },C\right) $ and $\Gamma
_{2}\left( Z,Z^{\prime },D\right) $ of (\ref{GMN}), (\ref{GMT}) are gaussian
functions derived in \cite{GLr}:%
\begin{equation}
\Gamma _{1}\left( Z,Z^{\prime },C\right) =\exp \left( -\frac{a_{C}\left(
Z\right) }{8\sigma _{C}^{2}}\left( C-C\left( \theta \right) \right)
^{2}\right)  \label{GMTH}
\end{equation}%
and:%
\begin{equation}
\Gamma _{2}\left( Z,Z^{\prime },D\right) =\exp \left( -\frac{a_{D}\left(
Z\right) }{8\sigma _{D}^{2}}\left( D-D\left( \theta \right) \right)
^{2}\right)  \label{GMTl}
\end{equation}%
Moreover, $\Gamma \left( T,\hat{T},\theta ,Z,Z^{\prime }\right) $ factors in
first approximation as:%
\begin{equation}
\Gamma \left( T,\hat{T},\theta ,Z,Z^{\prime }\right) =\Gamma _{0}\left(
T,\theta ,Z,Z^{\prime }\right) \Gamma _{0}\left( \hat{T},\theta ,Z,Z^{\prime
}\right)  \label{FCt}
\end{equation}

This allows to proceed as for the background state for $C$ and $D$. and to
minimize:%
\begin{equation}
S_{\Gamma }^{\left( 2\right) }=\Gamma _{0}^{\dag }\left( T,\theta
,Z,Z^{\prime }\right) \left( \frac{\rho h_{C}\left( \omega \left( \theta
,Z\right) \right) \left\vert \bar{\Psi}\left( \theta ,Z,Z^{\prime }\right)
\right\vert ^{2}}{\omega \left( \theta ,Z,\left\vert \Psi \right\vert
^{2}\right) }\right) \Gamma _{0}\left( T,\theta ,Z,Z^{\prime }\right)
\label{rst}
\end{equation}%
and:%
\begin{equation*}
S_{\Gamma }^{\left( 1\right) }=\Gamma _{0}^{\dag }\left( T,\theta
,Z,Z^{\prime }\right) \left( \frac{\sigma _{T}^{2}}{2}\nabla _{T}^{2}-\frac{1%
}{2\sigma _{T}^{2}}\left( \left( \frac{1}{\tau \omega }\left( T-\left\langle
T\right\rangle \right) \right) \right) ^{2}-\frac{1}{2\tau \omega \left(
Z\right) }\right) \Gamma _{0}\left( T,\theta ,Z,Z^{\prime }\right)
\end{equation*}
under the constraints (\ref{GMTH}) and (\ref{GMTl}). This corresponds to
project $\Gamma \left( T,\hat{T},\theta ,Z,Z^{\prime }\right) $ on the
background state:

\begin{eqnarray}
&&\Gamma _{0}\left( T,\hat{T},\theta ,Z,Z^{\prime }\right)  \label{GRTH} \\
&=&\Gamma _{0}\left( T,\theta ,Z,Z^{\prime }\right) \Gamma _{0}\left( \hat{T}%
,\theta ,Z,Z^{\prime }\right)  \notag \\
&=&\Gamma _{0}\left( T,\theta ,Z\right) \exp \left( -\frac{\rho h_{C}\left(
\omega \left( \theta ,Z\right) \right) \left\vert \bar{\Psi}\left( \theta
,Z,Z^{\prime }\right) \right\vert ^{2}}{4\sigma _{\hat{T}}^{2}\omega \left(
\theta ,Z,\left\vert \Psi \right\vert ^{2}\right) }\left( \left( \hat{T}%
-\left\langle \hat{T}\right\rangle \right) \right) ^{2}\right)  \notag
\end{eqnarray}%
We find ultimately the action for $\Gamma _{0}\left( T,\theta ,Z,Z^{\prime
}\right) $ quoted in the text:%
\begin{eqnarray}
S &&\left( \Gamma _{0}\right) =\int \Gamma _{0}^{\dag }\left( T,\theta
,Z,Z^{\prime }\right) \left( \frac{\sigma _{T}^{2}}{2}\nabla _{T}^{2}-\frac{1%
}{2\sigma _{T}^{2}}\left( \left( \frac{1}{\tau \omega }\left( T-\left\langle
T\right\rangle \right) \right) \left\vert \Psi \left( \theta ,Z\right)
\right\vert ^{2}\right) ^{2}-\frac{1}{2\tau \omega \left( Z\right) }\right)
\Gamma _{0}\left( T,\theta ,Z,Z^{\prime }\right) \\
&&-\Gamma _{0}^{\dag }\left( T,\theta ,Z,Z^{\prime }\right) \left(
a_{C}\left( Z\right) +a_{D}\left( Z\right) +\frac{\rho h_{C}\left( \omega
\left( \theta ,Z\right) \right) \left\vert \bar{\Psi}\left( \theta
,Z,Z^{\prime }\right) \right\vert ^{2}}{\omega \left( \theta ,Z,\left\vert
\Psi \right\vert ^{2}\right) }\right) \Gamma _{0}\left( T,\theta
,Z,Z^{\prime }\right)  \notag
\end{eqnarray}%
where:%
\begin{equation*}
\left\vert \bar{\Psi}\left( \theta ,Z,Z^{\prime }\right) \right\vert ^{2}=%
\frac{C_{Z,Z^{\prime }}\left( \theta \right) \left\vert \Psi \left( \theta
,Z\right) \right\vert ^{2}h_{C}\left( \omega \left( \theta ,Z\right) \right)
+D_{Z,Z^{\prime }}\left( \theta \right) \left\vert \Psi \left( \theta -\frac{%
\left\vert Z-Z^{\prime }\right\vert }{c},Z^{\prime }\right) \right\vert
^{2}h_{D}\left( \omega \left( \theta -\frac{\left\vert Z-Z^{\prime
}\right\vert }{c},Z^{\prime }\right) \right) }{h_{C}\left( \omega \left(
\theta ,Z\right) \right) }
\end{equation*}
This action results from the successive projections of partial background
states and becomes a function of $T$ only. This form will be used to
describe the static background states.

\subsection*{A3.2 Full system background state}

We first study a quasi-static first approximation, corresponding to an
averaging over individual neurons sgnls time scale. We consider the
classical approxmation. Assuming some static solution to the minimization
equation of $S\left( \Psi \right) $ (see \cite{GL}), We obtained the
background in \cite{GLr}:

\begin{equation}
\left\vert \Psi \left( Z\right) \right\vert ^{2}=\frac{2T\left( Z\right)
\left\langle \left\vert \Psi _{0}\left( Z^{\prime }\right) \right\vert
^{2}\right\rangle _{Z}}{\left( 1+\sqrt{1+4\left( \frac{\lambda \tau \nu
c-T\left( Z\right) }{\left( \frac{1}{\tau _{D}\alpha _{D}}+\frac{1}{\tau
_{C}\alpha _{C}}+\Omega \right) T\left( Z\right) -\frac{1}{\tau _{D}\alpha
_{D}}\lambda \tau \nu c}\right) ^{2}T\left( Z\right) \left\langle \left\vert
\Psi _{0}\left( Z^{\prime }\right) \right\vert ^{2}\right\rangle _{Z}}%
\right) }
\end{equation}%
$T\left( Z\right) $ is an averaged connectivity at $Z$ given by averaging
the connectivities reaching $Z$. The various parameters $\tau _{D}$, $\alpha
_{D}$,.. are defined below and describe the connectivity field equations. $%
\left\langle \left\vert \Psi _{0}\left( Z^{\prime }\right) \right\vert
^{2}\right\rangle _{Z}$ is the weighted average of the field defined $%
\left\vert \Psi _{0}\left( Z^{\prime }\right) \right\vert ^{2}$ above, the
weights being defined by connectivities.

Equation (\ref{M}) has itself static solutns $\omega _{0}\left( Z\right) $
describng some static activity at each point of the thread. They satisfy:%
\begin{equation*}
\omega ^{-1}\left( Z,\left\vert \Psi \right\vert ^{2}\right) =G\left( \int 
\frac{\kappa }{N}\frac{\omega \left( J,\Psi \right) T\left( Z,\theta
,Z_{1}\right) }{\omega \left( J,Z,\left\vert \Psi \right\vert ^{2}\right) }%
\left\vert \Psi \left( Z_{1}\right) \right\vert ^{2}dZ_{1}\right)
\end{equation*}

Replacing $\omega ^{-1}\left( J,Z,\left\vert \Psi \right\vert ^{2}\right) $
and $\left\vert \Psi \left( Z\right) \right\vert ^{2}$ allowed to minimize $%
\sum_{i}S_{\Gamma }^{\left( 1\right) }$. The background state for
connectivity functions may have two forms, for activated connections or
unactivated ones (those with $\left\langle T\right\rangle =0$). In
quasi-static approximation, we find:

\begin{eqnarray}
&&\left\vert \Gamma \right\vert _{a}^{2}\left( T,\hat{T},\theta ,C,D\right)
\label{gmv} \\
&\simeq &\left\{ \mathcal{N}\exp \left( -\frac{a_{C}\left( Z,Z^{\prime
}\right) }{2}\left( C-C\left( \theta \right) \right) ^{2}\right) \exp \left(
-\frac{a_{D}\left( Z,Z^{\prime }\right) }{2}\left( D-D\left( \theta \right)
\right) ^{2}\right) \right.  \notag \\
&&\times \exp \left( -\frac{\rho \left\vert \bar{\Psi}\left( \theta
,Z,Z^{\prime }\right) \right\vert ^{2}}{2}\left( \hat{T}-\left\langle \hat{T}%
\right\rangle \right) ^{2}\right)  \notag \\
&&\times \left. \left\Vert \Gamma _{0}\left( \theta ,Z,Z^{\prime }\right)
\right\Vert \exp \left( -\frac{\left\vert \Psi \left( \theta ,Z\right)
\right\vert ^{2}}{2\tau \omega }\left( T-\left\langle T\right\rangle \right)
^{2}\right) \right\} _{\left( Z,Z^{\prime }\right) ,\left\langle T\left(
Z,Z^{\prime }\right) \right\rangle \neq 0}  \notag
\end{eqnarray}%
\begin{eqnarray*}
&&\left\vert \Gamma \right\vert _{u}^{2}\left( T,\hat{T},\theta ,C,D\right)
\\
&\simeq &\left\{ \mathcal{N}\exp \left( -\frac{a_{C}\left( Z,Z^{\prime
}\right) }{2}\left( C-C\left( \theta \right) \right) ^{2}\right) \exp \left(
-\frac{a_{D}\left( Z,Z^{\prime }\right) }{2}\left( D-\left\langle
D\right\rangle \right) ^{2}\right) \right. \\
&&\left. \times \exp \left( -\frac{\rho \left\vert \bar{\Psi}\left( \theta
,Z,Z^{\prime }\right) \right\vert ^{2}}{2}\left( \hat{T}-\left\langle \hat{T}%
\right\rangle \right) ^{2}\right) \times \delta \left( T\right) \right\}
_{\left( Z,Z^{\prime }\right) ,\left\langle T\left( Z,Z^{\prime }\right)
\right\rangle \neq 0}
\end{eqnarray*}%
where $\mathcal{N}$ is a normalization factor ensuring that the constraint
over the number of connections is satisfied and where:%
\begin{eqnarray*}
\left\vert \bar{\Psi}\left( \theta ,Z,Z^{\prime }\right) \right\vert ^{2} &=&%
\frac{C_{Z,Z^{\prime }}\left( \theta \right) \left\vert \Psi \left( \theta
,Z\right) \right\vert ^{2}h_{C}\left( \omega \left( \theta ,Z\right) \right)
+D_{Z,Z^{\prime }}\left( \theta \right) \left\vert \Psi \left( \theta -\frac{%
\left\vert Z-Z^{\prime }\right\vert }{c},Z^{\prime }\right) \right\vert
^{2}h_{D}\left( \omega \left( \theta -\frac{\left\vert Z-Z^{\prime
}\right\vert }{c},Z^{\prime }\right) \right) }{h_{C}\left( \omega \left(
\theta ,Z\right) \right) } \\
a_{C}\left( Z,Z^{\prime }\right) &=&\frac{1}{\tau _{C}\omega }+\alpha _{C}%
\frac{\omega ^{\prime }\left\vert \Psi \left( \theta -\frac{\left\vert
Z-Z^{\prime }\right\vert }{c},Z^{\prime }\right) \right\vert ^{2}}{\omega }
\\
a_{D}\left( Z,Z^{\prime }\right) &=&\frac{1}{\tau _{D}\omega }+\alpha
_{D}\left\vert \Psi \left( \theta ,Z\right) \right\vert ^{2}
\end{eqnarray*}%
$\left\vert \bar{\Psi}\left( \theta ,Z,Z^{\prime }\right) \right\vert ^{2}$
is a weighted sum of the values of field $\left\vert \Psi \left( \theta
,Z\right) \right\vert ^{2}$ and $\left\vert \Psi \left( \theta -\frac{%
\left\vert Z-Z^{\prime }\right\vert }{c},Z^{\prime }\right) \right\vert ^{2}$%
. These quantities depend on the accumulation of input outpout spiks with
averages $\left\langle C\left( \theta \right) \right\rangle $ and $%
\left\langle D\left( \theta \right) \right\rangle $.

The averages for $C$ and $D$ in states $\Gamma _{1}\left( Z,Z^{\prime
},C\right) $ and $\Gamma _{2}\left( Z,Z^{\prime },D\right) $ respectively,
are: 
\begin{eqnarray}
C &\rightarrow &\left\langle C\left( \theta \right) \right\rangle =\frac{%
\alpha _{C}\frac{\omega ^{\prime }\left\vert \Psi \left( \theta -\frac{%
\left\vert Z-Z^{\prime }\right\vert }{c},Z^{\prime }\right) \right\vert ^{2}%
}{\omega }}{\frac{1}{\tau _{C}\omega }+\alpha _{C}\frac{\omega ^{\prime
}\left\vert \Psi \left( \theta -\frac{\left\vert Z-Z^{\prime }\right\vert }{c%
},Z^{\prime }\right) \right\vert ^{2}}{\omega }}=\frac{\alpha _{C}\omega
^{\prime }\left\vert \Psi \left( \theta -\frac{\left\vert Z-Z^{\prime
}\right\vert }{c},Z^{\prime }\right) \right\vert ^{2}}{\frac{1}{\tau _{C}}%
+\alpha _{C}\omega ^{\prime }\left\vert \Psi \left( \theta -\frac{\left\vert
Z-Z^{\prime }\right\vert }{c},Z^{\prime }\right) \right\vert ^{2}}\equiv
C\left( \theta \right)  \label{vrG} \\
D &\rightarrow &\left\langle D\left( \theta \right) \right\rangle =\frac{%
\alpha _{D}\omega \left\vert \Psi \left( \theta ,Z\right) \right\vert ^{2}}{%
\frac{1}{\tau _{D}}+\alpha _{D}\omega \left\vert \Psi \left( \theta
,Z\right) \right\vert ^{2}}\equiv D\left( \theta \right)  \label{vRG}
\end{eqnarray}

The average connectivities, and average variation of connectivities in the
background states satisf:%
\begin{equation*}
\left\langle T\left( Z,Z^{\prime }\right) \right\rangle =\lambda \tau
\left\langle \hat{T}\left( Z,Z^{\prime }\right) \right\rangle =\frac{\lambda
\tau \left( h\left( Z,Z^{\prime }\right) C_{Z,Z^{\prime }}\left( \theta
\right) \left\vert \Psi \left( \theta ,Z\right) \right\vert ^{2}\right) }{%
\left\vert \bar{\Psi}\left( \theta ,Z,Z^{\prime }\right) \right\vert ^{2}}
\end{equation*}%
These quantities depend on the accumulation of input outpt spks with
averages $\left\langle C\left( \theta \right) \right\rangle $ and $%
\left\langle D\left( \theta \right) \right\rangle $

Averages $\left\langle T\left( Z,Z^{\prime }\right) \right\rangle $, $%
\lambda \tau \left\langle \hat{T}\left( Z,Z^{\prime }\right) \right\rangle $
and $\left\vert \bar{\Psi}\left( \theta ,Z,Z^{\prime }\right) \right\vert
^{2}$ are quite constant if $\omega \left( \theta ,Z\right) $ have the same
freqnc, since $\left\vert \Psi \left( \theta ,Z\right) \right\vert ^{2}\sim
\nabla \omega \left( \theta ,Z\right) $ see Appendix 4.

The equations for the averages in the static background states reduce to:%
\begin{equation}
\left\langle T\left( Z,Z^{\prime }\right) \right\rangle =\lambda \tau
\left\langle \hat{T}\left( Z,Z^{\prime }\right) \right\rangle =\frac{\lambda
\tau h\left( Z,Z^{\prime }\right) \left\langle C_{Z,Z^{\prime }}\left(
\theta \right) \right\rangle \left\vert \Psi \left( Z\right) \right\vert ^{2}%
}{\left\vert \bar{\Psi}\left( Z,Z^{\prime }\right) \right\vert ^{2}}
\end{equation}%
with:%
\begin{equation*}
\left\vert \bar{\Psi}\left( \theta ,Z,Z^{\prime }\right) \right\vert ^{2}=%
\frac{\left\langle C_{Z,Z^{\prime }}\left( \theta \right) \right\rangle
\left\vert \Psi \left( Z\right) \right\vert ^{2}h_{C}\left( \omega \left(
\theta ,Z\right) \right) +\left\langle D_{Z,Z^{\prime }}\left( \theta
\right) \right\rangle \left\vert \Psi \left( Z^{\prime }\right) \right\vert
^{2}h_{D}\left( \omega \left( Z^{\prime }\right) \right) }{h_{C}\left(
\omega \left( Z\right) \right) }
\end{equation*}

\begin{eqnarray}
&&\omega ^{-1}\left( J,\theta ,Z,\left\vert \Psi \right\vert ^{2}\right) \\
&=&G\left( J\left( \theta ,Z\right) +\int \frac{\kappa }{N}\frac{\omega
\left( Z^{\prime }\right) T\left( Z,Z_{1}\right) }{\omega \left( Z\right) }%
\left( \mathcal{G}_{0}+\left\vert \Psi \left( Z_{1}\right) \right\vert
^{2}\right) dZ_{1}\right)  \notag
\end{eqnarray}%
\begin{equation}
T\left( Z,Z_{1}\right) =\int T\left\vert \Gamma \left( T,\hat{T},\omega
,\omega _{1},Z,Z_{1},C,D\right) \right\vert ^{2}dTd\hat{T}dCdD\equiv T\left(
Z,Z_{1}\right)
\end{equation}%
\begin{equation}
\left\langle T\left( Z,Z^{\prime }\right) \right\rangle =\frac{\lambda \tau
\exp \left( -\frac{\left\vert Z-Z^{\prime }\right\vert }{\nu c}\right) }{1+%
\frac{\alpha _{D}\omega h_{D}}{\alpha _{C}\omega ^{\prime }h_{C}}\frac{\frac{%
1}{\tau _{C}}+\alpha _{C}\omega ^{\prime }\left\vert \Psi \left( Z^{\prime
}\right) \right\vert ^{2}}{\frac{1}{\tau _{D}}+\alpha _{D}\omega \left\vert
\Psi \left( Z\right) \right\vert ^{2}}}
\end{equation}

We showed that under some approximations, the average values in this
background states present several possible patterns:%
\begin{eqnarray}
T\left( Z_{-},Z_{+}^{\prime }\right) &=&\frac{\lambda \tau \exp \left( -%
\frac{\left\vert Z-Z^{\prime }\right\vert }{\nu c}\right) \left( \frac{1}{%
\tau _{D}\alpha _{D}}+\frac{1}{b\bar{T}^{2}\left\langle \left\vert \Psi
_{0}\left( Z^{\prime }\right) \right\vert ^{2}\right\rangle _{Z}}\right) }{%
\frac{1}{\tau _{D}\alpha _{D}}+\frac{1}{\alpha _{C}\tau _{C}}+\frac{1}{b\bar{%
T}^{2}\left\langle \left\vert \Psi _{0}\left( Z^{\prime }\right) \right\vert
^{2}\right\rangle _{Z}}+b\bar{T}\left( \bar{T}\left\langle \left\vert \Psi
_{0}\left( Z^{\prime }\right) \right\vert ^{2}\right\rangle _{Z^{\prime
}}^{2}\right) ^{2}}\simeq 0  \label{LV} \\
T\left( Z_{+},Z_{+}^{\prime }\right) &=&\frac{\lambda \tau \exp \left( -%
\frac{\left\vert Z-Z^{\prime }\right\vert }{\nu c}\right) \left( \frac{1}{%
\tau _{D}\alpha _{D}}+b\bar{T}\left( \bar{T}\left\langle \left\vert \Psi
_{0}\left( Z^{\prime }\right) \right\vert ^{2}\right\rangle _{Z}^{2}\right)
^{2}\right) }{\frac{1}{\tau _{D}\alpha _{D}}+\frac{1}{\alpha _{C}\tau _{C}}+b%
\bar{T}\left( \bar{T}\left\langle \left\vert \Psi _{0}\left( Z^{\prime
}\right) \right\vert ^{2}\right\rangle _{Z}^{2}\right) ^{2}+b\bar{T}\left( 
\bar{T}\left\langle \left\vert \Psi _{0}\left( Z^{\prime }\right)
\right\vert ^{2}\right\rangle _{Z^{\prime }}^{2}\right) ^{2}}\simeq \frac{%
\lambda \tau \exp \left( -\frac{\left\vert Z-Z^{\prime }\right\vert }{\nu c}%
\right) }{2}  \notag \\
T\left( Z_{+},Z_{-}^{\prime }\right) &=&\frac{\lambda \tau \exp \left( -%
\frac{\left\vert Z-Z^{\prime }\right\vert }{\nu c}\right) \left( \frac{1}{%
\tau _{D}\alpha _{D}}+b\bar{T}\left( \bar{T}\left\langle \left\vert \Psi
_{0}\left( Z^{\prime }\right) \right\vert ^{2}\right\rangle _{Z}^{2}\right)
^{2}\right) }{\frac{1}{\tau _{D}\alpha _{D}}+\frac{1}{\alpha _{C}\tau _{C}}+b%
\bar{T}\left( \bar{T}\left\langle \left\vert \Psi _{0}\left( Z^{\prime
}\right) \right\vert ^{2}\right\rangle _{Z}^{2}\right) ^{2}+\frac{1}{b\bar{T}%
^{2}\left\langle \left\vert \Psi _{0}\left( Z^{\prime }\right) \right\vert
^{2}\right\rangle _{Z^{\prime }}}}\simeq \lambda \tau \exp \left( -\frac{%
\left\vert Z-Z^{\prime }\right\vert }{\nu c}\right)  \notag \\
T\left( Z_{-},Z_{-}^{\prime }\right) &\simeq &\frac{\lambda \tau \exp \left(
-\frac{\left\vert Z-Z^{\prime }\right\vert }{\nu c}\right) +\frac{1}{b\bar{T}%
^{2}\left\langle \left\vert \Psi _{0}\left( Z^{\prime }\right) \right\vert
^{2}\right\rangle _{Z}}}{1+\frac{\tau _{D}\alpha _{D}}{\alpha _{C}\tau _{C}}+%
\frac{1}{b\bar{T}^{2}\left\langle \left\vert \Psi _{0}\left( Z^{\prime
}\right) \right\vert ^{2}\right\rangle _{Z}}+\frac{1}{b\bar{T}%
^{2}\left\langle \left\vert \Psi _{0}\left( Z^{\prime }\right) \right\vert
^{2}\right\rangle _{Z^{\prime }}}}\simeq \frac{\lambda \tau \exp \left( -%
\frac{\left\vert Z-Z^{\prime }\right\vert }{\nu c}\right) }{2}  \notag
\end{eqnarray}

These result show the possbility of some emerging collective bound states.
These possible collective states are quite stationary for $\alpha _{C}\tau
_{C}<<1$ given the equation for connections. This ensures the stability of
such state at the time scale of activity oscillations.

Consequently, the field equation and the oscillations in activities can be
considered in a collective state of connections that shapes these
oscillations. This is the topic of Appendix 4.

\section*{Appendix 4. Minimization of $S_{e}\left( \Psi \right) $ for
dynamic fields. Oscillating activities around given background values.
Signal-induced oscillations}

\subsection*{A4.1 Classical level}

Equation (\ref{M}) is:%
\begin{equation*}
\omega ^{-1}\left( \theta ,Z,\left\vert \Psi \right\vert ^{2}\right)
=G\left( \int \frac{\kappa }{N}\frac{\omega \left( J,\theta -\frac{%
\left\vert Z-Z_{1}\right\vert }{c},Z_{1},\Psi \right) T\left( Z,\theta
,Z_{1},\theta -\frac{\left\vert Z-Z_{1}\right\vert }{c}\right) }{\omega
\left( J,\theta ,Z,\left\vert \Psi \right\vert ^{2}\right) }\left\vert \Psi
\left( \theta -\frac{\left\vert Z-Z_{1}\right\vert }{c},Z_{1}\right)
\right\vert ^{2}dZ_{1}\right)
\end{equation*}%
We can consider time dependent variations around the static equlbrm, due for
exampl to externl pertutbations by defining the deviations around the
equilibrium $\Omega \left( \theta ,Z\right) =\omega \left( \theta
,Z,\left\vert \Psi \right\vert ^{2}\right) -\omega _{0}\left( Z\right) $. We
show that this deviation satisfies some differential equation: 
\begin{equation}
G^{-1}\left( \omega \left( J\left( \theta \right) ,\theta \right) \right)
-G^{-1}\left( \omega _{0}\right) =J\left( \theta \right) +\frac{\hat{f}%
_{1}\nabla _{\theta }\omega \left( \theta ,Z\right) }{\omega \left( \theta
,Z\right) }+\frac{\hat{f}_{3}\nabla _{\theta }^{2}\omega \left( \theta
,Z\right) }{\omega \left( \theta ,Z\right) }+c^{2}\hat{f}_{3}\frac{\nabla
_{Z}^{2}\omega \left( \theta ,Z\right) }{\omega \left( \theta ,Z\right) }%
+T\Psi _{0}\delta \Psi \left( \theta ,Z\right)  \label{TS}
\end{equation}%
where:%
\begin{eqnarray*}
\hat{f}_{1} &=&\frac{W^{\prime }\left( 1\right) -W\left( 1\right) }{c}\Gamma
_{1}\text{, }\hat{f}_{3}=\frac{\left( W\left( 1\right) -W^{\prime }\left(
1\right) \right) \Gamma _{2}}{c^{2}} \\
\Gamma _{1} &=&\frac{\kappa }{NX_{r}}\int \left\vert Z-Z_{1}\right\vert
T\left( Z,Z_{1}\right) \mathcal{\bar{G}}_{0}\left( 0,Z_{1}\right) dZ_{1} \\
\Gamma _{2} &=&\frac{\kappa }{2NX_{r}}\int \left( Z-Z_{1}\right) ^{2}T\left(
Z,Z_{1}\right) \mathcal{\bar{G}}_{0}\left( 0,Z_{1}\right) dZ_{1}
\end{eqnarray*}%
The function $W\left( \frac{\omega \left( \theta ,Z\right) }{\omega \left(
\theta -\frac{\left\vert Z-Z_{1}\right\vert }{c},Z_{1}\right) }\right) $
computes the dependncy of $\left\langle T\left\vert \Gamma \left( T,\hat{T}%
,\theta ,Z,Z_{1}\right) \right\vert ^{2}\right\rangle $ in frequencies and $%
\mathcal{\bar{G}}_{0}\left( 0,Z_{1}\right) $ is some Green function
describing the signal propagation between two points of the thread. The
function represents some external perturbation inducing the deviation.

Disregarding the term $T\Psi _{0}\delta \Psi \left( \theta ,Z\right) $, for
every $Z$\ these equations (\ref{TS}) correspond to dampenng activity. This
dampening behaviour is the usual features for models of neural activities.
However considering the implications of statistical field theory allows to
refine and correct this result. Considering a static background field is
only an approximation.\ In a dynamic context, cells field in the background
oscillates and compensats dampning. Detrmined by solving field equation
frequencs. We show in \cite{GL} that such deviation modify the field $\Psi
\left( \theta ,Z\right) $ measuring the number of firing cells at pnt $Z$
and in turn the activity $\omega \left( J\left( \theta \right) ,\theta
,Z\right) $. In fact, the modification of the background is proportional to
the time variation of the deviation $\Omega \left( \theta ,Z\right) $, see
Appendix 5 for the following approximate dynamical solution:%
\begin{eqnarray}
\delta \Psi \left( \theta ,Z\right) &\simeq &\frac{\nabla _{\theta }\omega
\left( J,Z,\left\vert \Psi \right\vert ^{2}\right) }{V^{\prime \prime
}\left( \Psi _{0}\left( Z\right) \right) \omega _{0}^{2}\left(
J,Z,\left\vert \Psi _{0}\right\vert ^{2}\right) }\Psi _{0} \\
&\sim &\nabla _{\theta }\Omega \left( \theta ,Z\right)  \notag
\end{eqnarray}%
tht is:%
\begin{equation*}
\delta \Psi \left( \theta ,Z\right) \sim \nabla _{\theta }\Omega \left(
\theta ,Z\right)
\end{equation*}%
the proportionality depending on the caracteristic of the background and on
the potential $V$ controling the activity in the thread. Thus leads to
rewrite the last term in (\ref{TS}):%
\begin{eqnarray*}
T\delta \Psi \left( \theta ,Z\right) &\simeq &\delta \Psi \left( \theta
,Z\right) -\Gamma _{1}\nabla _{\theta }\delta \Psi \left( \theta ,Z\right) \\
&\simeq &N_{1}\nabla _{\theta }\omega _{0}\left( J,Z,\left\vert \Psi
_{0}\right\vert ^{2}\right) -N_{2}\nabla _{\theta }\omega _{0}\left(
J,Z,\left\vert \Psi _{0}\right\vert ^{2}\right)
\end{eqnarray*}%
This implies that (\ref{TS}) becomes:%
\begin{eqnarray*}
\hat{f}_{1} &=&\frac{W^{\prime }\left( 1\right) -W\left( 1\right) }{c}\Gamma
_{1}\text{, }\hat{f}_{3}=\frac{\left( W\left( 1\right) -W^{\prime }\left(
1\right) \right) \Gamma _{2}}{c^{2}} \\
\Gamma _{1} &=&\frac{\kappa }{NX_{r}}\int \left\vert Z-Z_{1}\right\vert
T\left( Z,Z_{1}\right) \mathcal{\bar{G}}_{0}\left( 0,Z_{1}\right) dZ_{1} \\
\Gamma _{2} &=&\frac{\kappa }{2NX_{r}}\int \left( Z-Z_{1}\right) ^{2}T\left(
Z,Z_{1}\right) \mathcal{\bar{G}}_{0}\left( 0,Z_{1}\right) dZ_{1}
\end{eqnarray*}%
\begin{equation}
f\Omega \left( \theta ,Z\right) =J\left( \theta ,Z\right) +\left( \frac{\hat{%
f}_{1}}{\omega \left( \theta ,Z\right) }+N_{1}\right) \nabla _{\theta
}\Omega \left( \theta ,Z\right) +\left( \frac{\hat{f}_{3}}{\omega \left(
\theta ,Z\right) }-N_{2}\right) \nabla _{\theta }^{2}\Omega \left( \theta
,Z\right) +\frac{c^{2}\hat{f}_{3}}{\omega \left( \theta ,Z\right) }\nabla
_{Z}^{2}\Omega \left( \theta ,Z\right)  \label{Ts}
\end{equation}%
with:%
\begin{eqnarray*}
N_{1} &=&\frac{\Psi _{0}\left( Z\right) }{U^{\prime \prime }\left(
X_{0}\right) \omega ^{2}\left( J\left( \theta \right) ,\theta ,Z,\mathcal{G}%
_{0}\right) } \\
N_{2} &=&\frac{\Gamma _{1}\Psi _{0}\left( Z\right) }{U^{\prime \prime
}\left( X_{0}\right) \omega ^{2}\left( J\left( \theta \right) ,\theta ,Z,%
\mathcal{G}_{0}\right) }
\end{eqnarray*}

Thus, the dynamic aspect of the background modify indirectly the activity
equations through coeffcients $N_{1}$ and $N_{2}$. This indirect
modifications stabilizes the fluctuations in activity. It results in some
non linear traveling solutions. These solutions may be replaced in first
approximation by linear solutions.

This stabilization is obtained by considering the coefficients $\frac{\hat{f}%
_{1}}{\omega \left( \theta ,Z\right) }+N_{1}$ responsible for dampened or
explosive solution. This coefficient varies with $\omega \left( J\left(
\theta \right) ,\theta ,Z,\mathcal{G}_{0}\right) $.

We show in \cite{GL} that the behavior of the solutions of (\ref{Ts})
depends on a threshold frequency $\omega _{1}$ the value of $\omega \left(
\theta ,Z\right) $ such that $\frac{\hat{f}_{1}}{\omega _{1}}+N_{1}=0$. When 
$\omega _{1}>\omega _{0}$, then $\omega _{0}$\ is a stable point since it
belongs to the domain in which $\frac{\hat{f}_{1}}{\omega \left( \theta
,Z\right) }+N_{1}<0$. Oscillatory patterns will dampen towards $\omega _{0}$%
. On the contrry, when $\omega _{1}<\omega _{0}$, the frequency $\omega _{0}$%
\ is an unstable point and an analyss shows tht the resulting dynamics thus
presents stable oscillations that are irregular in amplitude. The system
does not converge toward $\omega _{0}$ which remains unstable, but presents
the characteristic of some non linear travelling wave. This reslt wll be
useful while considering transitions of collective states.

Note that this result is more general than the one obtained in the linear
approximation. In our context, the possibility of travelling stable
oscillation is obtained for a whole range of parameters whereas the linear
approximation implies more restrictive condition. However, once the
possibility of travelling wave is understood, one can replace (\ref{TS}) by
its linear version (\ref{TS})reduces to:%
\begin{equation}
f\Omega \left( \theta ,Z\right) =J\left( \theta \right) +\left( \frac{\hat{f}%
_{1}}{\omega _{0}}+N_{1}\right) \nabla _{\theta }\Omega \left( \theta
,Z\right) +\left( \frac{\hat{f}_{3}}{\omega _{0}}-N_{2}\right) \nabla
_{\theta }^{2}\Omega \left( \theta ,Z\right) +\frac{c^{2}\hat{f}_{3}}{\omega
_{0}}\nabla _{Z}^{2}\Omega \left( \theta ,Z\right)  \label{wp}
\end{equation}
where $\omega _{0}$ is the average static solution, and where:%
\begin{equation}
\frac{\hat{f}_{1}}{\omega _{0}}+N_{1}=0  \label{nl}
\end{equation}%
is assumed to be satisfied to ensure persistent wave solution .

\subsection*{A4.2 Perturbative corrections}

The stabilizing effect previously describd is reinforced by including
perturbative corrections to the action through the effective action of the
system (see \cite{GL}). We show that this modifies the dynamic equations,
and introduces some non linear stabilizing effect. In (\ref{Ts}), the
perturbative corrections modify the coefficients $\frac{\hat{f}_{1}}{\omega
\left( \theta ,Z\right) }$ by adding some activity dependent term that
stabilize the equation.

vrll, dampening oscillations in activities modify the density of firing
neurons $\left\vert \Psi \left( \theta ,Z\right) \right\vert ^{2}$ that in
turn compensate this dampening.

The precise analysis is performd by starting with the effective action (\ref%
{gmf}). Equation (\ref{mG}) yields the corrective terms to $\omega
^{-1}\left( J\left( \theta \right) ,\theta ,Z,\mathcal{G}_{0}+\left\vert
\Psi \right\vert ^{2}\right) $ induced by $\Gamma _{p}\left( \Psi ^{\dagger
},\Psi \right) $. We focus on the weak field approximation to ensure
corrections of small magnitude.\ Since $p_{l}+\sum_{i}p_{l}^{i}\geqslant 2$,
the lowest order correction is for $m=1$ and $p_{l}+\sum_{i}p_{l}^{i}%
\geqslant 2=2$, and \cite{GL} shows that it is possible to define an
effective activity that can be identified as:%
\begin{equation}
\omega _{e}^{-1}\left( J\left( \theta \right) ,\theta ,Z,\mathcal{G}%
_{0}+\left\vert \Psi \right\vert ^{2}\right) =\omega ^{-1}\left( J\left(
\theta \right) ,\theta ,Z,\mathcal{G}_{0}+\left\vert \Psi \right\vert
^{2}\right) +Z
\end{equation}%
where $\omega \left( J\left( \theta \right) ,\theta ,Z,\mathcal{\bar{G}}%
_{0}+\left\vert \Psi \right\vert ^{2}\right) $ is the solution of:%
\begin{eqnarray*}
\omega ^{-1}\left( J,\theta ,Z,\left\vert \Psi \right\vert ^{2}\right)
&=&G\left( J\left( \theta ,Z\right) +\int \frac{\kappa }{N}\frac{\omega
\left( J,\theta -\frac{\left\vert Z-Z_{1}\right\vert }{c},Z_{1},\Psi \right)
T\left( Z,\theta ,Z_{1},\theta -\frac{\left\vert Z-Z_{1}\right\vert }{c}%
\right) }{\omega \left( J,\theta ,Z,\left\vert \Psi \right\vert ^{2}\right) }%
\right. \\
&&\times \left. \left( \mathcal{\bar{G}}_{0}\left( 0,Z_{1}\right)
+\left\vert \Psi \left( \theta -\frac{\left\vert Z-Z_{1}\right\vert }{c}%
,Z_{1}\right) \right\vert ^{2}\right) dZ_{1}\right)
\end{eqnarray*}%
Which is the classical activity equation, up to the inclusion of the Green
function $\mathcal{\bar{G}}_{0}\left( 0,Z_{1}\right) $ nd where:%
\begin{eqnarray}
Z &=&\int^{\theta }d\theta \sum_{\substack{ j\geqslant 1  \\ m\geqslant 1}}%
\sum_{\substack{ p_{l},\left( p_{l}^{i}\right) _{m\times j}  \\ %
p_{l}+\sum_{i}p_{l}^{i}\geqslant 2}}\left( \prod\limits_{i=1}^{m}\frac{%
\sharp _{j+1,m}\left( \left( p_{m},\left( p_{l}^{m}\right) \right) \right) }{%
4\bar{\sharp}_{j+1,m}\left( \left( p_{i},\left( p_{l}^{m}\right) \right)
\right) }\right) \frac{a_{j,m}}{2}  \label{Zqt} \\
&&\times \int \prod\limits_{i=1}^{m}\left\{ \frac{\delta
^{\sum_{l}p_{l}^{i}}\left( \nabla _{\theta }\omega ^{-1}\left( \theta
,Z,\left\vert \Psi \right\vert ^{2}\right) \right) }{\prod\limits_{l=1}^{j}%
\delta ^{p_{l}^{i}}\left\vert \Psi \left( \theta ^{\left( l\right)
},Z_{_{l}}\right) \right\vert ^{2}}\right\} \frac{\delta
^{\sum_{l}p_{l}}\left( \nabla _{\theta }\omega ^{-1}\left( \theta
,Z,\left\vert \Psi \right\vert ^{2}\right) \right) }{\prod\limits_{l=1}^{j}%
\delta ^{p_{l}}\left\vert \Psi \left( \theta ^{\left( l\right)
},Z_{_{l}}\right) \right\vert ^{2}}\prod\limits_{l=1}^{j}\left\vert \Psi
\left( \theta ^{\left( l\right) },Z_{l}\right) \right\vert ^{2}d\theta
^{\left( l\right) }dZ_{l}  \notag
\end{eqnarray}%
This series take into account the interaction between the frequencies and
the background field.

To find detailed results, we limit ourselves to the lowest order
corrections. We show that these corrections have the form:%
\begin{eqnarray}
\omega _{e}^{-1}\left( J\left( \theta \right) ,\theta ,Z,\mathcal{G}%
_{0}+\left\vert \Psi \right\vert ^{2}\right) &=&\omega ^{-1}\left( J\left(
\theta \right) ,\theta ,Z,\mathcal{G}_{0}+\left\vert \Psi \right\vert
^{2}\right)  \label{mge} \\
&&-\frac{f}{4\left( \frac{\hat{f}_{3}}{\omega \left( \theta ,Z\right) }%
-N_{2}\right) }\int \int^{\theta }\left( \frac{\delta \left( \omega
^{-1}\left( J\left( \theta \right) ,\theta ,Z,\mathcal{G}_{0}+\left\vert
\Psi \right\vert ^{2}\right) \right) }{\delta \left\vert \Psi \left( \theta
^{\left( l\right) },Z_{_{l}}\right) \right\vert ^{2}}\right) ^{2}  \notag \\
&&+\int \frac{1}{4}\nabla _{\theta }\left( \frac{\delta \left( \omega
^{-1}\left( J\left( \theta \right) ,\theta ,Z,\mathcal{G}_{0}+\left\vert
\Psi \right\vert ^{2}\right) \right) }{\delta \left\vert \Psi \left( \theta
^{\left( l\right) },Z_{_{l}}\right) \right\vert ^{2}}\right) ^{2}\left\vert
\Psi \left( \theta ^{\left( l\right) },Z_{_{l}}\right) \right\vert ^{2} 
\notag \\
&&+\frac{\frac{\hat{f}_{1}}{\omega \left( \theta ,Z\right) }+N_{1}}{4\left( 
\frac{\hat{f}_{3}}{\omega \left( \theta ,Z\right) }-N_{2}\right) }\int
\left( \frac{\delta \left( \omega ^{-1}\left( J\left( \theta \right) ,\theta
,Z,\mathcal{G}_{0}+\left\vert \Psi \right\vert ^{2}\right) \right) }{\delta
\left\vert \Psi \left( \theta ^{\left( l\right) },Z_{_{l}}\right)
\right\vert ^{2}}\right) ^{2}\left\vert \Psi \left( \theta ^{\left( l\right)
},Z_{_{l}}\right) \right\vert ^{2}  \notag
\end{eqnarray}%
In (\ref{mge}), the second and the third contributions:%
\begin{eqnarray}
&&-\frac{f}{4\left( \frac{\hat{f}_{3}}{\omega \left( \theta ,Z\right) }%
-N_{2}\right) }\int \int^{\theta }\left( \frac{\delta \left( \omega
^{-1}\left( J\left( \theta \right) ,\theta ,Z,\mathcal{G}_{0}+\left\vert
\Psi \right\vert ^{2}\right) \right) }{\delta \left\vert \Psi \left( \theta
^{\left( l\right) },Z_{_{l}}\right) \right\vert ^{2}}\right) ^{2}\left\vert
\Psi \left( \theta ^{\left( l\right) },Z_{_{l}}\right) \right\vert ^{2}
\label{ctB} \\
&&+\int \frac{1}{4}\nabla _{\theta }\left( \frac{\delta \left( \omega
^{-1}\left( J\left( \theta \right) ,\theta ,Z,\mathcal{G}_{0}+\left\vert
\Psi \right\vert ^{2}\right) \right) }{\delta \left\vert \Psi \left( \theta
^{\left( l\right) },Z_{_{l}}\right) \right\vert ^{2}}\right) ^{2}\left\vert
\Psi \left( \theta ^{\left( l\right) },Z_{_{l}}\right) \right\vert ^{2} 
\notag
\end{eqnarray}%
\bigskip describe the influence of the collective state defined by $%
\left\vert \Psi \left( \theta ^{\left( l\right) },Z_{_{l}}\right)
\right\vert ^{2}$ on the frequency at position $Z$ and time $\theta $. Under
our previous assumption that $\frac{\hat{f}_{3}}{\omega \left( \theta
,Z\right) }-N_{2}$ $<0$, and given that $f>0$, the first term in (\ref{ctB})
is positive, and thus, this term reduces $\omega \left( J\left( \theta
\right) ,\theta ,Z,\mathcal{G}_{0}+\left\vert \Psi \right\vert ^{2}\right) $%
. The higher the sensitivity $\frac{\delta \left( \omega ^{-1}\left( J\left(
\theta \right) ,\theta ,Z,\mathcal{G}_{0}+\left\vert \Psi \right\vert
^{2}\right) \right) }{\delta \left\vert \Psi \left( \theta ^{\left( l\right)
},Z_{_{l}}\right) \right\vert ^{2}}$ of the frequency to the collective
state, the more the frequency of the wave is reduced. The effect of this
smooting is cumulative in time, as shown by the integral over time arising
in this term. \ The second contribution in (\ref{ctB}) amplifies this
smoothing. Actually, this term is positive when $\frac{\delta \left( \omega
^{-1}\left( J\left( \theta \right) ,\theta ,Z,\mathcal{G}_{0}+\left\vert
\Psi \right\vert ^{2}\right) \right) }{\delta \left\vert \Psi \left( \theta
^{\left( l\right) },Z_{_{l}}\right) \right\vert ^{2}}$, i.e. the sensitivity
of frequency to the background field, increases in absolute value.\ As a
consequence, it reduces the oscillations of $\omega \left( J\left( \theta
\right) ,\theta ,Z,\mathcal{G}_{0}+\left\vert \Psi \right\vert ^{2}\right) $
when the frequency's dependency in the background field increases at
position $Z$ and time $\theta $.

The fourth term in (\ref{mge}):%
\begin{equation*}
\frac{\frac{\hat{f}_{1}}{\omega \left( \theta ,Z\right) }+N_{1}}{4\left( 
\frac{\hat{f}_{3}}{\omega \left( \theta ,Z\right) }-N_{2}\right) }\int
\left( \frac{\delta \left( \omega ^{-1}\left( J\left( \theta \right) ,\theta
,Z,\mathcal{G}_{0}+\left\vert \Psi \right\vert ^{2}\right) \right) }{\delta
\left\vert \Psi \left( \theta ^{\left( l\right) },Z_{_{l}}\right)
\right\vert ^{2}}\right) ^{2}\left\vert \Psi \left( \theta ^{\left( l\right)
},Z_{_{l}}\right) \right\vert ^{2}
\end{equation*}%
reinforces the mechanism of oscillation stabilization described in section
8.3. It has the sign of $-\left( \frac{\hat{f}_{1}}{\omega \left( \theta
,Z\right) }+N_{1}\right) $, given our assumption $\frac{\hat{f}_{3}}{\omega
\left( \theta ,Z\right) }-N_{2}$ $<0$ ensuring oscillatory behavior of $%
\omega \left( J\left( \theta \right) ,\theta ,Z,\mathcal{G}_{0}+\left\vert
\Psi \right\vert ^{2}\right) $. Thus, the correction to $\omega \left(
J\left( \theta \right) ,\theta ,Z,\mathcal{G}_{0}+\left\vert \Psi
\right\vert ^{2}\right) $ induced by this term has the sign of $\left( \frac{%
\hat{f}_{1}}{\omega \left( \theta ,Z\right) }+N_{1}\right) $: for $\frac{%
\hat{f}_{1}}{\omega \left( \theta ,Z\right) }+N_{1}>0$, the approximative
solution is oscillatory and explosive. Thus, the correction amplifies the
oscillations of $\omega \left( \theta ,Z\right) $ and the stabilization
mechanism applies. For $\omega \left( \theta ,Z\right) $ such that $\frac{%
\hat{f}_{1}}{\omega \left( \theta ,Z\right) }+N_{1}<0$, the correction term
turns negative and further decreases $\omega \left( \theta ,Z\right) $.

The series of higher corrections for (\ref{Zqt}) is computed in \cite{GL}.
This papr shows that, in the local approximation, the frequencies can be
described by a wave equation whose form depends on the stabilization
potential and the evolution of the background itself.

To sum up, the perturbative corrections account for interactions between the
classical solutions and the entire thread. Moreover these interactions
stabilize the traveling waves.

\subsection*{A4.3 General patterns of activities dynamics}

Several points sum-up the previous results and are useful for the rest of
the study.

1. The activities, and their oscillations wll impact the connectivity
states. They also depend on the connectivity, A state involving both
quantities corresponds to a state involving both systems and the stability
involve both variables.

2. Given the wave equation for activities, frequencies of activities can be
imposed by some signals. Some oscillatng solution exists built from these
signals. Both activities and connectivities states emerge and are built from
these signals.

3. Some interferences in signals leads to increased connectivit states.

\section*{Appendix 5. Stability}

We rewrite the saddle point equations, and then compute their first order
expansion around a particular saddle point characterizing a particular
collective state for a given group.

\subsection*{A5.1 Saddle-point equations}

\subsubsection*{A5.1.1 Derivation of the equations}

The saddle point equation for the effective action of field $\Psi \left(
\theta ,Z\right) $ is given by:%
\begin{eqnarray}
0 &\simeq &\frac{\delta \hat{S}_{cl}\left( \Psi ^{\dagger },\Psi \right) }{%
\delta \Psi ^{\dagger }\left( \theta ,Z\right) }+\Psi \left( \theta
,Z\right) \sum_{\substack{ j\geqslant 1  \\ m\geqslant 1}}\sum_{\substack{ %
\left( p_{l}^{i}\right) _{\left( m+1\right) \times j}  \\ %
\sum_{i}p_{l}^{i}\geqslant 2}}\sum_{k=1}^{m+1}a_{j,m}\int \int \left( \frac{1%
}{2}\frac{\delta ^{\sum_{l}p_{l}^{k}}\left( \nabla _{\theta }\omega
^{-1}\left( \theta ,Z,\left\vert \Psi \right\vert ^{2}\right) \right) }{%
\prod\limits_{l=1}^{j}\delta ^{\sum_{l}p_{l}^{k}}\left\vert \Psi \left(
\theta ^{\left( l\right) },Z_{_{l}}\right) \right\vert ^{2}}\right) \\
&&\times \prod\limits_{\substack{ i=1  \\ i\neq k}}^{m+1}\int \left\{ \frac{1%
}{2}\frac{\delta ^{\sum_{l}p_{l}^{i}}\left( \nabla _{\theta }\omega
^{-1}\left( \theta _{i},Z_{i},\left\vert \Psi \right\vert ^{2}\right)
\right) }{\prod\limits_{l=1}^{j}\delta ^{\sum_{l}p_{l}^{i}}\left\vert \Psi
\left( \theta ^{\left( l\right) },Z_{_{l}}\right) \right\vert ^{2}}%
\left\vert \Psi \left( \theta _{i},Z_{i}\right) \right\vert ^{2}\right\}
\left( \prod\limits_{l=1}^{j}\left\vert \Psi \left( \theta ^{\left( l\right)
},Z_{l}\right) \right\vert ^{2}\right) \prod\limits_{i=1}^{m}d\theta
_{i}dZ_{_{l}}\prod\limits_{l=1}^{j}d\theta ^{\left( l\right) }dZ_{l}  \notag
\\
&&+\Psi \left( \theta ,Z\right) \sum_{\substack{ j\geqslant 1  \\ m\geqslant
1 }}\sum_{\substack{ \left( p_{l}^{i}\right) _{\left( m+1\right) \times j} 
\\ \sum_{i}p_{l}^{i}\geqslant 2}}a_{j,m}\int \int
\sum_{k=1}^{j}\prod\limits_{i=1}^{m+1}\int \left\{ \frac{1}{2}\frac{\delta
^{\sum_{l}p_{l}^{i}}\left( \nabla _{\theta }\omega ^{-1}\left( \theta
_{i},Z_{i},\left\vert \Psi \right\vert ^{2}\right) \right) }{\delta
^{p_{k}^{i}}\left\vert \Psi \left( \theta ,Z\right) \right\vert
^{2}\prod\limits_{\substack{ l=1  \\ l\neq k}}^{j}\delta
^{\sum_{l}p_{l}^{i}}\left\vert \Psi \left( \theta ^{\left( l\right)
},Z_{_{l}}\right) \right\vert ^{2}}\left\vert \Psi \left( \theta
_{i},Z_{i}\right) \right\vert ^{2}\right\}  \notag \\
&&\times \left( \prod\limits_{\substack{ l=1  \\ l\neq k}}^{j}\left\vert
\Psi \left( \theta ^{\left( l\right) },Z_{l}\right) \right\vert ^{2}\right)
\prod\limits_{i=1}^{m}d\theta _{i}dZ_{_{l}}\prod\limits_{l=1}^{j}d\theta
^{\left( l\right) }dZ_{l}  \notag \\
&&+\Psi \left( \theta ,Z\right) \sum_{\substack{ j\geqslant 1  \\ m\geqslant
1 }}\sum_{\substack{ \left( p_{l}^{i}\right) _{\left( m+1\right) \times j} 
\\ \sum_{i}p_{l}^{i}\geqslant 2}}a_{j,m}\int \int \sum_{k=1}^{m+1}\left(
\int \Psi ^{\dagger }\left( \theta _{k},Z_{k}\right) \left\{ \frac{1}{2}%
\frac{\delta ^{1+\sum_{l}p_{l}^{k}}\left( \nabla _{\theta }\omega
^{-1}\left( \theta _{k},Z_{k},\left\vert \Psi \right\vert ^{2}\right)
\right) }{\delta \left\vert \Psi \left( \theta ,Z\right) \right\vert
^{2}\prod\limits_{l=1}^{j}\delta ^{\sum_{l}p_{l}^{k}}\left\vert \Psi \left(
\theta ^{\left( l\right) },Z_{_{l}}\right) \right\vert ^{2}}\right\} \Psi
\left( \theta _{k},Z_{k}\right) \right)  \notag \\
&&\times \prod\limits_{\substack{ i=1  \\ i\neq k}}^{m+1}\left\{ \frac{1}{2}%
\frac{\delta ^{\sum_{l}p_{l}^{i}}\left( \nabla _{\theta }\omega ^{-1}\left(
\theta _{i},Z_{i},\left\vert \Psi \right\vert ^{2}\right) \right) }{%
\prod\limits_{l=1}^{j}\delta ^{\sum_{l}p_{l}^{i}}\left\vert \Psi \left(
\theta ^{\left( l\right) },Z_{_{l}}\right) \right\vert ^{2}}\left\vert \Psi
\left( \theta _{i},Z_{i}\right) \right\vert ^{2}\right\} \left(
\prod\limits_{l=1}^{j}\left\vert \Psi \left( \theta ^{\left( l\right)
},Z_{l}\right) \right\vert ^{2}\right) \prod\limits_{i=1}^{m}d\theta
_{i}dZ_{_{l}}\prod\limits_{l=1}^{j}d\theta ^{\left( l\right) }dZ_{l}  \notag
\end{eqnarray}%
The several terms can be regrouped as:%
\begin{eqnarray}
0 &\simeq &\frac{\delta \hat{S}_{cl}\left( \Psi ^{\dagger },\Psi \right) }{%
\delta \Psi ^{\dagger }\left( \theta ,Z\right) }+\Psi \left( \theta
,Z\right) \sum_{\substack{ j\geqslant 1  \\ m\geqslant 1}}\sum_{\substack{ %
\left( p_{l}^{i}\right) _{\left( m+1\right) \times j}  \\ %
\sum_{i}p_{l}^{i}\geqslant 2}}\sum_{k=1}^{m+1}a_{j,m}  \label{Spt} \\
&&\times \int \int \left( \frac{1}{2}\frac{\delta ^{\sum_{l}p_{l}^{k}}\left[
\left( \nabla _{\theta }\omega ^{-1}\left( \theta ,Z,\left\vert \Psi
\right\vert ^{2}\right) +\nabla _{\theta }M\left( \left\{ \theta
_{k},Z_{k}\right\} ,\left\{ \theta ,Z\right\} \right) \right) \left\vert
\Psi \left( \theta _{k},Z_{k}\right) \right\vert ^{2}\right] }{%
\prod\limits_{l=1}^{j}\delta ^{\sum_{l}p_{l}^{k}}\left\vert \Psi \left(
\theta ^{\left( l\right) },Z_{_{l}}\right) \right\vert ^{2}}\right)  \notag
\\
&&\times \prod\limits_{\substack{ i=1  \\ i\neq k}}^{m+1}\int \left\{ \frac{1%
}{2}\frac{\delta ^{\sum_{l}p_{l}^{i}}\left( \nabla _{\theta }\omega
^{-1}\left( \theta _{i},Z_{i},\left\vert \Psi \right\vert ^{2}\right)
\right) }{\prod\limits_{l=1}^{j}\delta ^{\sum_{l}p_{l}^{i}}\left\vert \Psi
\left( \theta ^{\left( l\right) },Z_{_{l}}\right) \right\vert ^{2}}%
\left\vert \Psi \left( \theta _{i},Z_{i}\right) \right\vert ^{2}\right\}
\left( \prod\limits_{l=1}^{j}\left\vert \Psi \left( \theta ^{\left( l\right)
},Z_{l}\right) \right\vert ^{2}\right) \prod\limits_{i=1}^{m}d\theta
_{i}dZ_{_{l}}\prod\limits_{l=1}^{j}d\theta ^{\left( l\right) }dZ_{l}  \notag
\\
&&+\Psi \left( \theta ,Z\right) \sum_{\substack{ j\geqslant 1  \\ m\geqslant
1 }}\sum_{\substack{ \left( p_{l}^{i}\right) _{\left( m+1\right) \times j} 
\\ \sum_{i}p_{l}^{i}\geqslant 2}}a_{j,m}\int \int
\sum_{k=1}^{j}\prod\limits_{i=1}^{m+1}\int \left\{ \frac{1}{2}\frac{\delta
^{\sum_{l}p_{l}^{i}}\left( \nabla _{\theta }\omega ^{-1}\left( \theta
_{i},Z_{i},\left\vert \Psi \right\vert ^{2}\right) \right) }{\delta
^{p_{k}^{i}}\left\vert \Psi \left( \theta ,Z\right) \right\vert
^{2}\prod\limits_{\substack{ l=1  \\ l\neq k}}^{j}\delta
^{\sum_{l}p_{l}^{i}}\left\vert \Psi \left( \theta ^{\left( l\right)
},Z_{_{l}}\right) \right\vert ^{2}}\left\vert \Psi \left( \theta
_{i},Z_{i}\right) \right\vert ^{2}\right\}  \notag \\
&&\times \left( \prod\limits_{\substack{ l=1  \\ l\neq k}}^{j}\left\vert
\Psi \left( \theta ^{\left( l\right) },Z_{l}\right) \right\vert ^{2}\right)
\prod\limits_{i=1}^{m}d\theta _{i}dZ_{_{l}}\prod\limits_{l=1}^{j}d\theta
^{\left( l\right) }dZ_{l}  \notag
\end{eqnarray}%
We consider the notation:%
\begin{eqnarray*}
&&\prod\limits_{\substack{ i=1  \\ i\neq k}}^{m+1}\int \left\{ \frac{1}{2}%
\frac{\delta ^{\sum_{l}p_{l}^{i}}\left( \nabla _{\theta }\omega ^{-1}\left(
\theta _{i},Z_{i},\left\vert \Psi \right\vert ^{2}\right) \right) }{%
\prod\limits_{l=1}^{j}\delta ^{\sum_{l}p_{l}^{i}}\left\vert \Psi \left(
\theta ^{\left( l\right) },Z_{_{l}}\right) \right\vert ^{2}}\left\vert \Psi
\left( \theta _{i},Z_{i}\right) \right\vert ^{2}\right\} \\
&\rightarrow &\prod\limits_{\substack{ i=1  \\ i\neq k}}^{m+1}\frac{1}{2}%
\nabla _{\theta }M\left( \left( \theta _{i},Z_{i}\right) ,\left\{
p_{l}^{i},\left( \theta ^{\left( l\right) },Z_{_{l}}\right) \right\} \right)
\left\vert \Psi \left( \theta _{i},Z_{i}\right) \right\vert ^{2}
\end{eqnarray*}%
so that the first term writes:%
\begin{eqnarray*}
&&\frac{\delta \hat{S}_{cl}\left( \Psi ^{\dagger },\Psi \right) }{\delta
\Psi ^{\dagger }\left( \theta ,Z\right) } \\
&\rightarrow &-\frac{1}{2}\left( \nabla _{\theta }\left( \frac{\sigma
_{\theta }^{2}}{2}\nabla _{\theta }-\omega ^{-1}\left( J\left( \theta
\right) ,\theta ,Z,\mathcal{G}_{0}+\Psi \right) \right) \right) \Psi \left(
\theta ,Z\right) \\
&&+\frac{1}{2}\nabla _{\theta }M\left( \left( \theta _{1},Z_{1}\right)
,\left( \theta ,Z\right) \right) \left\vert \Psi \left( \theta
_{1},Z_{1}\right) \right\vert ^{2}\Psi \left( \theta ,Z\right) +U^{\prime
}\left( \left\vert \Psi \left( \theta ,Z\right) \right\vert ^{2}\right) \Psi
\left( \theta ,Z\right)
\end{eqnarray*}%
so that we have:%
\begin{eqnarray*}
&&\int \int \frac{1}{2}\frac{\delta ^{\sum_{l}p_{l}^{k}}\left[ \left( \nabla
_{\theta }\omega ^{-1}\left( \theta ,Z,\left\vert \Psi \right\vert
^{2}\right) +\nabla _{\theta }M\left( \left\{ \theta _{k},Z_{k}\right\}
,\left\{ \theta ,Z\right\} \right) \right) \left\vert \Psi \left( \theta
_{k},Z_{k}\right) \right\vert ^{2}\right] }{\prod\limits_{l=1}^{j}\delta
^{\sum_{l}p_{l}^{k}}\left\vert \Psi \left( \theta ^{\left( l\right)
},Z_{_{l}}\right) \right\vert ^{2}} \\
&\rightarrow &\frac{1}{2}\left( \nabla _{\theta }M\left( \left( \theta
,Z\right) ,\left\{ p_{l}^{k},\left( \theta ^{\left( l\right)
},Z_{_{l}}\right) \right\} \right) +\nabla _{\theta }M\left( \left( \theta
_{k},Z_{k}\right) ,\left\{ p_{l}^{k},\left( \theta ^{\left( l\right)
},Z_{_{l}}\right) \right\} ,\left\{ 1,\left( \theta ,Z\right) \right\}
\right) \left\vert \Psi \left( \theta _{k},Z_{k}\right) \right\vert
^{2}\right)
\end{eqnarray*}%
and:%
\begin{eqnarray*}
&&\prod\limits_{i=1}^{m+1}\int \left\{ \frac{1}{2}\frac{\delta
^{\sum_{l}p_{l}^{i}}\left( \nabla _{\theta }\omega ^{-1}\left( \theta
_{i},Z_{i},\left\vert \Psi \right\vert ^{2}\right) \right) }{\delta
^{p_{k}^{i}}\left\vert \Psi \left( \theta ,Z\right) \right\vert
^{2}\prod\limits_{\substack{ l=1  \\ l\neq k}}^{j}\delta
^{\sum_{l}p_{l}^{i}}\left\vert \Psi \left( \theta ^{\left( l\right)
},Z_{_{l}}\right) \right\vert ^{2}}\left\vert \Psi \left( \theta
_{i},Z_{i}\right) \right\vert ^{2}\right\} \\
&\rightarrow &\prod\limits_{i=1}^{m+1}\frac{1}{2}\nabla _{\theta }M\left(
\left( \theta _{i},Z_{i}\right) ,\left\{ p_{l}^{i},\left( \theta ^{\left(
l\right) },Z_{_{l}}\right) \right\} _{l\neq k},\left\{ p_{k}^{i},\left(
\theta ,Z\right) \right\} \right) \left\vert \Psi \left( \theta
_{i},Z_{i}\right) \right\vert ^{2}
\end{eqnarray*}%
The second term of (\ref{Spt}) becomes:%
\begin{eqnarray*}
&&\Psi \left( \theta ,Z\right) \sum_{k=1}^{m+1}\frac{1}{2}\left( \nabla
_{\theta }M\left( \left( \theta ,Z\right) ,\left\{ p_{l}^{k},\left( \theta
^{\left( l\right) },Z_{_{l}}\right) \right\} \right) +\nabla _{\theta
}M\left( \left( \theta _{k},Z_{k}\right) ,\left\{ p_{l}^{k},\left( \theta
^{\left( l\right) },Z_{_{l}}\right) \right\} ,\left\{ 1,\left( \theta
,Z\right) \right\} \right) \left\vert \Psi \left( \theta _{k},Z_{k}\right)
\right\vert ^{2}\right) \\
&&\times \left( \prod\limits_{\substack{ i=1  \\ i\neq k}}^{m+1}\frac{1}{2}%
\nabla _{\theta }M\left( \left( \theta _{i},Z_{i}\right) ,\left\{
p_{l}^{i},\left( \theta ^{\left( l\right) },Z_{_{l}}\right) \right\} \right)
\left\vert \Psi \left( \theta _{i},Z_{i}\right) \right\vert ^{2}\right)
\left( \prod\limits_{l=1}^{j}\left\vert \Psi \left( \theta ^{\left( l\right)
},Z_{l}\right) \right\vert ^{2}\right)
\end{eqnarray*}%
and the third term writes:%
\begin{eqnarray*}
&&\Psi \left( \theta ,Z\right) \sum_{k=1}^{j}\prod\limits_{i=1}^{m+1}\frac{1%
}{2}\nabla _{\theta }M\left( \left( \theta _{i},Z_{i}\right) ,\left\{
p_{l}^{i},\left( \theta ^{\left( l\right) },Z_{_{l}}\right) \right\} _{l\neq
k},\left\{ p_{k}^{i},\left( \theta ,Z\right) \right\} \right) \left\vert
\Psi \left( \theta _{i},Z_{i}\right) \right\vert ^{2} \\
&&\times \left( \prod\limits_{\substack{ l=1  \\ l\neq k}}^{j}\left\vert
\Psi \left( \theta ^{\left( l\right) },Z_{l}\right) \right\vert ^{2}\right)
\end{eqnarray*}%
leading ultimately to the saddle point equation:%
\begin{eqnarray}
0 &\rightarrow &-\frac{1}{2}\left( \nabla _{\theta }\left( \frac{\sigma
_{\theta }^{2}}{2}\nabla _{\theta }-\omega ^{-1}\left( J\left( \theta
\right) ,\theta ,Z,\mathcal{G}_{0}+\Psi \right) \right) \right) \Psi \left(
\theta ,Z\right)  \label{qsn} \\
&&+\frac{1}{2}\nabla _{\theta }M\left( \left( \theta _{1},Z_{1}\right)
,\left( \theta ,Z\right) \right) \left\vert \Psi \left( \theta
_{1},Z_{1}\right) \right\vert ^{2}\Psi \left( \theta ,Z\right) +U^{\prime
}\left( \left\vert \Psi \left( \theta ,Z\right) \right\vert ^{2}\right) \Psi
\left( \theta ,Z\right)  \notag \\
&&+\Psi \left( \theta ,Z\right) \sum_{\substack{ j\geqslant 1  \\ m\geqslant
1 }}\sum_{\substack{ \left( p_{l}^{i}\right) _{\left( m+1\right) \times j} 
\\ \sum_{i}p_{l}^{i}\geqslant 2}}\sum_{k=1}^{m+1}a_{j,m}\sum_{k=1}^{m+1}%
\frac{1}{2}\left( \nabla _{\theta }M\left( \left( \theta ,Z\right) ,\left\{
p_{l}^{k},\left( \theta ^{\left( l\right) },Z_{_{l}}\right) \right\} \right)
\right.  \notag \\
&&\left. +\nabla _{\theta }M\left( \left( \theta _{k},Z_{k}\right) ,\left\{
p_{l}^{k},\left( \theta ^{\left( l\right) },Z_{_{l}}\right) \right\}
,\left\{ 1,\left( \theta ,Z\right) \right\} \right) \left\vert \Psi \left(
\theta _{k},Z_{k}\right) \right\vert ^{2}\right)  \notag \\
&&\times \left( \prod\limits_{\substack{ i=1  \\ i\neq k}}^{m+1}\frac{1}{2}%
\nabla _{\theta }M\left( \left( \theta _{i},Z_{i}\right) ,\left\{
p_{l}^{i},\left( \theta ^{\left( l\right) },Z_{_{l}}\right) \right\} \right)
\left\vert \Psi \left( \theta _{i},Z_{i}\right) \right\vert ^{2}\right)
\prod\limits_{l=1}^{j}\left\vert \Psi \left( \theta ^{\left( l\right)
},Z_{l}\right) \right\vert ^{2}  \notag \\
&&+\Psi \left( \theta ,Z\right) \sum_{k=1}^{j}\prod\limits_{i=1}^{m+1}\frac{1%
}{2}\nabla _{\theta }M\left( \left( \theta _{i},Z_{i}\right) ,\left\{
p_{l}^{i},\left( \theta ^{\left( l\right) },Z_{_{l}}\right) \right\} _{l\neq
k},\left\{ p_{k}^{i},\left( \theta ,Z\right) \right\} \right) \left\vert
\Psi \left( \theta _{i},Z_{i}\right) \right\vert ^{2}\prod\limits 
_{\substack{ l=1  \\ l\neq k}}^{j}\left\vert \Psi \left( \theta ^{\left(
l\right) },Z_{l}\right) \right\vert ^{2}  \notag
\end{eqnarray}

\subsubsection*{A5.1.2 Approximate solution for $\Psi \left( \protect\theta %
,Z\right) $}

\bigskip around $\Psi _{0}\left( \theta ,Z\right) $ and setting $V=1$,
yields at the second order:%
\begin{eqnarray*}
\Gamma \left( \Psi ,\Psi ^{\dag }\right) &=&-\frac{1}{2}\int \delta \Psi
^{\dagger }\left( \theta ,Z\right) \left( \nabla _{\theta }\left( \frac{%
\sigma _{\theta }^{2}}{2}\nabla _{\theta }-\omega ^{-1}\left( J\left( \theta
\right) ,\theta ,Z,\mathcal{G}_{0}+\left\vert \Psi \right\vert ^{2}\right)
\right) \right) X_{0} \\
&&-\frac{1}{2}\int \delta \Psi ^{\dagger }\left( \theta ,Z\right) \left(
\nabla _{\theta }\left( \frac{\sigma _{\theta }^{2}}{2}\nabla _{\theta
}-\omega ^{-1}\left( J\left( \theta \right) ,\theta ,Z,\mathcal{G}%
_{0}+\left\vert \Psi \right\vert ^{2}\right) \right) \right) \delta \Psi
\left( \theta ,Z\right) \\
&&+\frac{1}{2}\int \delta \Psi ^{\dagger }\left( \theta ,Z\right) U^{\prime
\prime }\left( X_{0}\right) \delta \Psi \left( \theta ,Z\right)
\end{eqnarray*}%
with $\left\vert \Psi \right\vert ^{2}=X_{0}+\sqrt{X_{0}}\left( \delta
\left( \Psi ^{\dagger }+\delta \Psi \right) \right) $. This leads to the
first order condition for $\delta \Psi \left( \theta _{1},Z_{1}\right) $:%
\begin{eqnarray*}
0 &=&\frac{1}{2}\delta \Psi ^{\dagger }\left( \theta ,Z\right) \left(
-\nabla _{\theta }\left( \frac{\sigma _{\theta }^{2}}{2}\nabla _{\theta
}-\omega ^{-1}\left( J\left( \theta \right) ,\theta ,Z,\mathcal{G}%
_{0}+X_{0}\right) \right) +U^{\prime \prime }\left( X_{0}\right) \right) \\
&&-\frac{1}{2}\int \delta \Psi ^{\dagger }\left( \theta _{1},Z_{1}\right) 
\sqrt{X_{0}}\left( \nabla _{\theta }\frac{\delta \omega ^{-1}\left( J\left(
\theta _{1}\right) ,\theta _{1},Z_{1},\mathcal{G}_{0}+X_{0}\right) }{\delta
\left\vert \Psi \left( \theta ,Z\right) \right\vert ^{2}}\right)
X_{0}d\theta _{1}dZ_{1}
\end{eqnarray*}%
with solution $\delta \Psi ^{\dagger }\left( \theta ,Z\right) =0$. This
implies that the first order condition for $\delta \Psi ^{\dag }\left(
\theta ,Z\right) $ becomes: 
\begin{eqnarray}
&&0=-\frac{1}{2}\left( \nabla _{\theta }\left( \frac{\sigma _{\theta }^{2}}{2%
}\nabla _{\theta }-\omega ^{-1}\left( J\left( \theta \right) ,\theta ,Z,%
\mathcal{G}_{0}+\left\vert \Psi \right\vert ^{2}\right) \right) \right) X_{0}
\label{sDP} \\
&&-\frac{1}{2}\left( \nabla _{\theta }\left( \frac{\sigma _{\theta }^{2}}{2}%
\nabla _{\theta }-\omega ^{-1}\left( J\left( \theta \right) ,\theta ,Z,%
\mathcal{G}_{0}+\left\vert \Psi \right\vert ^{2}\right) \right) \right)
\delta \Psi \left( \theta ,Z\right)  \notag \\
&&+\frac{1}{2}U^{\prime \prime }\left( X_{0}\right) \delta \Psi \left(
\theta ,Z\right)  \notag
\end{eqnarray}%
Equation (\ref{sDP}) also rewrites:%
\begin{equation}
\left( -\left( \nabla _{\theta }\left( \frac{\sigma _{\theta }^{2}}{2}\nabla
_{\theta }-\omega ^{-1}\left( J\left( \theta \right) ,\theta ,Z,\mathcal{G}%
_{0}+\left\vert \Psi \right\vert ^{2}\right) \right) \right) +U^{\prime
\prime }\left( X_{0}\right) \right) \left( \delta \Psi \left( \theta
,Z\right) +X_{0}\right) =U^{\prime \prime }\left( X_{0}\right) X_{0}
\label{sDT}
\end{equation}%
Equation (\ref{sDT}) can be used to write $\delta \Psi \left( \theta
,Z\right) $ as a function of $\omega ^{-1}\left( J\left( \theta \right)
,\theta ,Z,\mathcal{G}_{0}+\left\vert \Psi \right\vert ^{2}\right) $:%
\begin{eqnarray}
\delta \Psi \left( \theta ,Z\right) &=&\left( \frac{\left( \nabla _{\theta
}\left( \frac{\sigma _{\theta }^{2}}{2}\nabla _{\theta }-\omega ^{-1}\left(
J\left( \theta \right) ,\theta ,Z,\mathcal{G}_{0}+\left\vert \Psi
\right\vert ^{2}\right) \right) \right) }{U^{\prime \prime }\left(
X_{0}\right) -\left( \nabla _{\theta }\left( \frac{\sigma _{\theta }^{2}}{2}%
\nabla _{\theta }-\omega ^{-1}\left( J\left( \theta \right) ,\theta ,Z,%
\mathcal{G}_{0}+\left\vert \Psi \right\vert ^{2}\right) \right) \right) }%
\right) X_{0}  \label{psv} \\
&=&-\frac{\nabla _{\theta }\left( \omega ^{-1}\left( J\left( \theta \right)
,\theta ,Z,\mathcal{G}_{0}+\left\vert \Psi \right\vert ^{2}\right) \right) }{%
U^{\prime \prime }\left( X_{0}\right) -\left( \nabla _{\theta }\left( \frac{%
\sigma _{\theta }^{2}}{2}\nabla _{\theta }-\omega ^{-1}\left( J\left( \theta
\right) ,\theta ,Z,\mathcal{G}_{0}+\left\vert \Psi \right\vert ^{2}\right)
\right) \right) }X_{0}  \notag
\end{eqnarray}%
In first approximation, for $U^{\prime \prime }\left( X_{0}\right) >>1$ and $%
\sigma _{\theta }^{2}<<1$, this yields: 
\begin{eqnarray}
\delta \Psi \left( \theta ,Z\right) &\simeq &-\frac{\nabla _{\theta }\omega
^{-1}\left( J\left( \theta \right) ,\theta ,Z,\mathcal{G}_{0}+\left\vert
\Psi \right\vert ^{2}\right) }{U^{\prime \prime }\left( X_{0}\right) +\nabla
_{\theta }\omega ^{-1}\left( J\left( \theta \right) ,\theta ,Z,\mathcal{G}%
_{0}+\left\vert \Psi \right\vert ^{2}\right) }X_{0}  \label{psG} \\
&\simeq &-\frac{\nabla _{\theta }\omega ^{-1}\left( J\left( \theta \right)
,\theta ,Z,\mathcal{G}_{0}+\left\vert \Psi \right\vert ^{2}\right) }{%
U^{\prime \prime }\left( X_{0}\right) }X_{0}  \notag
\end{eqnarray}

\subsection*{A5.2 First order variation of saddle-point equations}

The first order variation of (\ref{qsn}) involvs the terms:%
\begin{eqnarray*}
&&\Psi \left( \theta ,Z\right) \sum_{k=1}^{m+1}\frac{1}{2}\left( \nabla
_{\theta }M\left( \left( \theta _{k},Z_{k}\right) ,\left\{ p_{l}^{k},\left(
\theta ^{\left( l\right) },Z_{_{l}}\right) \right\} ,\left\{ 1,\left( \theta
,Z\right) \right\} \right) \Psi ^{\dag }\left( \theta _{k},Z_{k}\right)
\delta \Psi \left( \theta _{k},Z_{k}\right) \right) \\
&&\times \left( \prod\limits_{\substack{ i=1  \\ i\neq k}}^{m+1}\frac{1}{2}%
\nabla _{\theta }M\left( \left( \theta _{i},Z_{i}\right) ,\left\{
p_{l}^{i},\left( \theta ^{\left( l\right) },Z_{_{l}}\right) \right\} \right)
\left\vert \Psi \left( \theta _{i},Z_{i}\right) \right\vert ^{2}\right)
\prod\limits_{l=1}^{j}\left\vert \Psi \left( \theta ^{\left( l\right)
},Z_{l}\right) \right\vert ^{2}
\end{eqnarray*}%
and:%
\begin{eqnarray*}
&&\Psi \left( \theta ,Z\right) \sum_{k=1}^{m+1}\frac{1}{2}\left( \nabla
_{\theta }M\left( \left( \theta ,Z\right) ,\left\{ p_{l}^{k},\left( \theta
^{\left( l\right) },Z_{_{l}}\right) \right\} \right) +\nabla _{\theta
}M\left( \left( \theta _{k},Z_{k}\right) ,\left\{ p_{l}^{k},\left( \theta
^{\left( l\right) },Z_{_{l}}\right) \right\} ,\left\{ 1,\left( \theta
,Z\right) \right\} \right) \left\vert \Psi \left( \theta _{k},Z_{k}\right)
\right\vert ^{2}\right) \\
&&\times \left\{ \sum_{k^{\prime }=1,k^{\prime }\neq k}^{m+1}\Psi ^{\dag
}\left( \theta _{k^{\prime }},Z_{k^{\prime }}\right) \delta \Psi \left(
\theta _{k^{\prime }},Z_{k^{\prime }}\right) \left( \prod\limits_{\substack{ %
i=1  \\ i\neq k,k^{\prime }}}^{m+1}\frac{1}{2}\nabla _{\theta }M\left(
\left( \theta _{i},Z_{i}\right) ,\left\{ p_{l}^{i},\left( \theta ^{\left(
l\right) },Z_{_{l}}\right) \right\} \right) \left\vert \Psi \left( \theta
_{i},Z_{i}\right) \right\vert ^{2}\right) \prod\limits_{l=1}^{j}\left\vert
\Psi \left( \theta ^{\left( l\right) },Z_{l}\right) \right\vert ^{2}\right.
\\
&&+\left. \left( \prod\limits_{\substack{ i=1  \\ i\neq k}}^{m+1}\frac{1}{2}%
\nabla _{\theta }M\left( \left( \theta _{i},Z_{i}\right) ,\left\{
p_{l}^{i},\left( \theta ^{\left( l\right) },Z_{_{l}}\right) \right\} \right)
\left\vert \Psi \left( \theta _{i},Z_{i}\right) \right\vert ^{2}\right)
\sum_{l^{\prime }=1}^{j}\Psi ^{\dag }\left( \theta _{l^{\prime
}},Z_{l^{\prime }}\right) \delta \Psi \left( \theta _{l^{\prime
}},Z_{l^{\prime }}\right) \prod\limits_{\substack{ l=1  \\ l\neq l^{\prime } 
}}^{j}\left\vert \Psi \left( \theta ^{\left( l\right) },Z_{l}\right)
\right\vert ^{2}\right\}
\end{eqnarray*}

\bigskip so that the first order equation becomes:%
\begin{eqnarray*}
0 &=&\frac{1}{2}\nabla _{\theta }M\left( \left( \theta _{1},Z_{1}\right)
,\left( \theta ,Z\right) \right) \Psi ^{\dag }\left( \theta
_{1},Z_{1}\right) \delta \Psi \left( \theta _{1},Z_{1}\right) \Psi \left(
\theta ,Z\right) \\
&&+\Psi \left( \theta ,Z\right) \sum_{\substack{ j\geqslant 1  \\ m\geqslant
1 }}\sum_{\substack{ \left( p_{l}^{i}\right) _{\left( m+1\right) \times j} 
\\ \sum_{i}p_{l}^{i}\geqslant 2}}\sum_{k=1}^{m+1}a_{j,m}\sum_{k=1}^{m+1}%
\frac{1}{2}\left( \nabla _{\theta }M\left( \left( \theta _{k},Z_{k}\right)
,\left\{ p_{l}^{k},\left( \theta ^{\left( l\right) },Z_{_{l}}\right)
\right\} ,\left\{ 1,\left( \theta ,Z\right) \right\} \right) \Psi ^{\dag
}\left( \theta _{k},Z_{k}\right) \delta \Psi \left( \theta _{k},Z_{k}\right)
\right) \\
&&\times \left( \prod\limits_{\substack{ i=1  \\ i\neq k}}^{m+1}\frac{1}{2}%
\nabla _{\theta }M\left( \left( \theta _{i},Z_{i}\right) ,\left\{
p_{l}^{i},\left( \theta ^{\left( l\right) },Z_{_{l}}\right) \right\} \right)
\left\vert \Psi \left( \theta _{i},Z_{i}\right) \right\vert ^{2}\right)
\prod\limits_{l=1}^{j}\left\vert \Psi \left( \theta ^{\left( l\right)
},Z_{l}\right) \right\vert ^{2} \\
&&+\Psi \left( \theta ,Z\right) \sum_{\substack{ j\geqslant 1  \\ m\geqslant
1 }}\sum_{\substack{ \left( p_{l}^{i}\right) _{\left( m+1\right) \times j} 
\\ \sum_{i}p_{l}^{i}\geqslant 2}}\sum_{k=1}^{m+1}a_{j,m} \\
&&\times \sum_{k=1}^{m+1}\frac{1}{2}\left( \nabla _{\theta }M\left( \left(
\theta ,Z\right) ,\left\{ p_{l}^{k},\left( \theta ^{\left( l\right)
},Z_{_{l}}\right) \right\} \right) +\nabla _{\theta }M\left( \left( \theta
_{k},Z_{k}\right) ,\left\{ p_{l}^{k},\left( \theta ^{\left( l\right)
},Z_{_{l}}\right) \right\} ,\left\{ 1,\left( \theta ,Z\right) \right\}
\right) \left\vert \Psi \left( \theta _{k},Z_{k}\right) \right\vert
^{2}\right) \\
&&\times \left\{ \sum_{k^{\prime }=1,k^{\prime }\neq k}^{m+1}\frac{\Psi
^{\dag }\left( \theta _{k^{\prime }},Z_{k^{\prime }}\right) }{\left\vert
\Psi \left( \theta _{k^{\prime }},Z_{k^{\prime }}\right) \right\vert ^{2}}%
\delta \Psi \left( \theta _{k^{\prime }},Z_{k^{\prime }}\right) \left(
\prod\limits_{\substack{ i=1  \\ i\neq k}}^{m+1}\frac{1}{2}\nabla _{\theta
}M\left( \left( \theta _{i},Z_{i}\right) ,\left\{ p_{l}^{i},\left( \theta
^{\left( l\right) },Z_{_{l}}\right) \right\} \right) \left\vert \Psi \left(
\theta _{i},Z_{i}\right) \right\vert ^{2}\right)
\prod\limits_{l=1}^{j}\left\vert \Psi \left( \theta ^{\left( l\right)
},Z_{l}\right) \right\vert ^{2}\right. \\
&&+\left. \left( \prod\limits_{\substack{ i=1  \\ i\neq k}}^{m+1}\frac{1}{2}%
\nabla _{\theta }M\left( \left( \theta _{i},Z_{i}\right) ,\left\{
p_{l}^{i},\left( \theta ^{\left( l\right) },Z_{_{l}}\right) \right\} \right)
\left\vert \Psi \left( \theta _{i},Z_{i}\right) \right\vert ^{2}\right)
\sum_{l^{\prime }=1}^{j}\Psi ^{\dag }\left( \theta _{l^{\prime
}},Z_{l^{\prime }}\right) \delta \Psi \left( \theta _{l^{\prime
}},Z_{l^{\prime }}\right) \prod\limits_{\substack{ l=1  \\ l\neq l^{\prime } 
}}^{j}\left\vert \Psi \left( \theta ^{\left( l\right) },Z_{l}\right)
\right\vert ^{2}\right\}
\end{eqnarray*}

Using the saddle point equation (\ref{Spt}), we can replace:%
\begin{eqnarray*}
&&-\frac{1}{2}\left( \nabla _{\theta }\left( \frac{\sigma _{\theta }^{2}}{2}%
\nabla _{\theta }-\omega ^{-1}\left( J\left( \theta \right) ,\theta ,Z,%
\mathcal{G}_{0}+\Psi \right) \right) \right) \delta \Psi \left( \theta
,Z\right) \\
&&+\frac{1}{2}\nabla _{\theta }M\left( \left( \theta _{1},Z_{1}\right)
,\left( \theta ,Z\right) \right) \left\vert \Psi \left( \theta
_{1},Z_{1}\right) \right\vert ^{2}\delta \Psi \left( \theta ,Z\right)
+U^{\prime }\left( \left\vert \Psi \left( \theta ,Z\right) \right\vert
^{2}\right) \delta \Psi \left( \theta ,Z\right) +U^{\prime \prime }\left(
\left\vert \Psi \left( \theta ,Z\right) \right\vert ^{2}\right) \left\vert
\Psi \left( \theta ,Z\right) \right\vert ^{2}\delta \Psi \left( \theta
,Z\right) \\
&&+\delta \Psi \left( \theta ,Z\right) \sum_{\substack{ j\geqslant 1  \\ %
m\geqslant 1}}\sum_{\substack{ \left( p_{l}^{i}\right) _{\left( m+1\right)
\times j}  \\ \sum_{i}p_{l}^{i}\geqslant 2}}\sum_{k=1}^{m+1}a_{j,m} \\
&&\times \sum_{k=1}^{m+1}\frac{1}{2}\left( \nabla _{\theta }M\left( \left(
\theta ,Z\right) ,\left\{ p_{l}^{k},\left( \theta ^{\left( l\right)
},Z_{_{l}}\right) \right\} \right) +\nabla _{\theta }M\left( \left( \theta
_{k},Z_{k}\right) ,\left\{ p_{l}^{k},\left( \theta ^{\left( l\right)
},Z_{_{l}}\right) \right\} ,\left\{ 1,\left( \theta ,Z\right) \right\}
\right) \left\vert \Psi \left( \theta _{k},Z_{k}\right) \right\vert
^{2}\right) \\
&&\times \left( \prod\limits_{\substack{ i=1  \\ i\neq k}}^{m+1}\frac{1}{2}%
\nabla _{\theta }M\left( \left( \theta _{i},Z_{i}\right) ,\left\{
p_{l}^{i},\left( \theta ^{\left( l\right) },Z_{_{l}}\right) \right\} \right)
\left\vert \Psi \left( \theta _{i},Z_{i}\right) \right\vert ^{2}\right)
\prod\limits_{l=1}^{j}\left\vert \Psi \left( \theta ^{\left( l\right)
},Z_{l}\right) \right\vert ^{2} \\
&&+\delta \Psi \left( \theta ,Z\right) \sum_{\substack{ j\geqslant 1  \\ %
m\geqslant 1}}\sum_{\substack{ \left( p_{l}^{i}\right) _{\left( m+1\right)
\times j}  \\ \sum_{i}p_{l}^{i}\geqslant 2}}\sum_{k=1}^{m+1}a_{j,m} \\
&&\times \sum_{k=1}^{j}\prod\limits_{i=1}^{m+1}\frac{1}{2}\nabla _{\theta
}M\left( \left( \theta _{i},Z_{i}\right) ,\left\{ p_{l}^{i},\left( \theta
^{\left( l\right) },Z_{_{l}}\right) \right\} _{l\neq k},\left\{
p_{k}^{i},\left( \theta ,Z\right) \right\} \right) \left\vert \Psi \left(
\theta _{i},Z_{i}\right) \right\vert ^{2}\prod\limits_{\substack{ l=1  \\ %
l\neq k}}^{j}\left\vert \Psi \left( \theta ^{\left( l\right) },Z_{l}\right)
\right\vert ^{2}
\end{eqnarray*}%
by:%
\begin{eqnarray}
&&\left( -\frac{1}{2}\left( \nabla _{\theta }\left( \frac{\sigma _{\theta
}^{2}}{2}\nabla _{\theta }-\omega ^{-1}\left( J\left( \theta \right) ,\theta
,Z,\mathcal{G}_{0}+\Psi \right) \right) \right) +A\left( \Psi \left( \theta
,Z\right) \right) \nabla _{\theta }+U^{\prime \prime }\left( \left\vert \Psi
\left( \theta ,Z\right) \right\vert ^{2}\right) \left\vert \Psi \left(
\theta ,Z\right) \right\vert ^{2}\right) \delta \Psi \left( \theta ,Z\right)
\notag \\
&&+\delta \Psi \left( \theta ,Z\right) \left( \frac{1}{2}\left( \nabla
_{\theta }\left( \frac{\sigma _{\theta }^{2}}{2}\nabla _{\theta }-\omega
^{-1}\left( J\left( \theta \right) ,\theta ,Z,\mathcal{G}_{0}+\Psi \right)
\right) \right) +A\left( \Psi \left( \theta ,Z\right) \right) \nabla
_{\theta }\right) \Psi \left( \theta ,Z\right)  \label{nb}
\end{eqnarray}%
where $A\left( \Psi \left( \theta ,Z\right) \right) \nabla _{\theta }$
combines all the derivatives of $\Psi \left( \theta ,Z\right) $ arising in
the saddle point equation. Setting:%
\begin{equation*}
\delta \Psi \left( \theta ,Z\right) =\frac{\delta \Psi \left( \theta
,Z\right) }{\Psi \left( \theta ,Z\right) }\Psi \left( \theta ,Z\right)
\end{equation*}%
expression (\ref{nb}) becomes:%
\begin{eqnarray*}
&&\Psi \left( \theta ,Z\right) \left( \frac{1}{2}\left( -\frac{\sigma
_{\theta }^{2}}{2}\nabla _{\theta }+\omega ^{-1}\left( J\left( \theta
\right) ,\theta ,Z,\mathcal{G}_{0}+\Psi \right) \right) \nabla _{\theta
}+\left( A\left( \Psi \left( \theta ,Z\right) \right) -\frac{\sigma _{\theta
}^{2}}{2}\right) \nabla _{\theta }\right) \frac{\delta \Psi \left( \theta
,Z\right) }{\Psi \left( \theta ,Z\right) } \\
&&+U^{\prime \prime }\left( \left\vert \Psi \left( \theta ,Z\right)
\right\vert ^{2}\right) \left\vert \Psi \left( \theta ,Z\right) \right\vert
^{2}\delta \Psi \left( \theta ,Z\right)
\end{eqnarray*}%
As a consequence,the first order variation writes as a sum of two
expressions:%
\begin{equation}
0=E_{1}+E_{2}  \label{Smv}
\end{equation}%
where:%
\begin{eqnarray*}
E_{1} &=&\Psi \left( \theta ,Z\right) \left( \frac{1}{2}\left( -\frac{\sigma
_{\theta }^{2}}{2}\nabla _{\theta }+\omega ^{-1}\left( J\left( \theta
\right) ,\theta ,Z,\mathcal{G}_{0}+\Psi \right) \right) \nabla _{\theta
}+\left( A\left( \Psi \left( \theta ,Z\right) \right) -\frac{\sigma _{\theta
}^{2}}{2}\right) \nabla _{\theta }\right) \frac{\delta \Psi \left( \theta
,Z\right) }{\Psi \left( \theta ,Z\right) } \\
&&+U^{\prime \prime }\left( \left\vert \Psi \left( \theta ,Z\right)
\right\vert ^{2}\right) \left\vert \Psi \left( \theta ,Z\right) \right\vert
^{2}\delta \Psi \left( \theta ,Z\right) \\
&&\frac{1}{2}\nabla _{\theta }\left( \delta \omega ^{-1}\left( J\left(
\theta \right) ,\theta ,Z,\mathcal{G}_{0}+\Psi \right) \right) \Psi \left(
\theta ,Z\right) +\frac{1}{2}\nabla _{\theta }\delta M\left( \left( \theta
_{1},Z_{1}\right) ,\left( \theta ,Z\right) \right) \left\vert \Psi \left(
\theta _{1},Z_{1}\right) \right\vert ^{2}\Psi \left( \theta ,Z\right) \\
&&+\Psi \left( \theta ,Z\right) \sum_{\substack{ j\geqslant 1  \\ m\geqslant
1 }}\sum_{\substack{ \left( p_{l}^{i}\right) _{\left( m+1\right) \times j} 
\\ \sum_{i}p_{l}^{i}\geqslant 2}}\sum_{k=1}^{m+1}a_{j,m} \\
&&\times \sum_{k=1}^{m+1}\frac{1}{2}\left( \nabla _{\theta }\delta M\left(
\left( \theta ,Z\right) ,\left\{ p_{l}^{k},\left( \theta ^{\left( l\right)
},Z_{_{l}}\right) \right\} \right) +\nabla _{\theta }\delta M\left( \left(
\theta _{k},Z_{k}\right) ,\left\{ p_{l}^{k},\left( \theta ^{\left( l\right)
},Z_{_{l}}\right) \right\} ,\left\{ 1,\left( \theta ,Z\right) \right\}
\right) \left\vert \Psi \left( \theta _{k},Z_{k}\right) \right\vert
^{2}\right) \\
&&\times \left( \prod\limits_{\substack{ i=1  \\ i\neq k}}^{m+1}\frac{1}{2}%
\nabla _{\theta }M\left( \left( \theta _{i},Z_{i}\right) ,\left\{
p_{l}^{i},\left( \theta ^{\left( l\right) },Z_{_{l}}\right) \right\} \right)
\left\vert \Psi \left( \theta _{i},Z_{i}\right) \right\vert ^{2}\right)
\prod\limits_{l=1}^{j}\left\vert \Psi \left( \theta ^{\left( l\right)
},Z_{l}\right) \right\vert ^{2} \\
&&+\frac{1}{2}\nabla _{\theta }M\left( \left( \theta _{1},Z_{1}\right)
,\left( \theta ,Z\right) \right) \Psi ^{\dag }\left( \theta
_{1},Z_{1}\right) \delta \Psi \left( \theta _{1},Z_{1}\right) \Psi \left(
\theta ,Z\right)
\end{eqnarray*}%
and:%
\begin{eqnarray*}
&&E_{2}=\Psi \left( \theta ,Z\right) \sum_{\substack{ j\geqslant 1  \\ %
m\geqslant 1}}\sum_{\substack{ \left( p_{l}^{i}\right) _{\left( m+1\right)
\times j}  \\ \sum_{i}p_{l}^{i}\geqslant 2}}\sum_{k=1}^{m+1}a_{j,m}%
\sum_{k=1}^{m+1}\frac{1}{2}\left( \nabla _{\theta }M\left( \left( \theta
_{k},Z_{k}\right) ,\left\{ p_{l}^{k},\left( \theta ^{\left( l\right)
},Z_{_{l}}\right) \right\} ,\left\{ 1,\left( \theta ,Z\right) \right\}
\right) \Psi ^{\dag }\left( \theta _{k},Z_{k}\right) \delta \Psi \left(
\theta _{k},Z_{k}\right) \right) \\
&&\times \left( \prod\limits_{\substack{ i=1  \\ i\neq k}}^{m+1}\frac{1}{2}%
\nabla _{\theta }M\left( \left( \theta _{i},Z_{i}\right) ,\left\{
p_{l}^{i},\left( \theta ^{\left( l\right) },Z_{_{l}}\right) \right\} \right)
\left\vert \Psi \left( \theta _{i},Z_{i}\right) \right\vert ^{2}\right)
\prod\limits_{l=1}^{j}\left\vert \Psi \left( \theta ^{\left( l\right)
},Z_{l}\right) \right\vert ^{2} \\
&&+\Psi \left( \theta ,Z\right) \sum_{k=1}^{m+1}\frac{1}{2}\left( \nabla
_{\theta }M\left( \left( \theta ,Z\right) ,\left\{ p_{l}^{k},\left( \theta
^{\left( l\right) },Z_{_{l}}\right) \right\} \right) +\nabla _{\theta
}M\left( \left( \theta _{k},Z_{k}\right) ,\left\{ p_{l}^{k},\left( \theta
^{\left( l\right) },Z_{_{l}}\right) \right\} ,\left\{ 1,\left( \theta
,Z\right) \right\} \right) \left\vert \Psi \left( \theta _{k},Z_{k}\right)
\right\vert ^{2}\right) \\
&&\times \left\{ \sum_{k^{\prime }=1,k^{\prime }\neq k}^{m+1}\frac{\delta
\Psi \left( \theta _{k^{\prime }},Z_{k^{\prime }}\right) }{\Psi \left(
\theta _{k^{\prime }},Z_{k^{\prime }}\right) }\left( \prod\limits 
_{\substack{ i=1  \\ i\neq k,k^{\prime }}}^{m+1}\frac{1}{2}\nabla _{\theta
}M\left( \left( \theta _{i},Z_{i}\right) ,\left\{ p_{l}^{i},\left( \theta
^{\left( l\right) },Z_{_{l}}\right) \right\} \right) \left\vert \Psi \left(
\theta _{i},Z_{i}\right) \right\vert ^{2}\right)
\prod\limits_{l=1}^{j}\left\vert \Psi \left( \theta ^{\left( l\right)
},Z_{l}\right) \right\vert ^{2}\right. \\
&&+\left. \left( \prod\limits_{\substack{ i=1  \\ i\neq k}}^{m+1}\frac{1}{2}%
\nabla _{\theta }M\left( \left( \theta _{i},Z_{i}\right) ,\left\{
p_{l}^{i},\left( \theta ^{\left( l\right) },Z_{_{l}}\right) \right\} \right)
\left\vert \Psi \left( \theta _{i},Z_{i}\right) \right\vert ^{2}\right)
\sum_{l^{\prime }=1}^{j}\Psi ^{\dag }\left( \theta _{l^{\prime
}},Z_{l^{\prime }}\right) \delta \Psi \left( \theta _{l^{\prime
}},Z_{l^{\prime }}\right) \prod\limits_{\substack{ l=1  \\ l\neq k}}%
^{j}\left\vert \Psi \left( \theta ^{\left( l\right) },Z_{l}\right)
\right\vert ^{2}\right\}
\end{eqnarray*}%
\bigskip

\subsection*{A5.3 Fluctuation equation for activities}

\bigskip To obtain the fluctuation around a state connections, activity and
field, we start with:

\begin{eqnarray*}
\omega ^{-1}\left( J,\theta ,Z,\left\vert \Psi \right\vert ^{2}\right)
&=&G\left( J\left( \theta ,Z\right) +\int \frac{\kappa }{N}\frac{\omega
\left( J,\theta -\frac{\left\vert Z-Z_{1}\right\vert }{c},Z_{1},\Psi \right)
T\left( Z,\theta ,Z_{1},\theta -\frac{\left\vert Z-Z_{1}\right\vert }{c}%
\right) }{\omega \left( J,\theta ,Z,\left\vert \Psi \right\vert ^{2}\right) }%
\right. \\
&&\times \left. \left( \mathcal{\bar{G}}_{0}\left( 0,Z_{1}\right)
+\left\vert \Psi \left( \theta -\frac{\left\vert Z-Z_{1}\right\vert }{c}%
,Z_{1}\right) \right\vert ^{2}\right) dZ_{1}\right)
\end{eqnarray*}%
also written as:%
\begin{eqnarray*}
&&G^{-1}\left( \omega ^{-1}\left( J,\theta ,Z,\left\vert \Psi \right\vert
^{2}\right) \right) \\
&=&J\left( \theta ,Z\right) +\int \frac{\kappa }{N}\frac{\omega \left(
J,\theta -\frac{\left\vert Z-Z_{1}\right\vert }{c},Z_{1},\Psi \right)
T\left( Z,\theta ,Z_{1},\theta -\frac{\left\vert Z-Z_{1}\right\vert }{c}%
\right) }{\omega \left( J,\theta ,Z,\left\vert \Psi \right\vert ^{2}\right) }%
\left( \mathcal{\bar{G}}_{0}\left( 0,Z_{1}\right) +\left\vert \Psi \left(
\theta -\frac{\left\vert Z-Z_{1}\right\vert }{c},Z_{1}\right) \right\vert
^{2}\right) dZ_{1}
\end{eqnarray*}%
We then obtain, the derivatives of $\omega \left( J,\theta ,Z,\left\vert
\Psi \right\vert ^{2}\right) $ nd $\omega ^{-1}\left( J,\theta ,Z,\left\vert
\Psi \right\vert ^{2}\right) $. They are the equations for the system of
fluctuations of activities.

\subsubsection*{A5.3.1 Derivation of $\protect\delta \protect\omega \left( 
\protect\theta ,Z,\left\vert \Psi \right\vert ^{2}\right) $}

We start from:%
\begin{eqnarray*}
&&G^{-1}\left( \omega ^{-1}\left( J,\theta ,Z,\left\vert \Psi \right\vert
^{2}\right) \right) \\
&=&J\left( \theta ,Z\right) +\int \frac{\kappa }{N}\frac{\omega \left(
J,\theta -\frac{\left\vert Z-Z_{1}\right\vert }{c},Z_{1},\Psi \right)
T\left( Z,\theta ,Z_{1},\theta -\frac{\left\vert Z-Z_{1}\right\vert }{c}%
\right) }{\omega \left( J,\theta ,Z,\left\vert \Psi \right\vert ^{2}\right) }%
\left( \mathcal{\bar{G}}_{0}\left( 0,Z_{1}\right) +\left\vert \Psi \left(
\theta -\frac{\left\vert Z-Z_{1}\right\vert }{c},Z_{1}\right) \right\vert
^{2}\right) dZ_{1}
\end{eqnarray*}%
that can be rewritten, using a first order expansion in the variable $\delta
\omega \left( \theta ,Z,\left\vert \Psi \right\vert ^{2}\right) $:

\begin{eqnarray*}
&&-\delta \omega \left( \theta ,Z,\left\vert \Psi \right\vert ^{2}\right) 
\frac{\left( \omega ^{-1}\left( \theta ,Z,\left\vert \Psi \right\vert
^{2}\right) \right) ^{2}}{G^{\prime }\left( \omega ^{-1}\left( J,\theta
,Z,\left\vert \Psi \right\vert ^{2}\right) \right) } \\
&=&\int \frac{\kappa }{N}\frac{\delta \omega \left( \theta -\frac{\left\vert
Z-Z_{1}\right\vert }{c},Z_{1},\Psi \right) T\left( Z,\theta ,Z_{1},\theta -%
\frac{\left\vert Z-Z_{1}\right\vert }{c}\right) }{\omega \left( J,\theta
,Z,\left\vert \Psi \right\vert ^{2}\right) }\left( \mathcal{\bar{G}}%
_{0}\left( 0,Z_{1}\right) +\left\vert \Psi \left( \theta -\frac{\left\vert
Z-Z_{1}\right\vert }{c},Z_{1}\right) \right\vert ^{2}\right) dZ_{1} \\
&&-\delta \omega \left( \theta ,Z,\left\vert \Psi \right\vert ^{2}\right)
\left( \omega ^{-1}\left( \theta ,Z,\left\vert \Psi \right\vert ^{2}\right)
\right) \int \frac{\kappa }{N}\frac{\omega \left( \theta -\frac{\left\vert
Z-Z_{1}\right\vert }{c},Z_{1},\Psi \right) T\left( Z,\theta ,Z_{1},\theta -%
\frac{\left\vert Z-Z_{1}\right\vert }{c}\right) }{\omega \left( J,\theta
,Z,\left\vert \Psi \right\vert ^{2}\right) } \\
&&\times \left( \mathcal{\bar{G}}_{0}\left( 0,Z_{1}\right) +\left\vert \Psi
\left( \theta -\frac{\left\vert Z-Z_{1}\right\vert }{c},Z_{1}\right)
\right\vert ^{2}\right) dZ_{1} \\
&&+\int \frac{\kappa }{N}\frac{\omega \left( \theta -\frac{\left\vert
Z-Z_{1}\right\vert }{c},Z_{1},\Psi \right) \delta T\left( Z,\theta
,Z_{1},\theta -\frac{\left\vert Z-Z_{1}\right\vert }{c}\right) }{\omega
\left( J,\theta ,Z,\left\vert \Psi \right\vert ^{2}\right) }\left( \mathcal{%
\bar{G}}_{0}\left( 0,Z_{1}\right) +\left\vert \Psi \left( \theta -\frac{%
\left\vert Z-Z_{1}\right\vert }{c},Z_{1}\right) \right\vert ^{2}\right)
dZ_{1} \\
&&+\int \frac{\kappa }{N}\frac{\delta \omega \left( \theta -\frac{\left\vert
Z-Z_{1}\right\vert }{c},Z_{1},\Psi \right) T\left( Z,\theta ,Z_{1},\theta -%
\frac{\left\vert Z-Z_{1}\right\vert }{c}\right) }{\omega \left( J,\theta
,Z,\left\vert \Psi \right\vert ^{2}\right) }\mathcal{\delta }\left\vert \Psi
\left( \theta -\frac{\left\vert Z-Z_{1}\right\vert }{c},Z_{1}\right)
\right\vert ^{2}dZ_{1}
\end{eqnarray*}

that is:\bigskip

\begin{eqnarray}
&&\left( G^{-1}\left( \omega ^{-1}\left( \theta ,Z,\left\vert \Psi
\right\vert ^{2}\right) \right) -\frac{\left( G^{-1}\right) ^{\prime }\left(
\omega ^{-1}\left( \theta ,Z,\left\vert \Psi \right\vert ^{2}\right) \right) 
}{\left( \omega \left( \theta ,Z,\left\vert \Psi \right\vert ^{2}\right)
\right) }\right) \frac{\delta \omega \left( \theta ,Z,\left\vert \Psi
\right\vert ^{2}\right) }{\left( \omega \left( \theta ,Z,\left\vert \Psi
\right\vert ^{2}\right) \right) }  \label{FET} \\
&=&\int \frac{\kappa }{N}\frac{\delta \omega \left( \theta -\frac{\left\vert
Z-Z_{1}\right\vert }{c},Z_{1},\Psi \right) T\left( Z,\theta ,Z_{1},\theta -%
\frac{\left\vert Z-Z_{1}\right\vert }{c}\right) }{\omega \left( J,\theta
,Z,\left\vert \Psi \right\vert ^{2}\right) }\left( \mathcal{\bar{G}}%
_{0}\left( 0,Z_{1}\right) +\left\vert \Psi \left( \theta -\frac{\left\vert
Z-Z_{1}\right\vert }{c},Z_{1}\right) \right\vert ^{2}\right) dZ_{1}  \notag
\\
&&+\int \frac{\kappa }{N}\frac{\omega \left( \theta -\frac{\left\vert
Z-Z_{1}\right\vert }{c},Z_{1},\Psi \right) \delta T\left( Z,\theta
,Z_{1},\theta -\frac{\left\vert Z-Z_{1}\right\vert }{c}\right) }{\omega
\left( J,\theta ,Z,\left\vert \Psi \right\vert ^{2}\right) }\left( \mathcal{%
\bar{G}}_{0}\left( 0,Z_{1}\right) +\left\vert \Psi \left( \theta -\frac{%
\left\vert Z-Z_{1}\right\vert }{c},Z_{1}\right) \right\vert ^{2}\right)
dZ_{1}  \notag \\
&&+\int \frac{\kappa }{N}\frac{\omega \left( \theta -\frac{\left\vert
Z-Z_{1}\right\vert }{c},Z_{1},\Psi \right) T\left( Z,\theta ,Z_{1},\theta -%
\frac{\left\vert Z-Z_{1}\right\vert }{c}\right) }{\omega \left( J,\theta
,Z,\left\vert \Psi \right\vert ^{2}\right) }\mathcal{\delta }\left\vert \Psi
\left( \theta -\frac{\left\vert Z-Z_{1}\right\vert }{c},Z_{1}\right)
\right\vert ^{2}dZ_{1}  \notag
\end{eqnarray}

The fluctuation $\delta \omega \left( \theta ,Z,\left\vert \Psi \right\vert
^{2}\right) $ thus writes:\bigskip

\begin{eqnarray*}
&&\left( G^{-1}\left( \omega ^{-1}\left( \theta ,Z,\left\vert \Psi
\right\vert ^{2}\right) \right) -\frac{\left( G^{-1}\right) ^{\prime }\left(
\omega ^{-1}\left( \theta ,Z,\left\vert \Psi \right\vert ^{2}\right) \right) 
}{\left( \omega \left( \theta ,Z,\left\vert \Psi \right\vert ^{2}\right)
\right) }\right) \delta \omega \left( \theta ,Z,\left\vert \Psi \right\vert
^{2}\right) \\
&&-\int \frac{\kappa }{N}\delta \omega \left( \theta -\frac{\left\vert
Z-Z_{1}\right\vert }{c},Z_{1},\Psi \right) T\left( Z,\theta ,Z_{1},\theta -%
\frac{\left\vert Z-Z_{1}\right\vert }{c}\right) \left( \mathcal{\bar{G}}%
_{0}\left( 0,Z_{1}\right) +\left\vert \Psi \left( \theta -\frac{\left\vert
Z-Z_{1}\right\vert }{c},Z_{1}\right) \right\vert ^{2}\right) \\
&=&\frac{\kappa }{N}\omega \left( \theta -\frac{\left\vert
Z-Z_{1}\right\vert }{c},Z_{1},\Psi \right) T\left( Z,\theta ,Z_{1},\theta -%
\frac{\left\vert Z-Z_{1}\right\vert }{c}\right) \left( \mathcal{\bar{G}}%
_{0}\left( 0,Z_{1}\right) +\left\vert \Psi \left( \theta -\frac{\left\vert
Z-Z_{1}\right\vert }{c},Z_{1}\right) \right\vert ^{2}\right) \\
&&\times \left( \frac{\delta T\left( Z,\theta ,Z_{1},\theta -\frac{%
\left\vert Z-Z_{1}\right\vert }{c}\right) }{T\left( Z,\theta ,Z_{1},\theta -%
\frac{\left\vert Z-Z_{1}\right\vert }{c}\right) }+\frac{\delta \left\vert
\Psi \left( \theta -\frac{\left\vert Z-Z_{1}\right\vert }{c},Z_{1}\right)
\right\vert ^{2}}{\mathcal{\bar{G}}_{0}\left( 0,Z_{1}\right) +\left\vert
\Psi \left( \theta -\frac{\left\vert Z-Z_{1}\right\vert }{c},Z_{1}\right)
\right\vert ^{2}}\right)
\end{eqnarray*}

\bigskip We obtain $\frac{\delta T\left( Z,\theta ,Z_{1},\theta -\frac{%
\left\vert Z-Z_{1}\right\vert }{c}\right) }{T\left( Z,\theta ,Z_{1},\theta -%
\frac{\left\vert Z-Z_{1}\right\vert }{c}\right) }$ starting with:%
\begin{equation*}
T\left( Z,\theta ,Z_{1},\theta -\frac{\left\vert Z-Z_{1}\right\vert }{c}%
\right) \rightarrow \frac{\lambda \tau \exp \left( -\frac{\left\vert
Z-Z^{\prime }\right\vert }{\nu c}\right) }{1+\frac{\alpha _{D}\omega h_{D}}{%
\alpha _{C}\omega ^{\prime }h_{C}}\frac{\frac{1}{\tau _{C}}+\alpha
_{C}\omega ^{\prime }\left\vert \Psi \left( \theta -\frac{\left\vert
Z-Z^{\prime }\right\vert }{c},Z^{\prime }\right) \right\vert ^{2}}{\frac{1}{%
\tau _{D}}+\alpha _{D}\omega \left\vert \Psi \left( \theta ,Z\right)
\right\vert ^{2}}}
\end{equation*}%
and:%
\begin{equation*}
\frac{\delta T\left( Z,\theta ,Z_{1},\theta -\frac{\left\vert
Z-Z_{1}\right\vert }{c}\right) }{T}\rightarrow -\frac{\delta \left( 1+\frac{%
\alpha _{D}\omega h_{D}}{\alpha _{C}\omega ^{\prime }h_{C}}\frac{\frac{1}{%
\tau _{C}}+\alpha _{C}\omega ^{\prime }\left\vert \Psi \left( \theta -\frac{%
\left\vert Z-Z^{\prime }\right\vert }{c},Z^{\prime }\right) \right\vert ^{2}%
}{\frac{1}{\tau _{D}}+\alpha _{D}\omega \left\vert \Psi \left( \theta
,Z\right) \right\vert ^{2}}\right) }{1+\frac{\alpha _{D}\omega h_{D}}{\alpha
_{C}\omega ^{\prime }h_{C}}\frac{\frac{1}{\tau _{C}}+\alpha _{C}\omega
^{\prime }\left\vert \Psi \left( \theta -\frac{\left\vert Z-Z^{\prime
}\right\vert }{c},Z^{\prime }\right) \right\vert ^{2}}{\frac{1}{\tau _{D}}%
+\alpha _{D}\omega \left\vert \Psi \left( \theta ,Z\right) \right\vert ^{2}}}
\end{equation*}%
In first approximation we have:%
\begin{eqnarray*}
&&\delta \left( 1+\frac{\alpha _{D}\omega h_{D}}{\alpha _{C}\omega ^{\prime
}h_{C}}\frac{\frac{1}{\tau _{C}}+\alpha _{C}\omega ^{\prime }\left\vert \Psi
\left( \theta -\frac{\left\vert Z-Z^{\prime }\right\vert }{c},Z^{\prime
}\right) \right\vert ^{2}}{\frac{1}{\tau _{D}}+\alpha _{D}\omega \left\vert
\Psi \left( \theta ,Z\right) \right\vert ^{2}}\right) \simeq \frac{\alpha
_{D}h_{D}}{\alpha _{C}h_{C}}\delta \left( \frac{\omega }{\omega ^{\prime }}%
\right) \\
&\rightarrow &\frac{\alpha _{D}h_{D}}{\alpha _{C}h_{C}}\left( \frac{\omega
^{\prime }\delta \omega -\omega \delta \omega ^{\prime }}{\left( \omega
^{\prime }\right) ^{2}}-\frac{\delta \omega \delta \omega ^{\prime }}{\left(
\omega ^{\prime }\right) ^{2}}+2\frac{\omega \delta \omega ^{\prime }\delta
\omega ^{\prime }}{\left( \omega ^{\prime }\right) ^{3}}\right)
\end{eqnarray*}%
and:%
\begin{eqnarray*}
&&\frac{\partial T\left( Z,\theta ,Z_{1},\theta -\frac{\left\vert
Z-Z_{1}\right\vert }{c}\right) }{\partial \omega \left( \theta -\frac{%
\left\vert Z-Z_{1}\right\vert }{c},Z_{1},\Psi \right) } \\
&=&-\frac{\lambda \tau \exp \left( -\frac{\left\vert Z-Z^{\prime
}\right\vert }{\nu c}\right) }{\omega \left( \theta -\frac{\left\vert
Z-Z_{1}\right\vert }{c},Z_{1},\Psi \right) \left( 1+\frac{\alpha _{D}\omega
\left( J,\theta ,Z,\left\vert \Psi \right\vert ^{2}\right) h_{D}}{\alpha
_{C}\omega \left( \theta -\frac{\left\vert Z-Z_{1}\right\vert }{c}%
,Z_{1},\Psi \right) h_{C}}\frac{\frac{1}{\tau _{C}}+\alpha _{C}\omega \left(
\theta -\frac{\left\vert Z-Z_{1}\right\vert }{c},Z_{1},\Psi \right)
\left\vert \Psi \left( \theta -\frac{\left\vert Z-Z^{\prime }\right\vert }{c}%
,Z^{\prime }\right) \right\vert ^{2}}{\frac{1}{\tau _{D}}+\alpha _{D}\omega
\left( J,\theta ,Z,\left\vert \Psi \right\vert ^{2}\right) \left\vert \Psi
\left( \theta ,Z\right) \right\vert ^{2}}\right) ^{2}}
\end{eqnarray*}%
\begin{eqnarray*}
&&\frac{\partial T\left( Z,\theta ,Z_{1},\theta -\frac{\left\vert
Z-Z_{1}\right\vert }{c}\right) }{\partial \omega \left( \theta ,Z,\left\vert
\Psi \right\vert ^{2}\right) } \\
&=&\frac{\lambda \tau \exp \left( -\frac{\left\vert Z-Z^{\prime }\right\vert 
}{\nu c}\right) \omega \left( J,\theta ,Z,\left\vert \Psi \right\vert
^{2}\right) }{\omega ^{2}\left( \theta -\frac{\left\vert Z-Z_{1}\right\vert 
}{c},Z_{1},\Psi \right) \left( 1+\frac{\alpha _{D}\omega \left( J,\theta
,Z,\left\vert \Psi \right\vert ^{2}\right) h_{D}}{\alpha _{C}\omega \left(
\theta -\frac{\left\vert Z-Z_{1}\right\vert }{c},Z_{1},\Psi \right) h_{C}}%
\frac{\frac{1}{\tau _{C}}+\alpha _{C}\omega \left( \theta -\frac{\left\vert
Z-Z_{1}\right\vert }{c},Z_{1},\Psi \right) \left\vert \Psi \left( \theta -%
\frac{\left\vert Z-Z^{\prime }\right\vert }{c},Z^{\prime }\right)
\right\vert ^{2}}{\frac{1}{\tau _{D}}+\alpha _{D}\omega \left( J,\theta
,Z,\left\vert \Psi \right\vert ^{2}\right) \left\vert \Psi \left( \theta
,Z\right) \right\vert ^{2}}\right) ^{2}}
\end{eqnarray*}%
We then deduce that:%
\begin{eqnarray*}
\frac{\delta T\left( Z,\theta ,Z_{1},\theta -\frac{\left\vert
Z-Z_{1}\right\vert }{c}\right) }{T} &\rightarrow &-\frac{\frac{\alpha
_{D}h_{D}}{\alpha _{C}h_{C}}\frac{\omega ^{\prime }\delta \omega -\omega
\delta \omega ^{\prime }}{\left( \omega ^{\prime }\right) ^{2}}}{1+\frac{%
\alpha _{D}\omega h_{D}}{\alpha _{C}\omega ^{\prime }h_{C}}\frac{\frac{1}{%
\tau _{C}}+\alpha _{C}\omega ^{\prime }\left\vert \Psi \left( \theta -\frac{%
\left\vert Z-Z^{\prime }\right\vert }{c},Z^{\prime }\right) \right\vert ^{2}%
}{\frac{1}{\tau _{D}}+\alpha _{D}\omega \left\vert \Psi \left( \theta
,Z\right) \right\vert ^{2}}} \\
&\rightarrow &-\frac{\alpha _{D}h_{D}}{\alpha _{C}h_{C}}\frac{\omega
^{\prime }\delta \omega -\omega \delta \omega ^{\prime }}{\left( \omega
^{\prime }\right) ^{2}}\frac{T\left( Z,\theta ,Z_{1},\theta -\frac{%
\left\vert Z-Z_{1}\right\vert }{c}\right) }{\lambda \tau \exp \left( -\frac{%
\left\vert Z-Z^{\prime }\right\vert }{\nu c}\right) }
\end{eqnarray*}%
which is inserted in the fluctuation equation (\ref{FET}):%
\begin{eqnarray*}
&&\left( G^{-1}\left( \omega ^{-1}\left( \theta ,Z,\left\vert \Psi
\right\vert ^{2}\right) \right) -\frac{\left( G^{-1}\right) ^{\prime }\left(
\omega ^{-1}\left( \theta ,Z,\left\vert \Psi \right\vert ^{2}\right) \right) 
}{\left( \omega \left( \theta ,Z,\left\vert \Psi \right\vert ^{2}\right)
\right) }\right) \delta \omega \left( \theta ,Z,\left\vert \Psi \right\vert
^{2}\right) \\
&&-\int \frac{\kappa }{N}T\left( Z,\theta ,Z_{1},\theta -\frac{\left\vert
Z-Z_{1}\right\vert }{c}\right) \left( \mathcal{\bar{G}}_{0}\left(
0,Z_{1}\right) +\left\vert \Psi \left( \theta -\frac{\left\vert
Z-Z_{1}\right\vert }{c},Z_{1}\right) \right\vert ^{2}\right) S\left( \theta
,\theta -\frac{\left\vert Z-Z_{1}\right\vert }{c}\right) \delta \omega
\left( \theta ,Z,\left\vert \Psi \right\vert ^{2}\right) dZ_{1} \\
&&+\int dZ_{1}\frac{\alpha _{D}h_{D}}{\alpha _{C}h_{C}}\frac{T\left(
Z,\theta ,Z_{1},\theta -\frac{\left\vert Z-Z_{1}\right\vert }{c}\right) }{%
\lambda \tau \exp \left( -\frac{\left\vert Z-Z^{\prime }\right\vert }{\nu c}%
\right) }\frac{\omega \left( \theta -\frac{\left\vert Z-Z_{1}\right\vert }{c}%
,Z_{1},\Psi \right) -\omega \left( \theta ,Z,\left\vert \Psi \right\vert
^{2}\right) S\left( \theta ,\theta -\frac{\left\vert Z-Z_{1}\right\vert }{c}%
\right) }{\left( \omega \left( \theta -\frac{\left\vert Z-Z_{1}\right\vert }{%
c},Z_{1},\Psi \right) \right) ^{2}}\delta \omega \left( \theta ,Z,\left\vert
\Psi \right\vert ^{2}\right) \\
&&\times \frac{\kappa }{N}\omega \left( \theta -\frac{\left\vert
Z-Z_{1}\right\vert }{c},Z_{1},\Psi \right) T\left( Z,\theta ,Z_{1},\theta -%
\frac{\left\vert Z-Z_{1}\right\vert }{c}\right) \left( \mathcal{\bar{G}}%
_{0}\left( 0,Z_{1}\right) +\left\vert \Psi \left( \theta -\frac{\left\vert
Z-Z_{1}\right\vert }{c},Z_{1}\right) \right\vert ^{2}\right) \\
&=&\frac{\kappa }{N}\int dZ_{1}\omega \left( \theta -\frac{\left\vert
Z-Z_{1}\right\vert }{c},Z_{1},\Psi \right) T\left( Z,\theta ,Z_{1},\theta -%
\frac{\left\vert Z-Z_{1}\right\vert }{c}\right) \delta \left\vert \Psi
\left( \theta -\frac{\left\vert Z-Z_{1}\right\vert }{c},Z_{1}\right)
\right\vert ^{2}
\end{eqnarray*}%
This can be regrouped as:%
\begin{eqnarray*}
&&\delta \omega \left( \theta ,Z,\left\vert \Psi \right\vert ^{2}\right)
\left( G^{-1}\left( \omega ^{-1}\left( \theta ,Z,\left\vert \Psi \right\vert
^{2}\right) \right) -\frac{\left( G^{-1}\right) ^{\prime }\left( \omega
^{-1}\left( \theta ,Z,\left\vert \Psi \right\vert ^{2}\right) \right) }{%
\left( \omega \left( \theta ,Z,\left\vert \Psi \right\vert ^{2}\right)
\right) }\right. \\
&&\left. +\int dZ_{1}\frac{\alpha _{D}h_{D}}{\alpha _{C}h_{C}}\frac{T\left(
Z,\theta ,Z_{1},\theta -\frac{\left\vert Z-Z_{1}\right\vert }{c}\right) }{%
\lambda \tau \exp \left( -\frac{\left\vert Z-Z^{\prime }\right\vert }{\nu c}%
\right) }T\left( Z,\theta ,Z_{1},\theta -\frac{\left\vert Z-Z_{1}\right\vert 
}{c}\right) \left( \mathcal{\bar{G}}_{0}\left( 0,Z_{1}\right) +\left\vert
\Psi \left( \theta -\frac{\left\vert Z-Z_{1}\right\vert }{c},Z_{1}\right)
\right\vert ^{2}\right) \right) \\
&&-\int \frac{\kappa }{N}T\left( Z,\theta ,Z_{1},\theta -\frac{\left\vert
Z-Z_{1}\right\vert }{c}\right) \left( \mathcal{\bar{G}}_{0}\left(
0,Z_{1}\right) +\left\vert \Psi \left( \theta -\frac{\left\vert
Z-Z_{1}\right\vert }{c},Z_{1}\right) \right\vert ^{2}\right) S\left( \theta
,\theta -\frac{\left\vert Z-Z_{1}\right\vert }{c}\right) \delta \omega
\left( \theta ,Z,\left\vert \Psi \right\vert ^{2}\right) dZ_{1} \\
&&+\int dZ_{1}\frac{\alpha _{D}h_{D}}{\alpha _{C}h_{C}}\frac{T\left(
Z,\theta ,Z_{1},\theta -\frac{\left\vert Z-Z_{1}\right\vert }{c}\right) }{%
\lambda \tau \exp \left( -\frac{\left\vert Z-Z^{\prime }\right\vert }{\nu c}%
\right) }\frac{-\omega \left( \theta ,Z,\left\vert \Psi \right\vert
^{2}\right) S\left( \theta ,\theta -\frac{\left\vert Z-Z_{1}\right\vert }{c}%
\right) }{\omega \left( \theta -\frac{\left\vert Z-Z_{1}\right\vert }{c}%
,Z_{1},\Psi \right) }\delta \omega \left( \theta ,Z,\left\vert \Psi
\right\vert ^{2}\right) \\
&&\times \frac{\kappa }{N}T\left( Z,\theta ,Z_{1},\theta -\frac{\left\vert
Z-Z_{1}\right\vert }{c}\right) \left( \mathcal{\bar{G}}_{0}\left(
0,Z_{1}\right) +\left\vert \Psi \left( \theta -\frac{\left\vert
Z-Z_{1}\right\vert }{c},Z_{1}\right) \right\vert ^{2}\right) \\
&=&\frac{\kappa }{N}\int dZ_{1}\omega \left( \theta -\frac{\left\vert
Z-Z_{1}\right\vert }{c},Z_{1},\Psi \right) T\left( Z,\theta ,Z_{1},\theta -%
\frac{\left\vert Z-Z_{1}\right\vert }{c}\right) \delta \left\vert \Psi
\left( \theta -\frac{\left\vert Z-Z_{1}\right\vert }{c},Z_{1}\right)
\right\vert ^{2}
\end{eqnarray*}%
and this yields:%
\begin{eqnarray*}
&&\delta \omega \left( \theta ,Z,\left\vert \Psi \right\vert ^{2}\right) \\
&=&\int \frac{\kappa }{N}\hat{T}_{n}\left( Z,\theta ,Z_{1},\theta -\frac{%
\left\vert Z-Z_{1}\right\vert }{c}\right) \left( \mathcal{\bar{G}}_{0}\left(
0,Z_{1}\right) +\left\vert \Psi \left( \theta -\frac{\left\vert
Z-Z_{1}\right\vert }{c},Z_{1}\right) \right\vert ^{2}\right) \delta \omega
\left( \theta -\frac{\left\vert Z-Z_{1}\right\vert }{c},Z_{1},\Psi \right)
dZ_{1} \\
&&+\frac{\kappa }{N}\int dZ_{1}\omega \left( \theta -\frac{\left\vert
Z-Z_{1}\right\vert }{c},Z_{1},\Psi \right) \hat{T}\left( Z,\theta
,Z_{1},\theta -\frac{\left\vert Z-Z_{1}\right\vert }{c}\right) \delta
\left\vert \Psi \left( \theta -\frac{\left\vert Z-Z_{1}\right\vert }{c}%
,Z_{1}\right) \right\vert ^{2}
\end{eqnarray*}%
with:%
\begin{eqnarray*}
&&\hat{T}\left( Z,\theta ,Z_{1},\theta -\frac{\left\vert Z-Z_{1}\right\vert 
}{c}\right) \\
&\rightarrow &\frac{T\left( Z,\theta ,Z_{1},\theta -\frac{\left\vert
Z-Z_{1}\right\vert }{c}\right) }{G^{-1}\left( \omega ^{-1}\left( \theta
,Z,\left\vert \Psi \right\vert ^{2}\right) \right) -\frac{\left(
G^{-1}\right) ^{\prime }\left( \omega ^{-1}\left( \theta ,Z,\left\vert \Psi
\right\vert ^{2}\right) \right) }{\left( \omega \left( \theta ,Z,\left\vert
\Psi \right\vert ^{2}\right) \right) }+\tilde{T}_{1}}
\end{eqnarray*}%
\begin{equation*}
\tilde{T}_{1}=\int dZ_{1}\frac{\alpha _{D}h_{D}}{\alpha _{C}h_{C}}\frac{%
T\left( Z,\theta ,Z_{1},\theta -\frac{\left\vert Z-Z_{1}\right\vert }{c}%
\right) T\left( Z,\theta ,Z_{1},\theta -\frac{\left\vert Z-Z_{1}\right\vert 
}{c}\right) \left( \mathcal{\bar{G}}_{0}\left( 0,Z_{1}\right) +\left\vert
\Psi \left( \theta -\frac{\left\vert Z-Z_{1}\right\vert }{c},Z_{1}\right)
\right\vert ^{2}\right) }{\lambda \tau \exp \left( -\frac{\left\vert
Z-Z^{\prime }\right\vert }{\nu c}\right) }
\end{equation*}%
and:%
\begin{equation*}
\hat{T}_{n}\left( Z,\theta ,Z_{1},\theta -\frac{\left\vert
Z-Z_{1}\right\vert }{c}\right) =\hat{T}\left( Z,\theta ,Z_{1},\theta -\frac{%
\left\vert Z-Z_{1}\right\vert }{c}\right) \left( 1+\frac{\alpha _{D}h_{D}}{%
\alpha _{C}h_{C}}\frac{T\left( Z,\theta ,Z_{1},\theta -\frac{\left\vert
Z-Z_{1}\right\vert }{c}\right) }{\lambda \tau \exp \left( -\frac{\left\vert
Z-Z^{\prime }\right\vert }{\nu c}\right) }\frac{\omega \left( \theta
,Z,\left\vert \Psi \right\vert ^{2}\right) }{\omega \left( \theta -\frac{%
\left\vert Z-Z_{1}\right\vert }{c},Z_{1},\Psi \right) }\right)
\end{equation*}

\subsubsection*{A5.3.2 Derivation of $\protect\delta \protect\omega %
^{-1}\left( J,\protect\theta ,Z,\left\vert \Psi \right\vert ^{2}\right) $}

We start again from:%
\begin{eqnarray*}
&&G^{-1}\left( \omega ^{-1}\left( J,\theta ,Z,\left\vert \Psi \right\vert
^{2}\right) \right) \\
&=&J\left( \theta ,Z\right) +\int \frac{\kappa }{N}\frac{\omega \left(
J,\theta -\frac{\left\vert Z-Z_{1}\right\vert }{c},Z_{1},\Psi \right)
T\left( Z,\theta ,Z_{1},\theta -\frac{\left\vert Z-Z_{1}\right\vert }{c}%
\right) }{\omega \left( J,\theta ,Z,\left\vert \Psi \right\vert ^{2}\right) }%
\left( \mathcal{\bar{G}}_{0}\left( 0,Z_{1}\right) +\left\vert \Psi \left(
\theta -\frac{\left\vert Z-Z_{1}\right\vert }{c},Z_{1}\right) \right\vert
^{2}\right) dZ_{1}
\end{eqnarray*}%
that can be rewritten, using this time a first order expansion in $\delta
\omega ^{-1}\left( \theta ,Z,\left\vert \Psi \right\vert ^{2}\right) $:%
\begin{eqnarray*}
&&\left( \left( G^{-1}\right) ^{\prime }\left( \omega ^{-1}\left( \theta
,Z,\left\vert \Psi \right\vert ^{2}\right) \right) -\frac{G^{-1}\left(
\omega ^{-1}\left( \theta ,Z,\left\vert \Psi \right\vert ^{2}\right) \right) 
}{\omega ^{-1}\left( \theta ,Z,\left\vert \Psi \right\vert ^{2}\right) }%
\right) \delta \omega ^{-1}\left( \theta ,Z,\left\vert \Psi \right\vert
^{2}\right) \\
&=&\int \frac{\kappa }{N}\frac{\delta \omega \left( \theta -\frac{\left\vert
Z-Z_{1}\right\vert }{c},Z_{1},\Psi \right) T\left( Z,\theta ,Z_{1},\theta -%
\frac{\left\vert Z-Z_{1}\right\vert }{c}\right) }{\omega \left( J,\theta
,Z,\left\vert \Psi \right\vert ^{2}\right) }\left( \mathcal{\bar{G}}%
_{0}\left( 0,Z_{1}\right) +\left\vert \Psi \left( \theta -\frac{\left\vert
Z-Z_{1}\right\vert }{c},Z_{1}\right) \right\vert ^{2}\right) dZ_{1} \\
&&+\int \frac{\kappa }{N}\frac{\omega \left( \theta -\frac{\left\vert
Z-Z_{1}\right\vert }{c},Z_{1},\Psi \right) \delta T\left( Z,\theta
,Z_{1},\theta -\frac{\left\vert Z-Z_{1}\right\vert }{c}\right) }{\omega
\left( J,\theta ,Z,\left\vert \Psi \right\vert ^{2}\right) }\left( \mathcal{%
\bar{G}}_{0}\left( 0,Z_{1}\right) +\left\vert \Psi \left( \theta -\frac{%
\left\vert Z-Z_{1}\right\vert }{c},Z_{1}\right) \right\vert ^{2}\right)
dZ_{1} \\
&&+\int \frac{\kappa }{N}\frac{\omega \left( \theta -\frac{\left\vert
Z-Z_{1}\right\vert }{c},Z_{1},\Psi \right) T\left( Z,\theta ,Z_{1},\theta -%
\frac{\left\vert Z-Z_{1}\right\vert }{c}\right) }{\omega \left( J,\theta
,Z,\left\vert \Psi \right\vert ^{2}\right) }\mathcal{\delta }\left\vert \Psi
\left( \theta -\frac{\left\vert Z-Z_{1}\right\vert }{c},Z_{1}\right)
\right\vert ^{2}dZ_{1}
\end{eqnarray*}%
Using again that:%
\begin{eqnarray*}
\frac{\delta T\left( Z,\theta ,Z_{1},\theta -\frac{\left\vert
Z-Z_{1}\right\vert }{c}\right) }{T} &\rightarrow &-\frac{\frac{\alpha
_{D}h_{D}}{\alpha _{C}h_{C}}\frac{\omega ^{\prime }\delta \omega -\omega
\delta \omega ^{\prime }}{\left( \omega ^{\prime }\right) ^{2}}}{1+\frac{%
\alpha _{D}\omega h_{D}}{\alpha _{C}\omega ^{\prime }h_{C}}\frac{\frac{1}{%
\tau _{C}}+\alpha _{C}\omega ^{\prime }\left\vert \Psi \left( \theta -\frac{%
\left\vert Z-Z^{\prime }\right\vert }{c},Z^{\prime }\right) \right\vert ^{2}%
}{\frac{1}{\tau _{D}}+\alpha _{D}\omega \left\vert \Psi \left( \theta
,Z\right) \right\vert ^{2}}} \\
&\rightarrow &-\frac{\alpha _{D}h_{D}}{\alpha _{C}h_{C}}\frac{\omega
^{\prime }\delta \omega -\omega \delta \omega ^{\prime }}{\left( \omega
^{\prime }\right) ^{2}}\frac{T\left( Z,\theta ,Z_{1},\theta -\frac{%
\left\vert Z-Z_{1}\right\vert }{c}\right) }{\lambda \tau \exp \left( -\frac{%
\left\vert Z-Z^{\prime }\right\vert }{\nu c}\right) }
\end{eqnarray*}%
and:%
\begin{eqnarray*}
&&\frac{\omega \left( \theta -\frac{\left\vert Z-Z_{1}\right\vert }{c}%
,Z_{1},\Psi \right) \frac{\delta \omega ^{-1}\left( \theta ,Z,\left\vert
\Psi \right\vert ^{2}\right) }{\left( \omega ^{-1}\left( \theta
,Z,\left\vert \Psi \right\vert ^{2}\right) \right) ^{2}}-\omega \left(
\theta ,Z,\left\vert \Psi \right\vert ^{2}\right) \frac{\delta \omega
^{-1}\left( \theta -\frac{\left\vert Z-Z_{1}\right\vert }{c},Z_{1},\Psi
\right) }{\left( \omega ^{-1}\left( \theta -\frac{\left\vert
Z-Z_{1}\right\vert }{c},Z_{1},\Psi \right) \right) ^{2}}}{\omega \left(
\theta -\frac{\left\vert Z-Z_{1}\right\vert }{c},Z_{1},\Psi \right) } \\
&=&\frac{\delta \omega ^{-1}\left( \theta ,Z,\left\vert \Psi \right\vert
^{2}\right) -\frac{\omega ^{-1}\left( \theta ,Z,\left\vert \Psi \right\vert
^{2}\right) }{\omega ^{-1}\left( \theta -\frac{\left\vert Z-Z_{1}\right\vert 
}{c},Z_{1},\Psi \right) }\delta \omega ^{-1}\left( \theta -\frac{\left\vert
Z-Z_{1}\right\vert }{c},Z_{1},\Psi \right) }{\left( \omega ^{-1}\left(
\theta ,Z,\left\vert \Psi \right\vert ^{2}\right) \right) ^{2}}
\end{eqnarray*}%
the fluctuation equation for $\frac{\delta \omega ^{-1}\left( \theta
,Z,\left\vert \Psi \right\vert ^{2}\right) }{\omega ^{-1}\left( \theta
,Z,\left\vert \Psi \right\vert ^{2}\right) }$ writes: 
\begin{eqnarray*}
&&\left( \left( G^{-1}\right) ^{\prime }\left( \omega ^{-1}\left( \theta
,Z,\left\vert \Psi \right\vert ^{2}\right) \right) -\frac{G^{-1}\left(
\omega ^{-1}\left( \theta ,Z,\left\vert \Psi \right\vert ^{2}\right) \right) 
}{\omega ^{-1}\left( \theta ,Z,\left\vert \Psi \right\vert ^{2}\right) }%
\right) \frac{\delta \omega ^{-1}\left( \theta ,Z,\left\vert \Psi
\right\vert ^{2}\right) }{\omega ^{-1}\left( \theta ,Z,\left\vert \Psi
\right\vert ^{2}\right) } \\
&&+\int \frac{\kappa }{N}\frac{T\left( Z,\theta ,Z_{1},\theta -\frac{%
\left\vert Z-Z_{1}\right\vert }{c}\right) }{\left( \omega ^{-1}\left( \theta
-\frac{\left\vert Z-Z_{1}\right\vert }{c},Z_{1},\Psi \right) \right) ^{2}}%
\left( \mathcal{\bar{G}}_{0}\left( 0,Z_{1}\right) +\left\vert \Psi \left(
\theta -\frac{\left\vert Z-Z_{1}\right\vert }{c},Z_{1}\right) \right\vert
^{2}\right) \delta \omega ^{-1}\left( \theta -\frac{\left\vert
Z-Z_{1}\right\vert }{c},Z_{1},\Psi \right) dZ_{1} \\
&&+\int dZ_{1}\frac{\alpha _{D}h_{D}}{\alpha _{C}h_{C}}\frac{T\left(
Z,\theta ,Z_{1},\theta -\frac{\left\vert Z-Z_{1}\right\vert }{c}\right) }{%
\lambda \tau \exp \left( -\frac{\left\vert Z-Z^{\prime }\right\vert }{\nu c}%
\right) }\frac{\omega \left( \theta -\frac{\left\vert Z-Z_{1}\right\vert }{c}%
,Z_{1},\Psi \right) -\omega \left( \theta ,Z,\left\vert \Psi \right\vert
^{2}\right) S\left( \theta ,\theta -\frac{\left\vert Z-Z_{1}\right\vert }{c}%
\right) }{\left( \omega \left( \theta -\frac{\left\vert Z-Z_{1}\right\vert }{%
c},Z_{1},\Psi \right) \right) ^{2}}\delta \omega \left( \theta ,Z,\left\vert
\Psi \right\vert ^{2}\right) \\
&&\times \frac{\kappa }{N}\omega \left( \theta -\frac{\left\vert
Z-Z_{1}\right\vert }{c},Z_{1},\Psi \right) T\left( Z,\theta ,Z_{1},\theta -%
\frac{\left\vert Z-Z_{1}\right\vert }{c}\right) \left( \mathcal{\bar{G}}%
_{0}\left( 0,Z_{1}\right) +\left\vert \Psi \left( \theta -\frac{\left\vert
Z-Z_{1}\right\vert }{c},Z_{1}\right) \right\vert ^{2}\right) \\
&=&\frac{\kappa }{N}\int dZ_{1}\omega \left( \theta -\frac{\left\vert
Z-Z_{1}\right\vert }{c},Z_{1},\Psi \right) T\left( Z,\theta ,Z_{1},\theta -%
\frac{\left\vert Z-Z_{1}\right\vert }{c}\right) \delta \left\vert \Psi
\left( \theta -\frac{\left\vert Z-Z_{1}\right\vert }{c},Z_{1}\right)
\right\vert ^{2}
\end{eqnarray*}%
which is regrouped as:%
\begin{eqnarray*}
&&\left( \left( G^{-1}\right) ^{\prime }\left( \omega ^{-1}\left( \theta
,Z,\left\vert \Psi \right\vert ^{2}\right) \right) -\frac{G^{-1}\left(
\omega ^{-1}\left( \theta ,Z,\left\vert \Psi \right\vert ^{2}\right) \right) 
}{\omega ^{-1}\left( \theta ,Z,\left\vert \Psi \right\vert ^{2}\right) }%
\right) \frac{\delta \omega ^{-1}\left( \theta ,Z,\left\vert \Psi
\right\vert ^{2}\right) }{\omega ^{-1}\left( \theta ,Z,\left\vert \Psi
\right\vert ^{2}\right) } \\
&&+\int \frac{\kappa }{N}\frac{T\left( Z,\theta ,Z_{1},\theta -\frac{%
\left\vert Z-Z_{1}\right\vert }{c}\right) }{\left( \omega ^{-1}\left( \theta
-\frac{\left\vert Z-Z_{1}\right\vert }{c},Z_{1},\Psi \right) \right) ^{2}}%
\left( \mathcal{\bar{G}}_{0}\left( 0,Z_{1}\right) +\left\vert \Psi \left(
\theta -\frac{\left\vert Z-Z_{1}\right\vert }{c},Z_{1}\right) \right\vert
^{2}\right) \delta \omega ^{-1}\left( \theta -\frac{\left\vert
Z-Z_{1}\right\vert }{c},Z_{1},\Psi \right) dZ_{1} \\
&&-\int dZ_{1}\frac{\alpha _{D}h_{D}}{\alpha _{C}h_{C}}\frac{T\left(
Z,\theta ,Z_{1},\theta -\frac{\left\vert Z-Z_{1}\right\vert }{c}\right) }{%
\lambda \tau \exp \left( -\frac{\left\vert Z-Z^{\prime }\right\vert }{\nu c}%
\right) }\left( \frac{\delta \omega ^{-1}\left( \theta ,Z,\left\vert \Psi
\right\vert ^{2}\right) -\frac{\omega ^{-1}\left( \theta ,Z,\left\vert \Psi
\right\vert ^{2}\right) }{\omega ^{-1}\left( \theta -\frac{\left\vert
Z-Z_{1}\right\vert }{c},Z_{1},\Psi \right) }\delta \omega ^{-1}\left( \theta
-\frac{\left\vert Z-Z_{1}\right\vert }{c},Z_{1},\Psi \right) }{\left( \omega
^{-1}\left( \theta ,Z,\left\vert \Psi \right\vert ^{2}\right) \right) ^{2}}%
\right) \\
&&\times \frac{\kappa }{N}T\left( Z,\theta ,Z_{1},\theta -\frac{\left\vert
Z-Z_{1}\right\vert }{c}\right) \left( \mathcal{\bar{G}}_{0}\left(
0,Z_{1}\right) +\left\vert \Psi \left( \theta -\frac{\left\vert
Z-Z_{1}\right\vert }{c},Z_{1}\right) \right\vert ^{2}\right) \\
&=&\frac{\kappa }{N}\int dZ_{1}\omega \left( \theta -\frac{\left\vert
Z-Z_{1}\right\vert }{c},Z_{1},\Psi \right) T\left( Z,\theta ,Z_{1},\theta -%
\frac{\left\vert Z-Z_{1}\right\vert }{c}\right) \delta \left\vert \Psi
\left( \theta -\frac{\left\vert Z-Z_{1}\right\vert }{c},Z_{1}\right)
\right\vert ^{2}
\end{eqnarray*}%
We replace:%
\begin{equation*}
\left( \frac{\omega ^{-1}\left( \theta ,Z,\left\vert \Psi \right\vert
^{2}\right) }{\omega ^{-1}\left( \theta -\frac{\left\vert Z-Z_{1}\right\vert 
}{c},Z_{1},\Psi \right) }\right) ^{2}T\left( Z,\theta ,Z_{1},\theta -\frac{%
\left\vert Z-Z_{1}\right\vert }{c}\right) \rightarrow T^{\prime }\left(
Z,\theta ,Z_{1},\theta -\frac{\left\vert Z-Z_{1}\right\vert }{c}\right)
\end{equation*}%
which leads to:%
\begin{eqnarray*}
&&\delta \omega ^{-1}\left( \theta ,Z,\left\vert \Psi \right\vert
^{2}\right) \left( G^{-1}\omega ^{-1}\left( \theta ,Z,\left\vert \Psi
\right\vert ^{2}\right) -\omega ^{-1}\left( \theta ,Z,\left\vert \Psi
\right\vert ^{2}\right) \left( G^{-1}\right) ^{\prime }\omega ^{-1}\left(
\theta ,Z,\left\vert \Psi \right\vert ^{2}\right) \right. \\
&&\left. +\int dZ_{1}\frac{\alpha _{D}h_{D}}{\alpha _{C}h_{C}}\frac{T\left(
Z,\theta ,Z_{1},\theta -\frac{\left\vert Z-Z_{1}\right\vert }{c}\right) }{%
\lambda \tau \exp \left( -\frac{\left\vert Z-Z^{\prime }\right\vert }{\nu c}%
\right) }\frac{\kappa }{N}T^{\prime }\left( Z,\theta ,Z_{1},\theta -\frac{%
\left\vert Z-Z_{1}\right\vert }{c}\right) \left( \mathcal{\bar{G}}_{0}\left(
0,Z_{1}\right) +\left\vert \Psi \left( \theta -\frac{\left\vert
Z-Z_{1}\right\vert }{c},Z_{1}\right) \right\vert ^{2}\right) \right) \\
&=&\int \frac{\kappa }{N}T^{\prime }\left( Z,\theta ,Z_{1},\theta -\frac{%
\left\vert Z-Z_{1}\right\vert }{c}\right) \left( \mathcal{\bar{G}}_{0}\left(
0,Z_{1}\right) +\left\vert \Psi \left( \theta -\frac{\left\vert
Z-Z_{1}\right\vert }{c},Z_{1}\right) \right\vert ^{2}\right) \\
&&\times \left( 1+\frac{\alpha _{D}h_{D}}{\alpha _{C}h_{C}}\frac{T\left(
Z,\theta ,Z_{1},\theta -\frac{\left\vert Z-Z_{1}\right\vert }{c}\right) }{%
\lambda \tau \exp \left( -\frac{\left\vert Z-Z^{\prime }\right\vert }{\nu c}%
\right) }\frac{\omega ^{-1}\left( \theta ,Z,\left\vert \Psi \right\vert
^{2}\right) }{\omega ^{-1}\left( \theta -\frac{\left\vert Z-Z_{1}\right\vert 
}{c},Z_{1},\Psi \right) }\right) \delta \omega ^{-1}\left( \theta -\frac{%
\left\vert Z-Z_{1}\right\vert }{c},Z_{1},\Psi \right) dZ_{1} \\
&&-\frac{\kappa }{N}\int dZ_{1}\omega ^{-1}\left( \theta -\frac{\left\vert
Z-Z_{1}\right\vert }{c},Z_{1},\Psi \right) T^{\prime }\left( Z,\theta
,Z_{1},\theta -\frac{\left\vert Z-Z_{1}\right\vert }{c}\right) \delta
\left\vert \Psi \left( \theta -\frac{\left\vert Z-Z_{1}\right\vert }{c}%
,Z_{1}\right) \right\vert ^{2}
\end{eqnarray*}%
and this can be rewriten in the following form:%
\begin{eqnarray}
&&\delta \omega ^{-1}\left( \theta ,Z,\left\vert \Psi \right\vert ^{2}\right)
\label{dmg} \\
&=&\int \check{T}\left( Z,\theta ,Z_{1},\theta -\frac{\left\vert
Z-Z_{1}\right\vert }{c}\right) \left( \mathcal{\bar{G}}_{0}\left(
0,Z_{1}\right) +\left\vert \Psi \left( \theta -\frac{\left\vert
Z-Z_{1}\right\vert }{c},Z_{1}\right) \right\vert ^{2}\right) \delta \omega
^{-1}\left( \theta -\frac{\left\vert Z-Z_{1}\right\vert }{c},Z_{1},\Psi
\right) dZ_{1}  \notag \\
&&-\int dZ_{1}\omega ^{-1}\left( \theta -\frac{\left\vert Z-Z_{1}\right\vert 
}{c},Z_{1},\Psi \right) \frac{\check{T}\left( Z,\theta ,Z_{1},\theta -\frac{%
\left\vert Z-Z_{1}\right\vert }{c}\right) }{1+\frac{\alpha _{D}h_{D}}{\alpha
_{C}h_{C}}\frac{T\left( Z,\theta ,Z_{1},\theta -\frac{\left\vert
Z-Z_{1}\right\vert }{c}\right) }{\lambda \tau \exp \left( -\frac{\left\vert
Z-Z^{\prime }\right\vert }{\nu c}\right) }\frac{\omega ^{-1}\left( \theta
,Z,\left\vert \Psi \right\vert ^{2}\right) }{\omega ^{-1}\left( \theta -%
\frac{\left\vert Z-Z_{1}\right\vert }{c},Z_{1},\Psi \right) }}\delta
\left\vert \Psi \left( \theta -\frac{\left\vert Z-Z_{1}\right\vert }{c}%
,Z_{1}\right) \right\vert ^{2}  \notag
\end{eqnarray}%
where we define the effective connectivity $\check{T}\left( Z,\theta
,Z_{1},\theta -\frac{\left\vert Z-Z_{1}\right\vert }{c}\right) $ by: 
\begin{eqnarray*}
&&\check{T}\left( Z,\theta ,Z_{1},\theta -\frac{\left\vert
Z-Z_{1}\right\vert }{c}\right) \\
&=&\frac{\frac{\kappa }{N}T^{\prime }\left( Z,\theta ,Z_{1},\theta -\frac{%
\left\vert Z-Z_{1}\right\vert }{c}\right) \left( 1+\frac{\alpha _{D}h_{D}}{%
\alpha _{C}h_{C}}\frac{T\left( Z,\theta ,Z_{1},\theta -\frac{\left\vert
Z-Z_{1}\right\vert }{c}\right) }{\lambda \tau \exp \left( -\frac{\left\vert
Z-Z^{\prime }\right\vert }{\nu c}\right) }\frac{\omega ^{-1}\left( \theta
,Z,\left\vert \Psi \right\vert ^{2}\right) }{\omega ^{-1}\left( \theta -%
\frac{\left\vert Z-Z_{1}\right\vert }{c},Z_{1},\Psi \right) }\right) }{%
G^{-1}\left( \omega ^{-1}\left( \theta ,Z,\left\vert \Psi \right\vert
^{2}\right) \right) -\omega ^{-1}\left( \theta ,Z,\left\vert \Psi
\right\vert ^{2}\right) \left( G^{-1}\right) ^{\prime }\left( \omega
^{-1}\left( \theta ,Z,\left\vert \Psi \right\vert ^{2}\right) \right) +%
\tilde{T}_{1}} \\
&=&\frac{\left( \omega ^{-1}\left( \theta ,Z,\left\vert \Psi \right\vert
^{2}\right) \right) ^{2}T\left( Z,\theta ,Z_{1},\theta -\frac{\left\vert
Z-Z_{1}\right\vert }{c}\right) \left( 1+\frac{\alpha _{D}h_{D}}{\alpha
_{C}h_{C}}\frac{T\left( Z,\theta ,Z_{1},\theta -\frac{\left\vert
Z-Z_{1}\right\vert }{c}\right) }{\lambda \tau \exp \left( -\frac{\left\vert
Z-Z^{\prime }\right\vert }{\nu c}\right) }\frac{\omega ^{-1}\left( \theta
,Z,\left\vert \Psi \right\vert ^{2}\right) }{\omega ^{-1}\left( \theta -%
\frac{\left\vert Z-Z_{1}\right\vert }{c},Z_{1},\Psi \right) }\right) }{%
\left( \omega ^{-1}\left( \theta -\frac{\left\vert Z-Z_{1}\right\vert }{c}%
,Z_{1},\Psi \right) \right) ^{2}\left( G^{-1}\left( \omega ^{-1}\left(
\theta ,Z,\left\vert \Psi \right\vert ^{2}\right) \right) -\omega
^{-1}\left( \theta ,Z,\left\vert \Psi \right\vert ^{2}\right) \left(
G^{-1}\right) ^{\prime }\left( \omega ^{-1}\left( \theta ,Z,\left\vert \Psi
\right\vert ^{2}\right) \right) +\tilde{T}_{2}\right) }
\end{eqnarray*}

\bigskip where:%
\begin{equation*}
\tilde{T}_{1}=\frac{\kappa }{N}\int dZ_{1}\frac{\alpha _{D}h_{D}}{\alpha
_{C}h_{C}}\frac{T\left( Z,\theta ,Z_{1},\theta -\frac{\left\vert
Z-Z_{1}\right\vert }{c}\right) }{\lambda \tau \exp \left( -\frac{\left\vert
Z-Z^{\prime }\right\vert }{\nu c}\right) }T^{\prime }\left( Z,\theta
,Z_{1},\theta -\frac{\left\vert Z-Z_{1}\right\vert }{c}\right) \left( 
\mathcal{\bar{G}}_{0}\left( 0,Z_{1}\right) +\left\vert \Psi \left( \theta -%
\frac{\left\vert Z-Z_{1}\right\vert }{c},Z_{1}\right) \right\vert ^{2}\right)
\end{equation*}%
and:%
\begin{eqnarray*}
\tilde{T}_{2} &=&\frac{\kappa }{N}\int dZ_{1}\frac{\alpha _{D}h_{D}}{\alpha
_{C}h_{C}}\frac{T\left( Z,\theta ,Z_{1},\theta -\frac{\left\vert
Z-Z_{1}\right\vert }{c}\right) }{\lambda \tau \exp \left( -\frac{\left\vert
Z-Z^{\prime }\right\vert }{\nu c}\right) } \\
&&\times \frac{\left( \omega ^{-1}\left( \theta ,Z,\left\vert \Psi
\right\vert ^{2}\right) \right) ^{2}T\left( Z,\theta ,Z_{1},\theta -\frac{%
\left\vert Z-Z_{1}\right\vert }{c}\right) \left( \mathcal{\bar{G}}_{0}\left(
0,Z_{1}\right) +\left\vert \Psi \left( \theta -\frac{\left\vert
Z-Z_{1}\right\vert }{c},Z_{1}\right) \right\vert ^{2}\right) }{\left( \omega
^{-1}\left( \theta -\frac{\left\vert Z-Z_{1}\right\vert }{c},Z_{1},\Psi
\right) \right) ^{2}}
\end{eqnarray*}

\subsection*{A5.3 Higher order derivatives of activities}

The first and higher order derivatives of $\omega ^{-1}\left( \theta
,Z,\left\vert \Psi \right\vert ^{2}\right) $ are needed to express the
saddle point equation (\ref{Smv}) for fluctuations of the background field $%
\frac{\delta \Psi \left( \theta ,Z\right) }{\Psi \left( \theta ,Z\right) }$.
This is done by using an auxiliary path integral representation for $\omega
^{-1}\left( \theta ,Z,\left\vert \Psi \right\vert ^{2}\right) $.

\subsubsection*{A5.3.1 Computation of the first order derivatives of $%
\protect\omega ^{-1}\left( J,\protect\theta ,Z\right) $}

Expression (\ref{dmg}) yields:%
\begin{eqnarray}
\frac{\delta \omega ^{-1}\left( J,\theta ,Z\right) }{\delta \left\vert \Psi
\left( \theta -l_{1},Z_{1}\right) \right\vert ^{2}} &=&-\sum_{n=1}^{\infty }%
\frac{1}{\left\vert \Psi \left( \theta -l_{1},Z_{1}\right) \right\vert ^{2}}%
\int \frac{\omega ^{-1}\left( J,\theta -\sum_{l=1}^{n}\frac{\left\vert
Z^{\left( l-1\right) }-Z^{\left( l\right) }\right\vert }{c},Z_{1}\right) }{%
1+T_{0}\left( Z,\theta ,Z_{1},\theta -\sum_{l=1}^{n}\frac{\left\vert
Z^{\left( l-1\right) }-Z^{\left( l\right) }\right\vert }{c}\right) }
\label{dvtr} \\
&&\times \prod\limits_{l=1}^{n}\check{T}\left( \theta -\sum_{j=1}^{l-1}\frac{%
\left\vert Z^{\left( j-1\right) }-Z^{\left( j\right) }\right\vert }{c}%
,Z^{\left( l-1\right) },Z^{\left( l\right) },\omega ,\Psi \right) \delta
\left( l_{1}-\sum_{l=1}^{n}\frac{\left\vert Z^{\left( l-1\right) }-Z^{\left(
l\right) }\right\vert }{c}\right) \prod\limits_{l=1}^{n-1}dZ^{\left(
l\right) }  \notag
\end{eqnarray}%
with:%
\begin{eqnarray*}
&&T_{0}\left( Z,\theta ,Z_{1},\theta -\sum_{l=1}^{n}\frac{\left\vert
Z^{\left( l-1\right) }-Z^{\left( l\right) }\right\vert }{c}\right) \\
&=&\frac{\alpha _{D}h_{D}}{\alpha _{C}h_{C}}\frac{T\left( Z,\theta
,Z_{1},\theta -\sum_{l=1}^{n}\frac{\left\vert Z^{\left( l-1\right)
}-Z^{\left( l\right) }\right\vert }{c}\right) }{\lambda \tau \exp \left( -%
\frac{\left\vert Z-Z^{\prime }\right\vert }{\nu c}\right) }\frac{\omega
^{-1}\left( \theta -\sum_{l=1}^{n-1}\frac{\left\vert Z^{\left( l-1\right)
}-Z^{\left( l\right) }\right\vert }{c},Z,\left\vert \Psi \right\vert
^{2}\right) }{\omega ^{-1}\left( \theta -\sum_{l=1}^{n}\frac{\left\vert
Z^{\left( l-1\right) }-Z^{\left( l\right) }\right\vert }{c},Z_{1},\Psi
\right) } \\
&\simeq &\frac{\alpha _{D}h_{D}}{\alpha _{C}h_{C}}\left\langle \frac{T\left(
Z,\theta ,Z_{1},\theta -\sum_{l=1}^{n}\frac{\left\vert Z^{\left( l-1\right)
}-Z^{\left( l\right) }\right\vert }{c}\right) }{\lambda \tau \exp \left( -%
\frac{\left\vert Z-Z^{\prime }\right\vert }{\nu c}\right) }\right\rangle _{Z}
\\
&\rightarrow &T_{0}\left( Z_{1},\theta -\sum_{l=1}^{n}\frac{\left\vert
Z^{\left( l-1\right) }-Z^{\left( l\right) }\right\vert }{c}\right)
\end{eqnarray*}

\subsubsection*{A5.3.2 Expression for $\protect\omega ^{-1}\left( J,\protect%
\theta ,Z\right) $ in terms of an auxiliary path integral}

Equation (\ref{dvtr}) can be used to obtain an expression for $\omega
^{-1}\left( J,\theta ,Z\right) $, see \cite{GL}. The $\omega ^{-1}\left(
J,\theta ,Z\right) $ can be written with an auxiliary path integral (Some
details of this formulation are given in Appendix 7):%
\begin{eqnarray}
&&\omega ^{-1}\left( J,\theta ,Z\right) \\
&=&\omega _{0}^{-1}\left( J,\theta ,Z\right) +\sum_{n=1}^{\infty }\frac{1}{n!%
}\frac{\int \check{T}\Lambda ^{\dag }\left( Z,\theta \right) \int
\prod\limits_{i=1}^{n}\left( -\frac{\omega _{0}^{-1}\left( \theta ^{\left(
l\right) },Z_{_{l}}\right) }{1+T_{0}\left( \theta _{l},Z_{l},\right) }%
\right) \left\vert \Psi \left( J,\theta _{i},Z_{i}\right) \right\vert
^{2}\Lambda \left( Z_{i},\theta _{i}\right) d\left( Z_{i},\theta _{i}\right)
\exp \left( -S\left( \Lambda \right) \right) \mathcal{D}\Lambda }{\exp
\left( -S\left( \Lambda \right) \right) \mathcal{D}\Lambda }  \notag \\
&=&\omega _{0}^{-1}\left( J,\theta ,Z\right) +\frac{\int \check{T}\Lambda
^{\dag }\left( Z,\theta \right) \exp \left( -S\left( \Lambda \right) -\int
\Lambda \left( X,\theta \right) \frac{\omega _{0}\left( J,\theta ,Z\right) }{%
1+T_{0}\left( \theta _{l},Z_{l},\right) }\left\vert \Psi \left( J,\theta
,Z\right) \right\vert ^{2}d\left( X,\theta \right) \right) \mathcal{D}%
\Lambda }{\int \exp \left( -S\left( \Lambda \right) \right) \mathcal{D}%
\Lambda }  \notag
\end{eqnarray}%
with:%
\begin{eqnarray}
S\left( \Lambda \right) &=&\int \Lambda \left( X,\theta \right) \left(
1-\left( 1+\left\vert \Psi \right\vert ^{2}\right) \check{T}\right) \Lambda
^{\dag }\left( X,\theta \right) d\left( X,\theta \right)  \label{Fnc} \\
&&-\int \Lambda \left( Z^{\left( 1\right) },\theta ^{\left( 1\right)
}\right) \sum_{k}\frac{\delta ^{k}\left( \check{T}\left( \theta ^{\left(
1\right) }-\frac{\left\vert Z^{\left( 1\right) }-Z^{\left( 2\right)
}\right\vert }{c},Z^{\left( 1\right) },Z^{\left( 2\right) },\omega
_{0}^{-1}\right) \right) }{k!\prod\limits_{l=3}^{k+2}\delta ^{k}\omega
_{0}\left( J,\theta ^{\left( l\right) },Z^{\left( l\right) }\right) }\Lambda
^{\dag }\left( Z^{\left( 2\right) },\theta ^{\left( 1\right) }-\frac{%
\left\vert Z^{\left( 1\right) }-Z^{\left( 2\right) }\right\vert }{c}\right) 
\notag \\
&&\times \prod\limits_{l=3}^{k+2}\check{T}\left( \theta ^{\left( 1\right) }-%
\frac{\left\vert Z^{\left( 1\right) }-Z^{\left( l\right) }\right\vert }{c}%
,Z^{\left( 1\right) },Z^{\left( l\right) },\omega _{0}^{-1}\right) \Lambda
^{\dag }\left( \theta ^{\left( l\right) },Z^{\left( l\right) }\right)
d\theta ^{\left( 1\right) }\prod\limits_{l=1}^{k+2}dZ^{\left( l\right) } 
\notag \\
&=&\int \Lambda \left( Z,\theta \right) \left( 1-\left\vert \Psi \right\vert
^{2}\check{T}\right) \Lambda ^{\dag }\left( Z,\theta \right) d\left(
Z,\theta \right) -\int \Lambda \left( Z,\theta \right) \check{T}\left(
\theta -\frac{\left\vert Z-Z^{\left( 1\right) }\right\vert }{c},Z,Z^{\left(
1\right) },\omega _{0}^{-1}+\check{T}\Lambda ^{\dag }\right)  \notag \\
&&\times \Lambda ^{\dag }\left( Z^{\left( 1\right) },\theta -\frac{%
\left\vert Z-Z^{\left( 1\right) }\right\vert }{c}\right) dZdZ^{\left(
1\right) }d\theta  \notag
\end{eqnarray}%
where:%
\begin{eqnarray*}
&&\check{T}\left( \theta -\frac{\left\vert Z^{\left( 1\right) }-Z\right\vert 
}{c},Z^{\left( 1\right) },Z,\omega _{0}^{-1}+\check{T}\Lambda ^{\dag }\right)
\\
&=&\check{T}\left( \theta -\frac{\left\vert Z^{\left( 1\right)
}-Z\right\vert }{c},Z^{\left( 1\right) },Z,\omega _{0}^{-1}\left( Z,\theta
\right) +\int \check{T}\left( \theta -\frac{\left\vert Z-Z^{\left( 1\right)
}\right\vert }{c},Z^{\left( 1\right) },Z,\omega _{0}^{-1}\right) \Lambda
^{\dag }\left( Z^{\left( 1\right) },\theta -\frac{\left\vert Z-Z^{\left(
1\right) }\right\vert }{c}\right) dZ^{\left( 1\right) }\right)
\end{eqnarray*}

\subsubsection*{A5.3.3 Computation of the higher order derivatives of $%
\protect\omega ^{-1}\left( J,\protect\theta ,Z\right) $}

This allow to compute the successive derivatives of $\omega ^{-1}\left(
J,\theta ,Z\right) $ as:%
\begin{eqnarray*}
&&\frac{\delta ^{\sum_{l}p_{l}^{i}}}{\delta ^{\sum_{l}p_{l}^{i}}\left\vert
\Psi \left( \theta ^{\left( l\right) },Z_{_{l}}\right) \right\vert ^{2}}%
\omega ^{-1}\left( J,\theta ,Z\right) \\
&\rightarrow &\frac{\delta ^{\sum_{l}p_{l}^{i}}}{\delta
^{\sum_{l}p_{l}^{i}}\left\vert \Psi \left( \theta ^{\left( l\right)
},Z_{_{l}}\right) \right\vert ^{2}}\frac{\int \check{T}\Lambda ^{\dag
}\left( Z,\theta \right) \exp \left( -S\left( \Lambda \right) -\int \Lambda
\left( X,\theta \right) \frac{\omega _{0}^{-1}\left( J,\theta ,Z\right) }{%
1+T_{0}\left( \theta ,Z\right) }\left\vert \Psi \left( J,\theta ,Z\right)
\right\vert ^{2}d\left( X,\theta \right) \right) \mathcal{D}\Lambda }{\int
\exp \left( -S\left( \Lambda \right) \right) \mathcal{D}\Lambda }
\end{eqnarray*}%
The derivative of the action yields a contribution:%
\begin{eqnarray*}
&&\exp \left( \int \left( \Lambda \left( X,\theta \right) \left\vert \Psi
\right\vert ^{2}\check{T}\Lambda ^{\dag }\left( X,\theta \right) -\Lambda
\left( X,\theta \right) \omega _{0}^{-1}\left( J,\theta ,Z\right) \left\vert
\Psi \left( J,\theta ,Z\right) \right\vert ^{2}\right) d\left( X,\theta
\right) \right) \\
&\rightarrow &\left( \Lambda \left( \theta ^{\left( l\right)
},Z_{_{l}}\right) \check{T}\left( \theta _{l}-\frac{\left\vert Z^{\left(
1\right) }-Z_{l}\right\vert }{c},Z^{\left( 1\right) },Z_{l},\omega
_{0}\right) \Lambda ^{\dag }\left( Z^{\left( 1\right) },\theta _{l}-\frac{%
\left\vert Z^{\left( 1\right) }-Z_{l}\right\vert }{c}\right) -\Lambda \left(
\theta ^{\left( l\right) },Z_{_{l}}\right) \frac{\omega _{0}^{-1}\left(
\theta ^{\left( l\right) },Z_{_{l}}\right) }{1+T_{0}\left( \theta
_{l},Z_{l},\right) }\right)
\end{eqnarray*}%
and the successive derivatives have the form:%
\begin{eqnarray*}
&&\left\{ \int \mathcal{D}\Lambda \frac{\exp \left( -S\left( \Lambda \right)
\right) \mathcal{D}\Lambda }{\int \exp \left( -S\left( \Lambda \right)
\right) \mathcal{D}\Lambda }\check{T}\Lambda ^{\dag }\left( Z,\theta \right)
\right. \\
&&\left. \prod \left( \Lambda \left( \theta ^{\left( l\right)
},Z_{_{l}}\right) \check{T}\left( \theta _{l}-\frac{\left\vert Z^{\left(
1\right) }-Z_{l}\right\vert }{c},Z^{\left( 1\right) },Z_{l},\omega
_{0}\right) \Lambda ^{\dag }\left( Z^{\left( 1\right) },\theta _{l}-\frac{%
\left\vert Z^{\left( 1\right) }-Z_{l}\right\vert }{c}\right) -\Lambda \left(
\theta ^{\left( l\right) },Z_{_{l}}\right) \frac{\omega _{0}^{-1}\left(
\theta ^{\left( l\right) },Z_{_{l}}\right) }{1+T_{0}\left( \theta
_{l},Z_{l},\right) }\right) \right\} _{\left\vert \Psi \right\vert ^{2}=0}
\end{eqnarray*}%
and we find:%
\begin{eqnarray*}
&&\frac{\delta ^{n}\omega ^{-1}\left( J,\theta ,Z\right) }{%
\prod\limits_{i=1}^{n}\delta \left\vert \Psi \left( \theta
-l_{i},Z_{i}\right) \right\vert ^{2}} \\
&\rightarrow &\int \mathcal{D}\Lambda \frac{\exp \left( -S\left( \Lambda
\right) \right) \mathcal{D}\Lambda }{\int \exp \left( -S\left( \Lambda
\right) \right) \mathcal{D}\Lambda }\check{T}\Lambda ^{\dag }\left( Z,\theta
\right) \\
&&\prod \left( \Lambda \left( \theta ^{\left( l\right) },Z_{_{l}}\right) 
\check{T}\left( \theta _{l}-\frac{\left\vert Z^{\left( 1\right)
}-Z_{l}\right\vert }{c},Z^{\left( 1\right) },Z_{l},\omega _{0}^{-1}\right)
\Lambda ^{\dag }\left( Z^{\left( 1\right) },\theta _{l}-\frac{\left\vert
Z^{\left( 1\right) }-Z_{l}\right\vert }{c}\right) -\Lambda \left( \theta
^{\left( l\right) },Z_{_{l}}\right) \frac{\omega _{0}^{-1}\left( \theta
^{\left( l\right) },Z_{_{l}}\right) }{1+T_{0}\left( \theta _{l},Z_{l}\right) 
}\right)
\end{eqnarray*}

\bigskip Expanding the action functional (\ref{Fnc}) yields:%
\begin{eqnarray}
&&\frac{\delta ^{n}\omega ^{-1}\left( J,\theta ,Z\right) }{%
\prod\limits_{i=1}^{n}\delta \left\vert \Psi \left( \theta
-l_{i},Z_{i}\right) \right\vert ^{2}}  \label{Fnd} \\
&\rightarrow &\int \mathcal{D}\Lambda \exp \left( -\left( \int \Lambda
\left( X,\theta \right) \left( 1-\check{T}\right) \Lambda ^{\dag }\left(
X,\theta \right) d\left( X,\theta \right) \right) \right) \mathcal{D}\Lambda
\notag \\
&&\exp \left( \int \Lambda \left( Z^{\left( 1\right) },\theta ^{\left(
1\right) }\right) \sum_{k}\frac{\delta ^{k}\left( \check{T}\left( \theta
^{\left( 1\right) }-\frac{\left\vert Z^{\left( 1\right) }-Z^{\left( 2\right)
}\right\vert }{c},Z^{\left( 1\right) },Z^{\left( 2\right) },\omega \right)
\right) }{k!\prod\limits_{l=3}^{k+2}\delta ^{k}\omega \left( J,\theta
^{\left( l\right) },Z^{\left( l\right) }\right) }\Lambda ^{\dag }\left(
Z^{\left( 2\right) },\theta ^{\left( 1\right) }-\frac{\left\vert Z^{\left(
1\right) }-Z^{\left( 2\right) }\right\vert }{c}\right) \right.  \notag \\
&&\left. \times \prod\limits_{l=3}^{k+2}\check{T}\left( \theta ^{\left(
1\right) }-\frac{\left\vert Z^{\left( 1\right) }-Z^{\left( l\right)
}\right\vert }{c},Z^{\left( 1\right) },Z^{\left( l\right) },\omega \right)
\Lambda ^{\dag }\left( \theta ^{\left( l\right) },Z^{\left( l\right)
}\right) d\theta ^{\left( 1\right) }\prod\limits_{l=1}^{k+2}dZ^{\left(
l\right) }\right) \check{T}\Lambda ^{\dag }\left( Z,\theta \right)  \notag \\
&&\prod \left( \Lambda \left( \theta ^{\left( l\right) },Z_{_{l}}\right) 
\check{T}\left( \theta _{l}-\frac{\left\vert Z^{\left( 1\right)
}-Z_{l}\right\vert }{c},Z^{\left( 1\right) },Z_{l},\omega \right) \Lambda
^{\dag }\left( Z^{\left( 1\right) },\theta _{l}-\frac{\left\vert Z^{\left(
1\right) }-Z_{l}\right\vert }{c}\right) -\Lambda \left( \theta ^{\left(
l\right) },Z_{_{l}}\right) \frac{\omega ^{-1}\left( \theta ^{\left( l\right)
},Z_{_{l}}\right) }{1+T_{0}\left( \theta _{l},Z_{l},\right) }\right)  \notag
\end{eqnarray}%
with vacuum diagram removd.

In the previous expression:%
\begin{equation*}
\frac{\delta ^{k_{l}}\left( \check{T}\left( \theta ^{\left( 1\right) }-\frac{%
\left\vert Z^{\left( 1\right) }-Z^{\left( 2\right) }\right\vert }{c}%
,Z^{\left( 1\right) },Z^{\left( 2\right) },\omega \right) \right) }{%
k_{l}!\prod\limits_{l=3}^{k+2}\delta ^{k_{l}}\omega \left( J,\theta ^{\left(
l\right) },Z^{\left( l\right) }\right) }
\end{equation*}%
satisfies:%
\begin{equation*}
\sum k_{l}\leqslant n
\end{equation*}%
and:%
\begin{equation*}
-\frac{\omega ^{-1}\left( \theta ^{\left( l\right) },Z_{_{l}}\right) }{%
1+T_{0}\left( \theta _{l},Z_{l},\right) }
\end{equation*}%
arises $\sum k_{l}$ times. Given the propagator in (\ref{Fnd}), the product:%
\begin{equation*}
\prod \Lambda \left( \theta ^{\left( l\right) },Z_{_{l}}\right) \check{T}%
\left( \theta _{l}-\frac{\left\vert Z^{\left( 1\right) }-Z_{l}\right\vert }{c%
},Z^{\left( 1\right) },Z_{l},\omega \right) \Lambda ^{\dag }\left( Z^{\left(
1\right) },\theta _{l}-\frac{\left\vert Z^{\left( 1\right)
}-Z_{l}\right\vert }{c}\right)
\end{equation*}%
is replaced by 
\begin{equation*}
\rightarrow \prod \frac{\check{T}}{1-\check{T}}\left( \theta _{l}-\frac{%
\left\vert Z^{\left( 1\right) }-Z_{l}\right\vert }{c},Z^{\left( 1\right)
},Z_{l},\omega \right)
\end{equation*}%
and the factor arises $n-\sum k_{l}$ times.

Thus, $\frac{\delta ^{n}\omega ^{-1}\left( J,\theta ,Z\right) }{%
\prod\limits_{i=1}^{n}\delta \left\vert \Psi \left( \theta
-l_{i},Z_{i}\right) \right\vert ^{2}}$ has the form:%
\begin{eqnarray}
&&\frac{\delta ^{n}\omega ^{-1}\left( J,\theta ,Z\right) }{%
\prod\limits_{i=1}^{n}\delta \left\vert \Psi \left( \theta
-l_{i},Z_{i}\right) \right\vert ^{2}}  \label{Smb} \\
&\rightarrow &\sum \sum_{k}\frac{\delta ^{k}\left( \check{T}\left( \theta
^{\left( 1\right) }-\frac{\left\vert Z^{\left( 1\right) }-Z^{\left( 2\right)
}\right\vert }{c},Z^{\left( 1\right) },Z^{\left( 2\right) },\omega \right)
\right) }{k!\prod\limits_{l=3}^{k+2}\delta ^{k}\omega \left( J,\theta
^{\left( l\right) },Z^{\left( l\right) }\right) }\prod \frac{\check{T}}{1-%
\check{T}}\prod \frac{\omega ^{-1}\left( \theta ^{\left( l\right)
},Z_{_{l}}\right) }{1+T_{0}\left( \theta _{l},Z_{l},\right) }  \notag
\end{eqnarray}%
where the sum without index describes the sum of graphs ending at the $\frac{%
\omega ^{-1}\left( \theta ^{\left( l\right) },Z_{_{l}}\right) }{%
1+T_{0}\left( \theta _{l},Z_{l},\right) }$, the vertices $\frac{\delta
^{k}\left( \check{T}\left( \theta ^{\left( 1\right) }-\frac{\left\vert
Z^{\left( 1\right) }-Z^{\left( 2\right) }\right\vert }{c},Z^{\left( 1\right)
},Z^{\left( 2\right) },\omega \right) \right) }{k!\prod\limits_{l=3}^{k+2}%
\delta ^{k}\omega \left( J,\theta ^{\left( l\right) },Z^{\left( l\right)
}\right) }$ and the end points being connected by the kernels $\frac{\check{T%
}}{1-\check{T}}$.

\subsubsection*{A5.3.4 Estimation af the variations of the derivatives of $%
\protect\omega ^{-1}\left( J,\protect\theta ,Z\right) $}

We derive the variations of $\omega ^{-1}\left( J,\theta ,Z\right) $ and its
derivatives. In fact, given (\ref{Smb}), we have to obtain expressions for $%
\delta \frac{\check{T}}{1-\check{T}}$ , and the derivatives $\frac{\delta
^{k}\left( \check{T}\left( \theta ^{\left( 1\right) }-\frac{\left\vert
Z^{\left( 1\right) }-Z^{\left( 2\right) }\right\vert }{c},Z^{\left( 1\right)
},Z^{\left( 2\right) },\omega \right) \right) }{k!\prod\limits_{l=3}^{k+2}%
\delta ^{k}\omega \left( J,\theta ^{\left( l\right) },Z^{\left( l\right)
}\right) }$.

\bigskip

The variations of: 
\begin{equation*}
T\left( \theta ^{\left( 1\right) }-\frac{\left\vert Z^{\left( 1\right)
}-Z^{\left( 2\right) }\right\vert }{c},Z^{\left( 1\right) },Z^{\left(
2\right) },\omega \right)
\end{equation*}%
have already been accounted for in the derivation of $\omega ^{-1}\left(
J,\theta ,Z\right) $ so that we only keep:%
\begin{eqnarray*}
\delta \check{T} &=&\partial _{\frac{\delta \omega _{0}\left( J,\theta
^{\left( 1\right) }-\frac{\left\vert Z^{\left( 1\right) }-Z^{\left( 2\right)
}\right\vert }{c},Z^{\left( 2\right) }\right) }{\omega _{0}\left( J,\theta
^{\left( 1\right) }-\frac{\left\vert Z^{\left( 1\right) }-Z^{\left( 2\right)
}\right\vert }{c},Z^{\left( 2\right) }\right) }}\check{T}\left( \theta
^{\left( 1\right) }-\frac{\left\vert Z^{\left( 1\right) }-Z^{\left( 2\right)
}\right\vert }{c},Z^{\left( 1\right) },Z^{\left( 2\right) },\omega \right) \\
&&\times \left( \frac{\delta \omega \left( J,\theta ^{\left( 1\right) }-%
\frac{\left\vert Z^{\left( 1\right) }-Z^{\left( 2\right) }\right\vert }{c}%
,Z^{\left( 2\right) }\right) }{\omega \left( J,\theta ^{\left( 1\right) }-%
\frac{\left\vert Z^{\left( 1\right) }-Z^{\left( 2\right) }\right\vert }{c}%
,Z^{\left( 2\right) }\right) }-\frac{\delta \omega \left( J,\theta ^{\left(
1\right) },Z^{\left( 1\right) }\right) }{\omega \left( J,\theta ^{\left(
1\right) },Z^{\left( 1\right) }\right) }\right) \\
&&+\frac{\delta \check{T}_{i}\left( \theta ^{\left( 1\right) }-\frac{%
\left\vert Z^{\left( 1\right) }-Z^{\left( 2\right) }\right\vert }{c}%
,Z^{\left( 1\right) },Z^{\left( 2\right) },\omega \right) }{\check{T}\left(
\theta ^{\left( 1\right) }-\frac{\left\vert Z^{\left( 1\right) }-Z^{\left(
2\right) }\right\vert }{c},Z^{\left( 1\right) },Z^{\left( 2\right) },\omega
\right) }\check{T}\left( \theta ^{\left( 1\right) }-\frac{\left\vert
Z^{\left( 1\right) }-Z^{\left( 2\right) }\right\vert }{c},Z^{\left( 1\right)
},Z^{\left( 2\right) },\omega \right)
\end{eqnarray*}%
where:%
\begin{equation*}
\delta \check{T}_{i}\left( \theta ^{\left( 1\right) }-\frac{\left\vert
Z^{\left( 1\right) }-Z^{\left( 2\right) }\right\vert }{c},Z^{\left( 1\right)
},Z^{\left( 2\right) },\omega \right)
\end{equation*}%
is an intrinsic variation. As a consequence: 
\begin{equation*}
\delta \frac{\check{T}}{1-\check{T}}\rightarrow \frac{1}{1-\check{T}}\delta 
\check{T}\frac{1}{1-\check{T}}
\end{equation*}%
with:%
\begin{equation*}
\delta \check{T}=\delta \check{T}_{i}\left( \theta ^{\left( 1\right) }-\frac{%
\left\vert Z^{\left( 1\right) }-Z^{\left( 2\right) }\right\vert }{c}%
,Z^{\left( 1\right) },Z^{\left( 2\right) },\omega \right)
\end{equation*}%
Gathering these results, we are led to:

\bigskip

\begin{eqnarray*}
&&\delta \frac{\delta ^{k}\left( \check{T}\left( \theta ^{\left( 1\right) }-%
\frac{\left\vert Z^{\left( 1\right) }-Z^{\left( 2\right) }\right\vert }{c}%
,Z^{\left( 1\right) },Z^{\left( 2\right) },\omega \right) \right) }{%
k!\prod\limits_{l=3}^{k+2}\delta ^{k}\omega \left( J,\theta ^{\left(
l\right) },Z^{\left( l\right) }\right) } \\
&\rightarrow &\frac{\partial \delta ^{k}\left( \check{T}\left( \theta
^{\left( 1\right) }-\frac{\left\vert Z^{\left( 1\right) }-Z^{\left( 2\right)
}\right\vert }{c},Z^{\left( 1\right) },Z^{\left( 2\right) },\omega \right)
\right) }{\partial _{\frac{\delta \omega _{0}\left( J,\theta ^{\left(
1\right) }-\frac{\left\vert Z^{\left( 1\right) }-Z^{\left( 2\right)
}\right\vert }{c},Z^{\left( 2\right) }\right) }{\omega _{0}\left( J,\theta
^{\left( 1\right) }-\frac{\left\vert Z^{\left( 1\right) }-Z^{\left( 2\right)
}\right\vert }{c},Z^{\left( 2\right) }\right) }}k!\prod\limits_{l=3}^{k+2}%
\delta ^{k}\omega \left( J,\theta ^{\left( l\right) },Z^{\left( l\right)
}\right) } \\
&&\times \left( \delta \omega \left( J,\theta ^{\left( 1\right) },Z^{\left(
1\right) }\right) -\frac{\omega \left( J,\theta ^{\left( 1\right)
},Z^{\left( 1\right) }\right) }{\omega _{0}\left( J,\theta ^{\left( 1\right)
}-\frac{\left\vert Z^{\left( 1\right) }-Z^{\left( 2\right) }\right\vert }{c}%
,Z^{\left( 2\right) }\right) }\delta \omega \left( J,\theta ^{\left(
1\right) }-\frac{\left\vert Z^{\left( 1\right) }-Z^{\left( 2\right)
}\right\vert }{c},Z^{\left( 2\right) }\right) \right) \\
&&+\frac{\delta \check{T}_{i}\left( \theta ^{\left( 1\right) }-\frac{%
\left\vert Z^{\left( 1\right) }-Z^{\left( 2\right) }\right\vert }{c}%
,Z^{\left( 1\right) },Z^{\left( 2\right) },\omega \right) }{\check{T}\left(
\theta ^{\left( 1\right) }-\frac{\left\vert Z^{\left( 1\right) }-Z^{\left(
2\right) }\right\vert }{c},Z^{\left( 1\right) },Z^{\left( 2\right) },\omega
\right) }\frac{\delta ^{k}\left( \check{T}\left( \theta ^{\left( 1\right) }-%
\frac{\left\vert Z^{\left( 1\right) }-Z^{\left( 2\right) }\right\vert }{c}%
,Z^{\left( 1\right) },Z^{\left( 2\right) },\omega \right) \right) }{%
k!\prod\limits_{l=3}^{k+2}\delta ^{k}\omega \left( J,\theta ^{\left(
l\right) },Z^{\left( l\right) }\right) } \\
&\rightarrow &\frac{\delta ^{k}\left( \check{T}\left( \theta ^{\left(
1\right) }-\frac{\left\vert Z^{\left( 1\right) }-Z^{\left( 2\right)
}\right\vert }{c},Z^{\left( 1\right) },Z^{\left( 2\right) },\omega \right)
\right) }{\left( k-1\right) !\prod\limits_{l=3}^{k+2}\delta ^{k}\omega
\left( J,\theta ^{\left( l\right) },Z^{\left( l\right) }\right) }\left( 
\frac{\delta \omega \left( J,\theta ^{\left( 1\right) }-\frac{\left\vert
Z^{\left( 1\right) }-Z^{\left( 2\right) }\right\vert }{c},Z^{\left( 2\right)
}\right) }{\omega \left( J,\theta ^{\left( 1\right) }-\frac{\left\vert
Z^{\left( 1\right) }-Z^{\left( 2\right) }\right\vert }{c},Z^{\left( 2\right)
}\right) }-\frac{\delta \omega \left( J,\theta ^{\left( 1\right) },Z^{\left(
1\right) }\right) }{\omega \left( J,\theta ^{\left( 1\right) },Z^{\left(
1\right) }\right) }\right) \\
&&+\frac{\delta \check{T}_{i}\left( \theta ^{\left( 1\right) }-\frac{%
\left\vert Z^{\left( 1\right) }-Z^{\left( 2\right) }\right\vert }{c}%
,Z^{\left( 1\right) },Z^{\left( 2\right) },\omega \right) }{\check{T}\left(
\theta ^{\left( 1\right) }-\frac{\left\vert Z^{\left( 1\right) }-Z^{\left(
2\right) }\right\vert }{c},Z^{\left( 1\right) },Z^{\left( 2\right) },\omega
\right) }\frac{\delta ^{k}\left( \check{T}\left( \theta ^{\left( 1\right) }-%
\frac{\left\vert Z^{\left( 1\right) }-Z^{\left( 2\right) }\right\vert }{c}%
,Z^{\left( 1\right) },Z^{\left( 2\right) },\omega \right) \right) }{%
k!\prod\limits_{l=3}^{k+2}\delta ^{k}\omega \left( J,\theta ^{\left(
l\right) },Z^{\left( l\right) }\right) }
\end{eqnarray*}%
Given formula (\ref{Smb}), we also have to consider the variation at the
ending points:%
\begin{equation*}
\sum \frac{\delta \omega \left( \theta ^{\left( l\right) },Z_{_{l}}\right) }{%
\omega \left( \theta ^{\left( l\right) },Z_{_{l}}\right) }\frac{\delta
^{\sum_{l}p_{l}^{i}}}{\delta ^{\sum_{l}p_{l}^{i}}\left\vert \Psi \left(
\theta ^{\left( l\right) },Z_{_{l}}\right) \right\vert ^{2}}\omega
^{-1}\left( J,\theta ,Z\right)
\end{equation*}%
and we find the variation of the derivative:

\begin{eqnarray*}
&&\delta \frac{\delta ^{\sum_{l}p_{l}^{i}}}{\delta
^{\sum_{l}p_{l}^{i}}\left\vert \Psi \left( \theta ^{\left( l\right)
},Z_{_{l}}\right) \right\vert ^{2}}\omega ^{-1}\left( J,\theta ,Z\right) \\
&\rightarrow &\sum_{l}\frac{\delta \omega ^{-1}\left( \theta ^{\left(
l\right) },Z_{_{l}}\right) }{\omega ^{-1}\left( \theta ^{\left( l\right)
},Z_{_{l}}\right) }\frac{\delta ^{\sum_{l}p_{l}^{i}}}{\delta
^{\sum_{l}p_{l}^{i}}\left\vert \Psi \left( \theta ^{\left( l\right)
},Z_{_{l}}\right) \right\vert ^{2}}\omega ^{-1}\left( J,\theta ,Z\right)
+N\left\langle \frac{1}{1-\check{T}}\delta \check{T}\right\rangle \frac{%
\delta ^{\sum_{l}p_{l}^{i}}}{\delta ^{\sum_{l}p_{l}^{i}}\left\vert \Psi
\left( \theta ^{\left( l\right) },Z_{_{l}}\right) \right\vert ^{2}}\omega
^{-1}\left( J,\theta ,Z\right) \\
&&+M\frac{\left\langle \delta \frac{\delta ^{k}\left( \check{T}\left( \theta
^{\left( 1\right) }-\frac{\left\vert Z^{\left( 1\right) }-Z^{\left( 2\right)
}\right\vert }{c},Z^{\left( 1\right) },Z^{\left( 2\right) },\omega
_{0}\right) \right) }{k!\prod\limits_{l=3}^{k+2}\delta ^{k}\omega _{0}\left(
J,\theta ^{\left( l\right) },Z^{\left( l\right) }\right) }\right\rangle }{%
\left\langle \frac{\delta ^{k}\left( \check{T}\left( \theta ^{\left(
1\right) }-\frac{\left\vert Z^{\left( 1\right) }-Z^{\left( 2\right)
}\right\vert }{c},Z^{\left( 1\right) },Z^{\left( 2\right) },\omega
_{0}\right) \right) }{k!\prod\limits_{l=3}^{k+2}\delta ^{k}\omega _{0}\left(
J,\theta ^{\left( l\right) },Z^{\left( l\right) }\right) }\right\rangle }%
\frac{\delta ^{\sum_{l}p_{l}^{i}}}{\delta ^{\sum_{l}p_{l}^{i}}\left\vert
\Psi \left( \theta ^{\left( l\right) },Z_{_{l}}\right) \right\vert ^{2}}%
\omega ^{-1}\left( J,\theta ,Z\right)
\end{eqnarray*}%
For random fluctuations, we can consider that in average:%
\begin{eqnarray*}
&&\left\langle \frac{1}{1-\check{T}}\delta \check{T}\right\rangle \\
&\rightarrow &\left\langle \frac{1}{1-\check{T}}\frac{\delta \check{T}%
_{i}\left( \theta ^{\left( 1\right) }-\frac{\left\vert Z^{\left( 1\right)
}-Z^{\left( 2\right) }\right\vert }{c},Z^{\left( 1\right) },Z^{\left(
2\right) },\omega _{0}\right) }{\check{T}\left( \theta ^{\left( 1\right) }-%
\frac{\left\vert Z^{\left( 1\right) }-Z^{\left( 2\right) }\right\vert }{c}%
,Z^{\left( 1\right) },Z^{\left( 2\right) },\omega _{0}\right) }\check{T}%
\left( \theta ^{\left( 1\right) }-\frac{\left\vert Z^{\left( 1\right)
}-Z^{\left( 2\right) }\right\vert }{c},Z^{\left( 1\right) },Z^{\left(
2\right) },\omega _{0}\right) \right\rangle \\
&&\left\langle \delta \frac{\delta ^{k}\left( \check{T}\left( \theta
^{\left( 1\right) }-\frac{\left\vert Z^{\left( 1\right) }-Z^{\left( 2\right)
}\right\vert }{c},Z^{\left( 1\right) },Z^{\left( 2\right) },\omega
_{0}\right) \right) }{k!\prod\limits_{l=3}^{k+2}\delta ^{k}\omega _{0}\left(
J,\theta ^{\left( l\right) },Z^{\left( l\right) }\right) }\right\rangle \\
&\rightarrow &\left\langle \frac{\delta \check{T}_{i}\left( \theta ^{\left(
1\right) }-\frac{\left\vert Z^{\left( 1\right) }-Z^{\left( 2\right)
}\right\vert }{c},Z^{\left( 1\right) },Z^{\left( 2\right) },\omega
_{0}\right) }{\check{T}\left( \theta ^{\left( 1\right) }-\frac{\left\vert
Z^{\left( 1\right) }-Z^{\left( 2\right) }\right\vert }{c},Z^{\left( 1\right)
},Z^{\left( 2\right) },\omega _{0}\right) }\frac{\delta ^{k}\left( \check{T}%
\left( \theta ^{\left( 1\right) }-\frac{\left\vert Z^{\left( 1\right)
}-Z^{\left( 2\right) }\right\vert }{c},Z^{\left( 1\right) },Z^{\left(
2\right) },\omega _{0}\right) \right) }{k!\prod\limits_{l=3}^{k+2}\delta
^{k}\omega _{0}\left( J,\theta ^{\left( l\right) },Z^{\left( l\right)
}\right) }\right\rangle
\end{eqnarray*}%
and as a consequence, the derivatives become:%
\begin{eqnarray*}
&&\delta \frac{\delta ^{\sum_{l}p_{l}^{i}}}{\delta
^{\sum_{l}p_{l}^{i}}\left\vert \Psi \left( \theta ^{\left( l\right)
},Z_{_{l}}\right) \right\vert ^{2}}\omega ^{-1}\left( J,\theta ,Z\right) \\
&\rightarrow &\left( \sum_{l}\frac{\delta \omega ^{-1}\left( \theta ^{\left(
l\right) },Z_{_{l}}\right) }{\omega ^{-1}\left( \theta ^{\left( l\right)
},Z_{_{l}}\right) }+N\left\langle \frac{1}{1-\check{T}}\delta \check{T}%
\right\rangle +M\frac{\left\langle \delta \frac{\delta ^{k}\left( \check{T}%
\left( \theta ^{\left( 1\right) }-\frac{\left\vert Z^{\left( 1\right)
}-Z^{\left( 2\right) }\right\vert }{c},Z^{\left( 1\right) },Z^{\left(
2\right) },\omega _{0}\right) \right) }{k!\prod\limits_{l=3}^{k+2}\delta
^{k}\omega _{0}\left( J,\theta ^{\left( l\right) },Z^{\left( l\right)
}\right) }\right\rangle }{\left\langle \frac{\delta ^{k}\left( \check{T}%
\left( \theta ^{\left( 1\right) }-\frac{\left\vert Z^{\left( 1\right)
}-Z^{\left( 2\right) }\right\vert }{c},Z^{\left( 1\right) },Z^{\left(
2\right) },\omega _{0}\right) \right) }{k!\prod\limits_{l=3}^{k+2}\delta
^{k}\omega _{0}\left( J,\theta ^{\left( l\right) },Z^{\left( l\right)
}\right) }\right\rangle }\right) \\
&&\times \frac{\delta ^{\sum_{l}p_{l}^{i}}}{\delta
^{\sum_{l}p_{l}^{i}}\left\vert \Psi \left( \theta ^{\left( l\right)
},Z_{_{l}}\right) \right\vert ^{2}}\omega ^{-1}\left( J,\theta ,Z\right) \\
&\rightarrow &\left( \sum_{l}\frac{\delta \omega ^{-1}\left( \theta ^{\left(
l\right) },Z_{_{l}}\right) }{\omega ^{-1}\left( \theta ^{\left( l\right)
},Z_{_{l}}\right) }+\left\langle \frac{\delta \check{T}_{i}\left( \theta
^{\left( 1\right) }-\frac{\left\vert Z^{\left( 1\right) }-Z^{\left( 2\right)
}\right\vert }{c},Z^{\left( 1\right) },Z^{\left( 2\right) },\omega
_{0}\right) }{\check{T}\left( \theta ^{\left( 1\right) }-\frac{\left\vert
Z^{\left( 1\right) }-Z^{\left( 2\right) }\right\vert }{c},Z^{\left( 1\right)
},Z^{\left( 2\right) },\omega _{0}\right) }\right\rangle \right) \frac{%
\delta ^{\sum_{l}p_{l}^{i}}}{\delta ^{\sum_{l}p_{l}^{i}}\left\vert \Psi
\left( \theta ^{\left( l\right) },Z_{_{l}}\right) \right\vert ^{2}}\omega
^{-1}\left( J,\theta ,Z\right)
\end{eqnarray*}%
with $M$ and $N$ accounting for multiplicity.

Assuming $N=M$ the derivatives write to the first approximation:%
\begin{eqnarray*}
&&\delta \frac{\delta ^{\sum_{l}p_{l}^{i}}}{\delta
^{\sum_{l}p_{l}^{i}}\left\vert \Psi \left( \theta ^{\left( l\right)
},Z_{_{l}}\right) \right\vert ^{2}}\omega ^{-1}\left( J,\theta ,Z\right) \\
&=&\left( \sum_{l}\frac{\delta \omega ^{-1}\left( \theta ^{\left( l\right)
},Z_{_{l}}\right) }{\omega ^{-1}\left( \theta ^{\left( l\right)
},Z_{_{l}}\right) }+\left\langle \frac{\delta \check{T}_{i}\left( \theta
^{\left( 1\right) }-\frac{\left\vert Z^{\left( 1\right) }-Z^{\left( 2\right)
}\right\vert }{c},Z^{\left( 1\right) },Z^{\left( 2\right) },\omega
_{0}\right) }{\check{T}\left( \theta ^{\left( 1\right) }-\frac{\left\vert
Z^{\left( 1\right) }-Z^{\left( 2\right) }\right\vert }{c},Z^{\left( 1\right)
},Z^{\left( 2\right) },\omega _{0}\right) }\right\rangle \right) \frac{%
\delta ^{\sum_{l}p_{l}^{i}}}{\delta ^{\sum_{l}p_{l}^{i}}\left\vert \Psi
\left( \theta ^{\left( l\right) },Z_{_{l}}\right) \right\vert ^{2}}\omega
^{-1}\left( J,\theta ,Z\right)
\end{eqnarray*}

We can replace:%
\begin{equation*}
\left\langle \frac{\delta \check{T}_{i}\left( \theta ^{\left( 1\right) }-%
\frac{\left\vert Z^{\left( 1\right) }-Z^{\left( 2\right) }\right\vert }{c}%
,Z^{\left( 1\right) },Z^{\left( 2\right) },\omega _{0}\right) }{\check{T}%
\left( \theta ^{\left( 1\right) }-\frac{\left\vert Z^{\left( 1\right)
}-Z^{\left( 2\right) }\right\vert }{c},Z^{\left( 1\right) },Z^{\left(
2\right) },\omega _{0}\right) }\right\rangle
\end{equation*}%
by its average along the paths ending at the points $\left( \theta ^{\left(
l\right) },Z_{_{l}}\right) $:%
\begin{equation*}
\sum_{l}\left\langle \frac{\delta \check{T}\left( \theta ^{\left( l\right) }-%
\frac{\left\vert Z^{\left( 1\right) }-Z^{\left( l\right) }\right\vert }{c}%
,Z^{\left( 1\right) },Z^{\left( l\right) }\right) }{\check{T}\left( \theta
^{\left( l\right) }-\frac{\left\vert Z^{\left( 1\right) }-Z^{\left( 2\right)
}\right\vert }{c},Z^{\left( 1\right) },Z^{\left( l\right) }\right) }%
\right\rangle _{Z_{1}}
\end{equation*}%
and:%
\begin{eqnarray*}
&&\delta \frac{\delta ^{\sum_{l}p_{l}^{i}}}{\delta
^{\sum_{l}p_{l}^{i}}\left\vert \Psi \left( \theta ^{\left( l\right)
},Z_{_{l}}\right) \right\vert ^{2}}\omega ^{-1}\left( J,\theta ,Z\right) \\
&\rightarrow &\left( \sum_{l}\left( \frac{\delta \omega ^{-1}\left( \theta
^{\left( l\right) },Z_{_{l}}\right) }{\omega ^{-1}\left( \theta ^{\left(
l\right) },Z_{_{l}}\right) }+\left\langle \frac{\delta \check{T}\left(
\theta ^{\left( l\right) }-\frac{\left\vert Z^{\left( 1\right) }-Z^{\left(
l\right) }\right\vert }{c},Z^{\left( 1\right) },Z^{\left( l\right) }\right) 
}{\check{T}\left( \theta ^{\left( l\right) }-\frac{\left\vert Z^{\left(
1\right) }-Z^{\left( 2\right) }\right\vert }{c},Z^{\left( 1\right)
},Z^{\left( l\right) }\right) }\right\rangle _{Z_{1}}\right) \right) \frac{%
\delta ^{\sum_{l}p_{l}^{i}}}{\delta ^{\sum_{l}p_{l}^{i}}\left\vert \Psi
\left( \theta ^{\left( l\right) },Z_{_{l}}\right) \right\vert ^{2}}\omega
^{-1}\left( J,\theta ,Z\right)
\end{eqnarray*}

If we include local intrinsic variations $\check{T}_{i}$ of $\check{T}$ this
moifies the previous expression, and we are led to:%
\begin{eqnarray*}
&&\delta \frac{\delta ^{\sum_{l}p_{l}^{i}}}{\delta
^{\sum_{l}p_{l}^{i}}\left\vert \Psi \left( \theta ^{\left( l\right)
},Z_{_{l}}\right) \right\vert ^{2}}\omega ^{-1}\left( J,\theta ,Z\right) \\
&\rightarrow &\sum_{l}\frac{\delta \omega ^{-1}\left( \theta ^{\left(
l\right) },Z_{_{l}}\right) }{\omega ^{-1}\left( \theta ^{\left( l\right)
},Z_{_{l}}\right) }\frac{\delta ^{\sum_{l}p_{l}^{i}}}{\delta
^{\sum_{l}p_{l}^{i}}\left\vert \Psi \left( \theta ^{\left( l\right)
},Z_{_{l}}\right) \right\vert ^{2}}\omega ^{-1}\left( J,\theta ,Z\right) \\
&&+\sum \left[ \frac{\delta ^{\sum_{l}p_{l}^{i}}}{\delta
^{\sum_{l}p_{l}^{i}}\left\vert \Psi \left( \theta ^{\left( l\right)
},Z_{_{l}}\right) \right\vert ^{2}}\right] _{B}\omega ^{-1}\left( J,\theta
,Z\right) \\
&&\times \frac{\omega ^{-1}\left( \theta ^{\left( l\right) },Z_{_{l}}\right)
\delta \check{T}_{i}\left( \theta ^{\left( 1\right) }-\frac{\left\vert
Z^{\left( 1\right) }-Z^{\left( 2\right) }\right\vert }{c},Z^{\left( 1\right)
},Z^{\left( 2\right) },\omega _{0}\right) }{\check{T}\left( \theta ^{\left(
1\right) }-\frac{\left\vert Z^{\left( 1\right) }-Z^{\left( 2\right)
}\right\vert }{c},Z^{\left( 1\right) },Z^{\left( 2\right) },\omega
_{0}\right) }\left[ \frac{\delta ^{\sum_{l}p_{l}^{i}}}{\delta
^{\sum_{l}p_{l}^{i}}\left\vert \Psi \left( \theta ^{\left( l\right)
},Z_{_{l}}\right) \right\vert ^{2}}\right] _{B}\omega ^{-1}\left( J,\theta
,Z\right)
\end{eqnarray*}%
where:

\begin{equation*}
\frac{\delta ^{\sum_{l}p_{l}^{i}}}{\delta ^{\sum_{l}p_{l}^{i}}\left\vert
\Psi \left( \theta ^{\left( l\right) },Z_{_{l}}\right) \right\vert ^{2}}=%
\left[ \frac{\delta ^{\sum_{l}p_{l}^{i}}}{\delta
^{\sum_{l}p_{l}^{i}}\left\vert \Psi \left( \theta ^{\left( l\right)
},Z_{_{l}}\right) \right\vert ^{2}}\right] _{B}\left[ \frac{\delta
^{\sum_{l}p_{l}^{i}}}{\delta ^{\sum_{l}p_{l}^{i}}\left\vert \Psi \left(
\theta ^{\left( l\right) },Z_{_{l}}\right) \right\vert ^{2}}\right] _{B}
\end{equation*}%
with%
\begin{equation*}
\check{T}\left( 1-\check{T}\right) ^{-1}
\end{equation*}%
inserted $n$ times.

\subsection*{A.5.4 Formula for the $\protect\delta M\left( \left( \protect%
\theta ,Z\right) ,\left\{ p_{l}^{k},\left( \protect\theta ^{\left( l\right)
},Z_{_{l}}\right) \right\} \right) $}

The fluctuations equation (\ref{Smv}) involves the fluctuations $\delta
M\left( \left( \theta _{1},Z_{1}\right) ,\left( \theta ,Z\right) \right) $.
Using that: 
\begin{eqnarray}
&&\delta M\left( \left( \theta _{1},Z_{1}\right) ,\left( \theta ,Z\right)
\right)  \label{Fu} \\
&\rightarrow &M\left( \left( \theta _{1},Z_{1}\right) ,\left( \theta
,Z\right) \right) \frac{\delta \omega ^{-1}\left( \theta ,Z\right) }{\omega
^{-1}\left( \theta ,Z\right) }  \notag
\end{eqnarray}%
we obtain:%
\begin{eqnarray*}
&&\delta M\left( \left( \theta ,Z\right) ,\left\{ p_{l}^{k},\left( \theta
^{\left( l\right) },Z_{_{l}}\right) \right\} \right) \\
&=&\sum M\left( \left( \theta ,Z\right) ,\left\{ p_{l}^{k},\left( \theta
^{\left( l\right) },Z_{_{l}}\right) \right\} \right) p_{l}^{k}\frac{\delta
\omega ^{-1}\left( \theta ,Z\right) }{\omega ^{-1}\left( \theta ,Z\right) }
\\
&&+\sum M\left( \left( \theta ,Z\right) ,\left[ \left\{ p_{l}^{k},\left(
\theta ^{\left( l\right) },Z_{_{l}}\right) \right\} \right] _{B}\right) \\
&&\times \frac{\omega ^{-1}\left( \theta ^{\left( 1\right) },Z_{_{1}}\right)
\delta \check{T}_{i}\left( \theta ^{\left( 1\right) }-\frac{\left\vert
Z^{\left( 1\right) }-Z^{\left( 2\right) }\right\vert }{c},Z^{\left( 1\right)
},Z^{\left( 2\right) },\omega _{0}\right) }{\check{T}\left( \theta ^{\left(
1\right) }-\frac{\left\vert Z^{\left( 1\right) }-Z^{\left( 2\right)
}\right\vert }{c},Z^{\left( 1\right) },Z^{\left( 2\right) },\omega
_{0}\right) }M\left( \left( \theta ,Z\right) ,\left[ \left\{
p_{l}^{k},\left( \theta ^{\left( l\right) },Z_{_{l}}\right) \right\} \right]
_{B^{\prime }}\right)
\end{eqnarray*}%
where $Z_{_{1}}$ belongs to $B$ and $Z_{_{2}}$ belongs to $B$.

\begin{eqnarray*}
&&\delta M\left( \left( \theta _{k},Z_{k}\right) ,\left\{ p_{l}^{k},\left(
\theta ^{\left( l\right) },Z_{_{l}}\right) \right\} ,\left\{ 1,\left( \theta
,Z\right) \right\} \right) \\
&=&\sum M\left( \left( \theta _{k},Z_{k}\right) ,\left\{ p_{l}^{k},\left(
\theta ^{\left( l\right) },Z_{_{l}}\right) \right\} ,\left\{ 1,\left( \theta
,Z\right) \right\} \right) p_{l}^{k}\frac{\delta \omega ^{-1}\left( \theta
,Z\right) }{\omega ^{-1}\left( \theta ,Z\right) } \\
&&+M\left( \left( \theta _{k},Z_{k}\right) ,\left\{ p_{l}^{k},\left( \theta
^{\left( l\right) },Z_{_{l}}\right) \right\} ,\left\{ 1,\left( \theta
,Z\right) \right\} \right) \frac{\delta \omega ^{-1}\left( \theta ,Z\right) 
}{\omega ^{-1}\left( \theta ,Z\right) } \\
&&+\sum M\left( \left( \theta ,Z\right) ,\left[ \left\{ p_{l}^{k},\left(
\theta ^{\left( l\right) },Z_{_{l}}\right) \right\} ,\left\{ 1,\left( \theta
,Z\right) \right\} \right] _{B}\right) \\
&&\frac{\omega ^{-1}\left( \theta ^{\left( 1\right) },Z_{_{1}}\right) \delta 
\check{T}_{i}\left( \theta ^{\left( 1\right) }-\frac{\left\vert Z^{\left(
1\right) }-Z^{\left( 2\right) }\right\vert }{c},Z^{\left( 1\right)
},Z^{\left( 2\right) },\omega _{0}\right) }{\check{T}\left( \theta ^{\left(
1\right) }-\frac{\left\vert Z^{\left( 1\right) }-Z^{\left( 2\right)
}\right\vert }{c},Z^{\left( 1\right) },Z^{\left( 2\right) },\omega
_{0}\right) }M\left( \left( \theta ,Z\right) ,\left[ \left\{
p_{l}^{k},\left( \theta ^{\left( l\right) },Z_{_{l}}\right) \right\}
,\left\{ 1,\left( \theta ,Z\right) \right\} \right] _{B^{\prime }}\right)
\end{eqnarray*}%
Using the approximatn:%
\begin{eqnarray}
\frac{\delta ^{n}\omega ^{-1}\left( J,\theta ,Z\right) }{\prod%
\limits_{i=1}^{n}\delta \left\vert \Psi \left( \theta -l_{i},Z_{i}\right)
\right\vert ^{2}} &\simeq &\frac{\exp \left( -cl_{n}-\alpha \left(
\sum_{i=1}^{n-1}\left( \left( c\left( l_{i}-l_{i+1}\right) \right)
^{2}-\left\vert Z_{i}-Z_{i+1}\right\vert ^{2}\right) \right) \right) }{D^{n}}
\\
&&\times H\left( cl_{n}-\sum_{i=1}^{n-1}\left\vert Z_{i}-Z_{i+1}\right\vert
\right) \prod\limits_{i=1}^{n}\frac{\omega _{0}^{-1}\left( J,\theta
-l_{i},Z_{i}\right) }{\mathcal{\bar{G}}_{0}\left( 0,Z_{i}\right) \left(
1+T_{0}\left( \theta -l_{i},Z_{l},\right) \right) }  \notag
\end{eqnarray}%
and products:%
\begin{eqnarray*}
&&M\left( \left( \theta ,Z\right) ,\left[ \left\{ p_{l}^{k},\left( \theta
^{\left( l\right) },Z_{_{l}}\right) \right\} ,\left\{ 1,\left( \theta
,Z\right) \right\} \right] _{B}\right) \\
&&\times M\left( \left( \theta ,Z\right) ,\left[ \left\{ p_{l}^{k},\left(
\theta ^{\left( l\right) },Z_{_{l}}\right) \right\} ,\left\{ 1,\left( \theta
,Z\right) \right\} \right] _{B^{\prime }}\right)
\end{eqnarray*}%
are in general negligible.

Thus:%
\begin{eqnarray}
&&\delta M\left( \left( \theta ,Z\right) ,\left\{ p_{l}^{k},\left( \theta
^{\left( l\right) },Z_{_{l}}\right) \right\} \right)  \label{Fd} \\
&=&\sum M\left( \left( \theta ,Z\right) ,\left\{ p_{l}^{k},\left( \theta
^{\left( l\right) },Z_{_{l}}\right) \right\} \right) p_{l}^{k}\frac{\delta
\omega ^{-1}\left( \theta ^{\left( l\right) },Z_{_{l}}\right) }{\omega
^{-1}\left( \theta ^{\left( l\right) },Z_{_{l}}\right) }  \notag
\end{eqnarray}%
and:%
\begin{eqnarray}
&&\delta M\left( \left( \theta _{k},Z_{k}\right) ,\left\{ p_{l}^{k},\left(
\theta ^{\left( l\right) },Z_{_{l}}\right) \right\} ,\left\{ 1,\left( \theta
,Z\right) \right\} \right)  \label{Ftr} \\
&=&\sum M\left( \left( \theta _{k},Z_{k}\right) ,\left\{ p_{l}^{k},\left(
\theta ^{\left( l\right) },Z_{_{l}}\right) \right\} ,\left\{ 1,\left( \theta
,Z\right) \right\} \right) p_{l}^{k}\frac{\delta \omega ^{-1}\left( \theta
^{\left( l\right) },Z_{_{l}}\right) }{\omega ^{-1}\left( \theta ^{\left(
l\right) },Z_{_{l}}\right) }  \notag \\
&&+M\left( \left( \theta _{k},Z_{k}\right) ,\left\{ p_{l}^{k},\left( \theta
^{\left( l\right) },Z_{_{l}}\right) \right\} ,\left\{ 1,\left( \theta
,Z\right) \right\} \right) \frac{\delta \omega ^{-1}\left( \theta ^{\left(
l\right) },Z_{_{l}}\right) }{\omega ^{-1}\left( \theta ^{\left( l\right)
},Z_{_{l}}\right) }  \notag
\end{eqnarray}%
to the lowest order.

\subsection*{A5.5 Derivation of the terms involved in (\protect\ref{Smv})}

Using (\ref{Fu}), (\ref{Fd}) and (\ref{Ftr}) the first contribution involves 
$\delta \omega ^{-1}$:%
\begin{eqnarray*}
&&\frac{1}{2}\nabla _{\theta }\left( \delta \omega ^{-1}\left( \theta ,Z,%
\mathcal{G}_{0}+\Psi \right) \right) \Psi \left( \theta ,Z\right) +\frac{1}{2%
}\nabla _{\theta }M\left( \left( \theta _{1},Z_{1}\right) ,\left( \theta
,Z\right) \right) \left\vert \Psi \left( \theta _{1},Z_{1}\right)
\right\vert ^{2}\Psi \left( \theta ,Z\right) \frac{\delta \omega ^{-1}\left(
\theta ,Z\right) }{\omega ^{-1}\left( \theta ,Z\right) } \\
&&+\Psi \left( \theta ,Z\right) \sum_{k=1}^{m+1}\frac{1}{2}\nabla _{\theta
}\left( M\left( \left( \theta ,Z\right) ,\left\{ p_{l}^{k},\left( \theta
^{\left( l\right) },Z_{_{l}}\right) \right\} \right) \sum p_{l}^{k}\frac{%
\delta \omega ^{-1}\left( \theta ^{\left( l\right) },Z_{_{l}}\right) }{%
\omega ^{-1}\left( \theta ^{\left( l\right) },Z_{_{l}}\right) }\right. \\
&&+M\left( \left( \theta _{k},Z_{k}\right) ,\left\{ p_{l}^{k},\left( \theta
^{\left( l\right) },Z_{_{l}}\right) \right\} ,\left\{ 1,\left( \theta
,Z\right) \right\} \right) \left\vert \Psi \left( \theta _{k},Z_{k}\right)
\right\vert ^{2}\times \\
&&\left. \times \left\{ \sum p_{l}^{k}\frac{\delta \omega ^{-1}\left( \theta
^{\left( l\right) },Z_{_{l}}\right) }{\omega ^{-1}\left( \theta ^{\left(
l\right) },Z_{_{l}}\right) }+\frac{\delta \omega ^{-1}\left( \theta
,Z\right) }{\omega ^{-1}\left( \theta ,Z\right) }\right\} \right) \\
&&\times \left( \prod\limits_{\substack{ i=1  \\ i\neq k}}^{m+1}\frac{1}{2}%
\nabla _{\theta }M\left( \left( \theta _{i},Z_{i}\right) ,\left\{
p_{l}^{i},\left( \theta ^{\left( l\right) },Z_{_{l}}\right) \right\} \right)
\left\vert \Psi \left( \theta _{i},Z_{i}\right) \right\vert ^{2}\right)
\prod\limits_{l=1}^{j}\left\vert \Psi \left( \theta ^{\left( l\right)
},Z_{l}\right) \right\vert ^{2}
\end{eqnarray*}%
where, for the sake of simplicity:%
\begin{equation*}
\frac{\delta \omega ^{-1}\left( \theta ^{\left( l\right) },Z_{_{l}}\right) }{%
\omega ^{-1}\left( \theta ^{\left( l\right) },Z_{_{l}}\right) }
\end{equation*}%
stands for:%
\begin{equation*}
\frac{\delta \omega ^{-1}\left( \theta ^{\left( l\right) },Z_{_{l}}\right) }{%
\omega ^{-1}\left( \theta ^{\left( l\right) },Z_{_{l}}\right) }+\left\langle 
\frac{\delta \check{T}\left( \theta ^{\left( l\right) }-\frac{\left\vert
Z^{\left( 1\right) }-Z^{\left( l\right) }\right\vert }{c},Z^{\left( 1\right)
},Z^{\left( l\right) }\right) }{\check{T}\left( \theta ^{\left( l\right) }-%
\frac{\left\vert Z^{\left( 1\right) }-Z^{\left( 2\right) }\right\vert }{c}%
,Z^{\left( 1\right) },Z^{\left( l\right) }\right) }\right\rangle _{Z_{1}}
\end{equation*}%
Dividing by $\Psi \left( \theta ,Z\right) $ and reintroducing the sums:%
\begin{eqnarray*}
&&\frac{1}{2}\nabla _{\theta }\left( \delta \omega ^{-1}\left( \theta ,Z,%
\mathcal{G}_{0}+\Psi \right) \right) +\frac{1}{2}\nabla _{\theta }M\left(
\left( \theta _{1},Z_{1}\right) ,\left( \theta ,Z\right) \right) \left\vert
\Psi \left( \theta _{1},Z_{1}\right) \right\vert ^{2}\frac{\delta \omega
^{-1}\left( \theta ,Z\right) }{\omega ^{-1}\left( \theta ,Z\right) } \\
&&+\sum_{\substack{ j\geqslant 1  \\ m\geqslant 1}}\sum_{\substack{ \left(
p_{l}^{i}\right) _{\left( m+1\right) \times j}  \\ \sum_{i}p_{l}^{i}%
\geqslant 2}}\sum_{k=1}^{m+1}a_{j,m}\frac{1}{2}\nabla _{\theta }\left(
M\left( \left( \theta ,Z\right) ,\left\{ p_{l}^{k},\left( \theta ^{\left(
l\right) },Z_{_{l}}\right) \right\} \right) \sum p_{l}^{k}\frac{\delta
\omega ^{-1}\left( \theta ^{\left( l\right) },Z_{_{l}}\right) }{\omega
^{-1}\left( \theta ^{\left( l\right) },Z_{_{l}}\right) }\right. \\
&&+M\left( \left( \theta _{k},Z_{k}\right) ,\left\{ p_{l}^{k},\left( \theta
^{\left( l\right) },Z_{_{l}}\right) \right\} ,\left\{ 1,\left( \theta
,Z\right) \right\} \right) \left\vert \Psi \left( \theta _{k},Z_{k}\right)
\right\vert ^{2}\times \\
&&\left. \times \left\{ \sum p_{l}^{k}\frac{\delta \omega ^{-1}\left( \theta
^{\left( l\right) },Z_{_{l}}\right) }{\omega ^{-1}\left( \theta ^{\left(
l\right) },Z_{_{l}}\right) }+\frac{\delta \omega ^{-1}\left( \theta
,Z\right) }{\omega ^{-1}\left( \theta ,Z\right) }\right\} \right) \\
&&\times \left( \prod\limits_{\substack{ i=1  \\ i\neq k}}^{m+1}\frac{1}{2}%
\nabla _{\theta }M\left( \left( \theta _{i},Z_{i}\right) ,\left\{
p_{l}^{i},\left( \theta ^{\left( l\right) },Z_{_{l}}\right) \right\} \right)
\left\vert \Psi \left( \theta _{i},Z_{i}\right) \right\vert ^{2}\right)
\prod\limits_{l=1}^{j}\left\vert \Psi \left( \theta ^{\left( l\right)
},Z_{l}\right) \right\vert ^{2}
\end{eqnarray*}%
and this is factored as:%
\begin{eqnarray*}
&&\frac{1}{2}\nabla _{\theta }\left( \delta \omega ^{-1}\left( \theta ,Z,%
\mathcal{G}_{0}+\Psi \right) \right) +\frac{1}{2}\nabla _{\theta }M\left(
\left( \theta _{1},Z_{1}\right) ,\left( \theta ,Z\right) \right) \left\vert
\Psi \left( \theta _{1},Z_{1}\right) \right\vert ^{2}\frac{\delta \omega
^{-1}\left( \theta ,Z\right) }{\omega ^{-1}\left( \theta ,Z\right) } \\
&&+\sum_{\substack{ j\geqslant 1  \\ m\geqslant 1}}\sum_{\substack{ \left(
p_{l}^{i}\right) _{\left( m+1\right) \times j}  \\ \sum_{i}p_{l}^{i}%
\geqslant 2}}\sum_{k=1}^{m+1}a_{j,m}\frac{1}{2}\left( \frac{\nabla _{\theta
}\left( M\left( \left( \theta ,Z\right) ,\left\{ p_{l}^{k},\left( \theta
^{\left( l\right) },Z_{_{l}}\right) \right\} \right) \sum p_{l}^{k}\frac{%
\delta \omega ^{-1}\left( \theta ^{\left( l\right) },Z_{_{l}}\right) }{%
\omega ^{-1}\left( \theta ^{\left( l\right) },Z_{_{l}}\right) }\right) }{%
\nabla _{\theta }M\left( \left( \theta _{k},Z_{k}\right) ,\left\{
p_{l}^{k},\left( \theta ^{\left( l\right) },Z_{_{l}}\right) \right\} \right)
\left\vert \Psi \left( \theta _{k},Z_{k}\right) \right\vert ^{2}}\right. \\
&&\left. +\frac{\nabla _{\theta }M\left( \left( \theta _{k},Z_{k}\right)
,\left\{ p_{l}^{k},\left( \theta ^{\left( l\right) },Z_{_{l}}\right)
\right\} ,\left\{ 1,\left( \theta ,Z\right) \right\} \right) \left\{ \sum
p_{l}^{k}\frac{\delta \omega ^{-1}\left( \theta ^{\left( l\right)
},Z_{_{l}}\right) }{\omega ^{-1}\left( \theta ^{\left( l\right)
},Z_{_{l}}\right) }+\frac{\delta \omega ^{-1}\left( \theta ,Z\right) }{%
\omega ^{-1}\left( \theta ,Z\right) }\right\} }{\nabla _{\theta }M\left(
\left( \theta _{k},Z_{k}\right) ,\left\{ p_{l}^{k},\left( \theta ^{\left(
l\right) },Z_{_{l}}\right) \right\} \right) }\right) \\
&&\times \left( \prod\limits_{i=1}^{m+1}\frac{1}{2}\nabla _{\theta }M\left(
\left( \theta _{i},Z_{i}\right) ,\left\{ p_{l}^{i},\left( \theta ^{\left(
l\right) },Z_{_{l}}\right) \right\} \right) \left\vert \Psi \left( \theta
_{i},Z_{i}\right) \right\vert ^{2}\right) \prod\limits_{l=1}^{j}\left\vert
\Psi \left( \theta ^{\left( l\right) },Z_{l}\right) \right\vert ^{2}
\end{eqnarray*}

The second type of contribution involve $\delta \Psi $. Some of these terms
are "posterior" to $\left( \theta ,Z\right) $:\bigskip 
\begin{eqnarray*}
&&\frac{1}{2}\nabla _{\theta }M\left( \left( \theta _{1},Z_{1}\right)
,\left( 1,\theta ,Z\right) \right) \Psi ^{\dag }\left( \theta
_{1},Z_{1}\right) \delta \Psi \left( \theta _{1},Z_{1}\right) \Psi \left(
\theta ,Z\right) \\
&&+\Psi \left( \theta ,Z\right) \sum_{k=1}^{m+1}\frac{1}{2}\left( \nabla
_{\theta }M\left( \left( \theta _{k},Z_{k}\right) ,\left\{ p_{l}^{k},\left(
\theta ^{\left( l\right) },Z_{_{l}}\right) \right\} ,\left\{ 1,\left( \theta
,Z\right) \right\} \right) \Psi ^{\dag }\left( \theta _{k},Z_{k}\right)
\delta \Psi \left( \theta _{k},Z_{k}\right) \right) \\
&&\times \left( \prod\limits_{\substack{ i=1  \\ i\neq k}}^{m+1}\frac{1}{2}%
\nabla _{\theta }M\left( \left( \theta _{i},Z_{i}\right) ,\left\{
p_{l}^{i},\left( \theta ^{\left( l\right) },Z_{_{l}}\right) \right\} \right)
\left\vert \Psi \left( \theta _{i},Z_{i}\right) \right\vert ^{2}\right)
\prod\limits_{l=1}^{j}\left\vert \Psi \left( \theta ^{\left( l\right)
},Z_{l}\right) \right\vert ^{2}
\end{eqnarray*}%
The other terms are "anterior" to $\left( \theta ,Z\right) $:

\begin{eqnarray*}
&&\Psi \left( \theta ,Z\right) \sum_{k=1}^{m+1}\frac{1}{2}\left( \nabla
_{\theta }M\left( \left( \theta ,Z\right) ,\left\{ p_{l}^{k},\left( \theta
^{\left( l\right) },Z_{_{l}}\right) \right\} \right) +\nabla _{\theta
}M\left( \left( \theta _{k},Z_{k}\right) ,\left\{ p_{l}^{k},\left( \theta
^{\left( l\right) },Z_{_{l}}\right) \right\} ,\left\{ 1,\left( \theta
,Z\right) \right\} \right) \left\vert \Psi \left( \theta _{k},Z_{k}\right)
\right\vert ^{2}\right) \\
&&\times \left\{ \sum_{k^{\prime }=1,k^{\prime }\neq k}^{m+1}\frac{\delta
\Psi \left( \theta _{k^{\prime }},Z_{k^{\prime }}\right) }{\Psi \left(
\theta _{k^{\prime }},Z_{k^{\prime }}\right) }\left( \prod\limits 
_{\substack{ i=1  \\ i\neq k}}^{m+1}\frac{1}{2}\nabla _{\theta }M\left(
\left( \theta _{i},Z_{i}\right) ,\left\{ p_{l}^{i},\left( \theta ^{\left(
l\right) },Z_{_{l}}\right) \right\} \right) \left\vert \Psi \left( \theta
_{i},Z_{i}\right) \right\vert ^{2}\right) \prod\limits_{l=1}^{j}\left\vert
\Psi \left( \theta ^{\left( l\right) },Z_{l}\right) \right\vert ^{2}\right.
\\
&&+\left. \left( \prod\limits_{\substack{ i=1  \\ i\neq k}}^{m+1}\frac{1}{2}%
\nabla _{\theta }M\left( \left( \theta _{i},Z_{i}\right) ,\left\{
p_{l}^{i},\left( \theta ^{\left( l\right) },Z_{_{l}}\right) \right\} \right)
\left\vert \Psi \left( \theta _{i},Z_{i}\right) \right\vert ^{2}\right)
\sum_{l^{\prime }=1}^{j}\frac{\delta \Psi \left( \theta _{l^{\prime
}},Z_{l^{\prime }}\right) }{\Psi \left( \theta _{l^{\prime }},Z_{l^{\prime
}}\right) }\prod\limits_{l=1}^{j}\left\vert \Psi \left( \theta ^{\left(
l\right) },Z_{l}\right) \right\vert ^{2}\right\} \\
&=&\Psi \left( \theta ,Z\right) \sum_{k=1}^{m+1}\frac{1}{2}\left( \nabla
_{\theta }M\left( \left( \theta ,Z\right) ,\left\{ p_{l}^{k},\left( \theta
^{\left( l\right) },Z_{_{l}}\right) \right\} \right) +\nabla _{\theta
}M\left( \left( \theta _{k},Z_{k}\right) ,\left\{ p_{l}^{k},\left( \theta
^{\left( l\right) },Z_{_{l}}\right) \right\} ,\left\{ 1,\left( \theta
,Z\right) \right\} \right) \left\vert \Psi \left( \theta _{k},Z_{k}\right)
\right\vert ^{2}\right) \\
&&\times \left( \sum_{k^{\prime }=1,k^{\prime }\neq k}^{m+1}\frac{\delta
\Psi \left( \theta _{k^{\prime }},Z_{k^{\prime }}\right) }{\Psi \left(
\theta _{k^{\prime }},Z_{k^{\prime }}\right) }+\sum_{l^{\prime }=1}^{j}\frac{%
\delta \Psi \left( \theta _{l^{\prime }},Z_{l^{\prime }}\right) }{\Psi
\left( \theta _{l^{\prime }},Z_{l^{\prime }}\right) }\right) \\
&&\times \left( \prod\limits_{\substack{ i=1  \\ i\neq k}}^{m+1}\frac{1}{2}%
\nabla _{\theta }M\left( \left( \theta _{i},Z_{i}\right) ,\left\{
p_{l}^{i},\left( \theta ^{\left( l\right) },Z_{_{l}}\right) \right\} \right)
\left\vert \Psi \left( \theta _{i},Z_{i}\right) \right\vert ^{2}\right)
\prod\limits_{l=1}^{j}\left\vert \Psi \left( \theta ^{\left( l\right)
},Z_{l}\right) \right\vert ^{2}
\end{eqnarray*}%
\bigskip

Gathering these terms and summing yields:%
\begin{eqnarray*}
&&\frac{1}{2}\nabla _{\theta }M\left( \left( \theta _{1},Z_{1}\right)
,\left( \theta ,Z\right) \right) \Psi ^{\dag }\left( \theta
_{1},Z_{1}\right) \delta \Psi \left( \theta _{1},Z_{1}\right) \Psi \left(
\theta ,Z\right) \\
&&+\Psi \left( \theta ,Z\right) \sum_{\substack{ j\geqslant 1  \\ m\geqslant
1 }}\sum_{\substack{ \left( p_{l}^{i}\right) _{\left( m+1\right) \times j} 
\\ \sum_{i}p_{l}^{i}\geqslant 2}}\sum_{k=1}^{m+1}a_{j,m}\frac{1}{2}\left(
\nabla _{\theta }M\left( \left( \theta _{k},Z_{k}\right) ,\left\{
p_{l}^{k},\left( \theta ^{\left( l\right) },Z_{_{l}}\right) \right\}
,\left\{ 1,\left( \theta ,Z\right) \right\} \right) \Psi ^{\dag }\left(
\theta _{k},Z_{k}\right) \delta \Psi \left( \theta _{k},Z_{k}\right) \right)
\\
&&\times \left( \prod\limits_{\substack{ i=1  \\ i\neq k}}^{m+1}\frac{1}{2}%
\nabla _{\theta }M\left( \left( \theta _{i},Z_{i}\right) ,\left\{
p_{l}^{i},\left( \theta ^{\left( l\right) },Z_{_{l}}\right) \right\} \right)
\left\vert \Psi \left( \theta _{i},Z_{i}\right) \right\vert ^{2}\right)
\prod\limits_{l=1}^{j}\left\vert \Psi \left( \theta ^{\left( l\right)
},Z_{l}\right) \right\vert ^{2} \\
&&+\Psi \left( \theta ,Z\right) \sum_{\substack{ j\geqslant 1  \\ m\geqslant
1 }}\sum_{\substack{ \left( p_{l}^{i}\right) _{\left( m+1\right) \times j} 
\\ \sum_{i}p_{l}^{i}\geqslant 2}}\sum_{k=1}^{m+1}a_{j,m} \\
&&\times \frac{1}{2}\left( \nabla _{\theta }M\left( \left( \theta ,Z\right)
,\left\{ p_{l}^{k},\left( \theta ^{\left( l\right) },Z_{_{l}}\right)
\right\} \right) +\nabla _{\theta }M\left( \left( \theta _{k},Z_{k}\right)
,\left\{ p_{l}^{k},\left( \theta ^{\left( l\right) },Z_{_{l}}\right)
\right\} ,\left\{ 1,\left( \theta ,Z\right) \right\} \right) \left\vert \Psi
\left( \theta _{k},Z_{k}\right) \right\vert ^{2}\right) \\
&&\times \left( \sum_{k^{\prime }=1,k^{\prime }\neq k}^{m+1}\frac{\delta
\Psi \left( \theta _{k^{\prime }},Z_{k^{\prime }}\right) }{\Psi \left(
\theta _{k^{\prime }},Z_{k^{\prime }}\right) }+\sum_{l^{\prime }=1}^{j}\frac{%
\delta \Psi \left( \theta _{l^{\prime }},Z_{l^{\prime }}\right) }{\Psi
\left( \theta _{l^{\prime }},Z_{l^{\prime }}\right) }\right) \\
&&\times \left( \prod\limits_{\substack{ i=1  \\ i\neq k}}^{m+1}\frac{1}{2}%
\nabla _{\theta }M\left( \left( \theta _{i},Z_{i}\right) ,\left\{
p_{l}^{i},\left( \theta ^{\left( l\right) },Z_{_{l}}\right) \right\} \right)
\left\vert \Psi \left( \theta _{i},Z_{i}\right) \right\vert ^{2}\right)
\prod\limits_{l=1}^{j}\left\vert \Psi \left( \theta ^{\left( l\right)
},Z_{l}\right) \right\vert ^{2}
\end{eqnarray*}%
which can be factored as:%
\begin{eqnarray*}
&&\frac{1}{2}\nabla _{\theta }M\left( \left( \theta _{1},Z_{1}\right)
,\left( \theta ,Z\right) \right) \left\vert \Psi \left( \theta
_{1},Z_{1}\right) \right\vert ^{2}\frac{\delta \Psi \left( \theta
_{1},Z_{1}\right) }{\Psi \left( \theta _{1},Z_{1}\right) } \\
&&+\sum_{\substack{ j\geqslant 1  \\ m\geqslant 1}}\sum_{\substack{ \left(
p_{l}^{i}\right) _{\left( m+1\right) \times j}  \\ \sum_{i}p_{l}^{i}%
\geqslant 2}}\sum_{k=1}^{m+1}a_{j,m}\left\{ \frac{\nabla _{\theta }M\left(
\left( \theta ,Z\right) ,\left\{ p_{l}^{k},\left( \theta ^{\left( l\right)
},Z_{_{l}}\right) \right\} \right) }{\nabla _{\theta }M\left( \left( \theta
_{k},Z_{k}\right) ,\left\{ p_{l}^{k},\left( \theta ^{\left( l\right)
},Z_{_{l}}\right) \right\} \right) \left\vert \Psi \left( \theta
_{k},Z_{k}\right) \right\vert ^{2}}\right. \\
&&\times \left( \sum_{k^{\prime }=1,k^{\prime }\neq k}^{m+1}\frac{\delta
\Psi \left( \theta _{k^{\prime }},Z_{k^{\prime }}\right) }{\Psi \left(
\theta _{k^{\prime }},Z_{k^{\prime }}\right) }+\sum_{l^{\prime }=1}^{j}\frac{%
\delta \Psi \left( \theta _{l^{\prime }},Z_{l^{\prime }}\right) }{\Psi
\left( \theta _{l^{\prime }},Z_{l^{\prime }}\right) }\right) \\
&&+\left. \frac{\nabla _{\theta }M\left( \left( \theta _{k},Z_{k}\right)
,\left\{ p_{l}^{k},\left( \theta ^{\left( l\right) },Z_{_{l}}\right)
\right\} ,\left\{ 1,\left( \theta ,Z\right) \right\} \right) }{\nabla
_{\theta }M\left( \left( \theta _{k},Z_{k}\right) ,\left\{ p_{l}^{k},\left(
\theta ^{\left( l\right) },Z_{_{l}}\right) \right\} \right) }\left( \frac{%
\delta \Psi \left( \theta _{k},Z_{k}\right) }{\Psi \left( \theta
_{k},Z_{k}\right) }+\sum_{k^{\prime }=1,k^{\prime }\neq k}^{m+1}\frac{\delta
\Psi \left( \theta _{k^{\prime }},Z_{k^{\prime }}\right) }{\Psi \left(
\theta _{k^{\prime }},Z_{k^{\prime }}\right) }+\sum_{l^{\prime }=1}^{j}\frac{%
\delta \Psi \left( \theta _{l^{\prime }},Z_{l^{\prime }}\right) }{\Psi
\left( \theta _{l^{\prime }},Z_{l^{\prime }}\right) }\right) \right\} \\
&&\times \left( \prod\limits_{i=1}^{m+1}\frac{1}{2}\nabla _{\theta }M\left(
\left( \theta _{i},Z_{i}\right) ,\left\{ p_{l}^{i},\left( \theta ^{\left(
l\right) },Z_{_{l}}\right) \right\} \right) \left\vert \Psi \left( \theta
_{i},Z_{i}\right) \right\vert ^{2}\right) \prod\limits_{l=1}^{j}\left\vert
\Psi \left( \theta ^{\left( l\right) },Z_{l}\right) \right\vert ^{2}
\end{eqnarray*}

\bigskip\ Adding the contrbution:%
\begin{equation*}
\left( \frac{1}{2}\left( -\frac{\sigma _{\theta }^{2}}{2}\nabla _{\theta
}+\omega ^{-1}\left( J\left( \theta \right) ,\theta ,Z\right) \right) \nabla
_{\theta }+\left( A\left( \Psi \left( \theta ,Z\right) \right) -\frac{\sigma
_{\theta }^{2}}{2}\right) \nabla _{\theta }+U^{\prime \prime }\left(
\left\vert \Psi \left( \theta ,Z\right) \right\vert ^{2}\right) \left\vert
\Psi \left( \theta ,Z\right) \right\vert ^{2}\right) \frac{\delta \Psi
\left( \theta ,Z\right) }{\Psi \left( \theta ,Z\right) }
\end{equation*}%
and gathering the different contributions leads to:\bigskip

\begin{eqnarray*}
0 &\rightarrow &\left( \frac{1}{2}\left( -\frac{\sigma _{\theta }^{2}}{2}%
\nabla _{\theta }+\omega ^{-1}\left( J\left( \theta \right) ,\theta
,Z\right) \right) \nabla _{\theta }+\left( A\left( \Psi \left( \theta
,Z\right) \right) -\frac{\sigma _{\theta }^{2}}{2}\right) \nabla _{\theta
}+U^{\prime \prime }\left( \left\vert \Psi \left( \theta ,Z\right)
\right\vert ^{2}\right) \left\vert \Psi \left( \theta ,Z\right) \right\vert
^{2}\right) \frac{\delta \Psi \left( \theta ,Z\right) }{\Psi \left( \theta
,Z\right) } \\
&&+\frac{1}{2}\nabla _{\theta }\left( \delta \omega ^{-1}\left( J\left(
\theta \right) ,\theta ,Z\right) \right) +\frac{1}{2}\nabla _{\theta
}M\left( \left( \theta _{1},Z_{1}\right) ,\left( \theta ,Z\right) \right)
\left\vert \Psi \left( \theta _{1},Z_{1}\right) \right\vert ^{2}\frac{\delta
\omega ^{-1}\left( \theta ,Z\right) }{\omega ^{-1}\left( \theta ,Z\right) }
\\
&&+\sum_{\substack{ j\geqslant 1  \\ m\geqslant 1}}\sum_{\substack{ \left(
p_{l}^{i}\right) _{\left( m+1\right) \times j}  \\ \sum_{i}p_{l}^{i}%
\geqslant 2}}\sum_{k=1}^{m+1}a_{j,m}\frac{1}{2}\left( \frac{\nabla _{\theta
}\left( M\left( \left( \theta ,Z\right) ,\left\{ p_{l}^{k},\left( \theta
^{\left( l\right) },Z_{_{l}}\right) \right\} \right) \sum p_{l}^{k}\frac{%
\delta \omega ^{-1}\left( \theta ^{\left( l\right) },Z_{_{l}}\right) }{%
\omega ^{-1}\left( \theta ^{\left( l\right) },Z_{_{l}}\right) }\right) }{%
\nabla _{\theta }M\left( \left( \theta _{k},Z_{k}\right) ,\left\{
p_{l}^{k},\left( \theta ^{\left( l\right) },Z_{_{l}}\right) \right\} \right)
\left\vert \Psi \left( \theta _{k},Z_{k}\right) \right\vert ^{2}}\right. \\
&&\left. +\frac{\nabla _{\theta }M\left( \left( \theta _{k},Z_{k}\right)
,\left\{ p_{l}^{k},\left( \theta ^{\left( l\right) },Z_{_{l}}\right)
\right\} ,\left\{ 1,\left( \theta ,Z\right) \right\} \right) \left\{ \sum
p_{l}^{k}\frac{\delta \omega ^{-1}\left( \theta ^{\left( l\right)
},Z_{_{l}}\right) }{\omega ^{-1}\left( \theta ^{\left( l\right)
},Z_{_{l}}\right) }+\frac{\delta \omega ^{-1}\left( \theta ,Z\right) }{%
\omega ^{-1}\left( \theta ,Z\right) }\right\} }{\nabla _{\theta }M\left(
\left( \theta _{k},Z_{k}\right) ,\left\{ p_{l}^{k},\left( \theta ^{\left(
l\right) },Z_{_{l}}\right) \right\} \right) }\right) \\
&&\times \left( \prod\limits_{i=1}^{m+1}\frac{1}{2}\nabla _{\theta }M\left(
\left( \theta _{i},Z_{i}\right) ,\left\{ p_{l}^{i},\left( \theta ^{\left(
l\right) },Z_{_{l}}\right) \right\} \right) \left\vert \Psi \left( \theta
_{i},Z_{i}\right) \right\vert ^{2}\right) \prod\limits_{l=1}^{j}\left\vert
\Psi \left( \theta ^{\left( l\right) },Z_{l}\right) \right\vert ^{2} \\
&&+\frac{1}{2}\nabla _{\theta }M\left( \left( \theta _{1},Z_{1}\right)
,\left( \theta ,Z\right) \right) \left\vert \Psi \left( \theta
_{1},Z_{1}\right) \right\vert ^{2}\frac{\delta \Psi \left( \theta
_{1},Z_{1}\right) }{\Psi \left( \theta _{1},Z_{1}\right) } \\
&&+\sum_{\substack{ j\geqslant 1  \\ m\geqslant 1}}\sum_{\substack{ \left(
p_{l}^{i}\right) _{\left( m+1\right) \times j}  \\ \sum_{i}p_{l}^{i}%
\geqslant 2}}\sum_{k=1}^{m+1}a_{j,m}\left\{ \frac{\nabla _{\theta }M\left(
\left( \theta ,Z\right) ,\left\{ p_{l}^{k},\left( \theta ^{\left( l\right)
},Z_{_{l}}\right) \right\} \right) }{\nabla _{\theta }M\left( \left( \theta
_{k},Z_{k}\right) ,\left\{ p_{l}^{k},\left( \theta ^{\left( l\right)
},Z_{_{l}}\right) \right\} \right) \left\vert \Psi \left( \theta
_{k},Z_{k}\right) \right\vert ^{2}}\right. \\
&&\times \left( \sum_{k^{\prime }=1,k^{\prime }\neq k}^{m+1}\frac{\delta
\Psi \left( \theta _{k^{\prime }},Z_{k^{\prime }}\right) }{\Psi \left(
\theta _{k^{\prime }},Z_{k^{\prime }}\right) }+\sum_{l^{\prime }=1}^{j}\frac{%
\delta \Psi \left( \theta _{l^{\prime }},Z_{l^{\prime }}\right) }{\Psi
\left( \theta _{l^{\prime }},Z_{l^{\prime }}\right) }\right) \\
&&+\left. \frac{\nabla _{\theta }M\left( \left( \theta _{k},Z_{k}\right)
,\left\{ p_{l}^{k},\left( \theta ^{\left( l\right) },Z_{_{l}}\right)
\right\} ,\left\{ 1,\left( \theta ,Z\right) \right\} \right) }{\nabla
_{\theta }M\left( \left( \theta _{k},Z_{k}\right) ,\left\{ p_{l}^{k},\left(
\theta ^{\left( l\right) },Z_{_{l}}\right) \right\} \right) }\left( \frac{%
\delta \Psi \left( \theta _{k},Z_{k}\right) }{\Psi \left( \theta
_{k},Z_{k}\right) }+\sum_{k^{\prime }=1,k^{\prime }\neq k}^{m+1}\frac{\delta
\Psi \left( \theta _{k^{\prime }},Z_{k^{\prime }}\right) }{\Psi \left(
\theta _{k^{\prime }},Z_{k^{\prime }}\right) }+\sum_{l^{\prime }=1}^{j}\frac{%
\delta \Psi \left( \theta _{l^{\prime }},Z_{l^{\prime }}\right) }{\Psi
\left( \theta _{l^{\prime }},Z_{l^{\prime }}\right) }\right) \right\} \\
&&\times \left( \prod\limits_{i=1}^{m+1}\frac{1}{2}\nabla _{\theta }M\left(
\left( \theta _{i},Z_{i}\right) ,\left\{ p_{l}^{i},\left( \theta ^{\left(
l\right) },Z_{_{l}}\right) \right\} \right) \left\vert \Psi \left( \theta
_{i},Z_{i}\right) \right\vert ^{2}\right) \prod\limits_{l=1}^{j}\left\vert
\Psi \left( \theta ^{\left( l\right) },Z_{l}\right) \right\vert ^{2}
\end{eqnarray*}%
and coming back to:%
\begin{equation*}
\frac{\delta \omega ^{-1}\left( \theta ^{\left( l\right) },Z_{_{l}}\right) }{%
\omega ^{-1}\left( \theta ^{\left( l\right) },Z_{_{l}}\right) }\rightarrow 
\frac{\delta \omega ^{-1}\left( \theta ^{\left( l\right) },Z_{_{l}}\right) }{%
\omega ^{-1}\left( \theta ^{\left( l\right) },Z_{_{l}}\right) }+\left\langle 
\frac{\delta \check{T}\left( \theta ^{\left( l\right) }-\frac{\left\vert
Z^{\left( 1\right) }-Z^{\left( l\right) }\right\vert }{c},Z^{\left( 1\right)
},Z^{\left( l\right) }\right) }{\check{T}\left( \theta ^{\left( l\right) }-%
\frac{\left\vert Z^{\left( 1\right) }-Z^{\left( 2\right) }\right\vert }{c}%
,Z^{\left( 1\right) },Z^{\left( l\right) }\right) }\right\rangle _{Z_{1}}
\end{equation*}%
this is:%
\begin{eqnarray*}
0 &\rightarrow &\left( \frac{1}{2}\left( -\frac{\sigma _{\theta }^{2}}{2}%
\nabla _{\theta }+\omega ^{-1}\left( J\left( \theta \right) ,\theta
,Z\right) \right) \nabla _{\theta }+\left( A\left( \Psi \left( \theta
,Z\right) \right) -\frac{\sigma _{\theta }^{2}}{2}\right) \nabla _{\theta
}+U^{\prime \prime }\left( \left\vert \Psi \left( \theta ,Z\right)
\right\vert ^{2}\right) \left\vert \Psi \left( \theta ,Z\right) \right\vert
^{2}\right) \frac{\delta \Psi \left( \theta ,Z\right) }{\Psi \left( \theta
,Z\right) } \\
&&+\frac{1}{2}\nabla _{\theta }\left( \delta \omega ^{-1}\left( J\left(
\theta \right) ,\theta ,Z\right) \right) +\frac{1}{2}\nabla _{\theta
}M\left( \left( \theta _{1},Z_{1}\right) ,\left( \theta ,Z\right) \right)
\left\vert \Psi \left( \theta _{1},Z_{1}\right) \right\vert ^{2}\frac{\delta
\omega ^{-1}\left( \theta ,Z\right) }{\omega ^{-1}\left( \theta ,Z\right) }
\\
&&+\frac{1}{2}\nabla _{\theta }M\left( \left( \theta _{1},Z_{1}\right)
,\left( \theta ,Z\right) \right) \left\vert \Psi \left( \theta
_{1},Z_{1}\right) \right\vert ^{2}\frac{\delta \Psi \left( \theta
_{1},Z_{1}\right) }{\Psi \left( \theta _{1},Z_{1}\right) } \\
&&+\sum_{\substack{ j\geqslant 1  \\ m\geqslant 1}}\sum_{\substack{ \left(
p_{l}^{i}\right) _{\left( m+1\right) \times j}  \\ \sum_{i}p_{l}^{i}%
\geqslant 2}}\sum_{k=1}^{m+1}a_{j,m}\frac{1}{2}\left( \frac{\nabla _{\theta
}\left( M\left( \left( \theta ,Z\right) ,\left\{ p_{l}^{k},\left( \theta
^{\left( l\right) },Z_{_{l}}\right) \right\} \right) S_{1}\right) }{\nabla
_{\theta }M\left( \left( \theta _{k},Z_{k}\right) ,\left\{ p_{l}^{k},\left(
\theta ^{\left( l\right) },Z_{_{l}}\right) \right\} \right) \left\vert \Psi
\left( \theta _{k},Z_{k}\right) \right\vert ^{2}}\right. \\
&&\left. +\frac{\nabla _{\theta }\left( M\left( \left( \theta
_{k},Z_{k}\right) ,\left\{ p_{l}^{k},\left( \theta ^{\left( l\right)
},Z_{_{l}}\right) \right\} ,\left\{ 1,\left( \theta ,Z\right) \right\}
\right) S_{2}\right) }{\nabla _{\theta }M\left( \left( \theta
_{k},Z_{k}\right) ,\left\{ p_{l}^{k},\left( \theta ^{\left( l\right)
},Z_{_{l}}\right) \right\} \right) }\right) \\
&&\times \left( \prod\limits_{i=1}^{m+1}\frac{1}{2}\nabla _{\theta }M\left(
\left( \theta _{i},Z_{i}\right) ,\left\{ p_{l}^{i},\left( \theta ^{\left(
l\right) },Z_{_{l}}\right) \right\} \right) \left\vert \Psi \left( \theta
_{i},Z_{i}\right) \right\vert ^{2}\right) \prod\limits_{l=1}^{j}\left\vert
\Psi \left( \theta ^{\left( l\right) },Z_{l}\right) \right\vert ^{2}
\end{eqnarray*}

\bigskip where the sums $S_{1}$ and $S_{2}$ are given by:%
\begin{equation*}
S_{1}=\sum p_{l}^{k}\left( \frac{\delta \omega ^{-1}\left( \theta ^{\left(
l\right) },Z_{_{l}}\right) }{\omega ^{-1}\left( \theta ^{\left( l\right)
},Z_{_{l}}\right) }+\left\langle \frac{\delta \check{T}\left( \theta
^{\left( l\right) }-\frac{\left\vert Z^{\left( 1\right) }-Z^{\left( l\right)
}\right\vert }{c},Z^{\left( 1\right) },Z^{\left( l\right) }\right) }{\check{T%
}\left( \theta ^{\left( l\right) }-\frac{\left\vert Z^{\left( 1\right)
}-Z^{\left( 2\right) }\right\vert }{c},Z^{\left( 1\right) },Z^{\left(
l\right) }\right) }\right\rangle _{Z_{1}}\right) +\sum_{k^{\prime
}=1,k^{\prime }\neq k}^{m+1}\frac{\delta \Psi \left( \theta _{k^{\prime
}},Z_{k^{\prime }}\right) }{\Psi \left( \theta _{k^{\prime }},Z_{k^{\prime
}}\right) }+\sum_{l^{\prime }=1}^{j}\frac{\delta \Psi \left( \theta
_{l^{\prime }},Z_{l^{\prime }}\right) }{\Psi \left( \theta _{l^{\prime
}},Z_{l^{\prime }}\right) }
\end{equation*}%
and:%
\begin{eqnarray*}
S_{2} &=&\sum p_{l}^{k}\left( \frac{\delta \omega ^{-1}\left( \theta
^{\left( l\right) },Z_{_{l}}\right) }{\omega ^{-1}\left( \theta ^{\left(
l\right) },Z_{_{l}}\right) }+\left\langle \frac{\delta \check{T}\left(
\theta ^{\left( l\right) }-\frac{\left\vert Z^{\left( 1\right) }-Z^{\left(
l\right) }\right\vert }{c},Z^{\left( 1\right) },Z^{\left( l\right) }\right) 
}{\check{T}\left( \theta ^{\left( l\right) }-\frac{\left\vert Z^{\left(
1\right) }-Z^{\left( 2\right) }\right\vert }{c},Z^{\left( 1\right)
},Z^{\left( l\right) }\right) }\right\rangle _{Z_{1}}\right) +\frac{\delta
\omega ^{-1}\left( \theta ,Z\right) }{\omega ^{-1}\left( \theta ,Z\right) }+%
\frac{\delta \Psi \left( \theta _{k},Z_{k}\right) }{\Psi \left( \theta
_{k},Z_{k}\right) } \\
&&+\sum_{k^{\prime }=1,k^{\prime }\neq k}^{m+1}\frac{\delta \Psi \left(
\theta _{k^{\prime }},Z_{k^{\prime }}\right) }{\Psi \left( \theta
_{k^{\prime }},Z_{k^{\prime }}\right) }+\sum_{l^{\prime }=1}^{j}\frac{\delta
\Psi \left( \theta _{l^{\prime }},Z_{l^{\prime }}\right) }{\Psi \left(
\theta _{l^{\prime }},Z_{l^{\prime }}\right) }
\end{eqnarray*}

\subsection*{A5.6 Matricial form of the fluctuation equations\protect\bigskip%
}

\subsubsection*{A5.6.1 Matricial form}

We first rewrite the anterior contributions to the equation. Setting:%
\begin{equation*}
\overline{\sum }=\sum_{\substack{ j\geqslant 1  \\ m\geqslant 1}}\sum 
_{\substack{ \left( p_{l}^{i}\right) _{\left( m+1\right) \times j}  \\ %
\sum_{i}p_{l}^{i}\geqslant 2}}\sum_{k=1}^{m+1}a_{j,m}
\end{equation*}%
the first one is:%
\begin{eqnarray*}
&&\overline{\sum }\sum_{k=1}^{m+1}\frac{1}{2}\frac{\nabla _{\theta }\left(
M\left( \left( \theta ,Z\right) ,\left\{ p_{l}^{k},\left( \theta ^{\left(
l\right) },Z_{_{l}}\right) \right\} \right) S_{1}\right) }{\nabla _{\theta
}M\left( \left( \theta _{k},Z_{k}\right) ,\left\{ p_{l}^{k},\left( \theta
^{\left( l\right) },Z_{_{l}}\right) \right\} \right) \left\vert \Psi \left(
\theta _{k},Z_{k}\right) \right\vert ^{2}} \\
&&\times \left( \prod\limits_{i=1}^{m+1}\frac{1}{2}\nabla _{\theta }M\left(
\left( \theta _{i},Z_{i}\right) ,\left\{ p_{l}^{i},\left( \theta ^{\left(
l\right) },Z_{_{l}}\right) \right\} \right) \left\vert \Psi \left( \theta
_{i},Z_{i}\right) \right\vert ^{2}\right) \prod\limits_{l=1}^{j}\left\vert
\Psi \left( \theta ^{\left( l\right) },Z_{l}\right) \right\vert ^{2} \\
&=&\overline{\sum }\sum_{k=1}^{m+1}\left\{ \frac{1}{2}\frac{\nabla _{\theta
}\left( M\left( \left( \theta ,Z\right) ,\left\{ p_{l}^{k},\left( \theta
^{\left( l\right) },Z_{_{l}}\right) \right\} \right) \right) }{\nabla
_{\theta }M\left( \left( \theta _{k},Z_{k}\right) ,\left\{ p_{l}^{k},\left(
\theta ^{\left( l\right) },Z_{_{l}}\right) \right\} \right) \left\vert \Psi
\left( \theta _{k},Z_{k}\right) \right\vert ^{2}}S_{1}\right. \\
&&\left. +\frac{1}{2}\frac{M\left( \left( \theta ,Z\right) ,\left\{
p_{l}^{k},\left( \theta ^{\left( l\right) },Z_{_{l}}\right) \right\} \right) 
}{\nabla _{\theta }M\left( \left( \theta _{k},Z_{k}\right) ,\left\{
p_{l}^{k},\left( \theta ^{\left( l\right) },Z_{_{l}}\right) \right\} \right)
\left\vert \Psi \left( \theta _{k},Z_{k}\right) \right\vert ^{2}}\nabla
_{\theta }S_{1}\right\} \\
&&\times \left( \prod\limits_{i=1}^{m+1}\frac{1}{2}\nabla _{\theta }M\left(
\left( \theta _{i},Z_{i}\right) ,\left\{ p_{l}^{i},\left( \theta ^{\left(
l\right) },Z_{_{l}}\right) \right\} \right) \left\vert \Psi \left( \theta
_{i},Z_{i}\right) \right\vert ^{2}\right) \prod\limits_{l=1}^{j}\left\vert
\Psi \left( \theta ^{\left( l\right) },Z_{l}\right) \right\vert ^{2}
\end{eqnarray*}%
and this writes:%
\begin{eqnarray*}
&\rightarrow &M_{\omega }^{\left( 1,r\right) }\left( \left( \theta ,Z\right)
,\left( \theta _{1},Z_{1}\right) \right) \left( \frac{\delta \omega
^{-1}\left( \theta _{1},Z_{1}\right) }{\omega ^{-1}\left( \theta
_{1},Z_{1}\right) }+\left\langle \frac{\delta \check{T}\left( \theta
^{\left( l\right) }-\frac{\left\vert Z^{\left( 1\right) }-Z^{\left( l\right)
}\right\vert }{c},Z^{\left( 1\right) },Z^{\left( l\right) }\right) }{\check{T%
}\left( \theta ^{\left( l\right) }-\frac{\left\vert Z^{\left( 1\right)
}-Z^{\left( 2\right) }\right\vert }{c},Z^{\left( 1\right) },Z^{\left(
l\right) }\right) }\right\rangle _{Z_{1}}\right) \\
&&+M_{\omega }^{\left( 2,r\right) }\left( \left( \theta ,Z\right) ,\left(
\theta _{1},Z_{1}\right) \right) \nabla _{\theta }\left( \frac{\delta \omega
^{-1}\left( \theta _{1},Z_{1}\right) }{\omega ^{-1}\left( \theta
_{1},Z_{1}\right) }+\left\langle \frac{\delta \check{T}\left( \theta
^{\left( l\right) }-\frac{\left\vert Z^{\left( 1\right) }-Z^{\left( l\right)
}\right\vert }{c},Z^{\left( 1\right) },Z^{\left( l\right) }\right) }{\check{T%
}\left( \theta ^{\left( l\right) }-\frac{\left\vert Z^{\left( 1\right)
}-Z^{\left( 2\right) }\right\vert }{c},Z^{\left( 1\right) },Z^{\left(
l\right) }\right) }\right\rangle _{Z_{1}}\right) \\
&&+M_{\Psi }^{\left( 1,r\right) }\left( \left( \theta ,Z\right) ,\left(
\theta _{1},Z_{1}\right) \right) \frac{\delta \Psi \left( \theta
_{1},Z_{1}\right) }{\Psi \left( \theta _{1},Z_{1}\right) }+M_{\Psi }^{\left(
2,r\right) }\left( \left( \theta ,Z\right) ,\left( \theta _{1},Z_{1}\right)
\right) \nabla _{\theta }\frac{\delta \Psi \left( \theta _{1},Z_{1}\right) }{%
\Psi \left( \theta _{1},Z_{1}\right) }
\end{eqnarray*}%
The first contributions are given by the $\delta \omega ^{-1}$
contributions. We find: 
\begin{eqnarray*}
M_{\omega }^{\left( 1,r\right) }\left( \left( \theta ,Z\right) ,\left(
\theta _{1},Z_{1}\right) \right) &=&\overline{\sum }\sum_{k=1}^{m+1}H\left(
c\left( \theta -\theta _{1}\right) -\left\vert Z-Z_{1}\right\vert \right)
\int d\widetilde{\left[ \theta ,Z\right] }\frac{1}{2}\frac{\nabla _{\theta
}\left( M\left( \left( \theta ,Z\right) ,\left\{ p_{l}^{k},\left( \theta
^{\left( l\right) },Z_{_{l}}\right) \right\} \right) \right) \left( \sum
p_{l}^{k}\right) }{\nabla _{\theta }M\left( \left( \theta _{k},Z_{k}\right)
,\left\{ p_{l}^{k},\left( \theta ^{\left( l\right) },Z_{_{l}}\right)
\right\} \right) \left\vert \Psi \left( \theta _{k},Z_{k}\right) \right\vert
^{2}} \\
&&\times \left( \prod\limits_{i=1}^{m+1}\frac{1}{2}\nabla _{\theta }M\left(
\left( \theta _{i},Z_{i}\right) ,\left\{ p_{l}^{i},\left( \theta ^{\left(
l\right) },Z_{_{l}}\right) \right\} \right) \left\vert \Psi \left( \theta
_{i},Z_{i}\right) \right\vert ^{2}\right) \prod\limits_{l=1}^{j}\left\vert
\Psi \left( \theta ^{\left( l\right) },Z_{l}\right) \right\vert ^{2}
\end{eqnarray*}%
with:%
\begin{equation*}
d\widetilde{\left[ \theta ,Z\right] }=\sum_{k=1}^{m+1}\delta \left( \left(
\theta _{1},Z_{1}\right) -\left( \theta ^{\left( k\right) },Z_{k}\right)
\right) \prod\limits_{i=1}^{m+1}d\left( \theta _{i},Z_{i}\right)
\prod\limits_{l=1}^{j}d\left( \theta ^{\left( l\right) },Z_{l}\right)
\end{equation*}%
\begin{eqnarray*}
M_{\omega }^{\left( 2,r\right) }\left( \left( \theta ,Z\right) ,\left(
\theta _{1},Z_{1}\right) \right) &=&\overline{\sum }\sum_{k=1}^{m+1}H\left(
c\left( \theta -\theta _{1}\right) -\left\vert Z-Z_{1}\right\vert \right)
\int d\widetilde{\left[ \theta ,Z\right] }\frac{1}{2}\frac{M\left( \left(
\theta ,Z\right) ,\left\{ p_{l}^{k},\left( \theta ^{\left( l\right)
},Z_{_{l}}\right) \right\} \right) \left( \sum p_{l}^{k}\right) }{\nabla
_{\theta }M\left( \left( \theta _{k},Z_{k}\right) ,\left\{ p_{l}^{k},\left(
\theta ^{\left( l\right) },Z_{_{l}}\right) \right\} \right) \left\vert \Psi
\left( \theta _{k},Z_{k}\right) \right\vert ^{2}} \\
&&\times \left( \prod\limits_{i=1}^{m+1}\frac{1}{2}\nabla _{\theta }M\left(
\left( \theta _{i},Z_{i}\right) ,\left\{ p_{l}^{i},\left( \theta ^{\left(
l\right) },Z_{_{l}}\right) \right\} \right) \left\vert \Psi \left( \theta
_{i},Z_{i}\right) \right\vert ^{2}\right) \prod\limits_{l=1}^{j}\left\vert
\Psi \left( \theta ^{\left( l\right) },Z_{l}\right) \right\vert ^{2}
\end{eqnarray*}

\bigskip The second anterior contribution to consider is given by the
termsproportional to $\delta \Psi $:

\begin{eqnarray*}
&&\overline{\sum }\sum_{k=1}^{m+1}\frac{1}{2}\frac{\nabla _{\theta }\left(
M\left( \left( \theta ,Z\right) ,\left\{ p_{l}^{k},\left( \theta ^{\left(
l\right) },Z_{_{l}}\right) \right\} \right) \right) }{\nabla _{\theta
}M\left( \left( \theta _{k},Z_{k}\right) ,\left\{ p_{l}^{k},\left( \theta
^{\left( l\right) },Z_{_{l}}\right) \right\} \right) \left\vert \Psi \left(
\theta _{k},Z_{k}\right) \right\vert ^{2}}\left( \sum_{k^{\prime
}=1,k^{\prime }\neq k}^{m+1}\frac{\delta \Psi \left( \theta _{k^{\prime
}},Z_{k^{\prime }}\right) }{\Psi \left( \theta _{k^{\prime }},Z_{k^{\prime
}}\right) }+\sum_{l^{\prime }=1}^{j}\frac{\delta \Psi \left( \theta
_{l^{\prime }},Z_{l^{\prime }}\right) }{\Psi \left( \theta _{l^{\prime
}},Z_{l^{\prime }}\right) }\right) \\
&=&\overline{\sum }\frac{\delta \Psi \left( \theta _{1},Z_{1}\right) }{\Psi
\left( \theta _{1},Z_{1}\right) }\left\{ \sum_{k^{\prime }=1}^{m+1}\delta
\left( \left( \theta _{k^{\prime }},Z_{k^{\prime }}\right) -\left( \theta
_{1},Z_{1}\right) \right) \sum_{k=1,k^{\prime }\neq k}^{m+1}\frac{1}{2}\frac{%
\nabla _{\theta }\left( M\left( \left( \theta ,Z\right) ,\left\{
p_{l}^{k},\left( \theta ^{\left( l\right) },Z_{_{l}}\right) \right\} \right)
\right) }{\nabla _{\theta }M\left( \left( \theta _{k},Z_{k}\right) ,\left\{
p_{l}^{k},\left( \theta ^{\left( l\right) },Z_{_{l}}\right) \right\} \right)
\left\vert \Psi \left( \theta _{k},Z_{k}\right) \right\vert ^{2}}\right. \\
&&\left. +\sum_{l^{\prime }=1}^{j}\delta \left( \left( \theta _{l^{\prime
}},Z_{l^{\prime }}\right) -\left( \theta _{1},Z_{1}\right) \right)
\sum_{k=1}^{m+1}\frac{1}{2}\frac{\nabla _{\theta }\left( M\left( \left(
\theta ,Z\right) ,\left\{ p_{l}^{k},\left( \theta ^{\left( l\right)
},Z_{_{l}}\right) \right\} \right) \right) }{\nabla _{\theta }M\left( \left(
\theta _{k},Z_{k}\right) ,\left\{ p_{l}^{k},\left( \theta ^{\left( l\right)
},Z_{_{l}}\right) \right\} \right) \left\vert \Psi \left( \theta
_{k},Z_{k}\right) \right\vert ^{2}}\right\} \\
&=&\overline{\sum }\frac{\delta \Psi \left( \theta _{1},Z_{1}\right) }{\Psi
\left( \theta _{1},Z_{1}\right) }\left\{ \left( \sum_{k^{\prime
}=1}^{m+1}\delta \left( \left( \theta _{k^{\prime }},Z_{k^{\prime }}\right)
-\left( \theta _{1},Z_{1}\right) \right) +\sum_{l^{\prime }=1}^{j}\delta
\left( \left( \theta _{l^{\prime }},Z_{l^{\prime }}\right) -\left( \theta
_{1},Z_{1}\right) \right) \right) \right. \\
&&\times \sum_{k=1}^{m+1}\frac{1}{2}\frac{\nabla _{\theta }\left( M\left(
\left( \theta ,Z\right) ,\left\{ p_{l}^{k},\left( \theta ^{\left( l\right)
},Z_{_{l}}\right) \right\} \right) \right) }{\nabla _{\theta }M\left( \left(
\theta _{k},Z_{k}\right) ,\left\{ p_{l}^{k},\left( \theta ^{\left( l\right)
},Z_{_{l}}\right) \right\} \right) \left\vert \Psi \left( \theta
_{k},Z_{k}\right) \right\vert ^{2}} \\
&&\left. -\sum_{k=1}^{m+1}\delta \left( \left( \theta _{k},Z_{k}\right)
-\left( \theta _{1},Z_{1}\right) \right) \frac{1}{2}\frac{\nabla _{\theta
}\left( M\left( \left( \theta ,Z\right) ,\left\{ p_{l}^{k},\left( \theta
^{\left( l\right) },Z_{_{l}}\right) \right\} \right) \right) }{\nabla
_{\theta }M\left( \left( \theta _{k},Z_{k}\right) ,\left\{ p_{l}^{k},\left(
\theta ^{\left( l\right) },Z_{_{l}}\right) \right\} \right) \left\vert \Psi
\left( \theta _{k},Z_{k}\right) \right\vert ^{2}}\right\}
\end{eqnarray*}

which yields the following matrices:%
\begin{eqnarray*}
&&M_{\Psi }^{\left( 1,r\right) }\left( \left( \theta ,Z\right) ,\left(
\theta _{1},Z_{1}\right) \right) \\
&=&\overline{\sum }H\left( c\left( \theta -\theta _{1}\right) -\left\vert
Z-Z_{1}\right\vert \right) \left\{ \int d\overline{\left[ \theta ,Z\right] }%
\frac{1}{2}\sum_{k=1}^{m+1}\frac{1}{2}\frac{\nabla _{\theta }\left( M\left(
\left( \theta ,Z\right) ,\left\{ p_{l}^{k},\left( \theta ^{\left( l\right)
},Z_{_{l}}\right) \right\} \right) \right) }{\nabla _{\theta }M\left( \left(
\theta _{k},Z_{k}\right) ,\left\{ p_{l}^{k},\left( \theta ^{\left( l\right)
},Z_{_{l}}\right) \right\} \right) \left\vert \Psi \left( \theta
_{k},Z_{k}\right) \right\vert ^{2}}\right. \\
&&\left. -\int d\widetilde{\left[ \theta ,Z\right] }\frac{1}{2}\frac{\nabla
_{\theta }\left( M\left( \left( \theta ,Z\right) ,\left\{ p_{l}^{k},\left(
\theta ^{\left( l\right) },Z_{_{l}}\right) \right\} \right) \right) }{\nabla
_{\theta }M\left( \left( \theta _{k},Z_{k}\right) ,\left\{ p_{l}^{k},\left(
\theta ^{\left( l\right) },Z_{_{l}}\right) \right\} \right) \left\vert \Psi
\left( \theta _{k},Z_{k}\right) \right\vert ^{2}}\right\} \\
&&\times \left( \prod\limits_{i=1}^{m+1}\frac{1}{2}\nabla _{\theta }M\left(
\left( \theta _{i},Z_{i}\right) ,\left\{ p_{l}^{i},\left( \theta ^{\left(
l\right) },Z_{_{l}}\right) \right\} \right) \left\vert \Psi \left( \theta
_{i},Z_{i}\right) \right\vert ^{2}\right) \prod\limits_{l=1}^{j}\left\vert
\Psi \left( \theta ^{\left( l\right) },Z_{l}\right) \right\vert ^{2}
\end{eqnarray*}%
\begin{eqnarray*}
&&M_{\Psi }^{\left( 2,r\right) }\left( \left( \theta ,Z\right) ,\left(
\theta _{1},Z_{1}\right) \right) \\
&=&\overline{\sum }H\left( c\left( \theta -\theta _{1}\right) -\left\vert
Z-Z_{1}\right\vert \right) \left\{ \int d\overline{\left[ \theta ,Z\right] }%
\frac{1}{2}\sum_{k=1}^{m+1}\frac{1}{2}\frac{\left( M\left( \left( \theta
,Z\right) ,\left\{ p_{l}^{k},\left( \theta ^{\left( l\right)
},Z_{_{l}}\right) \right\} \right) \right) }{\nabla _{\theta }M\left( \left(
\theta _{k},Z_{k}\right) ,\left\{ p_{l}^{k},\left( \theta ^{\left( l\right)
},Z_{_{l}}\right) \right\} \right) \left\vert \Psi \left( \theta
_{k},Z_{k}\right) \right\vert ^{2}}\right. \\
&&\left. -\int d\widetilde{\left[ \theta ,Z\right] }\frac{1}{2}\frac{\left(
M\left( \left( \theta ,Z\right) ,\left\{ p_{l}^{k},\left( \theta ^{\left(
l\right) },Z_{_{l}}\right) \right\} \right) \right) }{\nabla _{\theta
}M\left( \left( \theta _{k},Z_{k}\right) ,\left\{ p_{l}^{k},\left( \theta
^{\left( l\right) },Z_{_{l}}\right) \right\} \right) \left\vert \Psi \left(
\theta _{k},Z_{k}\right) \right\vert ^{2}}\right\} \\
&&\times \left( \prod\limits_{i=1}^{m+1}\frac{1}{2}\nabla _{\theta }M\left(
\left( \theta _{i},Z_{i}\right) ,\left\{ p_{l}^{i},\left( \theta ^{\left(
l\right) },Z_{_{l}}\right) \right\} \right) \left\vert \Psi \left( \theta
_{i},Z_{i}\right) \right\vert ^{2}\right) \prod\limits_{l=1}^{j}\left\vert
\Psi \left( \theta ^{\left( l\right) },Z_{l}\right) \right\vert ^{2}
\end{eqnarray*}%
with:%
\begin{equation*}
d\overline{\left[ \theta ,Z\right] }=\left( \sum_{k^{\prime }=1}^{m+1}\delta
\left( \left( \theta _{k^{\prime }},Z_{k^{\prime }}\right) -\left( \theta
_{1},Z_{1}\right) \right) +\sum_{l^{\prime }=1}^{j}\delta \left( \left(
\theta _{l^{\prime }},Z_{l^{\prime }}\right) -\left( \theta
_{1},Z_{1}\right) \right) \right) \prod\limits_{i=1}^{m+1}d\left( \theta
_{i},Z_{i}\right) \prod\limits_{l=1}^{j}d\left( \theta ^{\left( l\right)
},Z_{l}\right)
\end{equation*}

\bigskip Posterior contributions includes contributions depending on $\delta
\omega ^{-1}$ and $\delta \Psi $:%
\begin{eqnarray*}
&\rightarrow &M_{\omega }^{\left( 1,r\right) }\left( \left( \theta ,Z\right)
,\left( \theta _{1},Z_{1}\right) \right) \left( \frac{\delta \omega
^{-1}\left( \theta _{1},Z_{1}\right) }{\omega ^{-1}\left( \theta
_{1},Z_{1}\right) }+\left\langle \frac{\delta \check{T}\left( \theta
^{\left( l\right) }-\frac{\left\vert Z^{\left( 1\right) }-Z^{\left( l\right)
}\right\vert }{c},Z^{\left( 1\right) },Z^{\left( l\right) }\right) }{\check{T%
}\left( \theta ^{\left( l\right) }-\frac{\left\vert Z^{\left( 1\right)
}-Z^{\left( 2\right) }\right\vert }{c},Z^{\left( 1\right) },Z^{\left(
l\right) }\right) }\right\rangle _{Z_{1}}\right) \\
&&+M_{\omega }^{\left( 2,r\right) }\left( \left( \theta ,Z\right) ,\left(
\theta _{1},Z_{1}\right) \right) \nabla _{\theta }\left( \frac{\delta \omega
^{-1}\left( \theta _{1},Z_{1}\right) }{\omega ^{-1}\left( \theta
_{1},Z_{1}\right) }+\left\langle \frac{\delta \check{T}\left( \theta
^{\left( l\right) }-\frac{\left\vert Z^{\left( 1\right) }-Z^{\left( l\right)
}\right\vert }{c},Z^{\left( 1\right) },Z^{\left( l\right) }\right) }{\check{T%
}\left( \theta ^{\left( l\right) }-\frac{\left\vert Z^{\left( 1\right)
}-Z^{\left( 2\right) }\right\vert }{c},Z^{\left( 1\right) },Z^{\left(
l\right) }\right) }\right\rangle _{Z_{1}}\right) \\
&&+M_{\Psi }^{\left( 1,r\right) }\left( \left( \theta ,Z\right) ,\left(
\theta _{1},Z_{1}\right) \right) \frac{\delta \Psi \left( \theta
_{1},Z_{1}\right) }{\Psi \left( \theta _{1},Z_{1}\right) }+M_{\Psi }^{\left(
2,r\right) }\left( \left( \theta ,Z\right) ,\left( \theta _{1},Z_{1}\right)
\right) \nabla _{\theta }\frac{\delta \Psi \left( \theta _{1},Z_{1}\right) }{%
\Psi \left( \theta _{1},Z_{1}\right) }
\end{eqnarray*}

\begin{eqnarray*}
&&\frac{1}{2}\nabla _{\theta }M\left( \left( \theta _{1},Z_{1}\right)
,\left( \theta ,Z\right) \right) \frac{\omega _{0}^{-1}\left( J,\theta
,Z\right) }{\mathcal{\bar{G}}_{0}\left( 0,Z\right) \left( 1+T_{0}\left(
\theta ,Z\right) \right) }\left\vert \Psi \left( \theta _{1},Z_{1}\right)
\right\vert ^{2}\frac{\delta \Psi \left( \theta _{1},Z_{1}\right) }{\Psi
\left( \theta _{1},Z_{1}\right) } \\
&&+\overline{\sum }\frac{1}{2}\frac{\nabla _{\theta }M\left( \left( \theta
_{k},Z_{k}\right) ,\left\{ p_{l}^{k},\left( \theta ^{\left( l\right)
},Z_{_{l}}\right) \right\} ,\left\{ 1,\left( \theta ,Z\right) \right\}
\right) S_{2}}{\nabla _{\theta }M\left( \left( \theta _{k},Z_{k}\right)
,\left\{ p_{l}^{k},\left( \theta ^{\left( l\right) },Z_{_{l}}\right)
\right\} \right) } \\
&&\times \left( \prod\limits_{i=1}^{m+1}\frac{1}{2}\nabla _{\theta }M\left(
\left( \theta _{i},Z_{i}\right) ,\left\{ p_{l}^{i},\left( \theta ^{\left(
l\right) },Z_{_{l}}\right) \right\} \right) \left\vert \Psi \left( \theta
_{i},Z_{i}\right) \right\vert ^{2}\right) \prod\limits_{l=1}^{j}\left\vert
\Psi \left( \theta ^{\left( l\right) },Z_{l}\right) \right\vert ^{2} \\
&=&M_{\omega }^{\left( 1,a\right) }\left( \left( \theta ,Z\right) ,\left(
\theta _{1},Z_{1}\right) \right) \left( \frac{\delta \omega ^{-1}\left(
\theta _{1},Z_{1}\right) }{\omega ^{-1}\left( \theta _{1},Z_{1}\right) }%
+\left\langle \frac{\delta \check{T}\left( \theta ^{\left( l\right) }-\frac{%
\left\vert Z^{\left( 1\right) }-Z^{\left( l\right) }\right\vert }{c}%
,Z^{\left( 1\right) },Z^{\left( l\right) }\right) }{\check{T}\left( \theta
^{\left( l\right) }-\frac{\left\vert Z^{\left( 1\right) }-Z^{\left( 2\right)
}\right\vert }{c},Z^{\left( 1\right) },Z^{\left( l\right) }\right) }%
\right\rangle _{Z_{1}}\right) \\
&&+M_{\omega }^{\left( 2,a\right) }\left( \left( \theta ,Z\right) ,\left(
\theta _{1},Z_{1}\right) \right) \nabla _{\theta }\left( \frac{\delta \omega
^{-1}\left( \theta _{1},Z_{1}\right) }{\omega ^{-1}\left( \theta
_{1},Z_{1}\right) }+\left\langle \frac{\delta \check{T}\left( \theta
^{\left( l\right) }-\frac{\left\vert Z^{\left( 1\right) }-Z^{\left( l\right)
}\right\vert }{c},Z^{\left( 1\right) },Z^{\left( l\right) }\right) }{\check{T%
}\left( \theta ^{\left( l\right) }-\frac{\left\vert Z^{\left( 1\right)
}-Z^{\left( 2\right) }\right\vert }{c},Z^{\left( 1\right) },Z^{\left(
l\right) }\right) }\right\rangle _{Z_{1}}\right) \\
&&+M_{\Psi }^{\left( 1,a\right) }\left( \left( \theta ,Z\right) ,\left(
\theta _{1},Z_{1}\right) \right) \frac{\delta \Psi \left( \theta
_{1},Z_{1}\right) }{\Psi \left( \theta _{1},Z_{1}\right) }+M_{\Psi }^{\left(
2,a\right) }\left( \left( \theta ,Z\right) ,\left( \theta _{1},Z_{1}\right)
\right) \nabla _{\theta }\frac{\delta \Psi \left( \theta _{1},Z_{1}\right) }{%
\Psi \left( \theta _{1},Z_{1}\right) } \\
&&+\bar{N}_{\omega }^{\left( 1\right) }\left( \left( \theta ,Z\right)
,\left( \theta _{1},Z_{1}\right) \right) \frac{\delta \omega ^{-1}\left(
\theta ,Z\right) }{\omega ^{-1}\left( \theta ,Z\right) }+\bar{N}_{\omega
}^{\left( 2\right) }\left( \left( \theta ,Z\right) ,\left( \theta
_{1},Z_{1}\right) \right) \nabla _{\theta }\frac{\delta \omega ^{-1}\left(
\theta ,Z\right) }{\omega ^{-1}\left( \theta ,Z\right) }
\end{eqnarray*}%
\begin{eqnarray*}
M_{\omega }^{\left( 1,a\right) }\left( \left( \theta ,Z\right) ,\left(
\theta _{1},Z_{1}\right) \right) &=&H\left( c\left( \theta _{1}-\theta
\right) -\left\vert Z-Z_{1}\right\vert \right) \overline{\sum }\int d%
\widetilde{\left[ \theta ,Z\right] }\frac{1}{2}\frac{\nabla _{\theta }\left(
M\left( \left( \theta _{k},Z_{k}\right) ,\left\{ p_{l}^{k},\left( \theta
^{\left( l\right) },Z_{_{l}}\right) \right\} ,\left\{ 1,\left( \theta
,Z\right) \right\} \right) \right) }{\nabla _{\theta }M\left( \left( \theta
_{k},Z_{k}\right) ,\left\{ p_{l}^{k},\left( \theta ^{\left( l\right)
},Z_{_{l}}\right) \right\} \right) } \\
&&\times \left( \prod\limits_{i=1}^{m+1}\frac{1}{2}\nabla _{\theta }M\left(
\left( \theta _{i},Z_{i}\right) ,\left\{ p_{l}^{i},\left( \theta ^{\left(
l\right) },Z_{_{l}}\right) \right\} \right) \left\vert \Psi \left( \theta
_{i},Z_{i}\right) \right\vert ^{2}\right) \prod\limits_{l=1}^{j}\left\vert
\Psi \left( \theta ^{\left( l\right) },Z_{l}\right) \right\vert ^{2}
\end{eqnarray*}%
\begin{eqnarray*}
M_{\omega }^{\left( 2,a\right) }\left( \left( \theta ,Z\right) ,\left(
\theta _{1},Z_{1}\right) \right) &=&H\left( c\left( \theta _{1}-\theta
\right) -\left\vert Z-Z_{1}\right\vert \right) \overline{\sum }\int d%
\widetilde{\left[ \theta ,Z\right] }\frac{1}{2}\frac{\left( M\left( \left(
\theta _{k},Z_{k}\right) ,\left\{ p_{l}^{k},\left( \theta ^{\left( l\right)
},Z_{_{l}}\right) \right\} ,\left\{ 1,\left( \theta ,Z\right) \right\}
\right) \right) }{\nabla _{\theta }M\left( \left( \theta _{k},Z_{k}\right)
,\left\{ p_{l}^{k},\left( \theta ^{\left( l\right) },Z_{_{l}}\right)
\right\} \right) } \\
&&\times \left( \prod\limits_{i=1}^{m+1}\frac{1}{2}\nabla _{\theta }M\left(
\left( \theta _{i},Z_{i}\right) ,\left\{ p_{l}^{i},\left( \theta ^{\left(
l\right) },Z_{_{l}}\right) \right\} \right) \left\vert \Psi \left( \theta
_{i},Z_{i}\right) \right\vert ^{2}\right) \prod\limits_{l=1}^{j}\left\vert
\Psi \left( \theta ^{\left( l\right) },Z_{l}\right) \right\vert ^{2}
\end{eqnarray*}%
\begin{eqnarray*}
N_{\omega }^{\left( 1\right) }\left( \left( \theta ,Z\right) ,\left( \theta
_{1},Z_{1}\right) \right) &=&H\left( c\left( \theta _{1}-\theta \right)
-\left\vert Z-Z_{1}\right\vert \right) \overline{\sum }\int d\overline{\left[
\theta ,Z\right] }\frac{1}{2}\frac{\nabla _{\theta }\left( M\left( \left(
\theta _{k},Z_{k}\right) ,\left\{ p_{l}^{k},\left( \theta ^{\left( l\right)
},Z_{_{l}}\right) \right\} ,\left\{ 1,\left( \theta ,Z\right) \right\}
\right) \right) }{\nabla _{\theta }M\left( \left( \theta _{k},Z_{k}\right)
,\left\{ p_{l}^{k},\left( \theta ^{\left( l\right) },Z_{_{l}}\right)
\right\} \right) } \\
&&\times \left( \prod\limits_{i=1}^{m+1}\frac{1}{2}\nabla _{\theta }M\left(
\left( \theta _{i},Z_{i}\right) ,\left\{ p_{l}^{i},\left( \theta ^{\left(
l\right) },Z_{_{l}}\right) \right\} \right) \left\vert \Psi \left( \theta
_{i},Z_{i}\right) \right\vert ^{2}\right) \prod\limits_{l=1}^{j}\left\vert
\Psi \left( \theta ^{\left( l\right) },Z_{l}\right) \right\vert ^{2}
\end{eqnarray*}%
\begin{eqnarray*}
N_{\omega }^{\left( 2\right) }\left( \left( \theta ,Z\right) ,\left( \theta
_{1},Z_{1}\right) \right) &=&H\left( c\left( \theta _{1}-\theta \right)
-\left\vert Z-Z_{1}\right\vert \right) \overline{\sum }\int d\overline{\left[
\theta ,Z\right] }\frac{1}{2}\frac{\left( M\left( \left( \theta
_{k},Z_{k}\right) ,\left\{ p_{l}^{k},\left( \theta ^{\left( l\right)
},Z_{_{l}}\right) \right\} ,\left\{ 1,\left( \theta ,Z\right) \right\}
\right) \right) }{\nabla _{\theta }M\left( \left( \theta _{k},Z_{k}\right)
,\left\{ p_{l}^{k},\left( \theta ^{\left( l\right) },Z_{_{l}}\right)
\right\} \right) } \\
&&\times \left( \prod\limits_{i=1}^{m+1}\frac{1}{2}\nabla _{\theta }M\left(
\left( \theta _{i},Z_{i}\right) ,\left\{ p_{l}^{i},\left( \theta ^{\left(
l\right) },Z_{_{l}}\right) \right\} \right) \left\vert \Psi \left( \theta
_{i},Z_{i}\right) \right\vert ^{2}\right) \prod\limits_{l=1}^{j}\left\vert
\Psi \left( \theta ^{\left( l\right) },Z_{l}\right) \right\vert ^{2}
\end{eqnarray*}%
\begin{eqnarray*}
\bar{N}_{\omega }^{\left( 1\right) }\left( \theta ,Z\right) &=&\int
N_{\omega }^{\left( 1\right) }\left( \left( \theta ,Z\right) ,\left( \theta
_{1},Z_{1}\right) \right) d\left( \theta _{1},Z_{1}\right) +\overline{\sum }%
\int \frac{1}{2}\nabla _{\theta }M\left( \left( \theta _{1},Z_{1}\right)
,\left( \theta ,Z\right) \right) \left\vert \Psi \left( \theta
_{1},Z_{1}\right) \right\vert ^{2}d\left( \theta _{1},Z_{1}\right) \\
\bar{N}_{\omega }^{\left( 2\right) }\left( \theta ,Z\right) &=&\int
N_{\omega }^{\left( 2\right) }\left( \left( \theta ,Z\right) ,\left( \theta
_{1},Z_{1}\right) \right) d\left( \theta _{1},Z_{1}\right) +\overline{\sum }%
\int \frac{1}{2}M\left( \left( \theta _{1},Z_{1}\right) ,\left( \theta
,Z\right) \right) \left\vert \Psi \left( \theta _{1},Z_{1}\right)
\right\vert ^{2}d\left( \theta _{1},Z_{1}\right)
\end{eqnarray*}%
where:%
\begin{equation*}
d\left[ \theta ,Z\right] =\sum_{k=1}^{m+1}\delta \left( \left( \theta
_{1},Z_{1}\right) -\left( \theta ^{\left( k\right) },Z_{k}\right) \right)
\prod\limits_{i=1}^{m+1}d\left( \theta _{i},Z_{i}\right)
\prod\limits_{l=1}^{j}d\left( \theta ^{\left( l\right) },Z_{l}\right)
\end{equation*}

\bigskip 
\begin{eqnarray*}
&&M_{\Psi }^{\left( 1,a\right) }\left( \left( \theta ,Z\right) ,\left(
\theta _{1},Z_{1}\right) \right) \\
&=&H\left( c\left( \theta -\theta _{1}\right) -\left\vert Z-Z_{1}\right\vert
\right) \overline{\sum }\int d\overline{\left[ \theta ,Z\right] }\frac{1}{2}%
\sum_{k=1}^{m+1}\frac{1}{2}\frac{\nabla _{\theta }M\left( \left( \theta
_{k},Z_{k}\right) ,\left\{ p_{l}^{k},\left( \theta ^{\left( l\right)
},Z_{_{l}}\right) \right\} ,\left\{ 1,\left( \theta ,Z\right) \right\}
\right) }{\nabla _{\theta }M\left( \left( \theta _{k},Z_{k}\right) ,\left\{
p_{l}^{k},\left( \theta ^{\left( l\right) },Z_{_{l}}\right) \right\} \right)
\left\vert \Psi \left( \theta _{k},Z_{k}\right) \right\vert ^{2}} \\
&&\times \left( \prod\limits_{i=1}^{m+1}\frac{1}{2}\nabla _{\theta }M\left(
\left( \theta _{i},Z_{i}\right) ,\left\{ p_{l}^{i},\left( \theta ^{\left(
l\right) },Z_{_{l}}\right) \right\} \right) \left\vert \Psi \left( \theta
_{i},Z_{i}\right) \right\vert ^{2}\right) \prod\limits_{l=1}^{j}\left\vert
\Psi \left( \theta ^{\left( l\right) },Z_{l}\right) \right\vert ^{2}
\end{eqnarray*}%
\begin{eqnarray*}
&&M_{\Psi }^{\left( 2,a\right) }\left( \left( \theta ,Z\right) ,\left(
\theta _{1},Z_{1}\right) \right) \\
&=&H\left( c\left( \theta -\theta _{1}\right) -\left\vert Z-Z_{1}\right\vert
\right) \overline{\sum }\int d\overline{\left[ \theta ,Z\right] }\frac{1}{2}%
\sum_{k=1}^{m+1}\frac{1}{2}\frac{M\left( \left( \theta _{k},Z_{k}\right)
,\left\{ p_{l}^{k},\left( \theta ^{\left( l\right) },Z_{_{l}}\right)
\right\} ,\left\{ 1,\left( \theta ,Z\right) \right\} \right) }{\nabla
_{\theta }M\left( \left( \theta _{k},Z_{k}\right) ,\left\{ p_{l}^{k},\left(
\theta ^{\left( l\right) },Z_{_{l}}\right) \right\} \right) \left\vert \Psi
\left( \theta _{k},Z_{k}\right) \right\vert ^{2}} \\
&&\times \left( \prod\limits_{i=1}^{m+1}\frac{1}{2}\nabla _{\theta }M\left(
\left( \theta _{i},Z_{i}\right) ,\left\{ p_{l}^{i},\left( \theta ^{\left(
l\right) },Z_{_{l}}\right) \right\} \right) \left\vert \Psi \left( \theta
_{i},Z_{i}\right) \right\vert ^{2}\right) \prod\limits_{l=1}^{j}\left\vert
\Psi \left( \theta ^{\left( l\right) },Z_{l}\right) \right\vert ^{2}
\end{eqnarray*}%
In first approximation:%
\begin{eqnarray*}
N_{\omega }^{\left( 1\right) }\left( \left( \theta ,Z\right) ,\left( \theta
_{1},Z_{1}\right) \right) &\simeq &\frac{M_{\omega }^{\left( 1,a\right)
}\left( \left( \theta ,Z\right) ,\left( \theta _{1},Z_{1}\right) \right) }{%
\left\langle \sum p_{l}^{k}\right\rangle } \\
N_{\omega }^{\left( 2\right) }\left( \left( \theta ,Z\right) ,\left( \theta
_{1},Z_{1}\right) \right) &\simeq &\frac{M_{\omega }^{\left( 2,a\right)
}\left( \left( \theta ,Z\right) ,\left( \theta _{1},Z_{1}\right) \right) }{%
\left\langle \sum p_{l}^{k}\right\rangle }
\end{eqnarray*}

Adding the contribution:

\begin{eqnarray*}
&&\left( \frac{1}{2}\left( -\frac{\sigma _{\theta }^{2}}{2}\nabla _{\theta
}+\omega ^{-1}\left( J\left( \theta \right) ,\theta ,Z\right) \right) \nabla
_{\theta }+\left( A\left( \Psi \left( \theta ,Z\right) \right) -\frac{\sigma
_{\theta }^{2}}{2}\right) \nabla _{\theta }+U^{\prime \prime }\left(
\left\vert \Psi \left( \theta ,Z\right) \right\vert ^{2}\right) \left\vert
\Psi \left( \theta ,Z\right) \right\vert ^{2}\right) \\
&&\times \frac{\delta \Psi \left( \theta ,Z\right) }{\Psi \left( \theta
,Z\right) }+\frac{1}{2}\nabla _{\theta }\left( \delta \omega ^{-1}\left(
\theta ,Z\right) \right)
\end{eqnarray*}%
and setting:%
\begin{equation*}
\left\langle \frac{\delta \check{T}\left( \theta ^{\left( l\right) }-\frac{%
\left\vert Z^{\left( 1\right) }-Z^{\left( l\right) }\right\vert }{c}%
,Z^{\left( 1\right) },Z^{\left( l\right) }\right) }{\check{T}\left( \theta
^{\left( l\right) }-\frac{\left\vert Z^{\left( 1\right) }-Z^{\left( 2\right)
}\right\vert }{c},Z^{\left( 1\right) },Z^{\left( l\right) }\right) }%
\right\rangle _{Z_{1}}\rightarrow \frac{\delta \check{T}\left( \theta
^{\left( l\right) },Z^{\left( l\right) }\right) }{\check{T}\left( \theta
^{\left( l\right) },Z^{\left( l\right) }\right) }
\end{equation*}%
\begin{equation*}
\left( \frac{1}{2}\left( -\frac{\sigma _{\theta }^{2}}{2}\nabla _{\theta
}+\omega ^{-1}\left( J\left( \theta \right) ,\theta ,Z\right) \right) \nabla
_{\theta }+\left( A\left( \Psi \left( \theta ,Z\right) \right) -\frac{\sigma
_{\theta }^{2}}{2}\right) \nabla _{\theta }+U^{\prime \prime }\left(
\left\vert \Psi \left( \theta ,Z\right) \right\vert ^{2}\right) \left\vert
\Psi \left( \theta ,Z\right) \right\vert ^{2}\right) \rightarrow D_{\theta }
\end{equation*}

\bigskip leads to the fluctuations equation:%
\begin{eqnarray*}
&&0=D_{\theta }\frac{\delta \Psi \left( \theta ,Z\right) }{\Psi \left(
\theta ,Z\right) }+\frac{1}{2}\nabla _{\theta }\left( \delta \omega
^{-1}\left( \theta ,Z\right) \right) \\
&&+M_{\omega }^{\left( 1,r\right) }\left( \left( \theta ,Z\right) ,\left(
\theta _{1},Z_{1}\right) \right) \left( \frac{\delta \omega ^{-1}\left(
\theta _{1},Z_{1}\right) }{\omega ^{-1}\left( \theta _{1},Z_{1}\right) }+%
\frac{\delta \check{T}\left( \theta _{1},Z_{1}\right) }{\check{T}\left(
\theta _{1},Z_{1}\right) }\right) +M_{\omega }^{\left( 2,r\right) }\left(
\left( \theta ,Z\right) ,\left( \theta _{1},Z_{1}\right) \right) \nabla
_{\theta }\left( \frac{\delta \omega ^{-1}\left( \theta _{1},Z_{1}\right) }{%
\omega ^{-1}\left( \theta _{1},Z_{1}\right) }+\frac{\delta \check{T}\left(
\theta _{1},Z_{1}\right) }{\check{T}\left( \theta _{1},Z_{1}\right) }\right)
\\
&&+M_{\Psi }^{\left( 1,r\right) }\left( \left( \theta ,Z\right) ,\left(
\theta _{1},Z_{1}\right) \right) \frac{\delta \Psi \left( \theta
_{1},Z_{1}\right) }{\Psi \left( \theta _{1},Z_{1}\right) }+M_{\Psi }^{\left(
2,r\right) }\left( \left( \theta ,Z\right) ,\left( \theta _{1},Z_{1}\right)
\right) \nabla _{\theta }\frac{\delta \Psi \left( \theta _{1},Z_{1}\right) }{%
\Psi \left( \theta _{1},Z_{1}\right) } \\
&&+M_{\omega }^{\left( 1,a\right) }\left( \left( \theta ,Z\right) ,\left(
\theta _{1},Z_{1}\right) \right) \left( \frac{\delta \omega ^{-1}\left(
\theta _{1},Z_{1}\right) }{\omega ^{-1}\left( \theta _{1},Z_{1}\right) }+%
\frac{\delta \check{T}\left( \theta _{1},Z_{1}\right) }{\check{T}\left(
\theta _{1},Z_{1}\right) }\right) +M_{\omega }^{\left( 2,a\right) }\left(
\left( \theta ,Z\right) ,\left( \theta _{1},Z_{1}\right) \right) \nabla
_{\theta }\left( \frac{\delta \omega ^{-1}\left( \theta _{1},Z_{1}\right) }{%
\omega ^{-1}\left( \theta _{1},Z_{1}\right) }+\frac{\delta \check{T}\left(
\theta _{1},Z_{1}\right) }{\check{T}\left( \theta _{1},Z_{1}\right) }\right)
\\
&&+\bar{N}_{\omega }^{\left( 1\right) }\left( \theta ,Z\right) \frac{\delta
\omega ^{-1}\left( \theta ,Z\right) }{\omega ^{-1}\left( \theta ,Z\right) }+%
\bar{N}_{\omega }^{\left( 2\right) }\left( \theta ,Z\right) \nabla _{\theta }%
\frac{\delta \omega ^{-1}\left( \theta ,Z\right) }{\omega ^{-1}\left( \theta
,Z\right) } \\
&&+M_{\Psi }^{\left( 1,a\right) }\left( \left( \theta ,Z\right) ,\left(
\theta _{1},Z_{1}\right) \right) \frac{\delta \Psi \left( \theta
_{1},Z_{1}\right) }{\Psi \left( \theta _{1},Z_{1}\right) }+M_{\Psi }^{\left(
2,a\right) }\left( \left( \theta ,Z\right) ,\left( \theta _{1},Z_{1}\right)
\right) \nabla _{\theta }\frac{\delta \Psi \left( \theta _{1},Z_{1}\right) }{%
\Psi \left( \theta _{1},Z_{1}\right) }
\end{eqnarray*}%
The system is completed by considering the fluctuation equation for $\delta
\omega ^{-1}$:%
\begin{eqnarray*}
&&\delta \omega ^{-1}\left( \theta ,Z,\left\vert \Psi \right\vert
^{2}\right) =\int \check{T}\left( Z,\theta ,Z_{1},\theta -\frac{\left\vert
Z-Z_{1}\right\vert }{c}\right) \left( \mathcal{\bar{G}}_{0}\left(
0,Z_{1}\right) +\left\vert \Psi \left( \theta -\frac{\left\vert
Z-Z_{1}\right\vert }{c},Z_{1}\right) \right\vert ^{2}\right) \\
&&\times \delta \omega ^{-1}\left( \theta -\frac{\left\vert
Z-Z_{1}\right\vert }{c},Z_{1},\Psi \right) dZ_{1} \\
&&-\int dZ_{1}\omega ^{-1}\left( \theta -\frac{\left\vert Z-Z_{1}\right\vert 
}{c},Z_{1},\Psi \right) \frac{\check{T}\left( Z,\theta ,Z_{1},\theta -\frac{%
\left\vert Z-Z_{1}\right\vert }{c}\right) }{1+\frac{\alpha _{D}h_{D}}{\alpha
_{C}h_{C}}\frac{T\left( Z,\theta ,Z_{1},\theta -\frac{\left\vert
Z-Z_{1}\right\vert }{c}\right) }{\lambda \tau \exp \left( -\frac{\left\vert
Z-Z^{\prime }\right\vert }{\nu c}\right) }\frac{\omega ^{-1}\left( \theta
,Z,\left\vert \Psi \right\vert ^{2}\right) }{\omega ^{-1}\left( \theta -%
\frac{\left\vert Z-Z_{1}\right\vert }{c},Z_{1},\Psi \right) }}\delta
\left\vert \Psi \left( \theta -\frac{\left\vert Z-Z_{1}\right\vert }{c}%
,Z_{1}\right) \right\vert ^{2}
\end{eqnarray*}%
\begin{eqnarray*}
\delta \omega ^{-1}\left( \theta ,Z,\left\vert \Psi \right\vert ^{2}\right)
&=&\left( 1-\check{T}\left( Z,\theta ,Z_{1},\theta -\frac{\left\vert
Z-Z_{1}\right\vert }{c}\right) \left\vert \Psi \left( \theta -\frac{%
\left\vert Z-Z_{1}\right\vert }{c},Z_{1}\right) \right\vert ^{2}\right) \\
&&\frac{\check{T}\left( Z,\theta ,Z_{1},\theta -\frac{\left\vert
Z-Z_{1}\right\vert }{c}\right) }{1+\frac{\alpha _{D}h_{D}}{\alpha _{C}h_{C}}%
\frac{T\left( Z,\theta ,Z_{1},\theta -\frac{\left\vert Z-Z_{1}\right\vert }{c%
}\right) }{\lambda \tau \exp \left( -\frac{\left\vert Z-Z^{\prime
}\right\vert }{\nu c}\right) }\frac{\omega ^{-1}\left( \theta ,Z,\left\vert
\Psi \right\vert ^{2}\right) }{\omega ^{-1}\left( \theta -\frac{\left\vert
Z-Z_{1}\right\vert }{c},Z_{1},\Psi \right) }}\delta \left\vert \Psi \left(
\theta -\frac{\left\vert Z-Z_{1}\right\vert }{c},Z_{1}\right) \right\vert
^{2}
\end{eqnarray*}%
with $\check{T}\left( Z,\theta ,Z_{1},\theta -\frac{\left\vert
Z-Z_{1}\right\vert }{c}\right) $ evaluated by:%
\begin{eqnarray*}
&&\check{T}\left( Z,\theta ,Z_{1},\theta -\frac{\left\vert
Z-Z_{1}\right\vert }{c}\right) \\
&=&\frac{\left( \omega ^{-1}\left( \theta ,Z,\left\vert \Psi \right\vert
^{2}\right) \right) ^{2}T\left( Z,\theta ,Z_{1},\theta -\frac{\left\vert
Z-Z_{1}\right\vert }{c}\right) \left( 1+\frac{\alpha _{D}h_{D}}{\alpha
_{C}h_{C}}\frac{T\left( Z,\theta ,Z_{1},\theta -\frac{\left\vert
Z-Z_{1}\right\vert }{c}\right) }{\lambda \tau \exp \left( -\frac{\left\vert
Z-Z^{\prime }\right\vert }{\nu c}\right) }\frac{\omega ^{-1}\left( \theta
,Z,\left\vert \Psi \right\vert ^{2}\right) }{\omega ^{-1}\left( \theta -%
\frac{\left\vert Z-Z_{1}\right\vert }{c},Z_{1},\Psi \right) }\right) }{%
\left( \omega ^{-1}\left( \theta -\frac{\left\vert Z-Z_{1}\right\vert }{c}%
,Z_{1},\Psi \right) \right) ^{2}\left( G^{-1}\omega ^{-1}\left( \theta
,Z,\left\vert \Psi \right\vert ^{2}\right) -\omega ^{-1}\left( \theta
,Z,\left\vert \Psi \right\vert ^{2}\right) \left( G^{-1}\right) ^{\prime
}\left( \omega ^{-1}\left( \theta ,Z,\left\vert \Psi \right\vert ^{2}\right)
\right) +\tilde{T}_{2}\right) }
\end{eqnarray*}%
with:%
\begin{eqnarray*}
\tilde{T}_{2} &=&\frac{\kappa }{N}\int dZ_{1}\frac{\alpha _{D}h_{D}}{\alpha
_{C}h_{C}}\frac{T\left( Z,\theta ,Z_{1},\theta -\frac{\left\vert
Z-Z_{1}\right\vert }{c}\right) }{\lambda \tau \exp \left( -\frac{\left\vert
Z-Z^{\prime }\right\vert }{\nu c}\right) }\frac{\left( \omega ^{-1}\left(
\theta ,Z,\left\vert \Psi \right\vert ^{2}\right) \right) ^{2}}{\left(
\omega ^{-1}\left( \theta -\frac{\left\vert Z-Z_{1}\right\vert }{c}%
,Z_{1},\Psi \right) \right) ^{2}} \\
&&\times T\left( Z,\theta ,Z_{1},\theta -\frac{\left\vert Z-Z_{1}\right\vert 
}{c}\right) \left( \mathcal{\bar{G}}_{0}\left( 0,Z_{1}\right) +\left\vert
\Psi \left( \theta -\frac{\left\vert Z-Z_{1}\right\vert }{c},Z_{1}\right)
\right\vert ^{2}\right)
\end{eqnarray*}%
We also need the fluctuations for connections:%
\begin{eqnarray*}
&&\frac{\delta \check{T}\left( \theta ^{\left( l\right) }-\frac{\left\vert
Z^{\left( 1\right) }-Z^{\left( l\right) }\right\vert }{c},Z^{\left( 1\right)
},Z^{\left( l\right) }\right) }{\check{T}\left( \theta ^{\left( l\right) }-%
\frac{\left\vert Z^{\left( 1\right) }-Z^{\left( 2\right) }\right\vert }{c}%
,Z^{\left( 1\right) },Z^{\left( l\right) }\right) } \\
&=&\frac{\partial _{\frac{\delta \omega _{0}\left( J,\theta ^{\left(
1\right) }-\frac{\left\vert Z^{\left( 1\right) }-Z^{\left( 2\right)
}\right\vert }{c},Z^{\left( 2\right) }\right) }{\omega _{0}\left( J,\theta
^{\left( 1\right) }-\frac{\left\vert Z^{\left( 1\right) }-Z^{\left( 2\right)
}\right\vert }{c},Z^{\left( 2\right) }\right) }}\check{T}\left( \theta
^{\left( 1\right) }-\frac{\left\vert Z^{\left( 1\right) }-Z^{\left( 2\right)
}\right\vert }{c},Z^{\left( 1\right) },Z^{\left( 2\right) },\omega
_{0}\right) }{\check{T}\left( \theta ^{\left( 1\right) }-\frac{\left\vert
Z^{\left( 1\right) }-Z^{\left( 2\right) }\right\vert }{c},Z^{\left( 1\right)
},Z^{\left( 2\right) },\omega _{0}\right) } \\
&&\times \left( \frac{\delta \omega _{0}\left( J,\theta ^{\left( 1\right) }-%
\frac{\left\vert Z^{\left( 1\right) }-Z^{\left( 2\right) }\right\vert }{c}%
,Z^{\left( 2\right) }\right) }{\omega _{0}\left( J,\theta ^{\left( 1\right)
}-\frac{\left\vert Z^{\left( 1\right) }-Z^{\left( 2\right) }\right\vert }{c}%
,Z^{\left( 2\right) }\right) }-\frac{\delta \omega _{0}\left( J,\theta
^{\left( 1\right) },Z^{\left( 1\right) }\right) }{\omega _{0}\left( J,\theta
^{\left( 1\right) },Z^{\left( 1\right) }\right) }\right) \\
&&+\frac{\delta \check{T}_{i}\left( \theta ^{\left( 1\right) }-\frac{%
\left\vert Z^{\left( 1\right) }-Z^{\left( 2\right) }\right\vert }{c}%
,Z^{\left( 1\right) },Z^{\left( 2\right) },\omega _{0}\right) }{\check{T}%
\left( \theta ^{\left( 1\right) }-\frac{\left\vert Z^{\left( 1\right)
}-Z^{\left( 2\right) }\right\vert }{c},Z^{\left( 1\right) },Z^{\left(
2\right) },\omega _{0}\right) } \\
&\rightarrow &\frac{\delta \check{T}_{i}\left( \theta ^{\left( 1\right) }-%
\frac{\left\vert Z^{\left( 1\right) }-Z^{\left( 2\right) }\right\vert }{c}%
,Z^{\left( 1\right) },Z^{\left( 2\right) },\omega _{0}\right) }{\check{T}%
\left( \theta ^{\left( 1\right) }-\frac{\left\vert Z^{\left( 1\right)
}-Z^{\left( 2\right) }\right\vert }{c},Z^{\left( 1\right) },Z^{\left(
2\right) },\omega _{0}\right) }
\end{eqnarray*}%
Remark that if $\frac{\delta \check{T}_{i}\left( \theta ^{\left( 1\right) }-%
\frac{\left\vert Z^{\left( 1\right) }-Z^{\left( 2\right) }\right\vert }{c}%
,Z^{\left( 1\right) },Z^{\left( 2\right) },\omega _{0}\right) }{\check{T}%
\left( \theta ^{\left( 1\right) }-\frac{\left\vert Z^{\left( 1\right)
}-Z^{\left( 2\right) }\right\vert }{c},Z^{\left( 1\right) },Z^{\left(
2\right) },\omega _{0}\right) }$ is persistent, the following relation is
satisfied:%
\begin{equation*}
\frac{\delta \check{T}\left( \theta ,Z\right) }{\check{T}\left( \theta
,Z\right) }=\frac{\delta \check{T}\left( \theta _{1},Z_{1}\right) }{\check{T}%
\left( \theta _{1},Z_{1}\right) }
\end{equation*}%
The three equations can then be written in matricial form:

\begin{equation}
0=\left\{ A+B+C\right\} \left( 
\begin{array}{c}
\frac{\delta \omega ^{-1}\left( \theta _{1},Z_{1}\right) }{\omega
^{-1}\left( \theta _{1},Z_{1}\right) } \\ 
\frac{\delta \Psi \left( \theta _{1},Z_{1}\right) }{\Psi \left( \theta
_{1},Z_{1}\right) } \\ 
\frac{\delta \check{T}\left( \theta ^{\left( l\right) },Z^{\left( l\right)
}\right) }{\check{T}}%
\end{array}%
\right)  \label{3D}
\end{equation}%
where the matrices are defined by:%
\begin{equation*}
A=\left( 
\begin{array}{ccc}
1 & 0 & 0 \\ 
\bar{N}_{\omega }^{\left( 1\right) }\left( \theta ,Z\right) +\bar{N}_{\omega
}^{\left( 2\right) }\left( \theta ,Z\right) \nabla _{\theta }+\frac{1}{2}%
\nabla _{\theta }\omega ^{-1}\left( \theta ,Z\right) & D_{\theta } & 0 \\ 
0 & 0 & 1%
\end{array}%
\right)
\end{equation*}%
\medskip 
\begin{equation*}
B=\left( 
\begin{array}{ccc}
-\frac{\omega ^{-1}\left( \theta -\frac{\left\vert Z-Z_{1}\right\vert }{c}%
,Z_{1},\Psi \right) }{\omega ^{-1}\left( \theta ,Z\right) }\check{T} & -%
\frac{\frac{\omega ^{-1}\left( \theta -\frac{\left\vert Z-Z_{1}\right\vert }{%
c},Z_{1},\Psi \right) }{\omega ^{-1}\left( \theta ,Z\right) }\check{T}}{1+%
\frac{\alpha _{D}h_{D}}{\alpha _{C}h_{C}}\frac{\omega ^{-1}\left( \theta
,Z,\left\vert \Psi \right\vert ^{2}\right) }{\omega ^{-1}\left( \theta -%
\frac{\left\vert Z-Z_{1}\right\vert }{c},Z_{1},\Psi \right) }\frac{\check{T}%
}{\lambda \tau \exp \left( -\frac{\left\vert Z-Z^{\prime }\right\vert }{\nu c%
}\right) }} & 0 \\ 
\begin{array}{c}
M_{\omega }^{\left( 1,r\right) }\left( \left( \theta ,Z\right) ,\left(
\theta _{1},Z_{1}\right) \right) \\ 
+M_{\omega }^{\left( 2,r\right) }\left( \left( \theta ,Z\right) ,\left(
\theta _{1},Z_{1}\right) \right) \nabla _{\theta }%
\end{array}
& 
\begin{array}{c}
M_{\Psi }^{\left( 1,r\right) }\left( \left( \theta ,Z\right) ,\left( \theta
_{1},Z_{1}\right) \right) \\ 
+M_{\Psi }^{\left( 2,r\right) }\left( \left( \theta ,Z\right) ,\left( \theta
_{1},Z_{1}\right) \right) \nabla _{\theta }%
\end{array}
& 
\begin{array}{c}
M_{\omega }^{\left( 1,r\right) }\left( \left( \theta ,Z\right) ,\left(
\theta _{1},Z_{1}\right) \right) \\ 
+M_{\omega }^{\left( 2,r\right) }\left( \left( \theta ,Z\right) ,\left(
\theta _{1},Z_{1}\right) \right) \nabla _{\theta }%
\end{array}
\\ 
0 & 0 & 0%
\end{array}%
\right)
\end{equation*}%
\medskip and:%
\begin{equation*}
C=\left( 
\begin{array}{ccc}
-\frac{\omega ^{-1}\left( \theta -\frac{\left\vert Z-Z_{1}\right\vert }{c}%
,Z_{1},\Psi \right) }{\omega ^{-1}\left( \theta ,Z\right) }\check{T} & -%
\frac{\frac{\omega ^{-1}\left( \theta -\frac{\left\vert Z-Z_{1}\right\vert }{%
c},Z_{1},\Psi \right) }{\omega ^{-1}\left( \theta ,Z\right) }\check{T}}{1+%
\frac{\alpha _{D}h_{D}}{\alpha _{C}h_{C}}\frac{\omega ^{-1}\left( \theta
,Z,\left\vert \Psi \right\vert ^{2}\right) }{\omega ^{-1}\left( \theta -%
\frac{\left\vert Z-Z_{1}\right\vert }{c},Z_{1},\Psi \right) }\frac{\check{T}%
}{\lambda \tau \exp \left( -\frac{\left\vert Z-Z^{\prime }\right\vert }{\nu c%
}\right) }} & 0 \\ 
\begin{array}{c}
M_{\omega }^{\left( 1,a\right) }\left( \left( \theta ,Z\right) ,\left(
\theta _{1},Z_{1}\right) \right) \\ 
+M_{\omega }^{\left( 2,a\right) }\left( \left( \theta ,Z\right) ,\left(
\theta _{1},Z_{1}\right) \right) \nabla _{\theta }%
\end{array}
& 
\begin{array}{c}
M_{\Psi }^{\left( 1,a\right) }\left( \left( \theta ,Z\right) ,\left( \theta
_{1},Z_{1}\right) \right) \\ 
+M_{\Psi }^{\left( 2,a\right) }\left( \left( \theta ,Z\right) ,\left( \theta
_{1},Z_{1}\right) \right) \nabla _{\theta }%
\end{array}
& 
\begin{array}{c}
M_{\omega }^{\left( 1,a\right) }\left( \left( \theta ,Z\right) ,\left(
\theta _{1},Z_{1}\right) \right) \\ 
+M_{\omega }^{\left( 2,a\right) }\left( \left( \theta ,Z\right) ,\left(
\theta _{1},Z_{1}\right) \right) \nabla _{\theta }%
\end{array}
\\ 
0 & 0 & 0%
\end{array}%
\right)
\end{equation*}%
If disregard $\frac{\delta \check{T}\left( \theta ^{\left( l\right)
},Z^{\left( l\right) }\right) }{\check{T}}$, by considering that the
conections vary slowly compared to the (invrse) activities $\omega
^{-1}\left( \theta ,Z\right) $ and the background field $\Psi \left( \theta
,Z\right) $, the system is define by $2$ equations:%
\begin{eqnarray}
0 &=&\left\{ \left( 
\begin{array}{cc}
1-\frac{\omega ^{-1}\left( \theta -\frac{\left\vert Z-Z_{1}\right\vert }{c}%
,Z_{1},\Psi \right) }{\omega ^{-1}\left( \theta ,Z\right) }\check{T} & -%
\frac{\frac{\omega ^{-1}\left( \theta -\frac{\left\vert Z-Z_{1}\right\vert }{%
c},Z_{1},\Psi \right) }{\omega ^{-1}\left( \theta ,Z\right) }\check{T}}{1+%
\frac{\alpha _{D}h_{D}}{\alpha _{C}h_{C}}\frac{\omega ^{-1}\left( \theta
,Z,\left\vert \Psi \right\vert ^{2}\right) }{\omega ^{-1}\left( \theta -%
\frac{\left\vert Z-Z_{1}\right\vert }{c},Z_{1},\Psi \right) }\frac{\check{T}%
}{\lambda \tau \exp \left( -\frac{\left\vert Z-Z^{\prime }\right\vert }{\nu c%
}\right) }} \\ 
\begin{array}{c}
\bar{N}_{\omega }^{\left( 1\right) }\left( \theta ,Z\right) +\bar{N}_{\omega
}^{\left( 2\right) }\left( \theta ,Z\right) \nabla _{\theta }+\frac{1}{2}%
\nabla _{\theta }\omega ^{-1}\left( \theta ,Z\right) \\ 
+M_{\omega }^{\left( 1,r\right) }\left( \left( \theta ,Z\right) ,\left(
\theta _{1},Z_{1}\right) \right) +M_{\omega }^{\left( 2,r\right) }\left(
\left( \theta ,Z\right) ,\left( \theta _{1},Z_{1}\right) \right) \nabla
_{\theta }%
\end{array}
& 
\begin{array}{c}
D_{\theta }+M_{\Psi }^{\left( 1,r\right) }\left( \left( \theta ,Z\right)
,\left( \theta _{1},Z_{1}\right) \right) \\ 
+M_{\Psi }^{\left( 2,r\right) }\left( \left( \theta ,Z\right) ,\left( \theta
_{1},Z_{1}\right) \right) \nabla _{\theta }%
\end{array}%
\end{array}%
\right) \right.  \notag \\
&&+\left. \left( 
\begin{array}{cc}
0 & 0 \\ 
\begin{array}{c}
M_{\omega }^{\left( 1,a\right) }\left( \left( \theta ,Z\right) ,\left(
\theta _{1},Z_{1}\right) \right) \\ 
+M_{\omega }^{\left( 2,a\right) }\left( \left( \theta ,Z\right) ,\left(
\theta _{1},Z_{1}\right) \right) \nabla _{\theta }%
\end{array}
& 
\begin{array}{c}
M_{\Psi }^{\left( 1,a\right) }\left( \left( \theta ,Z\right) ,\left( \theta
_{1},Z_{1}\right) \right) \\ 
+M_{\Psi }^{\left( 2,a\right) }\left( \left( \theta ,Z\right) ,\left( \theta
_{1},Z_{1}\right) \right) \nabla _{\theta }%
\end{array}%
\end{array}%
\right) \right\} \left( 
\begin{array}{c}
\frac{\delta \omega ^{-1}\left( \theta _{1},Z_{1}\right) }{\omega
^{-1}\left( \theta _{1},Z_{1}\right) } \\ 
\frac{\delta \Psi \left( \theta _{1},Z_{1}\right) }{\Psi \left( \theta
_{1},Z_{1}\right) }%
\end{array}%
\right)  \label{2D}
\end{eqnarray}%
We will estimate the coefficients in the next paragraph.

\subsubsection*{A5.6.2 \protect\bigskip Estimation of the coefficients}

We showed in \cite{GL} that the coefficients $M\left( \left( \theta
,Z\right) ,\left\{ p_{l}^{k},\left( \theta ^{\left( l\right)
},Z_{_{l}}\right) \right\} \right) $ can be estimated as: 
\begin{eqnarray}
&&M\left( \left( \theta ,Z\right) ,\left\{ p_{l}^{k},\left( \theta ^{\left(
l\right) },Z_{_{l}}\right) \right\} \right) \\
&\simeq &\frac{\exp \left( -c\left( \theta -\theta ^{\left( j\right)
}\right) -\alpha \left( \sum_{l=1}^{j-1}\left( \left( c\left( \theta
^{\left( l\right) }-\theta ^{\left( l+1\right) }\right) \right)
^{2}-\left\vert Z_{l}-Z_{l+1}\right\vert ^{2}\right) \right) \right) }{D^{j}}
\notag \\
&&\times H\left( c\left( \theta -\theta ^{\left( j\right) }\right)
-\sum_{i=1}^{j-1}\left\vert Z_{l}-Z_{l+1}\right\vert \right)
\prod\limits_{l=1}^{j}\left( \frac{\omega _{0}^{-1}\left( J,\theta ^{\left(
l\right) },Z_{l}\right) }{\mathcal{\bar{G}}_{0}\left( 0,Z_{l}\right) \left(
1+T_{0}\left( \theta ^{\left( l\right) },Z_{l},\right) \right) }\right)
^{p_{l}^{k}}  \notag \\
&=&M\left( \left( \theta ,Z\right) ,\left\{ \left( \theta ^{\left( l\right)
},Z_{_{l}}\right) \right\} \right) \prod\limits_{l=1}^{j}\left( \frac{\omega
_{0}^{-1}\left( J,\theta ^{\left( l\right) },Z_{l}\right) }{\mathcal{\bar{G}}%
_{0}\left( 0,Z_{l}\right) \left( 1+T_{0}\left( \theta ^{\left( l\right)
},Z_{l},\right) \right) }\right) ^{p_{l}^{k}}  \notag
\end{eqnarray}%
We use that, in the lacal approximation, we can replace:%
\begin{equation*}
\exp \left( -c\left( \theta ^{\left( j\right) }-\theta \right) -\alpha
\left( \left( \left( c\left( \theta ^{\left( j\right) }-\theta \right)
\right) ^{2}-\left\vert Z_{j}-Z\right\vert ^{2}\right) \right) \right)
\rightarrow \delta \left( \theta ^{\left( j\right) }-\theta \right) \left( 1+%
\frac{1}{c}\partial _{\theta }+\frac{1}{2c^{2}}\partial _{\theta }^{2}+\frac{%
1}{2}\partial _{Z}^{2}\right)
\end{equation*}%
so that we write:%
\begin{eqnarray}
&&M\left( \left( \theta ,Z\right) ,\left\{ p_{l}^{k},\left( \theta ^{\left(
l\right) },Z_{_{l}}\right) \right\} \right) \\
&\simeq &H\left( c\left( \theta -\theta ^{\left( j\right) }\right)
-\sum_{i=1}^{j-1}\left\vert Z_{l}-Z_{l+1}\right\vert \right)  \notag \\
&&\times \frac{\delta \left( \theta ^{\left( j\right) }-\theta \right)
\left( 1+\frac{1}{c}\partial _{\theta }+\frac{1}{2c^{2}}\partial _{\theta
}^{2}+\frac{1}{2}\partial _{Z}^{2}\right) }{D^{j}}\prod\limits_{l=1}^{j}%
\left( \frac{\omega _{0}^{-1}\left( J,\left( \theta ,Z\right) \right) }{%
\mathcal{\bar{G}}_{0}\left( 0,Z\right) \left( 1+T_{0}\left( \theta ,Z\right)
\right) }\right) ^{p_{l}^{k}}  \notag
\end{eqnarray}%
and similarly:%
\begin{eqnarray*}
&&M\left( \left( \theta _{k},Z_{k}\right) ,\left\{ p_{l}^{k},\left( \theta
^{\left( l\right) },Z_{_{l}}\right) \right\} ,\left\{ 1,\left( \theta
,Z\right) \right\} \right) \\
&=&M\left( \left( \theta _{k},Z_{k}\right) ,\left\{ \left( \theta ^{\left(
l\right) },Z_{_{l}}\right) \right\} ,\left\{ \left( \theta ,Z\right)
\right\} \right) \prod\limits_{l=1}^{j}\left( \frac{\omega _{0}^{-1}\left(
J,\theta ^{\left( l\right) },Z_{l}\right) }{\mathcal{\bar{G}}_{0}\left(
0,Z_{l}\right) \left( 1+T_{0}\left( \theta ^{\left( l\right) },Z_{l},\right)
\right) }\right) ^{p_{l}^{k}}\frac{\omega _{0}^{-1}\left( J,\theta ,Z\right) 
}{\mathcal{\bar{G}}_{0}\left( 0,Z\right) \left( 1+T_{0}\left( \theta
,Z\right) \right) }
\end{eqnarray*}

This allows to compute the ratios:%
\begin{eqnarray*}
&&\frac{\nabla _{\theta }\left( M\left( \left( \theta ,Z\right) ,\left\{
p_{l}^{k},\left( \theta ^{\left( l\right) },Z_{_{l}}\right) \right\} \right)
\right) }{\nabla _{\theta }M\left( \left( \theta _{k},Z_{k}\right) ,\left\{
p_{l}^{k},\left( \theta ^{\left( l\right) },Z_{_{l}}\right) \right\} \right)
\left\vert \Psi \left( \theta _{k},Z_{k}\right) \right\vert ^{2}} \\
&\rightarrow &\frac{\exp \left( -c\left( \theta -\theta ^{\left( 1\right)
}\right) -\alpha \left( \left( \left( c\left( \theta ^{\left( 1\right)
}-\theta \right) \right) ^{2}-\left\vert Z_{1}-Z\right\vert ^{2}\right)
\right) \right) }{\exp \left( -c\left( \theta _{k}-\theta ^{\left( 1\right)
}\right) -\alpha \left( \left( \left( c\left( \theta ^{\left( 1\right)
}-\theta _{k}\right) \right) ^{2}-\left\vert Z_{1}-Z_{k}\right\vert
^{2}\right) \right) \right) \left\vert \Psi \left( \theta _{k},Z_{k}\right)
\right\vert ^{2}}
\end{eqnarray*}

\begin{eqnarray*}
&&\frac{M\left( \left( \theta ,Z\right) ,\left\{ p_{l}^{k},\left( \theta
^{\left( l\right) },Z_{_{l}}\right) \right\} \right) }{\nabla _{\theta
}M\left( \left( \theta _{k},Z_{k}\right) ,\left\{ p_{l}^{k},\left( \theta
^{\left( l\right) },Z_{_{l}}\right) \right\} \right) \left\vert \Psi \left(
\theta _{k},Z_{k}\right) \right\vert ^{2}} \\
&\rightarrow &\frac{\exp \left( -c\left( \theta -\theta ^{\left( 1\right)
}\right) -\alpha \left( \left( \left( c\left( \theta ^{\left( 1\right)
}-\theta \right) \right) ^{2}-\left\vert Z_{1}-Z\right\vert ^{2}\right)
\right) \right) }{c\exp \left( -c\left( \theta _{k}-\theta ^{\left( 1\right)
}\right) -\alpha \left( \left( \left( c\left( \theta ^{\left( 1\right)
}-\theta _{k}\right) \right) ^{2}-\left\vert Z_{1}-Z_{k}\right\vert
^{2}\right) \right) \right) \left\vert \Psi \left( \theta _{k},Z_{k}\right)
\right\vert ^{2}}
\end{eqnarray*}

\begin{eqnarray*}
&&\frac{\nabla _{\theta }\left( M\left( \left( \theta _{k},Z_{k}\right)
,\left\{ p_{l}^{k},\left( \theta ^{\left( l\right) },Z_{_{l}}\right)
\right\} ,\left\{ 1,\left( \theta ,Z\right) \right\} \right) \right) }{%
\nabla _{\theta }M\left( \left( \theta _{k},Z_{k}\right) ,\left\{
p_{l}^{k},\left( \theta ^{\left( l\right) },Z_{_{l}}\right) \right\} \right) 
} \\
&\rightarrow &\frac{\exp \left( -c\left( \theta ^{\left( j\right) }-\theta
\right) -\alpha \left( \left( \left( c\left( \theta ^{\left( j\right)
}-\theta \right) \right) ^{2}-\left\vert Z_{j}-Z\right\vert ^{2}\right)
\right) \right) \omega _{0}^{-1}\left( J,\theta ,Z\right) }{\mathcal{\bar{G}}%
_{0}\left( 0,Z_{l}\right) \left( 1+T_{0}\left( \theta ,Z\right) \right) }
\end{eqnarray*}

and:%
\begin{eqnarray*}
&&\frac{\left( M\left( \left( \theta _{k},Z_{k}\right) ,\left\{
p_{l}^{k},\left( \theta ^{\left( l\right) },Z_{_{l}}\right) \right\}
,\left\{ 1,\left( \theta ,Z\right) \right\} \right) \right) }{\nabla
_{\theta }M\left( \left( \theta _{k},Z_{k}\right) ,\left\{ p_{l}^{k},\left(
\theta ^{\left( l\right) },Z_{_{l}}\right) \right\} \right) } \\
&\rightarrow &\frac{\exp \left( -c\left( \theta ^{\left( j\right) }-\theta
\right) -\alpha \left( \left( \left( c\left( \theta ^{\left( j\right)
}-\theta \right) \right) ^{2}-\left\vert Z_{j}-Z\right\vert ^{2}\right)
\right) \right) \omega _{0}^{-1}\left( J,\theta ,Z\right) }{c\mathcal{\bar{G}%
}_{0}\left( 0,Z_{l}\right) \left( 1+T_{0}\left( \theta ,Z\right) \right) }
\end{eqnarray*}

\bigskip Consequently, we can estimate the coefficients arising in the
matricial system:

\begin{eqnarray*}
M_{\omega }^{\left( 1,r\right) }\left( \left( \theta ,Z\right) ,\left(
\theta _{1},Z_{1}\right) \right) &=&\overline{\sum }\sum_{k=1}^{m+1}H\left(
c\left( \theta -\theta _{1}\right) -\left\vert Z-Z_{1}\right\vert \right)
\int d\widetilde{\left[ \theta ,Z\right] }\frac{1}{2}\frac{\nabla _{\theta
}\left( M\left( \left( \theta ,Z\right) ,\left\{ p_{l}^{k},\left( \theta
^{\left( l\right) },Z_{_{l}}\right) \right\} \right) \right) \left( \sum
p_{l}^{k}\right) }{\nabla _{\theta }M\left( \left( \theta _{k},Z_{k}\right)
,\left\{ p_{l}^{k},\left( \theta ^{\left( l\right) },Z_{_{l}}\right)
\right\} \right) \left\vert \Psi \left( \theta _{k},Z_{k}\right) \right\vert
^{2}} \\
&&\times \left( \prod\limits_{i=1}^{m+1}\frac{1}{2}\nabla _{\theta }M\left(
\left( \theta _{i},Z_{i}\right) ,\left\{ p_{l}^{i},\left( \theta ^{\left(
l\right) },Z_{_{l}}\right) \right\} \right) \left\vert \Psi \left( \theta
_{i},Z_{i}\right) \right\vert ^{2}\right) \prod\limits_{l=1}^{j}\left\vert
\Psi \left( \theta ^{\left( l\right) },Z_{l}\right) \right\vert ^{2}
\end{eqnarray*}%
with:%
\begin{equation*}
d\widetilde{\left[ \theta ,Z\right] }=\sum_{k=1}^{m+1}\delta \left( \left(
\theta _{1},Z_{1}\right) -\left( \theta ^{\left( k\right) },Z_{k}\right)
\right) \prod\limits_{i=1}^{m+1}d\left( \theta _{i},Z_{i}\right)
\prod\limits_{l=1}^{j}d\left( \theta ^{\left( l\right) },Z_{l}\right)
\end{equation*}%
so that:%
\begin{eqnarray*}
&&\frac{\nabla _{\theta }M\left( \left( \theta ,Z\right) ,\left\{
p_{l}^{k},\left( \theta ^{\left( l\right) },Z_{_{l}}\right) \right\} \right)
\left( \sum p_{l}^{k}\right) }{\nabla _{\theta }M\left( \left( \theta
_{k},Z_{k}\right) ,\left\{ p_{l}^{k},\left( \theta ^{\left( l\right)
},Z_{_{l}}\right) \right\} \right) \left\vert \Psi \left( \theta
_{k},Z_{k}\right) \right\vert ^{2}} \\
&\rightarrow &\delta \left( \left( \theta _{1},Z_{1}\right) -\left( \theta
^{\left( k\right) },Z_{k}\right) \right) \frac{\exp \left( -c\left( \theta
-\theta ^{\left( 1\right) }\right) -\alpha \left( \left( \left( c\left(
\theta ^{\left( 1\right) }-\theta \right) \right) ^{2}-\left\vert
Z_{1}-Z\right\vert ^{2}\right) \right) \right) }{\exp \left( -c\left( \theta
_{k}-\theta ^{\left( 1\right) }\right) -\alpha \left( \left( \left( c\left(
\theta ^{\left( 1\right) }-\theta _{k}\right) \right) ^{2}-\left\vert
Z_{1}-Z_{k}\right\vert ^{2}\right) \right) \right) \left\vert \Psi \left(
\theta _{k},Z_{k}\right) \right\vert ^{2}} \\
&\rightarrow &\delta \left( \left( \theta _{1},Z_{1}\right) -\left( \theta
^{\left( k\right) },Z_{k}\right) \right) \frac{1}{\left\vert \Psi \left(
\theta _{k},Z_{k}\right) \right\vert ^{2}}
\end{eqnarray*}%
We will write:%
\begin{eqnarray*}
M_{\omega }^{\left( 1,r\right) }\left( \left( \theta ,Z\right) ,\left(
\theta _{1},Z_{1}\right) \right) &\rightarrow &H\left( c\left( \theta
-\theta _{1}\right) -\left\vert Z-Z_{1}\right\vert \right) \overline{\sum }%
\sum_{k=1}^{m+1}\int d\widetilde{\left[ \theta ,Z\right] }\frac{1}{2}\frac{%
\left( \sum p_{l}^{k}\right) }{\left\vert \Psi \left( \theta
_{1},Z_{1}\right) \right\vert ^{2}} \\
&&\times \left( \prod\limits_{i=1}^{m+1}\frac{1}{2}\nabla _{\theta }M\left(
\left( \theta _{i},Z_{i}\right) ,\left\{ p_{l}^{i},\left( \theta ^{\left(
l\right) },Z_{_{l}}\right) \right\} \right) \left\vert \Psi \left( \theta
_{i},Z_{i}\right) \right\vert ^{2}\right) \prod\limits_{l=1}^{j}\left\vert
\Psi \left( \theta ^{\left( l\right) },Z_{l}\right) \right\vert ^{2} \\
&=&H\left( c\left( \theta -\theta _{1}\right) -\left\vert Z-Z_{1}\right\vert
\right) \hat{M}_{\omega }^{\left( 1,r\right) }\left( \left( \theta ,Z\right)
,\left( \theta _{1},Z_{1}\right) \right)
\end{eqnarray*}%
The second coefficient to compute is:%
\begin{eqnarray*}
&&\hat{M}_{\omega }^{\left( 1,r\right) }\left( \left( \theta ,Z\right)
,\left( \theta _{1},Z_{1}\right) \right) \\
&=&\delta \left( \theta ^{\left( j\right) }-\theta \right) \left( 1+\frac{1}{%
c}\partial _{\theta }+\frac{1}{2c^{2}}\partial _{\theta }^{2}+\frac{1}{2}%
\partial _{Z}^{2}\right) \sum_{\substack{ j\geqslant 1  \\ m\geqslant 1}}%
\sum _{\substack{ \left( p_{l}^{i}\right) _{\left( m+1\right) \times j}  \\ %
\sum_{i}p_{l}^{i}\geqslant 2}}\sum_{k=1}^{m+1}a_{j,m}\prod\limits_{l=1}^{j}%
\frac{\left( \frac{\omega _{0}^{-1}\left( J,\left( \theta ,Z\right) \right) 
}{\mathcal{\bar{G}}_{0}\left( 0,Z\right) \left( 1+T_{0}\left( \theta
,Z\right) \right) }\right) ^{p_{l}^{k}}}{D^{p_{l}^{k}}}\frac{\left( \sum
p_{l}^{k}\right) }{\left\vert \Psi \left( \theta _{1},Z_{1}\right)
\right\vert ^{2}} \\
&&\times \int \left( \prod\limits_{i=1,i\neq k}^{m+1}\frac{1}{2}\nabla
_{\theta }M\left( \left( \theta _{i},Z_{i}\right) ,\left\{ p_{l}^{i},\left(
\theta ^{\left( l\right) },Z_{_{l}}\right) \right\} \right) \left\vert \Psi
\left( \theta _{i},Z_{i}\right) \right\vert ^{2}\right)
\prod\limits_{l=1}^{j}\left\vert \Psi \left( \theta ^{\left( l\right)
},Z_{l}\right) \right\vert ^{2} \\
&\rightarrow &\delta \left( \theta ^{\left( j\right) }-\theta \right) \left(
1+\frac{1}{c}\partial _{\theta }+\frac{1}{2c^{2}}\partial _{\theta }^{2}+%
\frac{1}{2}\partial _{Z}^{2}\right) \sum_{p}\frac{p\left( \frac{\omega
_{0}^{-1}\left( J,\left( \theta ,Z\right) \right) }{\mathcal{\bar{G}}%
_{0}\left( 0,Z\right) \left( 1+T_{0}\left( \theta ,Z\right) \right) }\right)
^{p}}{D^{p}} \\
&&\times \sum_{\substack{ j\geqslant 1  \\ m\geqslant 1}}\sum_{\substack{ %
\left( p_{l}^{i}\right) _{\left( m\right) \times j}  \\ \sum_{i}p_{l}^{i}%
\geqslant 1}}\sum_{k=1}^{m}\left( m+1\right) a_{j,m+1}\int \left(
\prod\limits_{i=1}^{m}\frac{1}{2}\nabla _{\theta }M\left( \left( \theta
_{i},Z_{i}\right) ,\left\{ p_{l}^{i},\left( \theta ^{\left( l\right)
},Z_{_{l}}\right) \right\} \right) \left\vert \Psi \left( \theta
_{i},Z_{i}\right) \right\vert ^{2}\right) \prod\limits_{l=1}^{j}\left\vert
\Psi \left( \theta ^{\left( l\right) },Z_{l}\right) \right\vert ^{2}
\end{eqnarray*}%
This rewrites:%
\begin{eqnarray*}
&&\delta \left( \theta ^{\left( j\right) }-\theta \right) \left( 1+\frac{1}{c%
}\partial _{\theta }+\frac{1}{2c^{2}}\partial _{\theta }^{2}+\frac{1}{2}%
\partial _{Z}^{2}\right) \\
&&\times \omega _{0}^{-1}\left( J,\left( \theta ,Z\right) \right) \partial
_{\omega _{0}^{-1}\left( J,\left( \theta ,Z\right) \right) }\left( \frac{1}{%
1-\frac{\omega _{0}^{-1}\left( J,\left( \theta ,Z\right) \right) }{\mathcal{%
\bar{G}}_{0}\left( 0,Z\right) \left( 1+T_{0}\left( \theta ,Z\right) \right) D%
}}\right) \\
&&\times \sum_{\substack{ j\geqslant 1  \\ m\geqslant 1}}\sum_{\substack{ %
\left( p_{l}^{i}\right) _{\left( m\right) \times j}  \\ \sum_{i}p_{l}^{i}%
\geqslant 1}}\sum_{k=1}^{m}\left( m+1\right) a_{j,m+1}\int \left(
\prod\limits_{i=1}^{m}\frac{1}{2}\nabla _{\theta }M\left( \left( \theta
_{i},Z_{i}\right) ,\left\{ p_{l}^{i},\left( \theta ^{\left( l\right)
},Z_{_{l}}\right) \right\} \right) \left\vert \Psi \left( \theta
_{i},Z_{i}\right) \right\vert ^{2}\right) \prod\limits_{l=1}^{j}\left\vert
\Psi \left( \theta ^{\left( l\right) },Z_{l}\right) \right\vert ^{2} \\
&\rightarrow &\frac{\omega _{0}^{-1}\left( J,\left( \theta ,Z\right) \right) 
}{\mathcal{\bar{G}}_{0}\left( 0,Z\right) \left( 1+T_{0}\left( \theta
,Z\right) \right) D}\left( \frac{1}{1-\frac{\omega _{0}^{-1}\left( J,\left(
\theta ,Z\right) \right) }{\mathcal{\bar{G}}_{0}\left( 0,Z\right) \left(
1+T_{0}\left( \theta ,Z\right) \right) D}}\right) ^{2} \\
&&\times \sum_{\substack{ j\geqslant 1  \\ m\geqslant 1}}\sum_{\substack{ %
\left( p_{l}^{i}\right) _{\left( m\right) \times j}  \\ \sum_{i}p_{l}^{i}%
\geqslant 1}}\sum_{k=1}^{m}\left( m+1\right) a_{j,m+1}\int \left(
\prod\limits_{i=1}^{m}\frac{1}{2}\nabla _{\theta }M\left( \left( \theta
_{i},Z_{i}\right) ,\left\{ p_{l}^{i},\left( \theta ^{\left( l\right)
},Z_{_{l}}\right) \right\} \right) \left\vert \Psi \left( \theta
_{i},Z_{i}\right) \right\vert ^{2}\right) \prod\limits_{l=1}^{j}\left\vert
\Psi \left( \theta ^{\left( l\right) },Z_{l}\right) \right\vert ^{2}
\end{eqnarray*}

\bigskip We then compute $M_{\omega }^{\left( 2,r\right) }\left( \left(
\theta ,Z\right) ,\left( \theta _{1},Z_{1}\right) \right) $:

\begin{eqnarray*}
&&M_{\omega }^{\left( 2,r\right) }\left( \left( \theta ,Z\right) ,\left(
\theta _{1},Z_{1}\right) \right) \\
&=&\overline{\sum }\sum_{k=1}^{m+1}H\left( c\left( \theta -\theta
_{1}\right) -\left\vert Z-Z_{1}\right\vert \right) \int d\widetilde{\left[
\theta ,Z\right] }\frac{1}{2}\frac{M\left( \left( \theta ,Z\right) ,\left\{
p_{l}^{k},\left( \theta ^{\left( l\right) },Z_{_{l}}\right) \right\} \right)
\left( \sum p_{l}^{k}\right) }{\nabla _{\theta }M\left( \left( \theta
_{k},Z_{k}\right) ,\left\{ p_{l}^{k},\left( \theta ^{\left( l\right)
},Z_{_{l}}\right) \right\} \right) \left\vert \Psi \left( \theta
_{k},Z_{k}\right) \right\vert ^{2}} \\
&&\times \left( \prod\limits_{i=1}^{m+1}\frac{1}{2}\nabla _{\theta }M\left(
\left( \theta _{i},Z_{i}\right) ,\left\{ p_{l}^{i},\left( \theta ^{\left(
l\right) },Z_{_{l}}\right) \right\} \right) \left\vert \Psi \left( \theta
_{i},Z_{i}\right) \right\vert ^{2}\right) \prod\limits_{l=1}^{j}\left\vert
\Psi \left( \theta ^{\left( l\right) },Z_{l}\right) \right\vert ^{2} \\
&\rightarrow &\frac{1}{c}M_{\omega }^{\left( 1,r\right) }\left( \left(
\theta ,Z\right) ,\left( \theta _{1},Z_{1}\right) \right)
\end{eqnarray*}

\bigskip and $M_{\omega }^{\left( 1,a\right) }\left( \left( \theta ,Z\right)
,\left( \theta _{1},Z_{1}\right) \right) $:%
\begin{eqnarray*}
&&M_{\omega }^{\left( 1,a\right) }\left( \left( \theta ,Z\right) ,\left(
\theta _{1},Z_{1}\right) \right) \\
&=&H\left( c\left( \theta _{1}-\theta \right) -\left\vert Z-Z_{1}\right\vert
\right) \overline{\sum }\sum_{k=1}^{m+1}\int d\widetilde{\left[ \theta ,Z%
\right] }\frac{1}{2}\frac{\nabla _{\theta }M\left( \left( \theta
_{k},Z_{k}\right) ,\left\{ p_{l}^{k},\left( \theta ^{\left( l\right)
},Z_{_{l}}\right) \right\} ,\left\{ 1,\left( \theta ,Z\right) \right\}
\right) }{\nabla _{\theta }M\left( \left( \theta _{k},Z_{k}\right) ,\left\{
p_{l}^{k},\left( \theta ^{\left( l\right) },Z_{_{l}}\right) \right\} \right) 
} \\
&&\times \left( \prod\limits_{i=1}^{m+1}\frac{1}{2}\nabla _{\theta }M\left(
\left( \theta _{i},Z_{i}\right) ,\left\{ p_{l}^{i},\left( \theta ^{\left(
l\right) },Z_{_{l}}\right) \right\} \right) \left\vert \Psi \left( \theta
_{i},Z_{i}\right) \right\vert ^{2}\right) \prod\limits_{l=1}^{j}\left\vert
\Psi \left( \theta ^{\left( l\right) },Z_{l}\right) \right\vert ^{2} \\
&\rightarrow &H\left( c\left( \theta _{1}-\theta \right) -\left\vert
Z-Z_{1}\right\vert \right) \overline{\sum }\sum_{k=1}^{m+1}\int d\widetilde{%
\left[ \theta ,Z\right] }\frac{1}{2}\frac{\omega _{0}^{-1}\left( J,\theta
,Z\right) }{\mathcal{\bar{G}}_{0}\left( 0,Z_{l}\right) \left( 1+T_{0}\left(
\theta ,Z\right) \right) } \\
&&\times \left( \prod\limits_{i=1}^{m+1}\frac{1}{2}\nabla _{\theta }M\left(
\left( \theta _{i},Z_{i}\right) ,\left\{ p_{l}^{i},\left( \theta ^{\left(
l\right) },Z_{_{l}}\right) \right\} \right) \left\vert \Psi \left( \theta
_{i},Z_{i}\right) \right\vert ^{2}\right) \prod\limits_{l=1}^{j}\left\vert
\Psi \left( \theta ^{\left( l\right) },Z_{l}\right) \right\vert ^{2} \\
&\rightarrow &H\left( c\left( \theta _{1}-\theta \right) -\left\vert
Z-Z_{1}\right\vert \right) \frac{\frac{\omega _{0}^{-1}\left( J,\theta
,Z\right) }{\mathcal{\bar{G}}_{0}\left( 0,Z_{l}\right) \left( 1+T_{0}\left(
\theta ,Z\right) \right) }}{\frac{\left( \sum p_{l}^{k}\right) }{\left\vert
\Psi \left( \theta _{1},Z_{1}\right) \right\vert ^{2}}}\hat{M}_{\omega
}^{\left( 1,r\right) }\left( \left( \theta ,Z\right) ,\left( \theta
_{1},Z_{1}\right) \right)
\end{eqnarray*}%
and this factors as:%
\begin{eqnarray*}
&&M_{\omega }^{\left( 1,a\right) }\left( \left( \theta ,Z\right) ,\left(
\theta _{1},Z_{1}\right) \right) \\
&\rightarrow &-\delta \left( \theta ^{\left( j\right) }-\theta \right)
\left( 1+\frac{1}{c}\partial _{\theta }+\frac{1}{2c^{2}}\partial _{\theta
}^{2}+\frac{1}{2}\partial _{Z}^{2}\right) \frac{\omega _{0}^{-1}\left(
J,\theta ,Z\right) }{\mathcal{\bar{G}}_{0}\left( 0,Z_{l}\right) \left(
1+T_{0}\left( \theta ,Z\right) \right) }\sum_{p}\frac{\left( \frac{\omega
_{0}^{-1}\left( J,\left( \theta ,Z\right) \right) }{\mathcal{\bar{G}}%
_{0}\left( 0,Z\right) \left( 1+T_{0}\left( \theta ,Z\right) \right) }\right)
^{p}\left\vert \Psi \left( \theta ,Z\right) \right\vert ^{2}}{D^{p}} \\
&&\times \sum_{\substack{ j\geqslant 1  \\ m\geqslant 1}}\sum_{\substack{ %
\left( p_{l}^{i}\right) _{\left( m\right) \times j}  \\ \sum_{i}p_{l}^{i}%
\geqslant 1}}\sum_{k=1}^{m}\left( m+1\right) a_{j,m+1}\int \left(
\prod\limits_{i=1}^{m}\frac{1}{2}\nabla _{\theta }M\left( \left( \theta
_{i},Z_{i}\right) ,\left\{ p_{l}^{i},\left( \theta ^{\left( l\right)
},Z_{_{l}}\right) \right\} \right) \left\vert \Psi \left( \theta
_{i},Z_{i}\right) \right\vert ^{2}\right) \prod\limits_{l=1}^{j}\left\vert
\Psi \left( \theta ^{\left( l\right) },Z_{l}\right) \right\vert ^{2}
\end{eqnarray*}%
which becomes, after summing the inner series:%
\begin{eqnarray*}
&&M_{\omega }^{\left( 1,a\right) }\left( \left( \theta ,Z\right) ,\left(
\theta _{1},Z_{1}\right) \right) \\
&=&-\delta \left( \theta ^{\left( j\right) }-\theta \right) \left( 1+\frac{1%
}{c}\partial _{\theta }+\frac{1}{2c^{2}}\partial _{\theta }^{2}+\frac{1}{2}%
\partial _{Z}^{2}\right) \frac{\omega _{0}^{-1}\left( J,\theta ,Z\right) }{%
\mathcal{\bar{G}}_{0}\left( 0,Z_{l}\right) \left( 1+T_{0}\left( \theta
,Z\right) \right) }\left( \frac{1}{1-\frac{\omega _{0}^{-1}\left( J,\left(
\theta ,Z\right) \right) }{\mathcal{\bar{G}}_{0}\left( 0,Z\right) \left(
1+T_{0}\left( \theta ,Z\right) \right) D}}\right) \\
&&\times \sum_{\substack{ j\geqslant 1  \\ m\geqslant 1}}\sum_{\substack{ %
\left( p_{l}^{i}\right) _{\left( m\right) \times j}  \\ \sum_{i}p_{l}^{i}%
\geqslant 1}}\sum_{k=1}^{m}\left( m+1\right) a_{j,m+1}\int \left(
\prod\limits_{i=1}^{m}\frac{1}{2}\nabla _{\theta }M\left( \left( \theta
_{i},Z_{i}\right) ,\left\{ p_{l}^{i},\left( \theta ^{\left( l\right)
},Z_{_{l}}\right) \right\} \right) \left\vert \Psi \left( \theta
_{i},Z_{i}\right) \right\vert ^{2}\right) \prod\limits_{l=1}^{j}\left\vert
\Psi \left( \theta ^{\left( l\right) },Z_{l}\right) \right\vert ^{2}
\end{eqnarray*}%
The coefficient $M_{\omega }^{\left( 2,a\right) }\left( \left( \theta
,Z\right) ,\left( \theta _{1},Z_{1}\right) \right) $ is given by:%
\begin{equation*}
M_{\omega }^{\left( 2,a\right) }\left( \left( \theta ,Z\right) ,\left(
\theta _{1},Z_{1}\right) \right) =\frac{1}{c}M_{\omega }^{\left( 1,a\right)
}\left( \left( \theta ,Z\right) ,\left( \theta _{1},Z_{1}\right) \right)
\end{equation*}%
The coefficients $N_{\omega }^{\left( 1\right) }\left( \left( \theta
,Z\right) ,\left( \theta _{1},Z_{1}\right) \right) $ and $N_{\omega
}^{\left( 2\right) }\left( \left( \theta ,Z\right) ,\left( \theta
_{1},Z_{1}\right) \right) $ are defined by: 
\begin{eqnarray*}
&&N_{\omega }^{\left( 1\right) }\left( \left( \theta ,Z\right) ,\left(
\theta _{1},Z_{1}\right) \right) \\
&=&H\left( c\left( \theta _{1}-\theta \right) -\left\vert Z-Z_{1}\right\vert
\right) \overline{\sum }\int d\overline{\left[ \theta ,Z\right] }\frac{1}{2}%
\frac{\nabla _{\theta }\left( M\left( \left( \theta _{k},Z_{k}\right)
,\left\{ p_{l}^{k},\left( \theta ^{\left( l\right) },Z_{_{l}}\right)
\right\} ,\left\{ 1,\left( \theta ,Z\right) \right\} \right) \right) }{%
\nabla _{\theta }M\left( \left( \theta _{k},Z_{k}\right) ,\left\{
p_{l}^{k},\left( \theta ^{\left( l\right) },Z_{_{l}}\right) \right\} \right) 
} \\
&&\times \left( \prod\limits_{i=1}^{m+1}\frac{1}{2}\nabla _{\theta }M\left(
\left( \theta _{i},Z_{i}\right) ,\left\{ p_{l}^{i},\left( \theta ^{\left(
l\right) },Z_{_{l}}\right) \right\} \right) \left\vert \Psi \left( \theta
_{i},Z_{i}\right) \right\vert ^{2}\right) \prod\limits_{l=1}^{j}\left\vert
\Psi \left( \theta ^{\left( l\right) },Z_{l}\right) \right\vert ^{2} \\
&\rightarrow &-H\left( c\left( \theta _{1}-\theta \right) -\left\vert
Z-Z_{1}\right\vert \right) \overline{\sum }\int d\overline{\left[ \theta ,Z%
\right] }\frac{1}{2}\frac{\omega _{0}^{-1}\left( J,\theta ,Z\right) }{%
\mathcal{\bar{G}}_{0}\left( 0,Z_{l}\right) \left( 1+T_{0}\left( \theta
,Z\right) \right) } \\
&&\times \left( \prod\limits_{i=1}^{m+1}\frac{1}{2}\nabla _{\theta }M\left(
\left( \theta _{i},Z_{i}\right) ,\left\{ p_{l}^{i},\left( \theta ^{\left(
l\right) },Z_{_{l}}\right) \right\} \right) \left\vert \Psi \left( \theta
_{i},Z_{i}\right) \right\vert ^{2}\right) \prod\limits_{l=1}^{j}\left\vert
\Psi \left( \theta ^{\left( l\right) },Z_{l}\right) \right\vert ^{2}
\end{eqnarray*}%
and:%
\begin{equation*}
N_{\omega }^{\left( 2\right) }\left( \left( \theta ,Z\right) ,\left( \theta
_{1},Z_{1}\right) \right) =\frac{1}{c}N_{\omega }^{\left( 1\right) }\left(
\left( \theta ,Z\right) ,\left( \theta _{1},Z_{1}\right) \right)
\end{equation*}%
As a consequence:%
\begin{eqnarray*}
&&\bar{N}_{\omega }^{\left( 1\right) }\left( \theta ,Z\right) \\
&=&\int N_{\omega }^{\left( 1\right) }\left( \left( \theta ,Z\right) ,\left(
\theta _{1},Z_{1}\right) \right) d\left( \theta _{1},Z_{1}\right) +\overline{%
\sum }\int \frac{1}{2}\nabla _{\theta }M\left( \left( \theta
_{1},Z_{1}\right) ,\left( \theta ,Z\right) \right) \left\vert \Psi \left(
\theta _{1},Z_{1}\right) \right\vert ^{2}d\left( \theta _{1},Z_{1}\right) \\
&\rightarrow &-\frac{\omega _{0}^{-1}\left( J,\theta ,Z\right) }{\mathcal{%
\bar{G}}_{0}\left( 0,Z_{l}\right) \left( 1+T_{0}\left( \theta ,Z\right)
\right) }\int H\left( c\left( \theta _{1}-\theta \right) -\left\vert
Z-Z_{1}\right\vert \right) \\
&&\times \left( \prod\limits_{i=1}^{m+1}\frac{1}{2}\nabla _{\theta }M\left(
\left( \theta _{i},Z_{i}\right) ,\left\{ p_{l}^{i},\left( \theta ^{\left(
l\right) },Z_{_{l}}\right) \right\} \right) \left\vert \Psi \left( \theta
_{i},Z_{i}\right) \right\vert ^{2}\right) \prod\limits_{l=1}^{j}\left\vert
\Psi \left( \theta ^{\left( l\right) },Z_{l}\right) \right\vert ^{2} \\
&&+\int \frac{1}{2}\nabla _{\theta }M\left( \left( \theta _{1},Z_{1}\right)
,\left( \theta ,Z\right) \right) \left\vert \Psi \left( \theta
_{1},Z_{1}\right) \right\vert ^{2}d\left( \theta _{1},Z_{1}\right)
\end{eqnarray*}%
and\bigskip :%
\begin{equation*}
\bar{N}_{\omega }^{\left( 2\right) }\left( \theta ,Z\right) =\int N_{\omega
}^{\left( 2\right) }\left( \left( \theta ,Z\right) ,\left( \theta
_{1},Z_{1}\right) \right) d\left( \theta _{1},Z_{1}\right) +\overline{\sum }%
\int \frac{1}{2}M\left( \left( \theta _{1},Z_{1}\right) ,\left( \theta
,Z\right) \right) \left\vert \Psi \left( \theta _{1},Z_{1}\right)
\right\vert ^{2}d\left( \theta _{1},Z_{1}\right) =\frac{1}{c}\bar{N}_{\omega
}^{\left( 1\right) }\left( \theta ,Z\right)
\end{equation*}

Ultimately, the coefficients $M_{\Psi }^{\left( 1,r\right) }\left( \left(
\theta ,Z\right) ,\left( \theta _{1},Z_{1}\right) \right) $, $M_{\Psi
}^{\left( 2,r\right) }\left( \left( \theta ,Z\right) ,\left( \theta
_{1},Z_{1}\right) \right) $, $M_{\Psi }^{\left( 1,a\right) }\left( \left(
\theta ,Z\right) ,\left( \theta _{1},Z_{1}\right) \right) $\ and $M_{\Psi
}^{\left( 2,a\right) }\left( \left( \theta ,Z\right) ,\left( \theta
_{1},Z_{1}\right) \right) $ are derived:

\begin{eqnarray*}
&&M_{\Psi }^{\left( 1,r\right) }\left( \left( \theta ,Z\right) ,\left(
\theta _{1},Z_{1}\right) \right) \\
&=&\overline{\sum }H\left( c\left( \theta -\theta _{1}\right) -\left\vert
Z-Z_{1}\right\vert \right) \left\{ \int d\overline{\left[ \theta ,Z\right] }%
\frac{1}{2}\sum_{k=1}^{m+1}\frac{1}{2}\frac{\nabla _{\theta }M\left( \left(
\theta ,Z\right) ,\left\{ p_{l}^{k},\left( \theta ^{\left( l\right)
},Z_{_{l}}\right) \right\} \right) }{\nabla _{\theta }M\left( \left( \theta
_{k},Z_{k}\right) ,\left\{ p_{l}^{k},\left( \theta ^{\left( l\right)
},Z_{_{l}}\right) \right\} \right) \left\vert \Psi \left( \theta
_{k},Z_{k}\right) \right\vert ^{2}}\right. \\
&&\left. -\int d\widetilde{\left[ \theta ,Z\right] }\frac{1}{2}\frac{\nabla
_{\theta }M\left( \left( \theta ,Z\right) ,\left\{ p_{l}^{k},\left( \theta
^{\left( l\right) },Z_{_{l}}\right) \right\} \right) }{\nabla _{\theta
}M\left( \left( \theta _{k},Z_{k}\right) ,\left\{ p_{l}^{k},\left( \theta
^{\left( l\right) },Z_{_{l}}\right) \right\} \right) \left\vert \Psi \left(
\theta _{k},Z_{k}\right) \right\vert ^{2}}\right\} \\
&&\times \left( \prod\limits_{i=1}^{m+1}\frac{1}{2}\nabla _{\theta }M\left(
\left( \theta _{i},Z_{i}\right) ,\left\{ p_{l}^{i},\left( \theta ^{\left(
l\right) },Z_{_{l}}\right) \right\} \right) \left\vert \Psi \left( \theta
_{i},Z_{i}\right) \right\vert ^{2}\right) \prod\limits_{l=1}^{j}\left\vert
\Psi \left( \theta ^{\left( l\right) },Z_{l}\right) \right\vert ^{2} \\
&\rightarrow &\overline{\sum }\left( \int d\overline{\left[ \theta ,Z\right] 
}\frac{1}{2}\sum_{k=1}^{m+1}\frac{1}{2}\frac{1}{\left\vert \Psi \left(
\theta _{k},Z_{k}\right) \right\vert ^{2}}-\sum_{k=1}^{m+1}\int d\widetilde{%
\left[ \theta ,Z\right] }\frac{1}{2}\frac{1}{\left\vert \Psi \left( \theta
_{k},Z_{k}\right) \right\vert ^{2}}\right) \\
&&\times \left( \prod\limits_{i=1}^{m+1}\frac{1}{2}\nabla _{\theta }M\left(
\left( \theta _{i},Z_{i}\right) ,\left\{ p_{l}^{i},\left( \theta ^{\left(
l\right) },Z_{_{l}}\right) \right\} \right) \left\vert \Psi \left( \theta
_{i},Z_{i}\right) \right\vert ^{2}\right) \prod\limits_{l=1}^{j}\left\vert
\Psi \left( \theta ^{\left( l\right) },Z_{l}\right) \right\vert ^{2} \\
&\rightarrow &\delta \left( \theta ^{\left( j\right) }-\theta \right) \left(
1+\frac{1}{c}\partial _{\theta }+\frac{1}{2c^{2}}\partial _{\theta }^{2}+%
\frac{1}{2}\partial _{Z}^{2}\right) \sum_{\substack{ j\geqslant 1  \\ %
m\geqslant 1}}\sum_{\substack{ \left( p_{l}^{i}\right) _{\left( m+1\right)
\times j}  \\ \sum_{i}p_{l}^{i}\geqslant 2}}\sum_{k=1}^{m+1}a_{j,m}m\prod%
\limits_{l=1}^{j}\frac{\left( \frac{\omega _{0}^{-1}\left( J,\left( \theta
,Z\right) \right) }{\mathcal{\bar{G}}_{0}\left( 0,Z\right) \left(
1+T_{0}\left( \theta ,Z\right) \right) }\right) ^{p_{l}^{k}}}{D^{p_{l}^{k}}}%
\frac{1}{\left\vert \Psi \left( \theta _{1},Z_{1}\right) \right\vert ^{2}} \\
&&\times \int \left( \prod\limits_{i=1,i\neq k}^{m+1}\frac{1}{2}\nabla
_{\theta }M\left( \left( \theta _{i},Z_{i}\right) ,\left\{ p_{l}^{i},\left(
\theta ^{\left( l\right) },Z_{_{l}}\right) \right\} \right) \left\vert \Psi
\left( \theta _{i},Z_{i}\right) \right\vert ^{2}\right)
\prod\limits_{l=1}^{j}\left\vert \Psi \left( \theta ^{\left( l\right)
},Z_{l}\right) \right\vert ^{2} \\
&&+H\left( c\left( \theta -\theta _{1}\right) -\left\vert Z-Z_{1}\right\vert
\right) \overline{\sum }\left( m+1\right) \int \sum_{l^{\prime
}=1}^{j}\delta \left( \left( \theta _{l^{\prime }},Z_{l^{\prime }}\right)
-\left( \theta _{1},Z_{1}\right) \right) \frac{1}{2}\frac{1}{\left\vert \Psi
\left( \theta _{1},Z_{1}\right) \right\vert ^{2}} \\
&&\times \left( \prod\limits_{i=1}^{m+1}\frac{1}{2}\nabla _{\theta }M\left(
\left( \theta _{i},Z_{i}\right) ,\left\{ p_{l}^{i},\left( \theta ^{\left(
l\right) },Z_{_{l}}\right) \right\} \right) \left\vert \Psi \left( \theta
_{i},Z_{i}\right) \right\vert ^{2}\right) \prod\limits_{l=1}^{j}\left\vert
\Psi \left( \theta ^{\left( l\right) },Z_{l}\right) \right\vert ^{2}
\end{eqnarray*}%
Using that:%
\begin{equation*}
d\overline{\left[ \theta ,Z\right] }=\left( \sum_{k^{\prime }=1}^{m+1}\delta
\left( \left( \theta _{k^{\prime }},Z_{k^{\prime }}\right) -\left( \theta
_{1},Z_{1}\right) \right) +\sum_{l^{\prime }=1}^{j}\delta \left( \left(
\theta _{l^{\prime }},Z_{l^{\prime }}\right) -\left( \theta
_{1},Z_{1}\right) \right) \right) \prod\limits_{i=1}^{m+1}d\left( \theta
_{i},Z_{i}\right) \prod\limits_{l=1}^{j}d\left( \theta ^{\left( l\right)
},Z_{l}\right)
\end{equation*}%
we find:%
\begin{eqnarray*}
&&M_{\Psi }^{\left( 1,r\right) }\left( \left( \theta ,Z\right) ,\left(
\theta _{1},Z_{1}\right) \right) \\
&\rightarrow &\delta \left( \theta ^{\left( j\right) }-\theta \right) \left(
1+\frac{1}{c}\partial _{\theta }+\frac{1}{2c^{2}}\partial _{\theta }^{2}+%
\frac{1}{2}\partial _{Z}^{2}\right) \sum_{p}\frac{\left( \frac{\omega
_{0}^{-1}\left( J,\left( \theta ,Z\right) \right) }{\mathcal{\bar{G}}%
_{0}\left( 0,Z\right) \left( 1+T_{0}\left( \theta ,Z\right) \right) }\right)
^{p}\left\vert \Psi \left( \theta ,Z\right) \right\vert ^{2}}{D^{p}} \\
&&\times \sum_{\substack{ j\geqslant 1  \\ m\geqslant 1}}\sum_{\substack{ %
\left( p_{l}^{i}\right) _{\left( m\right) \times j}  \\ \sum_{i}p_{l}^{i}%
\geqslant 1}}\sum_{k=1}^{m}\left( m+1\right) a_{j,m+1}\int \left(
\prod\limits_{i=1}^{m}\frac{1}{2}\nabla _{\theta }M\left( \left( \theta
_{i},Z_{i}\right) ,\left\{ p_{l}^{i},\left( \theta ^{\left( l\right)
},Z_{_{l}}\right) \right\} \right) \left\vert \Psi \left( \theta
_{i},Z_{i}\right) \right\vert ^{2}\right) \prod\limits_{l=1}^{j}\left\vert
\Psi \left( \theta ^{\left( l\right) },Z_{l}\right) \right\vert ^{2}
\end{eqnarray*}%
which writes also:%
\begin{eqnarray*}
&&M_{\Psi }^{\left( 1,r\right) }\left( \left( \theta ,Z\right) ,\left(
\theta _{1},Z_{1}\right) \right) \\
&\rightarrow &\delta \left( \theta ^{\left( j\right) }-\theta \right) \left(
1+\frac{1}{c}\partial _{\theta }+\frac{1}{2c^{2}}\partial _{\theta }^{2}+%
\frac{1}{2}\partial _{Z}^{2}\right) \sum_{p}\frac{\left( \frac{\omega
_{0}^{-1}\left( J,\left( \theta ,Z\right) \right) }{\mathcal{\bar{G}}%
_{0}\left( 0,Z\right) \left( 1+T_{0}\left( \theta ,Z\right) \right) }\right)
^{p}}{D^{p}} \\
&&\times \sum_{\substack{ j\geqslant 1  \\ m\geqslant 1}}\sum_{\substack{ %
\left( p_{l}^{i}\right) _{\left( m\right) \times j}  \\ \sum_{i}p_{l}^{i}%
\geqslant 1}}\sum_{k=1}^{m}m\left( m+1\right) a_{j,m+1}\int \left(
\prod\limits_{i=1}^{m}\frac{1}{2}\nabla _{\theta }M\left( \left( \theta
_{i},Z_{i}\right) ,\left\{ p_{l}^{i},\left( \theta ^{\left( l\right)
},Z_{_{l}}\right) \right\} \right) \left\vert \Psi \left( \theta
_{i},Z_{i}\right) \right\vert ^{2}\right) \prod\limits_{l=1}^{j}\left\vert
\Psi \left( \theta ^{\left( l\right) },Z_{l}\right) \right\vert ^{2} \\
&&+H\left( c\left( \theta -\theta _{1}\right) -\left\vert Z-Z_{1}\right\vert
\right) \sum_{\substack{ j\geqslant 1  \\ m\geqslant 1}}\sum_{\substack{ %
\left( p_{l}^{i}\right) _{\left( m\right) \times j}  \\ \sum_{i}p_{l}^{i}%
\geqslant 1}}\sum_{k=1}^{m}\left( m+1\right) ^{2}l^{\prime }a_{j+1,m} \\
&&\times \int \left( \prod\limits_{i=1}^{m+1}\frac{1}{2}\nabla _{\theta
}M\left( \left( \theta _{i},Z_{i}\right) ,\left\{ p_{l}^{i},\left( \theta
^{\left( l\right) },Z_{_{l}}\right) \right\} \right) \left\vert \Psi \left(
\theta _{i},Z_{i}\right) \right\vert ^{2}\right)
\prod\limits_{l=1}^{j}\left\vert \Psi \left( \theta ^{\left( l\right)
},Z_{l}\right) \right\vert ^{2}
\end{eqnarray*}%
The coefficient $M_{\Psi }^{\left( 2,r\right) }\left( \left( \theta
,Z\right) ,\left( \theta _{1},Z_{1}\right) \right) $ is given by: 
\begin{equation*}
M_{\Psi }^{\left( 2,r\right) }\left( \left( \theta ,Z\right) ,\left( \theta
_{1},Z_{1}\right) \right) =\frac{1}{c}M_{\Psi }^{\left( 1,r\right) }\left(
\left( \theta ,Z\right) ,\left( \theta _{1},Z_{1}\right) \right)
\end{equation*}%
The coefficient $M_{\Psi }^{\left( 1,a\right) }\left( \left( \theta
,Z\right) ,\left( \theta _{1},Z_{1}\right) \right) $ is: 
\begin{eqnarray*}
&&M_{\Psi }^{\left( 1,a\right) }\left( \left( \theta ,Z\right) ,\left(
\theta _{1},Z_{1}\right) \right) \\
&=&H\left( c\left( \theta -\theta _{1}\right) -\left\vert Z-Z_{1}\right\vert
\right) \int d\overline{\left[ \theta ,Z\right] }\frac{1}{2}\sum_{k=1}^{m+1}%
\frac{1}{2}\frac{\nabla _{\theta }\left( \left( \theta _{k},Z_{k}\right)
,\left\{ p_{l}^{k},\left( \theta ^{\left( l\right) },Z_{_{l}}\right)
\right\} ,\left\{ 1,\left( \theta ,Z\right) \right\} \right) }{\nabla
_{\theta }M\left( \left( \theta _{k},Z_{k}\right) ,\left\{ p_{l}^{k},\left(
\theta ^{\left( l\right) },Z_{_{l}}\right) \right\} \right) \left\vert \Psi
\left( \theta _{k},Z_{k}\right) \right\vert ^{2}} \\
&&\times \left( \prod\limits_{i=1}^{m+1}\frac{1}{2}\nabla _{\theta }M\left(
\left( \theta _{k},Z_{k}\right) ,\left\{ p_{l}^{k},\left( \theta ^{\left(
l\right) },Z_{_{l}}\right) \right\} ,\left\{ 1,\left( \theta ,Z\right)
\right\} \right) \left\vert \Psi \left( \theta _{i},Z_{i}\right) \right\vert
^{2}\right) \prod\limits_{l=1}^{j}\left\vert \Psi \left( \theta ^{\left(
l\right) },Z_{l}\right) \right\vert ^{2} \\
&\rightarrow &-H\left( c\left( \theta -\theta _{1}\right) -\left\vert
Z-Z_{1}\right\vert \right) \int d\overline{\left[ \theta ,Z\right] }\frac{1}{%
2}\sum_{k=1}^{m+1}\frac{1}{2}\frac{1}{\left\vert \Psi \left( \theta
_{k},Z_{k}\right) \right\vert ^{2}} \\
&&\times \left( \prod\limits_{i=1}^{m+1}\frac{1}{2}\nabla _{\theta }M\left(
\left( \theta _{i},Z_{i}\right) ,\left\{ p_{l}^{i},\left( \theta ^{\left(
l\right) },Z_{_{l}}\right) \right\} \right) \left\vert \Psi \left( \theta
_{i},Z_{i}\right) \right\vert ^{2}\right) \prod\limits_{l=1}^{j}\left\vert
\Psi \left( \theta ^{\left( l\right) },Z_{l}\right) \right\vert ^{2}
\end{eqnarray*}%
and this expands as:%
\begin{eqnarray*}
&&M_{\Psi }^{\left( 1,a\right) }\left( \left( \theta ,Z\right) ,\left(
\theta _{1},Z_{1}\right) \right) \\
&\rightarrow &-\delta \left( \theta ^{\left( j\right) }-\theta \right)
\left( 1+\frac{1}{c}\partial _{\theta }+\frac{1}{2c^{2}}\partial _{\theta
}^{2}+\frac{1}{2}\partial _{Z}^{2}\right) \sum_{p}\frac{\left( \frac{\omega
_{0}^{-1}\left( J,\left( \theta ,Z\right) \right) }{\mathcal{\bar{G}}%
_{0}\left( 0,Z\right) \left( 1+T_{0}\left( \theta ,Z\right) \right) }\right)
^{p}}{D^{p}} \\
&&\times \sum_{\substack{ j\geqslant 1  \\ m\geqslant 1}}\sum_{\substack{ %
\left( p_{l}^{i}\right) _{\left( m\right) \times j}  \\ \sum_{i}p_{l}^{i}%
\geqslant 1}}\sum_{k=1}^{m}\left( m+1\right) ^{2}a_{j,m+1}\int \left(
\prod\limits_{i=1}^{m}\frac{1}{2}\nabla _{\theta }M\left( \left( \theta
_{i},Z_{i}\right) ,\left\{ p_{l}^{i},\left( \theta ^{\left( l\right)
},Z_{_{l}}\right) \right\} \right) \left\vert \Psi \left( \theta
_{i},Z_{i}\right) \right\vert ^{2}\right) \prod\limits_{l=1}^{j}\left\vert
\Psi \left( \theta ^{\left( l\right) },Z_{l}\right) \right\vert ^{2} \\
&&-H\left( c\left( \theta -\theta _{1}\right) -\left\vert Z-Z_{1}\right\vert
\right) \sum_{\substack{ j\geqslant 1  \\ m\geqslant 1}}\sum_{\substack{ %
\left( p_{l}^{i}\right) _{\left( m\right) \times j}  \\ \sum_{i}p_{l}^{i}%
\geqslant 1}}\sum_{k=1}^{m}\left( m+1\right) ^{2}l^{\prime }a_{j+1,m} \\
&&\times \int \left( \prod\limits_{i=1}^{m+1}\frac{1}{2}\nabla _{\theta
}M\left( \left( \theta _{i},Z_{i}\right) ,\left\{ p_{l}^{i},\left( \theta
^{\left( l\right) },Z_{_{l}}\right) \right\} \right) \left\vert \Psi \left(
\theta _{i},Z_{i}\right) \right\vert ^{2}\right)
\prod\limits_{l=1}^{j}\left\vert \Psi \left( \theta ^{\left( l\right)
},Z_{l}\right) \right\vert ^{2}
\end{eqnarray*}%
and:%
\begin{equation*}
M_{\Psi }^{\left( 2,a\right) }\left( \left( \theta ,Z\right) ,\left( \theta
_{1},Z_{1}\right) \right) =\frac{1}{c}M_{\Psi }^{\left( 1,a\right) }\left(
\left( \theta ,Z\right) ,\left( \theta _{1},Z_{1}\right) \right)
\end{equation*}

\subsection*{A5.7 Rewriting the equation in terms of \protect\bigskip $\frac{%
\protect\delta \Psi \left( \protect\theta ,Z\right) }{\Psi \left( \protect%
\theta ,Z\right) }$}

Working with the $2$ dimensional system (\ref{2D}), this writes:%
\begin{eqnarray}
0 &=&\left( 
\begin{array}{cc}
1-\frac{\omega ^{-1}\left( \theta -\frac{\left\vert Z-Z_{1}\right\vert }{c}%
,Z_{1},\Psi \right) }{\omega ^{-1}\left( \theta ,Z\right) }\check{T} & -%
\frac{\frac{\omega ^{-1}\left( \theta -\frac{\left\vert Z-Z_{1}\right\vert }{%
c},Z_{1},\Psi \right) }{\omega ^{-1}\left( \theta ,Z\right) }\check{T}}{1+%
\frac{\alpha _{D}h_{D}}{\alpha _{C}h_{C}}\frac{\omega ^{-1}\left( \theta
,Z,\left\vert \Psi \right\vert ^{2}\right) }{\omega ^{-1}\left( \theta -%
\frac{\left\vert Z-Z_{1}\right\vert }{c},Z_{1},\Psi \right) }\frac{\check{T}%
}{\lambda \tau \exp \left( -\frac{\left\vert Z-Z^{\prime }\right\vert }{\nu c%
}\right) }} \\ 
\begin{array}{c}
\left( \bar{N}_{\omega }^{\left( 1\right) }\left( \theta ,Z\right)
+M_{\omega }^{\left( 1,r\right) }\left( \left( \theta ,Z\right) ,\left(
\theta _{1},Z_{1}\right) \right) \right. \\ 
\left. +M_{\omega }^{\left( 1,a\right) }\left( \left( \theta ,Z\right)
,\left( \theta _{1},Z_{1}\right) \right) \right) \left( 1+\frac{1}{c}\nabla
_{\theta }\right) +\frac{1}{2}\nabla _{\theta }\omega ^{-1}\left( \theta
,Z\right)%
\end{array}
& 
\begin{array}{c}
D_{\theta }+\left( M_{\Psi }^{\left( 1,a\right) }\left( \left( \theta
,Z\right) ,\left( \theta _{1},Z_{1}\right) \right) \right. \\ 
\left. +M_{\Psi }^{\left( 1,a\right) }\left( \left( \theta ,Z\right) ,\left(
\theta _{1},Z_{1}\right) \right) \right) \left( 1+\frac{1}{c}\nabla _{\theta
}\right)%
\end{array}%
\end{array}%
\right) \\
&&\times \left( 
\begin{array}{c}
\frac{\delta \omega ^{-1}\left( \theta _{1},Z_{1}\right) }{\omega
^{-1}\left( \theta _{1},Z_{1}\right) } \\ 
\frac{\delta \Psi \left( \theta _{1},Z_{1}\right) }{\Psi \left( \theta
_{1},Z_{1}\right) }%
\end{array}%
\right)  \notag
\end{eqnarray}

\bigskip In the local approximation, we will replace the operator $\frac{%
\omega ^{-1}\left( \theta -\frac{\left\vert Z-Z_{1}\right\vert }{c}%
,Z_{1},\Psi \right) }{\omega ^{-1}\left( \theta ,Z\right) }\check{T}$ by a
second order expansion of its kernel. Setting:%
\begin{eqnarray*}
&&\tilde{T} \\
&\rightarrow &\frac{\kappa }{N}\int dZ_{1}\frac{\alpha _{D}h_{D}}{\alpha
_{C}h_{C}}\frac{T\left( Z,\theta ,Z_{1},\theta -\frac{\left\vert
Z-Z_{1}\right\vert }{c}\right) }{\lambda \tau \exp \left( -\frac{\left\vert
Z-Z^{\prime }\right\vert }{\nu c}\right) }\frac{\left( \omega ^{-1}\left(
\theta ,Z,\left\vert \Psi \right\vert ^{2}\right) \right) ^{2}T\left(
Z,\theta ,Z_{1},\theta -\frac{\left\vert Z-Z_{1}\right\vert }{c}\right)
\left( \mathcal{\bar{G}}_{0}\left( 0,Z_{1}\right) +\left\vert \Psi \left(
\theta -\frac{\left\vert Z-Z_{1}\right\vert }{c},Z_{1}\right) \right\vert
^{2}\right) }{\left( \omega ^{-1}\left( \theta -\frac{\left\vert
Z-Z_{1}\right\vert }{c},Z_{1},\Psi \right) \right) ^{2}}
\end{eqnarray*}

\begin{eqnarray*}
&&\frac{\omega ^{-1}\left( \theta -\frac{\left\vert Z-Z_{1}\right\vert }{c}%
,Z_{1},\Psi \right) }{\omega ^{-1}\left( \theta ,Z\right) }\check{T} \\
&\rightarrow &\frac{\omega ^{-1}\left( \theta ,Z\right) }{\omega ^{-1}\left(
\theta -\frac{\left\vert Z-Z_{1}\right\vert }{c},Z_{1},\Psi \right) } \\
&&\times \frac{T\left( Z,\theta ,Z_{1},\theta -\frac{\left\vert
Z-Z_{1}\right\vert }{c}\right) }{G^{-1}\left( \omega ^{-1}\left( \theta
,Z,\left\vert \Psi \right\vert ^{2}\right) \right) -\omega ^{-1}\left(
\theta ,Z,\left\vert \Psi \right\vert ^{2}\right) \left( G^{-1}\right)
^{\prime }\left( \omega ^{-1}\left( \theta ,Z,\left\vert \Psi \right\vert
^{2}\right) \right) +\tilde{T}} \\
&\simeq &\frac{\omega ^{-1}\left( \theta ,Z\right) }{\omega ^{-1}\left(
\theta -\frac{\left\vert Z-Z_{1}\right\vert }{c},Z_{1},\Psi \right) }\frac{%
T\left( Z,\theta ,Z_{1},\theta -\frac{\left\vert Z-Z_{1}\right\vert }{c}%
\right) }{G^{-1}\left( \omega ^{-1}\left( \theta ,Z,\left\vert \Psi
\right\vert ^{2}\right) \right) -\omega ^{-1}\left( \theta ,Z,\left\vert
\Psi \right\vert ^{2}\right) \left( G^{-1}\right) ^{\prime }\left( \omega
^{-1}\left( \theta ,Z,\left\vert \Psi \right\vert ^{2}\right) \right) }
\end{eqnarray*}%
we then replace:%
\begin{equation*}
\omega ^{-1}\left( \theta ,Z\right) \simeq \eta +\exp \left( i\Upsilon
\theta \right)
\end{equation*}%
where $\eta $ represents the static part of the (inverse) activity, with $%
\eta >1$.

As a consequence: 
\begin{eqnarray*}
&&\frac{\omega ^{-1}\left( \theta -\frac{\left\vert Z-Z_{1}\right\vert }{c}%
,Z_{1},\Psi \right) }{\omega ^{-1}\left( \theta ,Z\right) }\check{T} \\
&\rightarrow &\frac{\eta +\exp \left( i\Upsilon \theta \right) }{\eta +\exp
\left( i\Upsilon \left( \theta -\frac{\left\vert Z-Z_{1}\right\vert }{c}%
\right) \right) }\exp \left( -\frac{\left\vert Z-Z^{\prime }\right\vert }{%
\nu c}\right) \\
&\simeq &\frac{1}{1+\frac{\exp \left( -i\Upsilon \frac{\left\vert
Z-Z_{1}\right\vert }{c}\right) }{\eta +\exp \left( i\Upsilon \theta \right) }%
}\frac{\exp \left( -\frac{\left\vert Z-Z^{\prime }\right\vert }{\nu c}%
\right) }{G^{-1}\left( \omega ^{-1}\left( \theta ,Z,\left\vert \Psi
\right\vert ^{2}\right) \right) -\omega ^{-1}\left( \theta ,Z,\left\vert
\Psi \right\vert ^{2}\right) \left( G^{-1}\right) ^{\prime }\left( \omega
^{-1}\left( \theta ,Z,\left\vert \Psi \right\vert ^{2}\right) \right) }
\end{eqnarray*}%
Then, approximating:%
\begin{eqnarray*}
&&\frac{\eta +\exp \left( i\Upsilon \theta \right) }{\eta +\exp \left(
i\Upsilon \left( \theta -\frac{\left\vert Z-Z_{1}\right\vert }{c}\right)
\right) }\exp \left( -\frac{\left\vert Z-Z^{\prime }\right\vert }{\nu c}%
\right) \\
&\simeq &\frac{1}{1+\frac{\exp \left( -i\Upsilon \frac{\left\vert
Z-Z_{1}\right\vert }{c}\right) }{\eta +\exp \left( i\Upsilon \theta \right) }%
}\exp \left( -\frac{\left\vert Z-Z^{\prime }\right\vert }{\nu c}\right) \\
&\simeq &\left( 1-\frac{\exp \left( -i\Upsilon \frac{\left\vert
Z-Z_{1}\right\vert }{c}\right) }{\eta ^{2}}\right) \exp \left( -\frac{%
\left\vert Z-Z^{\prime }\right\vert }{\nu c}\right) \\
&\rightarrow &\exp \left( -\frac{\left\vert Z-Z^{\prime }\right\vert }{\nu c}%
\right) -\frac{\exp \left( -i\Upsilon \frac{\left\vert Z-Z_{1}\right\vert }{c%
}\right) }{\eta ^{2}}\exp \left( -\frac{\left\vert Z-Z^{\prime }\right\vert 
}{\nu c}\right)
\end{eqnarray*}%
we can set:%
\begin{eqnarray*}
&&\frac{\omega ^{-1}\left( \theta -\frac{\left\vert Z-Z_{1}\right\vert }{c}%
,Z_{1},\Psi \right) }{\omega ^{-1}\left( \theta ,Z\right) }\check{T} \\
&=&\frac{\exp \left( -\frac{\left\vert Z-Z^{\prime }\right\vert }{\nu c}%
\right) -\frac{\exp \left( -i\Upsilon \frac{\left\vert Z-Z_{1}\right\vert }{c%
}\right) }{\eta ^{2}}\exp \left( -\frac{\left\vert Z-Z^{\prime }\right\vert 
}{\nu c}\right) }{G^{-1}\left( \omega ^{-1}\left( \theta ,Z,\left\vert \Psi
\right\vert ^{2}\right) \right) -\omega ^{-1}\left( \theta ,Z,\left\vert
\Psi \right\vert ^{2}\right) \left( G^{-1}\right) ^{\prime }\left( \omega
^{-1}\left( \theta ,Z,\left\vert \Psi \right\vert ^{2}\right) \right) }
\end{eqnarray*}%
To find the sign of the denominator, we use that:%
\begin{equation*}
G\left( x\right) \simeq \frac{\arctan x}{\sqrt{x}}
\end{equation*}%
Considering the equation:%
\begin{equation*}
y=\frac{\arctan x}{\sqrt{x}}
\end{equation*}%
for:%
\begin{equation*}
T<<1
\end{equation*}%
we have%
\begin{equation*}
x<<1
\end{equation*}%
and:%
\begin{equation*}
y\simeq \sqrt{x}
\end{equation*}%
In this case, we can consider that:%
\begin{equation*}
x=y^{2}
\end{equation*}%
so that:%
\begin{eqnarray*}
&&G^{-1}\left( \omega ^{-1}\left( \theta ,Z,\left\vert \Psi \right\vert
^{2}\right) \right) -\omega ^{-1}\left( \theta ,Z,\left\vert \Psi
\right\vert ^{2}\right) \left( G^{-1}\right) ^{\prime }\left( \omega
^{-1}\left( \theta ,Z,\left\vert \Psi \right\vert ^{2}\right) \right) \\
&\rightarrow &y^{2}-2y^{2}<0
\end{eqnarray*}%
On the contrary, when:%
\begin{equation*}
T>>1
\end{equation*}%
we have:%
\begin{equation*}
x>>1
\end{equation*}%
As a consequence:%
\begin{equation*}
y=\frac{\frac{\pi }{2}}{\sqrt{x}}
\end{equation*}%
and:%
\begin{equation*}
x=\left( \frac{\frac{\pi }{2}}{y}\right) ^{2}
\end{equation*}%
In this case:%
\begin{eqnarray*}
&&G^{-1}\left( \omega ^{-1}\left( \theta ,Z,\left\vert \Psi \right\vert
^{2}\right) \right) -\omega ^{-1}\left( \theta ,Z,\left\vert \Psi
\right\vert ^{2}\right) \left( G^{-1}\right) ^{\prime }\left( \omega
^{-1}\left( \theta ,Z,\left\vert \Psi \right\vert ^{2}\right) \right) \\
&\rightarrow &y+2\left( \frac{\frac{\pi }{2}}{y}\right) ^{2}>0
\end{eqnarray*}%
Since, we consider groups with enhanced connectivities, we will only
consider the case $T>>1$, so that:%
\begin{equation*}
G^{-1}\left( \omega ^{-1}\left( \theta ,Z,\left\vert \Psi \right\vert
^{2}\right) \right) -\omega ^{-1}\left( \theta ,Z,\left\vert \Psi
\right\vert ^{2}\right) \left( G^{-1}\right) ^{\prime }\left( \omega
^{-1}\left( \theta ,Z,\left\vert \Psi \right\vert ^{2}\right) \right) >0
\end{equation*}

Thus, neglecting first order derivatives in the expansion, we replace the
kernel of $\frac{\omega ^{-1}\left( \theta -\frac{\left\vert
Z-Z_{1}\right\vert }{c},Z_{1},\Psi \right) }{\omega ^{-1}\left( \theta
,Z\right) }\check{T}$ \ by:

\begin{equation*}
\frac{\omega ^{-1}\left( \theta -\frac{\left\vert Z-Z_{1}\right\vert }{c}%
,Z_{1},\Psi \right) }{\omega ^{-1}\left( \theta ,Z\right) }\check{T}%
\rightarrow \left( \frac{i\Upsilon }{c\eta ^{2}}-\frac{1}{\nu c}\left( 1-%
\frac{1}{\eta ^{2}}\right) \right) ^{2}\nabla _{Z}^{2}
\end{equation*}%
up to a normalization, kixed to $1$. Acting on fluctuations with wave vector 
$k_{Z}$, this correspond to replace:%
\begin{equation*}
\frac{\omega ^{-1}\left( \theta -\frac{\left\vert Z-Z_{1}\right\vert }{c}%
,Z_{1},\Psi \right) }{\omega ^{-1}\left( \theta ,Z\right) }\check{T}%
\rightarrow -\left( \frac{i\Upsilon }{c\eta ^{2}}-\frac{1}{\nu c}\left( 1-%
\frac{1}{\eta ^{2}}\right) \right) ^{2}k_{Z}^{2}
\end{equation*}%
The dynamicl systm rewrites:

\begin{equation}
0=\left( 
\begin{array}{cc}
1-\left( \frac{i\Upsilon }{c\eta ^{2}}-\frac{1}{\nu c}\left( 1-\frac{1}{\eta
^{2}}\right) \right) ^{2}\nabla _{Z}^{2} & \frac{\left( \frac{i\Upsilon }{%
c\eta ^{2}}-\frac{1}{\nu c}\left( 1-\frac{1}{\eta ^{2}}\right) \right)
^{2}\nabla _{Z}^{2}}{1+\frac{\alpha _{D}h_{D}}{\alpha _{C}h_{C}}\frac{\omega
^{-1}\left( \theta ,Z,\left\vert \Psi \right\vert ^{2}\right) }{\omega
^{-1}\left( \theta -\frac{\left\vert Z-Z_{1}\right\vert }{c},Z_{1},\Psi
\right) }\frac{\check{T}}{\lambda \tau \exp \left( -\frac{\left\vert
Z-Z^{\prime }\right\vert }{\nu c}\right) }} \\ 
\begin{array}{c}
\left( \bar{N}_{\omega }^{\left( 1\right) }\left( \theta ,Z\right)
+M_{\omega }^{\left( 1,r\right) }\left( \left( \theta ,Z\right) ,\left(
\theta _{1},Z_{1}\right) \right) \right. \\ 
\left. +M_{\omega }^{\left( 1,a\right) }\left( \left( \theta ,Z\right)
,\left( \theta _{1},Z_{1}\right) \right) \right) +\frac{1}{2}\nabla _{\theta
}\omega ^{-1}\left( \theta ,Z\right)%
\end{array}
& 
\begin{array}{c}
D_{\theta }+\left( M_{\Psi }^{\left( 1,a\right) }\left( \left( \theta
,Z\right) ,\left( \theta _{1},Z_{1}\right) \right) \right. \\ 
\left. +M_{\Psi }^{\left( 1,a\right) }\left( \left( \theta ,Z\right) ,\left(
\theta _{1},Z_{1}\right) \right) \right)%
\end{array}%
\end{array}%
\right) \left( 
\begin{array}{c}
\frac{\delta \omega ^{-1}\left( \theta _{1},Z_{1}\right) }{\omega
^{-1}\left( \theta _{1},Z_{1}\right) } \\ 
\frac{\delta \Psi \left( \theta _{1},Z_{1}\right) }{\Psi \left( \theta
_{1},Z_{1}\right) }%
\end{array}%
\right)  \label{Dynt}
\end{equation}%
with the coefficients:%
\begin{eqnarray*}
&&M_{\omega }^{\left( 1,r\right) }\left( \left( \theta ,Z\right) ,\left(
\theta _{1},Z_{1}\right) \right) +M_{\omega }^{\left( 1,a\right) }\left(
\left( \theta ,Z\right) ,\left( \theta _{1},Z_{1}\right) \right) \\
&\rightarrow &\delta \left( \theta ^{\left( j\right) }-\theta \right) \left(
1+\frac{1}{c}\partial _{\theta }+\frac{1}{2c^{2}}\partial _{\theta }^{2}+%
\frac{1}{2}\partial _{Z}^{2}\right) \left( \frac{\frac{\omega
_{0}^{-1}\left( J,\theta ,Z\right) }{\mathcal{\bar{G}}_{0}\left(
0,Z_{l}\right) \left( 1+T_{0}\left( \theta ,Z\right) \right) D}}{1-\frac{%
\omega _{0}^{-1}\left( J,\left( \theta ,Z\right) \right) }{\mathcal{\bar{G}}%
_{0}\left( 0,Z\right) \left( 1+T_{0}\left( \theta ,Z\right) \right) D}}%
\right) ^{2} \\
&&\times \sum_{\substack{ j\geqslant 1  \\ m\geqslant 1}}\sum_{\substack{ %
\left( p_{l}^{i}\right) _{\left( m\right) \times j}  \\ \sum_{i}p_{l}^{i}%
\geqslant 1}}\sum_{k=1}^{m}\left( m+1\right) a_{j,m+1}\int \left(
\prod\limits_{i=1}^{m}\frac{1}{2}\nabla _{\theta }M\left( \left( \theta
_{i},Z_{i}\right) ,\left\{ p_{l}^{i},\left( \theta ^{\left( l\right)
},Z_{_{l}}\right) \right\} \right) \left\vert \Psi \left( \theta
_{i},Z_{i}\right) \right\vert ^{2}\right) \prod\limits_{l=1}^{j}\left\vert
\Psi \left( \theta ^{\left( l\right) },Z_{l}\right) \right\vert ^{2}
\end{eqnarray*}%
\begin{eqnarray*}
&&\bar{N}_{\omega }^{\left( 1\right) }\left( \theta ,Z\right) \\
&=&\int N_{\omega }^{\left( 1\right) }\left( \left( \theta ,Z\right) ,\left(
\theta _{1},Z_{1}\right) \right) d\left( \theta _{1},Z_{1}\right) +\overline{%
\sum }\int \frac{1}{2}\nabla _{\theta }M\left( \left( \theta
_{1},Z_{1}\right) ,\left( \theta ,Z\right) \right) \left\vert \Psi \left(
\theta _{1},Z_{1}\right) \right\vert ^{2}d\left( \theta _{1},Z_{1}\right) \\
&\rightarrow &-\frac{\omega _{0}^{-1}\left( J,\theta ,Z\right) }{\mathcal{%
\bar{G}}_{0}\left( 0,Z_{l}\right) \left( 1+T_{0}\left( \theta ,Z\right)
\right) }\int H\left( c\left( \theta _{1}-\theta \right) -\left\vert
Z-Z_{1}\right\vert \right) \\
&&\times \left( \prod\limits_{i=1}^{m+1}\frac{1}{2}\nabla _{\theta }M\left(
\left( \theta _{i},Z_{i}\right) ,\left\{ p_{l}^{i},\left( \theta ^{\left(
l\right) },Z_{_{l}}\right) \right\} \right) \left\vert \Psi \left( \theta
_{i},Z_{i}\right) \right\vert ^{2}\right) \prod\limits_{l=1}^{j}\left\vert
\Psi \left( \theta ^{\left( l\right) },Z_{l}\right) \right\vert ^{2}
\end{eqnarray*}

\bigskip 
\begin{eqnarray*}
&&M_{\omega }^{\left( 1,r\right) }\left( \left( \theta ,Z\right) ,\left(
\theta _{1},Z_{1}\right) \right) +M_{\omega }^{\left( 1,a\right) }\left(
\left( \theta ,Z\right) ,\left( \theta _{1},Z_{1}\right) \right) +\bar{N}%
_{\omega }^{\left( 1\right) }\left( \theta ,Z\right) \\
&\rightarrow &\delta \left( \theta ^{\left( j\right) }-\theta \right) \left(
1+\frac{1}{c}\partial _{\theta }+\frac{1}{2c^{2}}\partial _{\theta }^{2}+%
\frac{1}{2}\partial _{Z}^{2}\right) \frac{\omega _{0}^{-1}\left( J,\theta
,Z\right) }{\mathcal{\bar{G}}_{0}\left( 0,Z_{l}\right) \left( 1+T_{0}\left(
\theta ,Z\right) \right) D}\sum_{\substack{ j\geqslant 1  \\ m\geqslant 1}}%
\sum_{\substack{ \left( p_{l}^{i}\right) _{\left( m\right) \times j}  \\ %
\sum_{i}p_{l}^{i}\geqslant 1}}\sum_{k=1}^{m} \\
&&\times \left[ \left( m+1\right) a_{j,m+1}\frac{\frac{\omega
_{0}^{-1}\left( J,\theta ,Z\right) }{\mathcal{\bar{G}}_{0}\left(
0,Z_{l}\right) \left( 1+T_{0}\left( \theta ,Z\right) \right) D}}{\left( 1-%
\frac{\omega _{0}^{-1}\left( J,\left( \theta ,Z\right) \right) }{\mathcal{%
\bar{G}}_{0}\left( 0,Z\right) \left( 1+T_{0}\left( \theta ,Z\right) \right) D%
}\right) ^{2}}-\left( 1-\delta _{1,m}\right) a_{j,m}\right] \\
&&\times \sum_{\substack{ j\geqslant 1  \\ m\geqslant 1}}\sum_{\substack{ %
\left( p_{l}^{i}\right) _{\left( m\right) \times j}  \\ \sum_{i}p_{l}^{i}%
\geqslant 1}}\sum_{k=1}^{m}\left( m+1\right) a_{j,m+1}\int \left(
\prod\limits_{i=1}^{m}\frac{1}{2}\nabla _{\theta }M\left( \left( \theta
_{i},Z_{i}\right) ,\left\{ p_{l}^{i},\left( \theta ^{\left( l\right)
},Z_{_{l}}\right) \right\} \right) \left\vert \Psi \left( \theta
_{i},Z_{i}\right) \right\vert ^{2}\right) \prod\limits_{l=1}^{j}\left\vert
\Psi \left( \theta ^{\left( l\right) },Z_{l}\right) \right\vert ^{2}
\end{eqnarray*}

and:\bigskip 
\begin{eqnarray*}
&&M_{\Psi }^{\left( 1,a\right) }\left( \left( \theta ,Z\right) ,\left(
\theta _{1},Z_{1}\right) \right) +M_{\Psi }^{\left( 1,a\right) }\left(
\left( \theta ,Z\right) ,\left( \theta _{1},Z_{1}\right) \right) \\
&=&-\delta \left( \theta ^{\left( j\right) }-\theta \right) \left( 1+\frac{1%
}{c}\partial _{\theta }+\frac{1}{2c^{2}}\partial _{\theta }^{2}+\frac{1}{2}%
\partial _{Z}^{2}\right) \frac{1}{1-\frac{\omega _{0}^{-1}\left( J,\left(
\theta ,Z\right) \right) }{\mathcal{\bar{G}}_{0}\left( 0,Z\right) \left(
1+T_{0}\left( \theta ,Z\right) \right) D}} \\
&&\times \sum_{\substack{ j\geqslant 1  \\ m\geqslant 1}}\sum_{\substack{ %
\left( p_{l}^{i}\right) _{\left( m\right) \times j}  \\ \sum_{i}p_{l}^{i}%
\geqslant 1}}\sum_{k=1}^{m}\left( m+1\right) a_{j,m+1}\int \left(
\prod\limits_{i=1}^{m}\frac{1}{2}\nabla _{\theta }M\left( \left( \theta
_{i},Z_{i}\right) ,\left\{ p_{l}^{i},\left( \theta ^{\left( l\right)
},Z_{_{l}}\right) \right\} \right) \left\vert \Psi \left( \theta
_{i},Z_{i}\right) \right\vert ^{2}\right) \prod\limits_{l=1}^{j}\left\vert
\Psi \left( \theta ^{\left( l\right) },Z_{l}\right) \right\vert ^{2}
\end{eqnarray*}

Using that:%
\begin{eqnarray}
&&\nabla _{\theta }M\left( \left( \theta ,Z\right) ,\left\{ p_{l}^{k},\left(
\theta ^{\left( l\right) },Z_{_{l}}\right) \right\} \right) \\
&\simeq &-cH\left( c\left( \theta -\theta ^{\left( j\right) }\right)
-\sum_{i=1}^{j-1}\left\vert Z_{l}-Z_{l+1}\right\vert \right)  \notag \\
&&\times \frac{\delta \left( \theta ^{\left( j\right) }-\theta \right)
\left( 1+\frac{1}{c}\partial _{\theta }+\frac{1}{2c^{2}}\partial _{\theta
}^{2}+\frac{1}{2}\partial _{Z}^{2}\right) }{D^{j}}\prod\limits_{l=1}^{j}%
\left( \frac{\omega _{0}^{-1}\left( J,\left( \theta ,Z\right) \right) }{%
\mathcal{\bar{G}}_{0}\left( 0,Z\right) \left( 1+T_{0}\left( \theta ,Z\right)
\right) }\right) ^{p_{l}^{k}}  \notag \\
&\rightarrow &-c\frac{\frac{\omega _{0}^{-1}\left( J,\left( \theta ,Z\right)
\right) }{\mathcal{\bar{G}}_{0}\left( 0,Z\right) \left( 1+T_{0}\left( \theta
,Z\right) \right) D}}{1-\frac{\omega _{0}^{-1}\left( J,\left( \theta
,Z\right) \right) }{\mathcal{\bar{G}}_{0}\left( 0,Z\right) \left(
1+T_{0}\left( \theta ,Z\right) \right) D}}  \notag
\end{eqnarray}%
we have:%
\begin{eqnarray*}
&&\sum_{\substack{ j\geqslant 1  \\ m\geqslant 1}}\sum_{\substack{ \left(
p_{l}^{i}\right) _{\left( m\right) \times j}  \\ \sum_{i}p_{l}^{i}\geqslant
1 }}\sum_{k=1}^{m}\left( m+1\right) a_{j,m+1}\int \left(
\prod\limits_{i=1}^{m}\frac{1}{2}\nabla _{\theta }M\left( \left( \theta
_{i},Z_{i}\right) ,\left\{ p_{l}^{i},\left( \theta ^{\left( l\right)
},Z_{_{l}}\right) \right\} \right) \left\vert \Psi \left( \theta
_{i},Z_{i}\right) \right\vert ^{2}\right) \prod\limits_{l=1}^{j}\left\vert
\Psi \left( \theta ^{\left( l\right) },Z_{l}\right) \right\vert ^{2} \\
&\rightarrow &\sum_{\substack{ j\geqslant 1  \\ m\geqslant 1}}%
\sum_{k=1}^{m}\left( m+1\right) a_{j,m+1}\int \left( \prod\limits_{i=1}^{m}%
\frac{\delta \left( \theta ^{\left( j\right) }-\theta \right) \left( 1+\frac{%
1}{c}\partial _{\theta }+\frac{1}{2c^{2}}\partial _{\theta }^{2}+\frac{1}{2}%
\partial _{Z}^{2}\right) }{D^{j}}\right) \\
&&\times \left( -c\left\langle \frac{\left\vert \Psi \left( \theta
_{i},Z_{i}\right) \right\vert ^{2}}{G^{i}\left( \left( \theta
_{i},Z_{i}\right) \right) }\right\rangle \right)
^{m}\prod\limits_{l=1}^{j}\left( \frac{\frac{\omega _{0}^{-1}\left( J,\left(
\theta ^{\left( l\right) },Z_{l}\right) \right) }{\mathcal{\bar{G}}%
_{0}\left( 0,Z\right) \left( 1+T_{0}\left( \theta ^{\left( l\right)
},Z_{l}\right) \right) D}}{1-\frac{\omega _{0}^{-1}\left( J,\left( \theta
^{\left( l\right) },Z_{l}\right) \right) }{\mathcal{\bar{G}}_{0}\left(
0,Z\right) \left( 1+T_{0}\left( \theta ^{\left( l\right) },Z_{l}\right)
\right) D}}\right) ^{m}\left\vert \Psi \left( \theta ^{\left( l\right)
},Z_{l}\right) \right\vert ^{2}
\end{eqnarray*}%
which is written as a function: 
\begin{eqnarray*}
&\rightarrow &\sum_{\substack{ j\geqslant 1  \\ m\geqslant 1}}%
\sum_{k=1}^{m}\left( m+1\right) a_{j,m+1}\int \left( \prod\limits_{i=1}^{m}%
\frac{\delta \left( \theta ^{\left( j\right) }-\theta \right) \left( 1+\frac{%
1}{c}\partial _{\theta }+\frac{1}{2c^{2}}\partial _{\theta }^{2}+\frac{1}{2}%
\partial _{Z}^{2}\right) }{D^{j}}\right) \\
&&\times \left( -c\left\langle \frac{\left\vert \Psi \left( \theta
_{i},Z_{i}\right) \right\vert ^{2}}{G^{i}\left( \left( \theta
_{i},Z_{i}\right) \right) }\right\rangle \right) ^{m}\left( \frac{\frac{%
\omega _{0}^{-1}\left( J,\left( \theta ^{\left( l\right) },Z_{l}\right)
\right) }{\mathcal{\bar{G}}_{0}\left( 0,Z\right) \left( 1+T_{0}\left( \theta
^{\left( l\right) },Z_{l}\right) \right) D}}{1-\frac{\omega _{0}^{-1}\left(
J,\left( \theta ^{\left( l\right) },Z_{l}\right) \right) }{\mathcal{\bar{G}}%
_{0}\left( 0,Z\right) \left( 1+T_{0}\left( \theta ^{\left( l\right)
},Z_{l}\right) \right) D}}\right) ^{mj}\left\langle \left\vert \Psi \left(
\theta ^{\left( l\right) },Z_{l}\right) \right\vert ^{2}\right\rangle ^{j} \\
&\rightarrow &\left( -cA_{0}\left\langle \frac{\left\vert \Psi \left( \theta
_{i},Z_{i}\right) \right\vert ^{2}}{G^{i}\left( \left( \theta
_{i},Z_{i}\right) \right) }\right\rangle \left\langle \left\vert \Psi \left(
\theta ^{\left( l\right) },Z_{l}\right) \right\vert ^{2}\right\rangle
\right) G\left( A_{0},\left\langle \left\vert \Psi \left( \theta ^{\left(
l\right) },Z_{l}\right) \right\vert ^{2}\right\rangle \right)
\end{eqnarray*}%
with:%
\begin{equation*}
G\left( 0,0\right) =1
\end{equation*}%
and:%
\begin{equation*}
A_{0}=\frac{\frac{\omega _{0}^{-1}\left( J,\left( \theta ^{\left( l\right)
},Z_{l}\right) \right) }{\mathcal{\bar{G}}_{0}\left( 0,Z\right) \left(
1+T_{0}\left( \theta ^{\left( l\right) },Z_{l}\right) \right) D}}{1-\frac{%
\omega _{0}^{-1}\left( J,\left( \theta ^{\left( l\right) },Z_{l}\right)
\right) }{\mathcal{\bar{G}}_{0}\left( 0,Z\right) \left( 1+T_{0}\left( \theta
^{\left( l\right) },Z_{l}\right) \right) D}}
\end{equation*}%
As a consequence:%
\begin{eqnarray*}
&&M_{\Psi }^{\left( 1,a\right) }\left( \left( \theta ,Z\right) ,\left(
\theta _{1},Z_{1}\right) \right) +M_{\Psi }^{\left( 1,a\right) }\left(
\left( \theta ,Z\right) ,\left( \theta _{1},Z_{1}\right) \right) \\
&=&\delta \left( \theta ^{\left( j\right) }-\theta \right) \left( 1+\frac{1}{%
c}\partial _{\theta }+\frac{1}{2c^{2}}\partial _{\theta }^{2}+\frac{1}{2}%
\partial _{Z}^{2}\right) \\
&&\times c\frac{A_{0}\left\langle \frac{\left\vert \Psi \left( \theta
_{i},Z_{i}\right) \right\vert ^{2}}{G^{i}\left( \left( \theta
_{i},Z_{i}\right) \right) }\right\rangle \left\langle \left\vert \Psi \left(
\theta ^{\left( l\right) },Z_{l}\right) \right\vert ^{2}\right\rangle }{1-%
\frac{\omega _{0}^{-1}\left( J,\left( \theta ^{\left( l\right)
},Z_{l}\right) \right) }{\mathcal{\bar{G}}_{0}\left( 0,Z\right) \left(
1+T_{0}\left( \theta ^{\left( l\right) },Z_{l}\right) \right) D}}G\left(
A_{0},\left\langle \left\vert \Psi \left( \theta ^{\left( l\right)
},Z_{l}\right) \right\vert ^{2}\right\rangle \right) \\
&\rightarrow &\delta \left( \theta ^{\left( j\right) }-\theta \right) \left(
1+\frac{1}{c}\partial _{\theta }+\frac{1}{2c^{2}}\partial _{\theta }^{2}+%
\frac{1}{2}\partial _{Z}^{2}\right) \Lambda _{G}
\end{eqnarray*}%
and;%
\begin{eqnarray*}
&&M_{\omega }^{\left( 1,r\right) }\left( \left( \theta ,Z\right) ,\left(
\theta _{1},Z_{1}\right) \right) +M_{\omega }^{\left( 1,a\right) }\left(
\left( \theta ,Z\right) ,\left( \theta _{1},Z_{1}\right) \right) +\bar{N}%
_{\omega }^{\left( 1\right) }\left( \theta ,Z\right) \\
&=&-\delta \left( \theta ^{\left( j\right) }-\theta \right) \left( 1+\frac{1%
}{c}\partial _{\theta }+\frac{1}{2c^{2}}\partial _{\theta }^{2}+\frac{1}{2}%
\partial _{Z}^{2}\right) \\
&&\times cA_{0}^{2}\left\langle \frac{\left\vert \Psi \left( \theta
_{i},Z_{i}\right) \right\vert ^{2}}{G^{i}\left( \left( \theta
_{i},Z_{i}\right) \right) }\right\rangle \left\langle \left\vert \Psi \left(
\theta ^{\left( l\right) },Z_{l}\right) \right\vert ^{2}\right\rangle
H\left( A_{0},\left\langle \left\vert \Psi \left( \theta ^{\left( l\right)
},Z_{l}\right) \right\vert ^{2}\right\rangle \right) \\
&\rightarrow &\delta \left( \theta ^{\left( j\right) }-\theta \right) \left(
1+\frac{1}{c}\partial _{\theta }+\frac{1}{2c^{2}}\partial _{\theta }^{2}+%
\frac{1}{2}\partial _{Z}^{2}\right) \Lambda _{H}
\end{eqnarray*}%
with:%
\begin{equation*}
H\left( 0,0\right) =1
\end{equation*}%
The matricial system (\ref{Dynt}) allows to write $\frac{\delta \omega
^{-1}\left( \theta _{1},Z_{1}\right) }{\omega ^{-1}\left( \theta
_{1},Z_{1}\right) }$ as a function of $\frac{\delta \Psi \left( \theta
_{1},Z_{1}\right) }{\Psi \left( \theta _{1},Z_{1}\right) }$: 
\begin{eqnarray*}
\frac{\delta \omega ^{-1}\left( \theta _{1},Z_{1}\right) }{\omega
^{-1}\left( \theta _{1},Z_{1}\right) } &=&\frac{\frac{\frac{\omega
^{-1}\left( \theta -\frac{\left\vert Z-Z_{1}\right\vert }{c},Z_{1},\Psi
\right) }{\omega ^{-1}\left( \theta ,Z\right) }\check{T}_{1}k_{Z}^{2}}{1+%
\frac{\alpha _{D}h_{D}}{\alpha _{C}h_{C}}\frac{\omega ^{-1}\left( \theta
,Z,\left\vert \Psi \right\vert ^{2}\right) }{\omega ^{-1}\left( \theta -%
\frac{\left\vert Z-Z_{1}\right\vert }{c},Z_{1},\Psi \right) }\frac{\check{T}%
}{\lambda \tau \exp \left( -\frac{\left\vert Z-Z^{\prime }\right\vert }{\nu c%
}\right) }}}{1+\frac{\omega ^{-1}\left( \theta -\frac{\left\vert
Z-Z_{1}\right\vert }{c},Z_{1},\Psi \right) }{\omega ^{-1}\left( \theta
,Z\right) }\check{T}_{1}k_{Z}^{2}}\frac{\delta \Psi \left( \theta
_{1},Z_{1}\right) }{\Psi \left( \theta _{1},Z_{1}\right) } \\
&\rightarrow &\frac{\check{T}_{1}k_{Z}^{2}}{\left( \frac{\omega ^{-1}\left(
\theta ,Z\right) }{\omega ^{-1}\left( \theta -\frac{\left\vert
Z-Z_{1}\right\vert }{c},Z_{1},\Psi \right) }+\check{T}_{1}k_{Z}^{2}\right)
\left( 1+\frac{\alpha _{D}h_{D}}{\alpha _{C}h_{C}}\frac{\omega ^{-1}\left(
\theta ,Z,\left\vert \Psi \right\vert ^{2}\right) }{\omega ^{-1}\left(
\theta -\frac{\left\vert Z-Z_{1}\right\vert }{c},Z_{1},\Psi \right) }\frac{%
\check{T}}{\lambda \tau \exp \left( -\frac{\left\vert Z-Z^{\prime
}\right\vert }{\nu c}\right) }\right) }\frac{\delta \Psi \left( \theta
_{1},Z_{1}\right) }{\Psi \left( \theta _{1},Z_{1}\right) }
\end{eqnarray*}%
Consequently, the system (\ref{Dynt})\bigskip\ becomes a differential system
for $\delta \Psi \left( \theta ,Z\right) $:

\begin{eqnarray*}
&&0=\left( 1-\left( \frac{i\Upsilon }{c\eta ^{2}}-\frac{1}{\nu c}\left( 1-%
\frac{1}{\eta ^{2}}\right) \right) ^{2}\nabla _{Z}^{2}\right) \left(
D_{\theta }+M_{\Psi }^{\left( 1,a\right) }\left( \left( \theta ,Z\right)
,\left( \theta _{1},Z_{1}\right) \right) +M_{\Psi }^{\left( 1,a\right)
}\left( \left( \theta ,Z\right) ,\left( \theta _{1},Z_{1}\right) \right)
\right) \left( 1+\frac{1}{c}\nabla _{\theta }\right) \delta \Psi \left(
\theta ,Z\right) \\
&&-\left( \bar{N}_{\omega }^{\left( 1\right) }\left( \theta ,Z\right)
+M_{\omega }^{\left( 1,r\right) }\left( \left( \theta ,Z\right) ,\left(
\theta _{1},Z_{1}\right) \right) +M_{\omega }^{\left( 1,a\right) }\left(
\left( \theta ,Z\right) ,\left( \theta _{1},Z_{1}\right) \right) \right) \\
&&\times \frac{\left( \frac{i\Upsilon }{c\eta ^{2}}-\frac{1}{\nu c}\left( 1-%
\frac{1}{\eta ^{2}}\right) \right) ^{2}\nabla _{Z}^{2}}{1+\frac{\alpha
_{D}h_{D}}{\alpha _{C}h_{C}}\frac{\omega ^{-1}\left( \theta ,Z,\left\vert
\Psi \right\vert ^{2}\right) }{\omega ^{-1}\left( \theta -\frac{\left\vert
Z-Z_{1}\right\vert }{c},Z_{1},\Psi \right) }\frac{\check{T}}{\lambda \tau
\exp \left( -\frac{\left\vert Z-Z^{\prime }\right\vert }{\nu c}\right) }}%
\left( 1+\frac{1}{c}\nabla _{\theta }\right) \delta \Psi \left( \theta
,Z\right)
\end{eqnarray*}

\subsection*{A5.8 Solutions and stability}

\bigskip Using the change of variable:%
\begin{equation*}
\delta \Psi \left( \theta ,Z\right) \rightarrow \int \exp \left( -\frac{%
\theta -\theta _{0}}{c}\right) \delta \Psi \left( \theta _{0},Z\right)
d\theta _{0}
\end{equation*}%
this becoms:

\bigskip 
\begin{eqnarray*}
&&0=\left( 1-\left( \frac{i\Upsilon }{c\eta ^{2}}-\frac{1}{\nu c}\left( 1-%
\frac{1}{\eta ^{2}}\right) \right) ^{2}\nabla _{Z}^{2}\right) \left(
D_{\theta }+M_{\Psi }^{\left( 1,a\right) }\left( \left( \theta ,Z\right)
,\left( \theta _{1},Z_{1}\right) \right) +M_{\Psi }^{\left( 1,a\right)
}\left( \left( \theta ,Z\right) ,\left( \theta _{1},Z_{1}\right) \right)
\right) \delta \Psi \left( \theta ,Z\right) \\
&&-\left( \bar{N}_{\omega }^{\left( 1\right) }\left( \theta ,Z\right)
+M_{\omega }^{\left( 1,r\right) }\left( \left( \theta ,Z\right) ,\left(
\theta _{1},Z_{1}\right) \right) +M_{\omega }^{\left( 1,a\right) }\left(
\left( \theta ,Z\right) ,\left( \theta _{1},Z_{1}\right) \right) \right) \\
&&\times \frac{\left( \frac{i\Upsilon }{c\eta ^{2}}-\frac{1}{\nu c}\left( 1-%
\frac{1}{\eta ^{2}}\right) \right) ^{2}\nabla _{Z}^{2}}{1+\frac{\alpha
_{D}h_{D}}{\alpha _{C}h_{C}}\frac{\omega ^{-1}\left( \theta ,Z,\left\vert
\Psi \right\vert ^{2}\right) }{\omega ^{-1}\left( \theta -\frac{\left\vert
Z-Z_{1}\right\vert }{c},Z_{1},\Psi \right) }\frac{\check{T}}{\lambda \tau
\exp \left( -\frac{\left\vert Z-Z^{\prime }\right\vert }{\nu c}\right) }}%
\delta \Psi \left( \theta ,Z\right)
\end{eqnarray*}

We replace:%
\begin{eqnarray*}
&&D_{\theta } \\
&\rightarrow &\left( \frac{1}{2}\left( -\frac{\sigma _{\theta }^{2}}{2}%
\nabla _{\theta }+\omega ^{-1}\left( J\left( \theta \right) ,\theta
,Z\right) \right) \nabla _{\theta }+\left( A\left( \Psi \left( \theta
,Z\right) \right) -\frac{\sigma _{\theta }^{2}}{2}\right) \nabla _{\theta
}+U^{\prime \prime }\left( \left\vert \Psi \left( \theta ,Z\right)
\right\vert ^{2}\right) \left\vert \Psi \left( \theta ,Z\right) \right\vert
^{2}\right)
\end{eqnarray*}%
and project the equation on the function with given wave vector $k_{Z}$,
that is, we replace:%
\begin{equation*}
\nabla _{Z}^{2}\rightarrow -k_{Z}^{2}
\end{equation*}%
and the equation writes:

\begin{eqnarray*}
&&0=\left( 1+\left( \frac{i\Upsilon }{c\eta ^{2}}-\frac{1}{\nu c}\left( 1-%
\frac{1}{\eta ^{2}}\right) \right) ^{2}k_{Z}^{2}\right) \\
&&\times \left( -\left( \frac{1}{4}\sigma _{\theta }^{2}+\frac{1}{\alpha }%
\Lambda _{G}\right) \nabla _{\theta }^{2}+\left( \frac{1}{2}\omega
^{-1}\left( J\left( \theta \right) ,\theta ,Z\right) +\left( A\left( \Psi
\left( \theta ,Z\right) \right) -\frac{\sigma _{\theta }^{2}}{2}\right)
\right) \nabla _{\theta }\right. \\
&&\left. +U^{\prime \prime }\left( \left\vert \Psi \left( \theta ,Z\right)
\right\vert ^{2}\right) \left\vert \Psi \left( \theta ,Z\right) \right\vert
^{2}+\left( 1-\frac{1}{\alpha }k_{Z}^{2}\right) \Lambda _{G}\right) \delta
\Psi \left( \theta ,Z\right) \\
&&+\left( 1-\frac{1}{\alpha }k_{Z}^{2}-\frac{1}{\alpha }\partial _{\theta
}^{2}\right) \Lambda _{H}\frac{\left( \frac{i\Upsilon }{c\eta ^{2}}-\frac{1}{%
\nu c}\left( 1-\frac{1}{\eta ^{2}}\right) \right) ^{2}k_{Z}^{2}}{1+\frac{%
\alpha _{D}h_{D}}{\alpha _{C}h_{C}}\frac{\omega ^{-1}\left( \theta
,Z,\left\vert \Psi \right\vert ^{2}\right) }{\omega ^{-1}\left( \theta -%
\frac{\left\vert Z-Z_{1}\right\vert }{c},Z_{1},\Psi \right) }\frac{\check{T}%
}{\lambda \tau \exp \left( -\frac{\left\vert Z-Z^{\prime }\right\vert }{\nu c%
}\right) }}\delta \Psi \left( \theta ,Z\right)
\end{eqnarray*}

This equation has the form%
\begin{equation*}
0=\left( -A\nabla _{\theta }^{2}+B\nabla _{\theta }+C\right) \delta \Psi
\left( \theta ,Z\right)
\end{equation*}%
with:%
\begin{equation*}
A=\left( \frac{1}{4}\sigma _{\theta }^{2}+\frac{1}{\alpha }\Lambda
_{G}\right) +\frac{\frac{1}{\alpha }\Lambda _{H}}{1+\frac{\alpha _{D}h_{D}}{%
\alpha _{C}h_{C}}\frac{\omega ^{-1}\left( \theta ,Z,\left\vert \Psi
\right\vert ^{2}\right) }{\omega ^{-1}\left( \theta -\frac{\left\vert
Z-Z_{1}\right\vert }{c},Z_{1},\Psi \right) }\frac{\check{T}}{\lambda \tau
\exp \left( -\frac{\left\vert Z-Z^{\prime }\right\vert }{\nu c}\right) }}%
\frac{\left( \frac{i\Upsilon }{c\eta ^{2}}-\frac{1}{\nu c}\left( 1-\frac{1}{%
\eta ^{2}}\right) \right) ^{2}k_{Z}^{2}}{1+\left( \frac{i\Upsilon }{c\eta
^{2}}-\frac{1}{\nu c}\left( 1-\frac{1}{\eta ^{2}}\right) \right)
^{2}k_{Z}^{2}}
\end{equation*}%
\begin{equation*}
B=\frac{1}{2}\omega ^{-1}\left( J\left( \theta \right) ,\theta ,Z\right)
+\left( A\left( \Psi \left( \theta ,Z\right) \right) -\frac{\sigma _{\theta
}^{2}}{2}\right)
\end{equation*}

\begin{eqnarray*}
C &=&\left( U^{\prime \prime }\left( \left\vert \Psi \left( \theta ,Z\right)
\right\vert ^{2}\right) \left\vert \Psi \left( \theta ,Z\right) \right\vert
^{2}+\left( 1-\frac{1}{\alpha }k_{Z}^{2}\right) \Lambda _{G}\right) \\
&&+\frac{\left( 1-\frac{1}{\alpha }k_{Z}^{2}\right) \Lambda _{H}}{1+\frac{%
\alpha _{D}h_{D}}{\alpha _{C}h_{C}}\frac{\omega ^{-1}\left( \theta
,Z,\left\vert \Psi \right\vert ^{2}\right) }{\omega ^{-1}\left( \theta -%
\frac{\left\vert Z-Z_{1}\right\vert }{c},Z_{1},\Psi \right) }\frac{\check{T}%
}{\lambda \tau \exp \left( -\frac{\left\vert Z-Z^{\prime }\right\vert }{\nu c%
}\right) }}\frac{\left( \frac{i\Upsilon }{c\eta ^{2}}-\frac{1}{\nu c}\left(
1-\frac{1}{\eta ^{2}}\right) \right) ^{2}k_{Z}^{2}}{1+\left( \frac{i\Upsilon 
}{c\eta ^{2}}-\frac{1}{\nu c}\left( 1-\frac{1}{\eta ^{2}}\right) \right)
^{2}k_{Z}^{2}}
\end{eqnarray*}%
and:%
\begin{equation*}
\left( 1-\frac{1}{\alpha }k_{Z}^{2}\right) \Lambda _{G}\rightarrow c\frac{%
\frac{\omega _{0}^{-1}\left( J,\left( \theta ^{\left( l\right)
},Z_{l}\right) \right) }{\mathcal{\bar{G}}_{0}\left( 0,Z\right) \left(
1+T_{0}\left( \theta ^{\left( l\right) },Z_{l}\right) \right) D}}{\left( 1-%
\frac{\omega _{0}^{-1}\left( J,\left( \theta ^{\left( l\right)
},Z_{l}\right) \right) }{\mathcal{\bar{G}}_{0}\left( 0,Z\right) \left(
1+T_{0}\left( \theta ^{\left( l\right) },Z_{l}\right) \right) D}\right) ^{2}}%
\left\langle \frac{\left\vert \Psi \left( \theta _{i},Z_{i}\right)
\right\vert ^{2}}{G^{i}\left( \left( \theta _{i},Z_{i}\right) \right) }%
\right\rangle \left\langle \left\vert \Psi \left( \theta ^{\left( l\right)
},Z_{l}\right) \right\vert ^{2}\right\rangle \left( 1-\frac{1}{\alpha }%
k_{Z}^{2}\right)
\end{equation*}%
\begin{equation*}
\left( 1-\frac{1}{\alpha }k_{Z}^{2}\right) \Lambda _{H}\rightarrow -c\frac{%
\left( \frac{\omega _{0}^{-1}\left( J,\left( \theta ^{\left( l\right)
},Z_{l}\right) \right) }{\mathcal{\bar{G}}_{0}\left( 0,Z\right) \left(
1+T_{0}\left( \theta ^{\left( l\right) },Z_{l}\right) \right) D}\right) ^{2}%
}{\left( 1-\frac{\omega _{0}^{-1}\left( J,\left( \theta ^{\left( l\right)
},Z_{l}\right) \right) }{\mathcal{\bar{G}}_{0}\left( 0,Z\right) \left(
1+T_{0}\left( \theta ^{\left( l\right) },Z_{l}\right) \right) D}\right) ^{2}}%
\left\langle \frac{\left\vert \Psi \left( \theta _{i},Z_{i}\right)
\right\vert ^{2}}{G^{i}\left( \left( \theta _{i},Z_{i}\right) \right) }%
\right\rangle \left\langle \left\vert \Psi \left( \theta ^{\left( l\right)
},Z_{l}\right) \right\vert ^{2}\right\rangle \left( 1-\frac{1}{\alpha }%
k_{Z}^{2}\right)
\end{equation*}%
\begin{equation*}
\left( \frac{1}{4}\sigma _{\theta }^{2}+\frac{1}{\alpha }\Lambda _{G}\right)
+\frac{\frac{1}{\alpha }\Lambda _{H}}{1+\frac{\alpha _{D}h_{D}}{\alpha
_{C}h_{C}}\frac{\omega ^{-1}\left( \theta ,Z,\left\vert \Psi \right\vert
^{2}\right) }{\omega ^{-1}\left( \theta -\frac{\left\vert Z-Z_{1}\right\vert 
}{c},Z_{1},\Psi \right) }\frac{\check{T}}{\lambda \tau \exp \left( -\frac{%
\left\vert Z-Z^{\prime }\right\vert }{\nu c}\right) }}\frac{\left( \frac{%
i\Upsilon }{c\eta ^{2}}-\frac{1}{\nu c}\left( 1-\frac{1}{\eta ^{2}}\right)
\right) ^{2}k_{Z}^{2}}{1+\left( \frac{i\Upsilon }{c\eta ^{2}}-\frac{1}{\nu c}%
\left( 1-\frac{1}{\eta ^{2}}\right) \right) ^{2}k_{Z}^{2}}
\end{equation*}%
The equation is solved, by considering its Fourier transform;%
\begin{equation*}
\left( A\Omega ^{2}+iB\Omega +C\right) \delta \Psi \left( \theta ,Z\right) =0
\end{equation*}%
that is:%
\begin{equation}
\left( A\left( \Omega +i\frac{B}{2A}\right) ^{2}+C+\frac{B^{2}}{4A}\right)
\delta \Psi \left( \theta ,Z\right) =0  \label{Pqnssss}
\end{equation}%
The solutions are considered as perturbation. Given an initial deviation,
the dynamics is obtained by computing the Green function associated to (\ref%
{Pqnssss}).

Thus, the solution has the form:%
\begin{eqnarray*}
\delta \Psi \left( \theta ,Z\right) &\rightarrow &\int \frac{1}{\left(
A\Omega +i\frac{B}{2A}\right) ^{2}+C+\frac{B^{2}}{4A^{2}}}\exp \left(
-i\Omega \left( \theta -\theta _{0}\right) \right) d\theta \\
&\rightarrow &1_{\sqrt{C+\frac{B^{2}}{4A^{2}}}-\frac{B}{2A}>0}\exp \left(
-\left( \sqrt{C+\frac{B^{2}}{4A^{2}}}-\frac{B}{2A}\right) \left\vert \theta
-\theta _{0}\right\vert \right) \\
&&+1_{\sqrt{C+\frac{B^{2}}{4A^{2}}}+\frac{B}{2A}>0}\exp \left( -\left( \sqrt{%
C+\frac{B^{2}}{4A^{2}}}+\frac{B}{2A}\right) \left\vert \theta -\theta
_{0}\right\vert \right)
\end{eqnarray*}%
Such solution is stable if:%
\begin{equation*}
A>0
\end{equation*}%
and:%
\begin{equation*}
C+\frac{B^{2}}{4A}>0
\end{equation*}%
This corresponds to the following condition:%
\begin{eqnarray*}
&&\left( U^{\prime \prime }\left( \left\vert \Psi \left( \theta ,Z\right)
\right\vert ^{2}\right) \left\vert \Psi \left( \theta ,Z\right) \right\vert
^{2}+\left( 1-\frac{1}{\alpha }k_{Z}^{2}\right) \Lambda _{G}\right) \\
&&+\frac{\left( 1-\frac{1}{\alpha }k_{Z}^{2}\right) \Lambda _{H}}{1+\frac{%
\alpha _{D}h_{D}}{\alpha _{C}h_{C}}\frac{\omega ^{-1}\left( \theta
,Z,\left\vert \Psi \right\vert ^{2}\right) }{\omega ^{-1}\left( \theta -%
\frac{\left\vert Z-Z_{1}\right\vert }{c},Z_{1},\Psi \right) }\frac{\check{T}%
}{\lambda \tau \exp \left( -\frac{\left\vert Z-Z^{\prime }\right\vert }{\nu c%
}\right) }}\frac{\left( \frac{i\Upsilon }{c\eta ^{2}}-\frac{1}{\nu c}\left(
1-\frac{1}{\eta ^{2}}\right) \right) ^{2}k_{Z}^{2}}{1+\left( \frac{i\Upsilon 
}{c\eta ^{2}}-\frac{1}{\nu c}\left( 1-\frac{1}{\eta ^{2}}\right) \right)
^{2}k_{Z}^{2}} \\
&&+\frac{\left( \frac{1}{2}\omega ^{-1}\left( J\left( \theta \right) ,\theta
,Z\right) +\left( A\left( \Psi \left( \theta ,Z\right) \right) -\frac{\sigma
_{\theta }^{2}}{2}\right) \right) ^{2}}{\sigma _{\theta }^{2}+4\frac{\Lambda
_{G}}{\alpha }+\frac{\frac{4}{\alpha }\Lambda _{H}}{1+\frac{\alpha _{D}h_{D}%
}{\alpha _{C}h_{C}}\frac{\omega ^{-1}\left( \theta ,Z,\left\vert \Psi
\right\vert ^{2}\right) }{\omega ^{-1}\left( \theta -\frac{\left\vert
Z-Z_{1}\right\vert }{c},Z_{1},\Psi \right) }\frac{\check{T}}{\lambda \tau
\exp \left( -\frac{\left\vert Z-Z^{\prime }\right\vert }{\nu c}\right) }}%
\frac{\left( \frac{i\Upsilon }{c\eta ^{2}}-\frac{1}{\nu c}\left( 1-\frac{1}{%
\eta ^{2}}\right) \right) ^{2}k_{Z}^{2}}{1+\left( \frac{i\Upsilon }{c\eta
^{2}}-\frac{1}{\nu c}\left( 1-\frac{1}{\eta ^{2}}\right) \right)
^{2}k_{Z}^{2}}}
\end{eqnarray*}%
For $k_{Z}^{2}>>1$, this can be approximated by:%
\begin{eqnarray*}
&&U^{\prime \prime }\left( \left\vert \Psi \left( \theta ,Z\right)
\right\vert ^{2}\right) \left\vert \Psi \left( \theta ,Z\right) \right\vert
^{2}+\left( 1-\frac{1}{\alpha }k_{Z}^{2}\right) \Lambda _{G}+\left( 1-\frac{1%
}{\alpha }k_{Z}^{2}\right) \Lambda _{H} \\
&&+\frac{\left( \frac{1}{2}\omega ^{-1}\left( J\left( \theta \right) ,\theta
,Z\right) +\left( A\left( \Psi \left( \theta ,Z\right) \right) -\frac{\sigma
_{\theta }^{2}}{2}\right) \right) ^{2}}{\sigma _{\theta }^{2}+4\frac{\Lambda
_{G}}{\alpha }+\frac{4}{\alpha }\Lambda _{H}}>0
\end{eqnarray*}%
which is, in an expanded form:%
\begin{eqnarray*}
&&U^{\prime \prime }\left( \left\vert \Psi \left( \theta ,Z\right)
\right\vert ^{2}\right) \left\vert \Psi \left( \theta ,Z\right) \right\vert
^{2}+\Lambda _{G}\left( 1-\frac{\omega _{0}^{-1}\left( J,\left( \theta
^{\left( l\right) },Z_{l}\right) \right) }{\mathcal{\bar{G}}_{0}\left(
0,Z\right) \left( 1+T_{0}\left( \theta ^{\left( l\right) },Z_{l}\right)
\right) D}\right) \\
&&+\frac{\left( \frac{1}{2}\omega ^{-1}\left( J\left( \theta \right) ,\theta
,Z\right) +\left( A\left( \Psi \left( \theta ,Z\right) \right) -\frac{\sigma
_{\theta }^{2}}{2}\right) \right) ^{2}}{\sigma _{\theta }^{2}+4\frac{\Lambda
_{G}}{\alpha }\left( 1-\frac{\omega _{0}^{-1}\left( J,\left( \theta ^{\left(
l\right) },Z_{l}\right) \right) }{\mathcal{\bar{G}}_{0}\left( 0,Z\right)
\left( 1+T_{0}\left( \theta ^{\left( l\right) },Z_{l}\right) \right) D}%
\right) }>\frac{1}{\alpha }k_{Z}^{2}\Lambda _{G}\left( 1-\frac{\omega
_{0}^{-1}\left( J,\left( \theta ^{\left( l\right) },Z_{l}\right) \right) }{%
\mathcal{\bar{G}}_{0}\left( 0,Z\right) \left( 1+T_{0}\left( \theta ^{\left(
l\right) },Z_{l}\right) \right) D}\right)
\end{eqnarray*}%
When:%
\begin{equation*}
\frac{\omega _{0}^{-1}\left( J,\left( \theta ^{\left( l\right)
},Z_{l}\right) \right) }{\mathcal{\bar{G}}_{0}\left( 0,Z\right) \left(
1+T_{0}\left( \theta ^{\left( l\right) },Z_{l}\right) \right) D}<1
\end{equation*}%
the stability condition is:%
\begin{equation*}
\frac{1}{\alpha }k_{Z}^{2}\Lambda _{G}<\frac{U^{\prime \prime }\left(
\left\vert \Psi \left( \theta ,Z\right) \right\vert ^{2}\right) \left\vert
\Psi \left( \theta ,Z\right) \right\vert ^{2}}{1-\frac{\omega
_{0}^{-1}\left( J,\left( \theta ^{\left( l\right) },Z_{l}\right) \right) }{%
\mathcal{\bar{G}}_{0}\left( 0,Z\right) \left( 1+T_{0}\left( \theta ^{\left(
l\right) },Z_{l}\right) \right) D}}+\Lambda _{G}+\frac{\left( \frac{1}{2}%
\omega ^{-1}\left( J\left( \theta \right) ,\theta ,Z\right) +\left( A\left(
\Psi \left( \theta ,Z\right) \right) -\frac{\sigma _{\theta }^{2}}{2}\right)
\right) ^{2}}{\left( 1-\frac{\omega _{0}^{-1}\left( J,\left( \theta ^{\left(
l\right) },Z_{l}\right) \right) }{\mathcal{\bar{G}}_{0}\left( 0,Z\right)
\left( 1+T_{0}\left( \theta ^{\left( l\right) },Z_{l}\right) \right) D}%
\right) \left( \sigma _{\theta }^{2}+4\frac{\Lambda _{G}}{\alpha }\left( 1-%
\frac{\omega _{0}^{-1}\left( J,\left( \theta ^{\left( l\right)
},Z_{l}\right) \right) }{\mathcal{\bar{G}}_{0}\left( 0,Z\right) \left(
1+T_{0}\left( \theta ^{\left( l\right) },Z_{l}\right) \right) D}\right)
\right) }
\end{equation*}%
that is, when the wave vector satisifes:

\begin{eqnarray*}
k_{Z}^{2} &<&\alpha +\alpha \frac{1-\frac{\omega _{0}^{-1}\left( J,\left(
\theta ^{\left( l\right) },Z_{l}\right) \right) }{\mathcal{\bar{G}}%
_{0}\left( 0,Z\right) \left( 1+T_{0}\left( \theta ^{\left( l\right)
},Z_{l}\right) \right) D}}{\frac{\omega _{0}^{-1}\left( J,\left( \theta
^{\left( l\right) },Z_{l}\right) \right) }{\mathcal{\bar{G}}_{0}\left(
0,Z\right) \left( 1+T_{0}\left( \theta ^{\left( l\right) },Z_{l}\right)
\right) D}\left\langle \frac{\left\vert \Psi \left( \theta _{i},Z_{i}\right)
\right\vert ^{2}}{G^{i}\left( \left( \theta _{i},Z_{i}\right) \right) }%
\right\rangle \left\langle \left\vert \Psi \left( \theta ^{\left( l\right)
},Z_{l}\right) \right\vert ^{2}\right\rangle c} \\
&&\times \left( U^{\prime \prime }\left( \left\vert \Psi \left( \theta
,Z\right) \right\vert ^{2}\right) \left\vert \Psi \left( \theta ,Z\right)
\right\vert ^{2}+\frac{\left( \frac{1}{2}\omega ^{-1}\left( J\left( \theta
\right) ,\theta ,Z\right) +\left( A\left( \Psi \left( \theta ,Z\right)
\right) -\frac{\sigma _{\theta }^{2}}{2}\right) \right) ^{2}}{\left( \sigma
_{\theta }^{2}+4\frac{\Lambda _{G}}{\alpha }\left( 1-\frac{\omega
_{0}^{-1}\left( J,\left( \theta ^{\left( l\right) },Z_{l}\right) \right) }{%
\mathcal{\bar{G}}_{0}\left( 0,Z\right) \left( 1+T_{0}\left( \theta ^{\left(
l\right) },Z_{l}\right) \right) D}\right) \right) }\right)
\end{eqnarray*}

the fluctuations are stable.

Note that, when:%
\begin{equation*}
\frac{\omega _{0}^{-1}\left( J,\left( \theta ^{\left( l\right)
},Z_{l}\right) \right) }{\mathcal{\bar{G}}_{0}\left( 0,Z\right) \left(
1+T_{0}\left( \theta ^{\left( l\right) },Z_{l}\right) \right) D}>1
\end{equation*}%
the various series of coefficients do not converge, and the system is
unstable.

\section*{Appendix 6. Self organized collective states: effective action for
connectivities and background states}

Rather than considering signal induced binding between states, we rather
consider that connectivities states may themselves interact. In this
effective description, the activities are intergrated to derive a full
action for $\Delta \Gamma $, the connectivity field abov the bckgrnd.
Effective action depends on an initial background for fld $\left\vert \Psi
_{0}\left( Z\right) \right\vert ^{2}$ and activities $\omega _{0}\left(
Z\right) $. It is obtained by expanding the fields for connectivities as:%
\begin{eqnarray*}
\Gamma \left( T,\hat{T},\theta ,Z,Z^{\prime },C,D\right) &=&\Gamma
_{0}\left( T,\hat{T},\theta ,Z,Z^{\prime },C,D\right) +\Delta \Gamma \left(
T,\hat{T},\theta ,Z,Z^{\prime },C,D\right) \\
\Gamma ^{\dag }\left( T,\hat{T},\theta ,Z,Z^{\prime },C,D\right) &=&\Gamma
_{0}^{\dag }\left( T,\hat{T},\theta ,Z,Z^{\prime },C,D\right) +\Delta \Gamma
^{\dagger }\left( T,\hat{T},\theta ,Z,Z^{\prime },C,D\right)
\end{eqnarray*}%
and we show that in the static approximation, the action writes:%
\begin{eqnarray}
&&S\left( \Gamma \left( T,\hat{T},\theta ,Z,Z^{\prime },C,D\right) \right)
\label{fcm} \\
&=&S\left( \Gamma _{0}\left( T,\hat{T},\theta ,Z,Z^{\prime },C,D\right)
\right) +S_{e}\left( \Delta \Gamma \left( T,\hat{T},\theta ,Z,Z^{\prime
},C,D\right) \right)  \notag
\end{eqnarray}%
with:%
\begin{eqnarray}
&&S_{e}\left( \Delta \Gamma \left( T,\hat{T},\theta ,Z,Z^{\prime
},C,D\right) \right)  \label{sft} \\
&=&\Delta \Gamma ^{\dag }\left( T,\hat{T},\theta ,Z,Z^{\prime },C,D\right) %
\left[ \nabla _{T}\left( \nabla _{T}-\frac{\left( \lambda \left( \hat{T}%
-\left\langle \hat{T}\right\rangle \right) -\left( T-\left\langle
T\right\rangle \right) \right) }{\tau \omega _{0}\left( Z\right) +\Delta
\omega _{0}\left( Z,\left\vert \Psi \right\vert ^{2}\right) }\left\vert \Psi
_{0}\left( Z\right) \right\vert ^{2}\right) \right.  \notag \\
&&+\nabla _{\hat{T}}\left( \nabla _{\hat{T}}+\rho \left( C\frac{\left\vert
\Psi _{0}\left( Z\right) \right\vert ^{2}h_{C}\left( \omega _{0}\left(
Z\right) +\Delta \omega _{0}\left( Z,\left\vert \Psi \right\vert ^{2}\right)
\right) }{\omega _{0}\left( Z\right) +\Delta \omega _{0}\left( Z,\left\vert
\Psi \right\vert ^{2}\right) }\right. \right.  \notag \\
&&\left. \left. +D\frac{\left\vert \Psi _{0}\left( Z^{\prime }\right)
\right\vert ^{2}h_{D}\left( \omega _{0}\left( Z^{\prime }\right) +\Delta
\omega _{0}\left( Z^{\prime },\left\vert \Psi \right\vert ^{2}\right)
\right) }{\omega _{0}\left( Z\right) +\Delta \omega _{0}\left( Z,\left\vert
\Psi \right\vert ^{2}\right) }\right) \left( \hat{T}-\left\langle \hat{T}%
\right\rangle \right) \right) \Delta \Gamma \left( T,\hat{T},\theta
,Z,Z^{\prime },C,D\right)  \notag
\end{eqnarray}%
As expected, the fluctuations around the background state modify the static
activity:%
\begin{equation*}
\omega _{0}\left( Z\right) \rightarrow \omega _{0}\left( Z\right) +\Delta
\omega _{0}\left( Z,\left\vert \Psi \right\vert ^{2}\right)
\end{equation*}%
However, to the firs tapproximatn, we can neglect $\Delta \omega _{0}\left(
Z,\left\vert \Psi \right\vert ^{2}\right) $ and the effective action has the
form: 
\begin{eqnarray}
&&S\left( \Delta \Gamma \left( T,\hat{T},\theta ,Z,Z^{\prime }\right) \right)
\label{fcp} \\
&=&-\Delta \Gamma ^{\dag }\left( T,\hat{T},\theta ,Z,Z^{\prime }\right)
\left( \nabla _{T}\left( \nabla _{T}+\frac{\Delta T-\lambda \Delta \hat{T}}{%
\tau \omega _{0}\left( Z\right) }\right) \right) \Delta \Gamma \left( T,\hat{%
T},\theta ,Z,Z^{\prime }\right)  \notag \\
&&-\Delta \Gamma ^{\dag }\left( T,\hat{T},\theta ,Z,Z^{\prime }\right)
\nabla _{\hat{T}}\left( \nabla _{\hat{T}}+\left\vert \bar{\Psi}_{0}\left(
Z,Z^{\prime }\right) \right\vert ^{2}\Delta \hat{T}\right) \Delta \Gamma
\left( T,\hat{T},\theta ,Z,Z^{\prime }\right) -\hat{V}\left( \Delta \Gamma
,\Delta \Gamma ^{\dag }\right)  \notag \\
&&+U_{\Delta \Gamma }\left( \left\Vert \Delta \Gamma \left( Z,Z^{\prime
}\right) \right\Vert ^{2}\right)  \notag
\end{eqnarray}%
with:%
\begin{equation*}
\left\vert \bar{\Psi}_{0}\left( Z,Z^{\prime }\right) \right\vert ^{2}=\frac{%
\rho \left( C\left( \theta \right) \left\vert \Psi _{0}\left( Z\right)
\right\vert ^{2}\omega _{0}\left( Z\right) +D\left( \theta \right) \hat{T}%
\left\vert \Psi _{0}\left( Z^{\prime }\right) \right\vert ^{2}\omega
_{0}\left( Z^{\prime }\right) \right) }{\omega _{0}\left( Z\right) }
\end{equation*}%
\begin{eqnarray}
&&\hat{V}\left( \Delta \Gamma ,\Delta \Gamma ^{\dag }\right) =\Delta \Gamma
^{\dag }\left( T,\hat{T},\theta ,Z,Z^{\prime }\right)  \label{pnl} \\
&&\times \left( \nabla _{\hat{T}}\left( \frac{\rho D\left( \theta \right)
\left\langle \hat{T}\right\rangle \left\vert \Psi _{0}\left( Z^{\prime
}\right) \right\vert ^{2}}{\omega _{0}\left( Z\right) }\check{T}\left(
1-\left( 1+\left\langle \left\vert \Psi _{\Gamma }\right\vert
^{2}\right\rangle \right) \check{T}\right) ^{-1}\left[ O\frac{\Delta
T\left\vert \Delta \Gamma \left( \theta _{1},Z_{1},Z_{1}^{\prime }\right)
\right\vert ^{2}}{T\Lambda ^{2}}\right] \right) \right) \Delta \Gamma \left(
T,\hat{T},\theta ,Z,Z^{\prime }\right)  \notag
\end{eqnarray}%
and:%
\begin{equation}
O\left( Z,Z^{\prime },Z_{1}\right) =-\frac{\left\vert Z-Z^{\prime
}\right\vert }{c}\nabla _{\theta _{1}}+\frac{\left( Z^{\prime }-Z\right) ^{2}%
}{2}\left( \frac{\nabla _{Z_{1}}^{2}}{2}+\frac{\nabla _{\theta _{1}}^{2}}{%
2c^{2}}-\frac{\nabla _{Z}^{2}\omega _{0}\left( Z\right) }{2}\right)
\end{equation}%
\begin{eqnarray*}
\Delta T &=&T-\left\langle T\right\rangle \\
\Delta \hat{T} &=&\hat{T}-\left\langle \hat{T}\right\rangle
\end{eqnarray*}%
with $\left\langle T\right\rangle $ and $\left\langle \hat{T}\right\rangle $
are averages in the background field. The potential%
\begin{equation*}
U_{\Delta \Gamma }\left( \left\Vert \Delta \Gamma \left( Z,Z^{\prime
}\right) \right\Vert ^{2}\right)
\end{equation*}%
is the second order expansion of $U\left( \left\{ \left\vert \Gamma \left(
\theta ,Z,Z^{\prime },C,D\right) \right\vert ^{2}\right\} \right) $ around $%
\Gamma _{0}$ and $\Gamma _{0}^{\dag }$. Potential (\ref{pnl}) was derived in
(\cite{GLt}). It accnts for the distant activation between connectivity
states. Depending on the potential $U_{\Delta \Gamma }$\ and the background,
distant cells with enhanced connectivity have the possiblity to interact and
increase their connectn.

\subsection*{A6.1 Static collective states}

In (\cite{GLt}), we derived the conditions of existence for collective
states of the effective action (\ref{fcp}). Such state are described as a
collection of shifted states. The values of these shifts depend both on the
characteristics of the background field and the potential $U_{\Delta \Gamma
} $. For shifted states, the average connectivities are modified: $%
\left\langle \hat{T}\right\rangle \rightarrow \left\langle \hat{T}%
\right\rangle +\underline{\Delta \left\langle \hat{T}\right\rangle }$ and $%
\left\langle T\right\rangle \rightarrow \left\langle T\right\rangle +%
\underline{\Delta \left\langle T\right\rangle }$, where the values of the
shift $\underline{\Delta \left\langle \hat{T}\right\rangle }$ and $%
\underline{\Delta \left\langle T\right\rangle }$\ are computed in (\cite{GLt}%
).

We consider the average activity unaffected in first approximation to
consider the condition of emerg stt. The additional activities will be taken
into account below.

\subsubsection*{A6.1.1 Formulas for shifted states}

The static collective states are described by a set $W$ of doublet, such
that if $\left( Z,Z^{\prime }\right) \in W$ \ The solutions to the saddle
point equation of (\ref{fcp}) at these points become:%
\begin{eqnarray}
&&\Delta \Gamma _{\delta }\left( T,\hat{T},\theta ,Z,Z^{\prime }\right)
\label{SF} \\
&=&\exp \left( -\frac{1}{2}\left( \mathbf{\Delta T-}\underline{\mathbf{%
\Delta T}}\right) ^{t}\hat{U}\left( \mathbf{\Delta T-}\underline{\mathbf{%
\Delta T}}\right) \right)  \notag \\
&&\times H_{p}\left( \left( \mathbf{\Delta T}^{\prime }\mathbf{-}\underline{%
\mathbf{\Delta T}}^{\prime }\right) _{2}\frac{\sigma _{T}\lambda _{+}}{2%
\sqrt{2}}\left( \mathbf{\Delta T}^{\prime }\mathbf{-}\underline{\mathbf{%
\Delta T}}^{\prime }\right) _{2}\right) H_{p-\delta }\left( \left( \mathbf{%
\Delta T}^{\prime }\mathbf{-}\underline{\mathbf{\Delta T}}^{\prime }\right)
_{2}\frac{\sigma _{\hat{T}}\lambda _{-}}{2\sqrt{2}}\left( \mathbf{\Delta T}%
^{\prime }\mathbf{-}\underline{\mathbf{\Delta T}}^{\prime }\right)
_{2}\right)  \notag
\end{eqnarray}%
and:%
\begin{eqnarray}
&&\Delta \Gamma _{\delta }^{\dagger }\left( T,\hat{T},\theta ,Z,Z^{\prime
}\right)  \label{CJ} \\
&=&H_{p}\left( \left( \mathbf{\Delta T}^{\prime }\mathbf{-}\underline{%
\mathbf{\Delta T}}^{\prime }\right) _{2}\frac{\sigma _{T}\lambda _{+}}{2%
\sqrt{2}}\left( \mathbf{\Delta T}^{\prime }\mathbf{-}\underline{\mathbf{%
\Delta T}}^{\prime }\right) _{2}\right) H_{p-\delta }\left( \left( \mathbf{%
\Delta T}^{\prime }\mathbf{-}\underline{\mathbf{\Delta T}}^{\prime }\right)
_{2}\frac{\sigma _{\hat{T}}\lambda _{-}}{2\sqrt{2}}\left( \mathbf{\Delta T}%
^{\prime }\mathbf{-}\underline{\mathbf{\Delta T}}^{\prime }\right)
_{2}\right)  \notag
\end{eqnarray}%
where $H_{p}$ and $H_{p-\delta }$ are Hermite polynomials.

The variables involved are:%
\begin{eqnarray}
\mathbf{\Delta T-}\underline{\mathbf{\Delta T}} &=&\left( 
\begin{array}{c}
\Delta T-\underline{\Delta \left\langle T\right\rangle } \\ 
\Delta \hat{T}-\underline{\Delta \left\langle \hat{T}\right\rangle }%
\end{array}%
\right)  \label{SFT} \\
\mathbf{\Delta T}^{\prime }\mathbf{-}\underline{\mathbf{\Delta T}}^{\prime }
&=&P^{-1}\left( \mathbf{\Delta T-}\underline{\mathbf{\Delta T}}\right) 
\notag
\end{eqnarray}%
The matrices and parameters involved are provided in (\cite{GLt}), along
with the formula for $\underline{\Delta \left\langle T\right\rangle }$, $%
\underline{\Delta \left\langle \hat{T}\right\rangle }$.

The density of connectivities between $Z$ and $Z^{\prime }$ is given by $%
\left\vert \Delta \Gamma _{\delta }\left( T,\hat{T},\theta ,Z,Z^{\prime
}\right) \right\vert ^{2}$and can be understood as follows: regardless of
how the system is interpreted, whether as a set of groups of simple cells or
single complex cells at each point, the stable backgrounds are not defined
with a specific connectivity value. On the contrary, the background states
are described by a distribution around some average value. In other words,
the cells or groups of axons/dendrites are connected with strength of
connectivities that are distributed around this average.

\subsubsection{A6.1.2 Conditions for shifted state}

Shifted connectivities emerge at certain points under stability conditions.
Depending on the potential and the activity of cells at these points,
connectivity may either be enhanced or reduced. We established the
conditions for the existence of a shift in a previous work (\cite{GLt}). The
existence conditions for such a set depend on the background $\left\vert
\Psi _{0}\left( Z\right) \right\vert ^{2}$ and the potential $U_{\Delta
\Gamma }^{\prime \prime }$:%
\begin{equation}
\left\vert \left( u+v\right) -\left\langle u+v\right\rangle \right\vert <%
\sqrt{-8U_{\Delta \Gamma }\left( \left\Vert \Delta \Gamma \left( Z,Z^{\prime
}\right) \right\Vert _{\min }^{2}\right) U_{\Delta \Gamma }^{\prime \prime
}\left( \left\Vert \Delta \Gamma \left( Z,Z^{\prime }\right) \right\Vert
_{\min }^{2}\right) }  \label{thr}
\end{equation}%
where $\left\Vert \Delta \Gamma \left( Z,Z^{\prime }\right) \right\Vert
_{\min }^{2}$ is the minimum of the potential $U_{\Delta \Gamma }$ at $%
\left( Z,Z^{\prime }\right) $ and:

\begin{eqnarray*}
u &=&\frac{\left\vert \Psi _{0}\left( Z\right) \right\vert ^{2}}{\tau \omega
_{0}\left( Z\right) } \\
v &=&\rho C\frac{\left\vert \Psi _{0}\left( Z\right) \right\vert
^{2}h_{C}\left( \omega _{0}\left( Z\right) \right) }{\omega _{0}\left(
Z\right) }+\rho D\frac{\left\vert \Psi _{0}\left( Z^{\prime }\right)
\right\vert ^{2}h_{D}\left( \omega _{0}\left( Z^{\prime }\right) \right) }{%
\omega _{0}\left( Z\right) } \\
s &=&-\frac{\lambda \left\vert \Psi _{0}\left( Z\right) \right\vert ^{2}}{%
\omega _{0}\left( Z\right) }
\end{eqnarray*}

The bracket $\left\langle u+v\right\rangle $ represents the average of $u+v$
over the entire space. Therefore, the condition (\ref{thr}) for a shift is
relative. The main factor for allowing the emergence of modified
connectivities is the relative level of activity and frequencies with
respect to the entire system.

This condition means that, in first approximation, $\left\vert \Psi
_{0}\left( Z\right) \right\vert ^{2}$ must be below a threshold provided by
the right-hand-side of (\ref{thr}) for a state with enhanced connectivities
to exist. Since $\left\vert \Psi _{0}\left( Z\right) \right\vert ^{2}$ and
the potential $U_{\Delta \Gamma }$ \ vary along the thread, we may expect
some regions to be activated, while some others are not. The avtivation of
the zone is reinforced by the fact that elements of this zone are bounded by
the activatn.

\subsection*{A6.2 Collective states with dynamical activities}

Our earlier results were obtained by averaging over the entire system to
identify the elements that would become activated when connectivity was
modified. However, the resulting states themselves were not extensively
studied, especially in terms of their group interactions or the interactions
between different potential collective states. Additionally, there was no
mention of the activities associated with these possible states.

In the current context, when a group of states experiences a shift in
connectivity, these states are expected to interact collectively due to
these changes. Furthermore, several group of states should interact with
each other. These interactions are contingent upon the activities of each
constituent element within the group. The primary objective is to describe
the characteristic activity patterns for such independent groups of
interconnected cells and investigate the role of these characteristics in
their interactions.

To achieve this, we need to introduce dynamic aspects into the description
of collective states and expand the formalism to encompass interacting
groups of collective states. This begins with a reworking of (\ref{fcp}) to
incorporate dynamic aspects of interactions between elements within the
activated group. In (\cite{GLt}), the effective action (\ref{fcp}) was
initially derived to ascertain conditions for the emergence of states, and
it was sufficient to work with the neurons background activity.
Consequently, we will assume that neuronal activity is given by:%
\begin{equation*}
\omega _{0}\left( Z\right) +\Delta \omega \left( \theta ,Z,\left\vert \Psi
\right\vert ^{2}\right)
\end{equation*}%
where $\Delta \omega \left( \theta ,Z,\left\vert \Psi \right\vert
^{2}\right) $ is the internal activity of the group. We will find the
formula for the possible values $\Delta \omega \left( \theta ,Z,\left\vert
\Psi \right\vert ^{2}\right) $ and this will describe the set of possible
activities for the emerging states. In fact we will look specifically for
stable\ oscillating forms for $\Delta \omega \left( \theta ,Z,\left\vert
\Psi \right\vert ^{2}\right) $. This choice is justified in detail in
appendices 2 and 3 based on (\cite{GL}), (\cite{GLr}), (\cite{GLs}). The
reason relies on a field-theoretic perturbative argument in (\cite{GL}): a
perturbation in activities modifies the background state $\Psi $ which
compensate any dampening or enhancement in activity oscillations, resulting
in stable patterns.

\subsection{A6.2.1 Effective action}

We assume the existence of states with finite\ set $S^{2}$ $=\left\{ \left(
Z,Z^{\prime }\right) \right\} $, with $\left\Vert \Delta \Gamma \left( T,%
\hat{T},\theta ,Z,Z^{\prime }\right) \right\Vert ^{2}\neq 0$\footnote{%
The condition of existence for such collective state in a dynamic context is
discussed below.}. Using the averages (\ref{SFT}):%
\begin{equation*}
\left( \underline{\Delta \left\langle T\right\rangle },\underline{\Delta
\left\langle \hat{T}\right\rangle }\right)
\end{equation*}%
we can rewrite the effective action (\ref{fcp}) for the elements of the
group of activated states as:

\begin{eqnarray}
&&S\left( \Delta \Gamma \left( T,\hat{T},\theta ,Z,Z^{\prime }\right) \right)
\label{TRM} \\
&=&-\Delta \Gamma ^{\dag }\left( T,\hat{T},\theta ,Z,Z^{\prime }\right)
\left( \nabla _{T}\left( \nabla _{T}+\frac{\left( \Delta T-\underline{\Delta
\left\langle T\right\rangle }\right) -\lambda \left( \Delta \hat{T}-%
\underline{\Delta \left\langle \hat{T}\right\rangle }\right) }{\tau \omega
_{0}\left( Z\right) }\right) \right) \Delta \Gamma \left( T,\hat{T},\theta
,Z,Z^{\prime }\right)  \notag \\
&&-\Delta \Gamma ^{\dag }\left( T,\hat{T},\theta ,Z,Z^{\prime }\right)
\nabla _{\hat{T}}\left( \nabla _{\hat{T}}+\left\vert \bar{\Psi}_{0}\left(
Z,Z^{\prime }\right) \right\vert ^{2}\left( \Delta \hat{T}-\underline{\Delta
\left\langle \hat{T}\right\rangle }\right) \right) \Delta \Gamma \left( T,%
\hat{T},\theta ,Z,Z^{\prime }\right) +U_{\Delta \Gamma }\left( \left\Vert
\Delta \Gamma \left( Z,Z^{\prime }\right) \right\Vert ^{2}\right)  \notag
\end{eqnarray}

Where the sum over $Z\in S$ and $Z^{\prime }\in S$ is implicit. However,
this formula relies on the averages background activities $\omega _{0}\left(
Z\right) $, since it was designed to find average conditions for emergence
of modified states. Once a set of cells are activated, their interaction
implies some additional activity frequency $\Delta \omega \left( \theta
,Z,\left\vert \Psi \right\vert ^{2}\right) $ inducing a modification of
action that becomes:%
\begin{equation}
\hat{S}\left( \Delta \Gamma \left( T,\hat{T},\theta ,Z,Z^{\prime }\right)
\right) =S\left( \Delta \Gamma \left( T,\hat{T},\theta ,Z,Z^{\prime }\right)
\right) -\Delta V\left( \Delta \Gamma ,\Delta \Gamma ^{\dag }\right)
\label{MFD}
\end{equation}%
where:%
\begin{eqnarray}
&&\Delta V\left( \Delta \Gamma ,\Delta \Gamma ^{\dag }\right) =\Delta \Gamma
^{\dag }\left( T,\hat{T},\theta ,Z,Z^{\prime }\right)  \label{TSN} \\
&&\times \left( \nabla _{\hat{T}}\left( \frac{\rho }{\omega _{0}\left(
Z\right) }\left( D\left( \theta \right) \left\langle \hat{T}\right\rangle
\left\vert \Psi _{0}\left( Z^{\prime }\right) \right\vert ^{2}\left( \left(
\left( Z-Z^{\prime }\right) \left( \nabla _{Z}+\nabla _{Z}\omega _{0}\left(
Z\right) \right) +\frac{\left\vert Z-Z^{\prime }\right\vert }{c}\right)
\Delta \omega \left( \theta ,Z,\left\vert \Psi \right\vert ^{2}\right)
\right) \right) \right) \right)  \notag \\
&&\times \Delta \Gamma \left( T,\hat{T},\theta ,Z,Z^{\prime }\right)  \notag
\end{eqnarray}%
This additional term is similar to the potential (\ref{pnl}) but accounts
for an additional interaction between the activated elements as implied by
the terms $\Delta \omega \left( \theta ,Z,\left\vert \Psi \right\vert
^{2}\right) $. Eventhough similar to (\ref{pnl}), this term was not present
in (\ref{fcp}) since this action describes the possibility of emerging
groups, not the internal dynamics of such states.

To consider \ the possible dynamics of the group as a whole, we first
consider its possible dynamic activities given its shape and connectivity
magnitude.

\subsubsection*{A6.2.2 Dynamic activities $\Delta \protect\omega \left( 
\protect\theta ,Z,\left\vert \Psi \right\vert ^{2}\right) $ of collective
states}

In a state with $S^{2}$ $=\left\{ \left( Z,Z^{\prime }\right) \right\} $, we
aim at finding the possible activity frequencies $\omega \left( \theta
,Z,\left\vert \Psi \right\vert ^{2}\right) $ associated to the state $\Delta
T$. \ We consider the state $\left\{ \Delta T\left( Z,Z^{\prime }\right)
\right\} $ as a system with its own associated activity $\Delta \omega
\left( \theta ,Z,\left\vert \Psi \right\vert ^{2}\right) $ in the given
background field $\left\vert \Psi \right\vert ^{2}$.

To find $\omega \left( \theta ,Z,\left\vert \Psi \right\vert ^{2}\right) $,
we start with the defining equation: 
\begin{equation*}
\Delta \omega ^{-1}\left( \theta ,Z,\left\vert \Psi \right\vert ^{2}\right)
=G\left( \int \frac{\kappa }{N}\frac{\Delta \omega \left( J,\theta -\frac{%
\left\vert Z-Z_{1}\right\vert }{c},Z_{1},\Psi \right) \Delta T\left(
Z,\theta ,Z_{1},\theta -\frac{\left\vert Z-Z_{1}\right\vert }{c}\right) }{%
\Delta \omega \left( J,\theta ,Z,\left\vert \Psi \right\vert ^{2}\right) }%
\left\vert \Psi \left( \theta -\frac{\left\vert Z-Z_{1}\right\vert }{c}%
,Z_{1}\right) \right\vert ^{2}dZ_{1}\right)
\end{equation*}

We show in (\cite{GLw}) that $\Delta \omega ^{-1}\left( \theta ,Z,\left\vert
\Psi \right\vert ^{2}\right) $ decomposes into a static part and a variable
part. The solution for the specific activties of the group is:

\begin{eqnarray}
&&\Delta \omega \left( \theta ,Z_{i},\mathbf{T},\left\vert \Psi \right\vert
^{2}\right)  \label{CTv} \\
&=&\overline{\Delta \omega }\left( \mathbf{Z},\mathbf{T},\left\vert \Psi
\right\vert ^{2}\right)  \notag \\
&&+A\left( Z_{1}\right) \left( 1,\left( 1-\Delta \mathbf{\hat{T}}\exp \left(
-i\Upsilon _{p}\left( \mathbf{\hat{T}}\right) \frac{\left\vert \mathbf{%
\Delta Z}\right\vert }{c}\right) \right) ^{-1}\Delta \hat{T}_{1}\left( 
\mathbf{Z}\right) \exp \left( -i\Upsilon _{p}\left( \mathbf{\hat{T}}\right) 
\frac{\left\vert \mathbf{\Delta Z}_{1}\right\vert }{c}\right) \right)
^{t}\exp \left( i\Upsilon _{p}\left( \mathbf{\hat{T}}\right) \theta \right) 
\notag
\end{eqnarray}%
The static part for a group of activtd states 
\begin{equation*}
\overline{\Delta \omega }\left( \mathbf{Z},\mathbf{T},\left\vert \Psi
\right\vert ^{2}\right)
\end{equation*}%
is a vector, whose component $i$ satisfies for each point $Z_{i}$ of this
group:%
\begin{equation}
\overline{\Delta \omega }^{-1}\left( Z_{i},\left\vert \Psi \right\vert
^{2}\right) =G\left( \sum_{j}\frac{\kappa }{N}\frac{\overline{\Delta \omega }%
\left( Z_{j},\Psi \right) \Delta T\left( Z_{i},Z_{j}\right) }{\overline{%
\Delta \omega }\left( Z_{i},\left\vert \Psi \right\vert ^{2}\right) }%
\left\vert \Psi \left( Z_{j}\right) \right\vert ^{2}\right)  \label{DM}
\end{equation}

The second contribution in (\ref{CTv}) describes an osicllatory behavior for
all elements of the groups with a set of possible frequencs, with:%
\begin{equation*}
\Delta \hat{T}\left( Z_{i},Z_{j}\right) =\frac{\kappa }{N}\frac{\Delta
T\left( Z_{i},Z_{j}\right) \overline{\Delta \omega }\left( J,Z_{j},\Psi
\right) \left\vert \Psi _{0}\left( Z_{j}\right) \right\vert ^{2}}{%
G^{-1}\left( \overline{\Delta \omega }^{-1}\left( J,Z_{i},\left\vert \Psi
\right\vert ^{2}\right) \right) -\overline{\Delta \omega }^{-1}\left(
J,Z_{i},\left\vert \Psi \right\vert ^{2}\right) }
\end{equation*}

and $\overline{\Delta \omega }\left( \theta ,\mathbf{Z},\left\vert \Psi
\right\vert ^{2}\right) $ and $\Delta \hat{T}_{1}\left( \mathbf{Z}\right) $\
are vectors with coordinates $\overline{\Delta \omega }\left( J,\theta
,Z_{i},\left\vert \Psi \right\vert ^{2}\right) $ and $\Delta \hat{T}\left(
Z_{1},Z_{j}\right) $ respectively. The point $Z_{1}$ is an arbitrary point
chosen in the grp nd $A\left( Z_{1}\right) $ is the amplitude of $\Delta
\omega $ at some gvn $Z_{1}$.

The matrix $\Delta \mathbf{\hat{T}}$ has elements $\Delta \hat{T}\left(
Z_{i},Z_{j}\right) $ and the frequencies $\Upsilon _{p}\left( \mathbf{T}%
\right) $ belong to a discrete set satisfying the equation:%
\begin{equation}
\det \left( 1-\Delta \hat{T}\left( Z_{i},Z_{j}\right) \exp \left( -i\Upsilon
_{p}\frac{\left\vert Z_{i}-Z_{j}\right\vert }{c}\right) \right) =0
\label{frc}
\end{equation}%
The possible osillatory activities associated to the assembly is thus given
by the sets:%
\begin{equation*}
\left\{ \left\{ A\left( Z_{i}\right) \right\} _{i=1,...n},\Upsilon
_{p}\left( \left\{ \Delta \hat{T}\left( Z_{i},Z_{j}\right) \right\} \right)
\right\} _{p}
\end{equation*}%
where $p$ refers to the frequencies $\Upsilon $ and the amplitudes slve:%
\begin{equation*}
A\left( Z_{i}\right) =\sum_{j\neq i}A\left( Z_{j}\right) \Delta \hat{T}%
\left( Z_{i},Z_{j}\right) \exp \left( -i\Upsilon _{p}\left( \left\{ \Delta 
\hat{T}\left( Z_{i},Z_{j}\right) \right\} \right) \frac{\left\vert
Z_{i}-Z_{j}\right\vert }{c}\right)
\end{equation*}

\subsection*{A6.2.3 averages computations for activated states}

Now, considering (\ref{MFD}) for the group of shifted states, we rewrite the
action taking into account their particular interactions. Having obtained
the frequencies, we can come back to the action minimization and compute the
connectivity states along with average connectivities. To do so, we replace
in (\ref{TSN}) (see (\cite{GLt})):%
\begin{eqnarray}
\left( \left( \left( Z-Z^{\prime }\right) \left( \nabla _{Z}+\nabla
_{Z}\omega _{0}\left( Z\right) \right) +\frac{\left\vert Z-Z^{\prime
}\right\vert }{c}\right) \Delta \omega \left( \theta ,Z,\left\vert \Psi
\right\vert ^{2}\right) \right) &\rightarrow &\frac{\left\vert Z-Z^{\prime
}\right\vert }{c}\Delta \omega \left( \theta ,Z,\left\vert \Psi \right\vert
^{2}\right) \\
&\rightarrow &g\left\vert Z-Z^{\prime }\right\vert \Delta \omega \left(
\theta ,Z,\left\vert \Psi \right\vert ^{2}\right)  \notag
\end{eqnarray}%
At the connectivity time scale we replace $\Delta \omega \left( \theta
,Z,\left\vert \Psi \right\vert ^{2}\right) $ by its static part $\overline{%
\Delta \omega }\left( Z,\left\vert \Psi \right\vert ^{2}\right) $. \
Replacing this formula in (\ref{TSN}), using (\ref{MFD}) and (\ref{TRM}),
leads to the following action:%
\begin{eqnarray}
&&\hat{S}\left( \Delta \Gamma \left( T,\hat{T},\theta ,Z,Z^{\prime }\right)
\right)  \label{hst} \\
&=&-\Delta \Gamma ^{\dag }\left( T,\hat{T},\theta ,Z,Z^{\prime }\right)
\left( \nabla _{T}\left( \nabla _{T}+\frac{\left( \Delta T-\Delta
\left\langle T\right\rangle \right) -\lambda \left( \Delta \hat{T}-\Delta
\left\langle \hat{T}\right\rangle \right) }{\tau \omega _{0}\left( Z\right) }%
\left\vert \Psi \left( \theta ,Z\right) \right\vert ^{2}\right) \right)
\Delta \Gamma \left( T,\hat{T},\theta ,Z,Z^{\prime }\right)  \notag \\
&&-\Delta \Gamma ^{\dag }\left( T,\hat{T},\theta ,Z,Z^{\prime }\right)
\nabla _{\hat{T}}\left( \nabla _{\hat{T}}+\left\vert \bar{\Psi}_{0}\left(
Z,Z^{\prime }\right) \right\vert ^{2}\left( \Delta \hat{T}-\left( \Delta
\left\langle \hat{T}\right\rangle \right) \right) \right) \Delta \Gamma
\left( T,\hat{T},\theta ,Z,Z^{\prime }\right)  \notag \\
&&+U_{\Delta \Gamma }\left( \left\Vert \Delta \Gamma \left( Z,Z^{\prime
}\right) \right\Vert ^{2}\right)  \notag
\end{eqnarray}%
where $\Delta \left\langle T\right\rangle $ and $\Delta \left\langle \hat{T}%
\right\rangle $ are solutions of:%
\begin{eqnarray}
\Delta \left\langle T\right\rangle &=&\underline{\Delta \left\langle
T\right\rangle }+\lambda \left\langle \Delta ^{\omega }\hat{T}\right\rangle
\label{lbr} \\
\Delta \left\langle \hat{T}\right\rangle &=&\underline{\Delta \left\langle 
\hat{T}\right\rangle }+\left\langle \Delta ^{\omega }\hat{T}\right\rangle 
\notag
\end{eqnarray}%
with:%
\begin{equation}
\Delta ^{\omega }\hat{T}=\frac{\left( D\left( \theta \right) \left\langle 
\hat{T}\right\rangle \left\vert \Psi _{0}\left( Z^{\prime }\right)
\right\vert ^{2}\left\vert Z-Z^{\prime }\right\vert g\overline{\Delta \omega 
}\left( Z,\mathbf{T},\left\vert \Psi \right\vert ^{2}\right) \right) }{%
\left( C\left( \theta \right) \left\vert \Psi _{0}\left( Z\right)
\right\vert ^{2}\omega _{0}\left( Z\right) +D\left( \theta \right)
\left\langle \hat{T}\right\rangle \left\vert \Psi _{0}\left( Z^{\prime
}\right) \right\vert ^{2}\omega _{0}\left( Z^{\prime }\right) \right) }
\end{equation}%
where $\overline{\Delta \omega }\left( Z_{i},\left\vert \Psi \right\vert
^{2}\right) $ satisfies:%
\begin{equation}
\overline{\Delta \omega }^{-1}\left( Z_{i},\left\vert \Psi \right\vert
^{2}\right) =G\left( \sum_{j}\frac{\kappa }{N}\frac{\overline{\Delta \omega }%
\left( Z_{j},\Psi \right) \Delta T\left( Z_{i},Z_{j}\right) }{\overline{%
\Delta \omega }\left( Z_{i},\left\vert \Psi \right\vert ^{2}\right) }%
\left\vert \Psi \left( Z_{j}\right) \right\vert ^{2}\right)
\end{equation}%
The first equation defines the set $\left\langle \Delta \mathbf{T}%
\right\rangle $ and the second one yields $\Delta \left\langle \mathbf{\hat{T%
}}\right\rangle $. There are several solutions:%
\begin{equation}
\left( \left\langle \Delta \mathbf{T}\right\rangle ^{\alpha },\left\langle
\Delta \mathbf{\hat{T}}\right\rangle ^{\alpha }\right)
\end{equation}%
For each of these solutions, a sequence of frequencies $\left( \Upsilon
_{p}^{\alpha }\right) $ satisfying (\ref{frc}) are compatible. The variable
contribution of activities is given by:%
\begin{equation*}
\Delta \omega _{p}^{\alpha }\left( \theta ,Z,\mathbf{\Delta T}\right) =%
\overline{\Delta \omega }\left( Z,\mathbf{T},\left\vert \Psi \right\vert
^{2}\right) +\left( \mathbf{N}_{p}^{\alpha }\right) ^{-1}\Delta \mathbf{%
\omega }_{0}\exp \left( -i\Upsilon _{p}^{\alpha }\frac{\left\vert \Delta 
\mathbf{Z}_{i}\right\vert }{c}\right)
\end{equation*}%
where:%
\begin{equation*}
\left[ \mathbf{N}_{p}^{\alpha }\right] _{\left( Z_{i},Z_{j}\right) }=\left(
\delta _{ij}-\left[ \mathbf{\Delta T}\right] _{\left( Z_{i},Z_{j}\right)
}\exp \left( -i\Upsilon _{p}\frac{\left\vert Z_{i}-Z_{j}\right\vert }{c}%
\right) \right) ^{-1}
\end{equation*}

\subsection*{A6.2.4 Form of the activated states}

We will now derive the formula for $\Delta \Gamma $ and its conjugate $%
\Delta \Gamma ^{\dag }$ in the activated state. Using (\ref{lbr}), we showed
in \cite{GLw} \ that, after a change of variables, the effective action (\ref%
{hst}) rewrites as:%
\begin{eqnarray}
&&\hat{S}\left( \Delta \Gamma \left( T,\hat{T},\theta ,Z,Z^{\prime }\right)
\right)  \label{SH} \\
&=&-\Delta \Gamma ^{\dag }\left( T,\hat{T},\theta ,Z,Z^{\prime }\right) 
\notag \\
&&\left( \sigma _{T}^{2}\nabla _{T}^{2}-\frac{1}{2\sigma _{T}^{2}}\left( 
\frac{\left( \Delta T-\left\langle \Delta T\left( Z,Z^{\prime }\right)
\right\rangle _{p}^{\alpha }\right) -\lambda \left( \Delta \hat{T}%
-\left\langle \Delta \hat{T}\left( Z,Z^{\prime }\right) \right\rangle
_{p}^{\alpha }\right) }{\tau \omega _{S}\left( Z\right) }\left\vert \Psi
_{0}\left( \theta ,Z\right) \right\vert ^{2}\right) ^{2}\right) \Delta
\Gamma \left( T,\hat{T},\theta ,Z,Z^{\prime }\right)  \notag \\
&&-\Delta \Gamma ^{\dag }\left( T,\hat{T},\theta ,Z,Z^{\prime }\right) 
\notag \\
&&\times \left( \sigma _{\hat{T}}^{2}\nabla _{\hat{T}}^{2}-\frac{\left(
D\left( Z,Z^{\prime }\right) \left( \Delta \hat{T}-\left\langle \Delta \hat{T%
}\left( Z,Z^{\prime }\right) \right\rangle _{p}^{\alpha }\right) +\mathbf{M}%
^{\alpha }\left( Z,Z^{\prime }\right) \left( \mathbf{\Delta T}-\left\langle 
\mathbf{\Delta T}\left( Z,Z^{\prime }\right) \right\rangle _{p}^{\alpha
}\right) \right) ^{2}}{2\sigma _{\hat{T}}^{2}}\right) \Delta \Gamma \left( T,%
\hat{T},\theta ,Z,Z^{\prime }\right)  \notag \\
&&+C\left( Z,Z^{\prime }\right) \left\Vert \Delta \Gamma \left( T,\hat{T}%
,\theta ,Z,Z^{\prime }\right) \right\Vert ^{2}  \notag
\end{eqnarray}%
with:%
\begin{equation*}
D\left( Z,Z^{\prime }\right) =\frac{\rho \left( C\left( \theta \right)
\left\vert \Psi _{0}\left( Z\right) \right\vert ^{2}\omega _{0}\left(
Z\right) +D\left( \theta \right) \hat{T}\left\vert \Psi _{0}\left( Z^{\prime
}\right) \right\vert ^{2}\omega _{0}\left( Z^{\prime }\right) \right) }{%
\omega _{0}\left( Z\right) }
\end{equation*}

\begin{equation*}
C\left( Z,Z^{\prime }\right) =\tau \omega _{S}\left( Z\right) +\frac{\rho
\left( C\left( \theta \right) \left\vert \Psi _{0}\left( Z^{\prime }\right)
\right\vert ^{2}\omega _{S}\left( Z\right) +D\left( \theta \right)
\left\vert \Psi _{0}\left( Z^{\prime }\right) \right\vert ^{2}\omega
_{S}\left( Z^{\prime }\right) \right) }{\omega _{S}\left( Z\right) }
\end{equation*}%
and:%
\begin{equation*}
\mathbf{M}^{\alpha }\left( Z,Z^{\prime }\right) =\left( \frac{\rho D\left(
\theta \right) \left\langle \hat{T}\right\rangle \left\vert \Psi _{0}\left(
Z^{\prime }\right) \right\vert ^{2}A\left\vert Z-Z^{\prime }\right\vert }{%
\omega _{S}\left( Z\right) }\left( \nabla _{\mathbf{\Delta T}_{\left(
Z_{1},Z_{1}^{\prime }\right) }}\left( \overline{\Delta \omega }\left(
Z,\left\langle \mathbf{\Delta T}\right\rangle \right) \right) _{\left(
\left\langle \Delta \mathbf{T}_{\left( Z_{1},Z_{1}^{\prime }\right)
}\right\rangle _{p}^{\alpha }\right) }\right) \right)
\end{equation*}

Then, this effective action (\ref{SH}) rewrites as in the text:%
\begin{eqnarray}
&&\hat{S}\left( \Delta \Gamma \left( T,\hat{T},\theta ,Z,Z^{\prime }\right)
\right) \\
&=&-\Delta \Gamma ^{\dag }\left( T,\hat{T},\theta ,Z,Z^{\prime }\right)
\left( \nabla _{T}^{2}+\nabla _{\hat{T}}^{2}-\frac{1}{2}\left( \mathbf{%
\Delta T-}\left\langle \mathbf{\Delta T}\right\rangle _{p}^{\alpha }\right)
^{t}\mathbf{A}_{p}^{\alpha }\left( \mathbf{\Delta T-}\left\langle \mathbf{%
\Delta T}\right\rangle _{p}^{\alpha }\right) \right) \Delta \Gamma \left( T,%
\hat{T},\theta ,Z,Z^{\prime }\right)  \notag \\
&&+C\left( Z,Z^{\prime }\right) \left\Vert \Delta \Gamma \left( T,\hat{T}%
,\theta ,Z,Z^{\prime }\right) \right\Vert ^{2}  \notag
\end{eqnarray}%
and we show in \cite{GLw} that the minimization equations leads to the
activated state:%
\begin{equation}
\Delta \Gamma =\prod\limits_{Z,Z^{\prime }}\left\vert \Delta T\left(
Z,Z^{\prime }\right) ,\Delta \hat{T}\left( Z,Z^{\prime }\right) ,\alpha
\left( Z,Z^{\prime }\right) ,p\left( Z,Z^{\prime }\right) \right\rangle
\equiv \left\vert \mathbf{\alpha },\mathbf{p},S^{2}\right\rangle
\end{equation}%
with $S^{2}=\left\{ \left( Z,Z^{\prime }\right) \right\} $ where the states
are activated. In a developped form, this state is given by formula similar
to (\ref{SF}) and (\ref{CJ}):%
\begin{eqnarray}
&&\left\vert \mathbf{\alpha },\mathbf{p},S^{2}\right\rangle \\
&=&\exp \left( -\frac{1}{2}\left( \mathbf{\Delta T-}\left\langle \mathbf{%
\Delta T}\right\rangle _{p}^{\alpha }\right) ^{t}\mathbf{A}_{p}^{\alpha
}\left( \mathbf{\Delta T-}\left\langle \mathbf{\Delta T}\right\rangle
_{p}^{\alpha }\right) \right) H_{p}\left( \frac{1}{2}\left( \mathbf{\Delta T-%
}\left\langle \mathbf{\Delta T}\right\rangle _{p}^{\alpha }\right) ^{t}%
\mathbf{A}_{p}^{\alpha }\left( \mathbf{\Delta T-}\left\langle \mathbf{\Delta
T}\right\rangle _{p}^{\alpha }\right) \right)  \notag \\
&&\times H_{p}\left( \left( \mathbf{\Delta T}^{\prime }\mathbf{-}%
\left\langle \mathbf{\Delta T}\right\rangle _{p}^{\alpha }\right)
_{1}^{t}\left( \mathbf{D}_{p}^{\alpha }\right) _{1}\left( \mathbf{\Delta T}%
^{\prime }\mathbf{-}\left\langle \mathbf{\Delta T}\right\rangle _{p}^{\alpha
}\right) _{1}^{t}\right) H_{p-\delta }\left( \left( \mathbf{\Delta T}%
^{\prime }\mathbf{-}\left\langle \mathbf{\Delta T}\right\rangle _{p}^{\alpha
}\right) _{2}^{t}\left( \mathbf{D}_{p}^{\alpha }\right) _{2}\left( \mathbf{%
\Delta T}^{\prime }\mathbf{-}\left\langle \mathbf{\Delta T}\right\rangle
_{p}^{\alpha }\right) _{2}^{t}\right)  \notag
\end{eqnarray}%
with the conjugate:%
\begin{eqnarray}
\Delta \Gamma ^{\dag } &=&\left\langle \mathbf{\alpha },\mathbf{p}%
,S^{2}\right\vert \\
&=&H_{p}\left( \left( \mathbf{\Delta T}^{\prime }\mathbf{-}\left\langle 
\mathbf{\Delta T}\right\rangle _{p}^{\alpha }\right) _{1}^{t}\left( \mathbf{D%
}_{p}^{\alpha }\right) _{1}\left( \mathbf{\Delta T}^{\prime }\mathbf{-}%
\left\langle \mathbf{\Delta T}\right\rangle _{p}^{\alpha }\right)
_{1}^{t}\right) H_{p-\delta }\left( \left( \mathbf{\Delta T}^{\prime }%
\mathbf{-}\left\langle \mathbf{\Delta T}\right\rangle _{p}^{\alpha }\right)
_{2}^{t}\left( \mathbf{D}_{p}^{\alpha }\right) _{2}\left( \mathbf{\Delta T}%
^{\prime }\mathbf{-}\left\langle \mathbf{\Delta T}\right\rangle _{p}^{\alpha
}\right) _{2}^{t}\right)  \notag
\end{eqnarray}

\section*{Appendix 7. Activities and frequencies for interacting collective
states as function of non interacting states}

We formulate the activity equations for a collective state by considering
separately the different type of collectives states involved in this state.
The activity equations for each type are influenced by the activities of
other types. The equations for activities of the $i$-th group as:%
\begin{eqnarray}
&&\omega _{i}\left( Z_{a_{i}},\theta \right)  \label{frQ} \\
&=&G\left( \sum_{j}\sum_{\left\{ b_{j}\right\} }\frac{\kappa }{N}g^{ij}\frac{%
\omega _{j}\left( \theta -\frac{\left\vert Z_{a_{i}}-Z_{b_{j}}\right\vert }{c%
},Z_{b_{j}}\right) \left( \left( T_{ij}+\Delta T_{ij}\right) \left(
Z_{a_{i}},Z_{b_{j}},\theta -\frac{\left\vert Z_{a_{i}}-Z_{b_{j}}\right\vert 
}{c}\right) \right) }{\Delta \omega _{i}\left( Z_{a_{i}},\theta \right) }%
\left\vert \Psi _{j}\left( \theta -\frac{\left\vert
Z_{a_{i}}-Z_{b_{j}}\right\vert }{c},Z_{b_{j}}\right) \right\vert ^{2}\right)
\notag
\end{eqnarray}%
where as in the previous section:%
\begin{equation*}
\omega _{i}^{-1}\left( Z_{a_{i}},\theta \right) =\omega _{i0}^{-1}\left(
Z_{a_{i}}\right) +\Delta \omega _{i}^{-1}\left( Z_{a_{i}},\theta \right)
\end{equation*}%
Said differently, this is the equation for the activities of the $i$-th
component of the whole state.

Assuming a weak interactions between the components:%
\begin{equation*}
g^{ij}\Delta T_{ij}<<\Delta T_{ii}
\end{equation*}%
we expand this equation (\ref{frQ}) to the first order in $g^{ij}\Delta
T_{ij}$. It leads to writing the static and dynamic parts of activities as
functions of the non-interacting ones. Details are provided in appendix 5.
This description will be useful for depicting transitions between
non-interacting and interacting states.

\subsection*{A7.1 Static part}

Equilibrium\ static activities satify:%
\begin{eqnarray}
&&\overline{\Delta \omega }_{i}\left( Z_{a_{i}}\right)  \label{bsc} \\
&=&\sum_{j,b_{j}}\left( \delta _{\left( i,a_{i}\right) \left( j,b_{j}\right)
}-G^{\prime }\left( G^{-1}\left( \left( \omega _{Sf}\right) _{i}\right)
\right) \right.  \notag \\
&&\times \left. \left( \frac{\kappa }{N}\frac{G^{ij}\left( \left( \omega
_{Sf}\right) _{j}\right) \left( Z_{b_{j}}\right) T_{ij}\left(
Z_{a_{i}},Z_{b_{j}}\right) }{\left( \left( \omega _{Sf}\right) _{j}\right)
\left( Z_{a_{i}}\right) }\left( \mathcal{G}_{j0}+\left\vert \Psi _{j}\left(
Z_{b_{j}}\right) \right\vert ^{2}\right) \right) _{j\neq i}\right) _{ij}^{-1}%
\overline{\left( \Delta \omega _{f}\right) }_{j}\left( Z_{b_{j}}\right) 
\notag
\end{eqnarray}%
where:%
\begin{equation*}
\left( \omega _{Sf}\right) _{i}=\omega _{0i}+\overline{\left( \Delta \omega
_{f}\right) }_{i}
\end{equation*}
are the static part of the activities without interaction. Here, the $%
\overline{\left( \Delta \omega _{f}\right) }_{i}\left( Z_{a_{i}}\right) $
are the additional activities in absence of interactions and $\omega _{0i}$
is the background activity for field $\Psi _{i}$.

The $\overline{\left( \Delta \omega _{f}\right) }_{i}\left( Z_{a_{i}}\right) 
$ satisfy an equation similar to (\ref{DM}):%
\begin{equation}
\left( \omega _{Sf}\right) _{i}\left( Z_{a_{i}}\right) =G\left( \sum_{\beta
_{i}}\frac{\kappa }{N}\frac{\left( \left( \omega _{Sf}\right) _{i}\right)
\left( Z_{\beta _{i}}\right) T_{ij}\left( Z_{a_{i}},Z_{b_{j}}\right) }{%
\left( \left( \omega _{Sf}\right) _{i}\right) \left( Z_{a_{i}},\theta
\right) }\left\vert \Psi _{j}\left( Z_{b_{j}}\right) \right\vert ^{2}\right)
\label{STN}
\end{equation}%
where we assume $G^{ii}=1$. Formula for $\overline{\Delta \omega }_{i}\left(
Z_{a_{i}}\right) $ are derived in appendix 5 at the first order in $G^{ij}$:%
\begin{eqnarray}
&&\overline{\Delta \omega }_{i}\left( Z_{a_{i}}\right)  \label{STP} \\
&=&\sum_{j,b_{j}}\left( \delta _{\left( i,a_{i}\right) \left( j,b_{j}\right)
}-G^{\prime }\left( G^{-1}\left( \left( \omega _{Sf}\right) _{i}\left(
Z\right) \right) \right) \right.  \notag \\
&&\times \left. \left( \frac{\kappa }{N}\frac{G^{ij}\left( \omega
_{Sf}\right) _{j}\left( Z_{b_{j}}\right) T_{ij}\left(
Z_{a_{i}},Z_{b_{j}}\right) }{\left( \omega _{Sf}\right) _{i}\left(
Z_{a_{i}}\right) }\left( \mathcal{G}_{j0}+\left\vert \Psi _{j}\left(
Z_{b_{j}}\right) \right\vert ^{2}\right) \right) _{j\neq i}\right) _{ij}^{-1}%
\overline{\left( \Delta \omega _{f}\right) }_{j}\left( Z_{b_{j}}\right) 
\notag
\end{eqnarray}

\subsection*{A7.2 Non-static part}

\subsubsection{A7.2.1 Activities without interactions}

As before, the non static part of $n$ non-interacting activities are
solutions of:%
\begin{equation}
\left( \Delta \omega _{Df}\right) _{i}\left( Z_{a_{i}},\theta \right)
=\sum_{a_{j}}\check{T}_{ii}\left( Z_{a_{i}},Z_{a_{j}}\right) \left( \Delta
\omega _{Df}\right) _{i}\left( \theta -\frac{\left\vert Z_{\alpha
_{i}}-Z_{a_{j}}\right\vert }{c},Z_{a_{j}}\right)  \label{Sr}
\end{equation}%
where:%
\begin{eqnarray}
&&\check{T}_{ij}\left( Z_{a_{i}},Z_{b_{j}}\right)  \label{St} \\
&=&G^{\prime }\left( G^{-1}\left( \left( \omega _{S}\right) _{i}^{-1}\left(
Z_{a_{i}},\left\vert \Psi \right\vert ^{2}\right) \right) \right)  \notag \\
&&\times \frac{\kappa }{N}g^{ij}\frac{\left( \omega _{S}\right) _{j}\left(
Z_{b_{j}}\right) T_{ij}\left( Z_{a_{i}},Z_{b_{j}}\right) }{\sum_{b_{j}}\frac{%
\kappa }{N}g^{ij}\left( \omega _{S}\right) _{j}\left( Z_{b_{j}}\right)
G^{\prime }\left( G^{-1}\left( \left( \omega _{S}\right) _{i}^{-1}\left(
Z_{a_{i}},\left\vert \Psi \right\vert ^{2}\right) \right) \right) +\left(
\omega _{S}\right) _{i}^{2}\left( Z_{a_{i}}\right) }\left\vert \Psi
_{j}\left( Z_{b_{j}}\right) \right\vert ^{2}  \notag
\end{eqnarray}%
Rewriting the solutions as a vector:%
\begin{equation*}
\left( \left( \Delta \omega _{Df}\right) ^{-1}\left( Z_{a_{i}},\theta
\right) \right) _{a_{i}}\equiv \Delta \mathbf{\omega }_{Df}^{-1}\left( Z_{%
\mathbf{a}},\theta \right)
\end{equation*}%
and looking for oscillatory solutions:%
\begin{equation*}
\Delta \mathbf{\omega }_{Df}^{-1}\left( Z_{\mathbf{a}},\theta \right)
=\Delta \mathbf{\omega }_{f}^{-1}\left( Z_{\mathbf{a}}\right) \exp \left(
i\Upsilon \theta \right)
\end{equation*}%
yields the following formula, similar to the single type of cells case:

\begin{eqnarray*}
&&\left( \Delta \omega _{Df}\right) _{i}\left( \theta ,\mathbf{Z},\left\vert
\Psi \right\vert ^{2}\right) \\
&=&A_{i}\left( \left( Z_{1}\right) _{i}\right) \left( 1,\left( 1-\mathbf{%
\check{T}}_{ii}\exp \left( -i\Upsilon _{ip}\frac{\left\vert \mathbf{\Delta Z}%
\right\vert }{c}\right) \right) ^{-1}\check{T}_{1ii}\left( \mathbf{Z}\right)
\exp \left( -i\Upsilon _{ip}\frac{\left\vert \mathbf{\Delta Z}%
_{1}\right\vert }{c}\right) \right) ^{t}\exp \left( i\Upsilon _{ip}\left( 
\mathbf{\check{T}}_{ii}\right) \theta \right)
\end{eqnarray*}%
As in the $1$-field case, $A_{i}\left( \left( Z_{1}\right) _{i}\right) $ is
the amplitude of activity at one given point $\left( Z_{1}\right) _{i}$ of
the group $i$ and $\check{T}_{1ii}\left( \mathbf{Z}\right) $ is a vector
with components $\left( \check{T}_{1ii}\left( \mathbf{Z}\right) \right)
_{a_{i}}=\check{T}_{ii}\left( Z_{a_{i}},\left( Z_{1}\right) _{i}\right) $.

The $\Upsilon _{ip}\left( \mathbf{\check{T}}_{ii}\right) $ are equilbrium
frequencies. They are solutions of:%
\begin{equation}
\det \left( 1-\check{T}_{ii}\left( Z,Z_{1}\right) \exp \left( -i\Upsilon _{p}%
\frac{\left\vert Z-Z_{j}\right\vert }{c}\right) \right) =0  \label{MTV}
\end{equation}%
We write the solutions $\Upsilon _{p}\left( \mathbf{\check{T}}_{ii}\right) $%
. By diagonalization of the matrix involved in (\ref{MTV}), these
frequencies satisfy:%
\begin{equation*}
\prod\limits_{k}\left( 1-f_{i,k}\left( \Upsilon _{p}\right) \right) =0
\end{equation*}%
for some functions $f_{i,k}$. Thus, the solutions form a set:%
\begin{equation*}
\Upsilon _{p}^{i}=\left\{ \gamma _{i,k}\right\} _{k}
\end{equation*}

\subsubsection*{A7.2.2 Activities with interactions}

Comparing (\ref{SR}) and (\ref{Sr}) leads to the lowest order:%
\begin{equation}
\left( \Delta \omega _{D}\right) _{i}\left( Z_{a_{i}},\theta \right) =\left(
\Delta \omega _{Df}\right) _{i}\left( Z_{a_{i}},\theta \right) +\sum_{j\neq
i,\left\{ \left\{ b_{j}\right\} \right\} }\check{T}_{ij}\left(
Z_{a_{i}},Z_{b_{j}}\right) \left( \Delta \omega _{Df}\right) _{j}\left(
\theta -\frac{\left\vert Z_{a_{i}}-Z_{b_{j}}\right\vert }{c},Z_{b_{j}}\right)
\label{SR}
\end{equation}%
with $\check{T}_{ij}\left( Z_{a_{i}},Z_{b_{j}}\right) $ given by (\ref{St}).

\subsection*{A7.3 Frequencies for sets in interactions}

\subsubsection*{A7.3.1 Solutions for similar groups}

Rewriting the solutions of (\ref{SR}) as a vector:%
\begin{equation*}
\left( \left( \Delta \omega _{D}\right) _{i}^{-1}\left( Z_{\alpha
_{i}},\theta \right) \right) _{i}\equiv \left( \Delta \mathbf{\omega }%
_{D}\right) ^{-1}\left( Z_{\mathbf{\alpha }},\theta \right)
\end{equation*}%
we look for oscillatory solutions. For a composed state to exist, we have to
consider non-destructive interactions and this imply that we have to look
for similar frequencies between the various groups. As a consequence, we
assume that the solutions have the following form:%
\begin{equation*}
\left( \Delta \mathbf{\omega }_{D}\right) \left( Z_{\mathbf{a}},\theta
\right) =\left( \Delta \mathbf{\omega }\right) \left( Z_{\mathbf{a}}\right)
\exp \left( i\Upsilon \theta \right)
\end{equation*}%
which implies a solution for:%
\begin{equation*}
\Delta \mathbf{\omega }\left( Z_{\mathbf{a}}\right) =M\Delta \mathbf{\omega }%
\left( Z_{\mathbf{a}}\right)
\end{equation*}%
with:%
\begin{equation*}
M_{\left( ia_{i}\right) ,\left( jb_{j}\right) }=\hat{T}_{ij}\left(
Z_{a_{i}},Z_{b_{j}}\right) \exp \left( -i\Upsilon \frac{\left\vert
Z_{a_{i}}-Z_{b_{j}}\right\vert }{c}\right)
\end{equation*}%
\begin{equation*}
M=\left( M_{\left( ia_{i}\right) ,\left( jb_{j}\right) }\right) +\left(
M_{\left( ia_{i}\right) ,\left( jb_{j}\right) }\right) _{i\neq j}
\end{equation*}

\subsubsection*{A7.3.2 Composed frequencies as functions of
without-interaction frequencies}

Without interactions, the frequencies are in some state $\gamma _{i,l_{i}}$
satisfying: 
\begin{equation*}
f_{i,l_{i}}\left( \gamma _{i,l_{i}}\right) =1
\end{equation*}%
and the solution for the vector $\Upsilon $ depends on these quantities:%
\begin{equation}
\Upsilon _{\left( i,l_{i}\right) }=\sum_{i,j\neq i}\left[ \left[ M\right] %
\right] _{i,j}\frac{\gamma _{i,k_{i}}+\gamma _{j,l_{j}}}{2}\pm \sqrt{\left(
\sum_{i,j\neq i}\frac{\left[ \left[ M\right] \right] _{i,j}}{\sum_{i,j\neq i}%
\left[ \left[ M\right] \right] _{i,j}}\frac{\gamma _{i,k_{i}}-\gamma
_{j,l_{j}}}{2}\right) ^{2}+\sum_{i,j\neq i}\left[ \left[ M\right] \right]
_{i,j}}  \label{ntr}
\end{equation}%
The notation $\Upsilon _{\left( i,l_{i}\right) }$ encapsulates that the
resulting frequency for the new structure depends on the states $\left(
\gamma _{i,l_{i}}\right) $ of initial ones. The expression $\left[ \left[ M%
\right] \right] _{i,j}$: 
\begin{equation*}
\left[ \left[ M\right] \right] _{i,j}=\frac{\left[ \hat{M}_{j,i}\right]
_{l_{j},k_{i}}\left[ \hat{M}_{i,j}\right] _{k_{i},l_{j}}}{\left( \frac{%
\partial }{\partial \gamma }f_{j,l_{j}}\left( \gamma \right) \right)
_{\gamma _{j,l_{j}}}\left( \frac{\partial }{\partial \gamma }%
f_{i,k_{i}}\left( \gamma \right) \right) _{\gamma _{i,k_{i}}}}
\end{equation*}%
depends on the matrix $\hat{M}_{j,i}$ obtained from $M_{ij}$ by a change of
basis (see appendix 5 n P4).

Note that the supprt of the new structre may be smaller than the union of
initial support due to condition (\ref{thr}). As a consequence, the
background surroundig the initial objcts enables or not the emergnce of new
object.

\subsection*{A7.4 Full activity for composed collective states}

Ultimately, the possible activities for a composed state are obtained by
gathering static and non-static parts. The modified activity for component $%
i $ 
\begin{equation*}
\overline{\Delta \omega }_{i}\left( Z_{a_{i}}\right) +\left( \Delta \omega
_{D}\right) _{i}\left( Z_{a_{i}}\right) \exp \left( i\Upsilon _{\left\{
i,l_{i}\right\} }\theta \right)
\end{equation*}%
wher $\Upsilon _{\left( i,l_{i}\right) }$ is defined in (\ref{ntr}).\bigskip

\section*{Appendix 8. Computation of activities induced by external sources}

\subsection*{A8.1 Series expansion without source}

We write the activity as a series expansion:%
\begin{eqnarray}
&&\omega ^{-1}\left( \theta ^{\left( i\right) },Z\right) -\omega ^{-1}\left(
\theta ^{\left( i\right) },Z\right) _{\left\vert \Psi \right\vert ^{2}=0}
\label{xf} \\
&=&\int \sum_{n}\left( \frac{\delta ^{n}\omega ^{-1}\left( J,\theta
,Z\right) }{\prod\limits_{i=1}^{n}\delta \left\vert \Psi \left( \theta
-l_{i},Z_{i}\right) \right\vert ^{2}}\right) _{\left\vert \Psi \right\vert
^{2}=0}\prod\limits_{i=1}^{n}\left\vert \Psi \left( \theta
-l_{i},Z_{i}\right) \right\vert ^{2}dl_{i}dZ_{i}  \notag
\end{eqnarray}%
where $\check{T}\left( \theta ,Z,Z_{1},\omega ,\Psi \right) $ depends on the
connectivitis and the activities:%
\begin{eqnarray}
&&\check{T}\left( \theta ,Z,Z_{1},\omega ,\Psi \right)  \label{vdR} \\
&=&-\frac{\frac{\kappa }{N}\omega \left( J,\theta ,Z\right) T\left(
Z,Z_{1},\theta \right) G^{\prime }\left[ J,\omega ,\theta ,Z,\Psi \right]
\delta \left( l_{1}-\frac{\left\vert Z-Z_{1}\right\vert }{c}\right) \left(
\left\vert \Psi _{0}\left( Z_{1}\right) \right\vert ^{2}+\left\vert \Psi
\left( \theta -\frac{\left\vert Z-Z_{1}\right\vert }{c},Z_{1}\right)
\right\vert ^{2}\right) }{1-\left( \int \frac{\kappa }{N}\omega \left(
J,\theta -\frac{\left\vert Z-Z^{\prime }\right\vert }{c},Z^{\prime }\right) 
\frac{\partial T\left( Z,Z^{\prime },\theta \right) }{\partial \omega \left(
J,\theta ,Z\right) }\left\vert \Psi \left( \theta -\frac{\left\vert
Z-Z^{\prime }\right\vert }{c},Z^{\prime }\right) \right\vert ^{2}dZ^{\prime
}\right) G^{\prime }\left[ J,\omega ,\theta ,Z,\Psi \right] }  \notag
\end{eqnarray}%
and we show in \cite{GL} that the successive derivatives of $\omega
^{-1}\left( J,\theta ,Z\right) $ in (\ref{xf}) have themselves an expansion
in fields that can be written in terms of grahs: 
\begin{eqnarray}
&&\left( \frac{\delta ^{n}\omega ^{-1}\left( J,\theta ,Z\right) }{%
\prod\limits_{i=1}^{n}\delta \left\vert \Psi \left( \theta
-l_{i},Z_{i}\right) \right\vert ^{2}}\right) _{\left\vert \Psi \right\vert
^{2}=0}\prod\limits_{i=1}^{n}\left\vert \Psi \left( \theta
-l_{i},Z_{i}\right) \right\vert ^{2}  \notag \\
&=&\left( \sum_{m=1}^{n}\sum_{i=1}^{m}\sum_{\left(
line_{1},...,line_{m}\right) }\prod\limits_{i}F\left( line_{i}\right)
\prod\limits_{B}F\left( B\right) \right) \prod\limits_{i=1}^{n}\left\vert
\Psi \left( \theta -l_{i},Z_{i}\right) \right\vert ^{2}  \label{rdt}
\end{eqnarray}%
The expansion (\ref{rdt}) is obtained as follows. We associate the squared
field $\left\vert \Psi \left( \theta -l_{i},Z_{i}\right) \right\vert ^{2}$
to each point $Z_{i}$ . For $m=1,...,n$, we draw $m$ lines. At least one of
them is starting from $Z$. These lines are composed of an arbitrary number
of segments and all the points $Z_{i}$ are crossed by one line. Each line
ends at a point $Z_{i}$. The starting points of the lines have to branch
either at $Z$, either at some point of an other line. There are $m$
branching points of valence $k$ including the starting point at $Z$ Apart
from $Z$ the branching points have valence $3,...,n-1$. To each line $i$ of
length $L_{i}$, we associate the factor:%
\begin{eqnarray}
F\left( line_{i}\right) &=&\prod\limits_{l=1}^{L_{i}}\frac{\frac{\kappa }{N}%
T\left( Z^{\left( l-1\right) },Z^{\left( l\right) }\right) G^{\prime }\left[
J,\omega _{0},\theta -\sum_{j=1}^{l-1}\frac{\left\vert Z^{\left( j-1\right)
}-Z^{\left( j\right) }\right\vert }{c},Z^{\left( l-1\right) }\right] 
\mathcal{\bar{G}}_{0}\left( 0,Z^{\left( l\right) }\right) }{\omega
_{0}^{-1}\left( J,\theta -\sum_{j=1}^{l-1}\frac{\left\vert Z^{\left(
j-1\right) }-Z^{\left( j\right) }\right\vert }{c},Z^{\left( l-1\right)
}\right) }  \label{lf} \\
&&\times \frac{-\omega _{0}^{-1}\left( J,\theta -\sum_{l=1}^{L_{i}}\frac{%
\left\vert Z^{\left( l-1\right) }-Z^{\left( l\right) }\right\vert }{c}%
,Z_{i}\right) }{\mathcal{\bar{G}}_{0}\left( 0,Z_{i}\right) }  \notag \\
&=&\prod\limits_{l=1}^{L_{i}}\check{T}\left( \theta -\sum_{j=1}^{l-1}\frac{%
\left\vert Z^{\left( j-1\right) }-Z^{\left( j\right) }\right\vert }{c}%
,Z^{\left( l-1\right) },Z^{\left( l\right) },\omega _{0},\Psi \right) \frac{%
-\omega _{0}^{-1}\left( J,\theta -\sum_{l=1}^{L_{i}}\frac{\left\vert
Z^{\left( l-1\right) }-Z^{\left( l\right) }\right\vert }{c},Z_{i}\right) }{%
\mathcal{\bar{G}}_{0}\left( 0,Z_{i}\right) }  \notag
\end{eqnarray}%
and to each branching point $\left( X,\theta \right) =B$ of valence $k+2$
arising in the expansion, we associate the factor:%
\begin{equation}
F\left( B\right) =\frac{\delta ^{k}\left( \frac{\check{T}\left( Z^{\left(
l-1\right) },Z^{\left( l\right) }\right) G^{\prime }\left[ J,\omega
_{0},Z^{\left( l\right) }\right] \left\vert \Psi _{0}\left( Z_{l}\right)
\right\vert ^{2}}{\omega _{0}^{-1}\left( J,\theta ,Z^{\left( l\right)
}\right) }\right) }{\delta ^{k}\omega _{0}^{-1}\left( J,\theta ,Z^{\left(
l\right) }\right) }  \label{rc}
\end{equation}%
We then show that the generating function for the graphs is equal to the
partition function for an auxiliary complex field $\digamma \left( X,\theta
\right) $ with free Green function equal to $\frac{1}{1-\left( 1+\left\vert
\Psi \right\vert ^{2}\right) \check{T}}$ and interaction terms generating
the various graphs with arbitrary number of vertices. Here $\check{T}$ is
the operator with kernl:%
\begin{equation*}
\check{T}\left( \theta ^{\left( 1\right) }-\frac{\left\vert Z^{\left(
1\right) }-Z^{\left( 2\right) }\right\vert }{c},Z^{\left( 1\right)
},Z^{\left( 2\right) },\omega _{0}\right)
\end{equation*}%
The free part of the action for $\digamma \left( X,\theta \right) $ is:%
\begin{equation*}
S_{1}=\int \digamma \left( X,\theta \right) \left( 1-\left( 1+\left\vert
\Psi \right\vert ^{2}\right) \check{T}\right) \digamma ^{\dag }\left(
X,\theta \right) d\left( X,\theta \right)
\end{equation*}%
and the interaction terms have to induce the graphs with factor (\ref{rc}).
The $k+2$ valence vertex, with $k\geqslant 1$ is thus described by a term
involving (\ref{rc}) and writes:%
\begin{eqnarray*}
S_{2} &=&\int \digamma \left( Z^{\left( 1\right) },\theta ^{\left( 1\right)
}\right) \frac{\delta ^{k}\left( \check{T}\left( \theta ^{\left( 1\right) }-%
\frac{\left\vert Z^{\left( 1\right) }-Z^{\left( 2\right) }\right\vert }{c}%
,Z^{\left( 1\right) },Z^{\left( 2\right) },\omega _{0}\right) \right) }{%
k!\prod\limits_{l=3}^{k+2}\delta ^{k}\omega _{0}^{-1}\left( J,\theta
^{\left( l\right) },Z^{\left( l\right) }\right) }\digamma ^{\dag }\left(
Z^{\left( 2\right) },\theta ^{\left( 1\right) }-\frac{\left\vert Z^{\left(
1\right) }-Z^{\left( 2\right) }\right\vert }{c}\right) \\
&&\times \prod\limits_{l=3}^{k+2}\check{T}\left( \theta ^{\left( 1\right) }-%
\frac{\left\vert Z^{\left( 1\right) }-Z^{\left( l\right) }\right\vert }{c}%
,Z^{\left( 1\right) },Z^{\left( l\right) },\omega _{0}\right) \digamma
^{\dag }\left( \theta ^{\left( l\right) },Z^{\left( l\right) }\right)
d\theta ^{\left( 1\right) }\prod\limits_{l=1}^{k+2}dZ^{\left( l\right) }
\end{eqnarray*}%
Having found the free part of the action and the required vertices, the sum
of all graphs (\ref{xf}) yields, for $\frac{\left\vert \Psi \left( J,\theta
_{i},Z_{i}\right) \right\vert ^{2}}{\mathcal{\bar{G}}_{0}\left(
0,Z_{i}\right) }\rightarrow \left\vert \Psi \left( J,\theta
_{i},Z_{i}\right) \right\vert ^{2}$:%
\begin{eqnarray}
&&\omega _{0}^{-1}\left( J,\theta ,Z\right) +\sum_{n=1}^{\infty }\frac{1}{n!}%
\frac{\int \check{T}\digamma ^{\dag }\left( Z,\theta \right) \int
\prod\limits_{i=1}^{n}\left( -\omega _{0}^{-1}\left( J,\theta
_{i},Z_{i}\right) \right) \left\vert \Psi \left( J,\theta _{i},Z_{i}\right)
\right\vert ^{2}\digamma \left( Z_{i},\theta _{i}\right) d\left(
Z_{i},\theta _{i}\right) \exp \left( -S\left( \digamma \right) \right) 
\mathcal{D}\digamma }{\exp \left( -S\left( \digamma \right) \right) \mathcal{%
D}\digamma }  \notag \\
&=&\omega _{0}^{-1}\left( J,\theta ,Z\right) +\frac{\int \check{T}\digamma
^{\dag }\left( Z,\theta \right) \exp \left( -S\left( \digamma \right) -\int
\digamma \left( X,\theta \right) \omega _{0}^{-1}\left( J,\theta ,Z\right)
\left\vert \Psi \left( J,\theta ,Z\right) \right\vert ^{2}d\left( X,\theta
\right) \right) \mathcal{D}\digamma }{\int \exp \left( -S\left( \digamma
\right) \right) \mathcal{D}\digamma }  \label{PX}
\end{eqnarray}%
with:%
\begin{equation*}
S\left( \digamma \right) =S_{1}+S_{2}
\end{equation*}%
The sum can be computed as a compact expression:%
\begin{eqnarray}
S\left( \digamma \right) &=&\int \digamma \left( Z,\theta \right) \left(
1-\left\vert \Psi \right\vert ^{2}\check{T}\right) \digamma ^{\dag }\left(
Z,\theta \right) d\left( Z,\theta \right)  \label{SL} \\
&&-\int \digamma \left( Z,\theta \right) \check{T}\left( \theta -\frac{%
\left\vert Z-Z^{\left( 1\right) }\right\vert }{c},Z,Z^{\left( 1\right)
},\omega _{0}^{-1}+\check{T}\digamma ^{\dag }\right)  \notag \\
&&\times \digamma ^{\dag }\left( Z^{\left( 1\right) },\theta -\frac{%
\left\vert Z-Z^{\left( 1\right) }\right\vert }{c}\right) dZdZ^{\left(
1\right) }d\theta  \notag
\end{eqnarray}%
where:%
\begin{eqnarray*}
&&\check{T}\left( \theta -\frac{\left\vert Z^{\left( 1\right) }-Z\right\vert 
}{c},Z^{\left( 1\right) },Z,\omega _{0}^{-1}+\check{T}\digamma ^{\dag
}\right) \\
&=&\check{T}\left( \theta -\frac{\left\vert Z^{\left( 1\right)
}-Z\right\vert }{c},Z^{\left( 1\right) },Z,\right. \\
&&\left. \omega _{0}\left( Z,\theta \right) +\int \check{T}\left( \theta -%
\frac{\left\vert Z-Z^{\left( 1\right) }\right\vert }{c},Z^{\left( 1\right)
},Z,\omega _{0}\right) \digamma ^{\dag }\left( Z^{\left( 1\right) },\theta -%
\frac{\left\vert Z-Z^{\left( 1\right) }\right\vert }{c}\right) dZ^{\left(
1\right) }\right)
\end{eqnarray*}%
\bigskip

Integral (\ref{PX}) can be computed in the saddle point approximation. It is
obtained by replacing $\digamma ^{\dag }$ and\ $\digamma $ with their values
minimizing action $S\left( \digamma \right) $ defined in (\ref{SL}). The
field $\left\vert \Psi \left( J,\theta ,Z\right) \right\vert ^{2}$ has to be
evaluated in the background state. Formaly this implies that we have to
consider the effective action associated to the path integrl:%
\begin{equation*}
\int \exp \left( \frac{1}{2}\Psi ^{\dagger }\left( \theta ,Z\right) \nabla
\left( \frac{\sigma _{\theta }^{2}}{2}\nabla -\omega ^{-1}\right) \Psi
\left( \theta ,Z\right) +V\left( \Psi \right) \right) D\Psi D\Psi ^{\dagger }
\end{equation*}%
wth $\omega ^{-1}$ given by the expansion (\ref{PX}). Without source, the
path integrl was performed previously approximativly by considering the
background field $\Psi $ as a fluctuation around the minimum of $V$.

However, to proceed when sources are taken int account, we use a
perturbation expansion of (\ref{PX}) to rewrite the source term:%
\begin{equation*}
-\int \digamma \left( X,\theta \right) \omega _{0}^{-1}\left( J,\theta
,Z\right) \left\vert \Psi \left( J,\theta ,Z\right) \right\vert ^{2}d\left(
X,\theta \right)
\end{equation*}%
as a function of the stimuli:%
\begin{equation*}
\sum_{i}a\left( Z_{i},\theta \right) \left\vert \Psi \left( Z_{i},\theta
\right) \right\vert ^{2}
\end{equation*}

\subsection*{A8.2 Series expansion including source terms}

For given connectivity functions, we want to compute the path integral $\int
\exp \left( -S\left( \Psi \right) \right) $ for $\Psi \left( \theta
,Z\right) $ given a series of signals through time at some particular points
and the activit given by the auxiliary path integral (\ref{PX}) . As
explained in the text \cite{GLs}, this amounts to introduce in the path
integral the factor:

\begin{equation*}
\int \exp \left( \sum_{i}a\left( Z_{i},\theta _{0}\right) \left\vert \Psi
\left( Z_{i},\theta _{0}\right) \right\vert ^{2}\right) d\theta _{0}
\end{equation*}%
The term $\sum_{i}a\left( Z_{i},\theta \right) \left\vert \Psi \left(
Z_{i},\theta \right) \right\vert ^{2}$ corresponds to create and cancel some
stimulation that makes the field $\Psi \left( Z_{i},\theta \right) $ to
deviate from the static equilibrium. The exponential factor stands for the
possibility of several similar stimuli at the same point. The sum over $%
\theta $ ensures the repetion of the signal through some period of time. The
perturbation is implicitely, tensored by:%
\begin{equation*}
\prod\limits_{Z\neq Z_{i}}\delta \left( \left\vert \Psi \left( Z,\theta
_{0}\right) \right\vert ^{2}\right)
\end{equation*}%
to ensure that the perturbation arises only at points $Z_{i}$.

The path integral to consider is thus:

\begin{eqnarray}
&&\int \exp \left( -S\left( \Psi \right) \right) \int \exp \left(
\sum_{i}a\left( Z_{i},\theta _{0}\right) \left\vert \Psi \left( Z_{i},\theta
_{0}\right) \right\vert ^{2}\right) d\theta _{0}  \notag \\
&=&\int \exp \left( \frac{1}{2}\Psi ^{\dagger }\left( \theta ,Z\right)
\nabla \left( \frac{\sigma _{\theta }^{2}}{2}\nabla -\omega ^{-1}\right)
\Psi \left( \theta ,Z\right) \right) \int \exp \left( \sum_{i}a\left(
Z_{i},\theta _{0}\right) \left\vert \Psi \left( Z_{i},\theta _{0}\right)
\right\vert ^{2}\right) d\theta _{0}  \label{pr}
\end{eqnarray}%
with $\omega ^{-1}$ given by the auxiliary path integral (\ref{PX}):%
\begin{equation*}
\omega ^{-1}=\omega _{0}^{-1}\left( J,Z\right) +\frac{\int \check{T}\digamma
^{\dag }\left( Z,\theta \right) \exp \left( -S\left( \digamma \right) -\int
\digamma \left( X,\theta \right) \omega _{0}^{-1}\left( J,\theta ,Z\right)
\left\vert \Psi \left( J,\theta ,Z\right) \right\vert ^{2}d\left( X,\theta
\right) \right) \mathcal{D}\digamma }{\int \exp \left( -S\left( \digamma
\right) \right) \mathcal{D}\digamma }
\end{equation*}%
Howvr, ths formula is not the activty in presence of source. To find the
corrections due to the source, we compute in \cite{GLs} the intgral over $%
\left\vert \Psi \left( J,\theta ,Z\right) \right\vert ^{2}$ wth weight $\exp
\left( -S\left( \Psi \right) \right) $ using Wick theorem.

We show in \cite{GL} tht th perturbation expansion (\ref{pr}) rewrites:%
\begin{eqnarray*}
&&\int \exp \left( \frac{1}{2}\Psi ^{\dagger }\left( \theta ,Z\right) \nabla
\left( \frac{\sigma _{\theta }^{2}}{2}\nabla -\omega ^{-1}\right) \Psi
\left( \theta ,Z\right) \right) \int \exp \left( \sum_{i}a\left(
Z_{i},\theta _{0}\right) \left\vert \Psi \left( Z_{i},\theta _{0}\right)
\right\vert ^{2}\right) d\theta _{0} \\
&=&\int d\theta _{0}\int \exp \left( \frac{1}{2}\Psi ^{\dagger }\left(
\theta ,Z\right) \nabla \left( \frac{\sigma _{\theta }^{2}}{2}\nabla -\omega
^{-1}\right) \Psi \left( \theta ,Z\right) \right)
\end{eqnarray*}%
with $\omega ^{-1}$ represents th activity with sourc:%
\begin{equation}
\omega ^{-1}\left( J,\theta ,Z\right) =\omega _{0}^{-1}\left( J,Z\right) +%
\frac{\int \check{T}\digamma ^{\dag }\left( Z,\theta \right) \exp \left(
-S\left( \digamma \right) -\sum_{i}a\left( Z_{i},\theta _{0}\right) \frac{%
\omega _{0}^{-1}\left( J,\theta _{0},Z_{i}\right) }{\Lambda ^{2}}\digamma
\left( Z_{i},\theta \right) \right) \mathcal{D}\digamma }{\int \exp \left(
-S\left( \digamma \right) \right) \mathcal{D}\digamma }  \label{QV}
\end{equation}%
and $S\left( \digamma \right) $ obtained by replacing $\left\vert \Psi
\left( \theta ,Z\right) \right\vert ^{2}$ with $\frac{1}{\digamma }$:%
\begin{eqnarray}
S\left( \digamma \right) &=&\int \digamma \left( Z,\theta \right) \left(
1-\left\vert \Psi \right\vert ^{2}\check{T}\right) \digamma ^{\dag }\left(
Z,\theta \right) d\left( Z,\theta \right)  \label{SLp} \\
&&-\int \digamma \left( Z,\theta \right) \check{T}\left( \theta -\frac{%
\left\vert Z-Z^{\left( 1\right) }\right\vert }{c},Z,Z^{\left( 1\right)
},\omega _{0}^{-1}+\check{T}\digamma ^{\dag }\right) \digamma ^{\dag }\left(
Z^{\left( 1\right) },\theta -\frac{\left\vert Z-Z^{\left( 1\right)
}\right\vert }{c}\right) dZdZ^{\left( 1\right) }d\theta  \notag
\end{eqnarray}

Formula (\ref{QV}) is the path intgral formula for activity with source. We
can compute this integral through saddl pth approximation, and we obtain:

\begin{eqnarray}
&&\omega ^{-1}\left( J,\theta ,Z\right) =\omega _{0}^{-1}\left( J,\theta
,Z\right)  \label{sdn} \\
&&+\int^{\theta _{i}}\check{T}\left( 1-\left( 1+\frac{1}{\Lambda _{1}\left(
\left( \omega _{0}\left( J,\theta ,Z_{i}\right) _{i}\right) \right) }\right) 
\check{T}\right) ^{-1}\left( Z,\theta ,Z_{i},\theta _{i}\right) \left[
-\sum_{i}a\left( Z_{i},\theta _{i}\right) \frac{\omega _{0}\left( J,\theta
_{i},Z_{i}\right) }{\Lambda ^{2}}d\theta _{i}\right]  \notag \\
&\equiv &\omega _{0}^{-1}\left( J,\theta ,Z\right) -\sum_{i}\int K\left(
Z,\theta ,Z_{i},\theta _{i}\right) \left\{ a\left( Z_{i},\theta _{i}\right) 
\frac{\omega _{0}\left( J,\theta _{i},Z_{i}\right) }{\Lambda ^{2}}\right\}
d\theta _{i}  \notag
\end{eqnarray}%
where $\frac{1}{\Lambda }$ reprsent the trace of the Green function:%
\begin{equation*}
G\left( Z,Z,\theta \right)
\end{equation*}%
of operator:%
\begin{equation*}
-\nabla \left( \frac{\sigma _{\theta }^{2}}{2}\nabla -\omega _{0}^{-1}\right)
\end{equation*}%
and: 
\begin{equation*}
\check{T}_{1}=\frac{\check{T}}{\left( 1-\left( 1+\frac{1}{\Lambda }\right) 
\check{T}\right) }
\end{equation*}%
\begin{equation*}
\frac{1}{\Lambda _{1}\left( \left( \omega _{0}\left( J,\theta ,Z_{i}\right)
_{i}\right) \right) }=\frac{1}{\Lambda }-\frac{\check{T}_{1}\left[
-\sum_{i}a\left( Z_{i},\theta \right) \frac{\ \omega _{0}^{-1}\left(
J,\theta ,Z_{i}\right) }{\Lambda ^{2}}\right] }{\omega _{0}\left( Z\right) +%
\check{T}_{1}\left[ -\sum_{i}a\left( Z_{i},\theta \right) \frac{\ \omega
_{0}^{-1}\left( J,\theta ,Z_{i}\right) }{\Lambda ^{2}}\right] }
\end{equation*}%
and:%
\begin{equation*}
K\left( Z,\theta ,Z_{i},\theta _{i}\right) \propto \exp \left( -cl-\alpha
\left( \left( cl\right) ^{2}-\left\vert Z-Z_{i}\right\vert ^{2}\right)
\right) \exp \left( i\frac{\varpi \left( l-\left\vert Z-Z_{i}\right\vert
\right) }{c}\right)
\end{equation*}%
Considering oscillating signals $a\left( Z_{i},\theta \right) \propto \exp
\left( i\varpi \theta \right) $, and assuming "quite" straight lines of
length $\left\vert Z-Z_{i}\right\vert $ from $Z$ to $Z_{i}$, due to the
exponential factor in the transitions, leads to a phase shift proportional
to $\exp \left( i\frac{\varpi \left\vert Z_{i}-Z_{0}\right\vert }{%
c\left\vert Z-Z_{0}\right\vert }\right) $ where $Z_{0}\in \left\{
Z_{i}\right\} $ is the closest point to $Z$. Taking into account corrections
due to the length around $\left\vert Z-Z_{i}\right\vert $ contributes to a
phase shift of $\exp \left( i\frac{\varpi \left( l-\left\vert
Z-Z_{i}\right\vert \right) }{c}\right) $ in the integral: 
\begin{equation*}
\int K\left( Z,\theta ,Z_{i},\theta _{i}\right) \left\{ \sum_{i}a\left(
Z_{i},\theta _{i}\right) \frac{\omega _{0}^{-1}\left( J,\theta
_{i},Z_{i}\right) }{\Lambda ^{2}}\right\} d\theta _{i}
\end{equation*}%
For $a\left( Z_{i},\theta _{i}\right) $ constant equal to $a$ this reduces
to:%
\begin{eqnarray}
&&\int K\left( Z,\theta ,Z_{i},\theta _{i}\right) \left\{ \sum_{i}a\left(
Z_{i},\theta _{i}\right) \frac{\omega _{0}^{-1}\left( J,\theta
_{i},Z_{i}\right) }{\Lambda ^{2}}\right\} d\theta _{i}  \label{KSGL} \\
&\simeq &\frac{a\exp \left( -\left\vert Z-Z_{0}\right\vert \right) }{c\sqrt{%
\left( 1+2\alpha \left\vert Z-Z_{0}\right\vert \right) ^{2}+\left( \frac{%
\varpi }{c}\right) ^{2}}}\exp \left( i\left( \frac{\varpi \left( \left\vert
Z-Z_{0}\right\vert \right) }{c}-\arctan \left( \frac{\varpi }{c\left(
1+2\alpha \left\vert Z-Z_{0}\right\vert \right) }\right) \right) \right)
\sum_{i}\exp \left( i\frac{\varpi \left\vert Z_{i}-Z_{0}\right\vert }{%
c\left\vert Z-Z_{0}\right\vert }\right)  \notag
\end{eqnarray}%
leading to interferences. For large number of points $Z_{i}$:%
\begin{equation*}
\sum_{i}\exp \left( i\frac{\varpi \left\vert Z_{i}-Z_{0}\right\vert }{%
c\left\vert Z-Z_{0}\right\vert }\right) \simeq 0
\end{equation*}%
except for the maxima of interferences with magnitude:%
\begin{equation*}
\frac{a\exp \left( -\left\vert Z-Z_{0}\right\vert \right) }{c\sqrt{\left(
1+2\alpha \left\vert Z-Z_{0}\right\vert \right) ^{2}+\left( \frac{\varpi }{c}%
\right) ^{2}}}
\end{equation*}%
This leads to the formula for $\omega \left( J,\theta ,Z\right) $ at frst
order:%
\begin{equation}
\omega \left( J,\theta ,Z\right) =\omega _{0}\left( J,\theta ,Z\right)
+S\left( Z,\varpi \right) \sum_{i}\exp \left( i\frac{\varpi \left\vert
Z_{i}-Z_{0}\right\vert }{c\left\vert Z-Z_{0}\right\vert }\right)
\end{equation}%
with:%
\begin{equation*}
S\left( Z,\varpi \right) =\frac{a\exp \left( -\left\vert Z-Z_{0}\right\vert
\right) \omega _{0}\left( J,\theta ,Z\right) }{c\sqrt{\left( 1+2\alpha
\left\vert Z-Z_{0}\right\vert \right) ^{2}+\left( \frac{\varpi }{c}\right)
^{2}}}\exp \left( i\left( \frac{\varpi \left( \left\vert Z-Z_{0}\right\vert
\right) }{c}-\arctan \left( \frac{\varpi }{c\left( 1+2\alpha \left\vert
Z-Z_{0}\right\vert \right) }\right) \right) \right)
\end{equation*}

\section*{Appendix 9 derivation of the effective action ${\protect\Huge S}%
\left( {\protect\Huge \Lambda }^{{\protect\Large \dag }},{\protect\Huge %
\Lambda }\right) $}

Starting with the classical action (\ref{sfr}):

\begin{eqnarray*}
&&{\Huge S}\left( {\Huge \Lambda }^{{\Large \dag }},{\Huge \Lambda }\right)
\\
&=&\sum_{r,r^{\prime }}\int \prod\limits_{r=1}^{v}{\Huge \Lambda }^{{\Large %
\dag }}\left[ \left( \frac{\mathbf{C,S}_{\mathbf{C}}}{\Sigma /G_{\Sigma }}%
\right) _{r}\right] \frac{1}{\nu !}\frac{1}{\nu ^{\prime }!}{\Large S}_{\nu
,\nu ^{\prime }}\left[ \left\{ \left( \frac{\mathbf{C,S}_{\mathbf{C}}}{%
\Sigma /G_{\Sigma }}\right) _{r}\right\} ,\left\{ \left( \frac{\mathbf{C}%
^{\prime }\mathbf{,S}_{\mathbf{C}}^{\prime }}{\Sigma ^{\prime }/G_{\Sigma
}^{\prime }}\right) _{r^{\prime }}\right\} \right] \prod\limits_{r^{\prime
}=1}^{v^{\prime }}{\Huge \Lambda }\left[ \left( \frac{\mathbf{C}^{\prime }%
\mathbf{,S}_{\mathbf{C}}^{\prime }}{\Sigma ^{\prime }/G_{\Sigma }^{\prime }}%
\right) _{r^{\prime }}\right]
\end{eqnarray*}

The series for the effective action is computed by considering the series of
1PI (one-particle irreducible) graphs. These graphs are constructed as a
certain series of vertices connected by propagators. Since the processes
defined by the transitions must be treated as time-ordered, we consider that
vertices are divided into groups: 
\begin{equation*}
\left\{ \left\{ {\Large S}_{l_{j_{r}},l_{j_{r+1}}}\left[ \left\{ \left( 
\frac{\mathbf{C,S}_{\mathbf{C}}}{\Sigma /G_{\Sigma }}\right)
_{i_{s_{j_{r}}}^{j_{r}}}\right\} ,\left\{ \left( \frac{\mathbf{C,S}_{\mathbf{%
C}}}{\Sigma /G_{\Sigma }}\right) _{i_{s_{j_{r+1}}}^{j_{r+1}}}\right\} \right]
\right\} _{j_{r}<p_{r}}\right\} _{r\leqslant n}
\end{equation*}%
each group for a givn $n$ includes $p_{r}$ vertices labeld by $%
j_{r}=1,..p_{r}$:%
\begin{equation*}
{\Large S}_{l_{j_{r}},l_{j_{r}}^{\prime }}\left[ \left\{ \left( \frac{%
\mathbf{C,S}_{\mathbf{C}}}{\Sigma /G_{\Sigma }}\right)
_{i_{s_{j_{r}}}^{j_{r}}}\right\} ,\left\{ \left( \frac{\mathbf{C,S}_{\mathbf{%
C}}}{\Sigma /G_{\Sigma }}\right) _{i_{s_{j_{r}^{\prime }}}^{j_{r}^{\prime
}}}\right\} \right]
\end{equation*}%
with valence $\left( l_{j_{r}},l_{j_{r}}^{\prime }\right) $.

The various groups $r=1,...n-1$ corresponds to time-increased orderd
vertices. Then to draw the various graphs, we define the following sets:

\begin{equation*}
I^{j_{r}\prime }=\left\{ i_{s_{j_{r}^{\prime }}}\right\} _{s_{j_{r}^{\prime
}}\leqslant l_{j_{r}}^{\prime }}
\end{equation*}%
that represent the incoming vertices for one vertex, and the totl incoming
vertices%
\begin{equation*}
I_{r}^{\prime }=\cup _{^{j_{r}\prime }}I^{j_{r}\prime }
\end{equation*}%
for group $r$. Similarly for outcoming vertices:%
\begin{equation*}
I^{j_{r}}=\left\{ i_{s_{j_{r}}}\right\} _{s_{j_{r}}\leqslant l_{j_{r}}}
\end{equation*}%
and:%
\begin{equation*}
I_{r}=\cup _{r}I^{j_{r}}
\end{equation*}

We draw the graphs by drawing propagators from a grp $r$ to its
predecessors. This is done by defining the maps:%
\begin{equation*}
\epsilon _{j_{r}}:I^{j_{r}}\rightarrow \cup _{s<r}I_{s}^{\prime }
\end{equation*}%
from vertex $j_{r}${} to its predecessors and the global map from grp $r$ to
the possible predecessors: 
\begin{equation*}
\epsilon _{r}:I_{r}\rightarrow \cup _{s<r}I_{s}^{\prime }
\end{equation*}%
We also define the full map:%
\begin{equation*}
\epsilon =\left( \epsilon _{r}\right) :\cup _{r}I_{r}\rightarrow \cup
_{s}I_{s}^{\prime }
\end{equation*}%
Note that th vertics mps th cndtn tht:%
\begin{equation*}
\mathring{\cup}_{r}\left( \epsilon _{r}\left( I_{r}\right) \cap
I_{s}^{\prime }\right) =I_{s}^{\prime }
\end{equation*}%
so that all valences are connected to some propagator.

The global map defines all possible connections between valences. The value
of the corresponding graph becomes:

\begin{equation}
\sum_{\left\{ \epsilon :\cup :I_{r}\rightarrow \cup I_{r}^{\prime }\right\}
}\prod\limits_{r=1}^{n}\prod\limits_{j_{r}=1}^{p_{r}}\left( {\Large S}%
_{l_{j_{r}},l_{j_{r}}^{\prime }}\left[ \left\{ \left( \frac{\mathbf{C,S}_{%
\mathbf{C}}}{\Sigma /G_{\Sigma }}\right) _{i_{s_{j_{r}}}}\right\} ,\left\{
\left( \frac{\mathbf{C,S}_{\mathbf{C}}}{\Sigma /G_{\Sigma }}\right)
_{i_{s_{j_{r}^{\prime }}}}\right\} \right] \right) \prod\limits_{w_{r}\in
I_{r}}\mathit{G}\left( \left( \frac{\mathbf{C,S}_{\mathbf{C}}}{\Sigma
/G_{\Sigma }}\right) _{\epsilon _{r}\left( w_{r}\right) },\left( \frac{%
\mathbf{C,S}_{\mathbf{C}}}{\Sigma /G_{\Sigma }}\right) _{w_{r}}\right)
\label{Grh}
\end{equation}%
This graph contributes to the effective action with the term:%
\begin{eqnarray}
&&\prod\limits_{k^{\prime }=1}^{n_{i}^{\prime }}{\Huge \Lambda }^{{\Large %
\dag }}\left[ \left( \frac{\mathbf{C,S}_{\mathbf{C}}}{\Sigma /G_{\Sigma }}%
\right) _{k^{\prime }}^{\prime }\right]  \label{Vgr} \\
&&\times \left\{ \sum_{\left\{ \epsilon :\cup :I_{r}\rightarrow \cup
I_{r}^{\prime }\right\}
}\prod\limits_{r=1}^{n}\prod\limits_{j_{r}=1}^{p_{r}}\left( {\Large S}%
_{l_{j_{r}},l_{j_{r}}^{\prime }}\left[ \left\{ \left( \frac{\mathbf{C,S}_{%
\mathbf{C}}}{\Sigma /G_{\Sigma }}\right) _{i_{s_{j_{r}}}}\right\} ,\left\{
\left( \frac{\mathbf{C,S}_{\mathbf{C}}}{\Sigma /G_{\Sigma }}\right)
_{i_{s_{j_{r}^{\prime }}}}\right\} \right] \right) \right.  \notag \\
&&\times \left. \prod\limits_{w_{r}\in I_{r}}\mathit{G}\left( \left( \frac{%
\mathbf{C,S}_{\mathbf{C}}}{\Sigma /G_{\Sigma }}\right) _{\epsilon _{r}\left(
w_{r}\right) },\left( \frac{\mathbf{C,S}_{\mathbf{C}}}{\Sigma /G_{\Sigma }}%
\right) _{w_{r}}\right) \right\} \prod\limits_{k=1}^{n_{f}}{\Huge \Lambda }%
\left[ \left( \frac{\mathbf{C,S}_{\mathbf{C}}}{\Sigma /G_{\Sigma }}\right)
_{k^{\prime }}^{\prime }\right]  \notag
\end{eqnarray}%
where $n_{i}^{\prime }$ states $\left( \frac{\mathbf{C,S}_{\mathbf{C}}}{%
\Sigma /G_{\Sigma }}\right) _{i_{s_{j_{1}}}}$ are set to $\left( \frac{%
\mathbf{C,S}_{\mathbf{C}}}{\Sigma /G_{\Sigma }}\right) _{k^{\prime
}}^{\prime }$, for $k^{\prime }=1,...n_{i}^{\prime }$ nd $n_{f}$ states $%
\left( \frac{\mathbf{C,S}_{\mathbf{C}}}{\Sigma /G_{\Sigma }}\right)
_{i_{s_{j_{n}}}}$ are set to $\left( \frac{\mathbf{C,S}_{\mathbf{C}}}{\Sigma
/G_{\Sigma }}\right) _{k}$, for $k=1,...n_{f}$. The valence of the graph
satisfies the constraint: 
\begin{equation*}
0=\sum_{r}\sum_{j_{r}}l_{j_{r}}+n_{f}-\sum_{r}\sum_{j_{r}^{\prime
}}l_{j_{r}^{\prime }}-n_{i}^{\prime }
\end{equation*}%
To define the effective action the graphs (\ref{Grh}) in the series have to
be constrained to be $1$PI graphs, that is graphs that remain connected if
we cut one internal leg.

This formula (\ref{Vgr}) can be rewritten in a more expanded form to account
for the graphs irreductibility condition. Actually, we can check that $1$PI
graphs are obtained by connecting some loops certain loops through vertices
of valence at least equal to $2$, ensuring the irreductibility of the
resulting contributions. These loops are then connected to terms like $%
{\Huge \Lambda }^{{\Large \dag }}\left[ \left( \frac{\mathbf{C,S}_{\mathbf{C}%
}}{\Sigma /G_{\Sigma }}\right) _{k^{\prime }}^{\prime }\right] $ and ${\Huge %
\Lambda }\left[ \left( \frac{\mathbf{C,S}_{\mathbf{C}}}{\Sigma /G_{\Sigma }}%
\right) _{k^{\prime }}^{\prime }\right] $, producing contributions to the
effective action.

This is realized by considering maximal loops only of arbitrary length $N$.
Given the time ordering of the process, such loops are defined by dividing $%
N=p+p^{\prime }$, where $p$ and $p^{\prime }$ defining the branches of
propagators:%
\begin{equation*}
\prod\limits_{n=1,p^{\prime }}\mathit{G}\left( \left( \frac{\mathbf{C,S}_{%
\mathbf{C}}}{\Sigma /G_{\Sigma }}\right) _{n}^{\prime },\left( \frac{\mathbf{%
C,S}_{\mathbf{C}}}{\Sigma /G_{\Sigma }}\right) _{n+1}\right) \prod\limits 
_{\substack{ n=p^{\prime }+1,p+p^{\prime }  \\ p^{\prime }+1=1\bmod p}}%
\mathit{G}\left( \left( \frac{\mathbf{C,S}_{\mathbf{C}}}{\Sigma /G_{\Sigma }}%
\right) _{n+1}^{\prime },\left( \frac{\mathbf{C,S}_{\mathbf{C}}}{\Sigma
/G_{\Sigma }}\right) _{n}\right)
\end{equation*}%
These branches represent the forward propagation from an initial state.
These branches cross vertices at each point of the loops. Such vertices $%
{\Large S}_{l_{i},l_{i}^{\prime }}$ have two valences:%
\begin{equation*}
\left( \frac{\mathbf{C,S}_{\mathbf{C}}}{\Sigma /G_{\Sigma }}\right)
_{i}^{\prime },\left( \frac{\mathbf{C,S}_{\mathbf{C}}}{\Sigma /G_{\Sigma }}%
\right) _{i}
\end{equation*}%
used by the incoming and outgoing propagators, except at the beginning of
the loop, where two outgoing valences are needed to open the loop:%
\begin{equation*}
\left\{ \left( \frac{\mathbf{C,S}_{\mathbf{C}}}{\Sigma /G_{\Sigma }}\right)
_{k^{\prime }}^{\prime }\right\} _{2\leqslant k^{\prime }\leqslant
a_{i}^{\prime }},\left\{ \left( \frac{\mathbf{C,S}_{\mathbf{C}}}{\Sigma
/G_{\Sigma }}\right) \right\} _{2\leqslant k\leqslant a_{i}}
\end{equation*}%
and at the end of the loop, where two incoming valences:%
\begin{equation*}
\left( \frac{\mathbf{C,S}_{\mathbf{C}}}{\Sigma /G_{\Sigma }}\right)
_{p},\left( \frac{\mathbf{C,S}_{\mathbf{C}}}{\Sigma /G_{\Sigma }}\right) _{n}
\end{equation*}%
ar needed to close the loop.

The remaining valences of a vertex ${\Large S}_{l_{i},l_{i}^{\prime }}$ are
divided into two groups.

The first group consists of none, or at least two incoming or outgoing
valences that are needed to connect to other loops:%
\begin{equation*}
\left\{ \left( \frac{\mathbf{C,S}_{\mathbf{C}}}{\Sigma /G_{\Sigma }}\right)
_{k^{\prime }}^{\prime }\right\} _{2\leqslant k^{\prime }\leqslant
a_{i}^{\prime }},\left\{ \left( \frac{\mathbf{C,S}_{\mathbf{C}}}{\Sigma
/G_{\Sigma }}\right) \right\} _{2\leqslant k\leqslant a_{i}}
\end{equation*}%
as incoming or outgoing loop. The rest of valences are connected to any type
of internal graph (\ref{Grh}) without the condition of $1$ particl
irreductibility.

Practically this amount to write the initial vertices as:%
\begin{equation*}
{\Large S}_{l_{1},l_{1}^{\prime }}\left[ 
\begin{array}{c}
\left( \frac{\mathbf{C,S}_{\mathbf{C}}}{\Sigma /G_{\Sigma }}\right)
_{1}^{\prime },\left( \frac{\mathbf{C,S}_{\mathbf{C}}}{\Sigma /G_{\Sigma }}%
\right) _{p+1}^{\prime },\left\{ \left( \frac{\mathbf{C,S}_{\mathbf{C}}}{%
\Sigma /G_{\Sigma }}\right) _{k^{\prime }}^{\prime }\right\} _{2\leqslant
k^{\prime }\leqslant a_{1}^{\prime }},\left\{ \left( \frac{\mathbf{C,S}_{%
\mathbf{C}}}{\Sigma /G_{\Sigma }}\right) \right\} _{2\leqslant k\leqslant
a_{1}} \\ 
\left\{ \left( \frac{\mathbf{C,S}_{\mathbf{C}}}{\Sigma /G_{\Sigma }}\right)
_{k^{\prime }}^{\prime }\right\} _{k^{\prime }\leqslant l_{1}^{\prime
}-2-a_{1}^{\prime }},\left\{ \left( \frac{\mathbf{C,S}_{\mathbf{C}}}{\Sigma
/G_{\Sigma }}\right) \right\} _{k\leqslant l_{1}-a_{1}}%
\end{array}%
\right]
\end{equation*}%
nd:%
\begin{equation*}
{\Large S}_{l_{p},l_{p}^{\prime }}\left[ 
\begin{array}{c}
\left( \frac{\mathbf{C,S}_{\mathbf{C}}}{\Sigma /G_{\Sigma }}\right)
_{p},\left( \frac{\mathbf{C,S}_{\mathbf{C}}}{\Sigma /G_{\Sigma }}\right)
_{n},\left\{ \left( \frac{\mathbf{C,S}_{\mathbf{C}}}{\Sigma /G_{\Sigma }}%
\right) _{k^{\prime }}^{\prime }\right\} _{2\leqslant k^{\prime }\leqslant
a_{p}^{\prime }},\left\{ \left( \frac{\mathbf{C,S}_{\mathbf{C}}}{\Sigma
/G_{\Sigma }}\right) \right\} _{2\leqslant k\leqslant a_{p}} \\ 
\left\{ \left( \frac{\mathbf{C,S}_{\mathbf{C}}}{\Sigma /G_{\Sigma }}\right)
_{k^{\prime }}^{\prime }\right\} _{k^{\prime }\leqslant l_{p}^{\prime
}-a_{p}^{\prime }},\left\{ \left( \frac{\mathbf{C,S}_{\mathbf{C}}}{\Sigma
/G_{\Sigma }}\right) \right\} _{k\leqslant l_{p}-2-a_{p}}%
\end{array}%
\right]
\end{equation*}%
and the internal vertices:%
\begin{equation*}
{\Large S}_{l_{i},l_{i}^{\prime }}\left[ 
\begin{array}{c}
\left( \frac{\mathbf{C,S}_{\mathbf{C}}}{\Sigma /G_{\Sigma }}\right)
_{i}^{\prime },\left( \frac{\mathbf{C,S}_{\mathbf{C}}}{\Sigma /G_{\Sigma }}%
\right) _{i},\left\{ \left( \frac{\mathbf{C,S}_{\mathbf{C}}}{\Sigma
/G_{\Sigma }}\right) _{k^{\prime }}^{\prime }\right\} _{2\leqslant k^{\prime
}\leqslant a_{i}^{\prime }},\left\{ \left( \frac{\mathbf{C,S}_{\mathbf{C}}}{%
\Sigma /G_{\Sigma }}\right) \right\} _{2\leqslant k\leqslant a_{i}} \\ 
\left\{ \left( \frac{\mathbf{C,S}_{\mathbf{C}}}{\Sigma /G_{\Sigma }}\right)
_{k^{\prime }}^{\prime }\right\} _{k^{\prime }\leqslant l_{p_{i}}^{\prime
}-1-a_{i}^{\prime }},\left\{ \left( \frac{\mathbf{C,S}_{\mathbf{C}}}{\Sigma
/G_{\Sigma }}\right) \right\} _{k\leqslant l_{p_{i}}-1-a_{i}}%
\end{array}%
\right]
\end{equation*}%
while the contribution of such loop writes:

\begin{eqnarray}
&&\left\{ \mathit{G}\left( L_{p,p^{\prime }}\right) \right\} ^{\left(
S\right) }\left[ \left\{ \left( \frac{\mathbf{C,S}_{\mathbf{C}}}{\Sigma
/G_{\Sigma }}\right) _{k^{\prime }}^{\prime }\right\} _{2\leqslant k^{\prime
}\leqslant a_{i}^{\prime }},\left\{ \left( \frac{\mathbf{C,S}_{\mathbf{C}}}{%
\Sigma /G_{\Sigma }}\right) \right\} _{2\leqslant k\leqslant a_{i}}\right]
\label{Lp} \\
&&  \notag \\
&=&\frac{1}{Sf}\sum_{L_{p}=\left\{ \left( \frac{\mathbf{C,S}_{\mathbf{C}}}{%
\Sigma /G_{\Sigma }}\right) _{n},\left( \frac{\mathbf{C,S}_{\mathbf{C}}}{%
\Sigma /G_{\Sigma }}\right) _{n}^{\prime }\right\} _{n=1,...,p}}{\Large S}%
_{l_{1},l_{1}^{\prime }}\left[ 
\begin{array}{c}
\left( \frac{\mathbf{C,S}_{\mathbf{C}}}{\Sigma /G_{\Sigma }}\right)
_{1}^{\prime },\left( \frac{\mathbf{C,S}_{\mathbf{C}}}{\Sigma /G_{\Sigma }}%
\right) _{p+1}^{\prime },\left\{ \left( \frac{\mathbf{C,S}_{\mathbf{C}}}{%
\Sigma /G_{\Sigma }}\right) _{k^{\prime }}^{\prime }\right\} _{2\leqslant
k^{\prime }\leqslant a_{1}^{\prime }},\left\{ \left( \frac{\mathbf{C,S}_{%
\mathbf{C}}}{\Sigma /G_{\Sigma }}\right) \right\} _{2\leqslant k\leqslant
a_{1}} \\ 
\left\{ \left( \frac{\mathbf{C,S}_{\mathbf{C}}}{\Sigma /G_{\Sigma }}\right)
_{k^{\prime }}^{\prime }\right\} _{k^{\prime }\leqslant l_{1}^{\prime
}-2-a_{1}^{\prime }},\left\{ \left( \frac{\mathbf{C,S}_{\mathbf{C}}}{\Sigma
/G_{\Sigma }}\right) \right\} _{k\leqslant l_{1}-a_{1}}%
\end{array}%
\right]  \notag \\
&&  \notag \\
&&\times \prod\limits_{n=1,p^{\prime }}\mathit{G}\left( \left( \frac{\mathbf{%
C,S}_{\mathbf{C}}}{\Sigma /G_{\Sigma }}\right) _{n}^{\prime },\left( \frac{%
\mathbf{C,S}_{\mathbf{C}}}{\Sigma /G_{\Sigma }}\right) _{n+1}\right)
\prod\limits_{\substack{ n=p^{\prime }+1,p+p^{\prime }  \\ p^{\prime }+1=1%
\bmod p}}\mathit{G}\left( \left( \frac{\mathbf{C,S}_{\mathbf{C}}}{\Sigma
/G_{\Sigma }}\right) _{n+1}^{\prime },\left( \frac{\mathbf{C,S}_{\mathbf{C}}%
}{\Sigma /G_{\Sigma }}\right) _{n}\right)  \notag \\
&&\times {\Large S}_{l_{p},l_{p}^{\prime }}\left[ 
\begin{array}{c}
\left( \frac{\mathbf{C,S}_{\mathbf{C}}}{\Sigma /G_{\Sigma }}\right)
_{p},\left( \frac{\mathbf{C,S}_{\mathbf{C}}}{\Sigma /G_{\Sigma }}\right)
_{n},\left\{ \left( \frac{\mathbf{C,S}_{\mathbf{C}}}{\Sigma /G_{\Sigma }}%
\right) _{k^{\prime }}^{\prime }\right\} _{2\leqslant k^{\prime }\leqslant
a_{p}^{\prime }},\left\{ \left( \frac{\mathbf{C,S}_{\mathbf{C}}}{\Sigma
/G_{\Sigma }}\right) \right\} _{2\leqslant k\leqslant a_{p}} \\ 
\left\{ \left( \frac{\mathbf{C,S}_{\mathbf{C}}}{\Sigma /G_{\Sigma }}\right)
_{k^{\prime }}^{\prime }\right\} _{k^{\prime }\leqslant l_{p}^{\prime
}-a_{p}^{\prime }},\left\{ \left( \frac{\mathbf{C,S}_{\mathbf{C}}}{\Sigma
/G_{\Sigma }}\right) \right\} _{k\leqslant l_{p}-2-a_{p}}%
\end{array}%
\right]  \notag \\
&&  \notag \\
&&\times \prod\limits_{i\in S\subset \left\{ 1,...,p\right\} /\left\{
1,p\right\} }{\Large S}_{l_{i},l_{i}^{\prime }}\left[ 
\begin{array}{c}
\left( \frac{\mathbf{C,S}_{\mathbf{C}}}{\Sigma /G_{\Sigma }}\right)
_{i}^{\prime },\left( \frac{\mathbf{C,S}_{\mathbf{C}}}{\Sigma /G_{\Sigma }}%
\right) _{i},\left\{ \left( \frac{\mathbf{C,S}_{\mathbf{C}}}{\Sigma
/G_{\Sigma }}\right) _{k^{\prime }}^{\prime }\right\} _{2\leqslant k^{\prime
}\leqslant a_{i}^{\prime }},\left\{ \left( \frac{\mathbf{C,S}_{\mathbf{C}}}{%
\Sigma /G_{\Sigma }}\right) \right\} _{2\leqslant k\leqslant a_{i}} \\ 
\left\{ \left( \frac{\mathbf{C,S}_{\mathbf{C}}}{\Sigma /G_{\Sigma }}\right)
_{k^{\prime }}^{\prime }\right\} _{k^{\prime }\leqslant l_{p_{i}}^{\prime
}-1-a_{i}^{\prime }},\left\{ \left( \frac{\mathbf{C,S}_{\mathbf{C}}}{\Sigma
/G_{\Sigma }}\right) \right\} _{k\leqslant l_{p_{i}}-1-a_{i}}%
\end{array}%
\right]  \notag \\
&&  \notag \\
&&\times \prod\limits_{i\in \left\{ S\subset \left\{ 1,...,p\right\}
\right\} /\left\{ 1,p\right\} }B\left[ 
\begin{array}{c}
\left( \frac{\mathbf{C,S}_{\mathbf{C}}}{\Sigma /G_{\Sigma }}\right)
_{1}^{\prime },\left( \frac{\mathbf{C,S}_{\mathbf{C}}}{\Sigma /G_{\Sigma }}%
\right) _{p+1}^{\prime },\left\{ \left( \frac{\mathbf{C,S}_{\mathbf{C}}}{%
\Sigma /G_{\Sigma }}\right) _{k^{\prime }}^{\prime }\right\} _{k^{\prime
}\leqslant l_{p_{i}}^{\prime }-1-\underline{a}_{i}^{\prime }} \\ 
\left( \frac{\mathbf{C,S}_{\mathbf{C}}}{\Sigma /G_{\Sigma }}\right)
_{p},\left( \frac{\mathbf{C,S}_{\mathbf{C}}}{\Sigma /G_{\Sigma }}\right)
_{n},\left\{ \left( \frac{\mathbf{C,S}_{\mathbf{C}}}{\Sigma /G_{\Sigma }}%
\right) \right\} _{k\leqslant l_{p_{i}}-1-\underline{a}_{i}}%
\end{array}%
\right]  \notag
\end{eqnarray}

where:%
\begin{eqnarray*}
\underline{a}_{1}^{\prime } &=&a_{1}^{\prime }+1 \\
\underline{a}_{i}^{\prime } &=&a_{i}^{\prime },i>1 \\
\underline{a}_{p} &=&a_{p}+1 \\
\underline{a}_{i} &=&a_{i},i<p
\end{eqnarray*}%
and:%
\begin{eqnarray*}
&&B\left[ 
\begin{array}{c}
\left( \frac{\mathbf{C,S}_{\mathbf{C}}}{\Sigma /G_{\Sigma }}\right)
_{1}^{\prime },\left( \frac{\mathbf{C,S}_{\mathbf{C}}}{\Sigma /G_{\Sigma }}%
\right) _{p+1}^{\prime },\left\{ \left( \frac{\mathbf{C,S}_{\mathbf{C}}}{%
\Sigma /G_{\Sigma }}\right) _{k^{\prime }}^{\prime }\right\} _{k^{\prime
}\leqslant l_{p_{i}}^{\prime }-1-\underline{a}_{i}^{\prime }} \\ 
\left( \frac{\mathbf{C,S}_{\mathbf{C}}}{\Sigma /G_{\Sigma }}\right)
_{p},\left( \frac{\mathbf{C,S}_{\mathbf{C}}}{\Sigma /G_{\Sigma }}\right)
_{n},\left\{ \left( \frac{\mathbf{C,S}_{\mathbf{C}}}{\Sigma /G_{\Sigma }}%
\right) \right\} _{k\leqslant l_{p_{i}}-1-\underline{a}_{i}}%
\end{array}%
\right] \\
&& \\
&=&\sum_{\left\{ \epsilon :\cup :I_{r}\rightarrow \cup I_{r}^{\prime
}\right\} }\prod\limits_{r=1}^{n}\prod\limits_{j_{r}=1}^{p_{r}}\left( 
{\Large S}_{l_{j_{r}},l_{j_{r}}^{\prime }}\left[ \left\{ \left( \frac{%
\mathbf{C,S}_{\mathbf{C}}}{\Sigma /G_{\Sigma }}\right)
_{i_{s_{j_{r}}}}\right\} ,\left\{ \left( \frac{\mathbf{C,S}_{\mathbf{C}}}{%
\Sigma /G_{\Sigma }}\right) _{i_{s_{j_{r}^{\prime }}}}\right\} \right]
\right) \prod\limits_{w_{r}\in I_{r}}\mathit{G}\left( \left( \frac{\mathbf{%
C,S}_{\mathbf{C}}}{\Sigma /G_{\Sigma }}\right) _{\epsilon _{r}\left(
w_{r}\right) },\left( \frac{\mathbf{C,S}_{\mathbf{C}}}{\Sigma /G_{\Sigma }}%
\right) _{w_{r}}\right)
\end{eqnarray*}%
with:%
\begin{equation*}
0=\sum_{r}\sum_{j_{r}}l_{j_{r}}+\sum \left( l_{p_{i}}-1-\underline{a}%
_{i}\right) -\sum_{r}\sum_{j_{r}^{\prime }}l_{j_{r}^{\prime }}-\sum \left(
l_{p_{i}}^{\prime }-1-\underline{a}_{i}^{\prime }\right)
\end{equation*}

Graphically, the expression (\ref{Lp}) represents a maximal loop: a loop
with $p+p^{\prime }$ propagators that cross several vertices ${\Large S}%
_{l_{1},l_{1}^{\prime }}$, ${\Large S}_{l_{p},l_{p}^{\prime }}$, ${\Large S}%
_{l_{i},l_{i}^{\prime }}$. The loop begins at ${\Large S}_{l_{1},l_{1}^{%
\prime }}$ and ends at ${\Large S}_{l_{p},l_{p}^{\prime }}$. The vertices in
the loop include both incoming and outgoing legs, as well as internal
incoming and outgoing vertices, which are connected with each other by any
type f diagram:%
\begin{equation*}
B\left[ 
\begin{array}{c}
\left( \frac{\mathbf{C,S}_{\mathbf{C}}}{\Sigma /G_{\Sigma }}\right)
_{1}^{\prime },\left( \frac{\mathbf{C,S}_{\mathbf{C}}}{\Sigma /G_{\Sigma }}%
\right) _{p+1}^{\prime },\left\{ \left( \frac{\mathbf{C,S}_{\mathbf{C}}}{%
\Sigma /G_{\Sigma }}\right) _{k^{\prime }}^{\prime }\right\} _{k^{\prime
}\leqslant l_{p_{i}}^{\prime }-1-\underline{a}_{i}^{\prime }} \\ 
\left( \frac{\mathbf{C,S}_{\mathbf{C}}}{\Sigma /G_{\Sigma }}\right)
_{p},\left( \frac{\mathbf{C,S}_{\mathbf{C}}}{\Sigma /G_{\Sigma }}\right)
_{n},\left\{ \left( \frac{\mathbf{C,S}_{\mathbf{C}}}{\Sigma /G_{\Sigma }}%
\right) \right\} _{k\leqslant l_{p_{i}}-1-\underline{a}_{i}}%
\end{array}%
\right]
\end{equation*}%
The factor $\frac{1}{Sf}${} represents the symmetry factor of the overall
graph. This factor accounts for the indistinguishability of certain
configurations within the graph, ensuring that overcounting is avoided when
summing contributions.

The contribution of such a loop to the effective action is obtained by
summing over vertices of valence corresponding to field inclusions. These
vertices represent the points where external propagators meet. This leads to
replace the three types of vertices:%
\begin{eqnarray*}
&&{\Large S}_{l_{1},l_{1}^{\prime }}\left[ 
\begin{array}{c}
\left( \frac{\mathbf{C,S}_{\mathbf{C}}}{\Sigma /G_{\Sigma }}\right)
_{1}^{\prime },\left( \frac{\mathbf{C,S}_{\mathbf{C}}}{\Sigma /G_{\Sigma }}%
\right) _{p+1}^{\prime },\left\{ \left( \frac{\mathbf{C,S}_{\mathbf{C}}}{%
\Sigma /G_{\Sigma }}\right) _{k^{\prime }}^{\prime }\right\} _{2\leqslant
k^{\prime }\leqslant a_{1}^{\prime }},\left\{ \left( \frac{\mathbf{C,S}_{%
\mathbf{C}}}{\Sigma /G_{\Sigma }}\right) \right\} _{2\leqslant k\leqslant
a_{1}} \\ 
\left\{ \left( \frac{\mathbf{C,S}_{\mathbf{C}}}{\Sigma /G_{\Sigma }}\right)
_{k^{\prime }}^{\prime }\right\} _{k^{\prime }\leqslant l_{1}^{\prime
}-2-a_{1}^{\prime }},\left\{ \left( \frac{\mathbf{C,S}_{\mathbf{C}}}{\Sigma
/G_{\Sigma }}\right) \right\} _{k\leqslant l_{1}-a_{1}}%
\end{array}%
\right] \\
&& \\
&\rightarrow &{\Large S}_{l_{1},l_{1}^{\prime }}\left[ 
\begin{array}{c}
\left( \frac{\mathbf{C,S}_{\mathbf{C}}}{\Sigma /G_{\Sigma }}\right)
_{1}^{\prime },\left( \frac{\mathbf{C,S}_{\mathbf{C}}}{\Sigma /G_{\Sigma }}%
\right) _{p+1}^{\prime },\left\{ \left( \frac{\mathbf{C,S}_{\mathbf{C}}}{%
\Sigma /G_{\Sigma }}\right) _{k^{\prime }}^{\prime }\right\} _{2\leqslant
k^{\prime }\leqslant a_{1}^{\prime }},\left\{ \left( \frac{\mathbf{C,S}_{%
\mathbf{C}}}{\Sigma /G_{\Sigma }}\right) \right\} _{2\leqslant k\leqslant
a_{1}} \\ 
\left\{ \left( \frac{\mathbf{C,S}_{\mathbf{C}}}{\Sigma /G_{\Sigma }}\right)
_{k^{\prime }}^{\prime }\right\} _{k^{\prime }\leqslant l_{1}^{\prime
}-2-a_{1}^{\prime }},\left\{ \left( \frac{\mathbf{C,S}_{\mathbf{C}}}{\Sigma
/G_{\Sigma }}\right) \right\} _{k\leqslant l_{1}-a_{1}},\left[ \left\{
\left( \frac{\mathbf{C,S}_{\mathbf{C}}}{\Sigma /G_{\Sigma }}\right)
_{k_{1}^{\prime }}^{\prime }\right\} ,\left\{ \left( \frac{\mathbf{C,S}_{%
\mathbf{C}}}{\Sigma /G_{\Sigma }}\right) _{k_{1}}\right\} \right]%
\end{array}%
\right]
\end{eqnarray*}%
and:%
\begin{eqnarray*}
&&{\Large S}_{l_{p},l_{p}^{\prime }}\left[ 
\begin{array}{c}
\left( \frac{\mathbf{C,S}_{\mathbf{C}}}{\Sigma /G_{\Sigma }}\right)
_{p},\left( \frac{\mathbf{C,S}_{\mathbf{C}}}{\Sigma /G_{\Sigma }}\right)
_{n},\left\{ \left( \frac{\mathbf{C,S}_{\mathbf{C}}}{\Sigma /G_{\Sigma }}%
\right) _{k^{\prime }}^{\prime }\right\} _{2\leqslant k^{\prime }\leqslant
a_{p}^{\prime }},\left\{ \left( \frac{\mathbf{C,S}_{\mathbf{C}}}{\Sigma
/G_{\Sigma }}\right) \right\} _{2\leqslant k\leqslant a_{p}} \\ 
\left\{ \left( \frac{\mathbf{C,S}_{\mathbf{C}}}{\Sigma /G_{\Sigma }}\right)
_{k^{\prime }}^{\prime }\right\} _{k^{\prime }\leqslant l_{p}^{\prime
}-a_{p}^{\prime }},\left\{ \left( \frac{\mathbf{C,S}_{\mathbf{C}}}{\Sigma
/G_{\Sigma }}\right) \right\} _{k\leqslant l_{p}-2-a_{p}}%
\end{array}%
\right] \\
&\rightarrow &{\Large S}_{l_{p},l_{p}^{\prime }}\left[ 
\begin{array}{c}
\left( \frac{\mathbf{C,S}_{\mathbf{C}}}{\Sigma /G_{\Sigma }}\right)
_{p},\left( \frac{\mathbf{C,S}_{\mathbf{C}}}{\Sigma /G_{\Sigma }}\right)
_{n},\left\{ \left( \frac{\mathbf{C,S}_{\mathbf{C}}}{\Sigma /G_{\Sigma }}%
\right) _{k^{\prime }}^{\prime }\right\} _{2\leqslant k^{\prime }\leqslant
a_{p}^{\prime }},\left\{ \left( \frac{\mathbf{C,S}_{\mathbf{C}}}{\Sigma
/G_{\Sigma }}\right) \right\} _{2\leqslant k\leqslant a_{p}} \\ 
\left\{ \left( \frac{\mathbf{C,S}_{\mathbf{C}}}{\Sigma /G_{\Sigma }}\right)
_{k^{\prime }}^{\prime }\right\} _{k^{\prime }\leqslant l_{p}^{\prime
}-a_{p}^{\prime }},\left\{ \left( \frac{\mathbf{C,S}_{\mathbf{C}}}{\Sigma
/G_{\Sigma }}\right) \right\} _{k\leqslant l_{p}-2-a_{p}},\left[ \left\{
\left( \frac{\mathbf{C,S}_{\mathbf{C}}}{\Sigma /G_{\Sigma }}\right)
_{k_{p}^{\prime }}^{\prime }\right\} ,\left\{ \left( \frac{\mathbf{C,S}_{%
\mathbf{C}}}{\Sigma /G_{\Sigma }}\right) _{k_{p}}\right\} \right]%
\end{array}%
\right]
\end{eqnarray*}%
and:%
\begin{eqnarray*}
&&{\Large S}_{l_{i},l_{i}^{\prime }}\left[ 
\begin{array}{c}
\left( \frac{\mathbf{C,S}_{\mathbf{C}}}{\Sigma /G_{\Sigma }}\right)
_{i}^{\prime },\left( \frac{\mathbf{C,S}_{\mathbf{C}}}{\Sigma /G_{\Sigma }}%
\right) _{i},\left\{ \left( \frac{\mathbf{C,S}_{\mathbf{C}}}{\Sigma
/G_{\Sigma }}\right) _{k^{\prime }}^{\prime }\right\} _{2\leqslant k^{\prime
}\leqslant a_{i}^{\prime }},\left\{ \left( \frac{\mathbf{C,S}_{\mathbf{C}}}{%
\Sigma /G_{\Sigma }}\right) \right\} _{2\leqslant k\leqslant a_{i}} \\ 
\left\{ \left( \frac{\mathbf{C,S}_{\mathbf{C}}}{\Sigma /G_{\Sigma }}\right)
_{k^{\prime }}^{\prime }\right\} _{k^{\prime }\leqslant l_{p_{i}}^{\prime
}-1-a_{i}^{\prime }},\left\{ \left( \frac{\mathbf{C,S}_{\mathbf{C}}}{\Sigma
/G_{\Sigma }}\right) \right\} _{k\leqslant l_{p_{i}}-1-a_{i}}%
\end{array}%
\right] \\
&& \\
&\rightarrow &{\Large S}_{l_{i},l_{i}^{\prime }}\left[ 
\begin{array}{c}
\left( \frac{\mathbf{C,S}_{\mathbf{C}}}{\Sigma /G_{\Sigma }}\right)
_{i}^{\prime },\left( \frac{\mathbf{C,S}_{\mathbf{C}}}{\Sigma /G_{\Sigma }}%
\right) _{i},\left\{ \left( \frac{\mathbf{C,S}_{\mathbf{C}}}{\Sigma
/G_{\Sigma }}\right) _{k^{\prime }}^{\prime }\right\} _{2\leqslant k^{\prime
}\leqslant a_{i}^{\prime }},\left\{ \left( \frac{\mathbf{C,S}_{\mathbf{C}}}{%
\Sigma /G_{\Sigma }}\right) \right\} _{2\leqslant k\leqslant a_{i}} \\ 
\left\{ \left( \frac{\mathbf{C,S}_{\mathbf{C}}}{\Sigma /G_{\Sigma }}\right)
_{k^{\prime }}^{\prime }\right\} _{k^{\prime }\leqslant l_{p_{i}}^{\prime
}-1-a_{i}^{\prime }},\left\{ \left( \frac{\mathbf{C,S}_{\mathbf{C}}}{\Sigma
/G_{\Sigma }}\right) \right\} _{k\leqslant l_{p_{i}}-1-a_{i}},\left[ \left\{
\left( \frac{\mathbf{C,S}_{\mathbf{C}}}{\Sigma /G_{\Sigma }}\right)
_{k_{i}^{\prime }}^{\prime }\right\} ,\left\{ \left( \frac{\mathbf{C,S}_{%
\mathbf{C}}}{\Sigma /G_{\Sigma }}\right) _{k_{i}^{\prime }}\right\} \right]%
\end{array}%
\right]
\end{eqnarray*}%
leading to rewrites:%
\begin{eqnarray*}
&&\left\{ \mathit{G}\left( L_{p,p^{\prime }}\right) \right\} \left[ \left\{
\left( \frac{\mathbf{C,S}_{\mathbf{C}}}{\Sigma /G_{\Sigma }}\right)
_{k^{\prime }}^{\prime }\right\} _{2\leqslant k^{\prime }\leqslant
a_{i}^{\prime }},\left\{ \left( \frac{\mathbf{C,S}_{\mathbf{C}}}{\Sigma
/G_{\Sigma }}\right) \right\} _{2\leqslant k\leqslant a_{i}}\right] \\
&& \\
&\rightarrow &\left\{ \mathit{G}\left( L_{p,p^{\prime }}\right) \right\} %
\left[ 
\begin{array}{c}
\left\{ \left( \frac{\mathbf{C,S}_{\mathbf{C}}}{\Sigma /G_{\Sigma }}\right)
_{k^{\prime }}^{\prime }\right\} _{2\leqslant k^{\prime }\leqslant
a_{i}^{\prime }},\left\{ \left( \frac{\mathbf{C,S}_{\mathbf{C}}}{\Sigma
/G_{\Sigma }}\right) \right\} _{2\leqslant k\leqslant a_{i}} \\ 
\left[ \left\{ \left( \frac{\mathbf{C,S}_{\mathbf{C}}}{\Sigma /G_{\Sigma }}%
\right) _{k_{i}^{\prime }}^{\prime }\right\} ,\left\{ \left( \frac{\mathbf{%
C,S}_{\mathbf{C}}}{\Sigma /G_{\Sigma }}\right) _{k_{i}}\right\} \right]%
\end{array}%
\right]
\end{eqnarray*}

The contributions to the effective action are constructed by linking these
loops to incoming and outgoing fields and interconnecting loops with each
other, subject to the condition every connection between two loops includes
at least two propagators to maintain the $1$PI irreducibility of the graph.
These graphs result in the contributions for $L$ loops:%
\begin{eqnarray}
&&\prod\limits_{k_{i}^{\prime }=1}^{n_{i}^{\prime }}{\Huge \Lambda }^{%
{\Large \dag }}\left[ \left( \frac{\mathbf{C,S}_{\mathbf{C}}}{\Sigma
/G_{\Sigma }}\right) _{k_{i}^{\prime }}^{\prime }\right]  \label{Ff} \\
&&\times \left[ \left\{ \mathit{G}\left( L_{p,p^{\prime }}\right) \right\} %
\left[ 
\begin{array}{c}
\left\{ \left( \frac{\mathbf{C,S}_{\mathbf{C}}}{\Sigma /G_{\Sigma }}\right)
_{k^{\prime }}^{\prime }\right\} _{2\leqslant k^{\prime }\leqslant
a_{i}^{\prime }},\left\{ \left( \frac{\mathbf{C,S}_{\mathbf{C}}}{\Sigma
/G_{\Sigma }}\right) \right\} _{2\leqslant k\leqslant a_{i}} \\ 
\left[ \left\{ \left( \frac{\mathbf{C,S}_{\mathbf{C}}}{\Sigma /G_{\Sigma }}%
\right) _{k_{i}^{\prime }}^{\prime }\right\} ,\left\{ \left( \frac{\mathbf{%
C,S}_{\mathbf{C}}}{\Sigma /G_{\Sigma }}\right) _{k_{i}}\right\} \right]%
\end{array}%
\right] \right] \prod\limits_{k_{i}=1}^{n_{i}}{\Huge \Lambda }\left[ \left( 
\frac{\mathbf{C,S}_{\mathbf{C}}}{\Sigma /G_{\Sigma }}\right) _{k_{i}^{\prime
}}\right]  \notag
\end{eqnarray}%
These terms have to be considered as depending on variables:%
\begin{equation*}
\left\{ \left( \frac{\mathbf{C,S}_{\mathbf{C}}}{\Sigma /G_{\Sigma }}\right)
_{k^{\prime }}^{\prime }\right\} _{2\leqslant k^{\prime }\leqslant
a_{i}^{\prime }},\left\{ \left( \frac{\mathbf{C,S}_{\mathbf{C}}}{\Sigma
/G_{\Sigma }}\right) \right\} _{2\leqslant k\leqslant a_{i}}
\end{equation*}%
The effective action is obtained by the convolution of multiple loops
through these vriables.

Writing (\ref{Ff}) for short:%
\begin{equation*}
\prod\limits_{k_{i}^{\prime }=1}^{n_{i}^{\prime }}{\Huge \Lambda }^{{\Large %
\dag }}\left[ \left\{ \mathit{G}\left( L_{p,p^{\prime }}\right) \right\} %
\left[ \left\{ \left( \frac{\mathbf{C,S}_{\mathbf{C}}}{\Sigma /G_{\Sigma }}%
\right) _{k^{\prime }}^{\prime }\right\} _{2\leqslant k^{\prime }\leqslant
a_{i}^{\prime }},\left\{ \left( \frac{\mathbf{C,S}_{\mathbf{C}}}{\Sigma
/G_{\Sigma }}\right) \right\} _{2\leqslant k\leqslant a_{i}}\right] \right]
\prod\limits_{k_{i}=1}^{n_{i}}{\Huge \Lambda }
\end{equation*}%
and labeling the loops by $\gamma $ the convolutions are given by:%
\begin{eqnarray}
&&\prod\limits_{l=1}^{L}\left[ \prod\limits_{k_{i}^{\prime }=1}^{\left(
n_{i}^{\prime }\right) _{\gamma _{l}}}{\Huge \Lambda }^{{\Large \dag }}\left[
\left\{ \mathit{G}\left( L_{p,p^{\prime }}\right) \right\} \left[ \left\{
\left( \frac{\mathbf{C,S}_{\mathbf{C}}}{\Sigma /G_{\Sigma }}\right)
_{k_{\gamma _{l}}^{\prime }}^{\prime }\right\} _{2\leqslant k_{\gamma
_{l}}^{\prime }\leqslant \left( a_{i}^{\prime }\right) _{\gamma
_{l}}},\left\{ \left( \frac{\mathbf{C,S}_{\mathbf{C}}}{\Sigma /G_{\Sigma }}%
\right) \right\} _{2\leqslant k_{\gamma _{l}}\leqslant \left( a_{i}\right)
_{\gamma _{l}}}\right] \right] \prod\limits_{k_{i}=1}^{\left( n_{i}\right)
_{\gamma _{l}}}{\Huge \Lambda }\right]  \notag \\
&&\times C\left[ \left\{ \left\{ \left( \frac{\mathbf{C,S}_{\mathbf{C}}}{%
\Sigma /G_{\Sigma }}\right) _{k_{\gamma _{l}}^{\prime }}^{\prime }\right\}
_{2\leqslant k_{\gamma _{l}}^{\prime }\leqslant \left( a_{i}^{\prime
}\right) _{\gamma _{l}}},\left\{ \left( \frac{\mathbf{C,S}_{\mathbf{C}}}{%
\Sigma /G_{\Sigma }}\right) \right\} _{2\leqslant k_{\gamma _{l}}\leqslant
\left( a_{i}\right) _{\gamma _{l}}}\right\} _{l}\right]  \label{Fcr}
\end{eqnarray}

where $C$ is the kernel of convolution, which is computed by decomposing:%
\begin{equation*}
\left\{ \left( \frac{\mathbf{C,S}_{\mathbf{C}}}{\Sigma /G_{\Sigma }}\right)
_{k_{\gamma _{l}}^{\prime }}^{\prime }\right\} _{2\leqslant k_{\gamma
_{l}}^{\prime }\leqslant \left( a_{i}^{\prime }\right) _{\gamma
_{l}}}=\left( B_{\gamma _{l}}^{\prime }\right) _{1}\cup \left( B_{\gamma
_{l}}^{\prime }\right) _{2}\cup \left( B_{\gamma _{l}}^{\prime }\right)
_{3}\cup ...
\end{equation*}%
and:%
\begin{equation*}
\left\{ \left( \frac{\mathbf{C,S}_{\mathbf{C}}}{\Sigma /G_{\Sigma }}\right)
_{k_{\gamma _{l}}}\right\} _{2\leqslant k_{\gamma _{l}}\leqslant \left(
a_{i}\right) _{\gamma _{l}}}=\left( B_{\gamma _{l}}\right) _{1}\cup \left(
B_{\gamma _{l}}\right) _{2}\cup \left( B_{\gamma _{l}}\right) _{3}\cup ..
\end{equation*}%
where each block contains at least two of the $\left( \frac{\mathbf{C,S}_{%
\mathbf{C}}}{\Sigma /G_{\Sigma }}\right) _{k_{\gamma _{l}}^{\prime
}}^{\prime }$ or of the $\left( \frac{\mathbf{C,S}_{\mathbf{C}}}{\Sigma
/G_{\Sigma }}\right) _{k_{\gamma _{l}}}$.

The convolution writes:%
\begin{eqnarray*}
&&C\left[ \left\{ \left\{ \left( B_{\gamma _{l}}^{\prime }\right)
_{s_{l}^{\prime }}\right\} _{s_{l}^{\prime }},\left\{ \left( B_{\gamma
_{l}}\right) _{s_{l}}\right\} _{s_{l}}\right\} _{l}\right] \\
&=&\sum_{P}\prod\limits_{s_{l}^{\prime },s_{m}}\mathit{G}_{B}\left( \left(
B_{\gamma _{l}}^{\prime }\right) _{s_{l}^{\prime }},\left( B_{\gamma
_{m}}\right) _{s_{m}}\right)
\end{eqnarray*}%
where the sum is over the partitions $P$: 
\begin{equation*}
\cup _{l}\left( \left\{ \left( B_{\gamma _{l}}^{\prime }\right)
_{s_{l}^{\prime }}\right\} _{s_{l}^{\prime }}\cup \left\{ \left( B_{\gamma
_{l}}\right) _{s_{l}}\right\} _{s_{l}}\right) =\cup _{l,m}\left( \left(
B_{\gamma _{l}}^{\prime }\right) _{s_{l}^{\prime }}\cup \left\{ \left(
B_{\gamma _{m}}\right) _{s_{m}}\right\} _{s_{m}}\right)
\end{equation*}%
under the condition that:%
\begin{equation*}
\sharp \left( B_{\gamma _{l}}^{\prime }\right) _{s_{l}^{\prime }}=\sharp
\left( B_{\gamma _{m}}\right) _{s_{m}}
\end{equation*}%
and:%
\begin{equation*}
\mathit{G}_{B}\left( \left( B_{\gamma _{l}}^{\prime }\right) _{s_{l}^{\prime
}},\left( B_{\gamma _{m}}\right) _{s_{m}}\right) =\sum_{\left( a,b\right)
}\prod\limits_{\substack{ \left( \frac{\mathbf{C,S}_{\mathbf{C}}}{\Sigma
/G_{\Sigma }}\right) _{a}^{\prime }\in \left( B_{\gamma _{l}}^{\prime
}\right) _{s_{l}^{\prime }}  \\ \left( \frac{\mathbf{C,S}_{\mathbf{C}}}{%
\Sigma /G_{\Sigma }}\right) _{b}\in \left( B_{\gamma _{m}}\right) _{s_{m}}}}%
\mathit{G}\left( \left( \frac{\mathbf{C,S}_{\mathbf{C}}}{\Sigma /G_{\Sigma }}%
\right) _{a}^{\prime },\left( \frac{\mathbf{C,S}_{\mathbf{C}}}{\Sigma
/G_{\Sigma }}\right) _{b}\right)
\end{equation*}

The complete form of the effective action is obtained by reordering the
terms in (\ref{Fcr}) which leads to the coefficients. We fix the incoming
and outgoing vertices:%
\begin{equation*}
\left\{ \left( \frac{\mathbf{C,S}_{\mathbf{C}}}{\Sigma /G_{\Sigma }}\right)
_{n}\right\} ,\left\{ \left( \frac{\mathbf{C}^{\prime }\mathbf{,S}_{\mathbf{C%
}}^{\prime }}{\Sigma ^{\prime }/G_{\Sigma }^{\prime }}\right) _{n^{\prime
}}\right\}
\end{equation*}%
to obtain the contribution to the effective action:%
\begin{equation*}
\prod {\Huge \Lambda }^{{\Large \dag }}\left( \left\{ \left( \frac{\mathbf{C}%
^{\prime }\mathbf{,S}_{\mathbf{C}}^{\prime }}{\Sigma ^{\prime }/G_{\Sigma
}^{\prime }}\right) _{n^{\prime }}\right\} \right) {\Large F}\left[ \left\{
\left( \frac{\mathbf{C,S}_{\mathbf{C}}}{\Sigma /G_{\Sigma }}\right)
_{n}\right\} ,\left\{ \left( \frac{\mathbf{C}^{\prime }\mathbf{,S}_{\mathbf{C%
}}^{\prime }}{\Sigma ^{\prime }/G_{\Sigma }^{\prime }}\right) _{n^{\prime
}}\right\} \right] \prod {\Huge \Lambda }\left( \left\{ \left( \frac{\mathbf{%
C,S}_{\mathbf{C}}}{\Sigma /G_{\Sigma }}\right) _{n}\right\} ,\right)
\end{equation*}%
with the vertex:%
\begin{equation*}
{\Large F}\left[ \left\{ \left( \frac{\mathbf{C,S}_{\mathbf{C}}}{\Sigma
/G_{\Sigma }}\right) _{n}\right\} ,\left\{ \left( \frac{\mathbf{C}^{\prime }%
\mathbf{,S}_{\mathbf{C}}^{\prime }}{\Sigma ^{\prime }/G_{\Sigma }^{\prime }}%
\right) _{n^{\prime }}\right\} \right]
\end{equation*}%
obtained by summing all possible combinations of products of connected loops:%
\begin{eqnarray*}
&&{\Large F}\left[ \left\{ \left( \frac{\mathbf{C,S}_{\mathbf{C}}}{\Sigma
/G_{\Sigma }}\right) _{n}\right\} ,\left\{ \left( \frac{\mathbf{C}^{\prime }%
\mathbf{,S}_{\mathbf{C}}^{\prime }}{\Sigma ^{\prime }/G_{\Sigma }^{\prime }}%
\right) _{n^{\prime }}\right\} \right] \\
&=&\sum_{\substack{ \left\{ \left( \frac{\mathbf{C}^{\prime }\mathbf{,S}_{%
\mathbf{C}}^{\prime }}{\Sigma ^{\prime }/G_{\Sigma }^{\prime }}\right)
_{n^{\prime }}\right\} =\cup \left( \frac{\mathbf{C,S}_{\mathbf{C}}}{\Sigma
/G_{\Sigma }}\right) _{k_{i,l}^{\prime }\leqslant \left( n_{i}\right)
_{\gamma _{l}}}^{\prime }  \\ \left\{ \left( \frac{\mathbf{C,S}_{\mathbf{C}}%
}{\Sigma /G_{\Sigma }}\right) _{n}\right\} =\cup \left( \frac{\mathbf{C,S}_{%
\mathbf{C}}}{\Sigma /G_{\Sigma }}\right) _{k_{i,l}\leqslant \left(
n_{f}\right) _{\gamma _{l}}}}}\mathit{G}\left( \mathcal{L}^{\left( L\right)
}\right) \left[ \left\{ \left( \frac{\mathbf{C,S}_{\mathbf{C}}}{\Sigma
/G_{\Sigma }}\right) _{k_{i,l}^{\prime }\leqslant \left( n_{i}\right)
_{\gamma _{l}}}^{\prime }\right\} _{i,l},\left\{ \left( \frac{\mathbf{C,S}_{%
\mathbf{C}}}{\Sigma /G_{\Sigma }}\right) _{k_{i,l}\leqslant \left(
n_{f}\right) _{\gamma _{l}}}\right\} _{i,l}\right]
\end{eqnarray*}%
where:%
\begin{eqnarray}
&&\mathit{G}\left( \mathcal{L}^{\left( L\right) }\right) \left[ \left\{
\left( \frac{\mathbf{C,S}_{\mathbf{C}}}{\Sigma /G_{\Sigma }}\right)
_{k_{i,l}^{\prime }\leqslant \left( n_{i}\right) _{\gamma _{l}}}^{\prime
}\right\} _{i,l},\left\{ \left( \frac{\mathbf{C,S}_{\mathbf{C}}}{\Sigma
/G_{\Sigma }}\right) _{k_{i,l}\leqslant \left( n_{f}\right) _{\gamma
_{l}}}\right\} _{i,l}\right] \\
&=&\prod\limits_{l=1}^{L}\left\{ \mathit{G}\left( L_{p,p^{\prime }}\right)
\right\} \left[ 
\begin{array}{c}
\left\{ \left( \frac{\mathbf{C,S}_{\mathbf{C}}}{\Sigma /G_{\Sigma }}\right)
_{k_{\gamma _{l}}^{\prime }}^{\prime }\right\} _{2\leqslant k_{\gamma
_{l}}^{\prime }\leqslant \left( a_{i}^{\prime }\right) _{\gamma
_{l}}},\left\{ \left( \frac{\mathbf{C,S}_{\mathbf{C}}}{\Sigma /G_{\Sigma }}%
\right) \right\} _{2\leqslant k_{\gamma _{l}}\leqslant \left( a_{i}\right)
_{\gamma _{l}}} \\ 
\left[ \left\{ \left( \frac{\mathbf{C,S}_{\mathbf{C}}}{\Sigma /G_{\Sigma }}%
\right) _{k_{i,l}^{\prime }\leqslant \left( n_{i}\right) _{\gamma
_{l}}}^{\prime }\right\} ,\left\{ \left( \frac{\mathbf{C,S}_{\mathbf{C}}}{%
\Sigma /G_{\Sigma }}\right) _{k_{i,l}^{\prime }\leqslant \left( n_{f}\right)
_{\gamma _{l}}}\right\} \right]%
\end{array}%
\right]  \notag \\
&&\times C\left[ \left\{ \left\{ \left( \frac{\mathbf{C,S}_{\mathbf{C}}}{%
\Sigma /G_{\Sigma }}\right) _{k_{\gamma _{l}}^{\prime }}^{\prime }\right\}
_{2\leqslant k_{\gamma _{l}}^{\prime }\leqslant \left( a_{i}^{\prime
}\right) _{\gamma _{l}}},\left\{ \left( \frac{\mathbf{C,S}_{\mathbf{C}}}{%
\Sigma /G_{\Sigma }}\right) \right\} _{2\leqslant k_{\gamma _{l}}\leqslant
\left( a_{i}\right) _{\gamma _{l}}}\right\} _{l}\right]  \notag
\end{eqnarray}

\end{document}